%% file: mainACMstyle.tex
\documentclass[acmsmall]{acmart}

\newcommand{\true}{true}
\newcommand{\false}{false}
\newcommand{\PaperLongVersion}{\true}
\newcommand{\PaperLong}[1]{\ifthenelse{\equal{\PaperLongVersion}{\true}}{ #1}{\!}}
\newcommand{\PaperShort}[1]{\ifthenelse{\equal{\PaperLongVersion}{\false}}{ #1}{}}
\newcommand{\revision}[1]{#1}
\newcommand{\pt}{\%\textit{pt}\xspace}

\usepackage{hyperref}
\usepackage{soul}
\usepackage{subcaption}
\usepackage{ifthen}
\usepackage[shortlabels]{enumitem}
\usepackage{stfloats}
\usepackage{cuted}
\usepackage{xcolor}
\newcommand{\stitle}[1]{\paragraph{#1}}
\newtheorem{definition}{Definition}
\usepackage[export]{adjustbox}
\usepackage{numprint}
\usepackage{graphicx,multirow}
\usepackage{pifont}
\usepackage{xspace}
\usepackage{booktabs}

\begin{document}

\title{The Effects of Data Quality on Machine Learning Performance on Tabular Data}

\author{Sedir Mohammed}
\email{sedir.mohammed@hpi.de}
\author{Lukas Budach}
\email{lukas.budach@student.hpi.de} 
\author{Moritz Feuerpfeil}
\email{Moritz.Feuerpfeil@student.hpi.de}
\author{Nina Ihde}
\email{nina.ihde@student.hpi.de}
\author{Andrea Nathansen}
\email{andrea.nathansen@student.hpi.de}
\author{Nele Noack}
\email{nele.noack@student.hpi.de}
\author{Hendrik Patzlaff}
\email{hendrik.patzlaff@student.hpi.de}
\author{Felix Naumann}
\email{felix.naumann@hpi.de}
\affiliation{%
  \institution{Hasso Plattner Institute}
  \city{Potsdam}
  \country{Germany}
}
\author{Hazar Harmouch}
\email{h.harmouch@uva.nl}
\affiliation{%
  \institution{University of Amsterdam}
  \city{Amsterdam}
  \country{The Netherlands}
}

\renewcommand{\shortauthors}{Mohammed et al.}



\input{00-abstract}
\settopmatter{printacmref=false}
\renewcommand\footnotetextcopyrightpermission[1]{} 
\pagestyle{plain} 
\maketitle
\thispagestyle{empty}


\medskip\noindent\textbf{Code availability:}
Open-source code is available at \url{https://github.com/HPI-Information-Systems/DQ4AI}.

\input{10-intro}
\input{20-relatedwork}

\input{30-dimensions_pollution}
\input{40-ML_tasks}

\input{50-setup}
\input{60-results/60-overview}
\input{70-conclusion}

\section*{Acknowledgements}
This research was performed partially in the context of the \href{https://www.kitqar.de/de}{KITQAR} project, supported in part by Denkfabrik Digitale Arbeitsgemeinschaft im Bundesministerium für Arbeit und Soziales (BMAS).

\bibliographystyle{ACM-Reference-Format}
\bibliography{references}


\PaperLong{
\newpage
\appendix
\input{appendix/appendix}
}

\end{document}

%% file: 00-abstract.tex
\begin{abstract}
Modern artificial intelligence~(AI) applications require large quantities of training and test data. 
This need creates critical challenges not only concerning the availability of such data, but also regarding its quality.
For example, incomplete, erroneous, or inappropriate training data can lead to unreliable models that ultimately produce poor decisions. 
Trustworthy AI applications require high-quality training and test data along many quality dimensions, such as accuracy, completeness, and consistency. 

We explore empirically the relationship between six data quality dimensions and the performance of \revision{19} popular machine learning algorithms covering the tasks of classification, regression, and clustering, with the goal of explaining their performance in terms of data quality.
Our experiments distinguish three scenarios based on the AI pipeline steps that were fed with polluted data: polluted training data, test data, or both.
We conclude the paper with an extensive discussion of our observations. 
\end{abstract}

%% file: 10-intro.tex
\section{Data Quality and AI}
\label{sec:intro}
The rapid advances in the field of artificial intelligence~(AI) represent a great opportunity for further enhancement in many industries and sectors, some of which are critical in nature, such as autonomous driving and medical diagnosis.
The potential for AI has been enhanced by the recent and future enormous growth of data. 
However, this precious data raises significant challenges, such as data quality assessment, and, according to~\cite{noaiwithoutdata} data management, data democratization and data provenance.

Until recently, both academia and industry were mainly engaged in introducing new or improving existing machine learning~(ML) models, rather than finding remedies for any data challenges that fall beyond trivial cleaning or preparation steps. 
Nevertheless, the performance of AI-enhanced systems in practice is proven to be bounded by the quality of the underlying training data~\cite{breck2019data}. 
Moreover, data have a long lifetime and their use is usually not limited to a specific task, but can continuously be fed into the development of new models to solve new tasks. 
These observations led to a shift in research focus from a model-centric to a data-centric approach for building AI systems~\cite{Re2021}. 
In~2021, two workshops emerged to discuss the potential of data-centric AI and to initiate an interdisciplinary field that needs expertise from both data management and ML communities\footnote{\url{https://datacentricai.org/neurips21/} and \url{https://hai.stanford.edu/events/data-centric-ai-virtual-workshop}}.

In the field of data management, data quality is a well-studied topic that has been a major concern of organizations for decades, leading to the introduction of standards and quality frameworks~\cite{WangS96,Batini06}.
The recent advances in AI have brought data quality back into the spotlight in the context of building ``data ecosystems'' that cope with emerging data challenges posed by AI-based systems in enterprises~\cite{noaiwithoutdata}. 
Researchers pointed out such challenges, including data quality issues~\cite{gudivada2017data}, data life cycle concerns~\cite{PolyzotisRWZ18}, the connection to ML-OPs~\cite{abs-2102-07750}, and model management~\cite{SchelterBJSSS18}.
Furthermore, some studies presented a vision of data quality assessment tools~\cite{abs-2108-05935}, an automation of data quality verification~\cite{SchelterLSCBG18} or a methodology to summarize the quality of a dataset as datasheets~\cite{GebruMVVWDC21}, nutritional labels~\cite{StoyanovichH19}, and data cards~\cite{tagliabue2021dag}.

In this work, under the umbrella of data-centric AI, we revisit six selected data quality dimensions, namely \textit{consistent representation}, \textit{completeness}, \textit{feature accuracy}, \textit{target accuracy}, \textit{uniqueness} and \textit{target class balance}.
Our ultimate aim is to observe and understand ML model behavior in terms of data quality.
We test a variety of commonly used ML algorithms for solving classification, clustering, or regression tasks.
We analyze the performance of~\revision{19} ML algorithms covering the spectrum from simple models to complex deep learning models. 

\begin{figure}[th]
        \centering
        \includegraphics[width=0.5\textwidth]{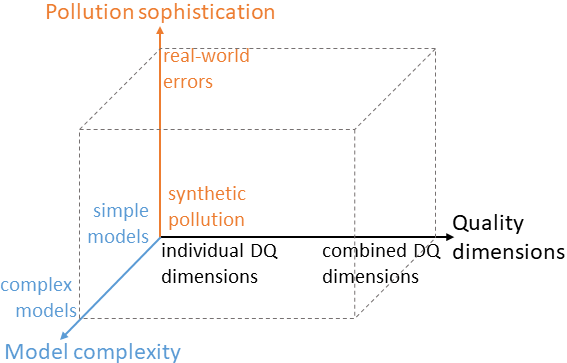}
        \caption{The wide scope of empirically studying the effect of data quality on ML algorithm performance.}
        \label{fig:scope}
    \end{figure}

Before diving further into the details of our experimental setup, we highlight the broad scope of plausible variations of such an empirical study,  showing the full perspective of any possible correlation between data quality and ML models.
As illustrated in Figure~\ref{fig:scope}, there are three main aspects: (1)~the model that can vary from simple models~(e.g., decision trees) to complex ones~(e.g., based on pre-trained embeddings); (2)~the pollution/error type can range from synthetically introduced problems to more difficult-to-detect real-world errors; (3)~data quality dimensions could be studied individually or by considering several dimensions at once assuming they are not independent.
Together, these dimensions span an enormous experimental space.
To gain the necessary basic understanding, we limit our experiments to full range of traditional and deep ML models, synthetic pollution, and individual data quality dimensions.
We plan to expand our study to cover the wider variations in future work.

Regarding data quality, we account for two aspects.
First, data plays a different role at different stages of the ML pipeline: Some systems use pre-trained models and thus the only available data is the ``test'' data; in many other cases, data scientists also need ``training'' data to build the models from scratch.
Second, training and testing data can be generated or collected by the same process from the same data source, so that they have similar quality.
In a more realistic case, training and testing data have different quality, especially when using pre-trained models and different sources or collection processes. 
To that end, we consider in this study three scenarios: Training and testing data have the same quality~(Scenario~3); the training data have high quality~(in terms of the studied quality dimensions) and lower quality testing data~(Scenario~2); and finally, the testing data have a high quality and the data used to build the models are of a lower quality~(Scenario~1). 

To vary data quality in each of these scenarios, we apply \emph{data pollution} or corruption to create degraded quality versions of the dataset at hand.
For each of the six quality dimensions, we designed a parameterized data polluter to introduce corresponding data errors. 
While we used real-world data only, for several of the datasets we had to manually create a clean version as a baseline to initiate the pollution process.
In these cases, we report the performance of ML models for both the ``original'' and the ``baseline'' datasets.

Research in the \revision{ML} community has studied the effects of label noise and missing values, and the data management community has studied the effects of data cleaning on classification, as we discuss in Section~\ref{sec:related}. 
Nevertheless, this paper is the first systematic study of the effects of data quality dimensions not only for classification, but also for clustering and regression tasks\revision{, while also considering various ML algorithms and different scenarios with training and test data of different data quality.}
Our work on real-world datasets with numerous experiments is a first step not only towards linking ML model performance to the underlying data quality, but also to understand and explain their connection. 

\medskip\noindent\textbf{Contributions.}
We present a comprehensive experimental study to understand the relation between data quality and ML model performance under the umbrella of data-centric AI, providing:
 \begin{itemize}
     \item A systematic empirical study that investigates the relation between six data quality dimensions and the performance of~\revision{19} ML algorithms.
     \item A simulation of real-life scenarios concerning data in ML pipelines. We perform a targeted analysis for cases where serving data, training data, test data, or both are of low quality. 
     \item Practical insights and learned lessons for data scientists. In addition, we raise several questions and point out possible directions for further research. 
     \item The open-source polluters, ML pipelines and all datasets as research artifacts are easily extendable with further quality dimensions, models, or datasets.
\end{itemize}

\medskip\noindent\textbf{Outline.} 
Next, we discuss related work in Section~\ref{sec:related}. Then, we formally define the six data quality dimensions together with a systematic pollution method for each in Section~\ref{sec:ddimentions}.
In Section~\ref{sec:MLalgorithms}, we briefly introduce the 19 ML algorithms for the three AI tasks of classification, regression, and clustering.
We describe our experimental setup in Section~\ref{sec:experimetal_setup}.
The results of the empirical evaluation, the core contribution of this paper, are discussed in Section~\ref{sec:results}.
Finally, we discuss our findings in Section~\ref{sec:discussion} and summarize this manuscript in Section~\ref{sec:conclusion}.

%% file: 20-relatedwork.tex
\section{Related Work}   
\label{sec:related}
\begin{table*}[!h]
\centering\small
\caption{\revision{Overview of related work.}}
\revision{
\begin{tabular}{p{2cm}|p{2.5cm}|p{2.9cm}|p{2.8cm}} 
\toprule
Reference & Data Quality \mbox{Dimensions} & ML Algorithms & Methodologies \\
\midrule
CleanML~\cite{LiRBZCZ21} & 
completeness, feature accuracy, uniqueness, consistent representation, target accuracy & 
Logistic Regression, KNN, Decision Tree, Random Forest, AdaBoost, Naive Bayes, XGBoost &
Considered different scenarios of dirty training and test data. \\ 
\midrule
Foroni \mbox{et al.~\cite{Foroni21}} & 
completeness, consistent representation, feature accuracy & 
Random Forest (classification), Linear Least Square (regression), k-means (clustering) & 
Introduces more errors to derive influence of low data quality. \\ 
\midrule
Neutatz \mbox{et al.~\cite{NeutatzCAYA22}} & 
uniqueness, consistent representation, feature accuracy, completeness & 
AutoSklearn builds model ensembles & 
Evaluation of state-of-the-art AutoML systems. \\ 
\midrule
Shah \mbox{et al.~\cite{10.14778/3648160.3648178}} & 
uniqueness & 
Random Forest, Logistic Regression, Neural Network, XGBoost & 
Analyzed the effect of duplicates. \\ 
\midrule
Fr{\'{e}}na and Verleysen~\cite{FrenayV14} & 
target accuracy & 
Referenced papers using, among others, KNN, Bayes Decision Tree, SVM, AdaBoost, Decision Tree &
Survey about the effect of dirty labels in training data. \\ 
\bottomrule
\end{tabular}
}
\label{tab:related_work}
\end{table*}
First, we report on the state of the art in data validation for ML\@.
Then, we discuss related work that studies the influence of data quality on ML models, namely by conducting an empirical evaluation, cleaning the data or by focusing on a specific error type like label noise.
\revision{We summarize related work in Table~\ref{tab:related_work}, highlighting the respective considered data quality dimensions, employed ML algorithms, and methodologies.}

\stitle{Data validation}
Several approaches have emerged to validate ML pipelines as well as the data fed to them, which includes training and serving data (data used in production).
These approaches use the concept of \emph{unit tests} to help engineers diagnose model-quality issues originating from data errors.
For instance, the validation system implemented by Breck et al.~\cite{breck2019data} and the similar system by Schelter et  al.~\cite{SchelterRB20} focus on validating serving data given a classification pipeline as a black box.

Generally, validation systems check, on the one hand, for traditional data quality dimensions, such as consistency and completeness, and on the other hand for ML dependent dimensions, such as model robustness and privacy~\cite{biessmann2021automated}. 
To help data scientists with the validation task, Schelter et al.\ introduced the experimental library JENGA~\cite{SchelterRB21}.
It enables testing ML model's robustness under data errors in serving data.
The authors use the concept of polluters or data corruptions as in our work.
However, they do not provide an extensive experimental study and their focus is on describing the framework. 

\stitle{Task-dependent data quality}
Foroni et al.\ argue that data quality assessment should not be performed in isolation from the task at hand~\cite{Foroni21}. 
Our results confirm this statement for ML models, as the same ``low'' quality data has a different effect when used to train different models. 
Their paper proposes a theoretical framework with a setup similar to our experiments, which evaluates the performance of a task given polluted datasets by various kinds of generated systematic noise.
\revision{The authors considered a multitude of errors in their work, which we can map to the data quality dimensions of our study as follows: completeness (missing values, generation of nulls), consistent representation (synonyms, abbreviations), and feature accuracy (spelling mistakes, permuted words, terms in different languages, numerical variations, scale modifications, arithmetic negations).}
The proposed framework then computes the variation effect factor or the sensitivity factor from the observed results of the task. 
Unlike our work, however, the authors focused only on polluted training and testing data~(Scenario~3 in our paper) and the experiments were conducted on a single dataset to evaluate only three models~(one per ML task) namely, Random Forest, $k$-means, and Linear Least Square. 
The authors made observations that agree with our findings, especially the fact that missing values (completeness) are a problem for all ML tasks.

\stitle{Data cleaning}
Li et al.\ investigated the impact of cleaning training data,~i.e., improving its quality, on the performance of classification algorithms~\cite{LiRBZCZ21}.
They obtained a clean version of the training data instead of systematically polluting it, as we did in this work.
Their effort yielded the CleanML benchmark.
They focused on five error types: missing values, outliers, duplicates, inconsistencies, and mislabels.
These error types are among the most popular error types, and thus some correspond to the data quality dimensions in our study. 
The authors observed that cleaning inconsistencies and duplicates is more likely to have low impact, while removing missing values and fixing mislabeled data is more likely to improve the classifier prediction. 
These observations align with our findings. 

The CleanML benchmark datasets have been recently used, among others, by Neutatz et al.~\cite{NeutatzCAYA22}.
The authors evaluate the ability of AutoML systems, such as AutoSklearn~\cite{FeurerKESBH15}, to produce a binary classification pipeline that can overcome the effect of the following types of errors in training data\revision{, which we again map to our considered data quality dimensions: uniqueness (duplicates), consistent representation (inconsistencies), feature accuracy (outliers), completeness (missing values), and target accuracy (mislabels)}.
The authors concluded that AutoML can handle duplicates, inconsistencies, and outliers, but not missing values.
The paper also points out that most current benchmark datasets contain only few real-world errors with insignificant impact on the ML performance even without any cleaning.
For this reason and as mentioned in Section~\ref{sec:intro}, our work uses synthetic errors to better characterize the correlation between ML models performance and data quality.

Clearly, our study aligns with data cleaning efforts but with a different goal. However,
our observations give data experts the understanding of the effects of data quality issues.
They can then use this knowledge to determine the robustness of their insights and decide which specific problems should to be tended to and when.
For an overview of the progress in cleaning for ML, we refer to~\cite{NeutatzCA021}.

The work of Shah et al.\ follows a similar direction as CleanML and ours~\cite{10.14778/3648160.3648178}.
However, the authors focus solely on investigating the influence of deduplication\revision{, which we can map to the data quality dimension uniqueness in our work,} on an underlying classification task.
They consider five different ML algorithms: random forest, logistic regression, a neural network, and XGBoost.
They also investigate the influence of different encodings on the underlying ML task.
The authors used 16 real-world datasets for the experiments, in which they manually annotated duplicates.
In addition to the complete removal of duplicates, the authors also look at the influence of a gradual introduction of duplicates into the data on the underlying ML task, similar to our setup.
In general, the authors have similar findings to ours that certain ML algorithms, such as logistic regression, are more robust against duplicates than others.

\stitle{Label noise}
The problem of label noise or mislabeling is one of the main concerns of the ML-community and has attracted much interest~\cite{FrenayV14}.
This problem is in essence a data quality problem. Fr{\'{e}}nay and Verleysen surveyed the literature related to classification using training data that suffers from \emph{label noise}, which is equivalent to the target accuracy dimension in our work~\cite{FrenayV14}.
They distinguish several sources of noise, discuss the potential ramifications, and categorize the methods 
into the classes ``noise-robust'', ``noise cleansing'', and ``noise-tolerant''. 
They conclude that label noise has diverse ramifications, including degrading classification accuracy, high complexity learning models, and difficulty in specifying relevant features.


In summary, we present the first systematic empirical study on how both training and test data quality affects not only classification but all three ML tasks.
We also provide a clear definition for each of the data quality dimensions and a respective method to systematically pollute the data.

%% file: 30-dimensions_pollution.tex
\section{Data quality dimensions and data pollution}
\label{sec:ddimentions}

We present the definition of the six selected data quality dimensions, along with our methods to systemically pollute a dataset along those dimensions. In this work, we use the ML terms \emph{feature} and \emph{sample} to refer to columns and rows, respectively. During pollution, we assume that features' data types and the placeholders that represent missing values in each feature are given.

\subsection{Consistent Representation}
\label{sec:consistent_representation}
A dataset is consistent in its representation if no feature has two or more unique values that are semantically equivalent.
I.e., each real-world entity or concept is referred to by only one representation. For example, in a feature ``city'', \texttt{New York} shall not be also represented as \texttt{NYC} or \texttt{NY}.
\revision{Consistent representation is different from uniqueness, which focuses on ensuring that no duplicate records exist in the dataset (see Section~\ref{subsection:uniqueness}).}
\begin{definition}
    The degree of \emph{inconsistency} of a feature~$c$, denoted as $\mathit{InCons}(c)$, is the ratio of the minimum number of replacement operations required to transform it into a consistent state and the number of samples in the dataset. 
    \end{definition}
This definition applies only to categorical features,~i.e., strings or integers that encode categorical values, whereas numerical features and dates are considered to be consistent and theirs~$\mathit{InCons}(c)=0$. 
\PaperLong{The degree of consistency of a feature cannot be derived by subtracting~$\lambda_{cr}$ from 1 because it also depends on the number of representations of an original value~(see Figure~\ref{fig:consistent-representation-pollution-quality}).
    \begin{figure}[htbp]
        \centering
        \includegraphics[width=0.6\textwidth]{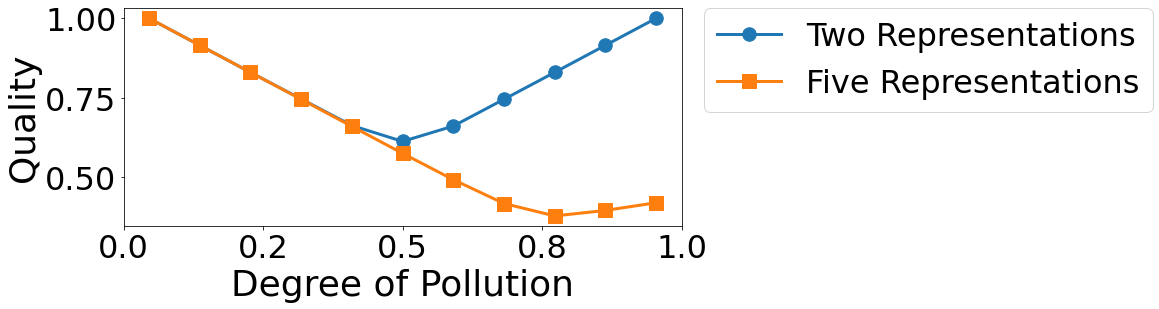}
        \caption{Relation of pollution \& quality for consistent representation with different numbers of representations per original value}
        \label{fig:consistent-representation-pollution-quality}
    \end{figure}}
\begin{definition}
We define the degree of \emph{consistency} of a dataset~$d$ with~$f$ features as follows.

\begin{equation}
    \label{eq:per-dataset-consistant}
        \mathit{Consistency}(d) = 1 - \frac{1}{f}\cdot\sum_{i \in \mathbf{1,...,f}} \mathit{InCons}(c_i)
\end{equation}
\end{definition}

\revision{
\stitle{Example}
Consider a dataset $d$ with a single categorical feature ``City'', containing the ten distinct values [\texttt{New York}, \texttt{NYC}, \texttt{Los Angeles}, \texttt{LA}, \texttt{San Francisco}, \texttt{SF}, \texttt{NY}, \texttt{Los Angeles}, \texttt{SF}, \texttt{San Francisco}].
This ``City'' feature includes multiple representations of the same cities: \texttt{New York} is also represented as \texttt{NYC} and \texttt{NY}, \texttt{Los Angeles} as \texttt{LA}, and \texttt{San Francisco} as \texttt{SF}.
To achieve consistency in this feature, \texttt{NYC} and \texttt{NY} could be replaced with \texttt{New York}, \texttt{LA} with \texttt{Los Angeles}, and \texttt{SF} with \texttt{San Francisco}.
The minimum number of replacement operations required to achieve this consistency is~5.
Thus, the inconsistency of the ``City'' feature is $\mathit{InCons}(City)= \frac{5}{10} = 0.5$ and the degree of consistency for the dataset is calculated as $\mathit{Consistency}(d) = 1 - \frac{1}{1} \cdot 0.5 = 0.5$.
}

\stitle{Pollution} 
We have two inputs: First, the percentage of samples to be polluted~$\lambda_{cr}$, defined by a value between~0 and~1, and second, for each unique value~$v$ of a pollutable feature, the number of representations~$k_{v}$ for that value~(including~$v$ itself).
For each categorical feature, we choose randomly the samples to be polluted.
Then, we generate~$k_{v}-1$ new representations for each unique value~$v$ of its values.
The new representations of a string value are produced as new non-existing values by appending a trailing ascending number to the end of the value, whereas for integers, new integers are added after the maximum existing one. 
These sample's entries at this feature are replaced by a randomly picked fresh representation of the original value.

\revision{When pollution introduces alternative representations in the data, it typically decreases the measured consistency according to Eq.~\ref{eq:per-dataset-consistant}.
However, as the quality measure is based on the number of unique representations rather than their semantic integrity (as defined by the clean data).
Thus, consistency can be achieved by aligning the correct values to match the current state of the data. 
If the data contains significant inconsistencies -- indicating substantial pollution -- fewer transformations may be required to align the correct values with the noisy ones than to correct the noisy values themselves. 
As a result, the calculated consistency, as per Eq.~\ref{eq:per-dataset-consistant}, may increase.
For consistency-related results, we show prediction accuracy curves only up to the point where measured consistency begins to increase during pollution.}

\subsection{Completeness}
\label{sec:completeness}
The problem of missing values exists in many real-world datasets. Some of these values are actually missing, e.g., missing readings due to a failure in a sensor, while others are represented by a placeholder, such as ``unknown'' or ``NaN''.
\PaperLong{For example, when medical sensors monitor temperature, blood pressure and other health information of a person and one sensor fails for a period of time, there are no values present for this sensor and time in the recorded dataset.
In the case of a survey form containing optional input fields, personal attributes like the employment status could be given as ``unknown''.
The respective value that has a missing value could potentially be absent. When processing data automatically, for example in a data frame, the value itself usually exists, but is not informative or useful for analysis and therefore equivalent to the value being missing, decreasing the completeness of the dataset.}
\begin{definition}
The \emph{completeness} of a feature~$c$ is the ratio of the number of non-missing values and the total number of samples~$n$ in this dataset. The completeness of a dataset~$d$ is defined as 
\begin{equation}
    \label{eq:per-dataset-completeness}
        \mathit{Completeness}(d) = 1-\frac{1}{f}\cdot \sum_{i \in \mathbf{1,...,f}} \mathit{missing}(c_i)
\end{equation}
\revision{where $\mathit{missing}(c_i)$ denotes the ratio of the total number of missing values in the feature $c_i$ and the number of samples in the dataset.}
\end{definition}
\PaperLong{A dataset with a completeness of~1 has no missing values in it. In case of a completeness of~0, the whole dataset consists only of missing values, except for the target feature.}
For \revision{ML}, samples with a missing value for the target feature are usually removed from the dataset, as they cannot be used for training.
Thus, we exclude the target feature while computing completeness.

\revision{\stitle{Example}
Consider a dataset $d$ that has two features (excluding the target feature) and four samples. 
This means that there are eight cells in total for computing the completeness.
If two values of each feature are missing in the dataset, we calculate $\mathit{Completeness}(d) = 1-\frac{1}{2} \cdot \left(\frac{2}{4}+\frac{2}{4}\right) = 0.5$.
}

\stitle{Pollution} 
We inject missing values ``completely at random''~\cite{MCAR-definition} according to a specified pollution percentage~$\lambda_{c}$ for each feature.
If there are already missing values in the feature, we account for those values and inject only the remaining number of missing values necessary to reach~$\lambda_{c}$.
A feature-specific placeholder is used to represent all missing values injected into this feature. 
The placeholder value does not carry information except that the value is missing.
A typical placeholder is ``Not a Number''~(NaN). 
Many implementations of ML algorithms cannot handle NaN values. 
For this reason, we choose a representation as placeholders that can be used in computations, but lie outside the usual domain of a feature and are still distinguishable from the actual non-missing values of the feature.
For example,~$-1$ for ``age'' feature, or the string ``empty'' for ``genre'' categorical feature in a movie table. 
We manually selected the used placeholders, as this task requires some domain knowledge to determine suitable values~\footnote{\revision{See https://github.com/HPI-Information-Systems/DQ4AI/blob/main/metadata.json for detailed setting of the placeholders.}}.
\PaperLong{A placeholder value representation can count as imputation~\cite{SchelterRB21}.
Most imputation methods, such as taking the mean, reconstruct some amount of information based on other observed values.
As the placeholder does not contain information related to the data and has no reconstruction involved, we still consider a placeholder representation as pollution.
Further, comparing different imputation methods would drift apart from the dimension completeness because it would interfere with other dimensions like accuracy.
We do not make any assumptions about the underlying distribution and dependencies of missing data in our datasets.
The probability of an value to have a missing value is not influenced by any other observed or unobserved value in the data.
In other terms, we only consider data that is ``Missing Completely at Random''~(MCAR)~\cite{MCAR-definition} in our experiments.}

\subsection{Feature Accuracy}
\label{subsection:facc_def}
Feature accuracy reflects to which extent feature values in a given dataset equal their respective ground truth values.\PaperLong{The more cells deviate from their actual value and the stronger pronounced this deviation is, the lower is the feature accuracy.} 
\begin{definition}    
     The \emph{feature accuracy} measures the deviation of a column's values from their respective ground truth values. For a categorical feature $c$, we define the feature accuracy as follows.     
 \begin{equation}
    \label{eq:per-feature-acc-cat}
        \mathit{cFAcc(c)} = 1 - \frac{\mathit{mismatches}(c)}{n}
\end{equation}
where~$\mathit{mismatches}(c)$ denotes the number of values in the feature~$c$ that are different from the ground truth, and~$n$ is the number of samples in the dataset.\PaperLong{The erroneous values' ratio is then subtracted from~1, meaning that~1 is the best possible quality and~0 is the worst possible quality.}
For numerical features, we define feature accuracy as follows. 
 \begin{equation}
    \label{eq:per-feature-acc-num}
       \mathit{nFAcc}(c) = 1 - \frac{\mathit{avg\_dist}(c)}{\mathit{mean\_gt}(c)}
    \end{equation}
\begin{equation}
    \label{eq:average-dist}
        \mathit{avg\_dist}(c_i) = \frac{1}{n} \cdot \sum\limits_{j=0}^{n-1} |gt_{i, j} - v_{i, j}|
    \end{equation}
where~$\mathit{avg\_dist(c)}$ is the average of the absolute distances between the ground truth and values in~$c$~(see Equation~\ref{eq:average-dist}) and~$\mathit{mean\_gt(c)}$ is the mean of the ground truth values of~$c$. 
\end{definition}
In Equation~\ref{eq:average-dist},~$j$ is used as the index of a specific sample. Hence,~$gt_{i, j}$ denotes the ground truth and~$v_{i, j}$ the value of the sample with index~$j$ in feature~$i$.
\PaperLong{As for the categorical features, a quality of~1 indicates a clean feature.} In contrast to the categorical feature quality measure, the numerical measure can fall below~0 and has no defined lower bound.
\PaperLong{However, we found that all datasets polluted reasonably also yield a numerical quality~$>0$.}

The feature accuracy of an entire dataset consists of two metrics: The average feature accuracy of all categorical features~$\mathit{cFAccuracy}$ and the average feature accuracy of all numerical features~$\mathit{nFAccuracy}$.
\PaperLong{This is caused by the fact that with numeric features all samples are polluted, and with categorical features only a certain percentage of the samples are polluted. 
Therefore, the feature accuracy of both feature types is calculated differently, which leads to both feature types having different accuracy ranges.

The feature accuracy quality measure of all categorical features~$\mathit{cFAccuracy}$ is defined as the average of the feature accuracy of all categorical features as can be seen in Equation~\ref{eq:quality-measure-fa-cat}.
Similarly, Equation~\ref{eq:quality-measure-fa-num} shows that the feature accuracy quality measure of all numeric features~$nFAccuracy$ is defined as the average of all per-feature accuracies. }The numbers of categorical and numeric features are given by~$n_{cat}$ and~$n_{num}$, respectively.
    \begin{equation}
    \label{eq:quality-measure-fa-cat}
        \mathit{cFAccuracy}(d) = \frac{1}{n_{\mathit{cat}}} \cdot \sum\limits_{i=0}^{n_{\mathit{cat}}-1} \mathit{cFAcc}(c_i)
    \end{equation}
    \begin{equation}
    \label{eq:quality-measure-fa-num}
         \mathit{nFAccuracy}(d)= \frac{1}{n_{\mathit{num}}} \cdot \sum\limits_{i=0}^{n_{\mathit{num}}-1} \mathit{nFAcc}(c_i)
    \end{equation}

\revision{\stitle{Example}
Consider a dataset $d$ with five samples and with the two features ``City'' (categorical) and ``Temperature'' (numerical).
In this example ``City'' has one mismatch, similar to the example for consistent representation, \texttt{LA} should be \texttt{Los Angeles}.
We would calculate the feature accuracy for ``City'', the categorical feature, as $\mathit{cFAccuracy}(d) = \frac{1}{1}\cdot\left(1-\frac{1}{5}\right) = 0.8$.}
    
\stitle{Pollution}
\PaperLong{The polluter takes three arguments. The first argument~$\lambda_{fa}$ is a dictionary that maps feature names to float numbers in the interval~$[0.0, 1.0]$. 
It describes the level of pollution that should be utilized for each given feature. 
The argument~$\lambda_{fa}$ can also be defined as a single float number, meaning that the same level of pollution is applied to all features.
The second and third polluter arguments each contain a complete list of the available categorical and numeric feature names.\\}
The pollution is executed differently, depending on the feature type. 
For categorical features, the level of pollution~$\lambda_{fa}$ for a specific feature~$c$ determines the percentage of samples to be polluted.
\PaperLong{The samples to pollute are chosen randomly. 
However, the seed for this selection is fixed to (1)~ensure reproducibility of the results and (2)~allow for a level of pollution~$\lambda_{fa}(c) = 0.2$ to be a direct extension of $\lambda_{fa}(c) = 0.1$.} The randomly selected samples are polluted by exchanging the current category with a random, but different category from feature $c$' domain.
\PaperLong{Consequently, a level of pollution~$\lambda_{fa}(c) = 1.0$ for a categorical feature~$c$ means that the categories of all samples are updated and~$\lambda_{fa}(c) = 0.0$ indicates that all categories of $c$ stay the same.}

For numeric features, we add normally distributed noise to all samples of the feature~$c$: 
$\textit{noise}(c)= X \cdot \mathit{mean\_gt}(c)$ where $X$ is a random sample drawn from the normal Gaussian distribution~$N(\mu, \sigma^2)$ with~$\mu = 0$ and~$\sigma^2 = \lambda_{fa}$. 
The level of pollution~$\lambda_{fa}$ determines the standard deviation of the normal distribution and thus denotes how wide it is spread.
\PaperLong{Again, it is ensured that the same seed is used per feature in consecutive pollution runs on the same dataset to keep the behavior consistent and comparable.} 

\subsection{Target Accuracy}
For each sample in a dataset, the target feature contains either a class/label in classification tasks or a numeric value in regression tasks. 
There might be some incorrect labels due to human or machine error,~e.g., a dog labeled as ``wolf''.
\PaperLong{\begin{definition}
     The \emph{target accuracy} of a dataset is the deviation of its target feature values from their respective ground truth values. For a categorical target, the target accuracy is the ratio of correct values in the target feature. 
     \begin{equation}\label{eq:target_accuracy_cat}
        \mathit{cTAccuracy}(d) = 1 - \frac{\mathit{mismatches}(target)}{n}
    \end{equation}
     For a target with numerical values, we define the target accuracy as follows.  
         \begin{equation}\label{eq:target_accuracy_num}
        \mathit{nTAccuracy}(d) = 1 - \frac{\mathit{avg\_dis}(\mathit{target})}{\mathit{mean\_gt}(\mathit{target})}
    \end{equation}
Where~$\mathit{avg\_dist}$ is the averaged sum of the absolute distances~(Manhattan distance) of the ground truth and target feature values and~$\mathit{mean\_gt}$ is the mean of the ground truth values.
\end{definition}}
The definition of the target accuracy of a dataset is equivalent to the definition of the feature accuracy of its target feature~(see previous section). Nevertheless, the target feature is the most important feature because of its influence on prediction performance. Thus, it is beneficial to study its accuracy separately.\PaperLong{By scaling with the mean, we obtain a measurement that is less dependent on the actual target domain and, thus, more comparable between different datasets. This, in theory, allows for a negative quality metric where on average every target value is more than one mean away from its ground truth. We disregard those cases and define~0 as the lowest possible quality metric in our experiments.} 
\revision{\stitle{Example}
As the target accuracy is calculated similarly to feature accuracy, see Section~\ref{subsection:facc_def} for an example.}

\stitle{Pollution} Naturally, we used the same pollution method as for feature accuracy, based on the target type.\PaperLong{In general, this polluter takes a single argument for degree of pollution~$\lambda_{ta}$. For categorical target, this value is interpreted as the fraction of data points that should be polluted. The pollution itself replaces the current label with a randomly chosen label that differs from the current one. For numerical targets,~$\lambda_{ta}$ is interpreted as the variance of normally distributed noise that is scaled by the mean of the original target value distribution and added onto the target values of the specified subset~(e.g., train or test data).} 
  
\subsection{Uniqueness}
\label{subsection:uniqueness}
Redundant data does not provide additional information to the ML model for the training process. 
Thus, deduplication is a common step in ML pipelines to avoid overfitting.
\PaperLong{Furthermore, there is a plethora of existing research on how to detect duplicates and how to remove them \cite{duplicates-quality, relational-deduplication}. 
In this report, we evaluate how the number of duplicates present in a dataset influences the performance of ML models. 
We investigate whether pre-processing datasets to remove duplicates is an essential step toward improved ML performance.
According to Chen et al.\, there are different definitions when to consider two samples as duplicates, such as having identical primary keys or the equality in every feature of the two samples~\cite{redundancy_def}. In practice, there are also often non-exact duplicates, where some features differ slightly, for example timestamps. Introducing non-exact duplicates in the polluter would not only interfere with the redundancy dimension, but also with consistent representation~(see Section~\ref{sec:consistent_representation}). }Exact duplicate rows are, in general, easy to detect. Yet, this step is still expensive, especially for large datasets.
\PaperLong{In this regard, ideally, there are as few duplicates as possible in a dataset. Therefore, we only consider fully equal samples as duplicates.} 
\begin{definition}      
        The \emph{uniqueness} of a dataset~$d$ is the fraction of unique samples within the dataset. We normalize the value as follows.   
    \begin{equation}\label{def_uniqueness}
           \mathit{Uniqueness}(d) = \frac{\mathit{unique\_samples}(d)-1}{n-1}
    \end{equation}
We subtract~1 from the denominator and numerator to allow a quality metric score of~0.
\end{definition}
\PaperLong{A dataset with the quality metric of~1 does not contain any duplicates, whereas the quality metric of~0 refers to a dataset containing only one unique record -- even if the dataset contains many records overall.}
\revision{\stitle{Example}
Consider a dataset $d$ with ten samples.
In this example, let three of the ten samples appear as exact duplicates of the other rows.
Thus, seven of the ten samples are unique and we calculate the uniqueness for this example as $\mathit{Uniqueness}(d) = \frac{\mathit{7-1}}{10-1} = \frac{2}{3}$.}

\stitle{Pollution} 
\PaperLong{Input datasets can contain duplicates themselves.} To pollute a dataset along the uniqueness dimension, we first remove all existing exact duplicates.\PaperLong{This allows to pollute the dataset in incremental fashion.} Then, we add exact duplicates of randomly selected samples to the dataset.\PaperLong{Actually, we increase the dataset size to avoid data loss that can affect ML models performance.} 
The number of the added duplicates is determined by the duplication factor~$\rho$: for each class~${cl}$ with~$n_{cl}$ samples, we add~$\mathit{dup}_{cl}=(\rho-1)\cdot n_{cl}$ duplicates to avoid changing the class balance (see Section~\ref{subsec:target-class-balance-polluter}).
Thus, the size of the polluted dataset is~$n\cdot\rho$ and its uniqueness is~$1/\rho$.\PaperLong{The duplication factor $\rho$ ranges from~1, meaning no pollution is applied, to potentially infinity. It is important to decide how often each sample appears in the polluted dataset. One trivial approach would be to duplicate each by the same factor. One issue of this approach is the limited applicability to real-world scenarios.
In reality, the number of duplicates per sample depends on the data domain.
Manually inserted form data, for example, probably contain a normally distributed number of duplicates due to human errors, with a mean of~2.
Working with sensor data, due to misconfiguration of sensors, the distribution of duplicate count could be uniform.
Analyzing web index data based on web traffic, the duplicates could be distributed according to the Zipf distribution.
Thus, the polluter needs to be flexible regarding the distribution of duplicate counts per sample.} For each randomly selected sample from a class~$cl$, we add~$x$ duplicates of this sample and then continue sampling and adding duplicates to reach~$\mathit{dup}_{cl}$.
We draw~$x$ from a pre-defined distribution: we apply uniform, normal, and Zipf distributions, in addition to adding a single duplicate of each selected sample. 
\PaperLong{The duplicate counts generated by the specified distribution function define only the number of duplicates to add for the respective sample each time the sample is randomly selected for duplication.
For this reason, the actual resulting distribution of duplicate cluster sizes after pollution likely differs from the specified distribution.
This can happen especially with large duplication factors, but is inevitable, as otherwise classes with a low number of sampled duplicate counts could limit the number of generated elements.}

\subsection{Target Class Balance}
\label{subsec:target-class-balance-polluter}
Many ML algorithms assume a relatively equal number of samples per target class, i.e., a balanced dataset, to achieve satisfactory performance.
Clustering or classification algorithms on top of an imbalanced dataset may fail at identifying structures or even miss smaller classes completely.
For example, the $k$-Means algorithm suffers from the ``uniform effect'', i.e., it recognizes clusters of approximately uniform sizes even if they are not present in the input data~\cite{kumar2015subset}.
\begin{definition} 
    \label{def_imbalance}
     Given a dataset~$d$ with~$m$ target classes $cl_1,..., cl_m$ of~$n_{cl_1},..., n_{cl_m}$ samples per class, respectively, and~$\forall i, j: 1 \leq i < j \leq m \iff n_{cl_i} \leq n_{cl_j}$, the target class \emph{imbalance} is defined as the sum of the pairwise differences between the number of samples per class: 
    \begin{equation}\label{eq_sum_pairwise}
        \mathit{ImBalance}(d) = \frac{1}{2} \cdot \sum_{i, j \in \mathbf{1,...,m}} |n_{cl_i} - n_{cl_j}| 
    \end{equation}
\end{definition}
As the worst imbalance case, we assume a maximal imbalanced dataset that has~$\lceil m/2 \rceil$ classes with~$n_{\mathit{cmax}}$ samples and the remaining classes have~0~samples, where~$n_{\mathit{cmax}}$ is the maximum number of samples that a class can have\PaperLong{(see Figure~\ref{fig:classbalance-worst-case})}. The target class imbalance of such a dataset is~$\varepsilon = \lceil m/2 \rceil \cdot \lfloor m/2 \rfloor \cdot n_{\mathit{cmax}}$.\PaperLong{This is clearly a hypothetical case, as no class exists if it has~$0$ samples; otherwise we could add infinite classes with~$0$ samples to each dataset.
However, this constructed the worst case allows us to define the target class balance quality measure.} 

\begin{definition} 
The target class \emph{balance} of a dataset~$d$ is the deviation from its imbalance score, normalized by the imbalance score of the worst case.
   \begin{equation}\label{eq_quality_balance}
        \mathit{Balance}(d) = 1 - \frac{\mathit{ImBalance}(d)}{\varepsilon}
    \end{equation}
\end{definition}
If all classes in the dataset have the same number of samples, then~$\mathit{Balance}(d)$ is maximal and equals~1. 
In contrast, $\mathit{Balance}(d)$ reaches its minimum ($\lim\limits_{d \to w} \mathit{Balance}(d) = 0$) if the balance of the classes in the dataset approaches the hypothetical worst case.
\PaperLong{\begin{figure}[htbp]
        \centering
        \includegraphics[width=0.5\textwidth]{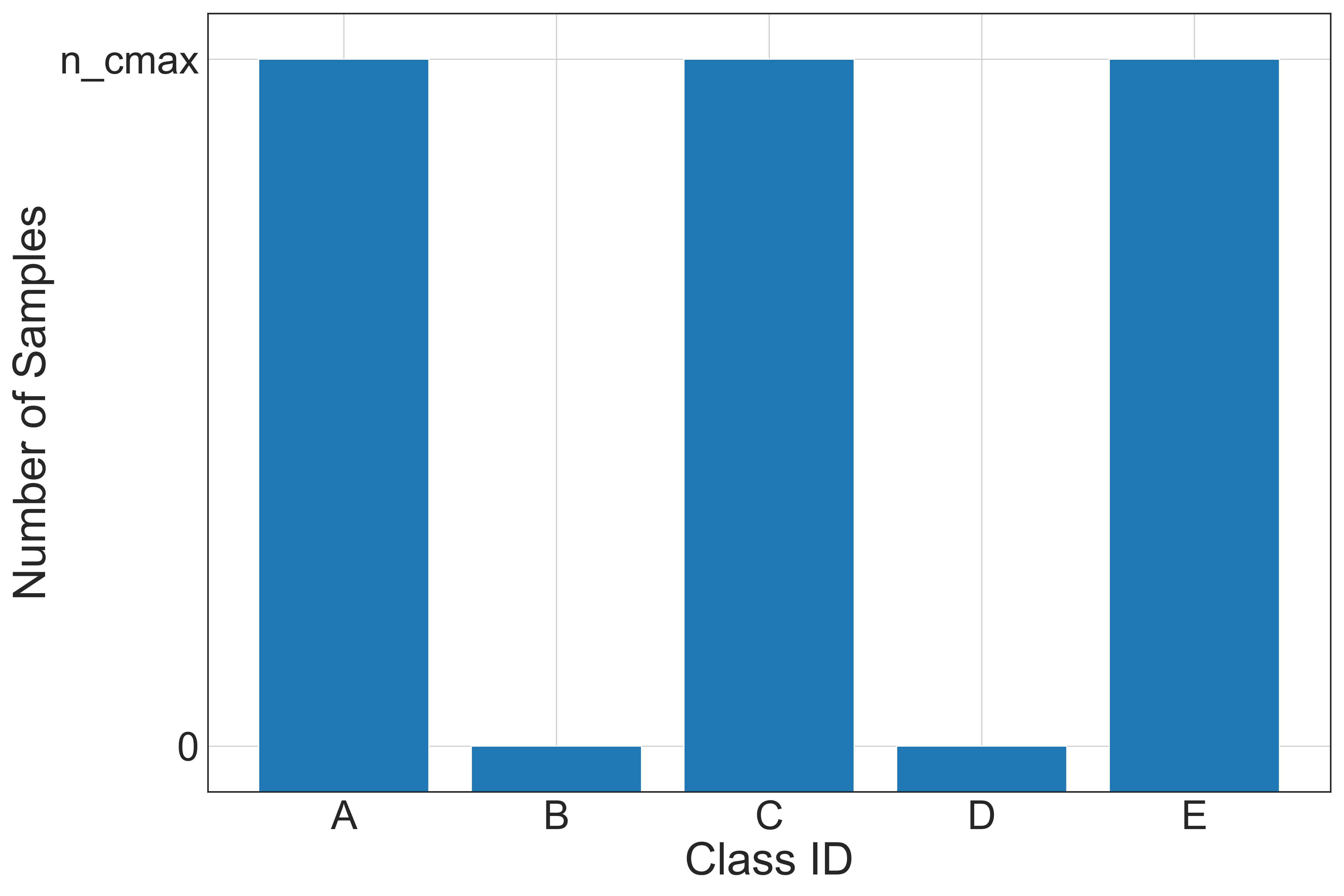}
        \caption{Example of worst-case target class balance. Plot starts below zero to clearly show classes with no samples.}
        \label{fig:classbalance-worst-case}
    \end{figure}}
\revision{\stitle{Example}
Consider a dataset $d$ with the three target classes \texttt{expensive}, \texttt{moderate}, and \texttt{cheap}. 
Let the number of samples in each respective class be 
$n_{expensive} = 10$, $n_{moderate} = 20$, and $n_{cheap} = 30$.
We calculate the target class imbalance using the sum of pairwise differences between the number of samples per class as $\mathit{ImBalance}(d) = \frac{1}{2} \cdot \left( |10 - 20| + |10 - 30| + |20 - 30| \right) = 20$.}

\revision{To compute the target class balance, we first need the imbalance score for the worst case. 
In the worst case scenario, two of the three classes have $n_{\mathit{cmax}} = 30$ samples, and one class has 0 samples. 
Thus, the worst-case imbalance score is $\varepsilon = \lceil 3/2 \rceil \cdot \lfloor 3/2 \rfloor \cdot n_{\mathit{cmax}} = 60$.}
\revision{Finally, we compute the target class balance as $\mathit{Balance}(d) = \frac{20}{60} = 1 - 0.333 = 0.667$.
}

\stitle{Pollution}
We have two inputs for pollution: The degree of imbalance~$\lambda_{cb}$ and the number of samples in the polluted version~$\tilde{n}$. 
We can choose~$\tilde{n}$ arbitrarily as a multiple of~$m$ or calculate it from the data as the number of samples from the original dataset, needed to produce the maximum pollution level.
\PaperLong{In both cases, its validity at the maximum imbalance level and the balanced dataset is checked again. 
In case of an invalid sample count, the next valid, smaller possible sample count is calculated and used while a warning is presented to the user.
For calculating~$\tilde{n}$, we also consider that each class must have a minimum number of samples in the original dataset to be able to produce this imbalance. 
If this requirement cannot be satisfied, a new number of total samples~$\tilde{n}$ in the imbalanced dataset is iteratively determined until it is possible to create the maximal imbalance.}
We use~$\lambda_{cb}$ to calculate the number of samples per class in the polluted version.
\PaperLong{It is a number in the interval~$[0, 1]$ and it is not directly linked to~$\mathit{Balance}(d)$: $\lambda_{cb} = 0$  creates a fully balanced dataset, which is to be used as a baseline for comparing all the imbalanced datasets to, as the original dataset maybe imbalanced itself;~$\lambda_{cb} = 1$ creates the most heavily imbalanced dataset.
Note that the dataset produced by~$\lambda_{cb} = 1$ is not the hypothetical worst case mentioned in the definition section above. 
Instead, it produces a dataset in which the smallest class has~$0\%$ of the samples of the largest class and where our restriction of constant changes in sample counts between the classes still stands.
While mathematically, this polluted dataset with one class completely removed would be the most imbalanced, this does not suit the purpose of examining the effects of class imbalance on the ML process.
Therefore, we restrict the most heavily imbalanced dataset to have a class~$c_m$ containing the maximal number of~$s_{\mathit{max}}$ samples and a class~$c_1$ containing the minimal number of~$s_{\mathit{min}} = \lceil0.01 \cdot s_{\mathit{max}}\rceil$ samples.
Due to this calculation, the class balance polluter works best if the class~$c_m$ has at least~$s_{\mathit{max}}=100$ samples. 
This state of imbalance is reached at a degree of imbalance~$\lambda_{cb} < 1$. 
However, anything above this degree is ignored by the polluter.}
The degree of imbalance~$\lambda_{cb}$ for any polluted dataset version satisfies: $\lambda_{cb} = \frac{\mathit{ImBalance}(\tilde{d})}{1/3 \cdot (m+1) \cdot \tilde{n}}$.

To pollute, we order the target classes by their sizes descending and for the classes of the same size, we use the class ID in ascending order.
Using the defined order, we create a class imbalance with an equal difference~$\Delta$ for each subsequent pair of classes.
The used samples of each class are randomly selected, i.e., in the polluted dataset, the following holds:
$\forall 1 < j \leq m: (\tilde{n}_{c_{j-1}} \leq \tilde{n}_{c_j}) \land (\tilde{n}_{c_j} - \tilde{n}_{c_{j-1}}) = \Delta$. 
\PaperLong{Figure~\ref{fig:classbalance-balance-examples} in the depicted \textit{Balance~2} shows an example of such a distribution.
This choice has been made to simplify the pollution process by removing all the other possible methods of creating an imbalance and making it more easily reproducible as no random component is needed to determine the per-class sample counts in the polluted dataset. 
Figure~\ref{fig:classbalance-balance-examples} shows the two per-class sample count distributions \textit{Balance~1} and \textit{Balance~3}, which exemplify the need to limit the pollution method as they differ in per-class sample counts but still produce the same data quality.
    \begin{figure}[htbp]
        \centering
        \includegraphics[width=0.5\textwidth]{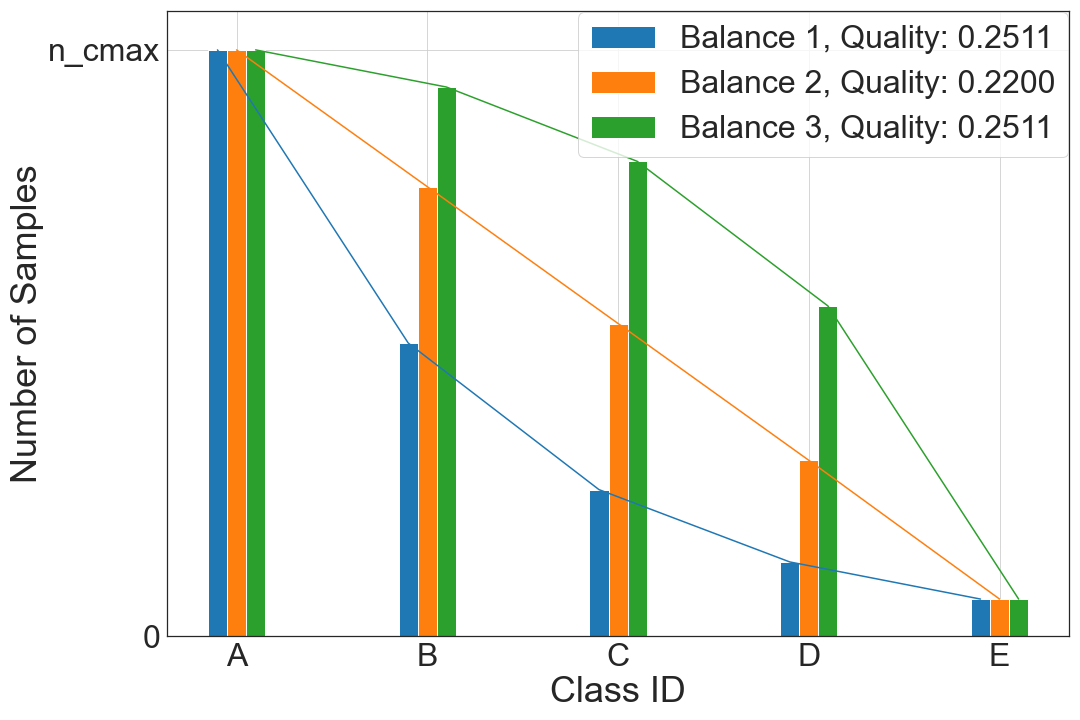}
        \caption{Examples of possible per-class sample distributions and their corresponding quality.}
        \label{fig:classbalance-balance-examples}
    \end{figure}}
    
To create a class imbalance, we calculate the class size of a balanced dataset~$\frac{\tilde{n}}{m}$ with~$\tilde{n}$ samples and~$m$ classes.
Then, we iteratively add/remove samples from the classes based on their order: We remove samples from all classes that are at indices~$\lfloor \frac{m}{2} \rfloor - 1$ and below, and add samples to all classes at indices above that, unless~$m$ is odd: then the size of the class at index~$\lfloor \frac{m}{2} \rfloor$ stays constant. 

%% file: 40-ML_tasks.tex
\section{Machine Learning tasks} 
\label{sec:MLalgorithms}
This section introduces the \revision{19} algorithms that we employed for the three machine learning (ML) tasks of \textit{classification}, \textit{regression} and \textit{clustering}. 
Our selection was guided to cover a broad range of methods families for each task. 

\stitle{Classification}
\revision{Classification is a supervised learning method that uses labeled data to learn the dependency between features and a discrete target variable. 
The goal is to assign samples to predefined classes (predict the correct class) by learning patterns in labeled training data.}
We selected a variety of classification algorithms that fall into different categories.
We included two linear classification models: Logistic regression~(LogR)~\cite{McCullaghN89} and support vector machine~(SVM)~\cite{cortes1995support}; two tree-based algorithms: a decision tree~(DT)~\cite{Breiman1983ClassificationAR} and a gradient boosting~(GB)~\cite{10.1214/aos/1013203451} classifier; a $k$-nearest neighbors~(KNN) classifier~\cite{altman1992introduction}; \revision{two} neural network-based \revision{models}: a multi-layer perceptron~(MLP), in the form of three variants, concretely with 1~(MLP-1), 5~(MLP-5) and 10-hidden layers~(MLP-10)~\cite{neural_networks_article} \revision{and TabNet~\cite{tabnet}~(TN), a transformer-based deep learning model}.

\stitle{Regression}
\revision{Regression is a supervised learning method that uses labelled data to learn the dependency between features and a \emph{continuous} target variable.
There is a plethora of regression algorithms.}
To evaluate how error-prone different categories of regression algorithms are and how more complex algorithms within one category perform, we compare \revision{seven} of the most widely used approaches for regression out of three categories of regression algorithms.
We selected two linear-regression-based algorithms: a linear regression~(LR)~\cite{montgomery2021introduction}, and a ridge regression~(RR)~\cite{hoerl1970ridge}; three tree-based algorithms: a decision tree~(DT)~\cite{Breiman1983ClassificationAR}, a random forest~(RF)~\cite{breiman2001random} and a gradient boosting~(GB) regression; and again \revision{two} deep-learning based approach\revision{es}: \revision{a} multi-layer perceptron~(MLP) with 1~(MLP-1), 5~(MLP-5) and 10-hidden layers~(MLP-10)~\cite{neural_networks_article} \revision{and TabNet~\cite{tabnet}~(TN)}.

\stitle{Clustering}
\revision{Clustering algorithms aim to find groups of samples with similar features in a given dataset. 
As clustering is an unsupervised task, our clustering models are not trained but directly receive all data as test data for the prediction of the target label. 
Depending on dataset properties, such as dimensionality, distribution or data types, current research recommends different clustering algorithms. 
These can be grouped into different categories or algorithm families.}
We decided to use one algorithm from each of the five most commonly used categories of clustering algorithms~\cite{rokach2005clustering}: the Gaussian mixture clustering algorithm~\cite{reynolds2009gaussian} from the distribution-based family; the $k$-means~\cite{sinaga2020unsupervised} and the $k$-prototypes~\cite{ji2013improved} algorithms from the centroid-based family; the agglomerative clustering algorithm~\cite{day1984efficient} from the hierarchical family; the \textit{ordering points to identify cluster structure} (OPTICS) algorithm~\cite{ankerst1999optics} from the density-based family; and a deep autoencoder neural network from deep learning-based family~\cite{song2013auto}.

%% file: 50-setup.tex
\section{Experimental setup}
\label{sec:experimetal_setup}
\begin{figure}[!htb]
\centering
\includegraphics[width=.9\linewidth]{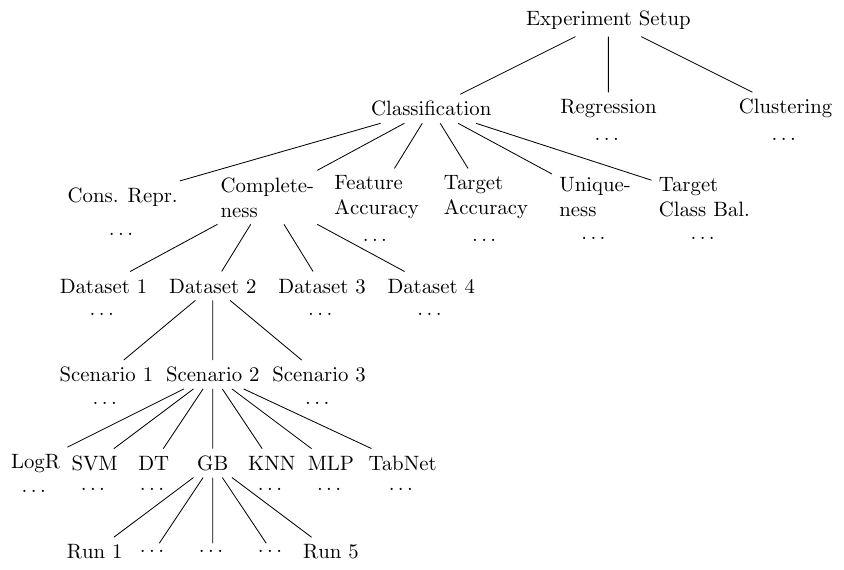}
\caption{\revision{Experimental setup yielding $(2\times6\times3\times3\times7\times5)+(1\times5\times3\times3\times7\times5)+(1\times6\times4\times1\times5\times5) = \numprint{4905}$ individual experiments.}}
\label{fig:experiment_setup_tree}
\end{figure}

This section gives an overview of our implementation and introduces our datasets together with the parameterization and performance measures of the analyzed ML models. The experiments sweep a wide scope of parameters as shown in Figure~\ref{fig:experiment_setup_tree}, including three tasks, six pollution types, \revision{four} datasets, three pollution scenarios (for clustering just one scenario), \revision{seven} approaches per task \revision{(five for clustering)}, and five runs to eliminate measurement variance \revision{(one for TabNet, to reduce extended training time)}. 
Multiplying, this yields \revision{\numprint{4905}} experimental runs to produce all results. 

\PaperLong{\subsection{Hardware}
We ran \revision{the majority of} our experiments on a DELTA D12z-M2-ZR machine. 
The server has AMD EPYC 7702P Xeon~(2.00GHz-3.35GHz, 64-Core) processor, 512\,GB DDR4-3200 DIMM RAM and runs Ubuntu 24.04 LTS Server Edition.
The server has two NVIDIA Quadro RTX GPUs~(5000/16\,GB, A6000/48\,GB).} 
\subsection{Implementation}\label{sec:scenarios}
The implementation of the polluters and ML pipelines are written in Python~3.10 using\PaperLong{the} scikit-learn\PaperLong{\cite{sklearn_main}}~1.3.2 and PyTorch\PaperLong{~\cite{pytorch_main}}~2.1.2\PaperLong{supporting NVIDIA CUDA~12.0~\cite{cuda_main} libraries}.
We evaluate the performance of the ML models in three scenarios: 
\begin{itemize}
    \item Scenario~1 -- polluted training set\label{text:scenario_1}
    \item Scenario~2 -- polluted test set\label{text:scenario_2}
    \item Scenario~3 -- polluted training and test sets\label{text:scenario_3}
\end{itemize}
These scenarios are considered only for classification and regression, as clustering does not have a separate training and test set.
To create the scenarios, we randomly split the data with a stratified~80:20 split into training and test sets. 

We then pollute the training and test sets separately, varying the ratio of pollution between~0 and~1 in increments of~0.1. For the consistent representation dimension, we tested~$k_{v}=2$ and~$k_{v}=5$, thus adding one or four new representations per~$v$ in the polluted dataset, respectively.
For uniqueness experiments, we varied~$\rho$ from~1 to~5 in steps to lead to linear quality decrease by~$0.1$ per step,~i.e.,~$\frac{10}{9},~\frac{10}{8}$, \ldots
\PaperLong{We stopped at a~$\rho$ of~5, because for lower quality~$\rho$ increases faster than linearly,~e.g., it would have to be~10 for a quality of~0.1, which would make the experiments much slower. 
Regarding duplicate count distribution functions, all samples receive a duplicate count of~1 and we used a normal distribution with mean~1 and standard deviation~5.}
Before applying any ML model, we one-hot encode the categorical features. 
Then, we measure the performance of the respective ML models, given the specific scenario for the specific dataset polluted with the specific polluter configuration.
We run each polluter configuration five times with a different random seed, obtaining five results per ML model in that setting, which we then aggregate by averaging.

For regression, we discretized the data with manually specified bin-step sizes before the stratified split. 
We use the discretized version also for the class balance and uniqueness polluters, as they require the target feature to consist of discrete classes. 
For all other polluters, we use the original continuous representation. The target feature distribution in our regression datasets is mostly close to a normal distribution. 

Applying the class balance polluter would lead to a heavily decreased dataset size of much less than~$50\%$ after balancing. 
That is why, when using the class balance polluter, we discard the discretized classes with very few samples, which would result in a balanced dataset with such a small size. 
To allow consistent comparisons with the original dataset, i.e., having the same classes contained in the data, we also discard those classes with few samples from the original dataset for the experiments with the class balance polluter.

\subsection{Datasets}
To investigate the correlation between the studied data quality dimensions and the chosen ML algorithms, we use the \revision{ten} datasets shown in Table~\ref{tab:datasets}.
We chose them for a variety of domains, sample sizes and characteristics. 
Our choice was also influenced by the ML task that the dataset is used for.
\begin{table*}[ht]
\centering
    \caption{Overview of the used datasets after pre-processing.}
    \begin{tabular}{l|r|r|r|r|r}
        \toprule
        Name & Samples & Features & Categorical & Numerical  & Classes \\
        \midrule
\multicolumn{6}{c}{Classification}\\   
        \midrule
        \textsf{Credit} & \numprint{1000} & 20 & 13 & 7 & 2 \\ 
        \hline
        \textsf{Contraceptive} & \numprint{1473} & 9 & 7 & 2 & 3 \\ 
        \hline
        \textsf{Telco} & \numprint{7032} & 19 & 16 & 3 & 2 \\ 
        \hline
        \revision{\textsf{COVID}} & \revision{\numprint{1025152}} & \revision{17} & \revision{16} & \revision{1} & \revision{2} \\ 
        \midrule
  \multicolumn{6}{c}{Regression}\\ 
        \midrule
        \textsf{Houses} & \numprint{1460} & 79 & 46 & 33 & - \\
        \hline
        \textsf{IMDB} & \numprint{5993} & 12 & 8 & 4 & - \\
        \hline
        \textsf{Cars} & \numprint{15157} & 8 & 3 & 5 & - \\
        \hline
        \revision{\textsf{COVID}} & \revision{\numprint{1025152}} & \revision{17} & \revision{17} & \revision{0} & \revision{-} \\ 
        \midrule
  \multicolumn{6}{c}{Clustering (all datasets were sampled)}\\ 
        \midrule
        \textsf{Bank} & \numprint{7500} & 3 & 2 & 1 & 6 \\ 
        \hline
        \textsf{Covertype} & \numprint{7504} & 54 & 44 & 10 & 7 \\ 
        \hline
        \textsf{Letter} & \numprint{7514} & 16 & 0 & 16 & 26 \\ 
        \hline
        \revision{\textsf{COVID}} & \revision{\numprint{25000}} & \revision{17} & \revision{16} & \revision{1} & \revision{4} \\ 
        \bottomrule
    \end{tabular}
    \label{tab:datasets}
\end{table*}

\stitle{Classification}
IBM's Telco Customer Churn~(\textsf{Telco}) dataset represents\PaperLong{7043} customers from a fictional telecommunications company~\cite{telco_dataset}.
\PaperLong{It contains personal information about customers~(e.g., gender and seniority) and their contracts~(e.g., type of contract and monthly charges).} 
The target variable \texttt{Churn} describes whether the customer cancelled their contract within the last month.
\PaperLong{We dropped the customerID sample because of its lack of information content for the classification task. The original German Credit dataset was donated to the UCI Machine Learning Repository in 1994 by Hofmann~\cite{credit_dataset}. 
Then a corrected version was introduced by Grömping who identified inconsistencies in the coding table and corrected them~\cite{credit_dataset_corrected}~(we use the corrected version).}
The \textsf{Credit} dataset contains\PaperLong{a stratified sample of~\numprint{1000}} credits between the years~1973 and~1975 from a southern German bank~\cite{credit_dataset_corrected}.
\PaperLong{It contains the personal data about people who applied for a credit~(e.g., marital status or age) and about the credit itself~(e.g., purpose or duration).} 
The target variable tells whether a customer \texttt{complied} with the conditions of the contract or not.

The \textsf{Contraceptive} dataset was part of the 1987 National Indonesia Contraceptive Prevalence Survey, asking non-pregnant married women about their contraceptive methods~\cite{contraceptive_dataset_uci}.
\PaperLong{The dataset consists of~\numprint{1473} samples containing personal information about the wife's and husband's education, age, the number of children and much more.}
The classification task is to determine the contraceptive method choice\PaperLong{out of \emph{No-use}, \emph{Long-term} and \emph{Short-term} with the target variable \texttt{Contraceptive method used}}.

\revision{The COVID-19 Dataset~(\textsf{COVID}) dataset, released by the Mexican government, contains anonymized data from patients either confirmed or suspected to be infected with the coronavirus~\cite{covid_dataset, covid_dataset_kaggle}.}
\PaperLong{The original dataset includes~21 features and~\numprint{1048576} patient records.
After performing data cleaning -- such as removing records with missing values --  we reduced the dataset to~17 features and~\numprint{1025152} records.}
\revision{The classification task is to predict patient \texttt{mortality}, i.e., whether a patient died during the hospital stay or afterwards (monitored by epidemiological surveillance units or health jurisdictions)~\cite{covid_dataset}.}


\stitle{Regression}
The \textsf{Houses} dataset\PaperLong{was created as a modern replacement of the widely used but outdated Boston Housing dataset~\cite{boston-house-prices}. 
Located in Ames, Iowa, it} contains features of houses in the city to determine their~\texttt{sales price}~\cite{ames-house-prices}. 
We use the dataset in the form it was presented in a Kaggle challenge~\cite{houses-dataset}.
\PaperLong{There, only the training set includes the sale prices, which is why we take the training set with~\numprint{1460} samples from the challenge. 
We removed the \texttt{Id} feature because its only purpose is to identify houses uniquely. 
This results in~79 features plus the target feature. 
There are missing values~(\texttt{NaN}) in five features, which we replaced by computationally usable placeholders. 
One numerical feature, the year a garage was built, contains information related to the categorical feature \texttt{garage type} and has missing values for observations where the \texttt{garage type} indicates that there is no garage anyway. 
In those cases, we set~$0$ as a placeholder for the garage year to represent the meaning in context with the garage type to distinguish them from the MCAR values that the polluter injects in our experiments. 
For the remaining four of the features with missing values, their occurrence is independent of other features. 
We treat them as MCAR values and represent them as placeholders outside the feature domains.
}

The \textsf{IMDB} dataset contains features and ratings for films and series that were retrieved from the IMDB website~\cite{imdb-dataset}.
\PaperLong{We excluded all samples with missing values in the \texttt{rating} attribute, as it serves as the target variable. 
For the remaining \numprint{5993} samples, we removed the \texttt{name} feature, as it merely identifies the films and series. 
This leaves~12 features, excluding the target feature, where two inherently numeric features contain missing values represented as text placeholders.
For the first feature, \texttt{duration}, the missing values are unrelated to other features.
To handle this, we assigned a numeric placeholder outside the domain, enabling the processing of these values as numerical rather than categorical.
For the second feature, \texttt{episodes}, there are only missing values for film elements. 
We do not count those as MCAR values because they are related to the fact that the element is a film and therefore contain information, which is why we set the placeholder to~0 here.

}
The \textsf{Cars} dataset collects listings for used cars of different manufacturers~\cite{cars-dataset}.
\PaperLong{It contains technical attributes of the cars, as well as \texttt{model}, \texttt{year}, and \texttt{tax}.} The~\texttt{sales price} is the feature that shall be predicted. 
The data is stored in files grouped by manufacturer. 
We use only the data of \textit{VW}\@.

\revision{We also used the \textsf{COVID} dataset, previously presented in the context of classification, for regression~\cite{covid_dataset, covid_dataset_kaggle}. 
In this case, the goal is to predict the \texttt{age} of the patients.}

\stitle{Clustering}
\label{subsubsection:clustering-datasets}
The \textsf{Letter} dataset contains\PaperLong{\numprint{20000} samples of different} statistical measures of character images~\cite{letter-dataset,uci-repo}.
\PaperLong{Each sample represents one of the 26 capital letters in the English alphabet, generated by randomly distorting images of the letters using approximately 20 different fonts.
The measured statistics were scaled, bounding the feature values to integers~$i$ within the range~$0 \leq i \leq 15$. 
Since this dataset lacks categorical features, the consistent representation polluter cannot be applied.
This is a deliberate decision, as the majority of existing clustering datasets do not contain categorical features, and we would like to examine the other data quality dimensions on a clustering-typical dataset.

}
The \textsf{Bank} dataset\PaperLong{was created through marketing campaigns of a banking institution conducted via phone calls. 
It} contains different characteristics of a person and has a binary target stating whether a term deposit is~\texttt{sub\-scribed} \cite{bank-dataset,moro2014data}.
\PaperLong{For clustering, however, having only two clusters is uncommon and, therefore not a representative use-case for our experiments.} 
We use a subset of this dataset, taking the education level as the target and keeping only three features related to it to have more than two clusters. 
We removed all classes with fewer than~\numprint{2000} samples to avoid significant data loss when applying the target class balance polluter.
The \textsf{Covertype} dataset\PaperLong{was created by the Colorado State University and} consists of descriptive information about forested areas\PaperLong{and thereby helps natural resource managers in their decision-making processes}~\cite{covertype-dataset,blackard1999comparative}.
\PaperLong{Each sample contains cartographic measures for an observation of a~30×30\,m cell. 
It includes numerical features such as \texttt{slope} or \texttt{elevation}, as well as categorical features like \texttt{wilderness area} or \texttt{soil type}, which are used to derive the \texttt{cover type},~i.e., the dominant forest type in the study area. 
Exemplary target classes are ``Spuce/Fir'' and ``Krummholz''.

} 

\revision{We also used the \textsf{COVID} dataset in the context of the clustering task, where the objective is to cluster patients based on their COVID-19 \texttt{test findings}~(three different degrees of severity or not infected)~\cite{covid_dataset, covid_dataset_kaggle}.
}

To evaluate clustering, we sampled all datasets as the last step of its preprocessing to reduce their size, as further processing would have required too much main memory.
\PaperLong{For the sampling process, a total of \numprint{7500} (for \textsf{COVID}, \numprint{25000}) samples were targeted for each dataset. 
To ensure a balanced dataset, an equal number of data points were selected for each class.} 
We slightly varied the number of samples selected to be a multiple of the dataset's class count larger than~\numprint{7500}.
Instead of sampling once, we decided to also use the same five random seeds used for any random operations in the polluters to create a total of five preprocessed datasets for each dataset. 
This choice was made to account for the potential impact the sampling may have on the data if done using only one seed. 

\subsection{Model Parameters}
This section documents our parameter settings. All parameters that are not mentioned are kept at their default values in scikit-learn.

\stitle{Classification} For LogR, we increase \textit{max\_iter} to~\numprint{2000} for better convergence. For a multi-class dataset, we fit a binary problem for each label. 
The maximum number of iterations for each MLP variant~(MLP-1, MLP-5, MLP-10) is~\numprint{1000}. 
For SVM, we use a linear kernel and scale the input data with the scikit-learn \textit{StandardScaler}\PaperLong{: the time to converge would otherwise make it infeasible given the large number of experiments we run}.
\revision{For TabNet (TN), we set the maximum number of iterations to~10.
Additionally we used used a patience value of~3 to stop earlier -- if the loss does not change further.}
\revision{For the COVID dataset, we used PyTorch for the experiments with MLP, SVM and KNN due to runtime considerations. 
In the case of MLP, we adjusted several hyperparameters, including reducing the maximum number of iterations to~10, as experiments showed no further change in loss beyond this point. 
Similar to TN, we introduced a patience value of three, along with a required loss rate change of~0.01.
For KNN and SVM, we chose the default values of scikit-learn.
}

\stitle{Regression} For LR and RF, we set \textit{n\_jobs} to~$-1$ to use all available processors. 
For all MLP variants, we increase the \textit{max\_iter} parameter that defines the maximum number of epochs to~\numprint{3000} \PaperLong{so that each MLP is able to converge when trained on the original datasets}.
\revision{Similar to the classification task, we set the maximum number of iterations for TabNet to~10, along with a patience value of three.}
\revision{For the COVID dataset, we again used PyTorch for MLP and also set the maximum number of iterations to~10, with a patience value of~3.}

\stitle{Clustering} The actual number of clusters is passed to the $k$-Means/$k$-Prototypes, Gaussian mixture, and agglomerative algorithms. 
If categorical features exist, the agglomerative algorithm uses Gower's distance measure~\cite{gower1971general}. 
We set the \textit{affinity} parameter to ``precomputed''~(a pre-calculated distance matrix is to be employed) and the \textit{linkage} parameter to ``average'', which defines the distance between two clusters as the average distance between all samples in the first and all samples in the second cluster. 
For OPTICS, we defined only the minimum cluster size parameter \textit{min\_cluster\_size}, which specifies how many samples need to be in a cluster for it to be classified as such, as~100.
\PaperLong{We chose this number based on a combination of experiments and knowledge about our data. 
We are certain that in our original datasets, each cluster contains substantially more samples than~100.} 
Our autoencoder is trained for~200 epochs and optimized using the Adam optimizer~\cite{kingma2014adam} with a learning rate of~0.003 and mean squared error loss. 
The dataset is split into~$80\%$ training and~20$\%$ test data and loaded in batches of~128 shuffled samples. 
For all models that need a random seed, we used~42 as a seed.

\subsection{Model Performance}
\label{subsec:performance-metrics}
\PaperLong{We briefly describe the respective measure chosen to evaluate the performance of the three different tasks.}

\stitle{Classification}
The most common performance metric in classification tasks is accuracy: the number of correct predictions over the number of total predictions. But accuracy can be misleading for imbalanced datasets.
For example, a majority class prediction baseline yields a~90\% accuracy on data that is naturally distributed among two classes with a ratio of~9:1.
Therefore, we use the $F_1$-\emph{score}: it better accounts for class imbalance, which does exist in our datasets.
For instance, the \textsf{Telco} and \textsf{Credit} datasets are unbalanced in their target classes, with a~70/30 split or worse.
Usually, the $F_1$-score is measured for a single target class, but as we do not make assumptions about the importance of the target class, we report the average of the $F_1$-scores over all target classes.

\stitle{Regression}
Mean squared error~(MSE) is the commonly used metric to evaluate regression algorithms. 
However, it is highly dependent on the data domain,~e.g., it is expected to be much larger for house prices than for movie ratings.
As this makes comparisons of algorithm performance across datasets difficult, we use the \emph{coefficient of determination}~$R^{2}$, which measures the fraction of variance in the data that is explained by the regression model~\cite{lewis2015applied}.
An~$R^{2}$ of~$1$ means that the model explains all variance, while a model that achieves an~$R^{2}$ of~$0$ is as good as one that always predicts the mean of the target feature regardless of the input.
If~$R^{2}$ is negative, the model's predictions are even less accurate than always predicting the mean.

\stitle{Clustering}
The mutual information~(MI) score describes how much information is shared between two clusters.
This metric applies only if the samples are grouped with the same others, regardless of the target labels.
We use an adapted version of MI called the \emph{adjusted mutual information}~(AMI)~\cite{NguyenEB09}.
This version corrects the MI score for random choice and normalizes its value to a range between~0 and~1, which is necessary as the MI score tends to increase as the number of clusters increases, regardless of the quality of the clustering produced~\cite{vinh2010information}. 

%% file: 60-results/60-overview.tex
\section{Results}
\label{sec:results}
We discuss our observations grouped by machine learning~(ML) task and data quality dimension. 
For all plots, the horizontal axis indicates the decreasing data quality~(training, test, or both) (Section~\ref{sec:ddimentions}) and the vertical axis indicates the increasing ML model performance metric~(Section~\ref{subsec:performance-metrics}). 
The data quality of the (unpolluted) baseline dataset is indicated by a dotted vertical line (``Original DQ''). 
For most polluters, the original and the baseline datasets are identical.
All values in the plots are an average of five runs for each algorithm per polluted dataset.

For the feature accuracy pollution, we plotted the average of the two metrics~$\mathit{cFAccuracy}(d)$ and~$\mathit{nFAccuracy}(d)$ described in Section~\ref{subsection:facc_def}. 
Also, due to the definition of consistent representation~(Section~\ref{sec:consistent_representation}), the data quality could increase again~(instead of decreasing) with a higher pollution level.
Thus, we add the degrees of pollution in the plots for such cases. 

We present and discuss the results for each of the scenarios introduced in Section~\ref{sec:scenarios}: Scenario~1 -- polluted training set; Scenario~2 -- polluted test set; and Scenario~3 -- polluted training and test sets.
\PaperShort{To limit the length of this paper, we include the plots for only one dataset per ML task.
The full set of results can be found in the extended work~\cite{ourtechreport}.
}
\input{60-results/61-classification}
\input{60-results/62-regression}
\input{60-results/63-clustering}

%% file: 60-results/61-classification.tex
\subsection{Classification}
\label{subsection:classification_results}

\PaperShort{\input{Latex_Figure/classification/summary_telco}}
We discuss the effect of polluting data along the six data quality dimensions of \revision{four} datasets, namely \textsf{Credit}, \textsf{Contraceptive}, \textsf{Telco}, \revision{and \textsf{COVID}}, on the performance of \revision{seven} classification algorithms, namely LogR, SVM, DT, GB, KNN, \revision{TN,} and MLP. 
We consider MLP in the form of three variants: MLP-1, MLP-5 and MLP-10\@.
\PaperLong{To better understand the behavior of the studied classification models,} we include two baselines trained and tested on clean data, namely a \textit{majority-class classifier} and a \textit{class ratio classifier}.
The first assigns all test dataset instances to the majority class in the training dataset.
The latter selects a label based on the labels' ratios in the training dataset.
The target accuracy/balance pollution would shift the class ratios of the training data, which is not reflected in the baseline performance because we report only on clean training and test data.
Only for \textsf{Contraceptive}, the classifiers are not binary.
\PaperShort{\revision{Figure~\ref{fig:classification-results-all-telco} shows the results only for \textsf{Telco}. 
However, in the description of the result per data quality dimension, respectively data pollution, we also include the results for the other considered datasets.}}

\revision{Although we do not focus on optimizing the prediction accuracy in this study, the results \PaperShort{in Figure~\ref{fig:classification-results-all-telco}} in the context of TN mostly show a significantly lower performance than the other considered classification algorithms. 
We see this effect for the smaller datasets.
As a transformer-based model, TN seems to be more data-hungry and requires larger datasets to fully exploit its potential~(\citet{tabnet} also show this trend in a dedicated experiment). 
However, for \textsf{COVID}, which is much larger than the other considered datasets, TN achieves similar performance compared to the other classification algorithms.}

\stitle{Consistent Representation} 
\label{subsubsection:classification_results_consistent}
Introducing new representations of the categorical values of the training dataset has a limited impact on the performance of the studied classification algorithms on all datasets\PaperLong{(see the first and third rows of Figure~\ref{fig:classification-results-all-ConsistentRepresentationk5})}\PaperShort{, as we see in Figure~\ref{fig:classification-results-all-ConsistentRepresentationk5-1-telco} for \textsf{Telco}}.
\revision{However, TN is slightly more erratic than the other algorithms.
Considering \textsf{Telco}, performance drops by almost 10 percentage points~(\pt) at 20\% pollution.
Later, at a pollution of at least 45\%, the performance increases, then suddenly drops again by around 8\pt.}
For Scenario~2\PaperLong{, in which the model is trained on original (clean) data and then is supposed to classify polluted ``real-world'' data}, we observe a slight and slow decrease in the performance of all algorithms with the decrease in test data consistency. 
Comparing \PaperShort{Figures~\ref{fig:classification-results-all-ConsistentRepresentationk5-2-telco} and~\ref{fig:classification-results-all-ConsistentRepresentationk5-3-telco}}\PaperLong{the second and third rows of Figure~\ref{fig:classification-results-all-ConsistentRepresentationk5}}, we can observe a better performance when both training and testing datasets suffer from the same inconsistent representations. 
In particular, MLP-1 shows a significantly higher susceptibility for pollution in Scenario~2 than MLP-5 and MLP-10.
Both variants~(MLP-5, MLP-10) show an almost constant performance even after~50\% pollution.
\PaperLong{\input{Latex_Figure/classification/Consistent_Representation_5}}
\revision{Also TN shows a constant performance, with only an initial fluctuation up to a pollution of 15\%.}
Considering inconsistent representations with only two representations per original value, we note less resilience by some of the algorithm's and a clear decrease in their performance after polluting~50\% of the values in Scenarios~1 and~2\PaperLong{(see Figure~\ref{fig:classification-results-all-ConsistentRepresentation})}. 

\stitle{Completeness} 
Our intuition was that compromising the completeness of the training data in Scenario~1 leads to classifiers that are increasingly biased towards the imputed placeholder values due to their sheer number.
\PaperLong{The first row in Figure~\ref{fig:classification-results-all-completeness}}\PaperShort{However, the results} show a surprisingly limited decline in the $F_1$-score, suggesting that the models are affected but not biased. 
\revision{The only exceptions on \textsf{Telco} are the SVM model, which drops drastically in performance once more than half the dataset is polluted, and the GB model, which also shows a significant performance drop after a pollution level of 80\%}, as \PaperShort{Figure~\ref{fig:classification-results-all-completeness-1-telco}}\PaperLong{\input{Latex_Figure/classification/Completeness}} shows.
In contrast, in Scenario~2\PaperShort{(see Figure~\ref{fig:classification-results-all-completeness-2-telco})}\PaperLong{(shown in the second row in Figure~\ref{fig:classification-results-all-completeness})}, the degradation in prediction performance is faster and ends below the performance of the majority class baseline.

Comparing the results of Scenario~3 to the other scenarios\PaperLong{in Figure~\ref{fig:classification-results-all-completeness}}, we notice for all datasets that the risk of classifying incomplete test data seems to be less if the training has already been carried out on incomplete data.

\stitle{Feature Accuracy} 
Training the classifiers on noisy data in Scenario~1 has a non-negligible impact on their performance that varies by the dataset\PaperLong{(as Figure~\ref{fig:classification-results-all-FeatureAccuracy} shows)}.
For \textsf{Contraceptive}\PaperLong{(see Figure~\ref{fig:classification-results-all-FeatureAccuracy-1-contra})}, the algorithms show a certain robustness until the quality reaches a threshold of~0.8, where the performance degrades more steeply and eventually falls below baseline performance between a quality of~0.4 and~0.2.
A similar robustness can be noticed for \textsf{Credit}\PaperLong{in Figure~\ref{fig:classification-results-all-FeatureAccuracy-1-credit}}, except for MLP\revision{-1}, which perform\revision{s} much worse~($>$10\pt drop in $F_1$-score) after introducing only a small amount of noise to the features.
The linear models~(SVM and LogR) and GB as a tree-based model seem very robust to degrading feature accuracy up to a certain point, where they suddenly lose performance rapidly until they meet the majority class baseline.
For the \textsf{COVID} dataset, we observe an initial sharp decline in performance across all algorithms as the feature accuracy decreases, reaching a stable $F_1$-score once the quality drops to 0.6.
Beyond this point, the $F_1$-score remains relatively constant until the quality falls below 0.3.
\PaperLong{The differences in the datasets do not allow us to make general statements besides that initial robustness. 
For more insights, one could investigate the influence of the separate feature accuracy qualities~(for categorical and numerical values) as the combination of them does not allow us to reason about their individual contribution to the overall loss in performance.}\PaperLong{\input{Latex_Figure/classification/Feature_Accurecy}}

For Scenario~2\PaperShort{(see Figure~\ref{fig:classification-results-all-FeatureAccuracy-2-telco})}\PaperLong{(see second row in Figure~\ref{fig:classification-results-all-FeatureAccuracy})}, the performances linearly decrease with reduced feature quality of the test data, while staying above baseline performance most of the time.
Reducing the quality of the test data to~0.5 causes a drop of about~10\pt in $F_1$-score for the linear models.
\revision{TN shows a slight upward trend in performance, with a 10\pt increase between the original performance and the performance at 100\% pollution.}
\PaperLong{There is almost no significant difference in performance for the KNN classifier in Figure~\ref{fig:classification-results-all-FeatureAccuracy-2-credit} which seems to be a dataset specific finding. 
Generally, our findings suggest that these algorithms are rather robust against reduced feature accuracy on the tabular data that we tested with.} 

For Scenario~3\PaperShort{(see Figure~\ref{fig:classification-results-all-FeatureAccuracy-3-telco})}\PaperLong{(see first and third row in Figure~\ref{fig:classification-results-all-FeatureAccuracy})}, we observe a similar behavior as in Scenario~1, but all algorithms show an increase in their performance for low feature accuracy quality.

\stitle{Target Accuracy} 
The target accuracy is especially relevant in the classification task, as it simulates labeling errors/noise in the data.
\PaperLong{Scenario~1 in the first row in Figure~\ref{fig:classification-results-all-TargetAccuracy} could depict a real life situation in which the training labels were collected by crowd workers~(noisy and inconsistent) and the test~labels were carefully handpicked by experts~(as close to the ground truth as possible). 
Therefore, the results show the connection between labeling errors in the training set and loss in real-world performance of the classifier.}
For Scenario~1, there is almost a linear decline in performance for \textsf{Contraceptive} and \textsf{Credit} in response to decreasing the training dataset target accuracy\PaperLong{, as  Figures~\ref{fig:classification-results-all-TargetAccuracy-1-contra} and~\ref{fig:classification-results-all-TargetAccuracy-1-credit} show}.
However, the performance of the MLP variants drops relatively faster and is more erratic than that of the other algorithms.
\revision{TN generally follows the trend of the other algorithms but is slightly more erratic considering the \textsf{Contraceptive} dataset.
With \textsf{COVID}~(see Figure~\ref{fig:classification-results-all-TargetAccuracy-1-covid}), all algorithms show a sharp drop in performance at 50\% pollution.}
Same applies for \textsf{Telco}, which shows in Figure~\ref{fig:classification-results-all-TargetAccuracy-1-telco} a similar behavior for \revision{all algorithms, except KNN and DT, which steeply decline in performance around a target accuracy of~0.5. 
Conversely, KNN and DT follow a more linear performance pattern.}

For all the datasets and all ML algorithms, we found that once the target accuracy of the training data is equal to or worse than 1 divided by the number of classes, then the prediction performance is below that of the class ratio baseline classifier. 
In addition, we note that up to~20\% of training labels could be flipped without a performance' losses of no more than~10\pt in $F_1$-score for most of the algorithms. 
The performance of all considered MLP variants and SVM\PaperLong{in Figures~\ref{fig:classification-results-all-TargetAccuracy-1-credit} and~\ref{fig:classification-results-all-TargetAccuracy-1-telco}} also showed a very high variance across the five different seeds, which indicates that they are the most sensitive to incorrectly labeled samples.

The results of Scenario~2 show a linear trend and is consistent across all datasets, with only slight differences in the slopes of the linear trend. 
20\% more incorrectly labeled samples in the test set leads to a prediction accuracy decrease of up to~10\pt, which shows \revision{the importance of carefully} labeling the test set.
\revision{TN is the exception for \textsf{Credit}, with a slightly increasing slope during pollution.}
\PaperLong{If the test set has many mislabeled samples, then this could result in scrapping the model as it performs below the baseline even though it might perform better for real-workd data~(see the cross-section of model performances and baselines,~e.g., in Figure~\ref{fig:classification-results-all-TargetAccuracy-2-telco}).}\PaperLong{\input{Latex_Figure/classification/Target_Accurecy} The second scenario in the second row in Figure~\ref{fig:classification-results-all-TargetAccuracy} shows a situation where the training data was labeled cautiously, but the test data contains mislabeled samples.}

For Scenario~3\PaperShort{(see Figure~\ref{fig:classification-results-all-TargetAccuracy-3-telco})}\PaperLong{(see third column in Figure~\ref{fig:classification-results-all-TargetAccuracy})}, we observed a similar behavior of the algorithms as in Scenario~1, with one difference: After polluting~1 divided by the number of classes of the samples both in the training and test data, the $F_1$-score starts to increase at different slopes for all algorithms regardless of the dataset.
\revision{This trend is mainly due to the fact that for datasets with two classes half of the labels have already been flipped at 50\% pollution, leading to fewer misclassifications by the algorithms.}

\stitle{Uniqueness} 
For all datasets, uniqueness does not have much of an impact on the performance of all classifiers in all scenarios,\PaperShort{as \revision{the} Figures~\ref{fig:classification-results-all-Uniqueness_dc1-1-telco}, \ref{fig:classification-results-all-Uniqueness_dc1-2-telco}, and~\ref{fig:classification-results-all-Uniqueness_dc1-3-telco} show.}\PaperLong{as the Figures~\ref{fig:classification-results-all-Uniqueness_dc1} show.}  
\revision{One of the} largest drop\revision{s} in the~$F_1$-score is observed for MLP-1 on \textsf{Credit} \revision{for Scenario~1 and~3}. 
This drastic drop in performance is attributed to the size of \textsf{Credit}, only~\numprint{1000} records, where generalization is already difficult enough, so introducing duplicates gives too much weight to single instances. 
On the one hand, it is interesting that MLP-1 performance already drops with~5\% pollution in Scenarios~1 and~3 on \textsf{Credit}\PaperLong{(see Figures~\ref{fig:classification-results-all-Uniqueness_dc1-1-credit} and \ref{fig:classification-results-all-Uniqueness_dc1-3-credit})}. 
This suggests that deduplication is an important pre-processing step before training an MLP on a small dataset. 
On the other hand, the results on the other datasets with linear models suggest that exact duplicates in both training and testing data do not significantly decrease classification performance.
\revision{However, we see a strong effect on the performance of SVM in the \textsf{COVID} dataset.
With a pollution of 20\%, there is a sudden change in performance, with the $F_1$-score briefly falling by over 15\pt.
The influence of duplicates on SVM may be more significant in large datasets, as SVM prioritizes the duplicated points rather than focusing on a broad distribution of points, causing the decision boundary to center more around these points than the overall data.}\PaperLong{\input{Latex_Figure/classification/Uniqueness}}

\stitle{Target Class Balance} 
\PaperLong{
In Figure~\ref{fig:classification-results-all-ClassBalance}, training begins on a balanced dataset (quality of~1.0), and as the quality decreases, the imbalance in the target variable progressively shifts toward the original majority class.
It is worth emphasizing that this shift toward the original majority class explains the performance improvement observed in most algorithms up to a certain point~(e.g., a quality of approximately~0.25 in Figure~\ref{fig:classification-results-all-ClassBalance-1-telco}).}
In Scenarios~1 and~3\PaperLong{(first and third row in Figure~\ref{fig:classification-results-all-ClassBalance})}, once the imbalance affects more than half the samples for the binary classification datasets, all algorithms' performance slowly drops towards the performance of the majority class baseline because they are trained on only a handful of samples from the minority class and therefore have no chance to actually learning the patterns of this class.
\revision{With \textsf{COVID}, this drop occurs later -- when over 90\% of the samples are polluted.
The performance then drops abruptly by over 10\pt for some algorithms, such as MLP-1 and TN.
SVM is an outlier in performance and shows erratic behavior even in the early stages of pollution.}\PaperShort{
Figures~\ref{fig:classification-results-all-ClassBalance-1-telco} and~\ref{fig:classification-results-all-ClassBalance-3-telco} show the described behavior on \textsf{Telco}.}
All the algorithms behave similarly until this point, with only a few exceptions~(mainly SVM and \revision{TN}). 
This suggests that the training data does not have to reflect the actual real-world class balance, as long as it is equally or more balanced. 
\PaperLong{The training data serves as the pivot in determining model robustness. 
If only the test data is imbalanced while the training data remains balanced (Scenario~2), the algorithms tend to show greater stability. 
Conversely, training a model on data imbalanced data, results in ML model robustness that is more dataset-specific.
For instance, as Figures~\ref{fig:classification-results-all-ClassBalance-1-credit} and~\ref{fig:classification-results-all-ClassBalance-3-credit} show, the performance of ML algorithms is more erratic when using the \textsf{Credit} dataset compared to the other datasets.}\PaperLong{\input{Latex_Figure/classification/Class_Balance}}

For Scenario~2\PaperShort{(see Figure~\ref{fig:classification-results-all-ClassBalance-2-telco}),}\PaperLong{(see second row in Figure~\ref{fig:classification-results-all-ClassBalance}), which reflects the real-world situation of training a ML model on an initial batch of real-world data and once the model reaches production, there is a distribution shift in the test data that enters the pipeline.} we observe that the models are \revision{quite} robust against the distribution shift, which moves towards a balance in class frequency, and they sometimes even increase their performance\PaperLong{(only the KNN in Figure~\ref{fig:classification-results-all-ClassBalance-2-credit} decreases significantly)}.
\revision{Only TN using \textsf{Telco} and SVM using \textsf{COVID} show a behavior against the trend of the other algorithms.}
\PaperLong{Similar to Scenario~1, the performance on class imbalance past the original class balance is dataset-dependent and, for example, much steeper for \textsf{Telco} than for the other two datasets, as we see in Figure~\ref{fig:classification-results-all-ClassBalance-2-telco}.} 

%% file: Latex_Figure/classification/summary_telco.tex
\begin{figure*}[!htbp]
    \centering
\begin{adjustbox}{minipage=\linewidth}
\begin{subfigure}[b]{0.32\textwidth}
\includegraphics[width=\textwidth]{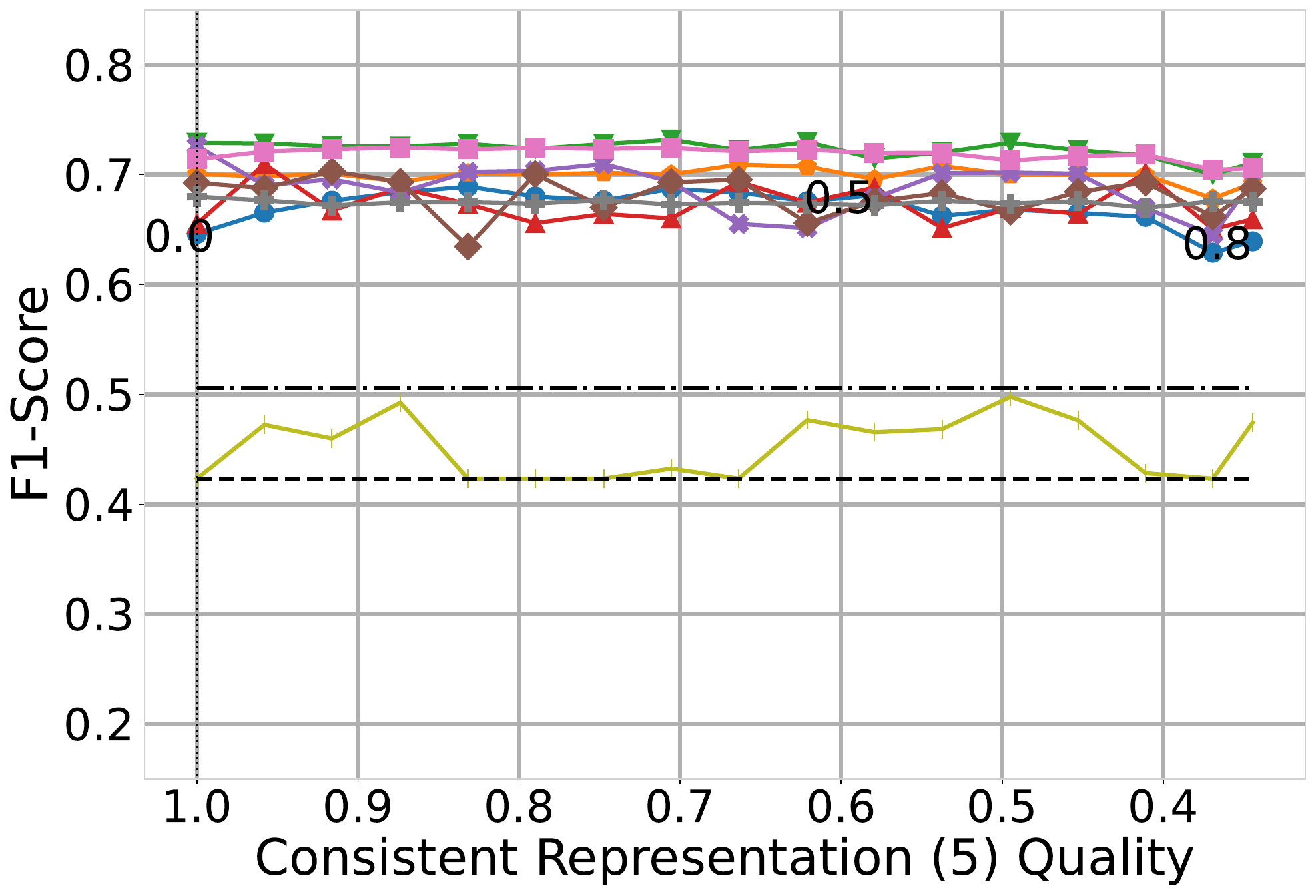}
\caption{Consistency $k_v = 5$ (Sce. 1)}
\label{fig:classification-results-all-ConsistentRepresentationk5-1-telco}
\end{subfigure}
\begin{subfigure}[b]{0.32\textwidth}
\includegraphics[width=\textwidth]{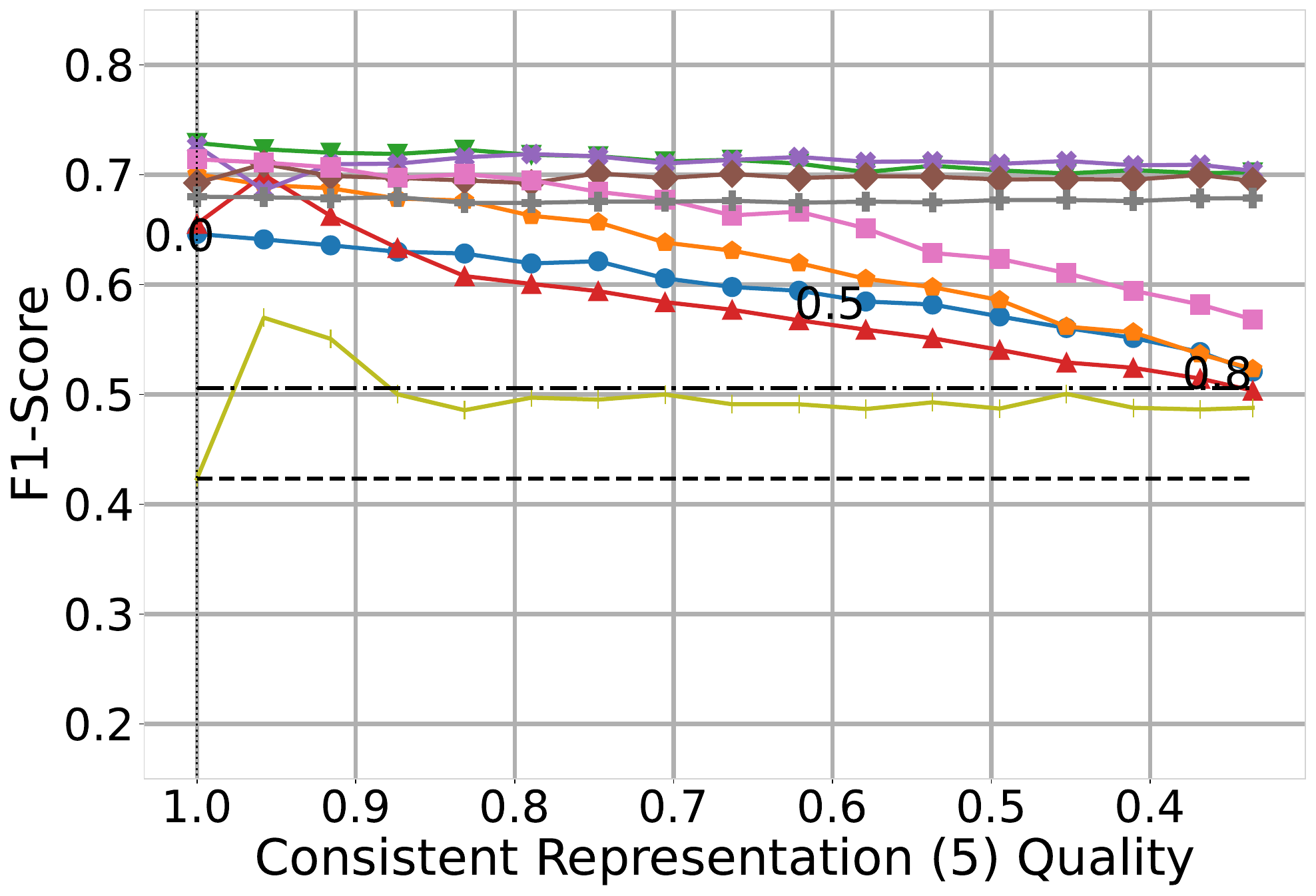}
\caption{Consistency $k_v = 5$ (Sce. 2)}
\label{fig:classification-results-all-ConsistentRepresentationk5-2-telco}
\end{subfigure}
\begin{subfigure}[b]{0.32\textwidth}
\includegraphics[width=\textwidth]{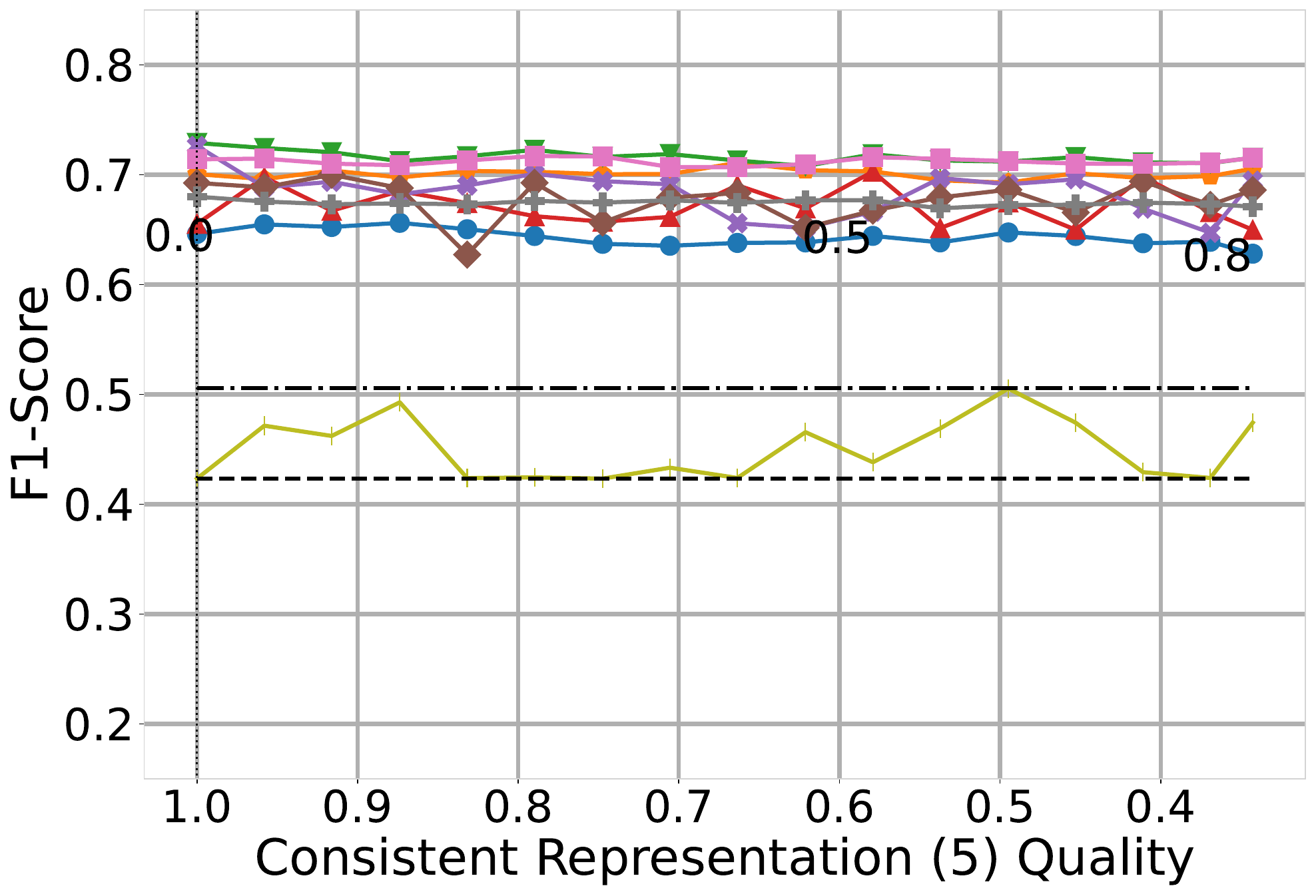}
\caption{Consistency $k_v = 5$ (Sce. 3)}
\label{fig:classification-results-all-ConsistentRepresentationk5-3-telco}
\end{subfigure}

    \begin{subfigure}[b]{0.32\textwidth}
        \includegraphics[width=\textwidth]{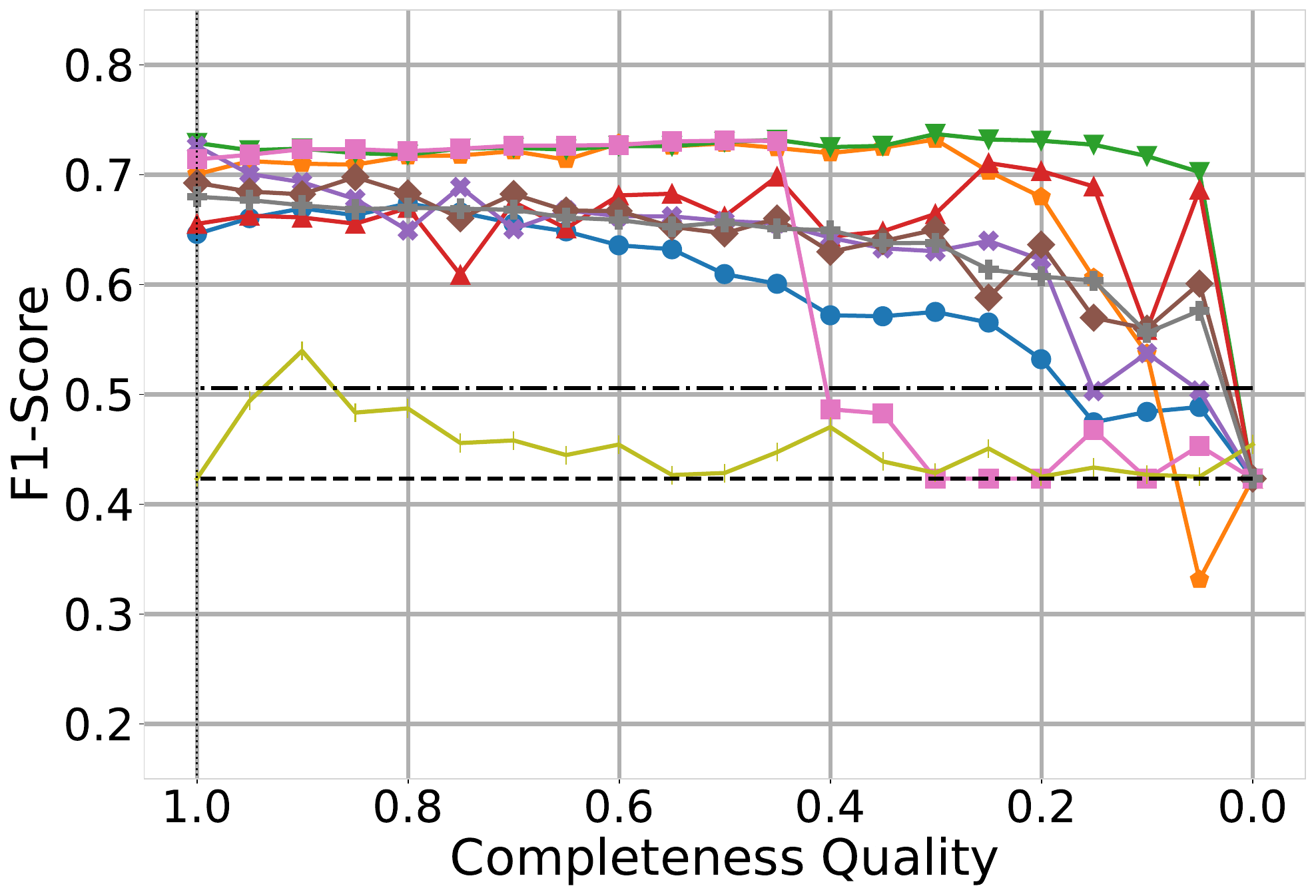}
        \caption{Completeness (Sce. 1)}
        \label{fig:classification-results-all-completeness-1-telco}
    \end{subfigure}
    \begin{subfigure}[b]{0.32\textwidth}
        \includegraphics[width=\textwidth]{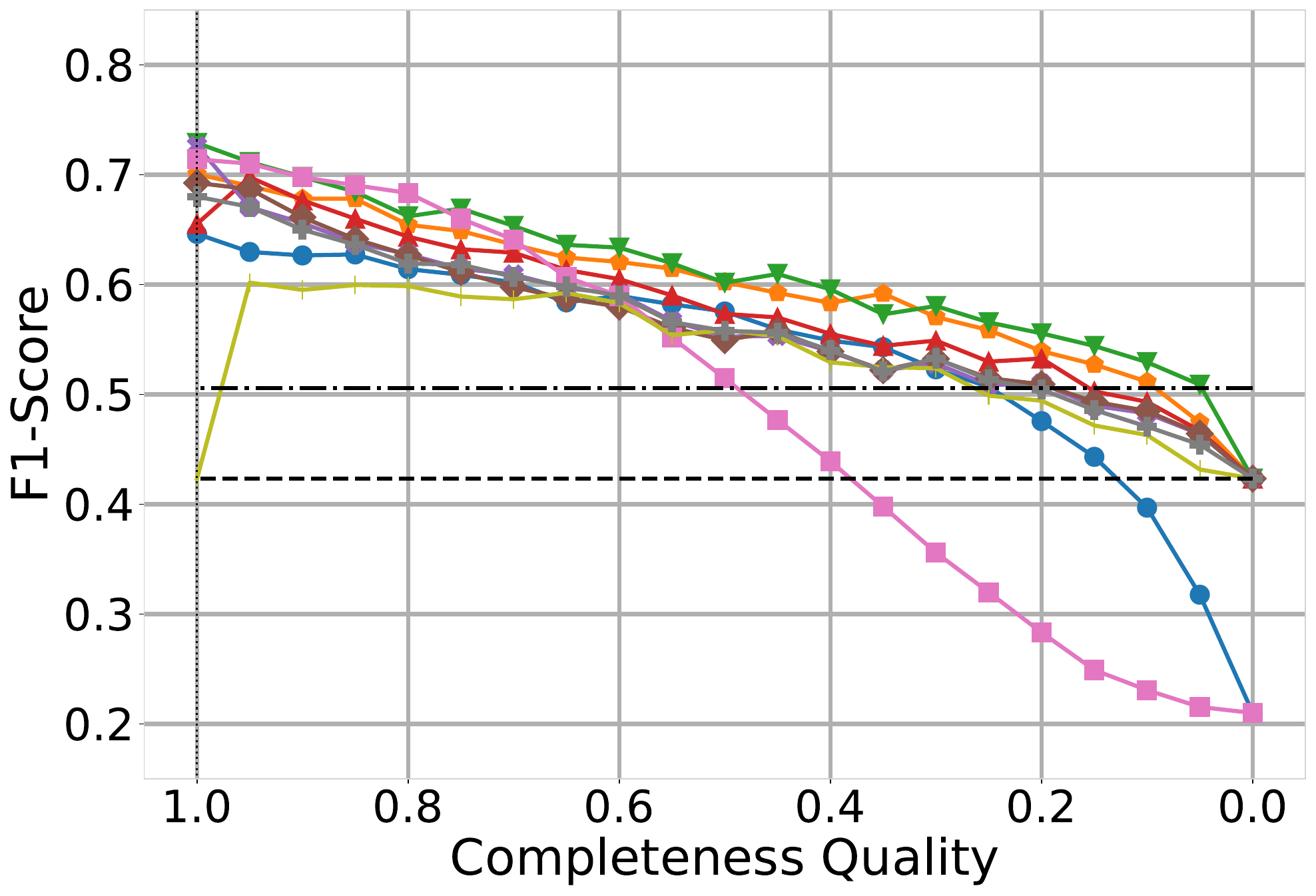}
        \caption{Completeness (Sce. 2)}
        \label{fig:classification-results-all-completeness-2-telco}
    \end{subfigure}
    \begin{subfigure}[b]{0.32\textwidth}
        \includegraphics[width=\textwidth]{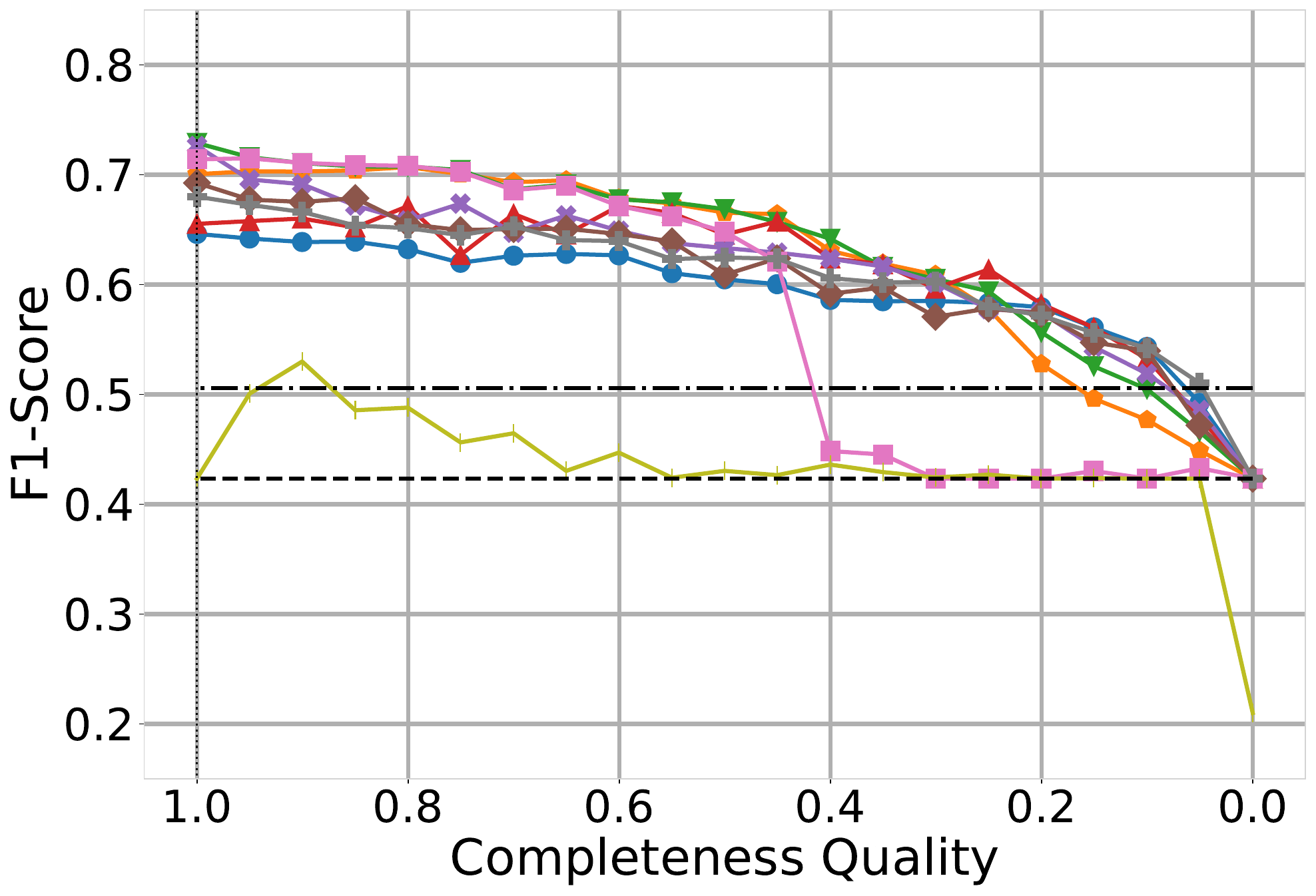}
        \caption{Completeness (Sce. 3)}
        \label{fig:classification-results-all-completeness-3-telco}
    \end{subfigure}

    \begin{subfigure}[b]{0.32\textwidth}
        \includegraphics[width=\textwidth]{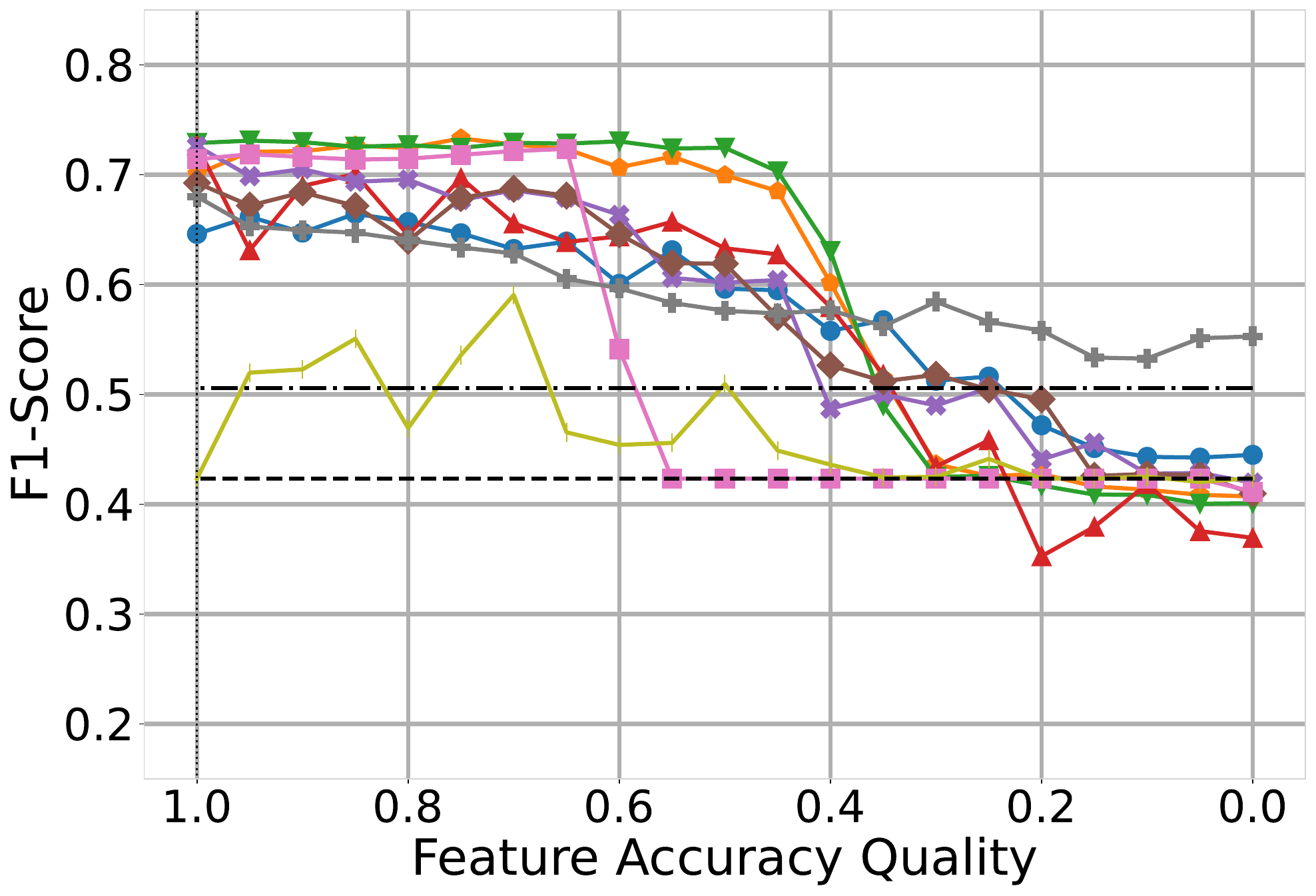}
        \caption{Feature Accuracy (Sce. 1)}
        \label{fig:classification-results-all-FeatureAccuracy-1-telco}
    \end{subfigure}
    \begin{subfigure}[b]{0.32\textwidth}
        \includegraphics[width=\textwidth]{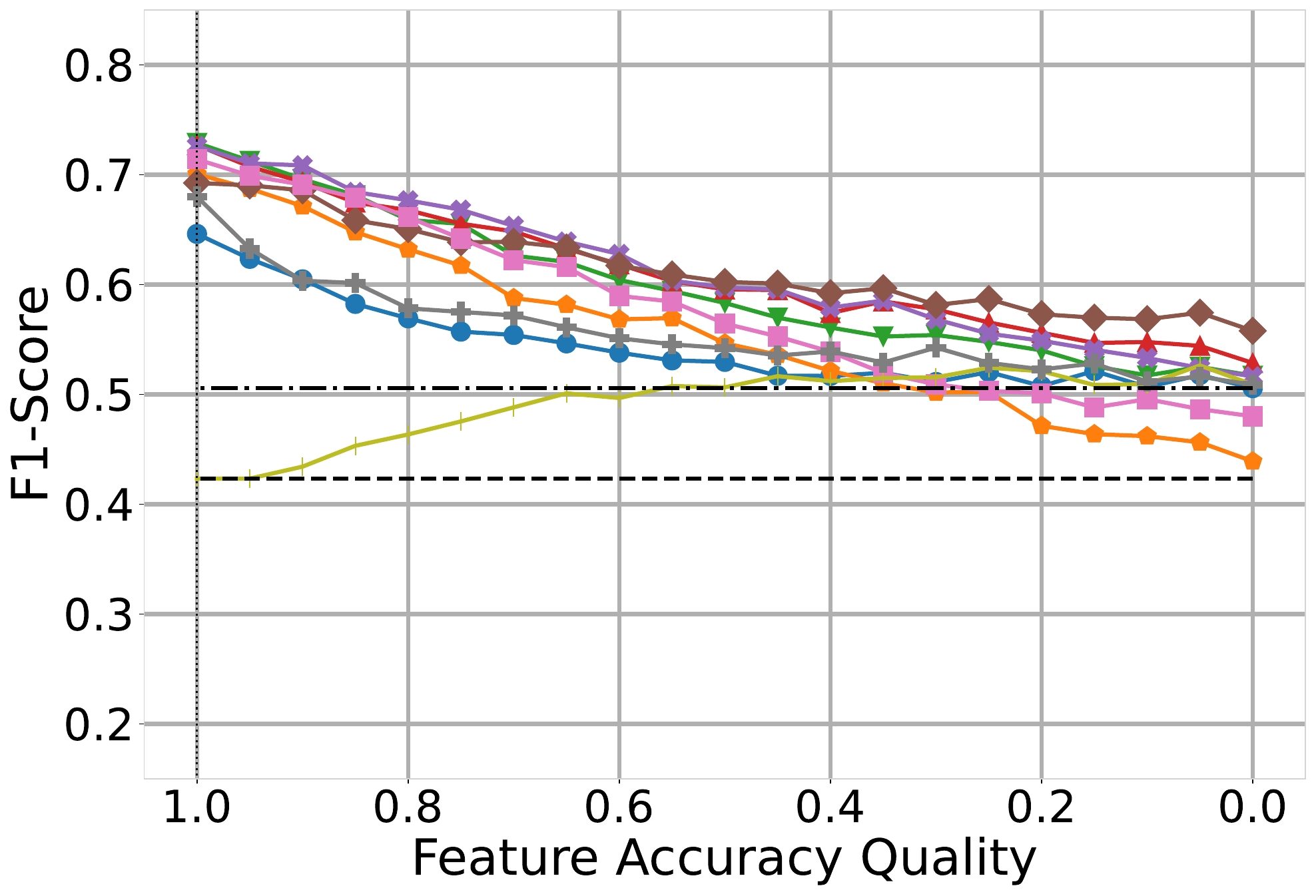}
        \caption{Feature Accuracy (Sce. 2)}
        \label{fig:classification-results-all-FeatureAccuracy-2-telco}
    \end{subfigure}
    \begin{subfigure}[b]{0.32\textwidth}
        \includegraphics[width=\textwidth]{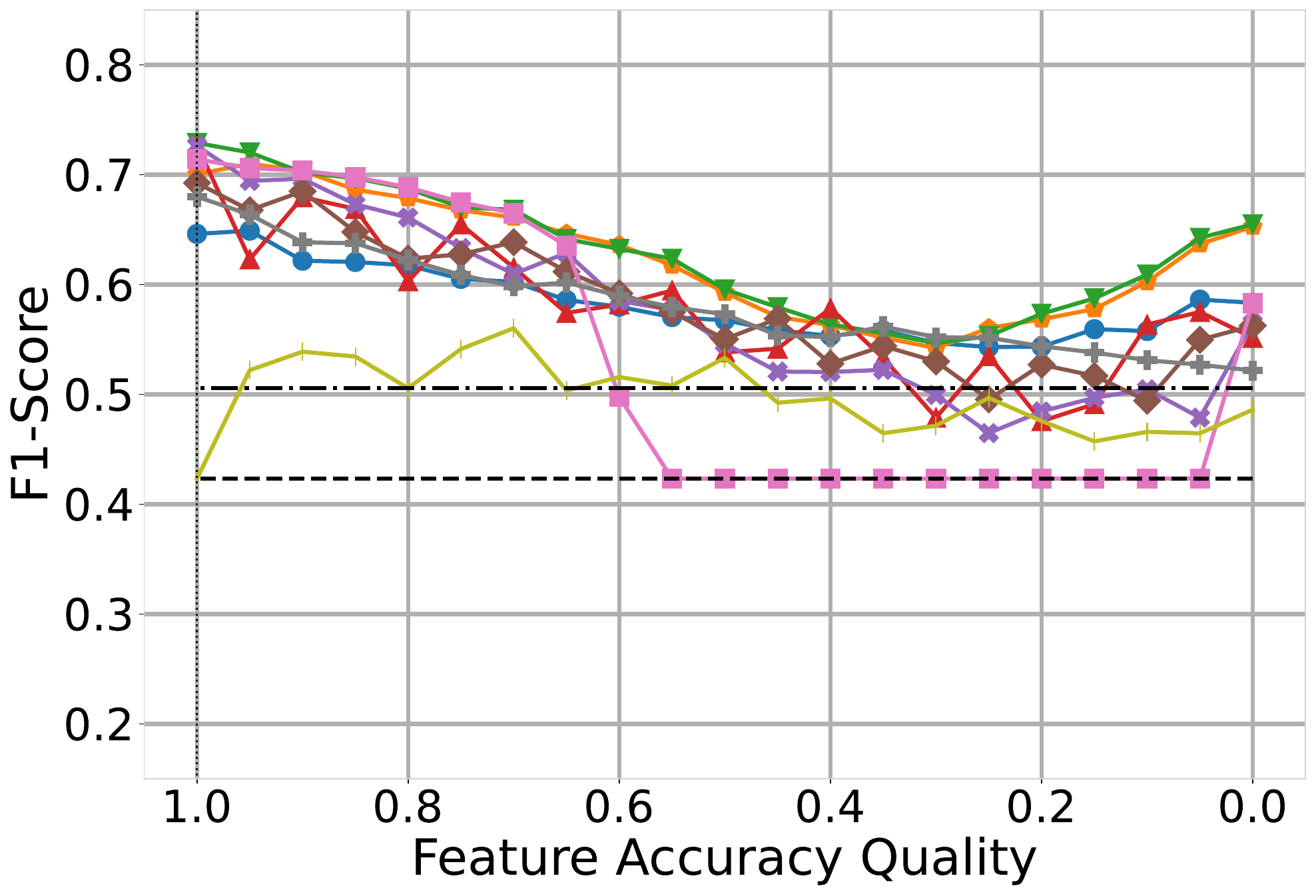}
        \caption{Feature Accuracy (Sce. 3)}
        \label{fig:classification-results-all-FeatureAccuracy-3-telco}
    \end{subfigure}

    \begin{subfigure}[b]{0.32\textwidth}
        \includegraphics[width=\textwidth]{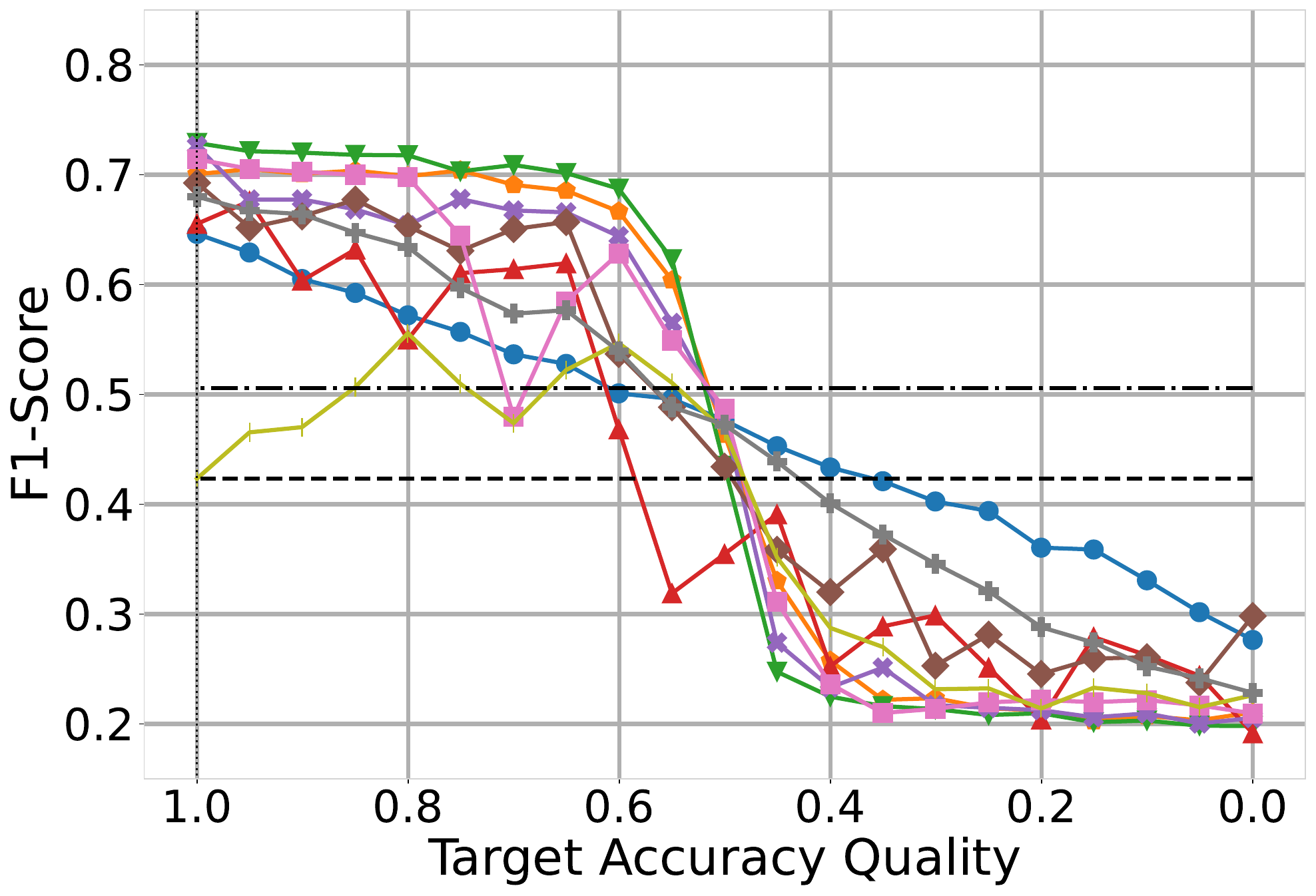}
        \caption{Target Accuracy (Sce. 1)}
        \label{fig:classification-results-all-TargetAccuracy-1-telco}
    \end{subfigure}
    \begin{subfigure}[b]{0.32\textwidth}
        \includegraphics[width=\textwidth]{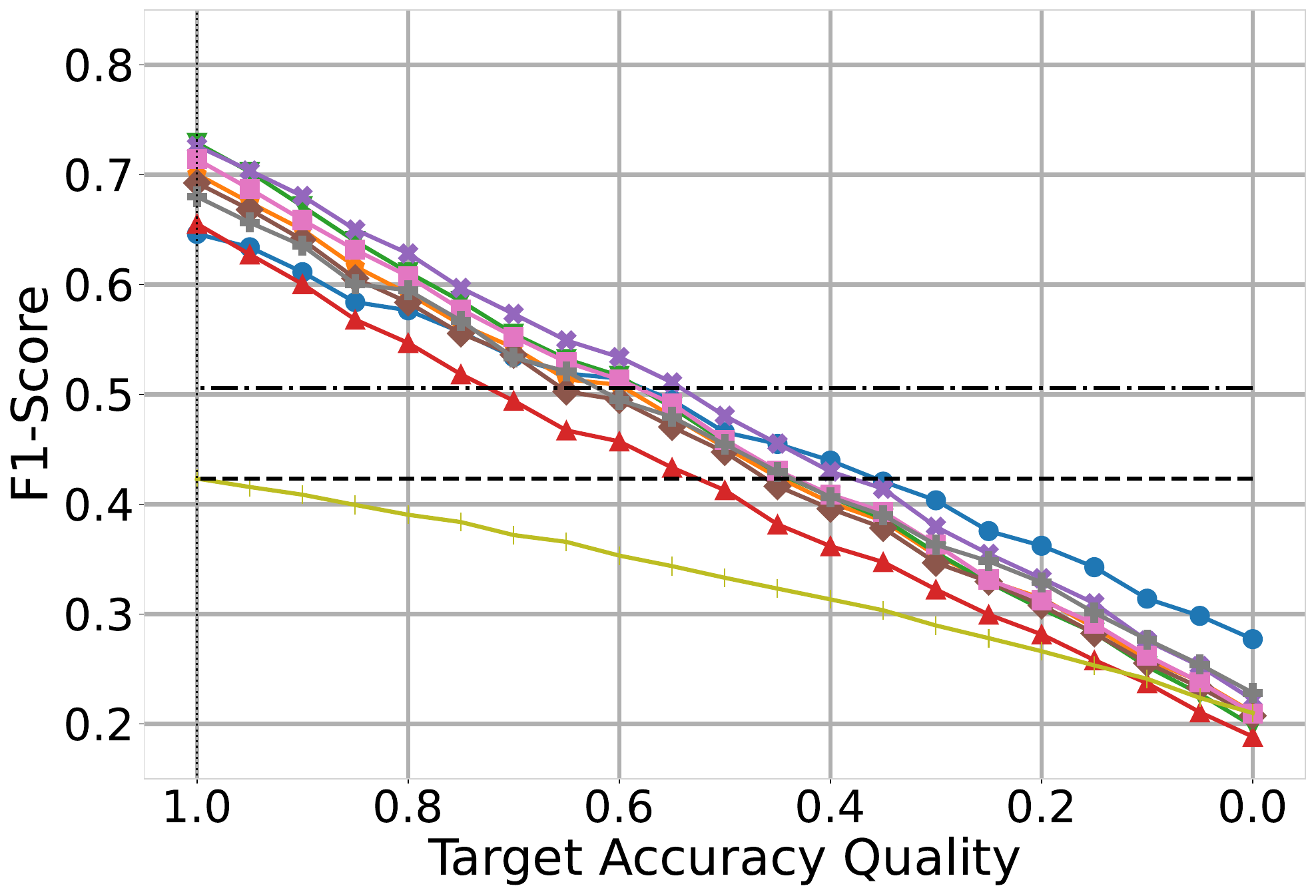}
        \caption{Target Accuracy (Sce. 2)}
        \label{fig:classification-results-all-TargetAccuracy-2-telco}
    \end{subfigure}
    \begin{subfigure}[b]{0.32\textwidth}
        \includegraphics[width=\textwidth]{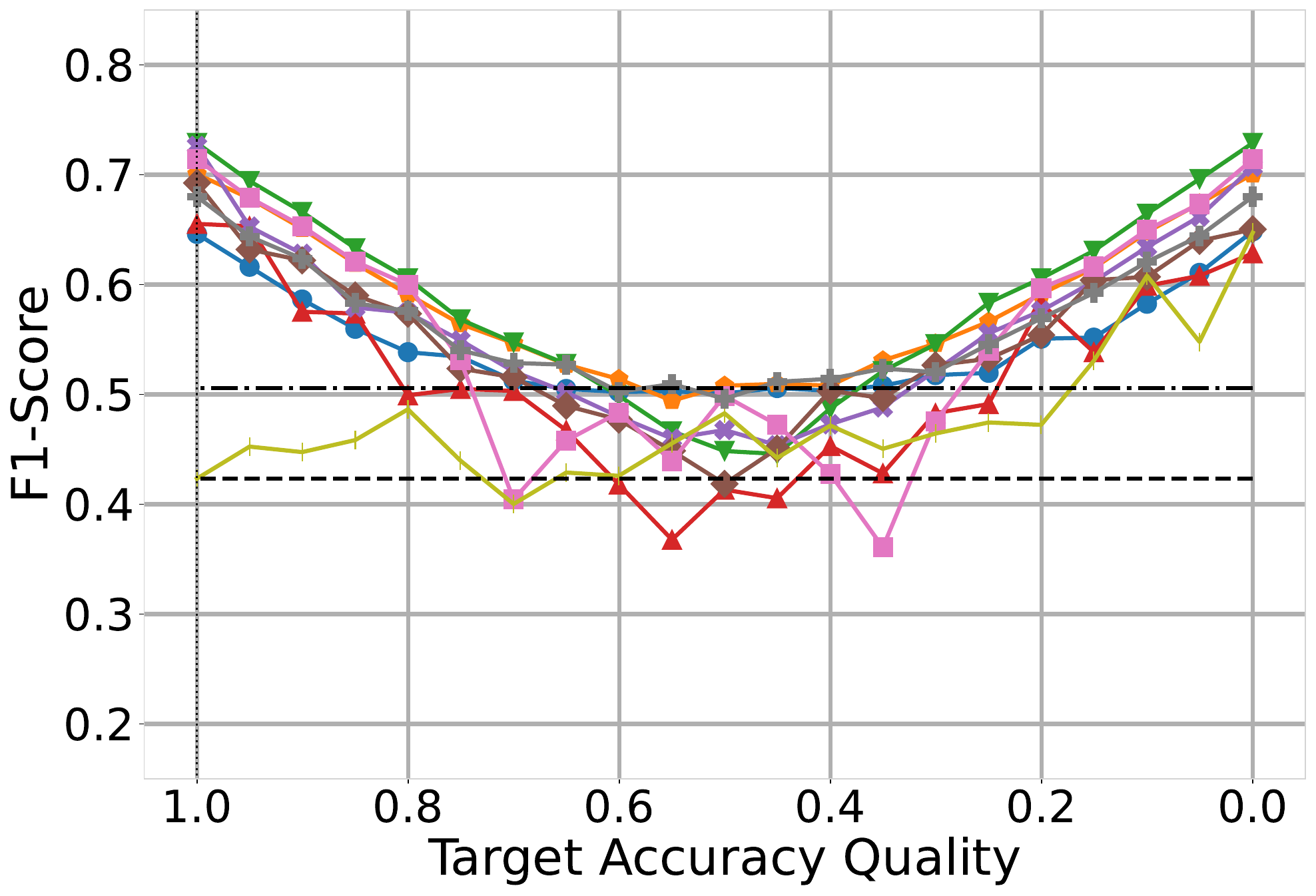}
        \caption{Target Accuracy (Sce. 3)}
        \label{fig:classification-results-all-TargetAccuracy-3-telco}
    \end{subfigure}
    
    \begin{subfigure}[b]{0.32\textwidth}
        \includegraphics[width=\textwidth]{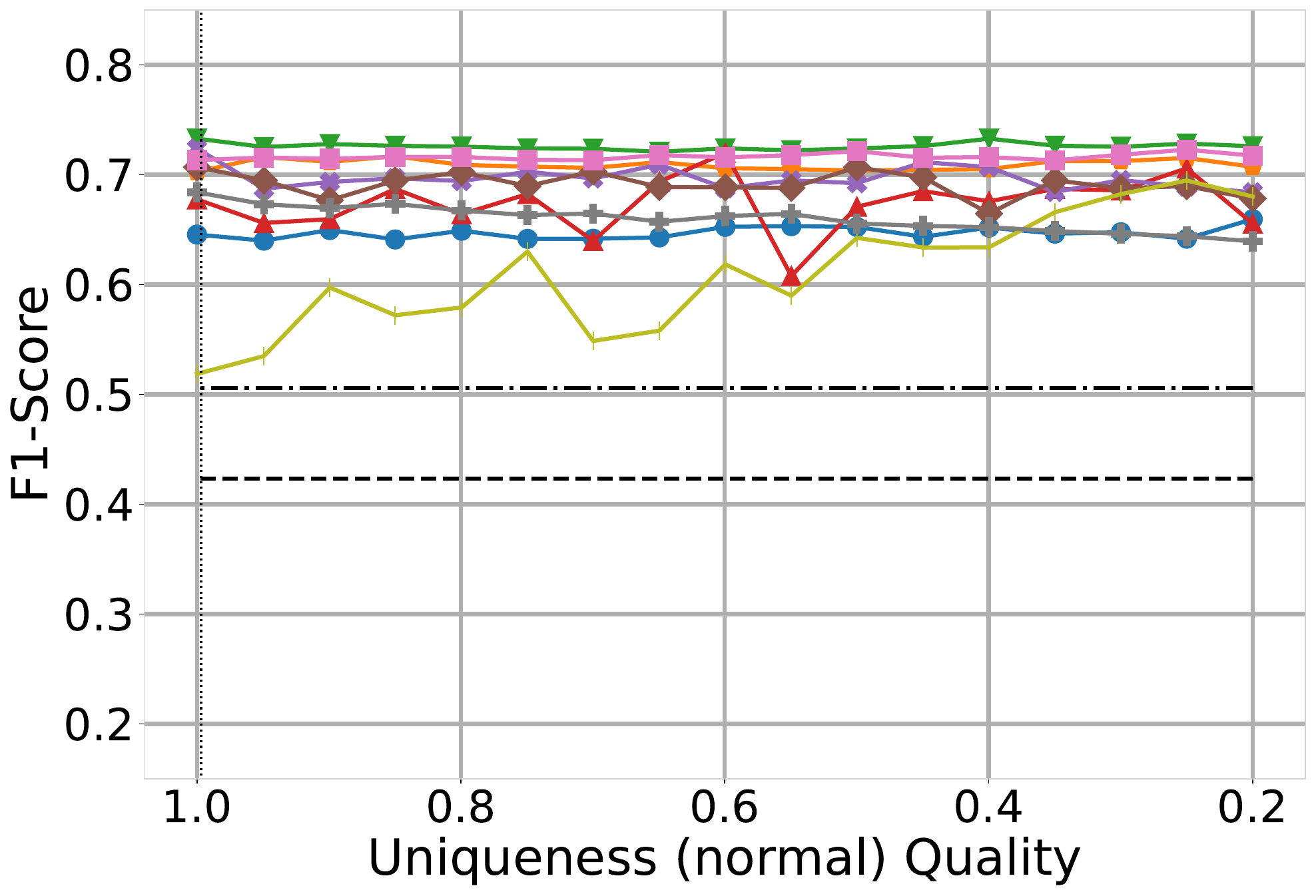}
        \caption{Uniqueness (Sce. 1)}
        \label{fig:classification-results-all-Uniqueness_dc1-1-telco}
    \end{subfigure}
    \begin{subfigure}[b]{0.32\textwidth}
        \includegraphics[width=\textwidth]{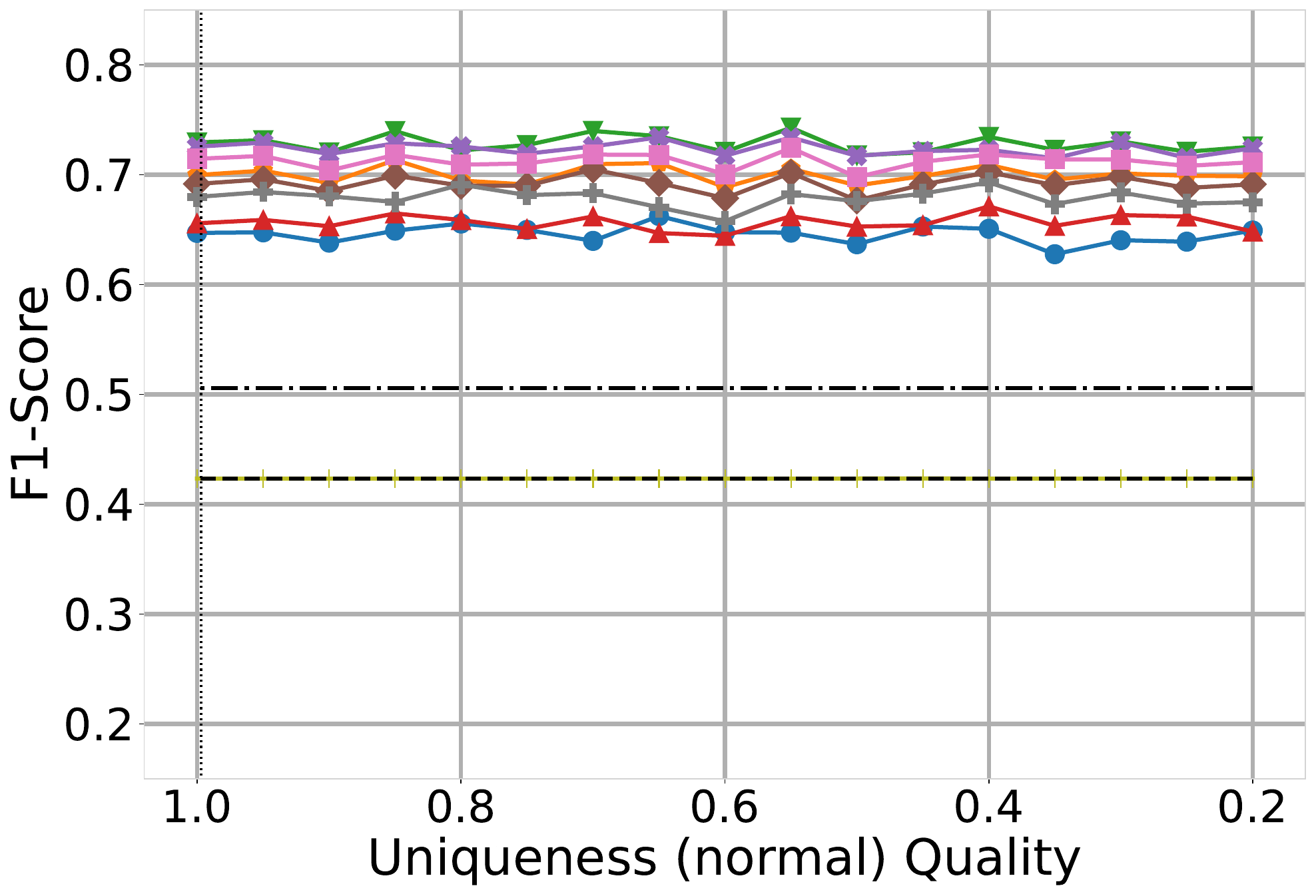}
        \caption{Uniqueness (Sce. 2)}
        \label{fig:classification-results-all-Uniqueness_dc1-2-telco}
    \end{subfigure}
    \begin{subfigure}[b]{0.32\textwidth}
        \includegraphics[width=\textwidth]{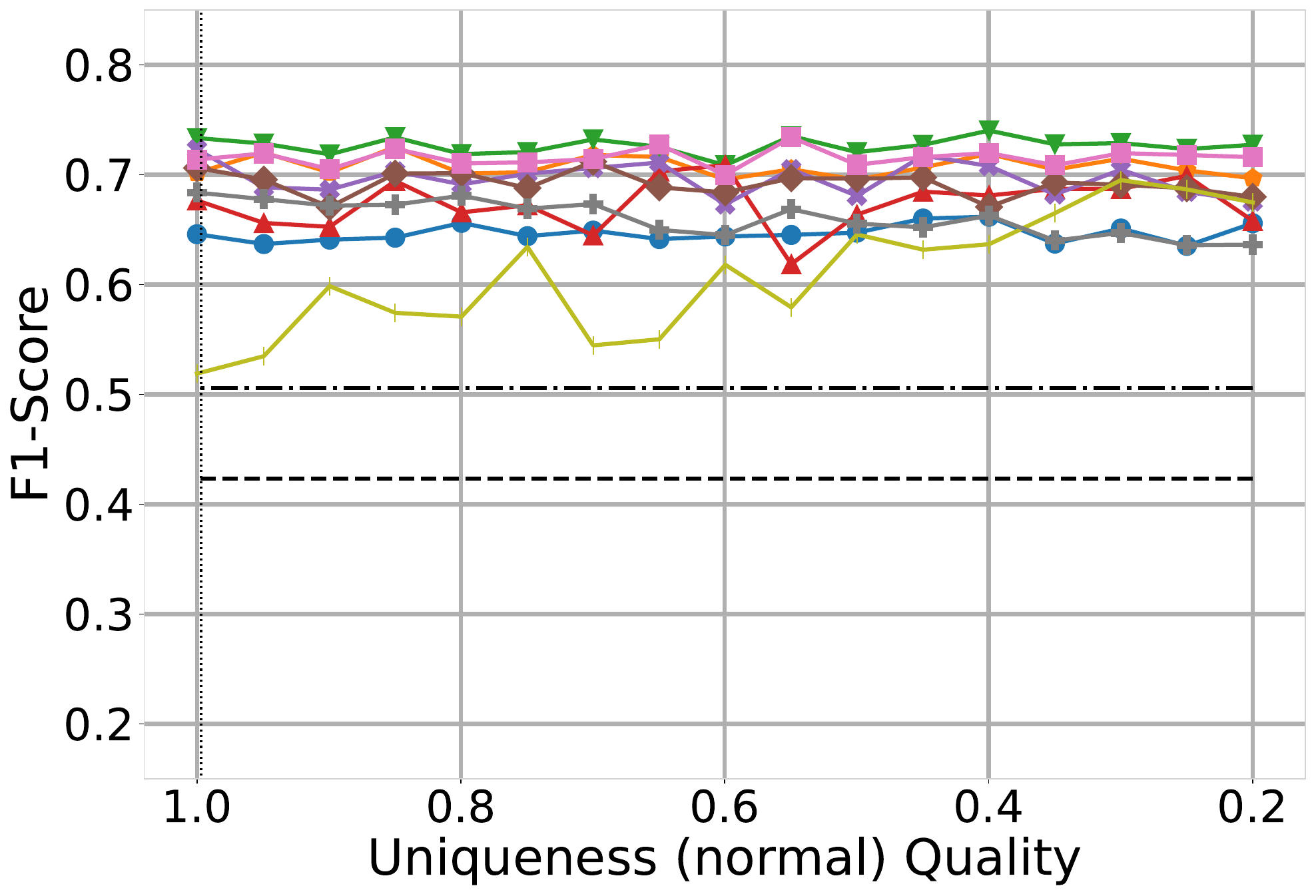}
        \caption{Uniqueness (Sce. 3)}
        \label{fig:classification-results-all-Uniqueness_dc1-3-telco}
    \end{subfigure}

 \begin{subfigure}[b]{0.32\textwidth}
        \includegraphics[width=\textwidth]{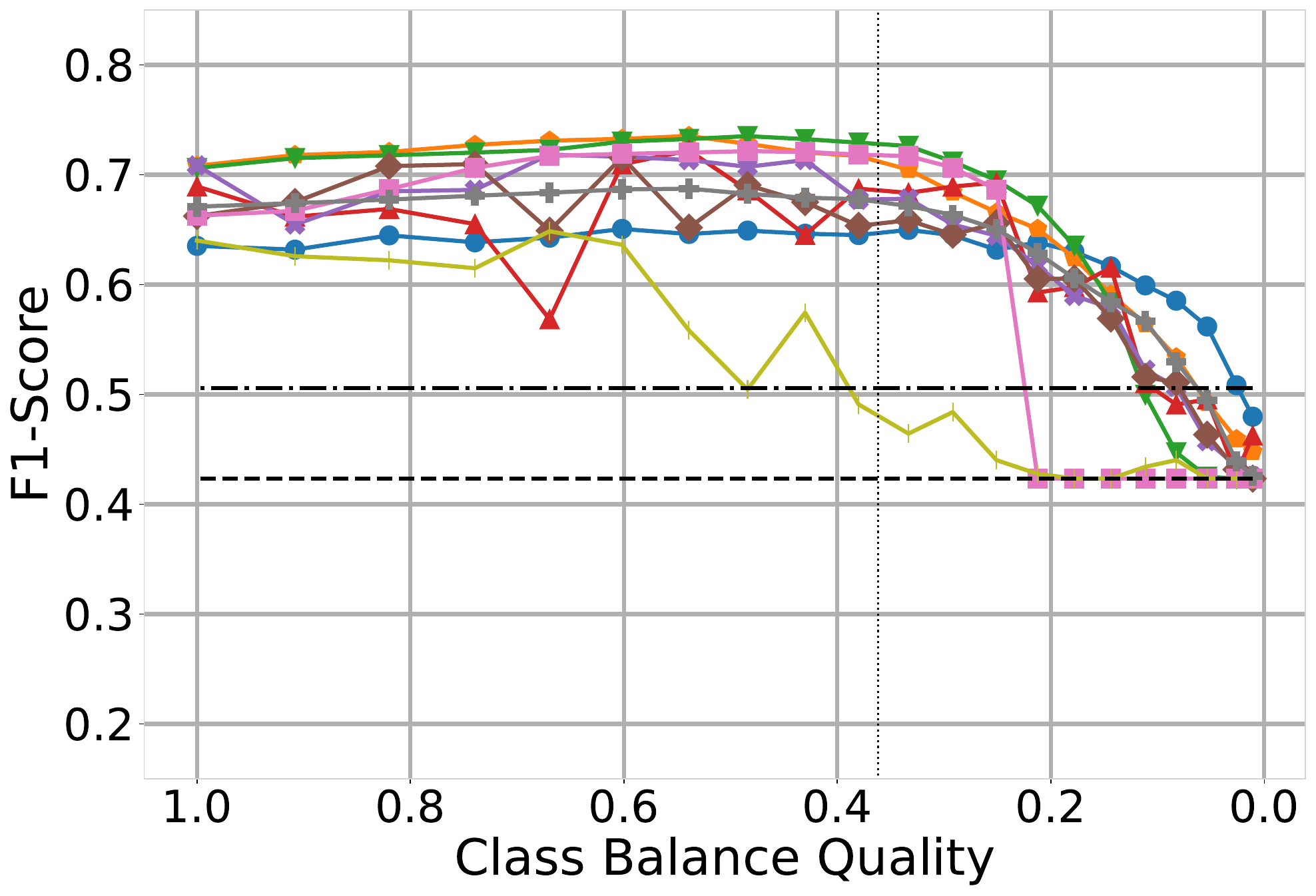}
        \caption{Class Balance (Sce. 1)}
        \label{fig:classification-results-all-ClassBalance-1-telco}
    \end{subfigure}
    \begin{subfigure}[b]{0.32\textwidth}
        \includegraphics[width=\textwidth]{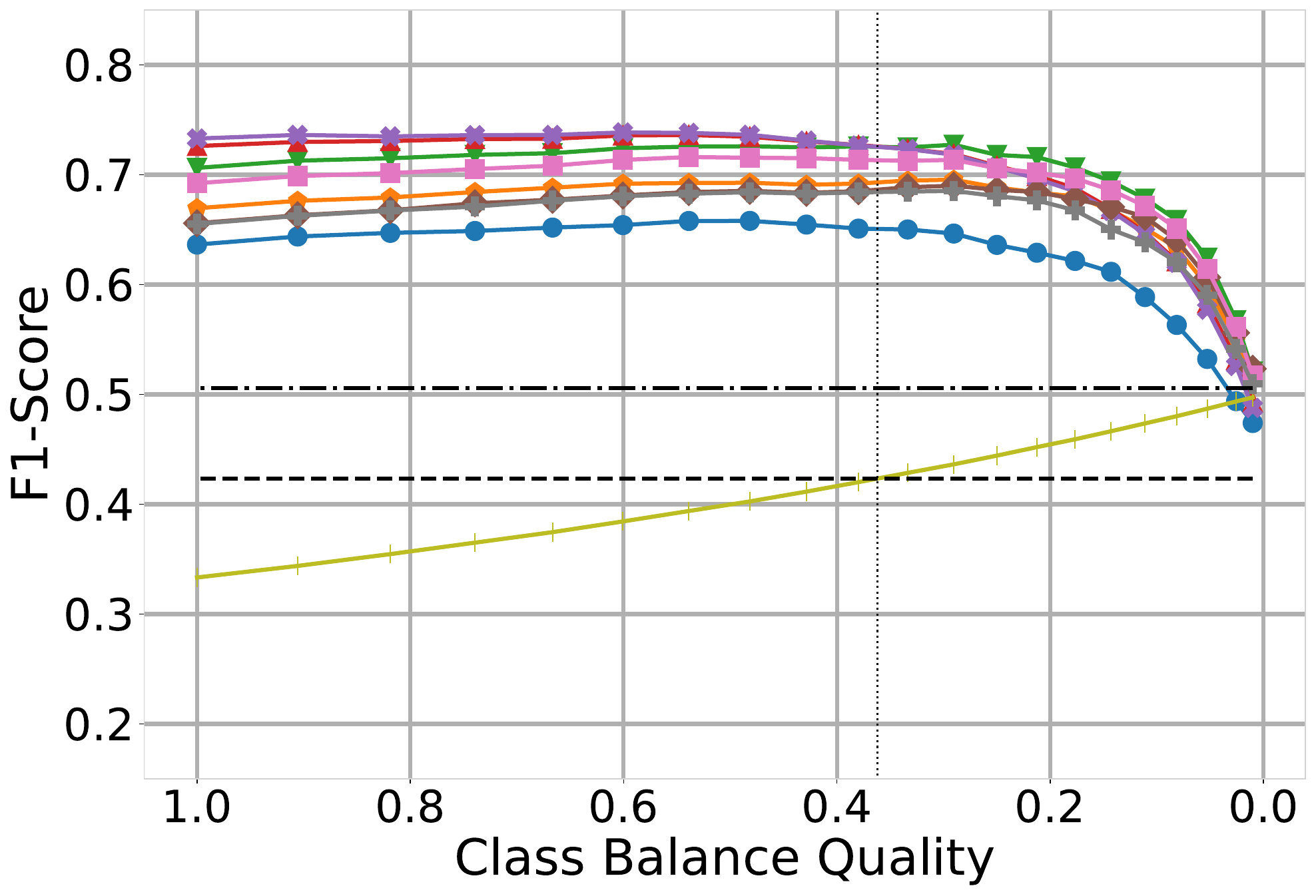}
        \caption{Class Balance (Sce. 2)}
        \label{fig:classification-results-all-ClassBalance-2-telco}
    \end{subfigure}
    \begin{subfigure}[b]{0.32\textwidth}
        \includegraphics[width=\textwidth]{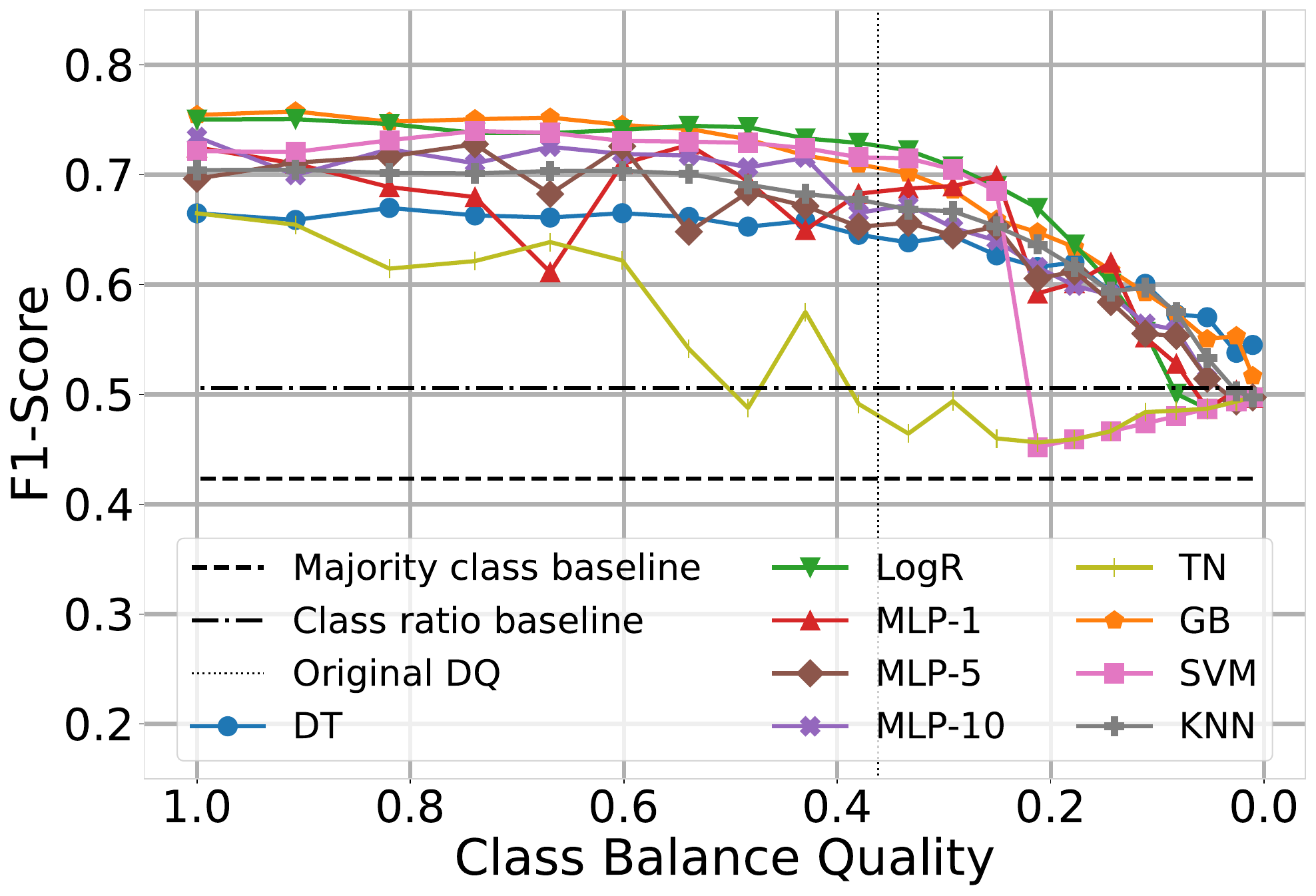}
        \caption{Class Balance (Sce. 3)}
        \label{fig:classification-results-all-ClassBalance-3-telco}
    \end{subfigure}
  \end{adjustbox}
\caption{\revision{$F_1$-scores of the classification algorithms for \textsf{Telco} dataset.}}
\label{fig:classification-results-all-telco}
\end{figure*}

%% file: Latex_Figure/classification/Consistent_Representation_5.tex
\begin{figure*}[t]
\centering
\raisebox{0.4\height}{\rotatebox{90}{Scenario 1}}\hspace{0.3em}
\begin{subfigure}[b]{0.23\linewidth}
\includegraphics[width=\linewidth]{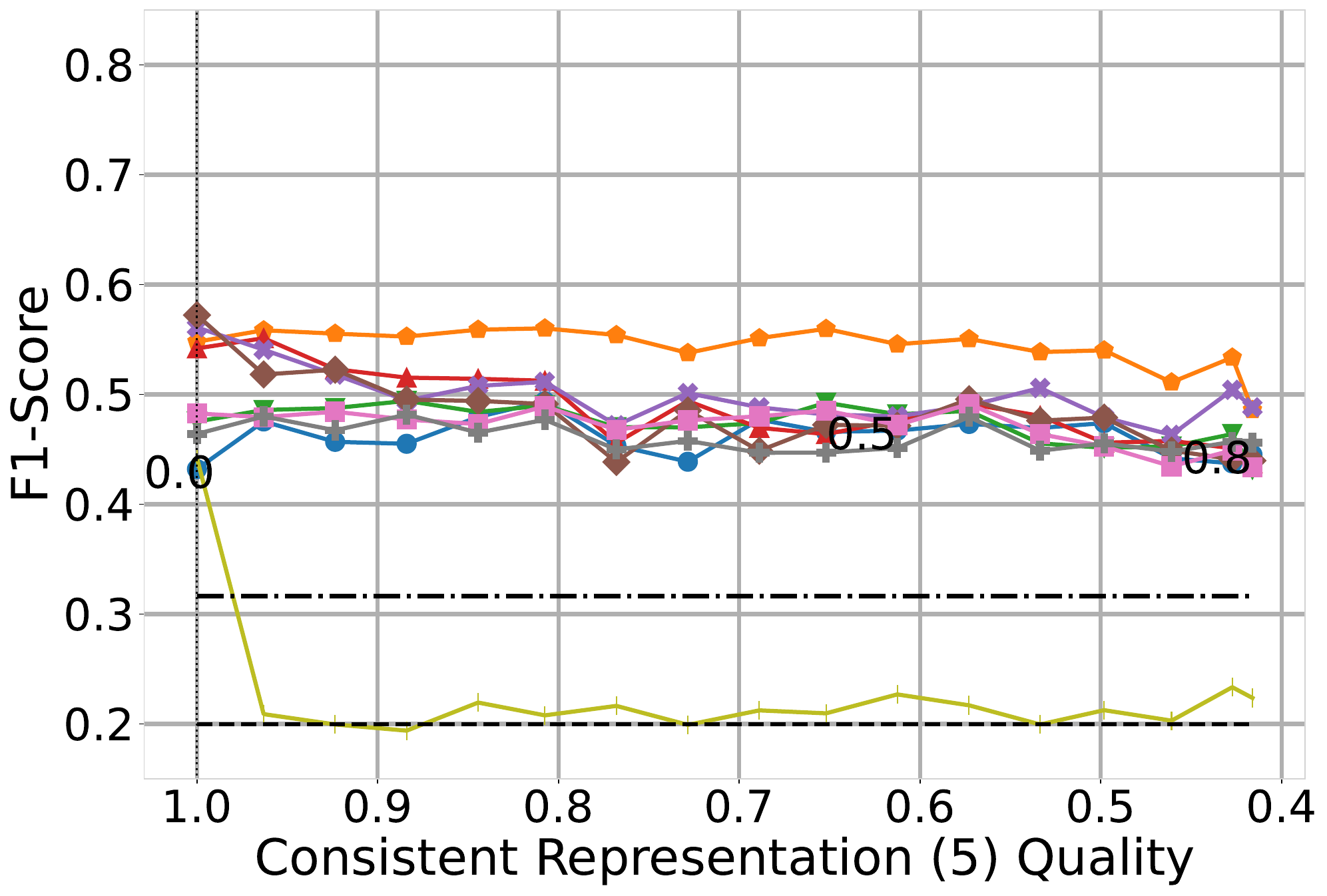}
\caption{\textsf{Contraceptive}}
\label{fig:classification-results-all-ConsistentRepresentationk5-1-contra}
\end{subfigure}
\begin{subfigure}[b]{0.23\linewidth}
\includegraphics[width=\linewidth]{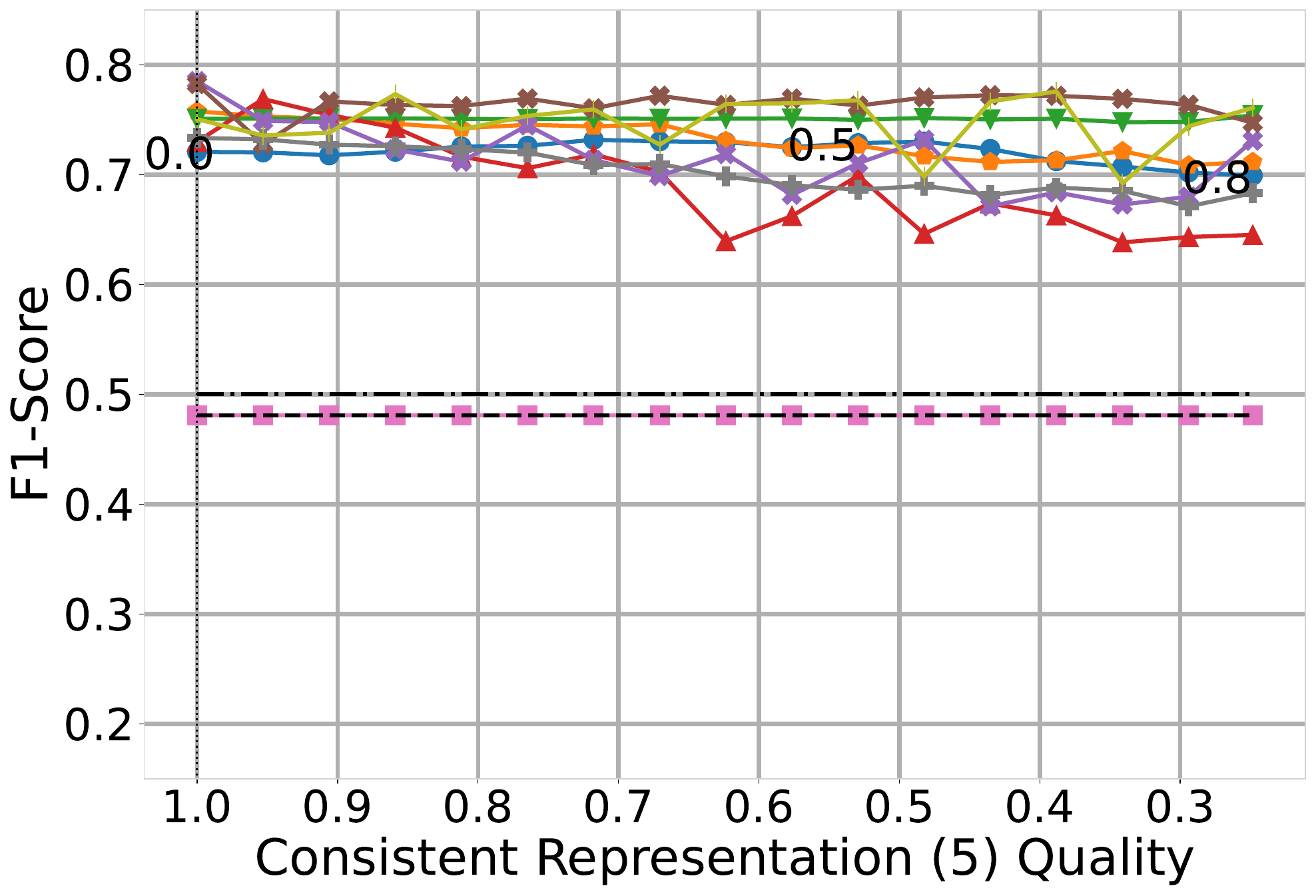}
\caption{\textsf{COVID}}
\label{fig:classification-results-all-ConsistentRepresentationk5-1-covid}
\end{subfigure}
\begin{subfigure}[b]{0.23\linewidth}
\includegraphics[width=\linewidth]{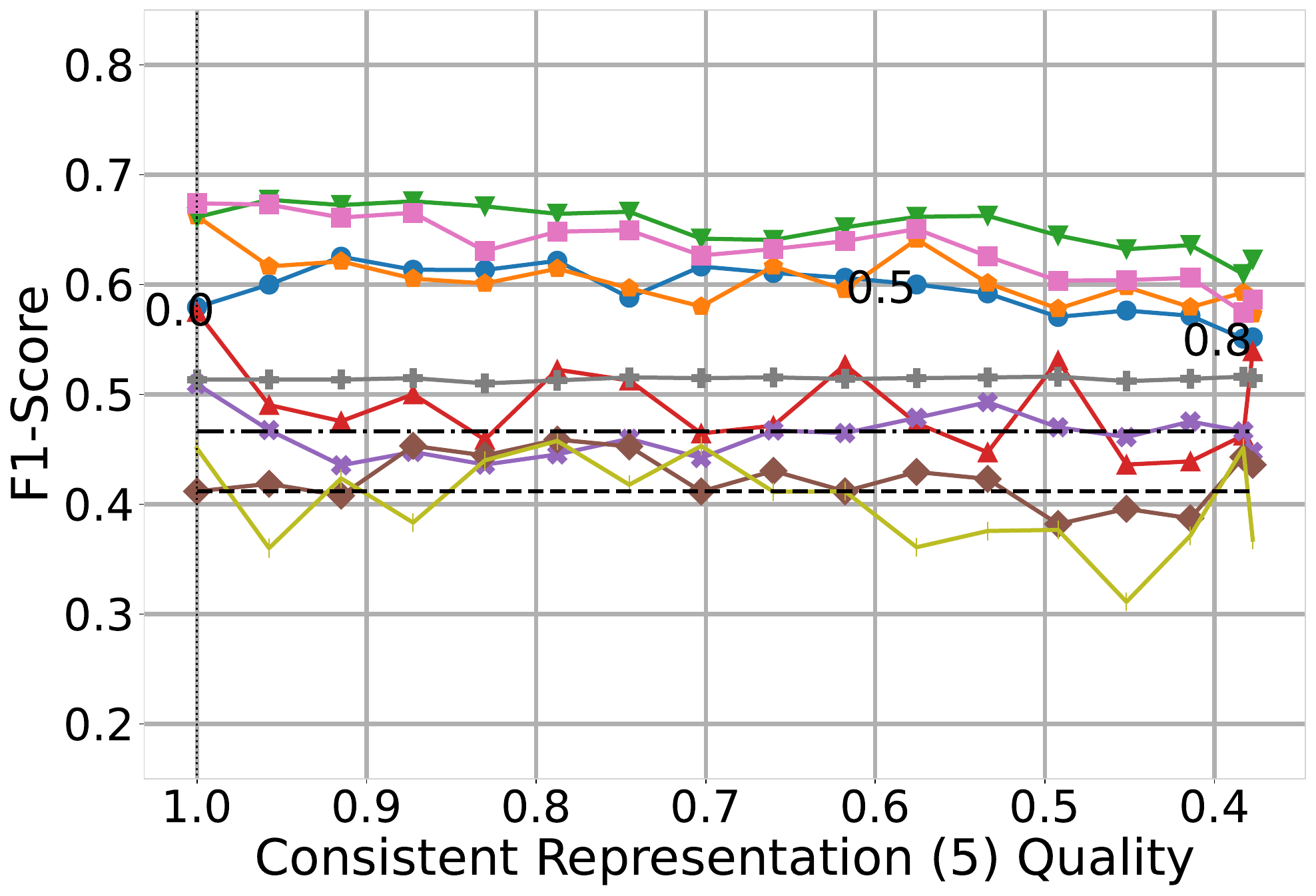}
\caption{\textsf{Credit}}
\label{fig:classification-results-all-ConsistentRepresentationk5-1-credit}
\end{subfigure}
\begin{subfigure}[b]{0.23\linewidth}
\includegraphics[width=\linewidth]{figures/classification/telco_train_polluted_test_clean_ConsistentRepresentationPolluter_five.pdf}
\caption{\textsf{Telco}}
\label{fig:classification-results-all-ConsistentRepresentationk5-1-telco}
\end{subfigure}

\raisebox{0.4\height}{\rotatebox{90}{Scenario 2}}\hspace{0.3em}
\begin{subfigure}[b]{0.23\linewidth}
\includegraphics[width=\linewidth]{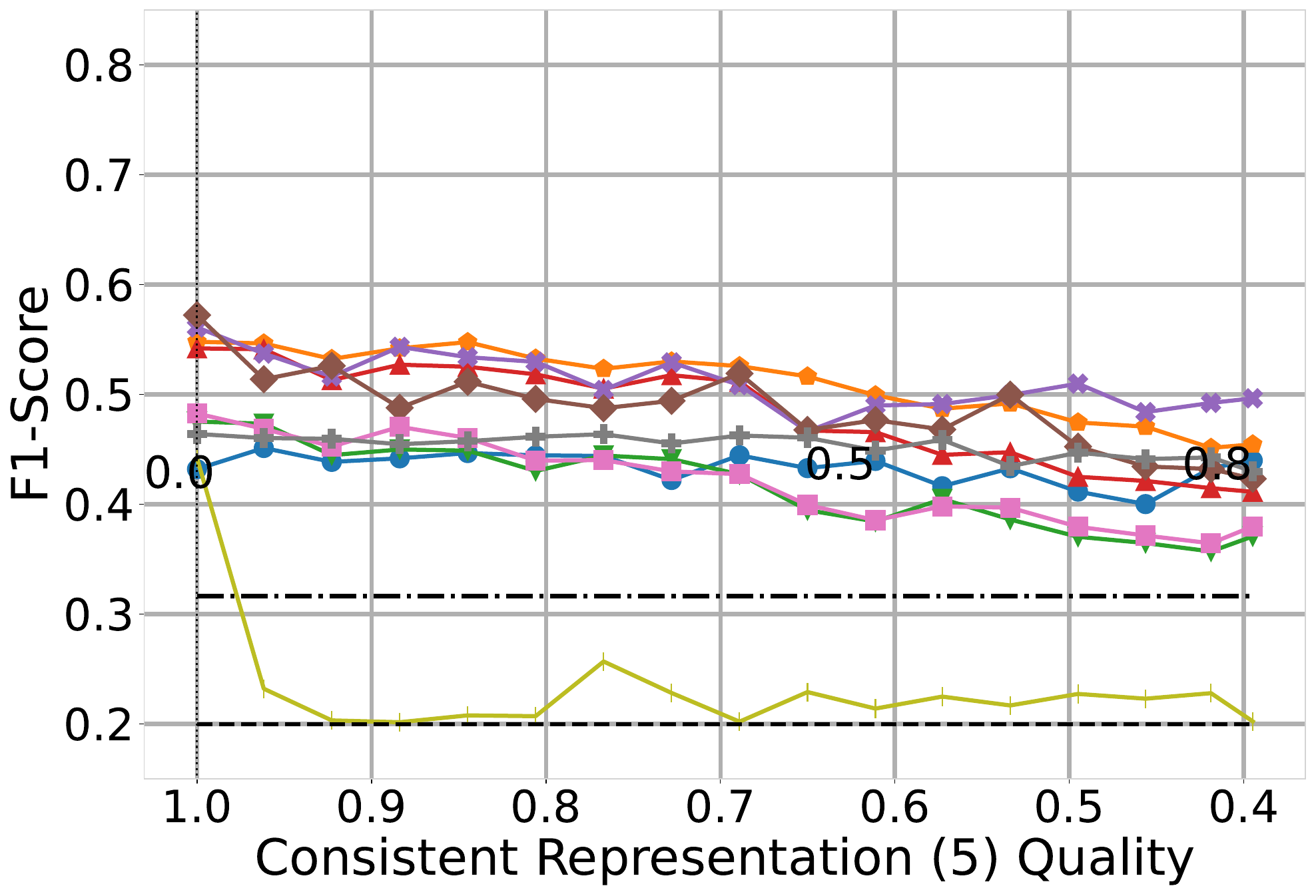}
\caption{\textsf{Contraceptive}}
\label{fig:classification-results-all-ConsistentRepresentationk5-2-contra}
\end{subfigure}
\begin{subfigure}[b]{0.23\linewidth}
\includegraphics[width=\linewidth]{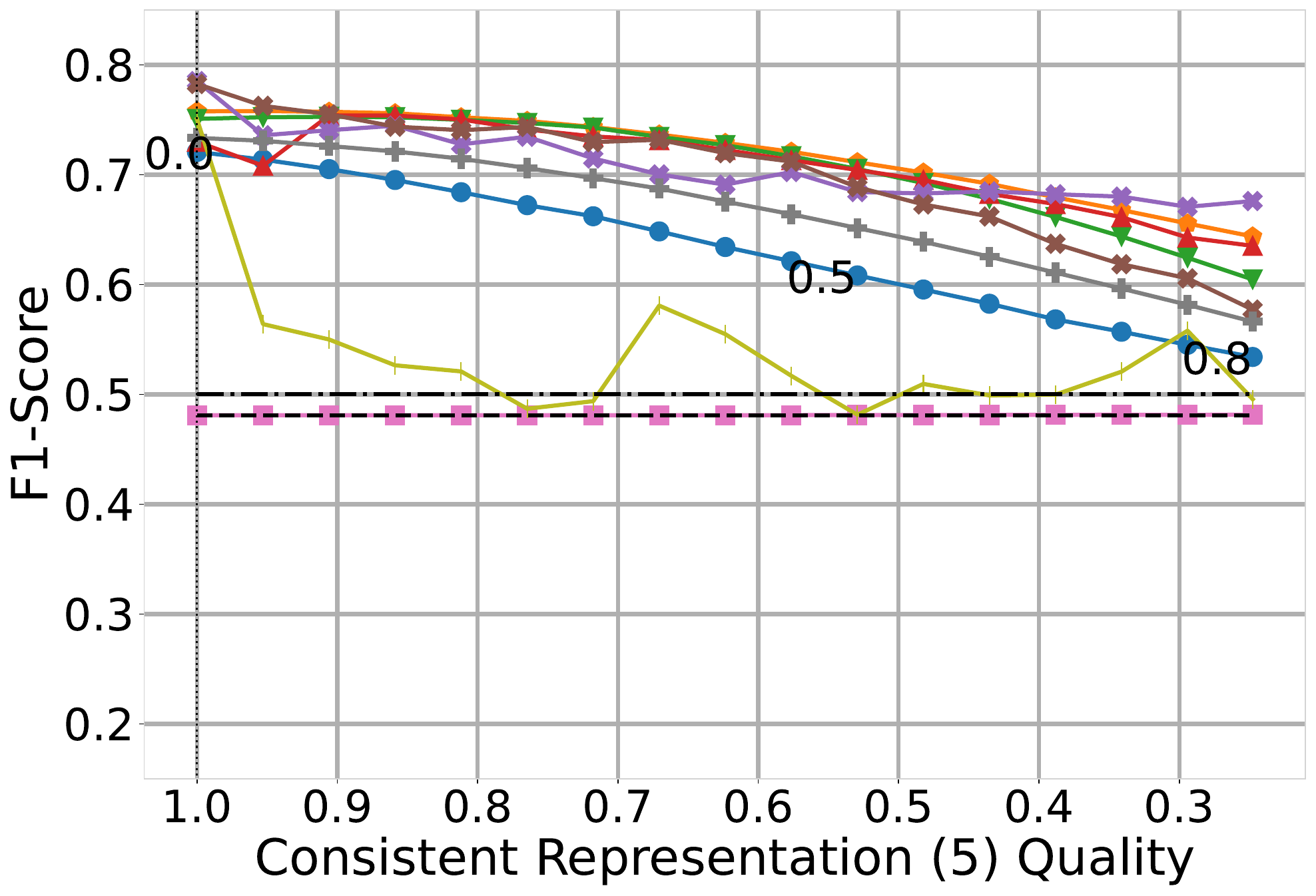}
\caption{\textsf{COVID}}
\label{fig:classification-results-all-ConsistentRepresentationk5-2-covid}
\end{subfigure}
\begin{subfigure}[b]{0.23\linewidth}
\includegraphics[width=\linewidth]{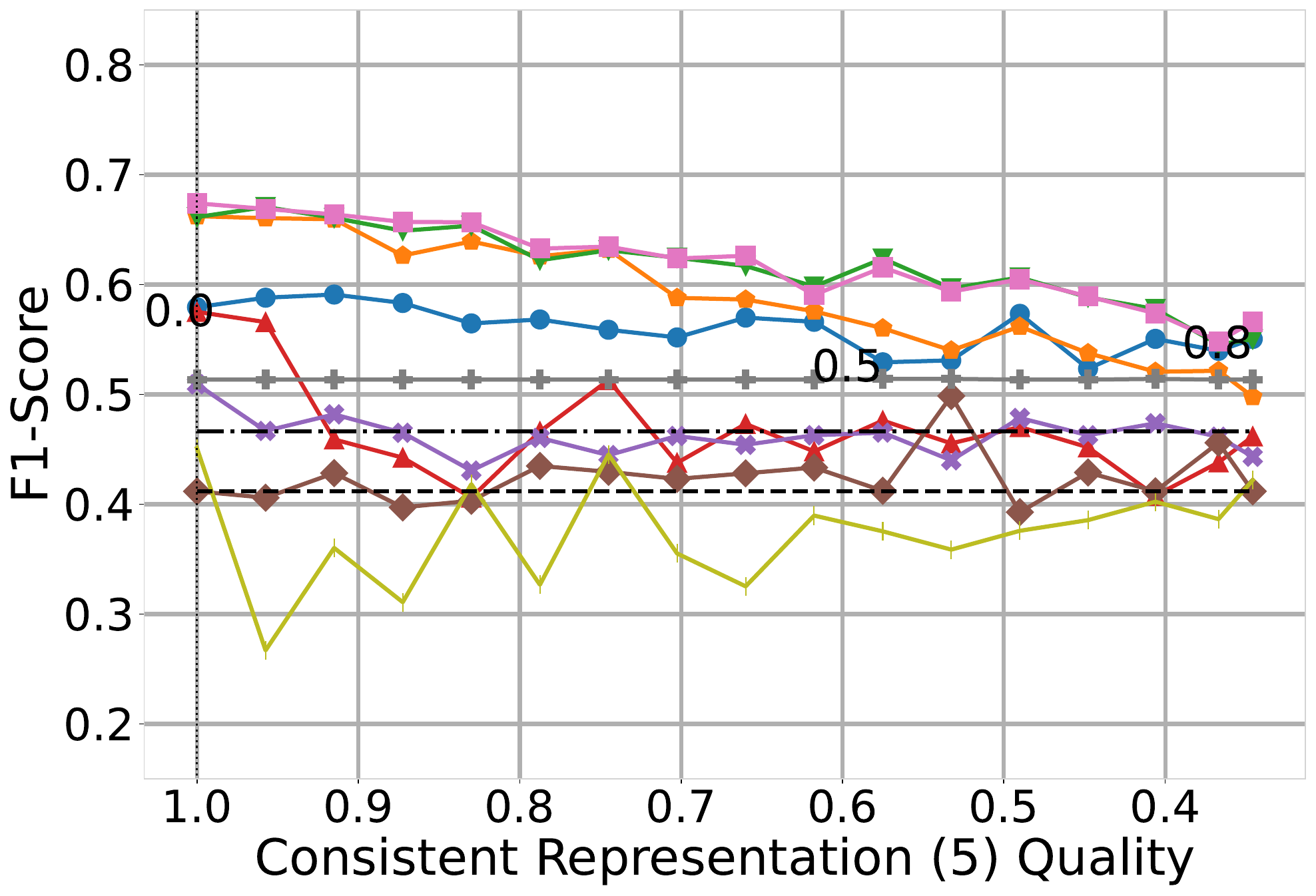}
\caption{\textsf{Credit}}
\label{fig:classification-results-all-ConsistentRepresentationk5-2-credit}
\end{subfigure}
\begin{subfigure}[b]{0.23\linewidth}
\includegraphics[width=\linewidth]{figures/classification/telco_train_clean_test_polluted_ConsistentRepresentationPolluter_five.pdf}
\caption{\textsf{Telco}}
\label{fig:classification-results-all-ConsistentRepresentationk5-2-telco}
\end{subfigure}

\raisebox{0.4\height}{\rotatebox{90}{Scenario 3}}\hspace{0.3em}
\begin{subfigure}[b]{0.23\linewidth}
\includegraphics[width=\linewidth]{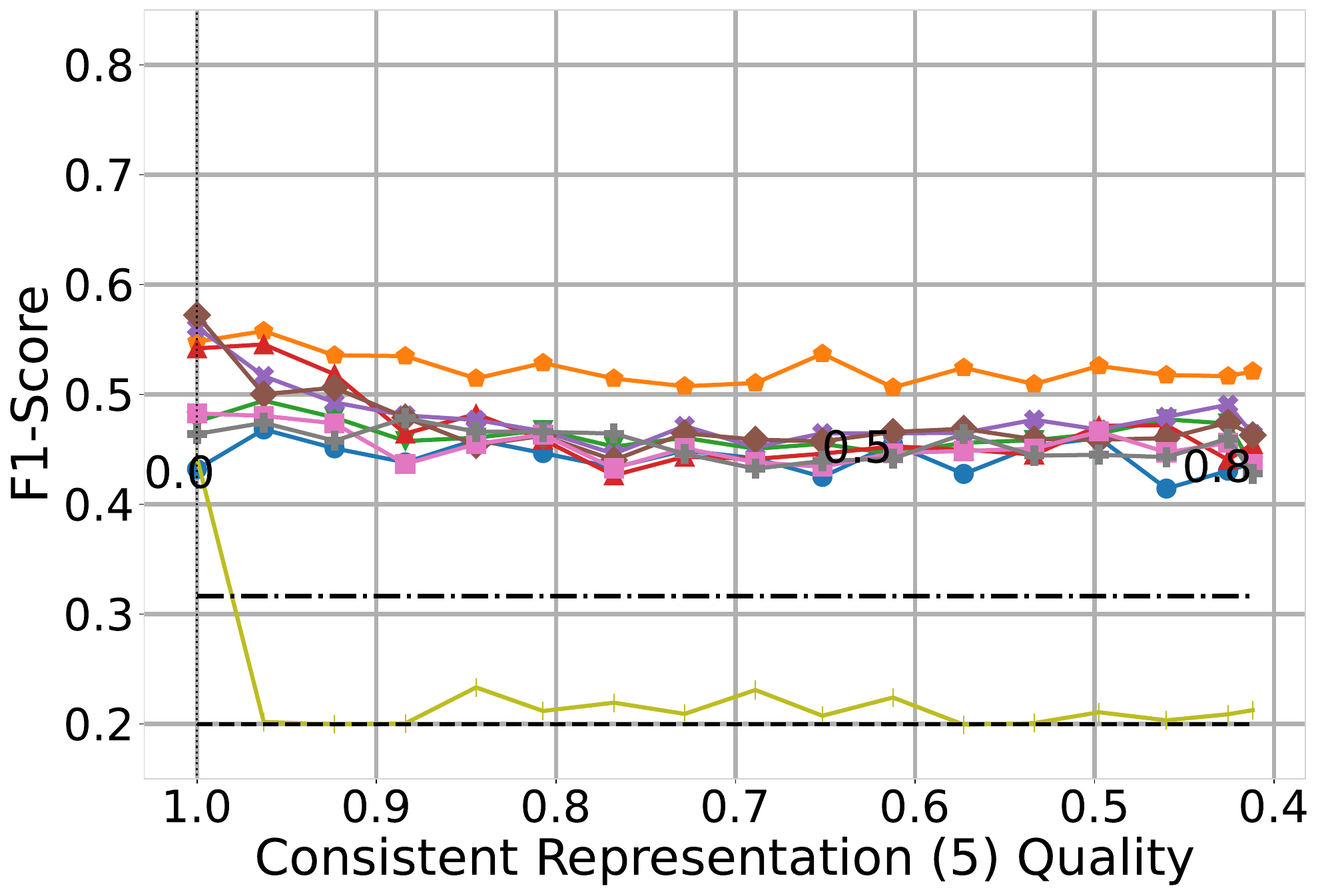}
\caption{\textsf{Contraceptive}}
\label{fig:classification-results-all-ConsistentRepresentationk5-3-contra}
\end{subfigure}
\begin{subfigure}[b]{0.23\linewidth}
\includegraphics[width=\linewidth]{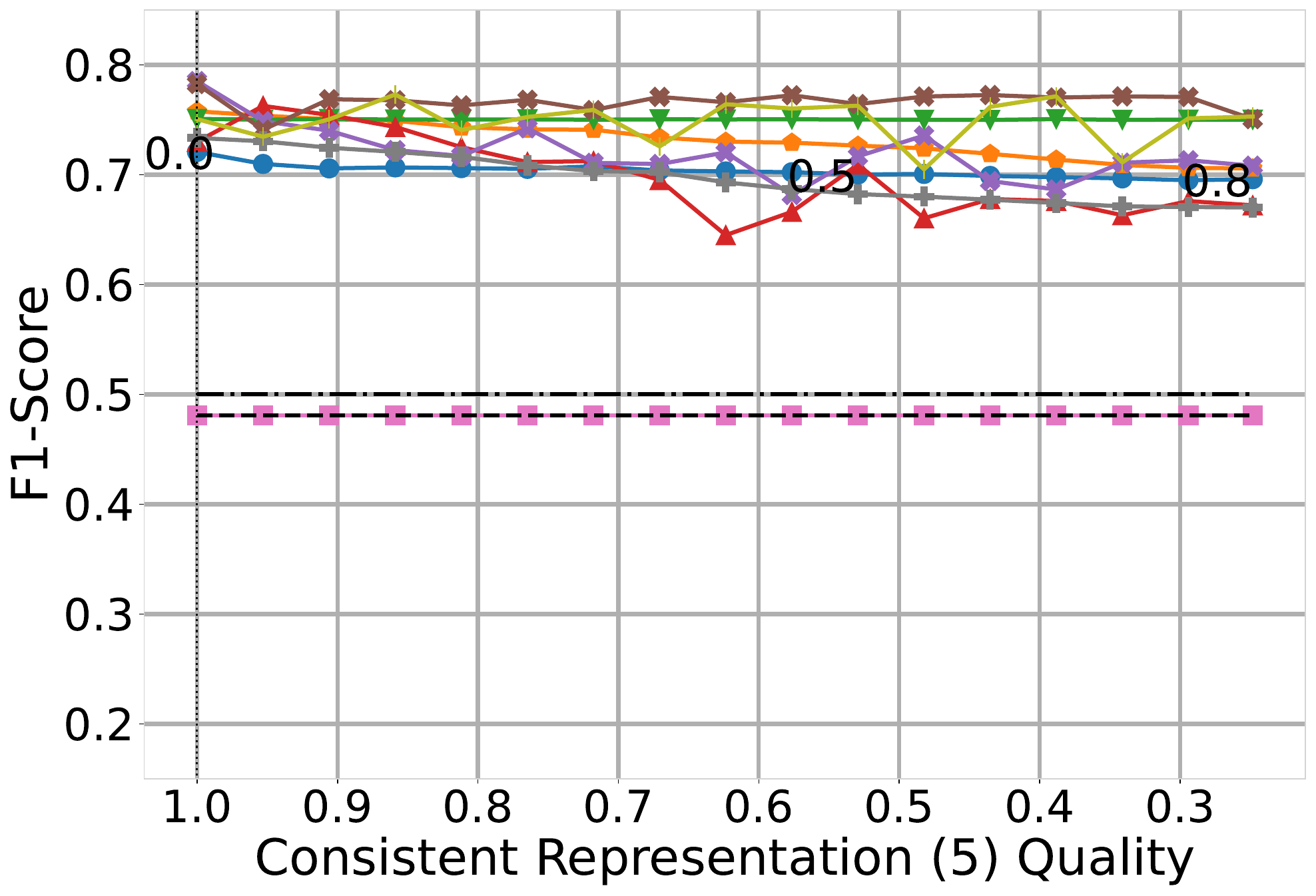}
\caption{\textsf{COVID}}
\label{fig:classification-results-all-ConsistentRepresentationk5-3-covid}
\end{subfigure}
\begin{subfigure}[b]{0.23\linewidth}
\includegraphics[width=\linewidth]{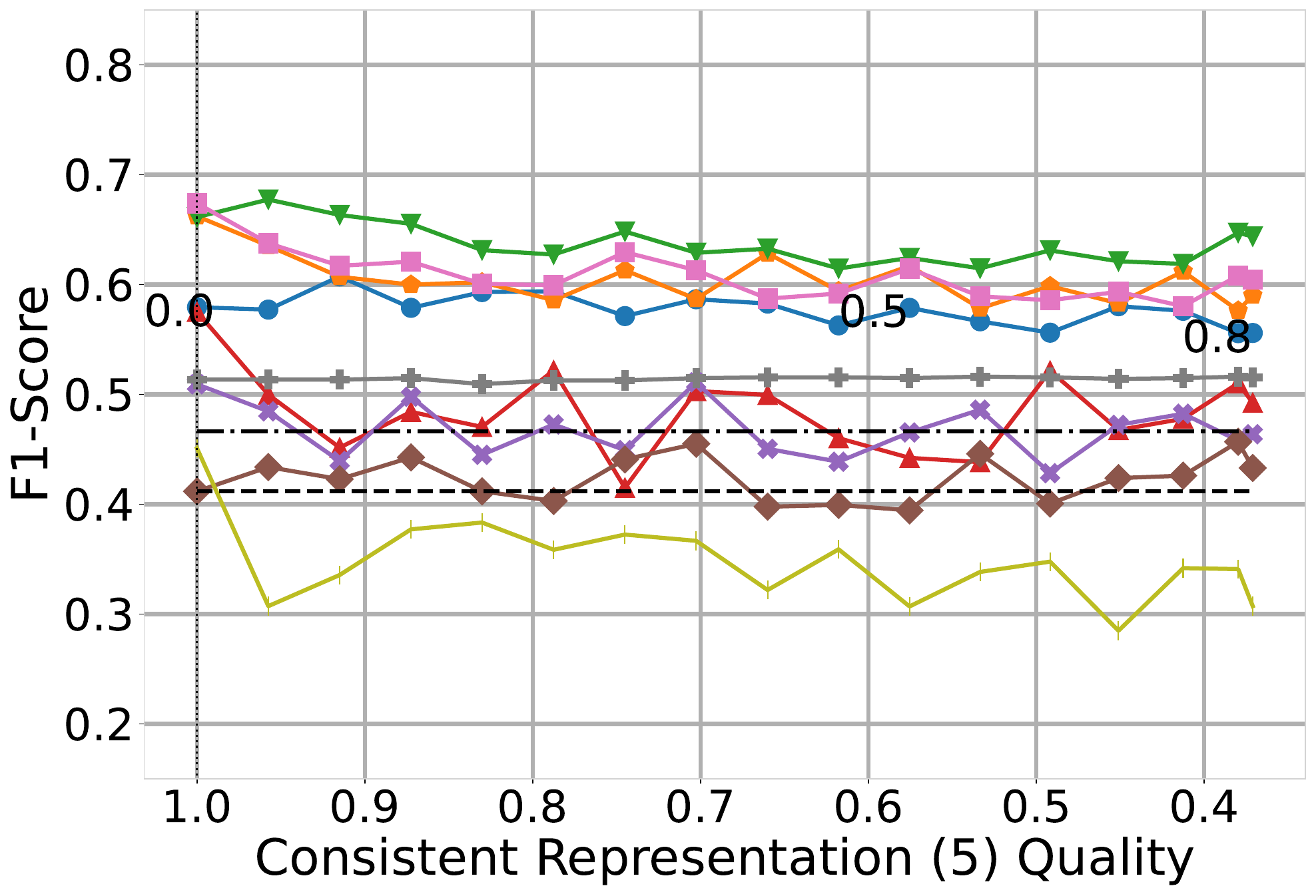}
\caption{\textsf{Credit}}
\label{fig:classification-results-all-ConsistentRepresentationk5-3-credit}
\end{subfigure}
\begin{subfigure}[b]{0.23\linewidth}
\includegraphics[width=\linewidth]{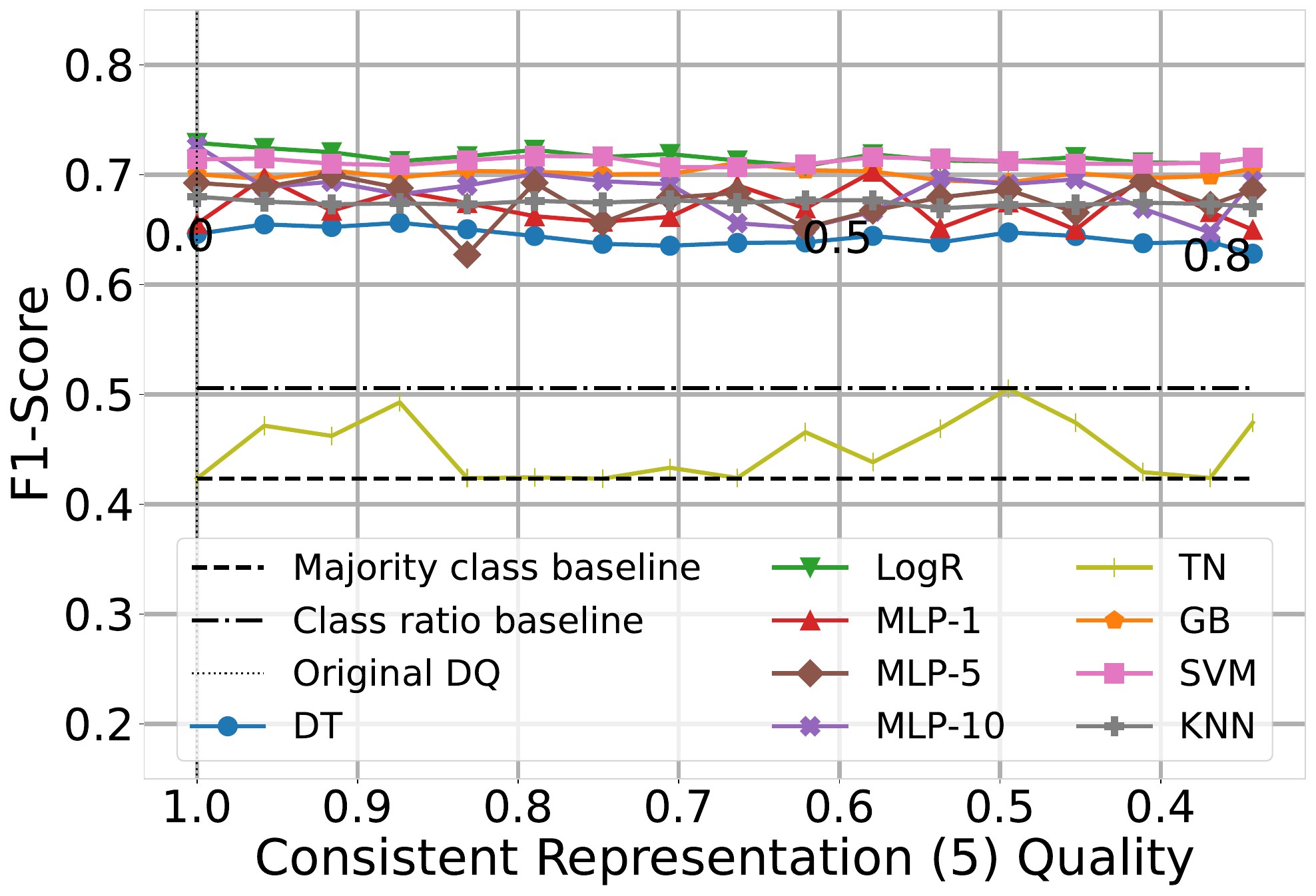}
\caption{\textsf{Telco}}
\label{fig:classification-results-all-ConsistentRepresentationk5-3-telco}
\end{subfigure}
\caption{$F_1$-scores of the classification algorithms for consistent representation with $k_v = 5$.}
\label{fig:classification-results-all-ConsistentRepresentationk5}
\end{figure*}

%% file: Latex_Figure/classification/Completeness.tex
\begin{figure*}[t]
    \centering
\raisebox{0.4\height}{\rotatebox{90}{Scenario 1}}\hspace{0.1em}
\begin{subfigure}[b]{0.23\linewidth}
        \includegraphics[width=\linewidth]{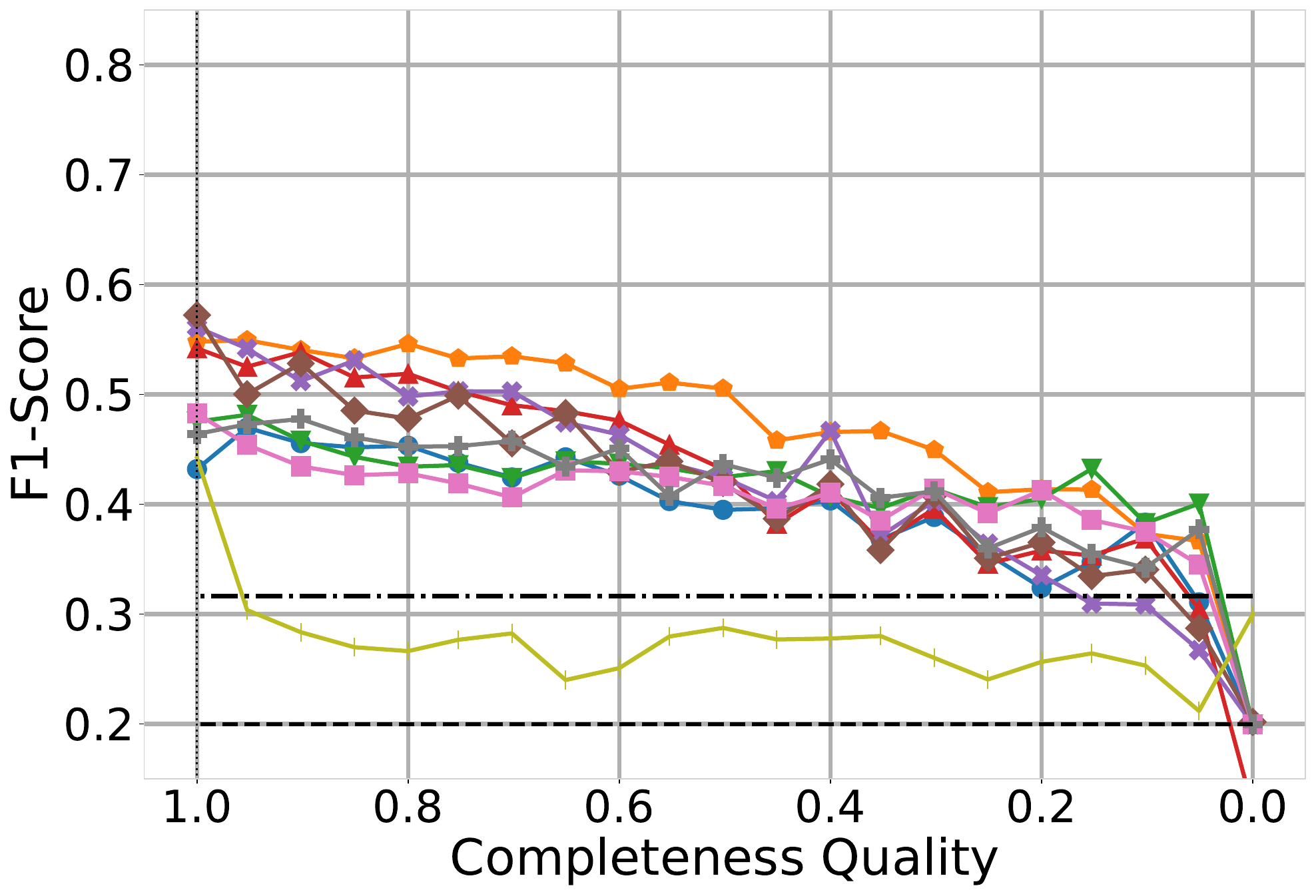}
        \caption{\textsf{Contraceptive}}
        \label{fig:classification-results-all-completeness-1-contra}
    \end{subfigure}
\begin{subfigure}[b]{0.23\linewidth}
        \includegraphics[width=\linewidth]{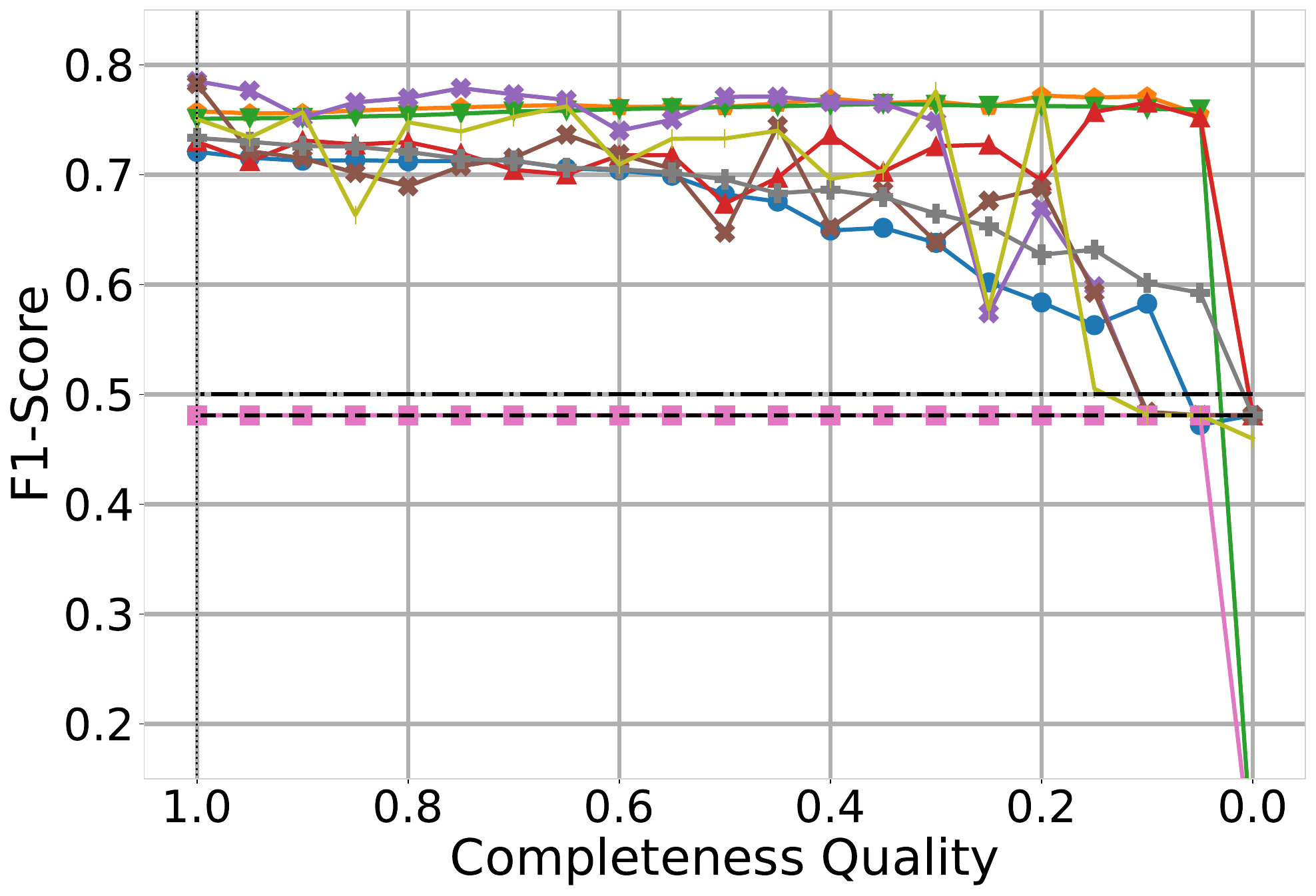}
        \caption{\textsf{COVID}}
        \label{fig:classification-results-all-completeness-1-covid}
    \end{subfigure}
\begin{subfigure}[b]{0.23\linewidth}
        \includegraphics[width=\linewidth]{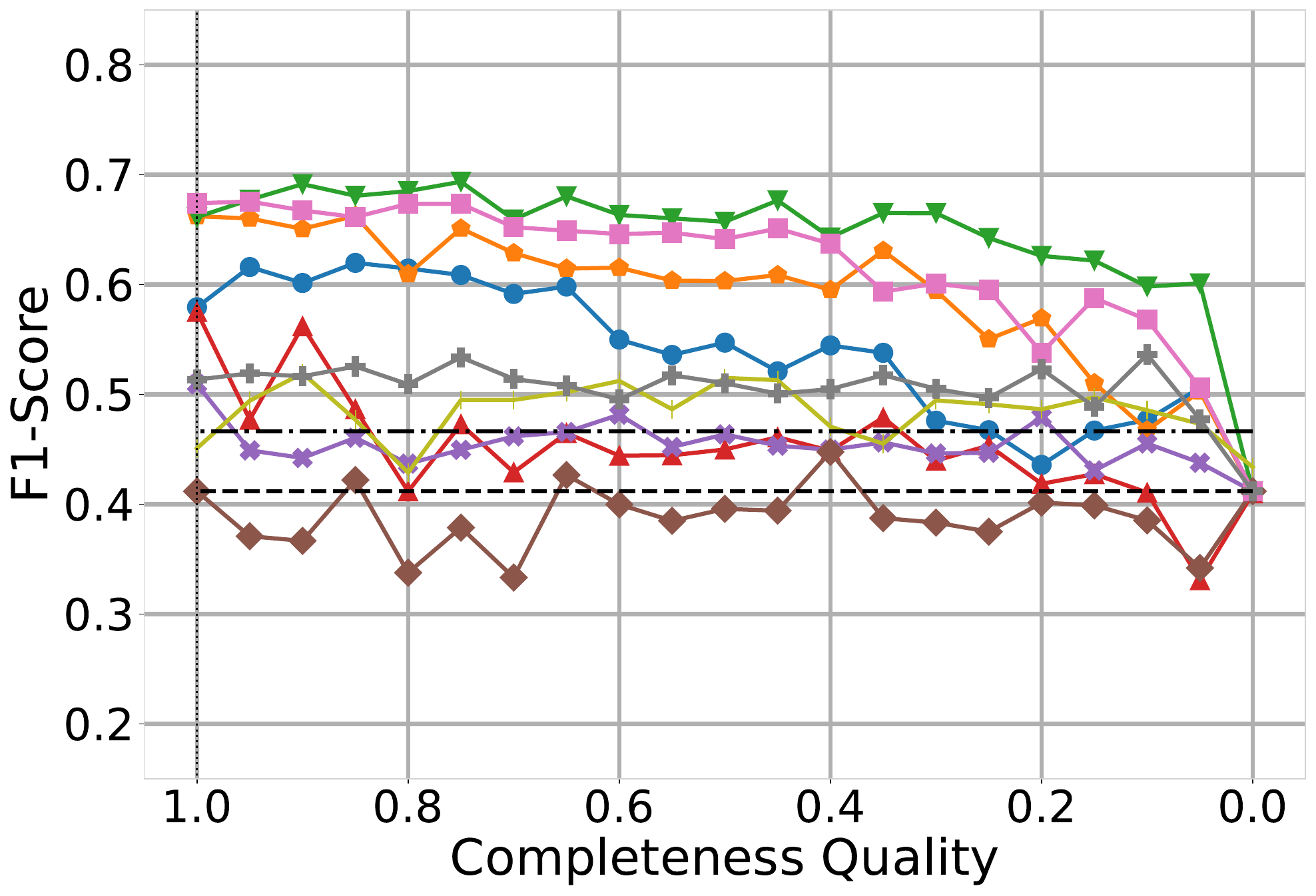}
        \caption{\textsf{Credit}}
        \label{fig:classification-results-all-completeness-1-credit}
    \end{subfigure}
\begin{subfigure}[b]{0.23\linewidth}
        \includegraphics[width=\linewidth]{figures/classification/telco_train_polluted_test_clean_CompletenessPolluter.pdf}
        \caption{\textsf{Telco}}
        \label{fig:classification-results-all-completeness-1-telco}
    \end{subfigure}

\raisebox{0.4\height}{\rotatebox{90}{Scenario 2}}\hspace{0.1em}
\begin{subfigure}[b]{0.23\linewidth}
        \includegraphics[width=\linewidth]{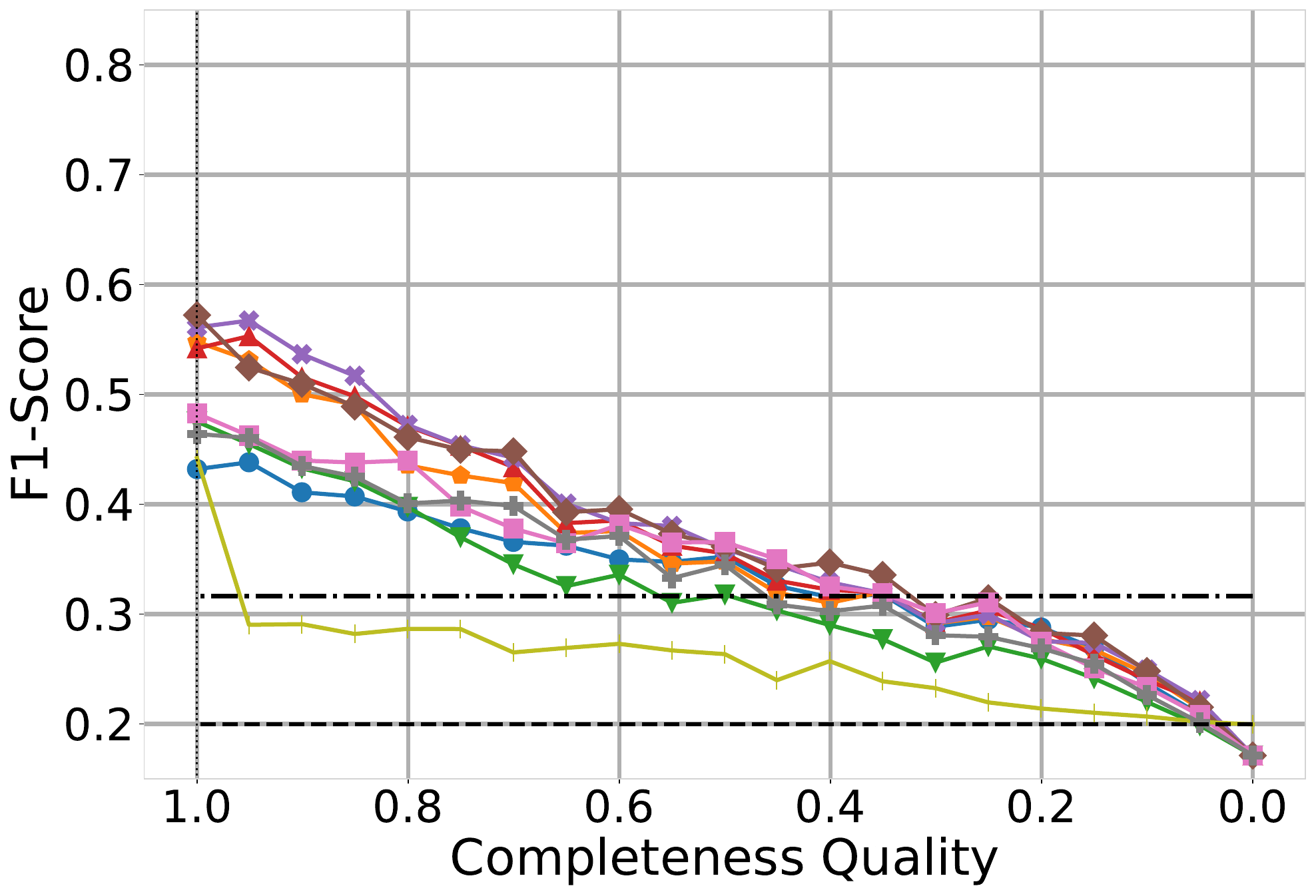}
        \caption{\textsf{Contraceptive}}
        \label{fig:classification-results-all-completeness-2-contra}
    \end{subfigure}
\begin{subfigure}[b]{0.23\linewidth}
        \includegraphics[width=\linewidth]{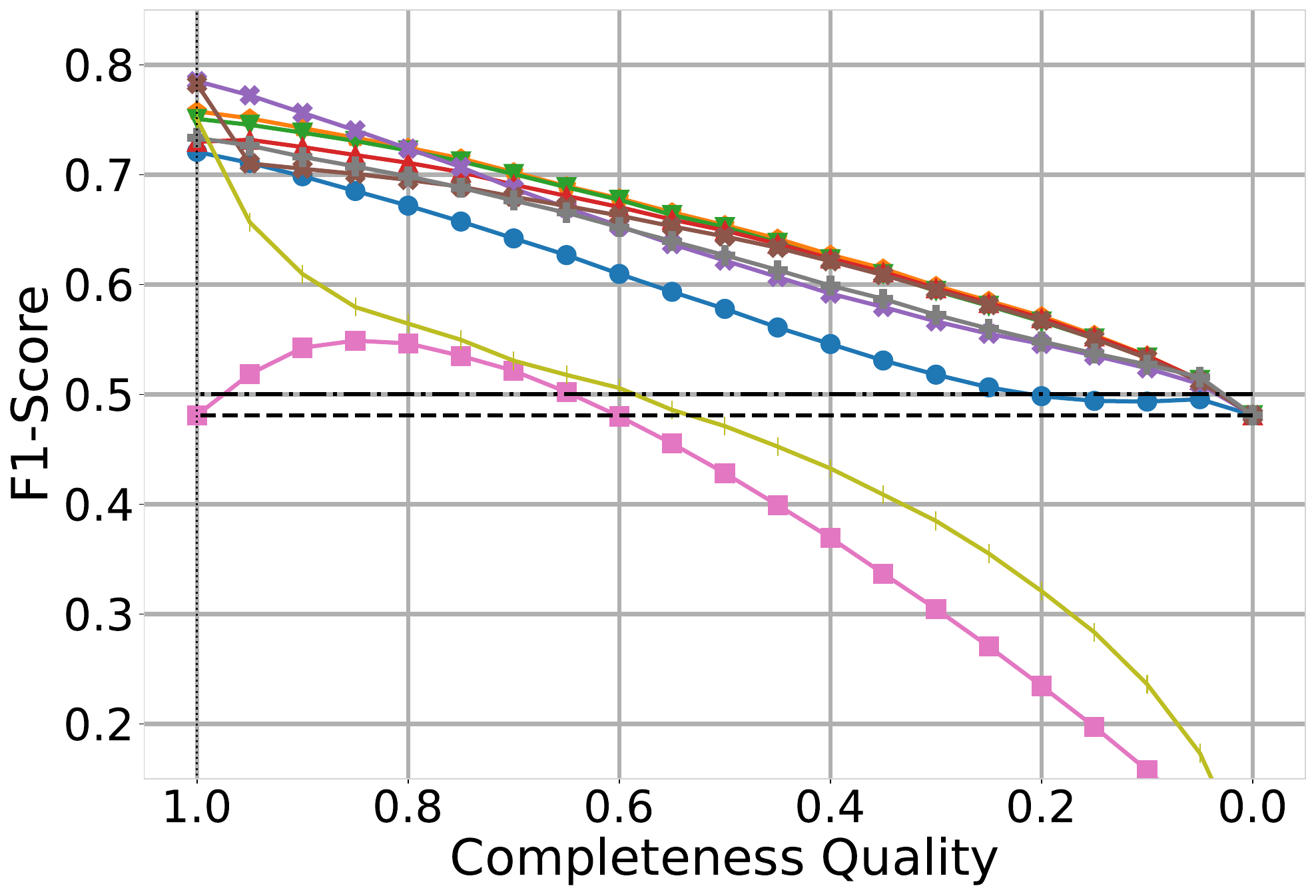}
        \caption{\textsf{COVID}}
        \label{fig:classification-results-all-completeness-2-covid}
    \end{subfigure}
\begin{subfigure}[b]{0.23\linewidth}
        \includegraphics[width=\linewidth]{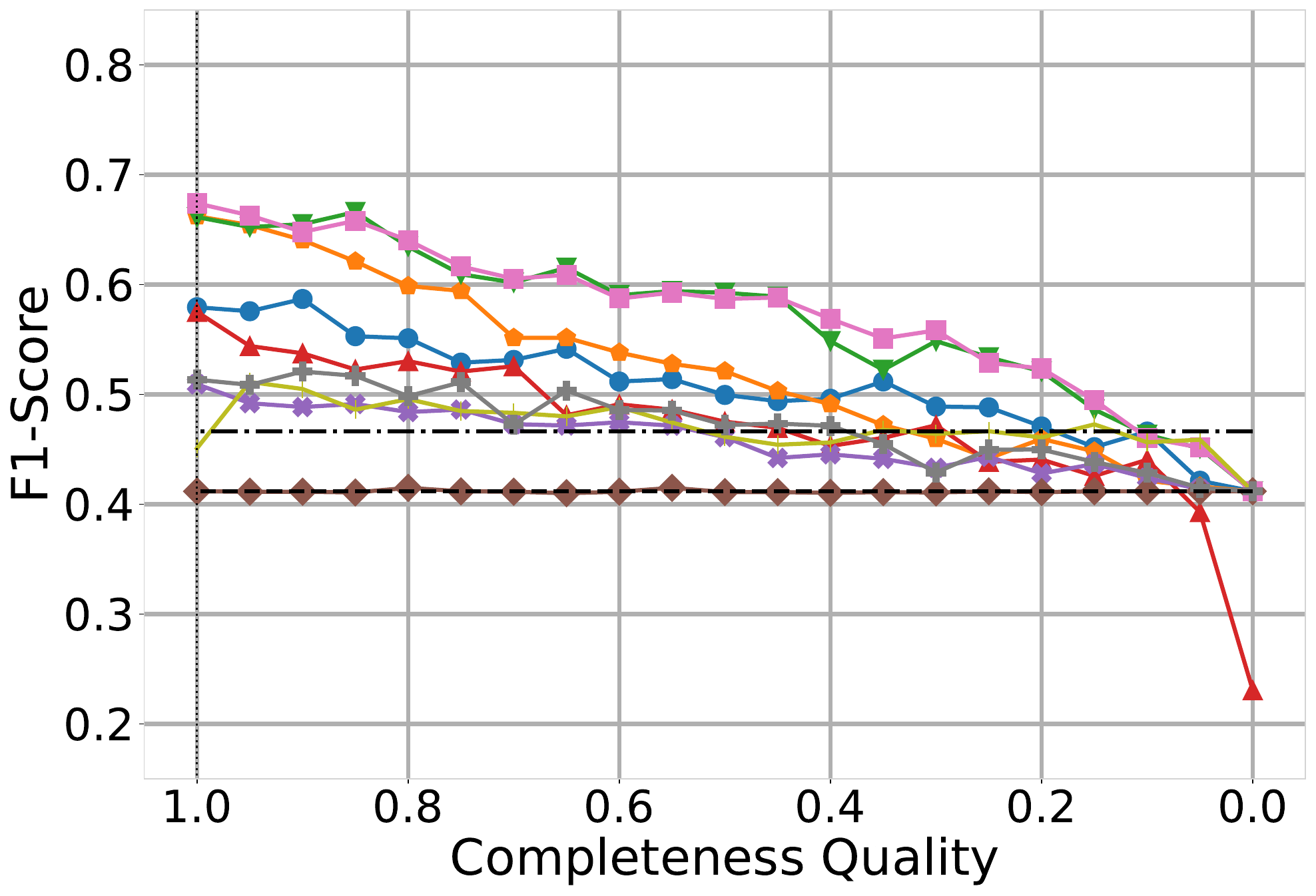}
        \caption{\textsf{Credit}}
        \label{fig:classification-results-all-completeness-2-credit}
    \end{subfigure}
\begin{subfigure}[b]{0.23\linewidth}
        \includegraphics[width=\linewidth]{figures/classification/telco_train_clean_test_polluted_CompletenessPolluter.pdf}
        \caption{\textsf{Telco}}
        \label{fig:classification-results-all-completeness-2-telco}
    \end{subfigure}

\raisebox{0.4\height}{\rotatebox{90}{Scenario 3}}\hspace{0.1em}
\begin{subfigure}[b]{0.23\linewidth}
        \includegraphics[width=\linewidth]{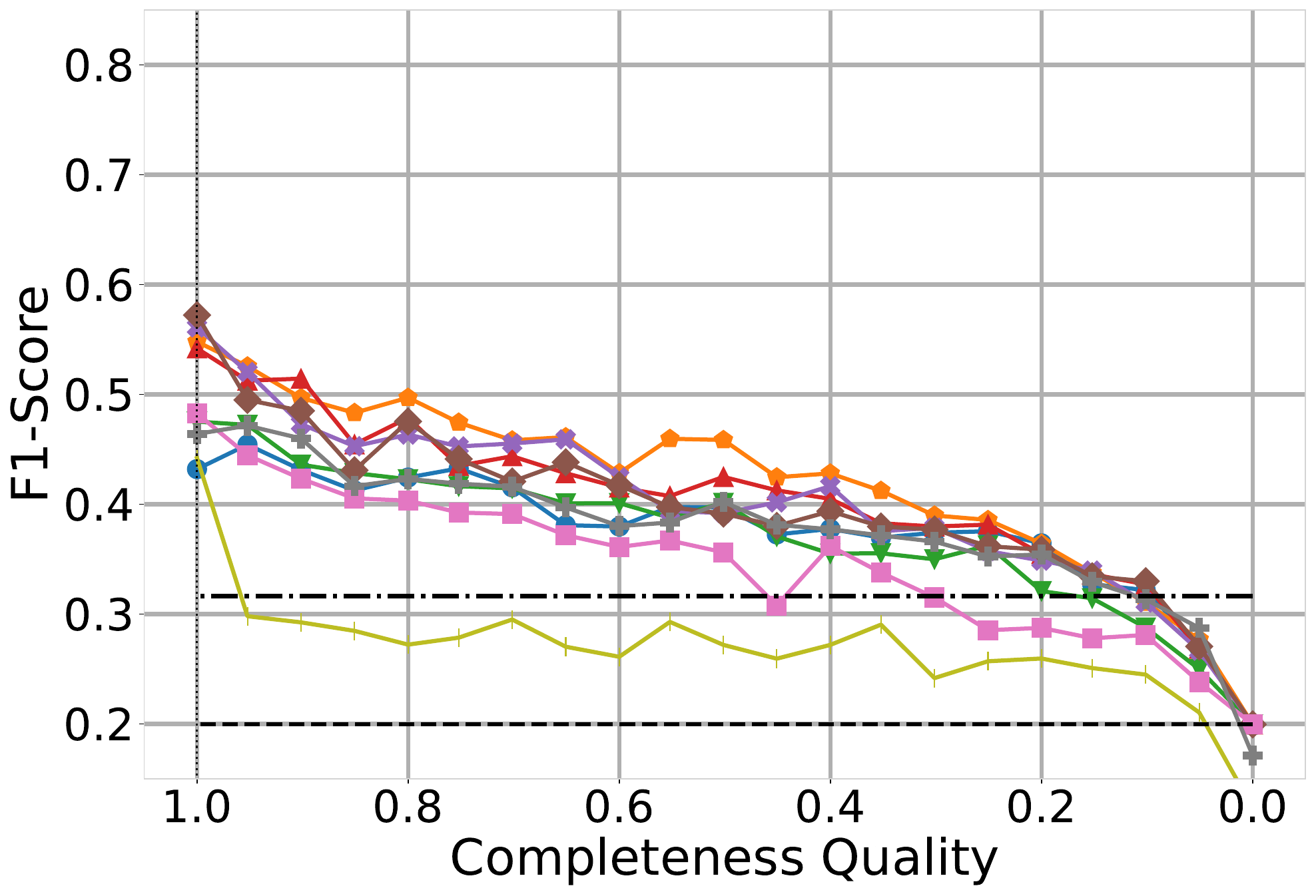}
        \caption{\textsf{Contraceptive}}
        \label{fig:classification-results-all-completeness-3-contra}
    \end{subfigure}
\begin{subfigure}[b]{0.23\linewidth}
        \includegraphics[width=\linewidth]{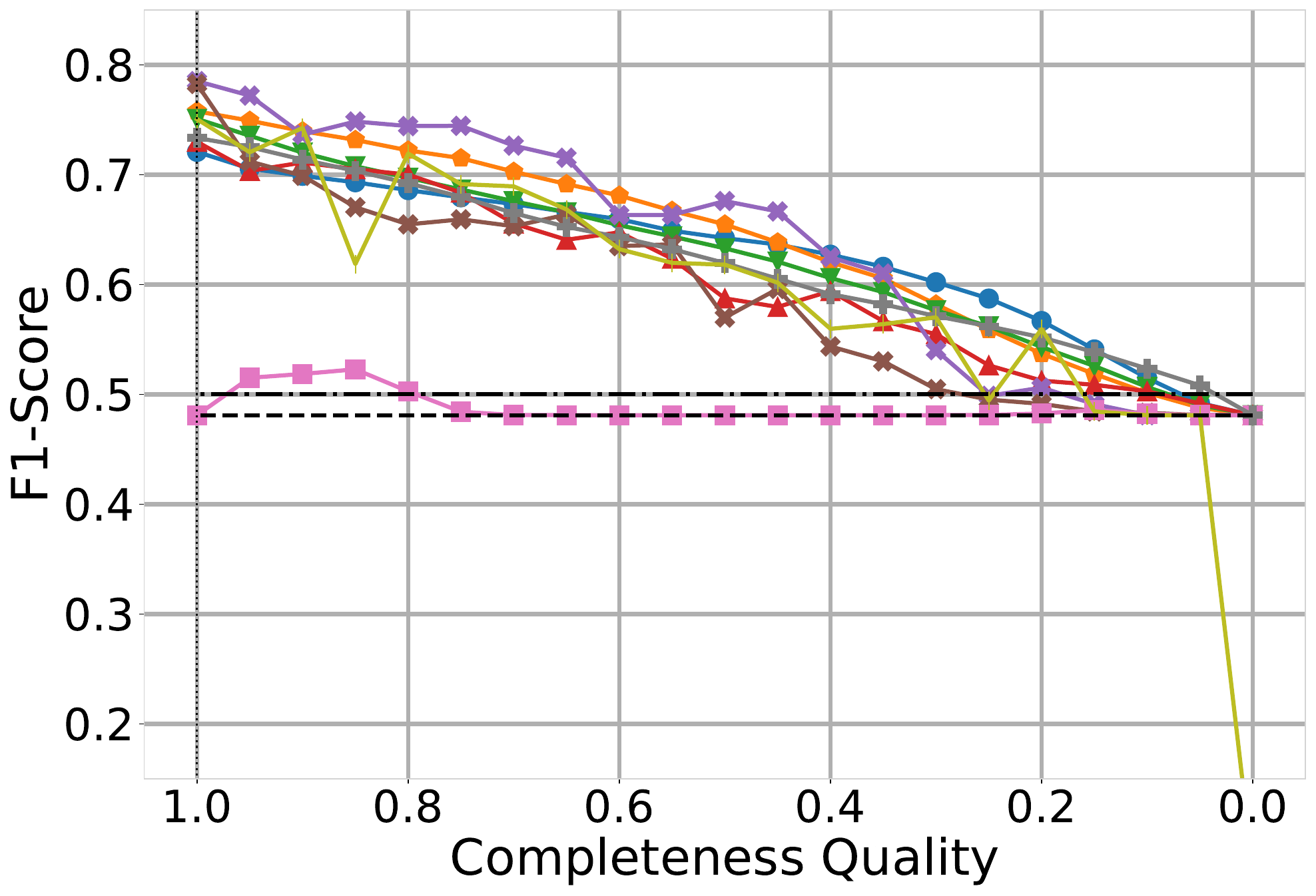}
        \caption{\textsf{COVID}}
        \label{fig:classification-results-all-completeness-3-covid}
    \end{subfigure}
\begin{subfigure}[b]{0.23\linewidth}
        \includegraphics[width=\linewidth]{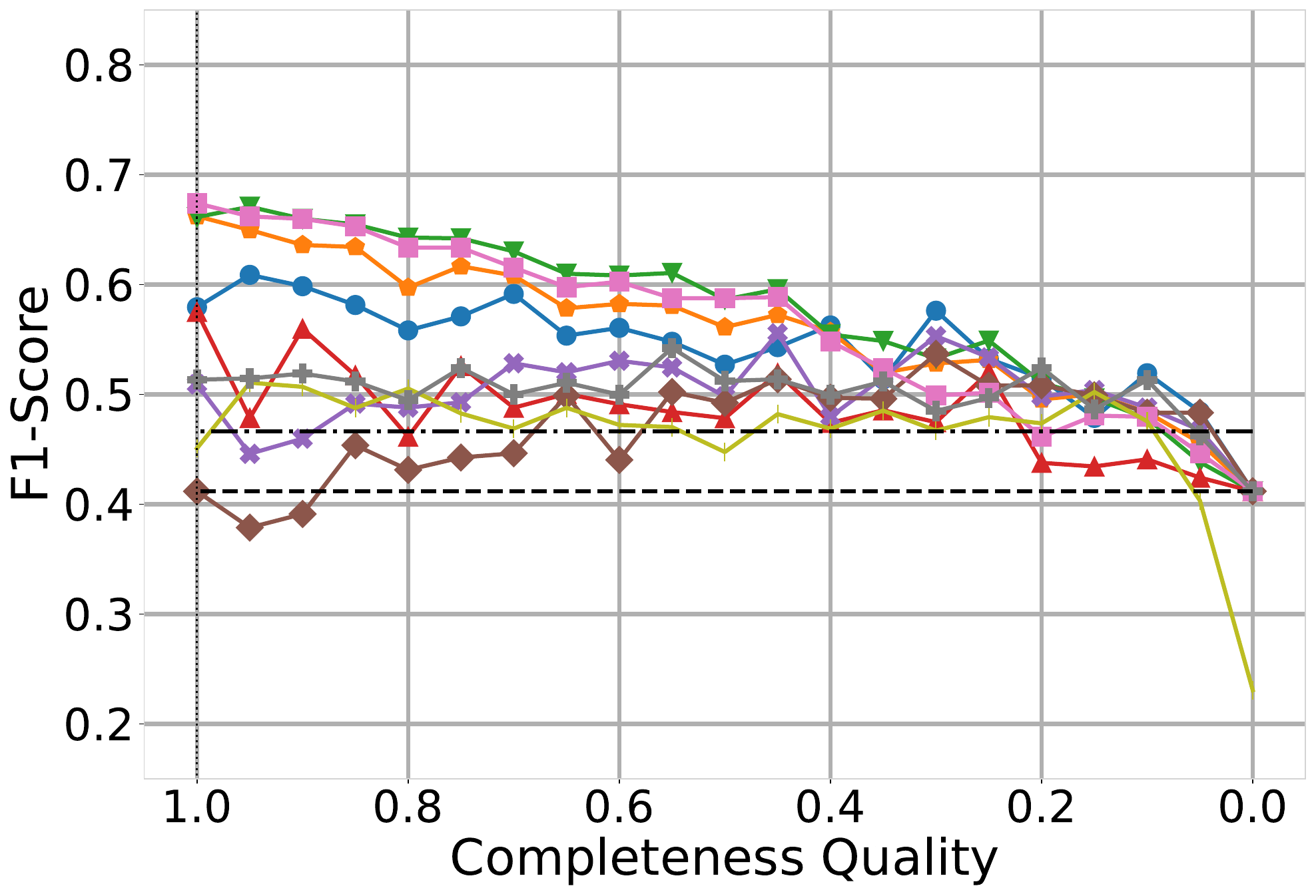}
        \caption{\textsf{Credit}}
        \label{fig:classification-results-all-completeness-3-credit}
    \end{subfigure}
\begin{subfigure}[b]{0.23\linewidth}
        \includegraphics[width=\linewidth]{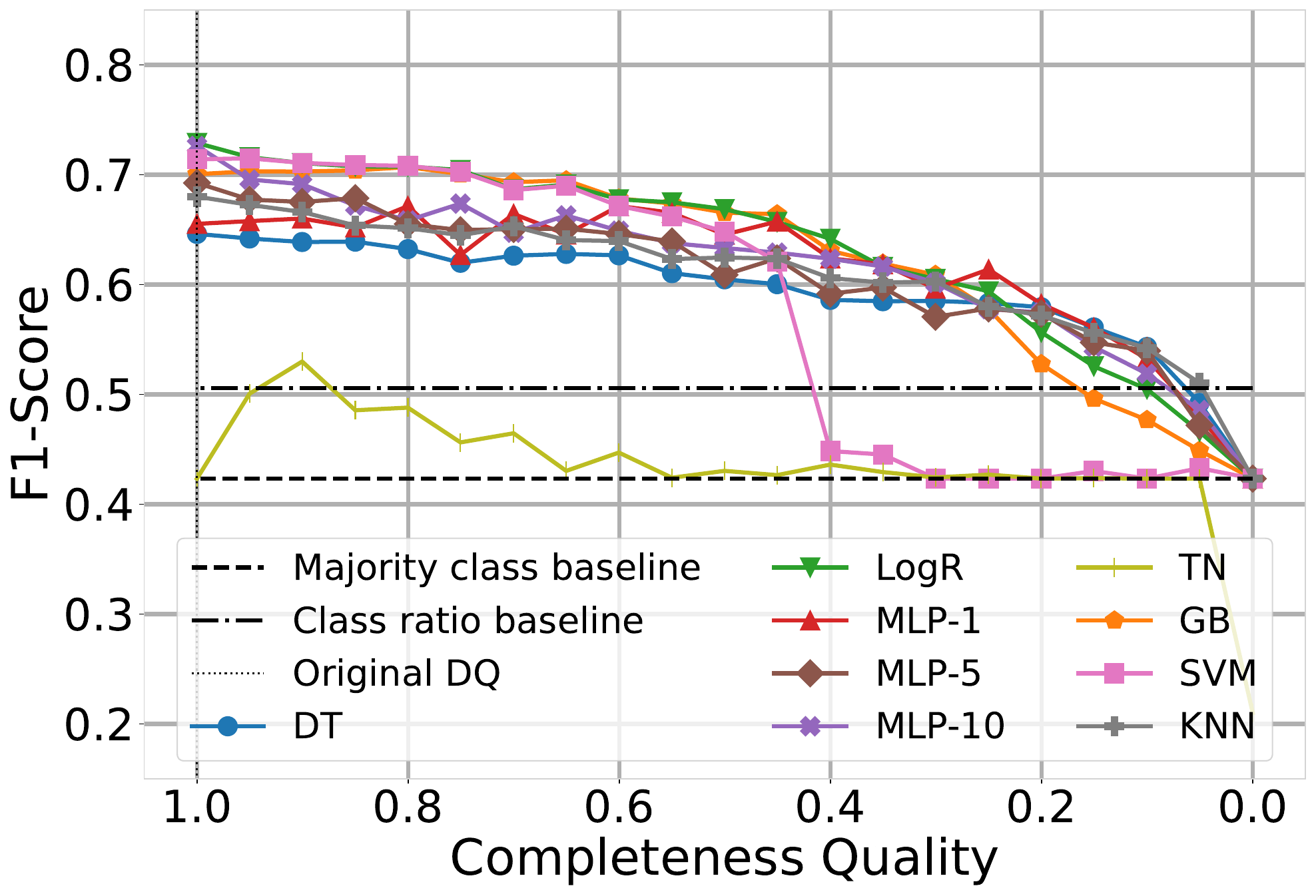}
        \caption{\textsf{Telco}}
        \label{fig:classification-results-all-completeness-3-telco}
    \end{subfigure}
    \caption{$F_1$-scores of the classification algorithms for completeness.}
    \label{fig:classification-results-all-completeness}
\end{figure*}

%% file: Latex_Figure/classification/Feature_Accurecy.tex
\begin{figure*}[t]
    \centering
\raisebox{0.4\height}{\rotatebox{90}{Scenario 1}}\hspace{0.1em}
\begin{subfigure}[b]{0.23\linewidth}
        \includegraphics[width=\linewidth]{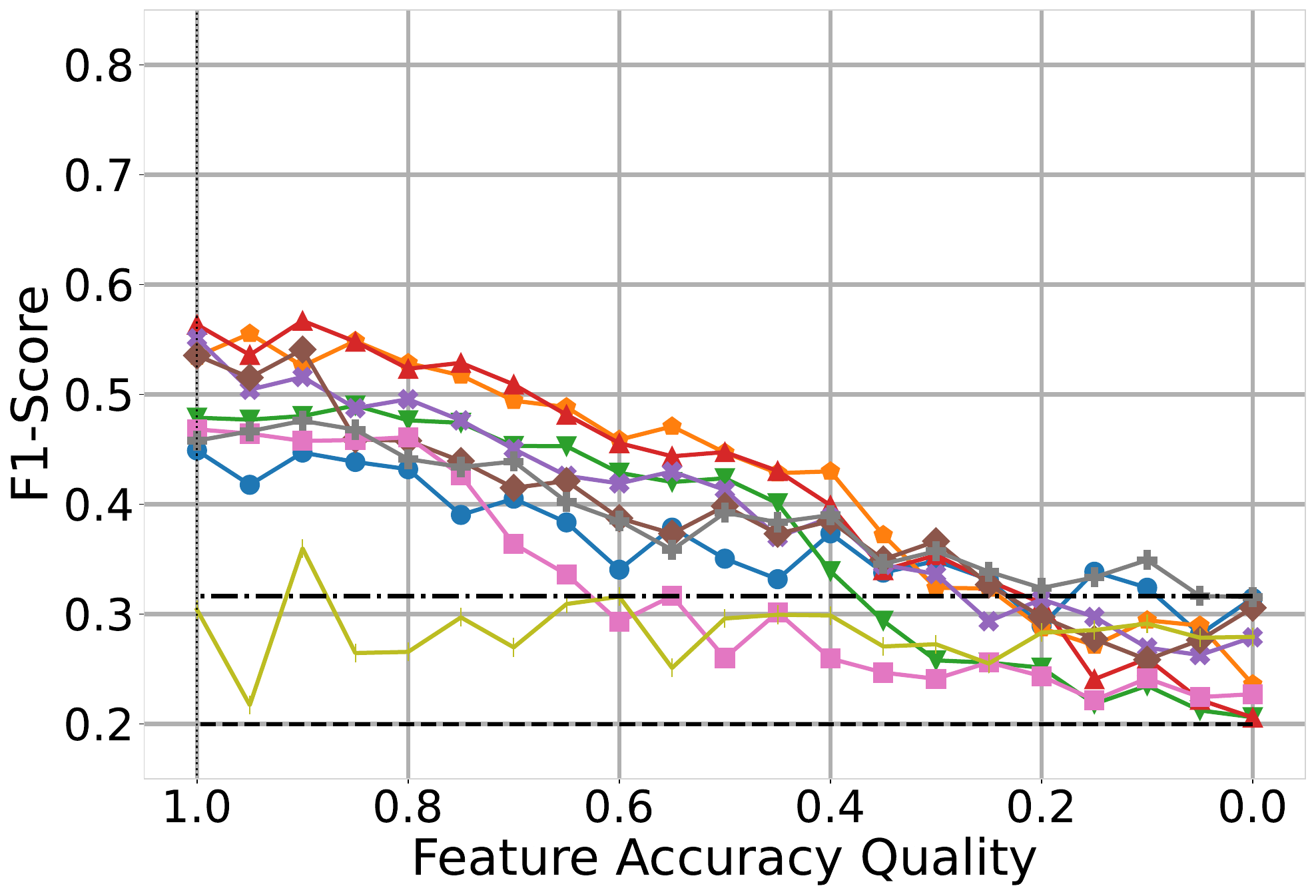}
        \caption{\textsf{Contraceptive}}
        \label{fig:classification-results-all-FeatureAccuracy-1-contra}
    \end{subfigure}
\begin{subfigure}[b]{0.23\linewidth}
        \includegraphics[width=\linewidth]{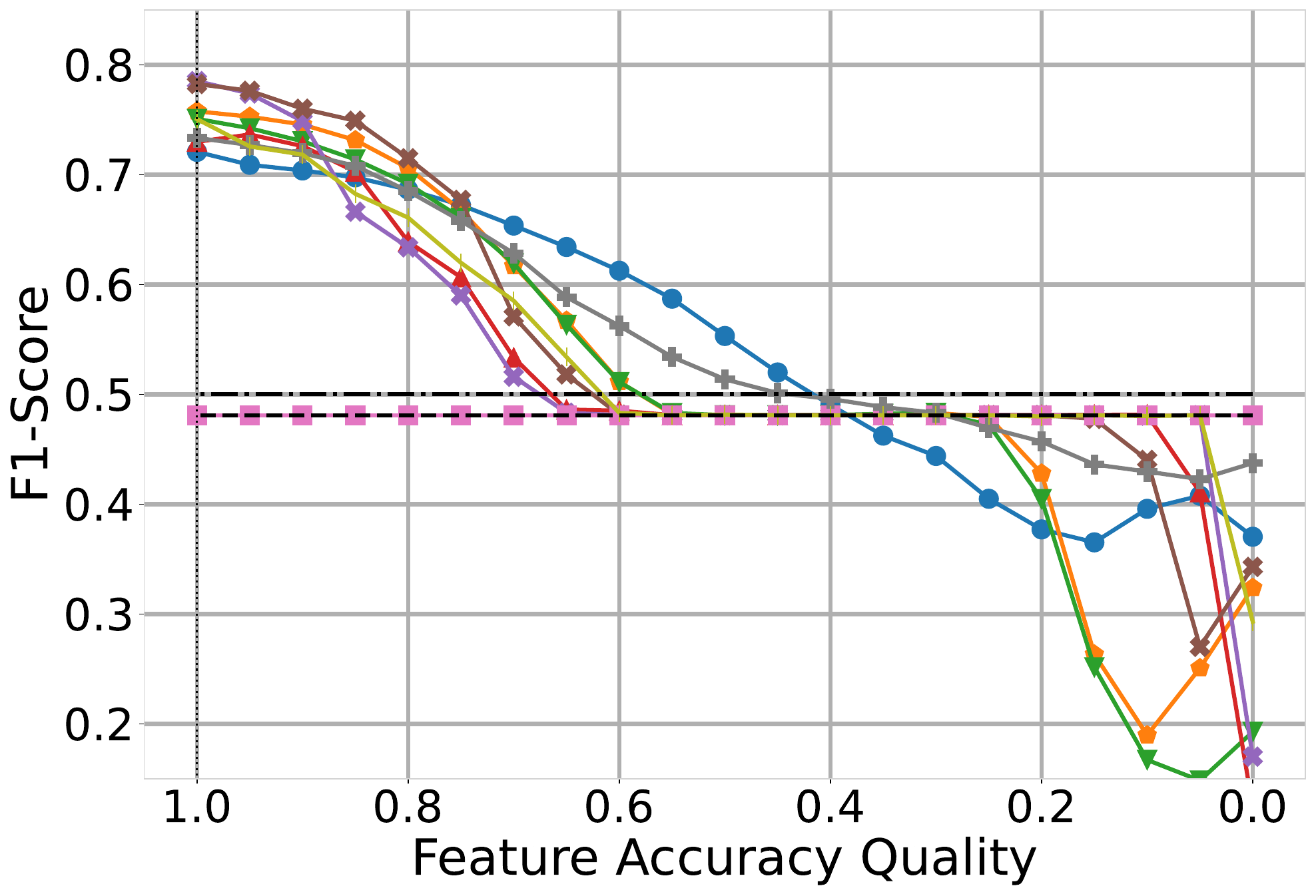}
        \caption{\textsf{COVID}}
        \label{fig:classification-results-all-FeatureAccuracy-1-covid}
    \end{subfigure}
\begin{subfigure}[b]{0.23\linewidth}
        \includegraphics[width=\linewidth]{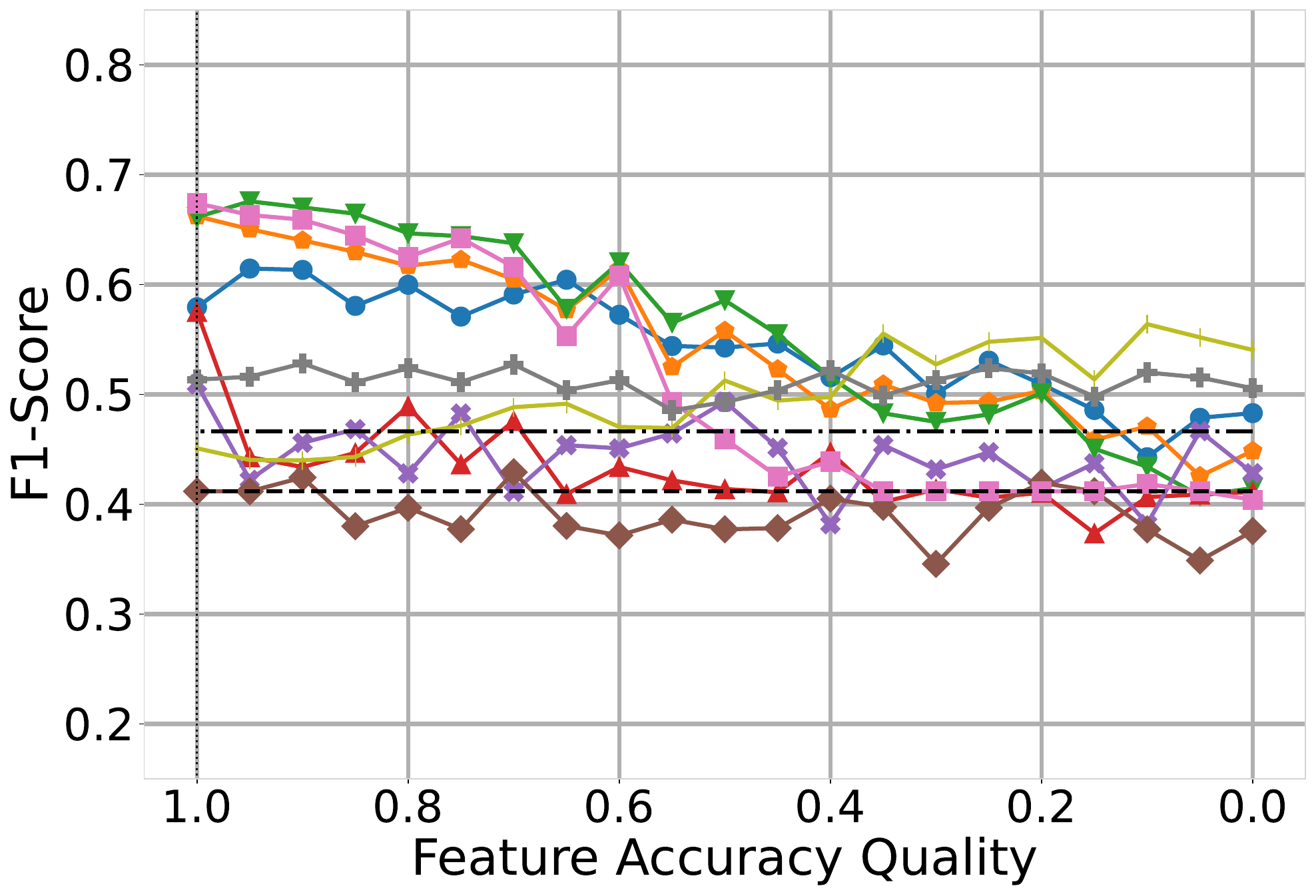}
        \caption{\textsf{Credit}}
        \label{fig:classification-results-all-FeatureAccuracy-1-credit}
    \end{subfigure}
\begin{subfigure}[b]{0.23\linewidth}
        \includegraphics[width=\linewidth]{figures/classification/telco_train_polluted_test_clean_FeatureAccuracyPolluter.pdf}
        \caption{\textsf{Telco}}
        \label{fig:classification-results-all-FeatureAccuracy-1-telco}
    \end{subfigure}

\raisebox{0.4\height}{\rotatebox{90}{Scenario 2}}\hspace{0.1em}
\begin{subfigure}[b]{0.23\linewidth}
        \includegraphics[width=\linewidth]{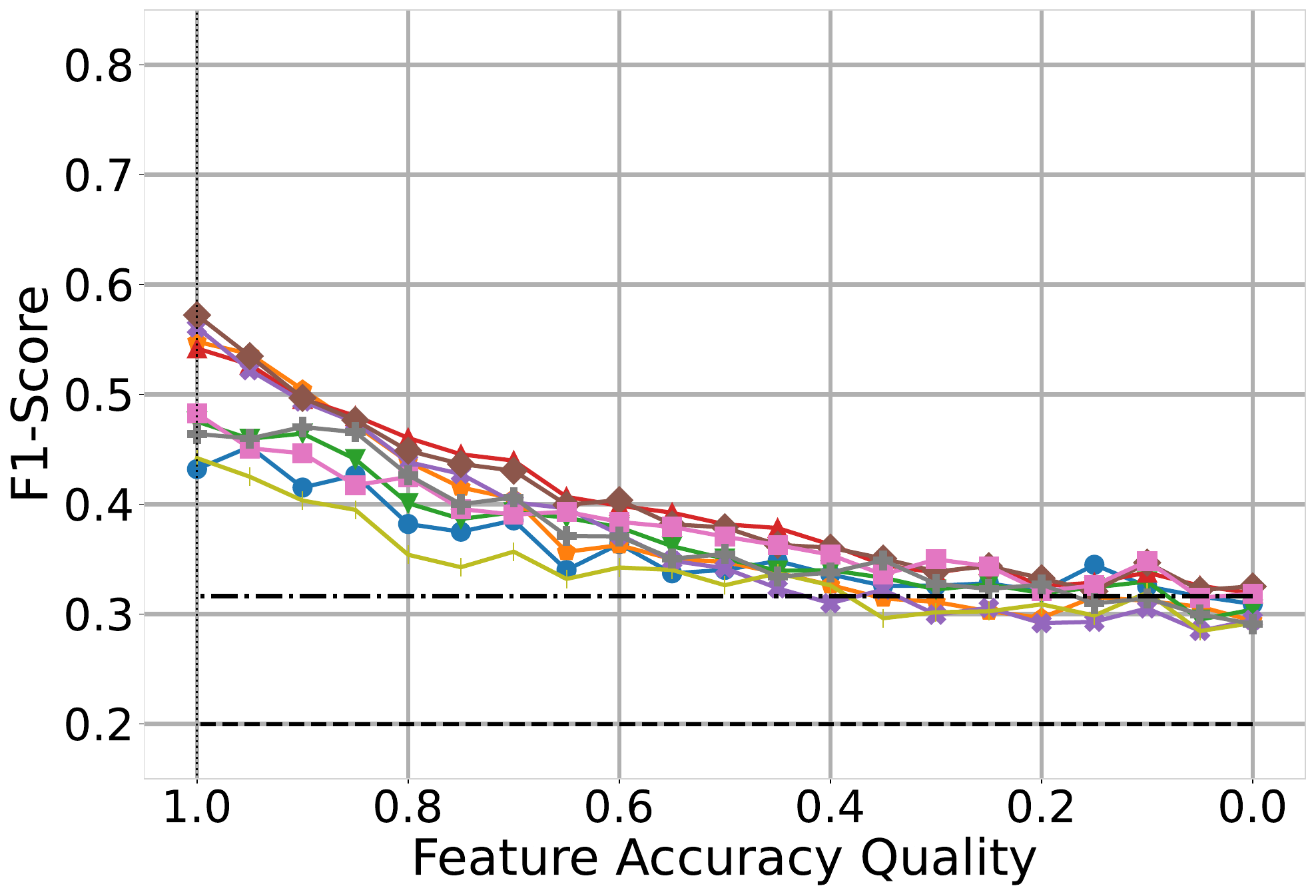}
        \caption{\textsf{Contraceptive}}
        \label{fig:classification-results-all-FeatureAccuracy-2-contra}
    \end{subfigure}
\begin{subfigure}[b]{0.23\linewidth}
        \includegraphics[width=\linewidth]{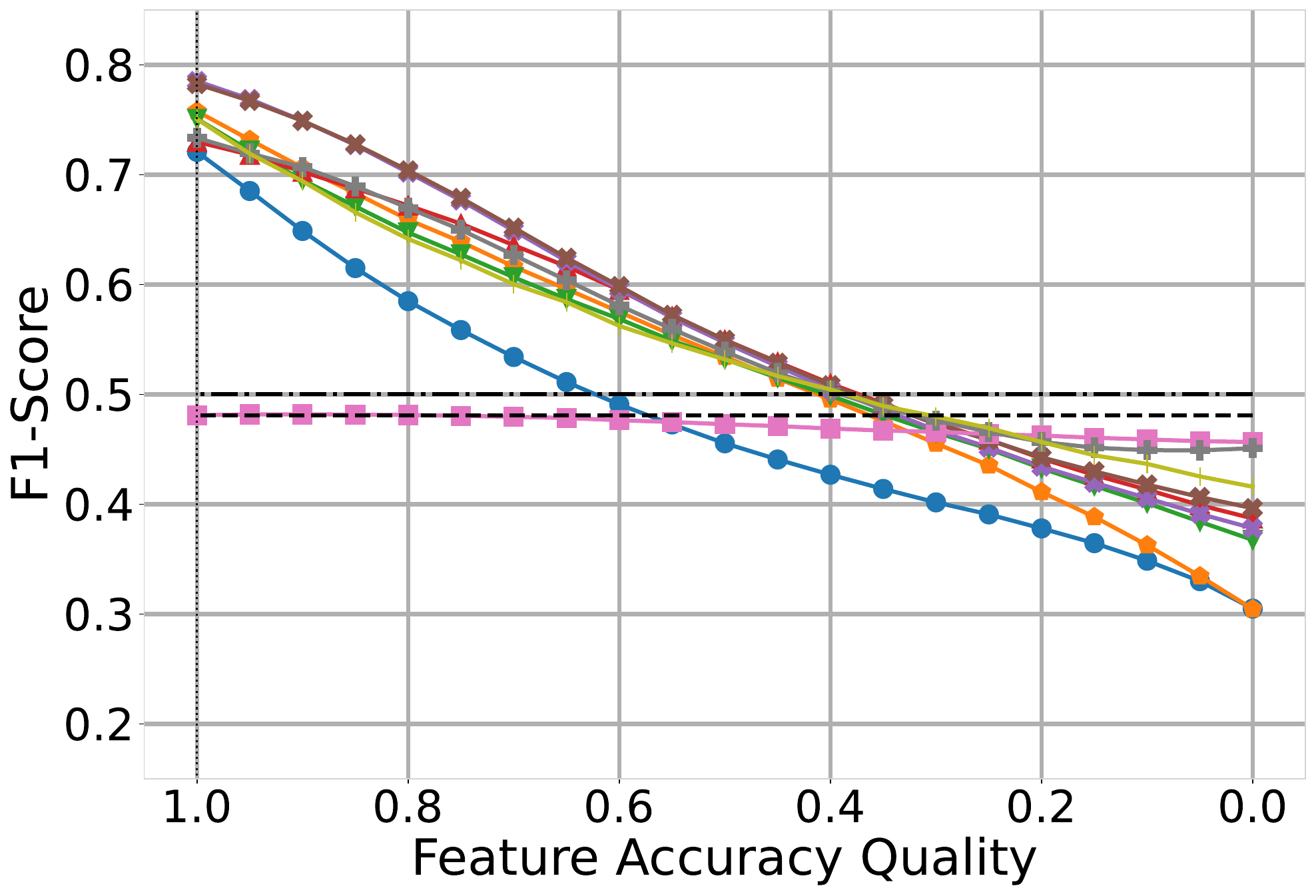}
        \caption{\textsf{COVID}}
        \label{fig:classification-results-all-FeatureAccuracy-2-covid}
    \end{subfigure}
\begin{subfigure}[b]{0.23\linewidth}
        \includegraphics[width=\linewidth]{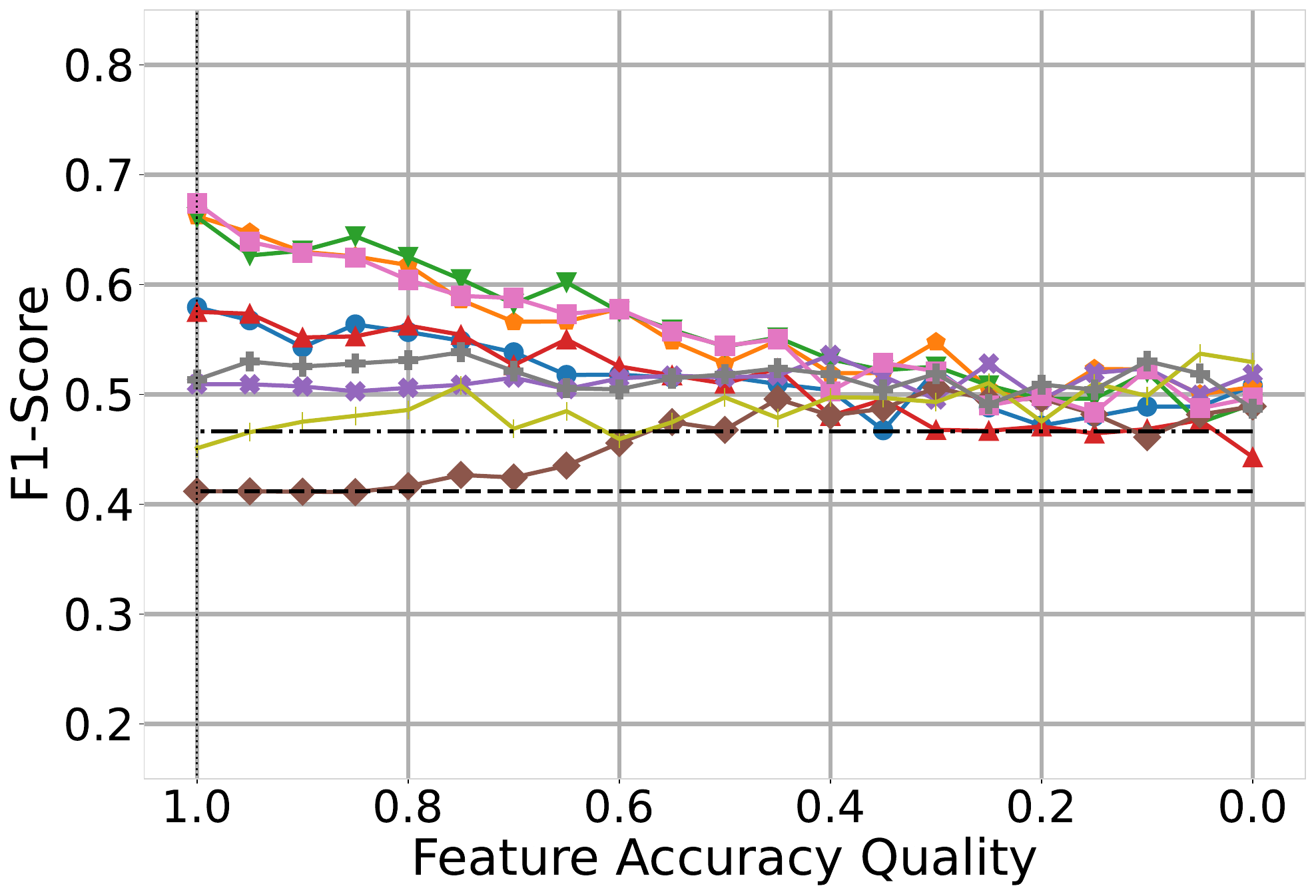}
        \caption{\textsf{Credit}}
        \label{fig:classification-results-all-FeatureAccuracy-2-credit}
    \end{subfigure}
\begin{subfigure}[b]{0.23\linewidth}
        \includegraphics[width=\linewidth]{figures/classification/telco_train_clean_test_polluted_FeatureAccuracyPolluter.pdf}
        \caption{\textsf{Telco}}
        \label{fig:classification-results-all-FeatureAccuracy-2-telco}
    \end{subfigure}

\raisebox{0.4\height}{\rotatebox{90}{Scenario 3}}\hspace{0.1em}
\begin{subfigure}[b]{0.23\linewidth}
        \includegraphics[width=\linewidth]{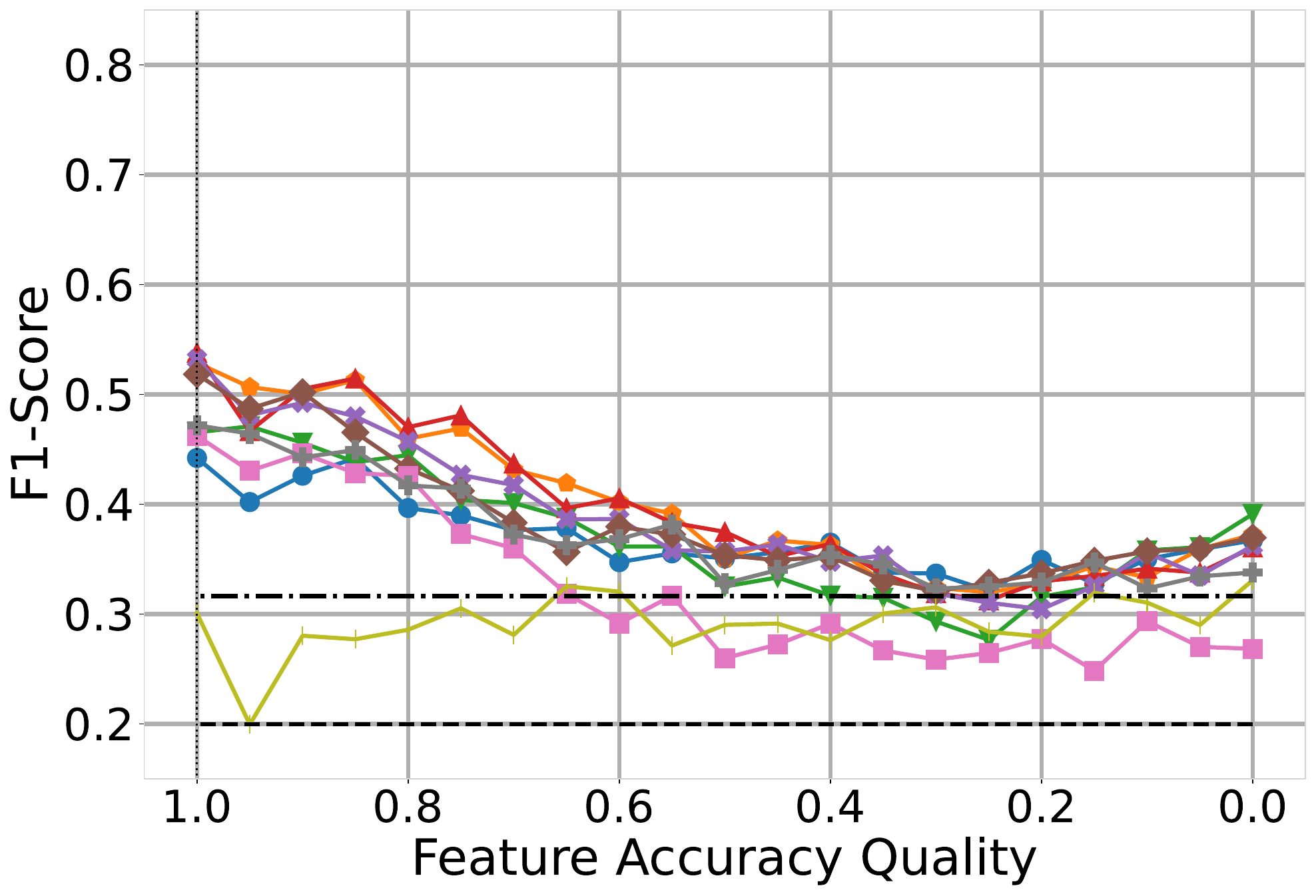}
        \caption{\textsf{Contraceptive}}
        \label{fig:classification-results-all-FeatureAccuracy-3-contra}
    \end{subfigure}
\begin{subfigure}[b]{0.23\linewidth}
        \includegraphics[width=\linewidth]{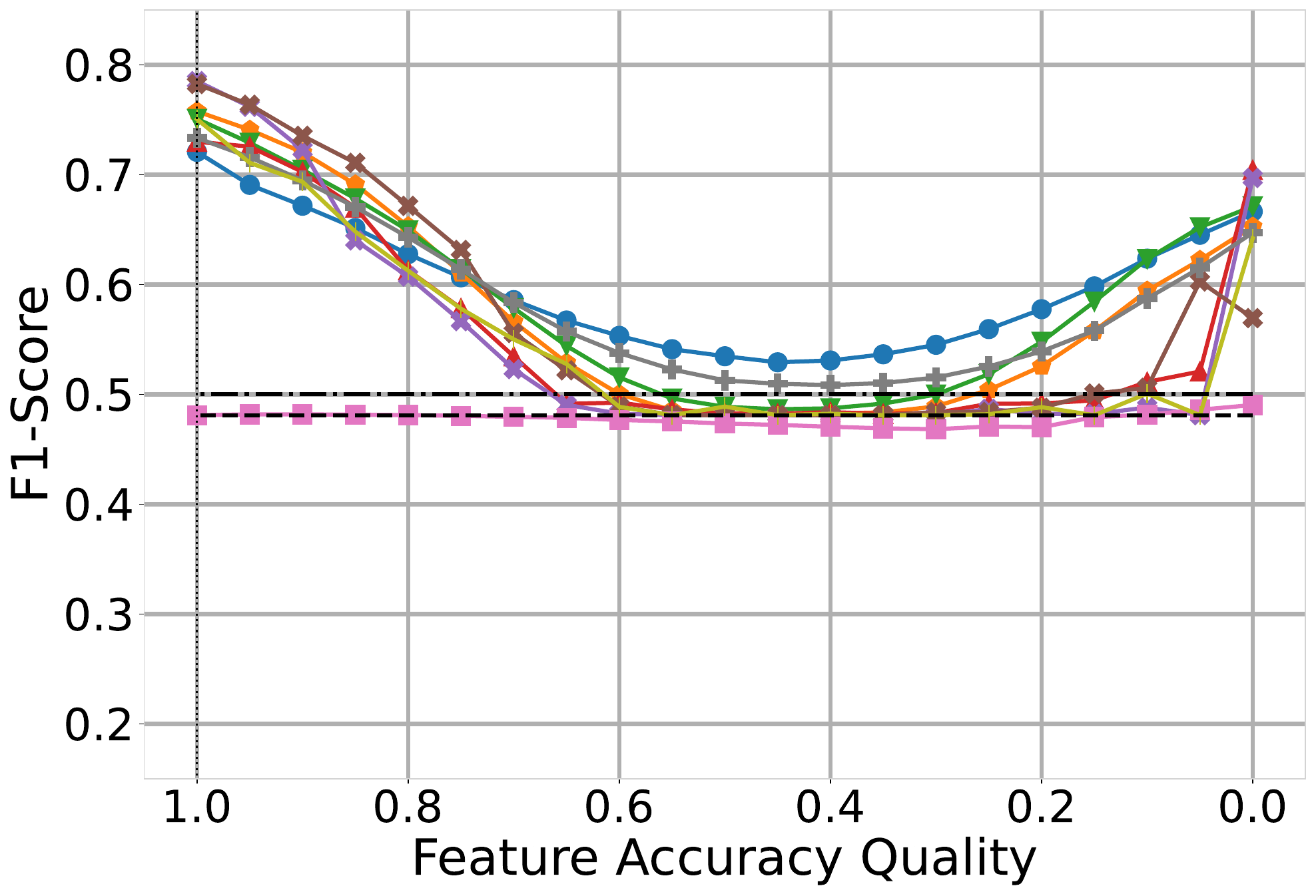}
        \caption{\textsf{COVID}}
        \label{fig:classification-results-all-FeatureAccuracy-3-covid}
    \end{subfigure}
\begin{subfigure}[b]{0.23\linewidth}
        \includegraphics[width=\linewidth]{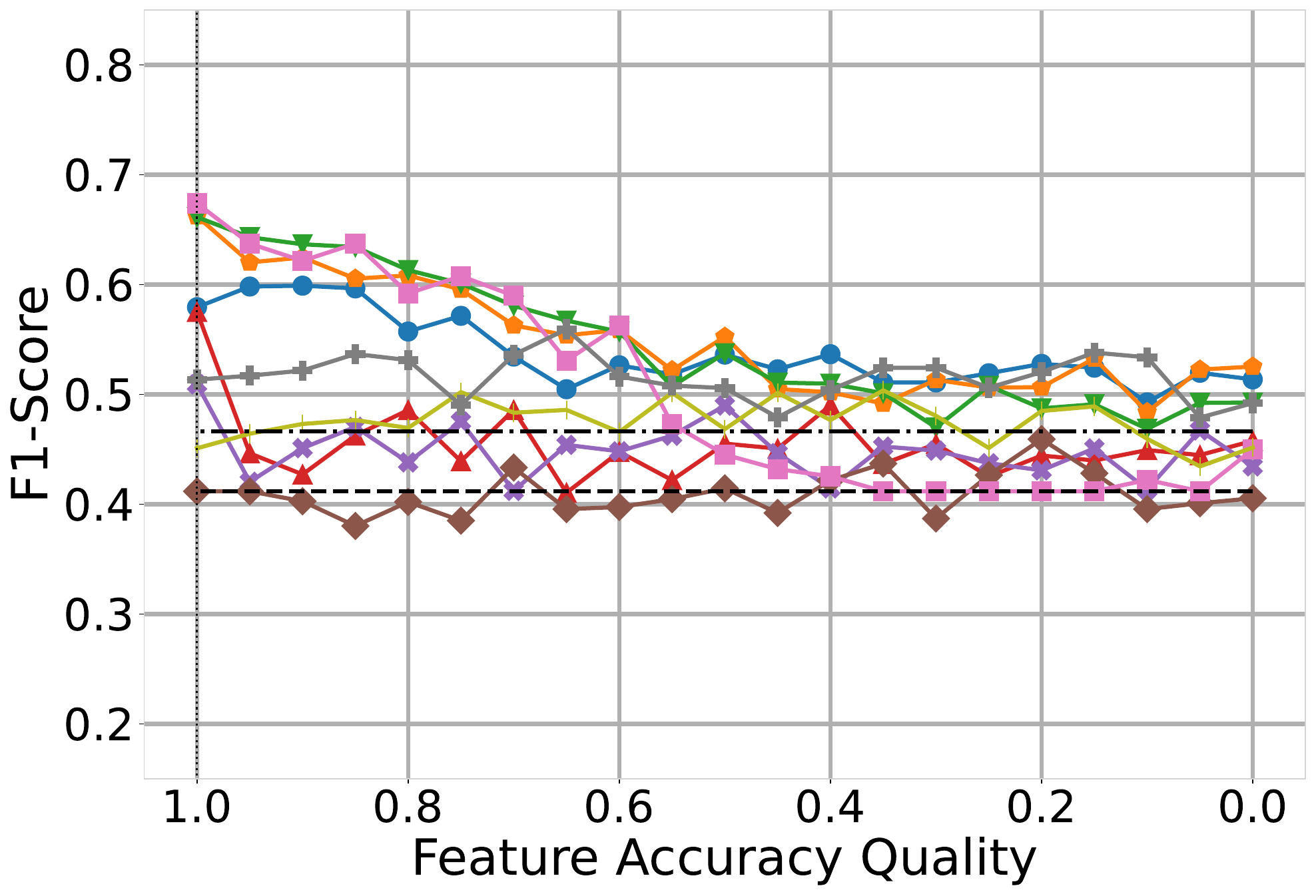}
        \caption{\textsf{Credit}}
        \label{fig:classification-results-all-FeatureAccuracy-3-credit}
    \end{subfigure}
\begin{subfigure}[b]{0.23\linewidth}
        \includegraphics[width=\linewidth]{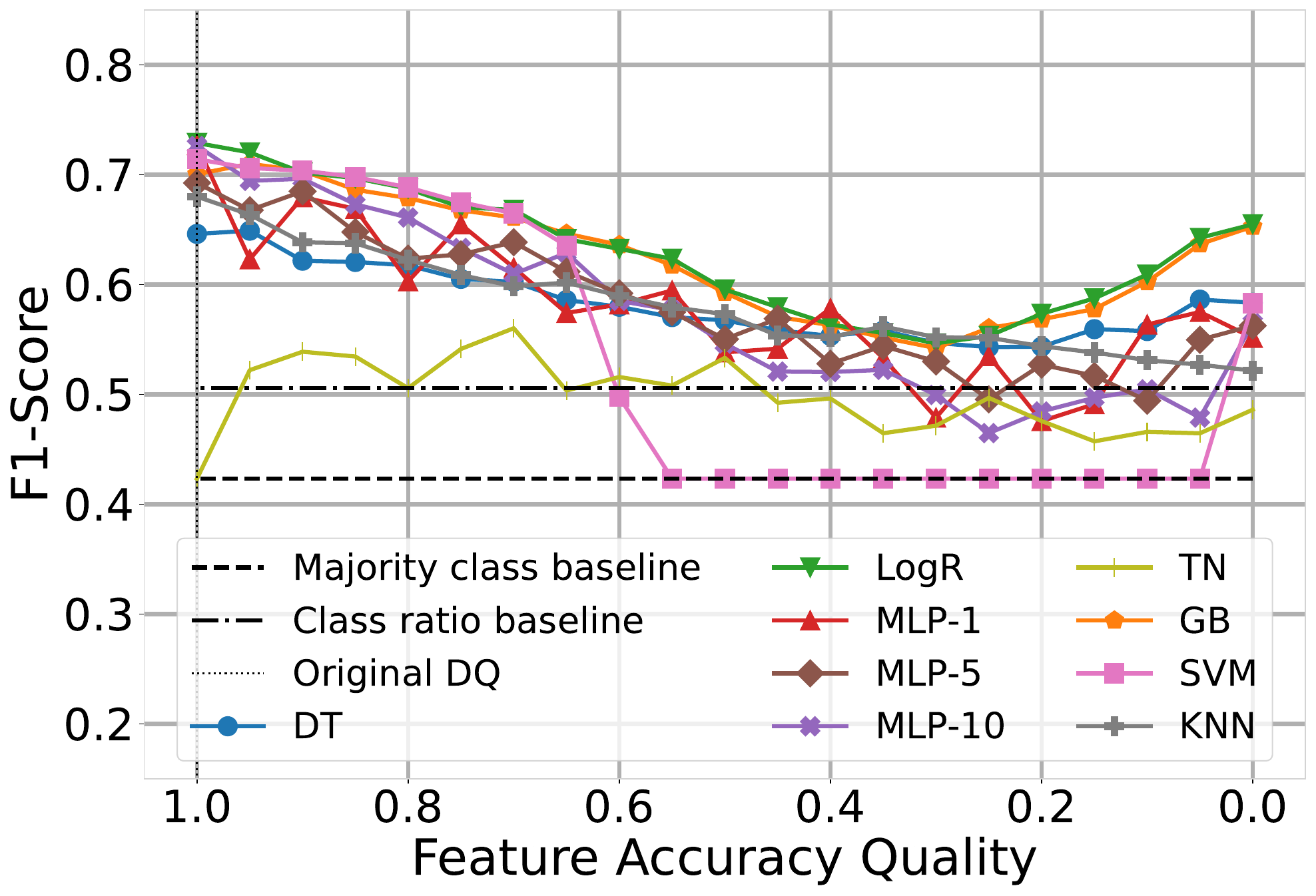}
        \caption{\textsf{Telco}}
        \label{fig:classification-results-all-FeatureAccuracy-3-telco}
    \end{subfigure}
    \caption{$F_1$-scores of the classification algorithms for feature accuracy.}
    \label{fig:classification-results-all-FeatureAccuracy}
\end{figure*}

%% file: Latex_Figure/classification/Target_Accurecy.tex
\begin{figure*}[t]
    \centering
\raisebox{0.4\height}{\rotatebox{90}{Scenario 1}}\hspace{0.3em}
\begin{subfigure}[b]{0.23\linewidth}
        \includegraphics[width=\linewidth]{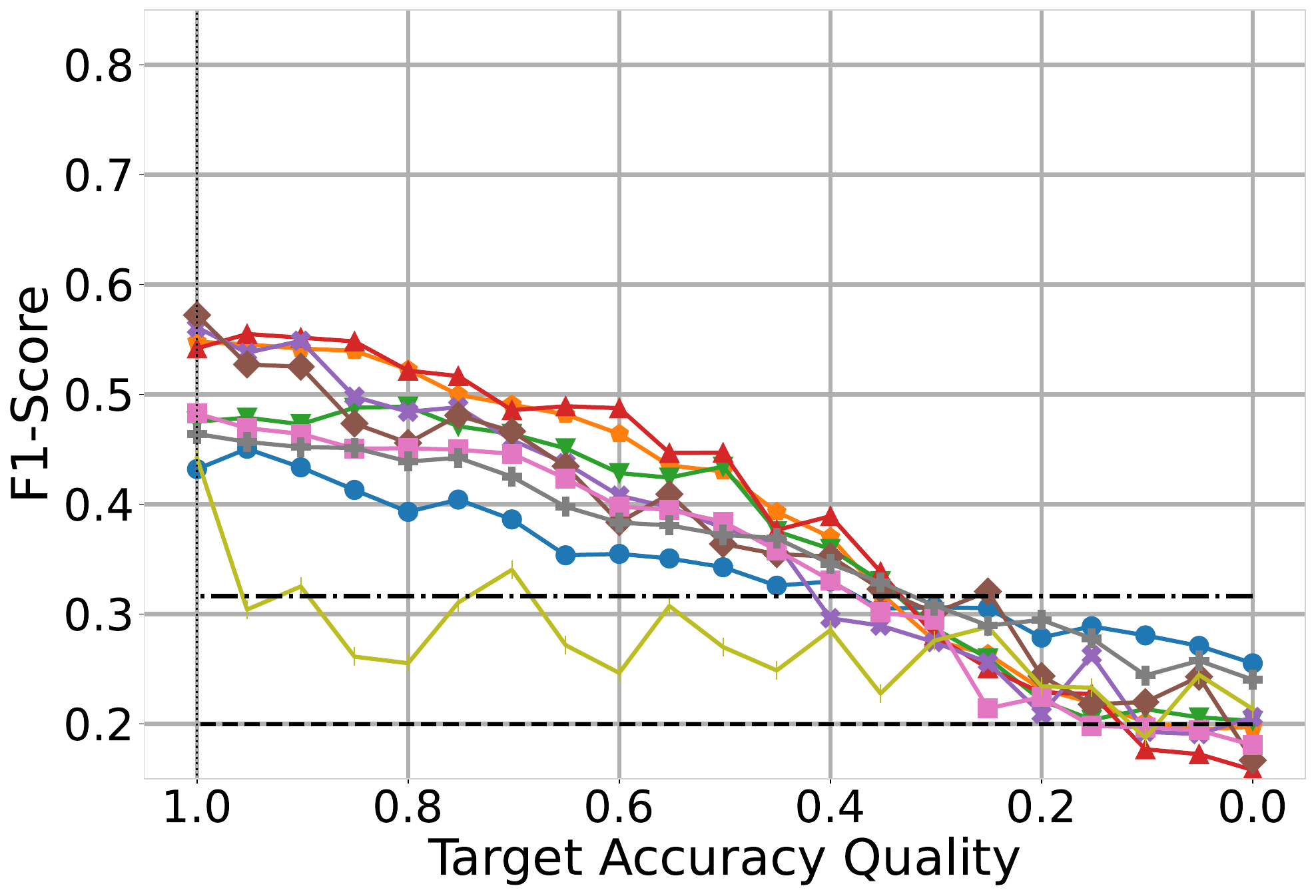}
        \caption{\textsf{Contraceptive}}
        \label{fig:classification-results-all-TargetAccuracy-1-contra}
    \end{subfigure}
\begin{subfigure}[b]{0.23\linewidth}
        \includegraphics[width=\linewidth]{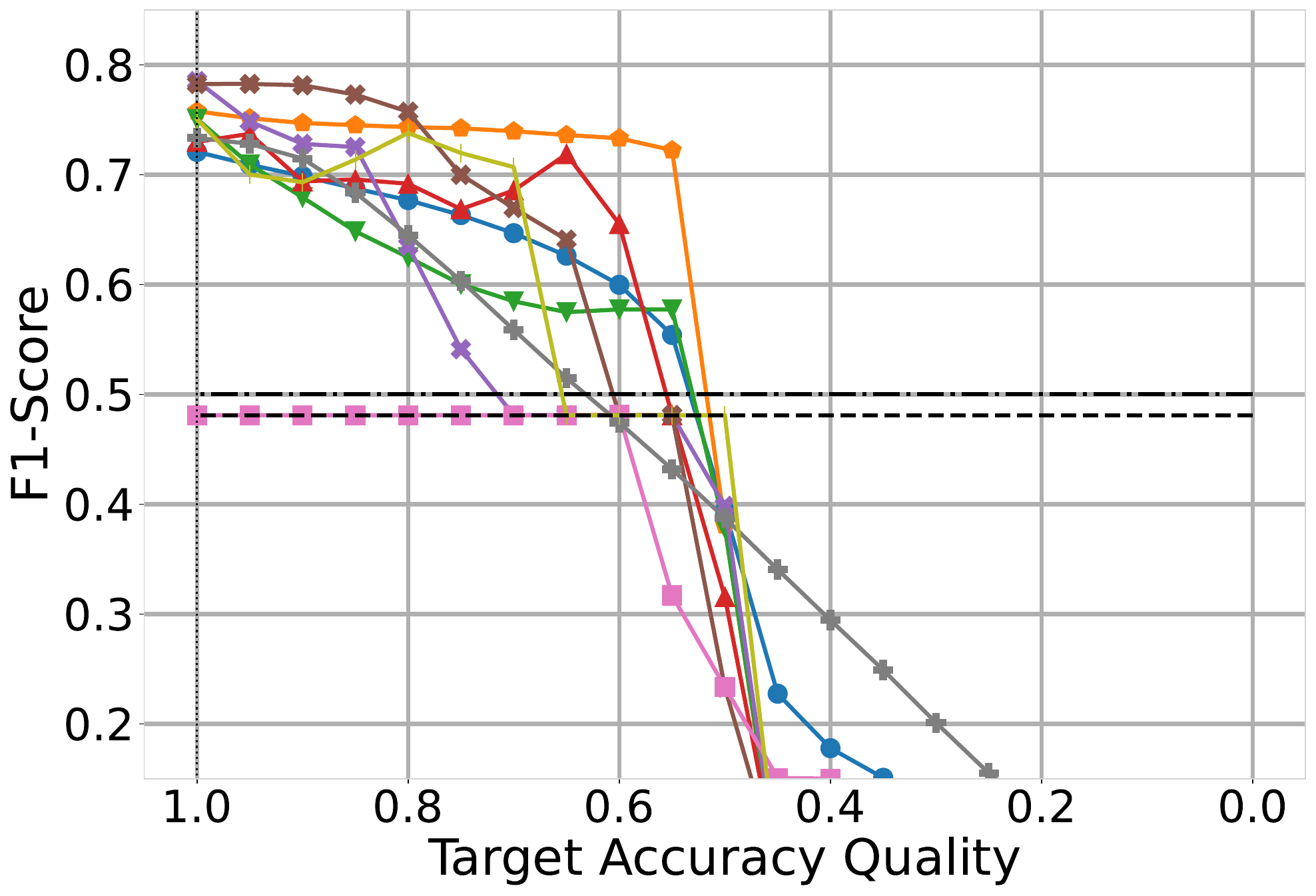}
        \caption{\textsf{COVID}}
        \label{fig:classification-results-all-TargetAccuracy-1-covid}
    \end{subfigure}
\begin{subfigure}[b]{0.23\linewidth}
        \includegraphics[width=\linewidth]{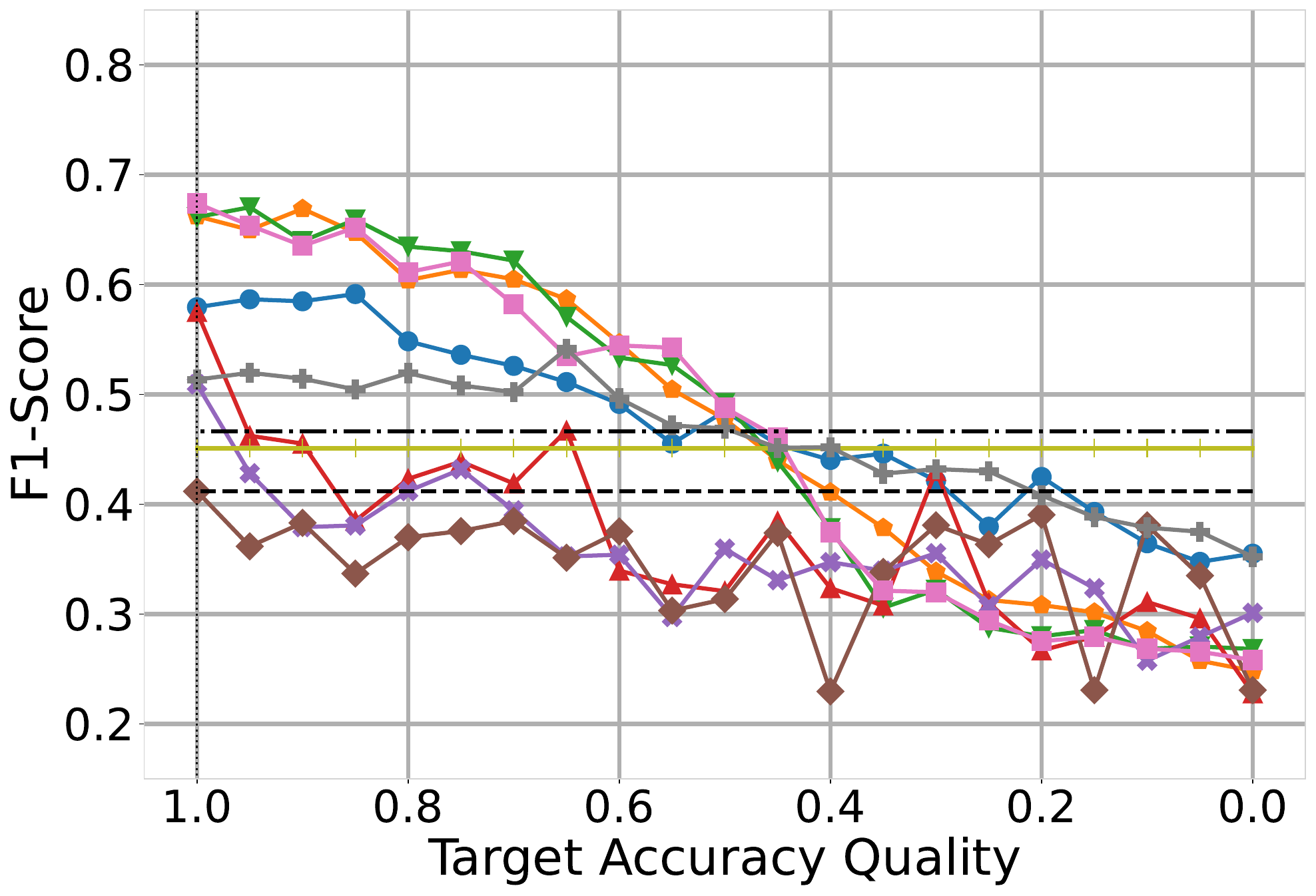}
        \caption{\textsf{Credit}}
        \label{fig:classification-results-all-TargetAccuracy-1-credit}
    \end{subfigure}
\begin{subfigure}[b]{0.23\linewidth}
        \includegraphics[width=\linewidth]{figures/classification/telco_train_polluted_test_clean_TargetAccuracyPolluter.pdf}
        \caption{\textsf{Telco}}
        \label{fig:classification-results-all-TargetAccuracy-1-telco}
    \end{subfigure}

\raisebox{0.4\height}{\rotatebox{90}{Scenario 2}}\hspace{0.3em}
\begin{subfigure}[b]{0.23\linewidth}
        \includegraphics[width=\linewidth]{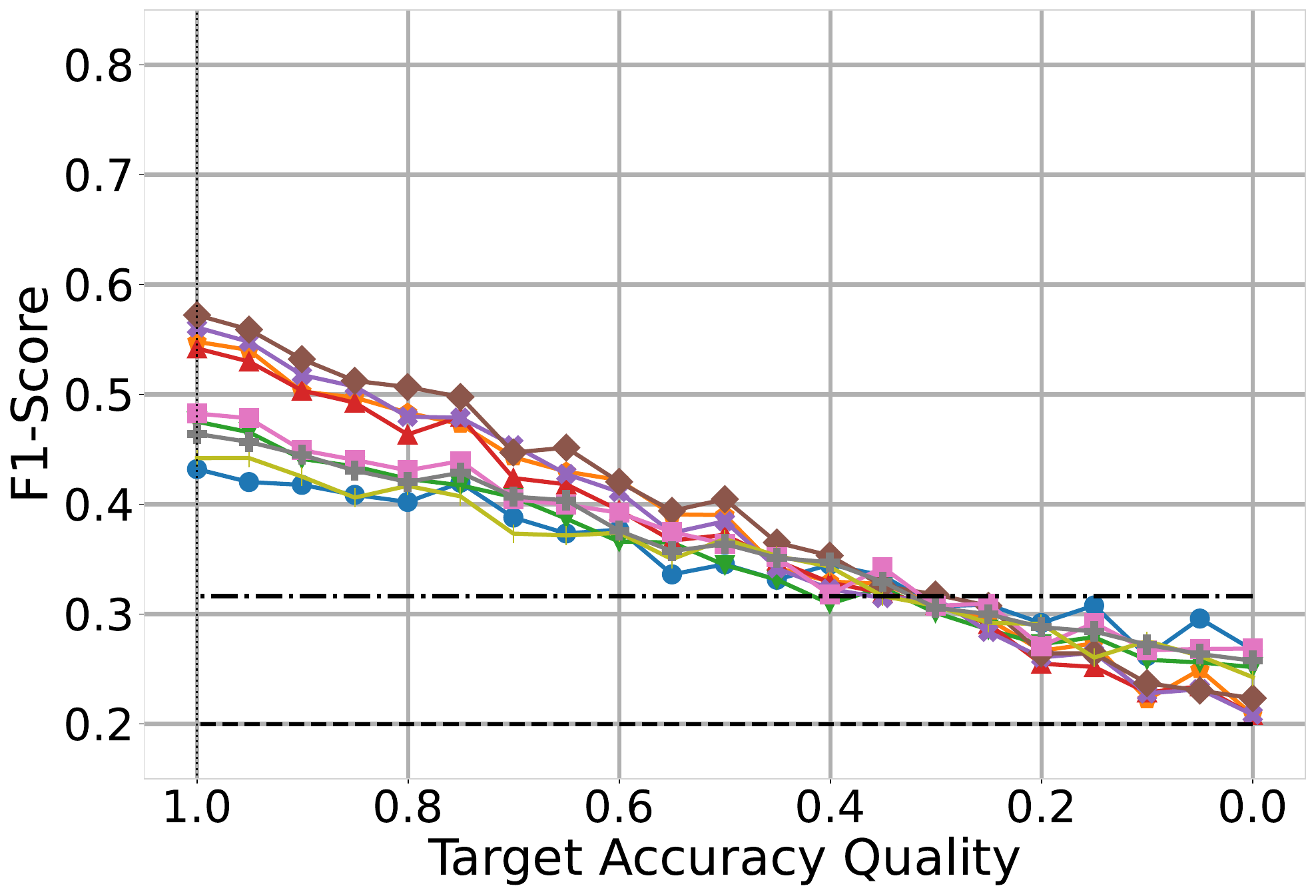}
        \caption{\textsf{Contraceptive}}
        \label{fig:classification-results-all-TargetAccuracy-2-contra}
    \end{subfigure}
\begin{subfigure}[b]{0.23\linewidth}
        \includegraphics[width=\linewidth]{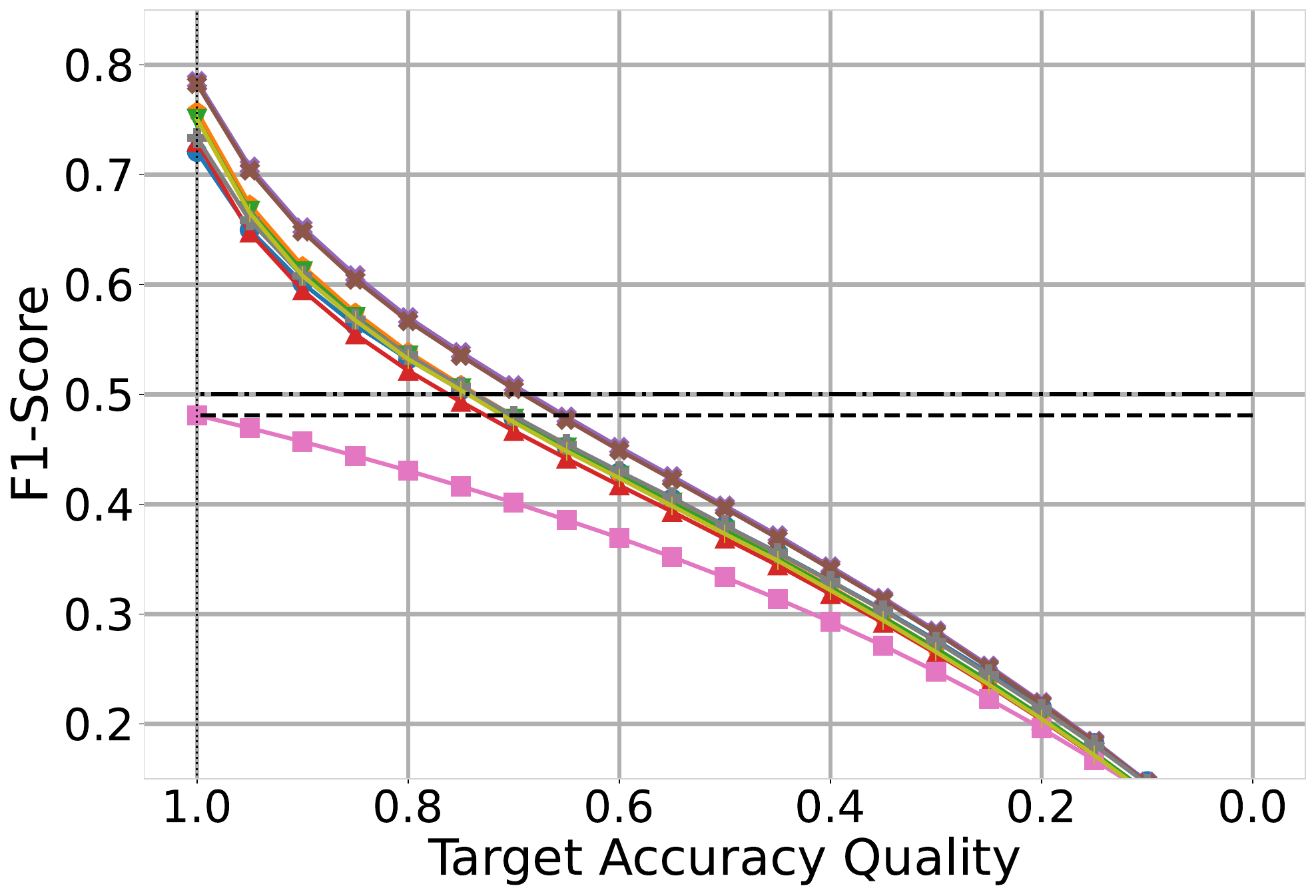}
        \caption{\textsf{COVID}}
        \label{fig:classification-results-all-TargetAccuracy-2-covid}
    \end{subfigure}
\begin{subfigure}[b]{0.23\linewidth}
        \includegraphics[width=\linewidth]{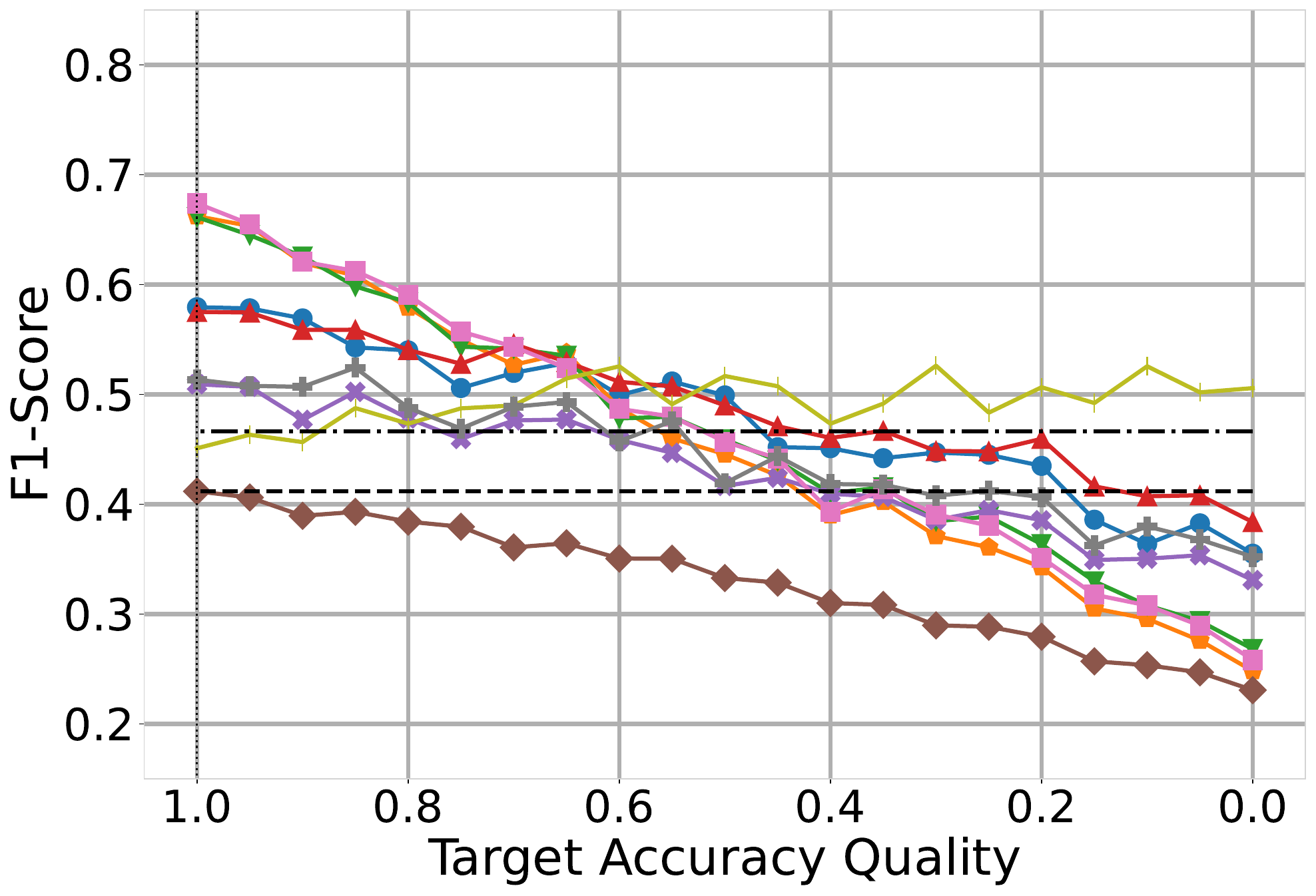}
        \caption{\textsf{Credit}}
        \label{fig:classification-results-all-TargetAccuracy-2-credit}
    \end{subfigure}
\begin{subfigure}[b]{0.23\linewidth}
        \includegraphics[width=\linewidth]{figures/classification/telco_train_clean_test_polluted_TargetAccuracyPolluter.pdf}
        \caption{\textsf{Telco}}
        \label{fig:classification-results-all-TargetAccuracy-2-telco}
    \end{subfigure}

\raisebox{0.4\height}{\rotatebox{90}{Scenario 3}}\hspace{0.3em}
\begin{subfigure}[b]{0.23\linewidth}
        \includegraphics[width=\linewidth]{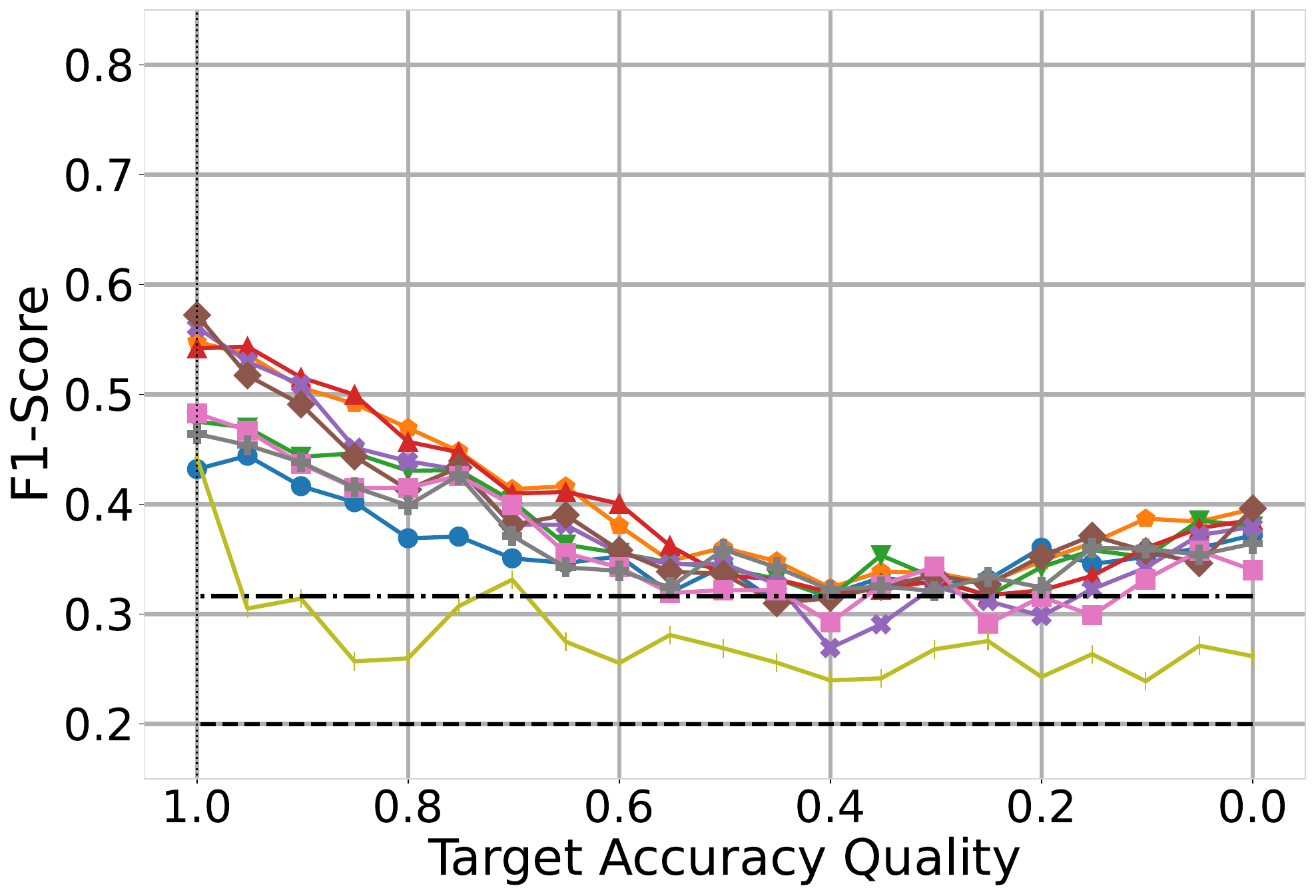}
        \caption{\textsf{Contraceptive}}
        \label{fig:classification-results-all-TargetAccuracy-3-contra}
    \end{subfigure}
\begin{subfigure}[b]{0.23\linewidth}
        \includegraphics[width=\linewidth]{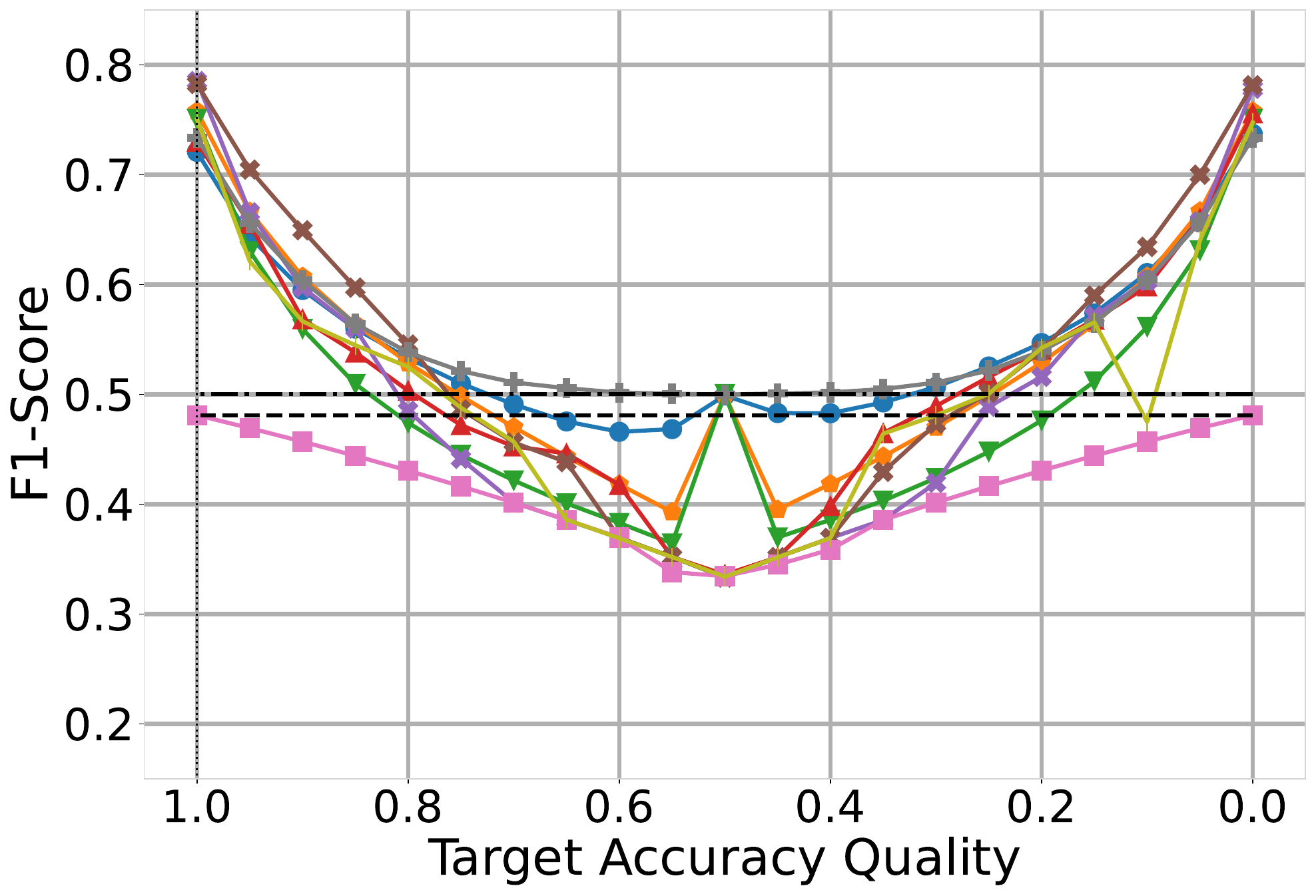}
        \caption{\textsf{COVID}}
        \label{fig:classification-results-all-TargetAccuracy-3-covid}
    \end{subfigure}
\begin{subfigure}[b]{0.23\linewidth}
        \includegraphics[width=\linewidth]{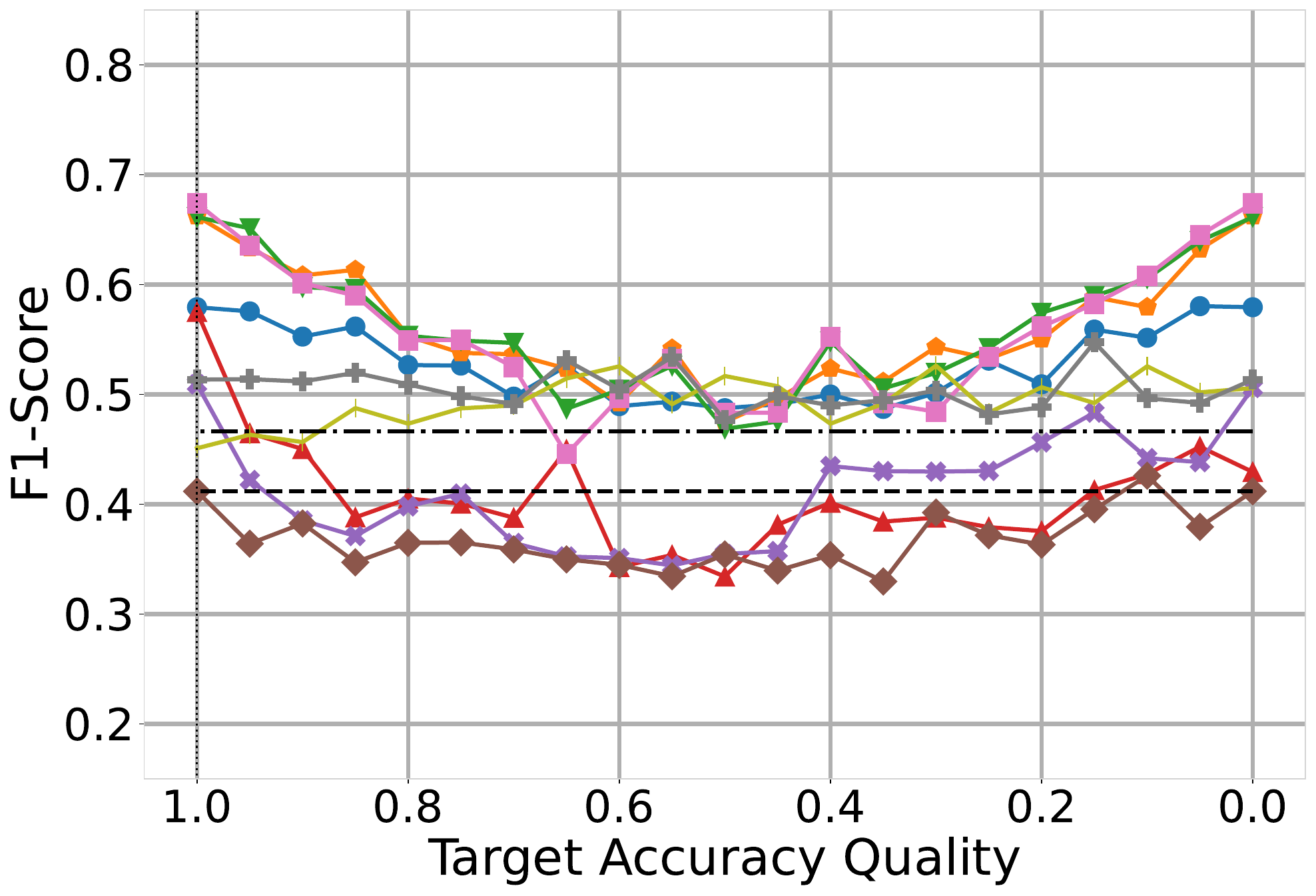}
        \caption{\textsf{Credit}}
        \label{fig:classification-results-all-TargetAccuracy-3-credit}
    \end{subfigure}
\begin{subfigure}[b]{0.23\linewidth}
        \includegraphics[width=\linewidth]{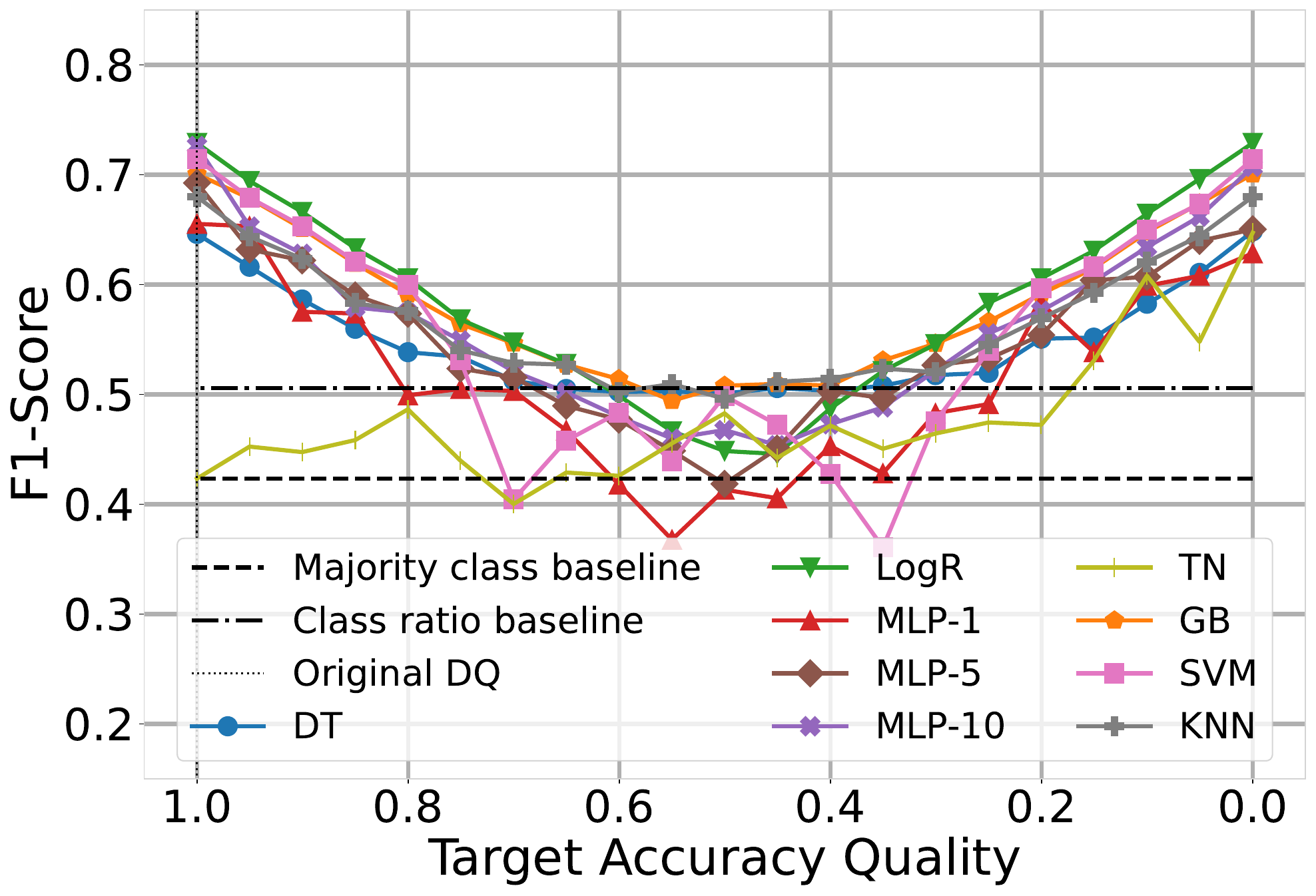}
        \caption{\textsf{Telco}}
        \label{fig:classification-results-all-TargetAccuracy-3-telco}
    \end{subfigure}
    \caption{$F_1$-scores of the classification algorithms for target accuracy.}
    \label{fig:classification-results-all-TargetAccuracy}
\end{figure*}

%% file: Latex_Figure/classification/Uniqueness.tex
\begin{figure*}[t]
    \centering
\raisebox{0.4\height}{\rotatebox{90}{Scenario 1}}\hspace{0.3em}
\begin{subfigure}[b]{0.23\linewidth}
        \includegraphics[width=\linewidth]{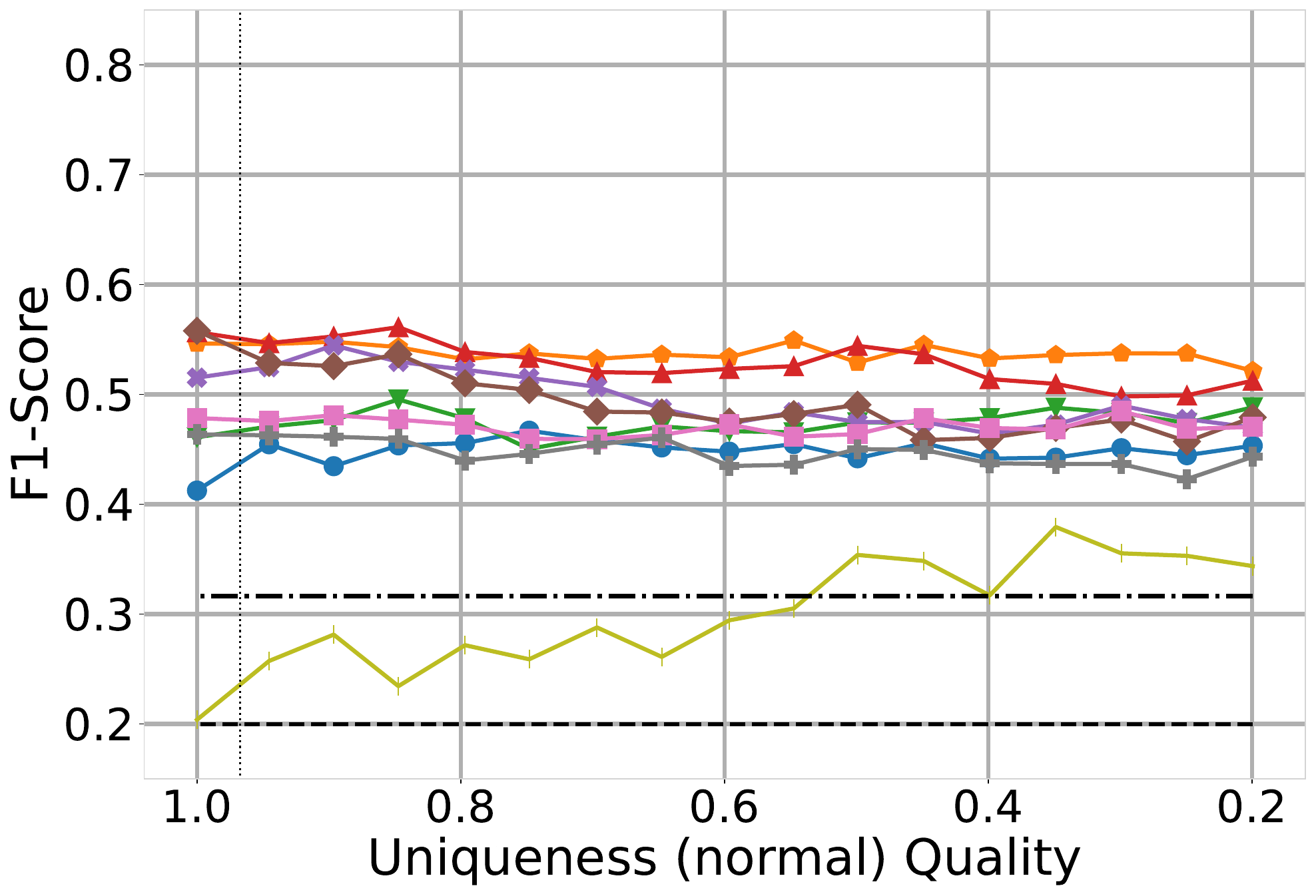}
        \caption{\textsf{Contraceptive}}
        \label{fig:classification-results-all-Uniqueness_dc1-1-contra}
    \end{subfigure}
\begin{subfigure}[b]{0.23\linewidth}
        \includegraphics[width=\linewidth]{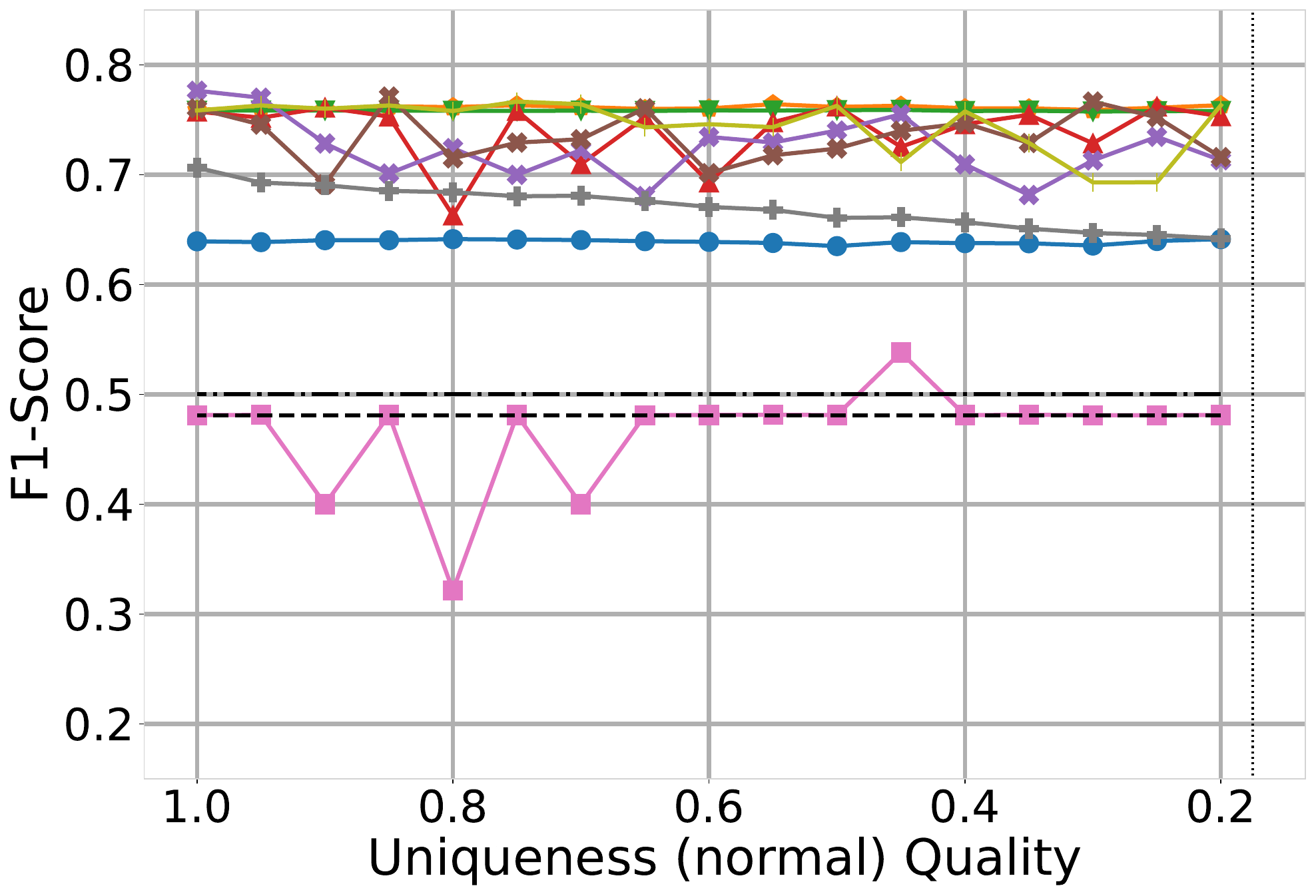}
        \caption{\textsf{COVID}}
        \label{fig:classification-results-all-Uniqueness_dc1-1-covid}
    \end{subfigure}
\begin{subfigure}[b]{0.23\linewidth}
        \includegraphics[width=\linewidth]{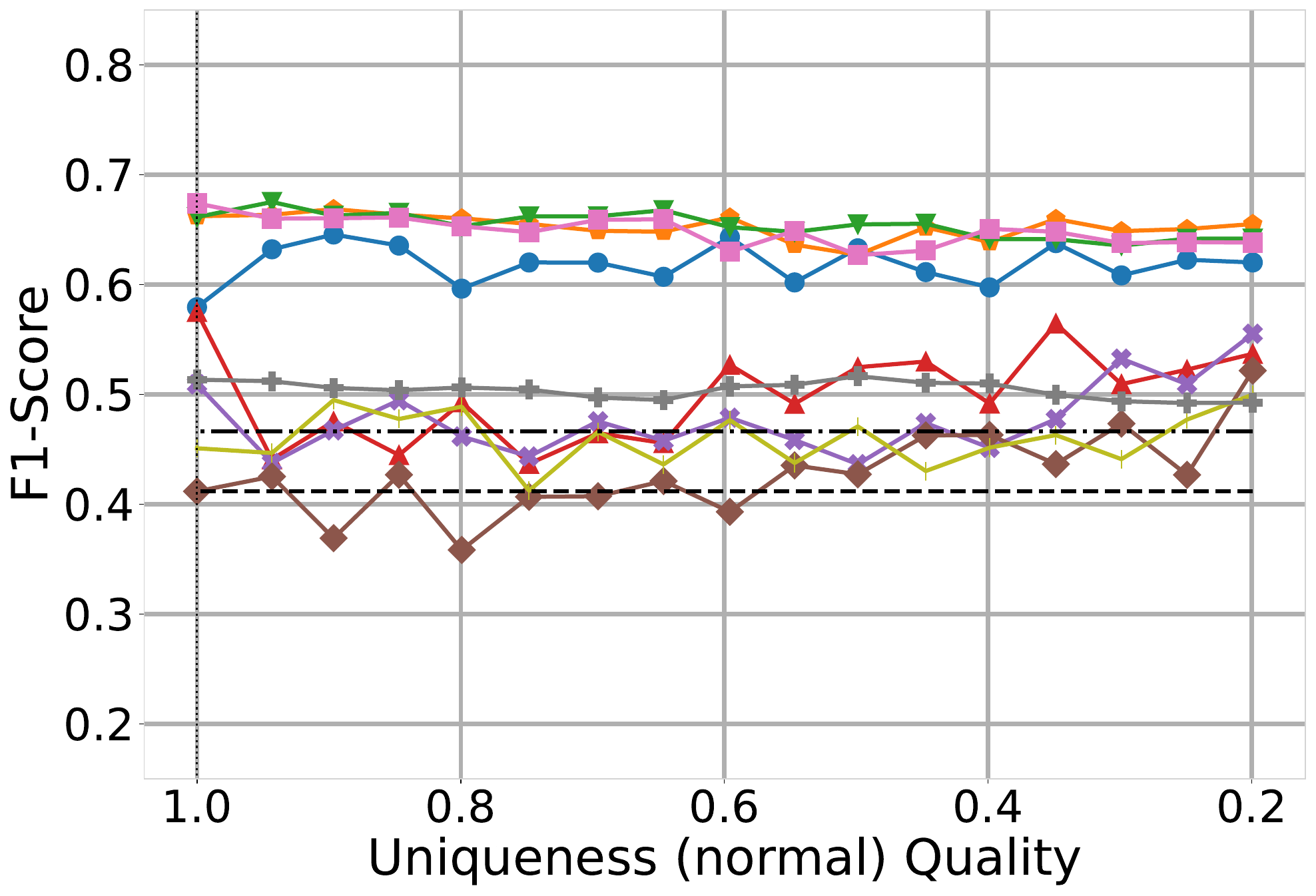}
        \caption{\textsf{Credit}}
        \label{fig:classification-results-all-Uniqueness_dc1-1-credit}
    \end{subfigure}
\begin{subfigure}[b]{0.23\linewidth}
        \includegraphics[width=\linewidth]{figures/classification/telco_train_polluted_test_clean_UniquenessPolluter_normal.pdf}
        \caption{\textsf{Telco}}
        \label{fig:classification-results-all-Uniqueness_dc1-1-telco}
    \end{subfigure}

\raisebox{0.4\height}{\rotatebox{90}{Scenario 2}}\hspace{0.3em}
\begin{subfigure}[b]{0.23\linewidth}
        \includegraphics[width=\linewidth]{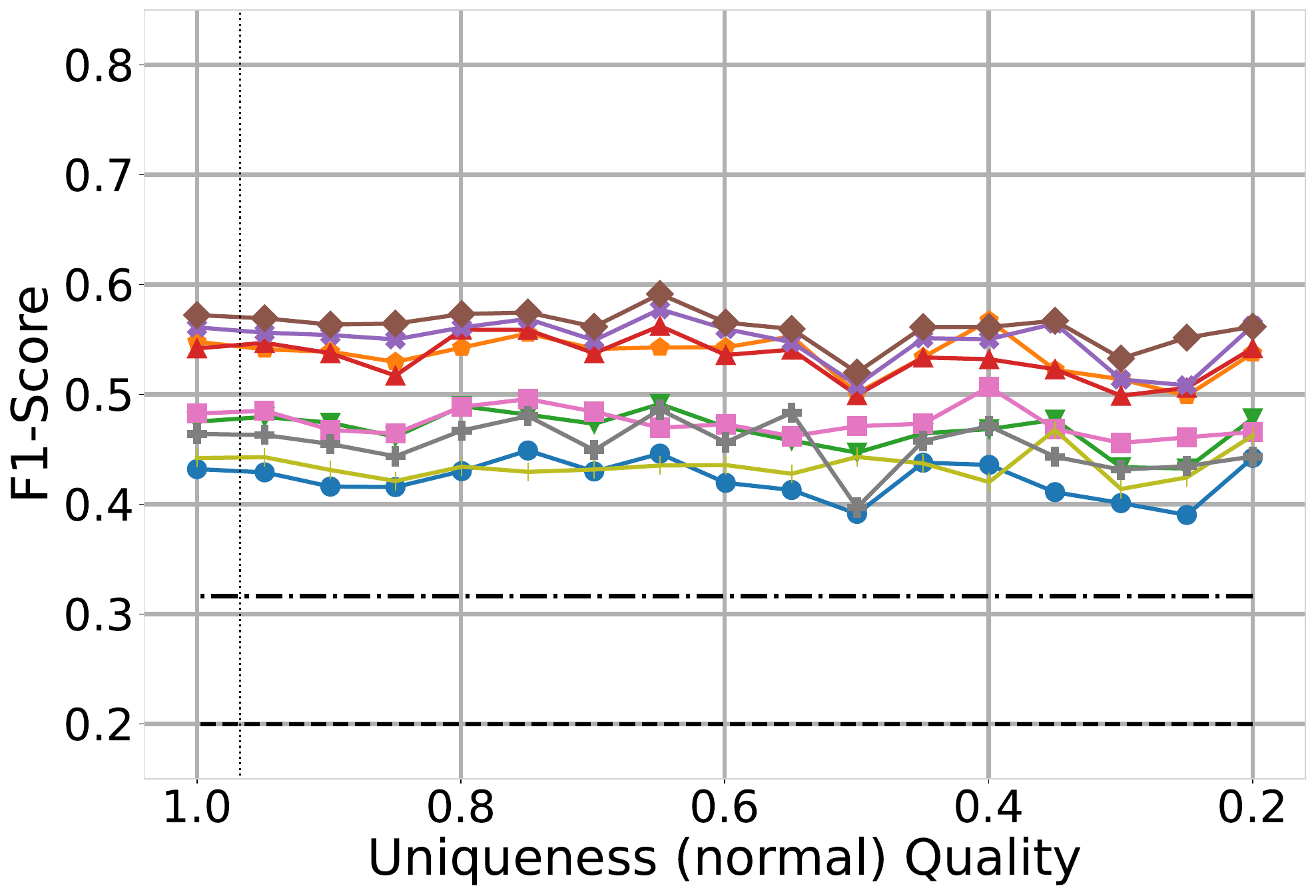}
        \caption{\textsf{Contraceptive}}
        \label{fig:classification-results-all-Uniqueness_dc1-2-contra}
    \end{subfigure}
\begin{subfigure}[b]{0.23\linewidth}
        \includegraphics[width=\linewidth]{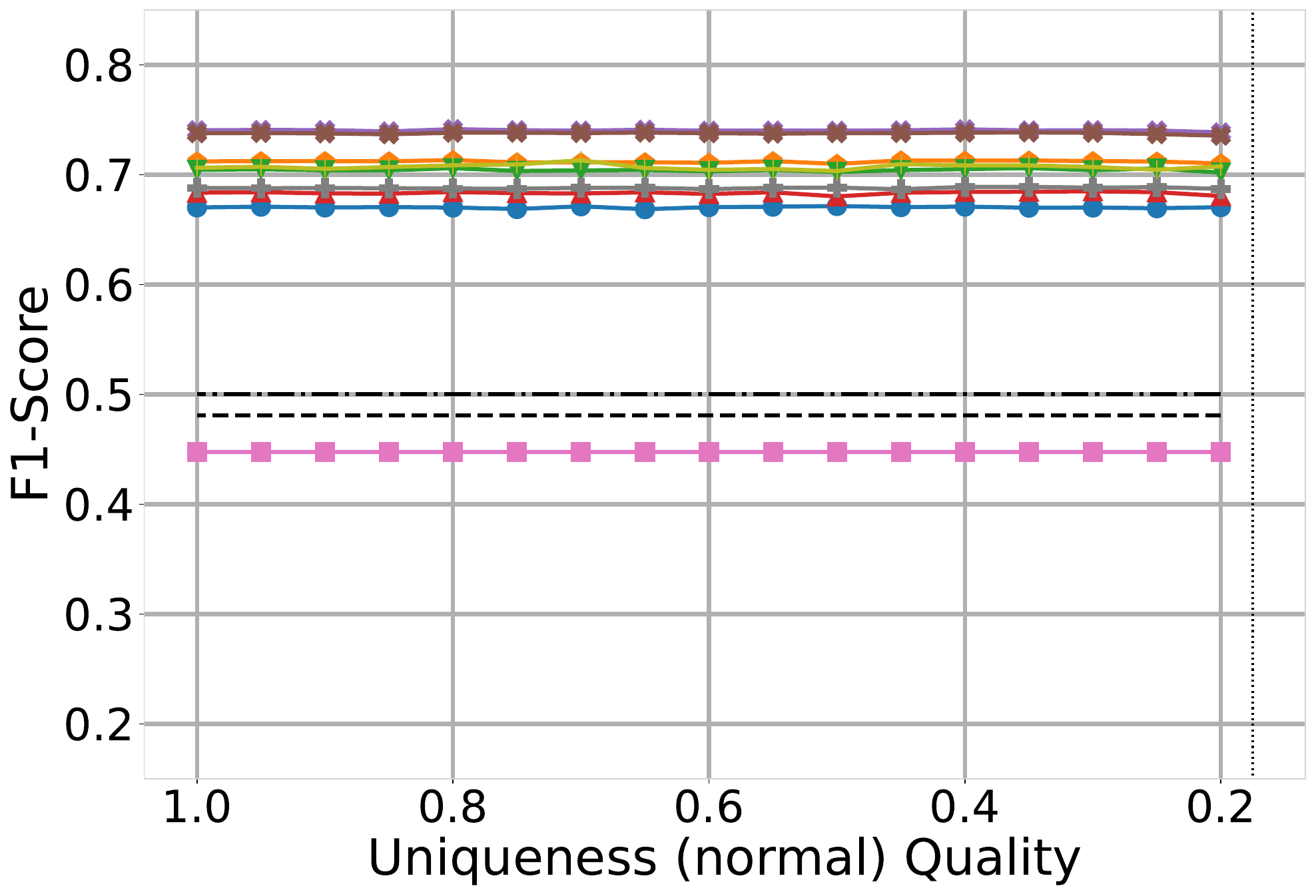}
        \caption{\textsf{COVID}}
        \label{fig:classification-results-all-Uniqueness_dc1-2-covid}
    \end{subfigure}
\begin{subfigure}[b]{0.23\linewidth}
        \includegraphics[width=\linewidth]{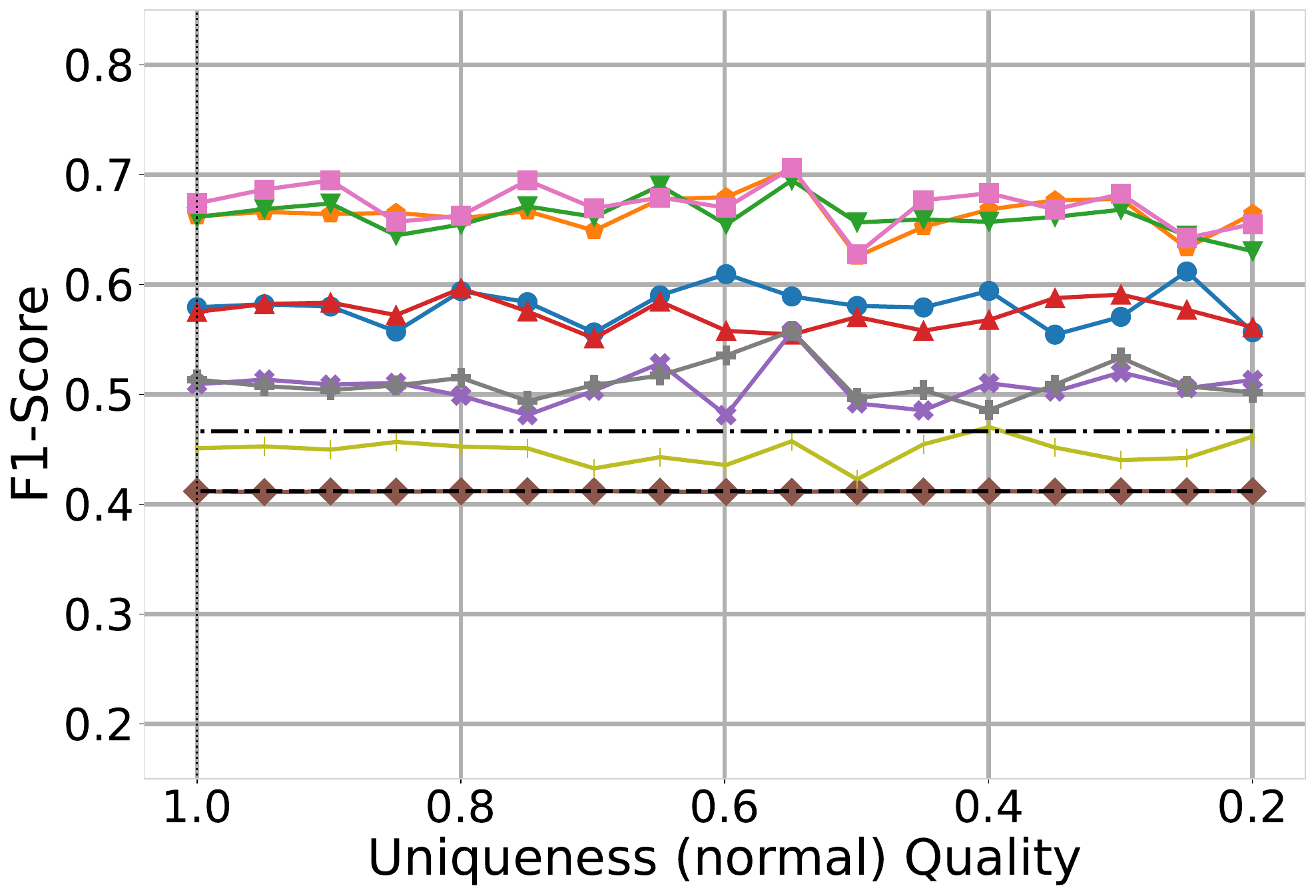}
        \caption{\textsf{Credit}}
        \label{fig:classification-results-all-Uniqueness_dc1-2-credit}
    \end{subfigure}
\begin{subfigure}[b]{0.23\linewidth}
        \includegraphics[width=\linewidth]{figures/classification/telco_train_clean_test_polluted_UniquenessPolluter_normal.pdf}
        \caption{\textsf{Telco}}
        \label{fig:classification-results-all-Uniqueness_dc1-2-telco}
    \end{subfigure}

\raisebox{0.4\height}{\rotatebox{90}{Scenario 3}}\hspace{0.3em}
\begin{subfigure}[b]{0.23\linewidth}
        \includegraphics[width=\linewidth]{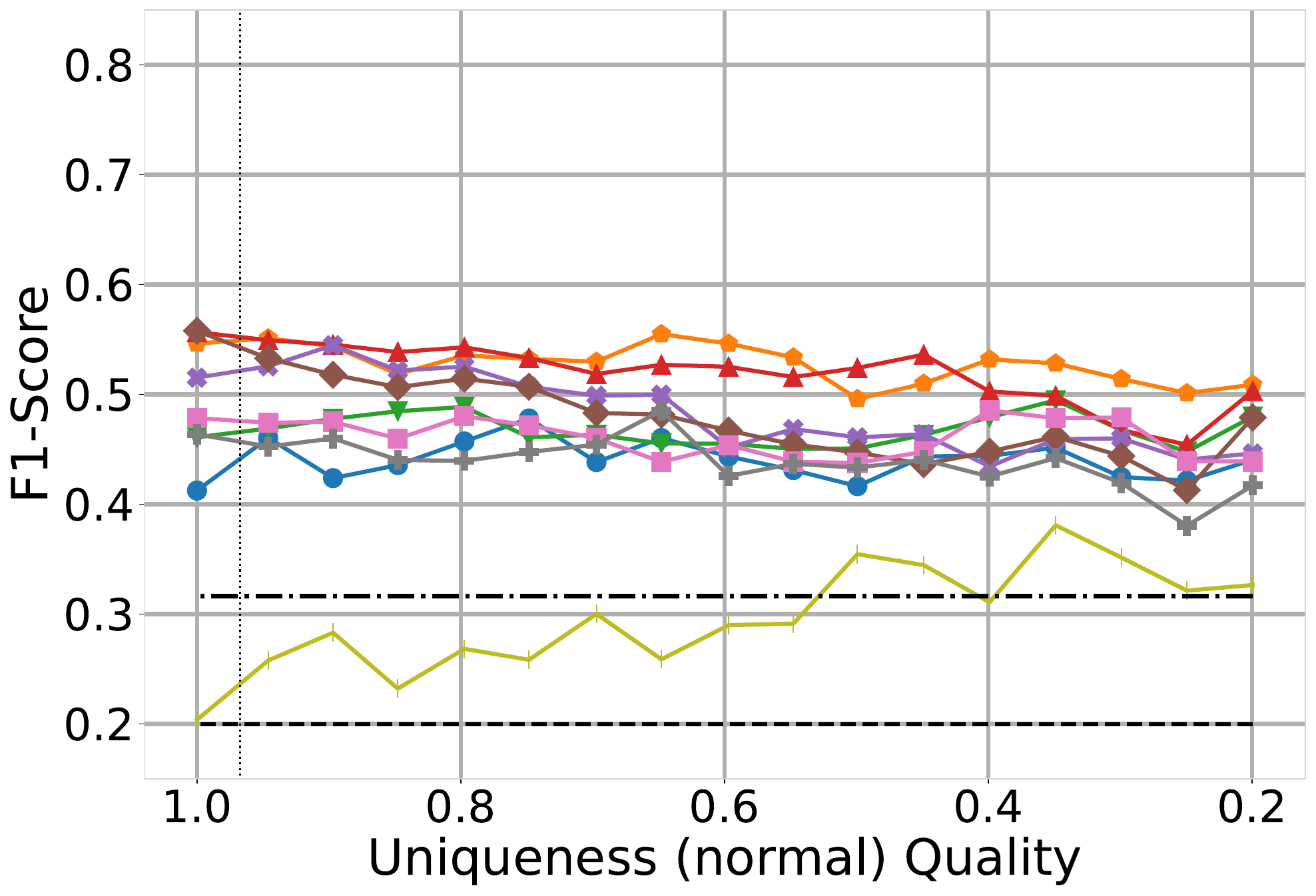}
        \caption{\textsf{Contraceptive}}
        \label{fig:classification-results-all-Uniqueness_dc1-3-contra}
    \end{subfigure}
\begin{subfigure}[b]{0.23\linewidth}
        \includegraphics[width=\linewidth]{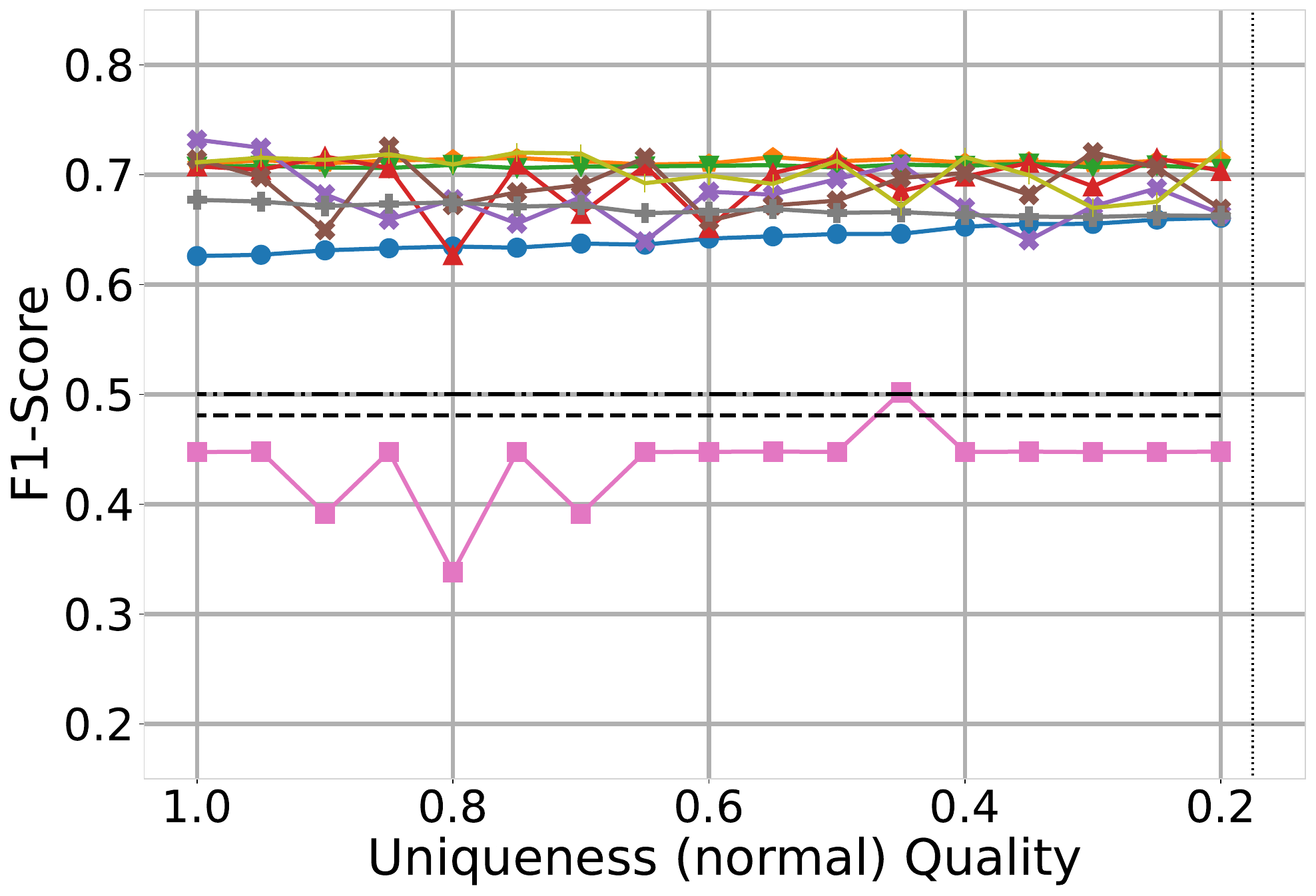}
        \caption{\textsf{COVID}}
        \label{fig:classification-results-all-Uniqueness_dc1-3-covid}
    \end{subfigure}
\begin{subfigure}[b]{0.23\linewidth}
        \includegraphics[width=\linewidth]{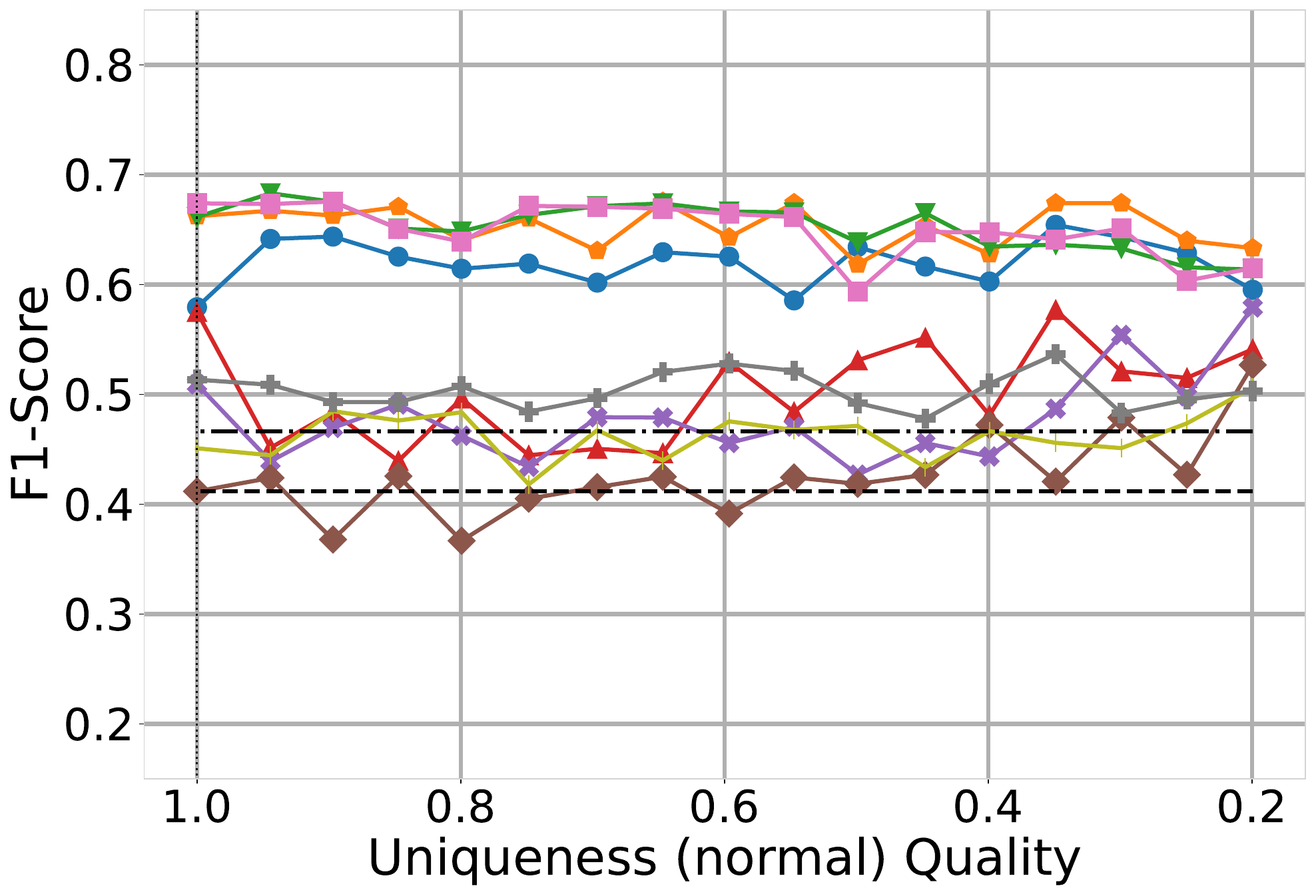}
        \caption{\textsf{Credit}}
        \label{fig:classification-results-all-Uniqueness_dc1-3-credit}
    \end{subfigure}
\begin{subfigure}[b]{0.23\linewidth}
        \includegraphics[width=\linewidth]{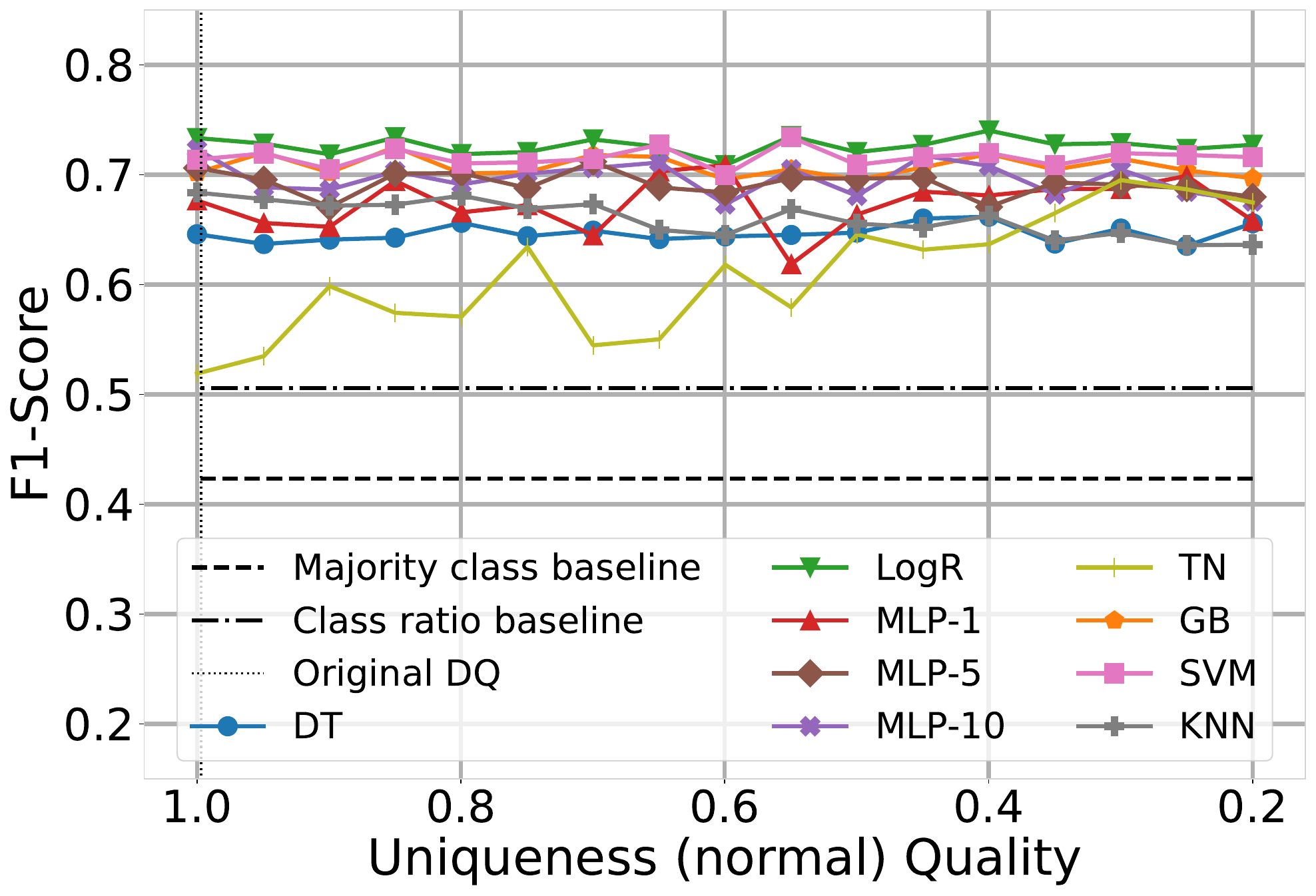}
        \caption{\textsf{Telco}}
        \label{fig:classification-results-all-Uniqueness_dc1-3-telco}
    \end{subfigure}
    \caption{$F_1$-scores of the classification algorithms for uniqueness.}
    \label{fig:classification-results-all-Uniqueness_dc1}
\end{figure*}

%% file: Latex_Figure/classification/Class_Balance.tex
\begin{figure*}[t]
    \centering
\raisebox{0.4\height}{\rotatebox{90}{Scenario 1}}\hspace{0.1em}
\begin{subfigure}[b]{0.23\linewidth}
        \includegraphics[width=\linewidth]      {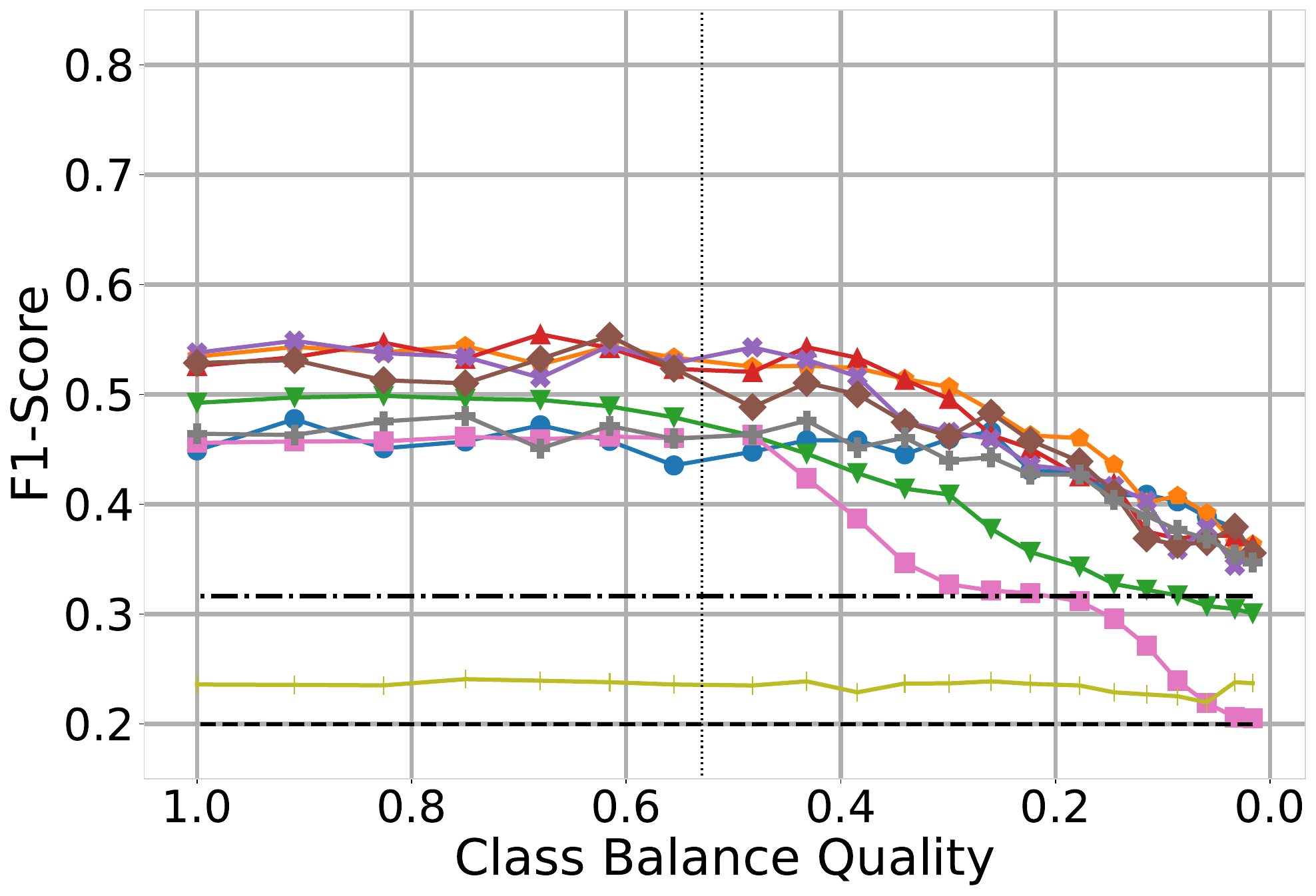}
        \caption{\textsf{Contraceptive}}
        \label{fig:classification-results-all-ClassBalance-1-contra}
\end{subfigure}
\begin{subfigure}[b]{0.23\linewidth}
        \includegraphics[width=\linewidth]{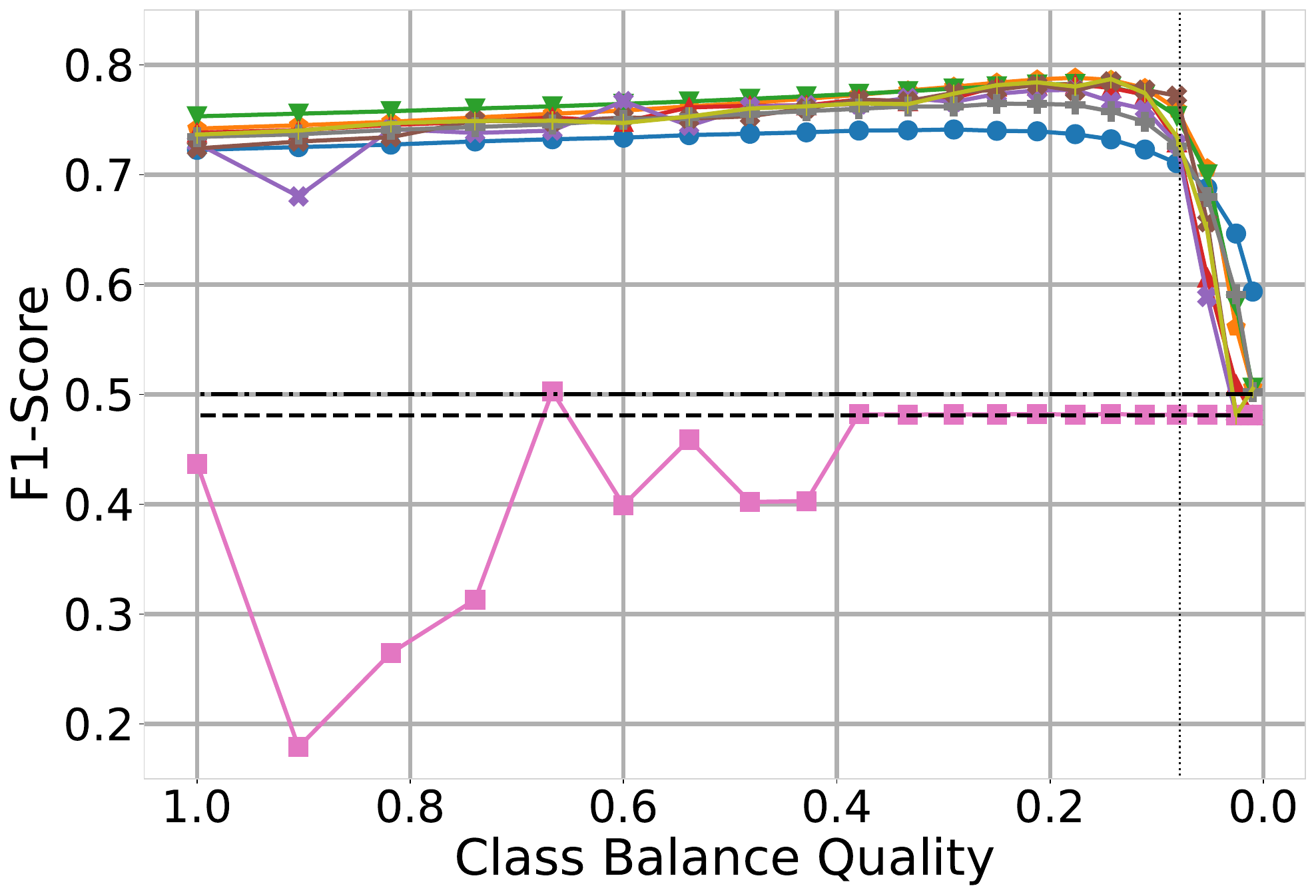}
        \caption{\textsf{COVID}}
        \label{fig:classification-results-all-ClassBalance-1-covid}
\end{subfigure}
\begin{subfigure}[b]{0.23\linewidth}
        \includegraphics[width=\linewidth]{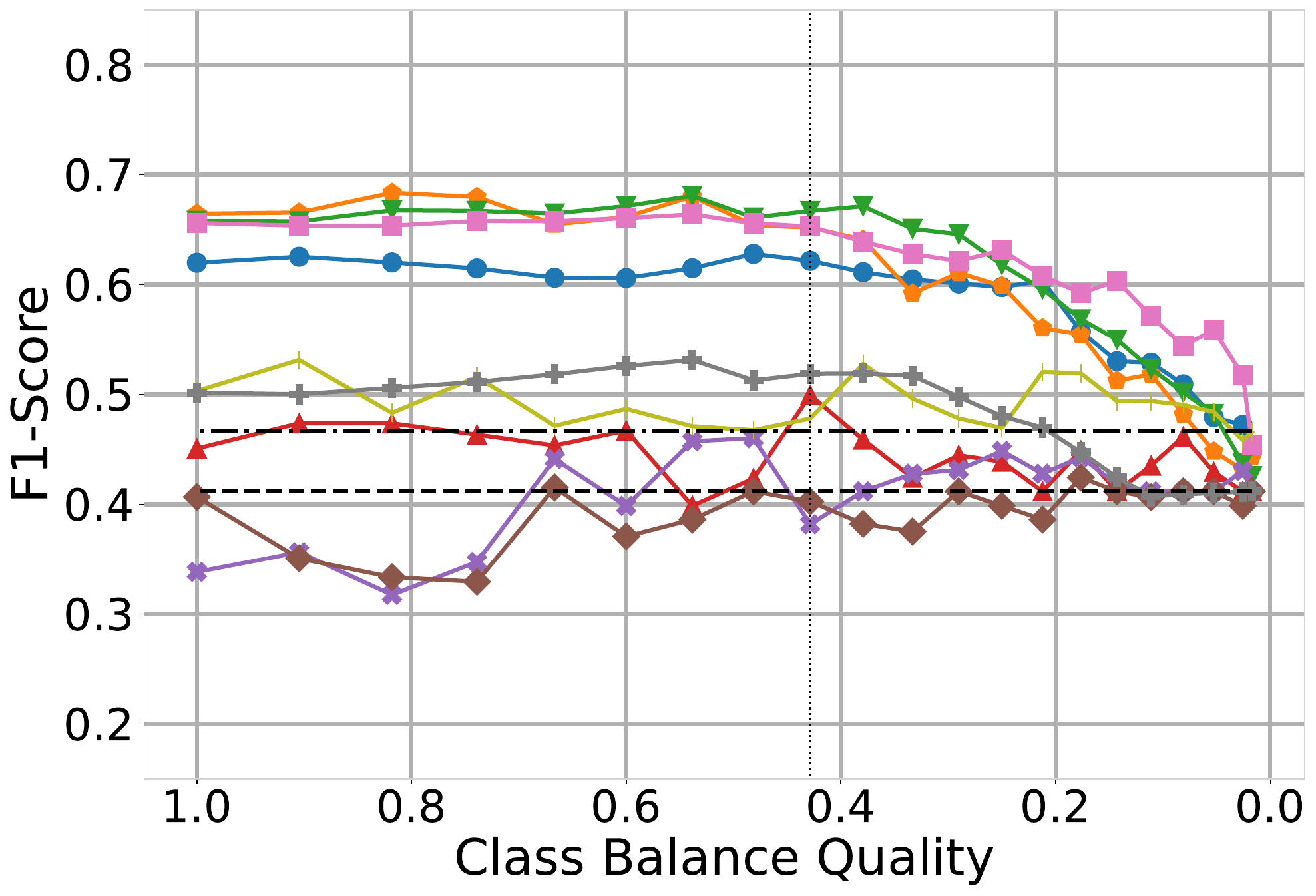}
        \caption{\textsf{Credit}}
        \label{fig:classification-results-all-ClassBalance-1-credit}
\end{subfigure}
\begin{subfigure}[b]{0.23\linewidth}
        \includegraphics[width=\linewidth]{figures/classification/telco_train_polluted_test_clean_ClassBalancePolluter.pdf}
        \caption{\textsf{Telco}}
        \label{fig:classification-results-all-ClassBalance-1-telco}
\end{subfigure}

\raisebox{0.4\height}{\rotatebox{90}{Scenario 2}}\hspace{0.3em}
\begin{subfigure}[b]{0.23\linewidth}
        \includegraphics[width=\linewidth]{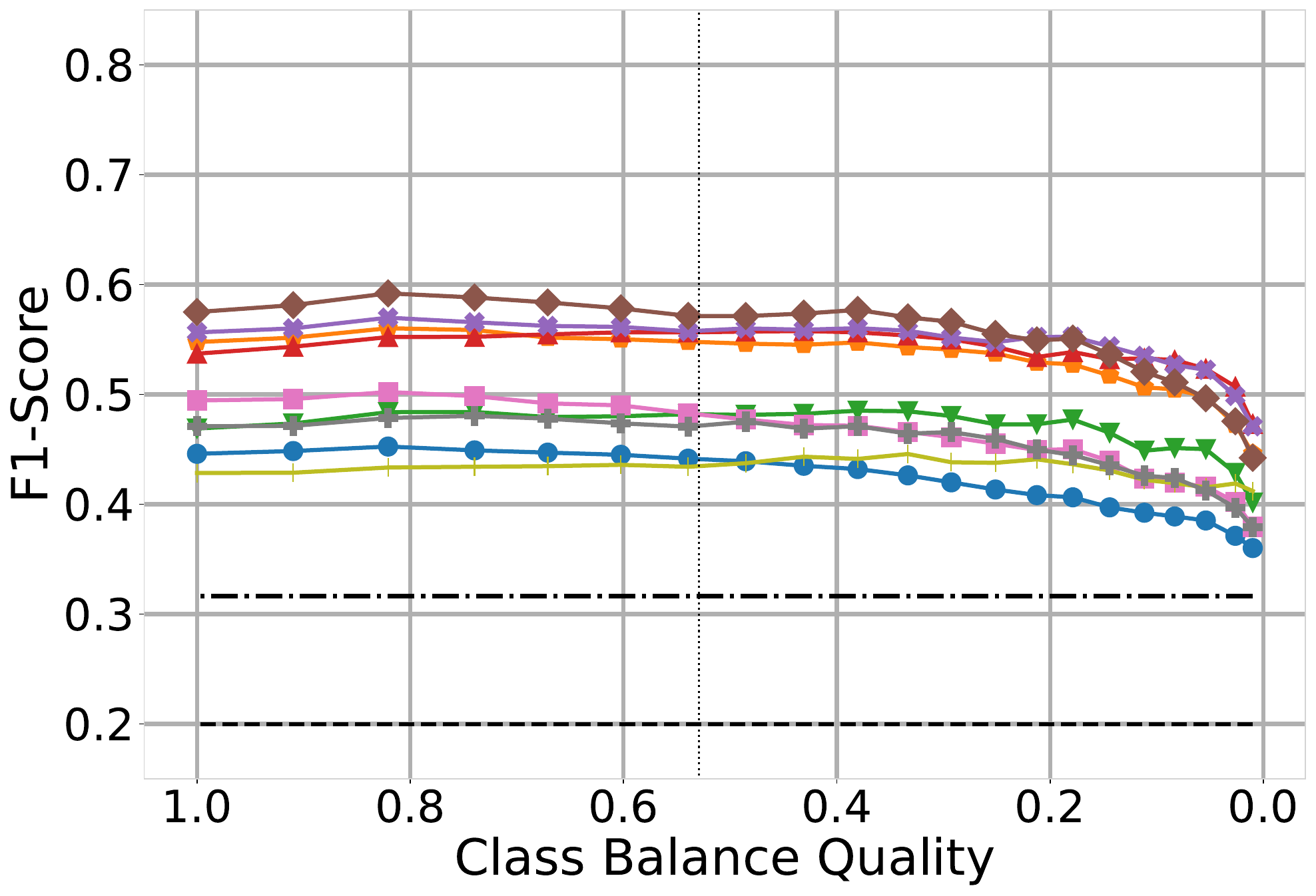}
        \caption{\textsf{Contraceptive}}
        \label{fig:classification-results-all-ClassBalance-2-contra}
\end{subfigure}
\begin{subfigure}[b]{0.23\linewidth}
        \includegraphics[width=\linewidth]{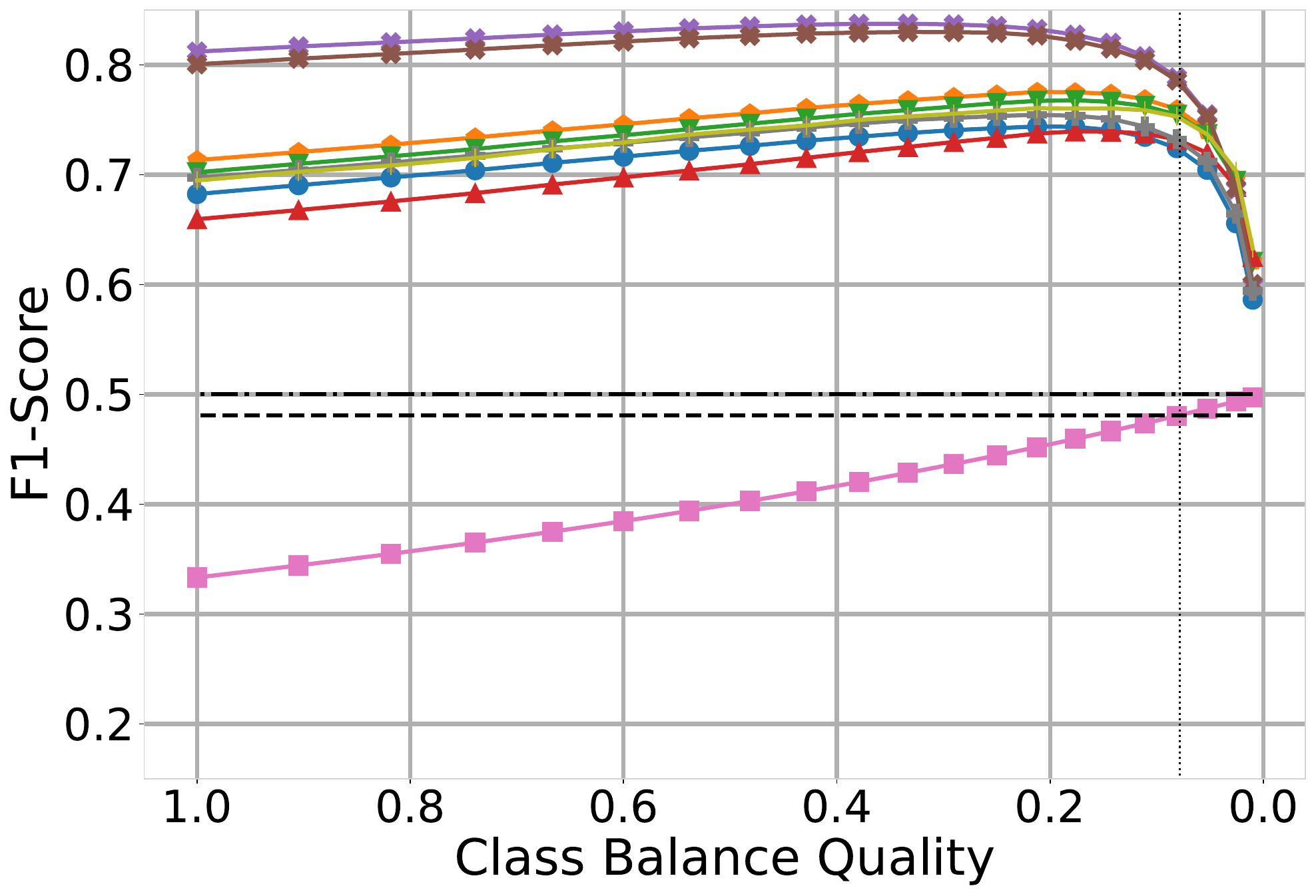}
        \caption{\textsf{COVID}}
        \label{fig:classification-results-all-ClassBalance-2-covid}
\end{subfigure}
\begin{subfigure}[b]{0.23\linewidth}
        \includegraphics[width=\linewidth]{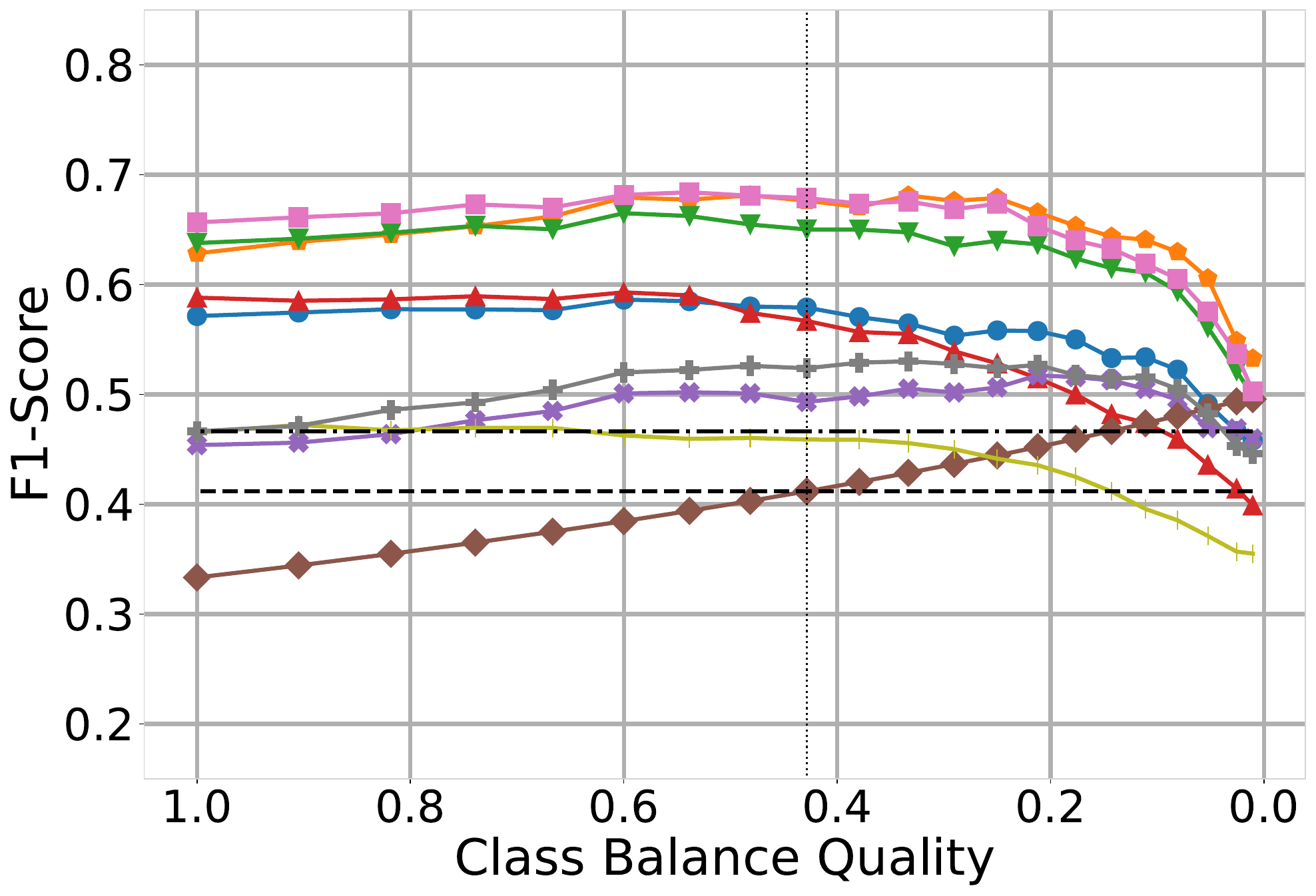}
        \caption{\textsf{Credit}}
        \label{fig:classification-results-all-ClassBalance-2-credit}
\end{subfigure}
\begin{subfigure}[b]{0.24\textwidth}
        \includegraphics[width=\textwidth]{figures/classification/telco_train_clean_test_polluted_ClassBalancePolluter.pdf}
        \caption{\textsf{Telco}}
        \label{fig:classification-results-all-ClassBalance-2-telco}
\end{subfigure}

\raisebox{0.4\height}{\rotatebox{90}{Scenario 3}}\hspace{0.3em}
\begin{subfigure}[b]{0.23\linewidth}
        \includegraphics[width=\linewidth]{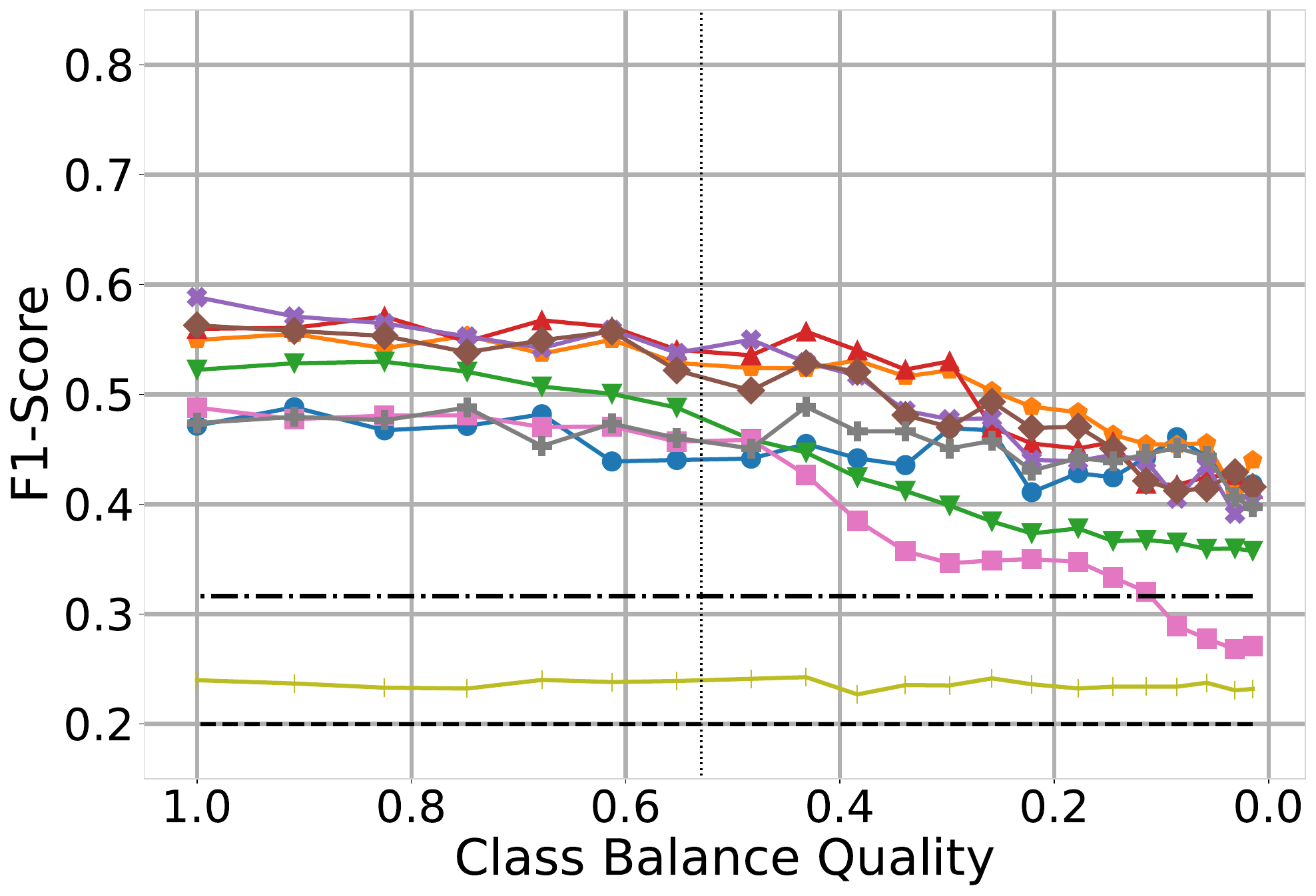}
        \caption{\textsf{Contraceptive}}
        \label{fig:classification-results-all-ClassBalance-3-contra}
\end{subfigure}
\begin{subfigure}[b]{0.23\linewidth}
        \includegraphics[width=\linewidth]{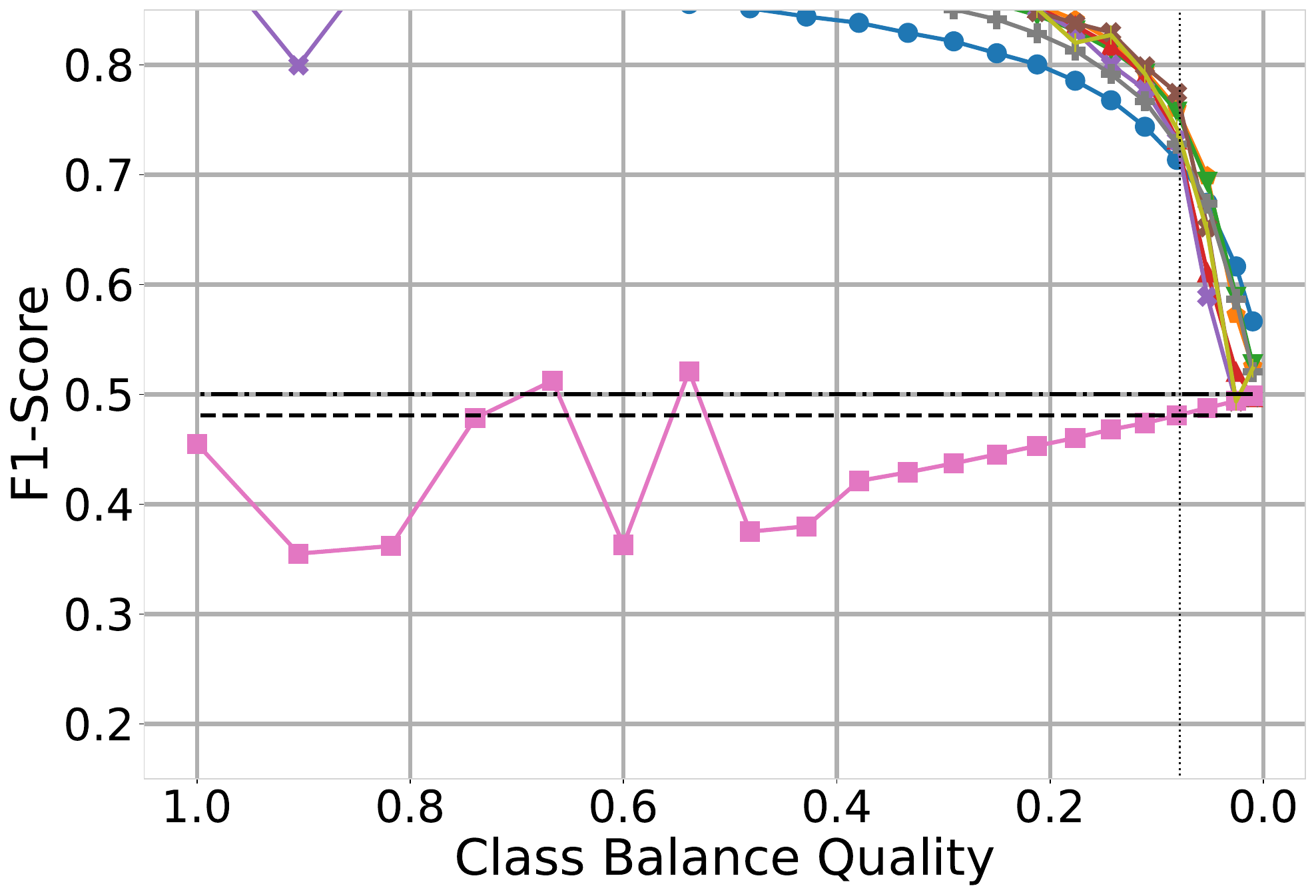}
        \caption{\textsf{COVID}}
        \label{fig:classification-results-all-ClassBalance-3-covid}
\end{subfigure}
\begin{subfigure}[b]{0.23\linewidth}
        \includegraphics[width=\linewidth]{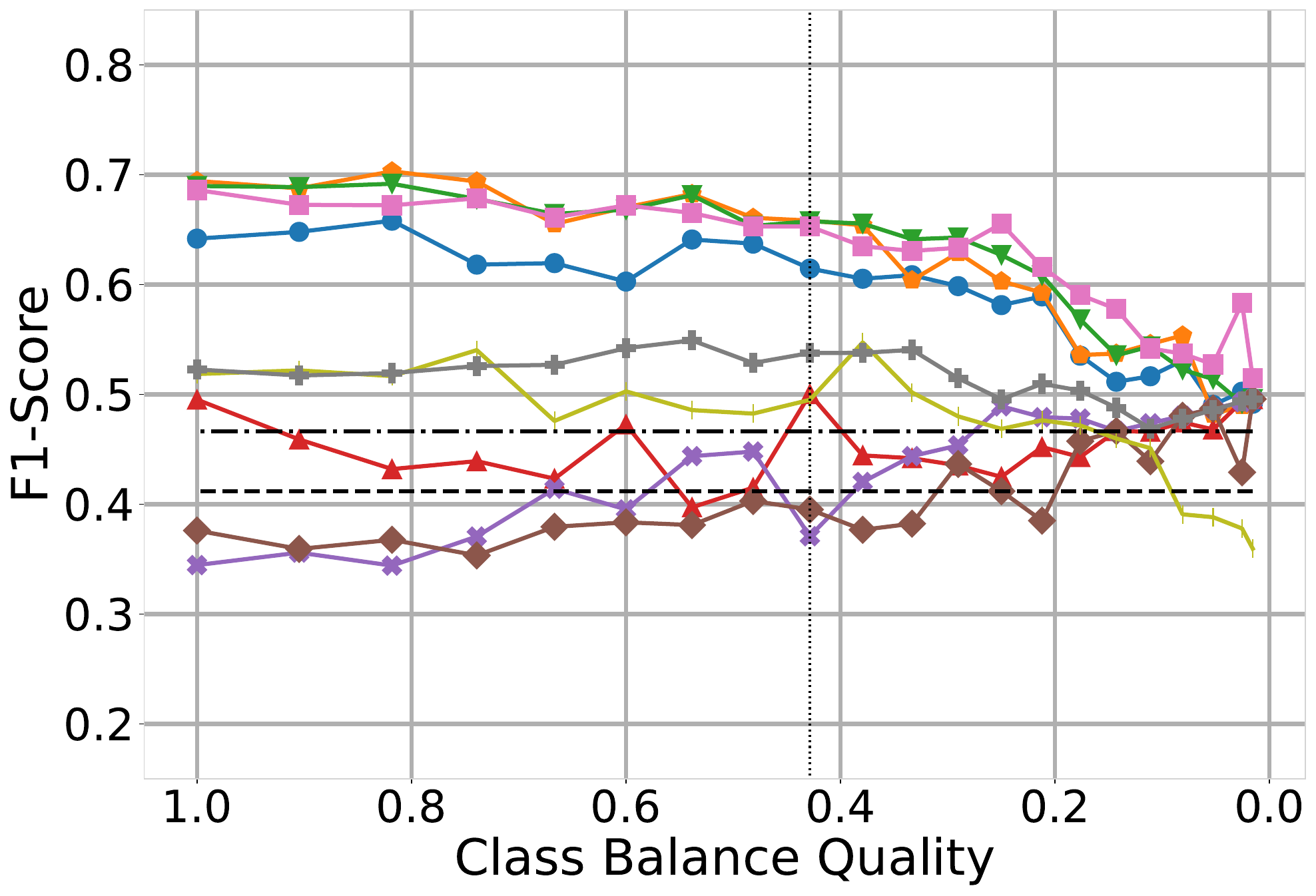}
        \caption{\textsf{Credit}}
        \label{fig:classification-results-all-ClassBalance-3-credit}
\end{subfigure}
\begin{subfigure}[b]{0.24\textwidth}
        \includegraphics[width=\textwidth]{figures/classification/telco_train_polluted_test_polluted_ClassBalancePolluter_legend.pdf}
        \caption{\textsf{Telco}}
        \label{fig:classification-results-all-ClassBalance-3-telco}
\end{subfigure}
    
    \caption{$F_1$-scores of the classification algorithms for target class balance.}
    \label{fig:classification-results-all-ClassBalance}
\end{figure*}

%% file: 60-results/62-regression.tex
\subsection{Regression}
\label{subsec:regression-results}
\PaperShort{\input{Latex_Figure/regression/summary_house}}
We discuss here the effect of degrading the six data quality dimensions of four datasets, namely: \textsf{Houses}, \textsf{IMDB}\revision{,} \textsf{Cars} \revision{and \textsf{COVID}}, on the performance of \revision{seven} regression algorithms, namely: LR, RR, DT, GB, RF, MLP\revision{, TN}.
\revision{Similar to the classification task}, we consider MLP in the form of three variants: MLP-1, MLP-5 and MLP-10\@.
As LR and RR differ only in the regularization employed in RR, their performance lines in the result plots often overlap.
\PaperLong{This means that in cases where the LR performance line is not visible at all in a plot, it is hidden behind the RR line.}
\PaperShort{\revision{Figure~\ref{fig:regression-results-all-house} shows the results of the experiments using the \textsf{Houses} dataset.}
We only show~$R^{2}$ of~$0$ or larger, since a negative~$R^{2}$ means that the model is worse than always predicting the mean. 
Thus, a model in the negative ranges would not be of interest and the resulting vertical axis scale would make it harder to analyze the behavior in the range of interest,~$0$ to~$1$.
\PaperShort{As the results of TN are not visible in the Figures due to TN's low overall performance using small datasets, we additionally included in Figure~\ref{fig:regression-results-all-covid} the results of the \textsf{COVID} dataset.
However, as for classification, we include the findings for the other datasets in the descriptions of the results per data quality dimension.}}
\PaperLong{As the MLP performance on \textsf{IMDB} is already negative without pollution, the MLP line is not visible in the plots for \textsf{IMDB}.
Same applies for TN for \textsf{Houses}.}

\stitle{Consistent Representation}
\PaperLong{\input{Latex_Figure/regression/Consistent_Representation_5}
For consistent representation, we show only a plot with adding 4~new representations per unique value,~i.e.,~$k_v = 5$~(the plots for~$k_v = 2$ can be found in Figure~\ref{fig:regression-results-all-ConsistentRepresentation}\PaperShort{in the extended version}). 
As described in the introduction of Section~\ref{sec:results}, a higher percentage of polluted samples can lead to an increase in quality again, which is why performance would rise again when more than~$80\%$ of samples are polluted for~$k_v = 5$.
However, we stop the line at this point.
We add some percentages next to the lines for better orientation.

}
An increase of the representations of values in categorical features leads to a decrease in~$R^{2}$ of all algorithms \revision{and datasets}\PaperLong{, as shown in Figure~\ref{fig:regression-results-all-ConsistentRepresentationk5}}. 
The severity of this decrease depends on the scenario. 
In Scenario~2\PaperShort{(see Figure~\ref{fig:regression-results-all-ConsistentRepresentationk5-2-houses}),}\PaperLong{(second row in Figure~\ref{fig:regression-results-all-ConsistentRepresentationk5}), when the inserted inconsistent representations have not been present during training,} the performance decrease is the largest, especially for RR and LR\@. 

Adding inconsistent representations during training but not during testing,~i.e., Scenario~1\PaperShort{(see Figure~\ref{fig:regression-results-all-ConsistentRepresentationk5-1-houses})}\PaperLong{(first row in Figure~\ref{fig:regression-results-all-ConsistentRepresentationk5})} has \revision{generally} a smaller effect \revision{on the performance}\revision{, except for DT}. 
The effect of the pollution is \revision{quite low} in Scenario~3 for \textsf{Houses}~(see Figure~\ref{fig:regression-results-all-ConsistentRepresentationk5-3-houses})\revision{,} \textsf{Cars}\PaperLong{(see Figure~\ref{fig:regression-results-all-ConsistentRepresentationk5-3-cars})} \revision{and \textsf{COVID}\PaperLong{(see Figure~\ref{fig:regression-results-all-ConsistentRepresentationk5-3-covid})}.}
\revision{F}or \textsf{IMDB} \PaperLong{(see Figure~\ref{fig:regression-results-all-ConsistentRepresentationk5-3-imdb})} it is more similar to \revision{its} Scenario~1\PaperLong{(see Figure~\ref{fig:regression-results-all-ConsistentRepresentationk5-1-imdb}).}
\PaperLong{
Regarding the algorithms, LR often shows considerable outliers when the training set is polluted}. 
Due to the one-hot encoding of categorical features\PaperLong{and LR learning a linear impact of each of those one-hot features}, LR is affected by the inherently discrete differences\PaperLong{in the one-hot features}. 
The regularization of RR fixes this extreme behavior.\PaperLong{LR and RR both show a larger performance drop than the three other algorithms in Scenario~2, probably due to the linear method poorly handling the one-hot features that were never~$1$ during training.}
The tree-based methods, RF and GB, are more stable in most cases, along with all considered MLP variants~(MLP-1, MLP-5, MLP-10).
\revision{In Scenarios~1 and~2 for \textsf{COVID}, TN shows sudden unstable behavior at different points.
In Scenario~1, TN's is stable over whole pollution increase, whereas in Scenario~2, performance decreases sharply with only a small amount of pollution introduced.}
The non-linearity of those methods weakens the effect of the inconsistencies\PaperLong{in the one-hot features.

The effect of inconsistency is larger on \textsf{IMDB} than on \textsf{Houses}, \textsf{Cars} and \textsf{COVID}~(see Figure~\ref{fig:regression-results-all-ConsistentRepresentationk5}). 

The observed trends are mostly similar when adding only one new representation during pollution, as shown in Figure~\ref{fig:regression-results-all-ConsistentRepresentation}, where the quality increases again after polluting~$50\%$ of samples. 
In Scenario~3, the algorithm performances using \textsf{Houses} rise for this increasing quality, which is expected, as the single new representation becomes the majority in both the training and test set.
}

\stitle{Completeness}
\label{subsubsection:results-regression-completeness}
Reducing completeness of a dataset leads to a heavy performance degradation on all algorithms in all scenarios, as shown\PaperShort{representatively in Figure~\ref{fig:regression-results-all-house} for \textsf{Houses}}\PaperLong{in Figure~\ref{fig:regression-results-all-completeness}}. 
The RF and GB performance \revision{generally} decreases the slowest\PaperLong{, especially much slower than DT in most cases}. 
These are ensemble methods, which makes them more robust against missing values.
\revision{Considering Scenarios~1 and~3, TN shows the most robust performance using the \textsf{COVID} data~(see Figure~\ref{fig:regression-results-all-ConsistentRepresentationk5-1-covid} and ~\ref{fig:regression-results-all-ConsistentRepresentationk5-3-covid}).}\PaperLong{LR and RR show similar behavior, with RR performing better or similar than LR. 
The performance of the MLP variants is mostly between that of LR/RR and DT.}

For all datasets, the strongest decrease happens in Scenario~2~\PaperShort{(see Figure~\ref{fig:regression-results-all-completeness-2-houses})}\PaperLong{(second row in Figure~\ref{fig:regression-results-all-completeness}), where the missing values only appear in the test set, not in the training set}. 
LR and RR are the most affected algorithms: the linear relation learned by those algorithms is easier to confuse with the newly inserted placeholders outside the features' domain.\PaperLong{The value chosen as placeholder could also have an effect,~e.g., whether choosing~$-1$ or~$-1000$. 
Different placeholders could lead to a different behavior, which could also hold for the other scenarios and algorithms.}

The best performance is achieved in Scenario~3\PaperShort{(see Figure~\ref{fig:regression-results-all-completeness-3-houses})}\PaperLong{(last row in Figure~\ref{fig:regression-results-all-completeness})}, indicating that models perform better on datasets with missing data if the missing data also exists during the training.
The effect of reducing the completeness differs per dataset.\PaperLong{For \textsf{Houses}, the performance decrease in Scenario~1 is larger than for \textsf{IMDB} and especially than for \textsf{COVID}. 
The performance in Scenario~3 of \textsf{Houses} first decreases slower than in Scenario~1 and then rapidly drops when the completeness is lower than~$0.2$. 
For \textsf{IMDB} and \textsf{COVID}, the performance decrease in Scenario~3 is more of a linear form and faster than in Scenario~1. 
Those differences between the datasets could have their origin in the size of the datasets.} 
For example, the small size of \textsf{Houses} causes the algorithms to overfit and produce better results when training and test sets contain missing values.
Whereas larger datasets could reduce this effect, making the amount of non-missing information in the test set the more important factor. 
\PaperLong{\input{Latex_Figure/regression/Completeness}}

\stitle{Feature Accuracy}
\PaperLong{\input{Latex_Figure/regression/Feature_Accurecy}} 
Decreasing feature accuracy causes a clear performance degradation of the regression algorithms in all scenarios, as shown\PaperShort{representatively in Figure~\ref{fig:regression-results-all-house} for \textsf{Houses}.}\PaperLong{in Figure~\ref{fig:regression-results-all-FeatureAccuracy}.} 

For Scenario~1, the performance decrease for all considered algorithms, except for DT, is slower in higher quality ranges and faster in lower quality ranges as the differences between training and test set become too large.
\revision{But also in Scenario~3, the majority of the algorithms, especially RR and GB, perform quite robustly against noise.
GB and RR benefit from the normal distribution of the noise.}
As for completeness, the~$R^{2}$ degradation is the highest in Scenario~2. 
In Scenario~3, \revision{the} algorithms show more linearly decreasing performance\revision{, with a stagnation or even improvement in ranges of very low quality~(especially for the \textsf{COVID} dataset using TN, GB and RR in Figure~\ref{fig:regression-results-all-FeatureAccuracy-3-covid}). 
Apparently, the inaccuracies in training and test set align when the pollution is high, leading to an increase after a performance minimum}.
The DT algorithm is more sensitive to feature accuracy drop than the other algorithms in Scenarios~1 and~3.
\PaperLong{We assume that this is due to DT being sensitive to changes in the data, without having the assumption of normally distributed noise like LR/RR.} DT's performance decreases especially fast on \textsf{IMDB} and \textsf{COVID} due to the high number of categorical features.
All considered MLP variants are generally more robust than DT regarding feature accuracy.
There are minimal performance differences between the different MLP variants, except in Scenario~3 in \textsf{Houses} -- MLP-5 and MLP-10 have a slightly faster performance drop than MLP-1.

\PaperLong{When comparing between datasets, it is noticeable that for all scenarios the effect of pollution is overall the strongest on \textsf{Cars}, which is the dataset with the smallest number of features, followed by \textsf{IMDB} and \textsf{COVID}.} 

\stitle{Target Accuracy}
\PaperLong{\input{Latex_Figure/regression/Target_Accurecy}}We observe a strong~$R^{2}$ degradation in response to decreasing target accuracy for all datasets and in all scenarios, like for feature accuracy and completeness, but at a weaker level. 

Scenario~1\PaperLong{~(first row in Figure~\ref{fig:regression-results-all-TargetAccuracy})} shows the strongest degradation with different speeds for different datasets. 
For \textsf{Cars}\PaperLong{(see Figure~\ref{fig:regression-results-all-TargetAccuracy-1-cars})}, having a low number of features and high number of records, we observe a heavy performance degradation for RF and DT, whereas the performance of GB and LR/RR stays relatively constant as the noise is normally distributed.
\PaperLong{Thus, it does not influence the regression lines to such an extent that the performance decreases.}
In contrast, there is a \revision{strong} degradation of LR and RR performance for \textsf{Houses}~(see Figure~\ref{fig:regression-results-all-TargetAccuracy-1-houses}) due to the large number of features with low sample size.
In general, all MLP variants are quite resilient for low target accuracy.
However, similar to feature accuracy, MLP-5 and MLP-10 are less robust than MLP-1 for \textsf{Houses} at a lower target accuracy.
\PaperLong{For \textsf{IMDB} (see Figure~\ref{fig:regression-results-all-TargetAccuracy-1-imdb}), it has a baseline performance below~0, and thus the effect of pollution is not considered here.
For \textsf{Houses}, the MLP-1 variant performs better than the other algorithms with increasing pollution.}
Comparing within the LR-based and the tree-based families: RR performs better than LR, and GB outperforms RF and DT.
\PaperLong{Thus, the improved version of the algorithm performs better than the more simple version across all datasets.

Contrasting Scenario~1 and Scenario~3~(third row in Figure~\ref{fig:regression-results-all-TargetAccuracy}), we see that the perceived performance, which is shown as Scenario~1, decreases faster than the actual performance, which is shown in Scenario~3, in higher quality ranges.} 

\revision{For all datasets, we can observe that} \revision{i}n lower quality ranges, the perceived performance in Scenario~3\PaperShort{(see Figure~\ref{fig:regression-results-all-TargetAccuracy-3-houses})} decreases slower or \revision{even} stagnates compared to Scenario~1.
\revision{Also in Scenario~2 all datasets follow the same trend:} \revision{T}he performance declines gradually with the decrease in target accuracy, as the algorithms increasingly fail in predicting the emerging patterns caused by polluting only the test data.
\PaperLong{Comparing the pollution effect on the datasets for Scenario~2, we see that the strongest degradation is present for \textsf{IMDB}. 
\textsf{Cars} and \textsf{Houses} dataset perform similar regarding this pollution method.}

\stitle{Uniqueness}
\PaperLong{\input{Latex_Figure/regression/Uniqueness_1}}
Decreasing uniqueness does not have a considerable effect on the performance of the algorithms, regardless of the scenario or dataset\revision{apart from few exceptions}. 
On the datasets with many samples, such as \textsf{COVID}, we see \revision{less effects in the performances, except for TN in Scenario~1, which behaves much more erratic than the simpler models (see Figure~\ref{fig:regression-results-all-Uniqueness_dc1-1-covid})}.
Datasets with a small sample size, like \textsf{Houses}~(see Figures~\ref{fig:regression-results-all-Uniqueness_dc1-1-houses} and~\ref{fig:regression-results-all-Uniqueness_dc1-3-houses}), show a slight performance decrease with decreased uniqueness in Scenarios~1 and~3 due to an exaggerated influence of few samples on the training dataset, as they occur several times as a result of duplication.

Comparing the uniqueness with duplicate count sampled by normal distribution~\PaperLong{(see Figure~\ref{fig:regression-results-all-Uniqueness_dctnormal})} and uniqueness with all samples having duplicate count of~1\PaperLong{(see Figure~\ref{fig:regression-results-all-Uniqueness_dc1})}, we do not see different behavior for \textsf{Cars} and \textsf{IMDB}. 
For \textsf{Houses} \revision{and \textsf{COVID}}, inserting duplicates following a normal distribution causes a larger degradation of all algorithms' performance, especially for Scenarios~2 and~3. 
As \textsf{Cars}, \textsf{IMDB}, and \textsf{COVID} have fewer features and more samples compared to \textsf{Houses}, this explains the higher resilience of the algorithms.

\stitle{Target Class Balance}
\PaperLong{\input{Latex_Figure/regression/Class_Balance}} 
Similar to the uniqueness dimension, we observe a low effect of the target class imbalance increase in all scenarios on all datasets \revision{for imbalance up to 50\%,}\PaperShort{as shown in last row in Figure~\ref{fig:regression-results-all-house} for \textsf{Houses} \revision{and Figure~\ref{fig:regression-results-all-covid} for \textsf{COVID}.}}\PaperLong{as shown in Figure~\ref{fig:regression-results-all-ClassBalance}.}
Two facts can justify this observation: Firstly, regression datasets usually have a continuous target variable, mostly normally distributed and, therefore, inherently unbalanced target variable. 
Secondly, we had to discretize the target feature and discard resulting classes with very few samples before applying pollution, as described in Section~\ref{sec:scenarios}.
Only with a high imbalance above~50\%, we start to recognize a degraded performance of all algorithms.

\PaperLong{
Even though there is no big influence of the pollution in general, we see similar behavior between algorithms. 
All seven algorithms show the same relative degradation when decreasing the target class balance, and thus handle target accuracy similarly. 
Comparing the datasets, we see similar responses to an increased pollution. 
The performance degradation on \textsf{IMDB} is faster than on other datasets, especially if the imbalance was introduced only to test data~(see Figure~\ref{fig:regression-results-all-ClassBalance-2-imdb}).
\textsf{COVID} shows in Scenario~1 and~2 and even increasing performance with increasing pollution.
Target class balance has the lowest influence on \textsf{Cars}.}

%% file: Latex_Figure/regression/summary_house.tex
\begin{figure*}[!htbp]
\captionsetup[subfigure]{aboveskip=-1pt,belowskip=-1pt}
    \centering
\begin{adjustbox}{minipage=\linewidth}
    \begin{subfigure}[b]{0.32\textwidth}
        \includegraphics[width=\textwidth]{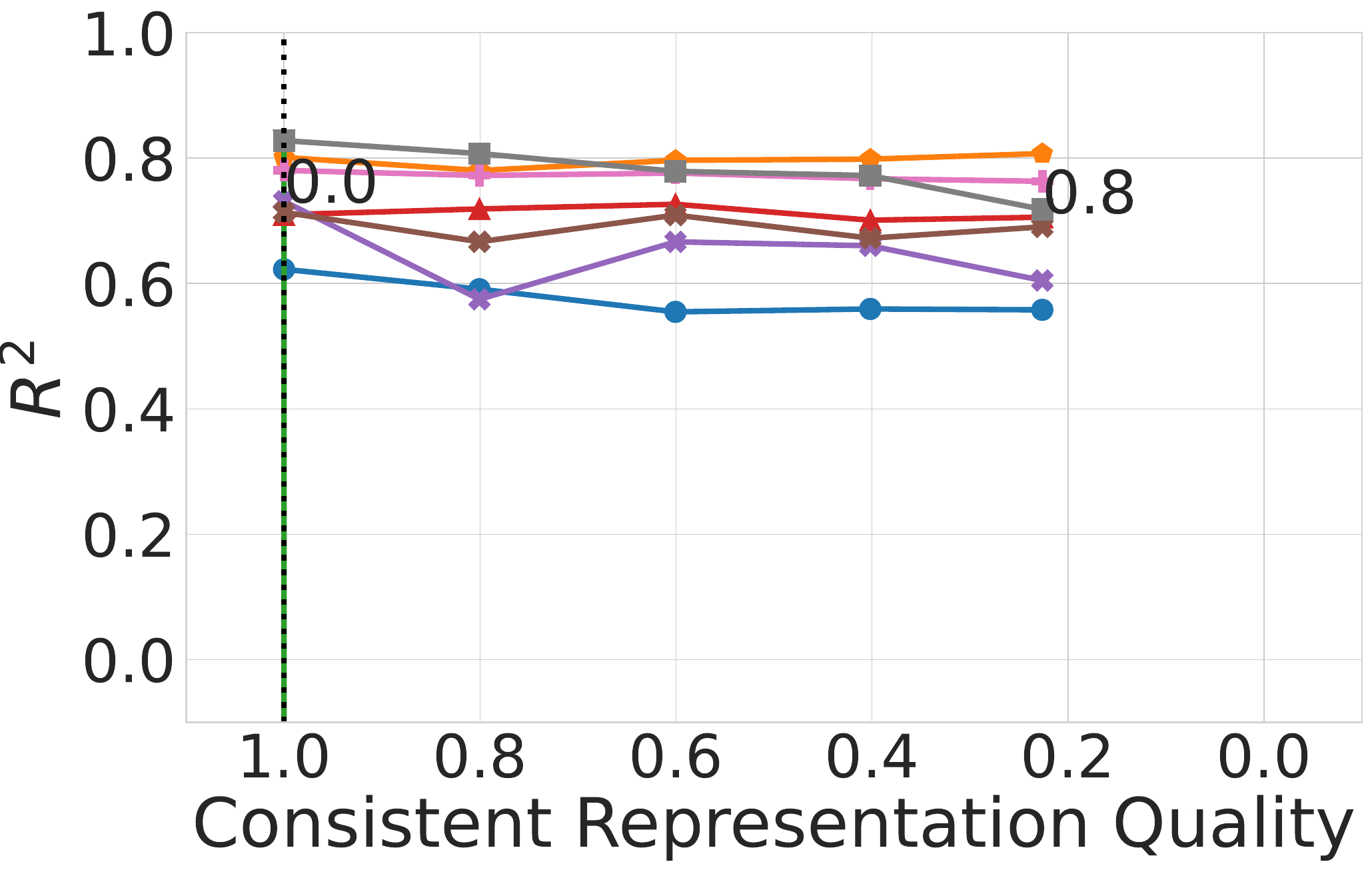}
        \caption{Consistency with $k_{v}=5$ (Sce. 1)}
        \label{fig:regression-results-all-ConsistentRepresentationk5-1-houses}
    \end{subfigure}
    \begin{subfigure}[b]{0.32\textwidth}
        \includegraphics[width=\textwidth]{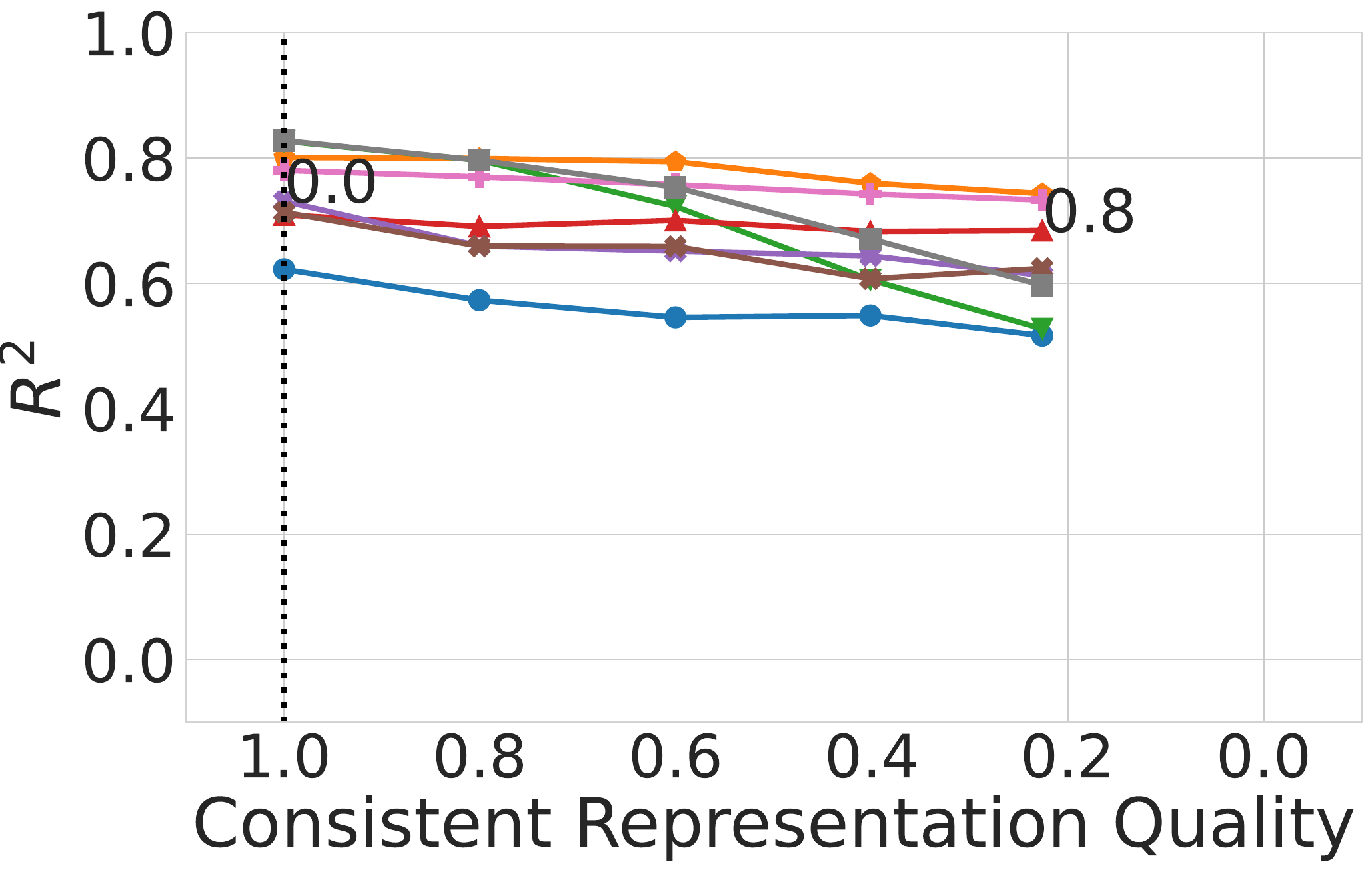}
        \caption{Consistency with $k_{v}=5$ (Sce. 2)}
        \label{fig:regression-results-all-ConsistentRepresentationk5-2-houses}
    \end{subfigure}
   \begin{subfigure}[b]{0.32\textwidth}
        \includegraphics[width=\textwidth]{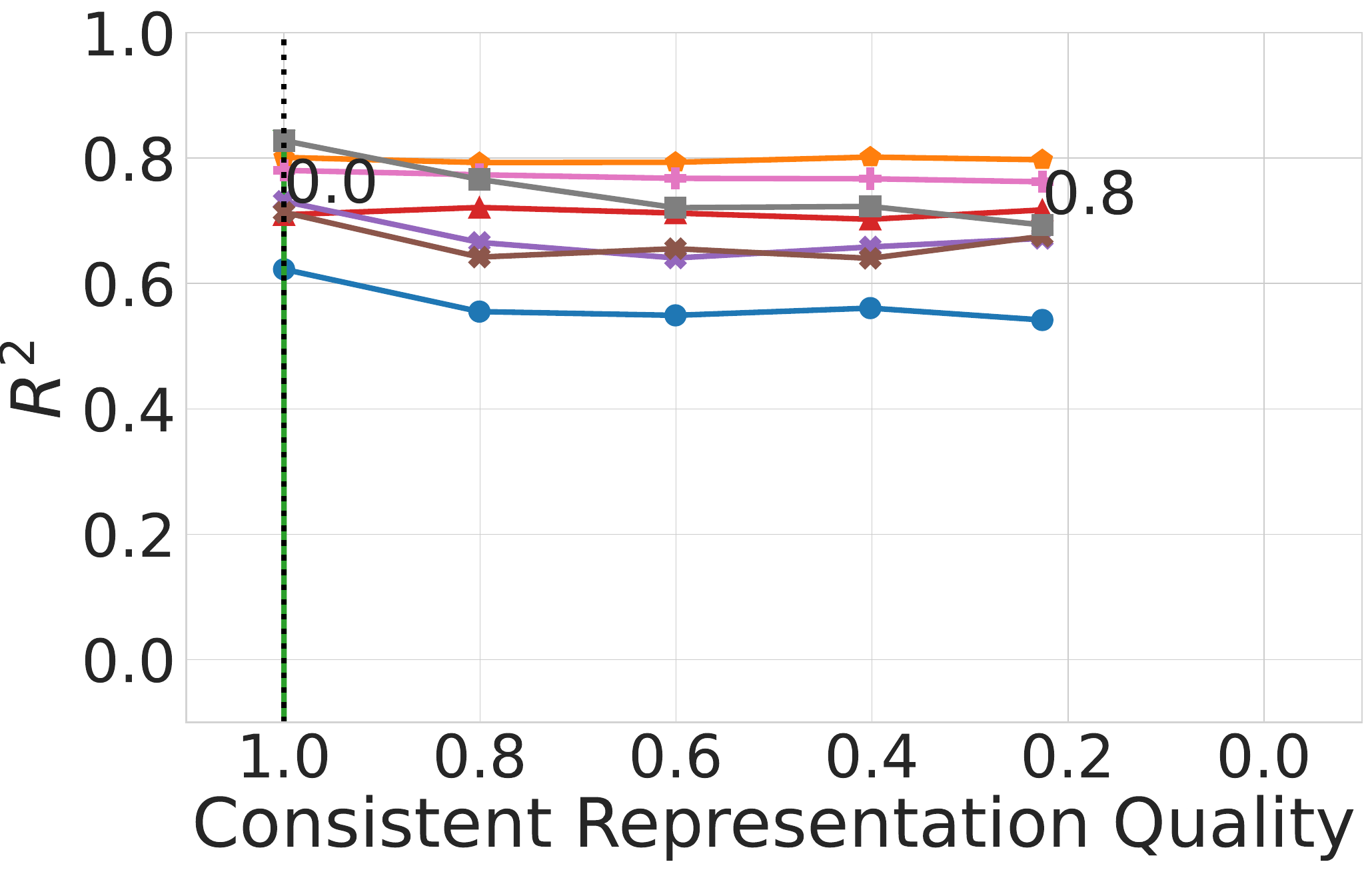}
        \caption{Consistency with $k_{v}=5$ (Sce. 3)}
        \label{fig:regression-results-all-ConsistentRepresentationk5-3-houses}
    \end{subfigure} \\
    \begin{subfigure}[b]{0.32\textwidth}
        \includegraphics[width=\textwidth]{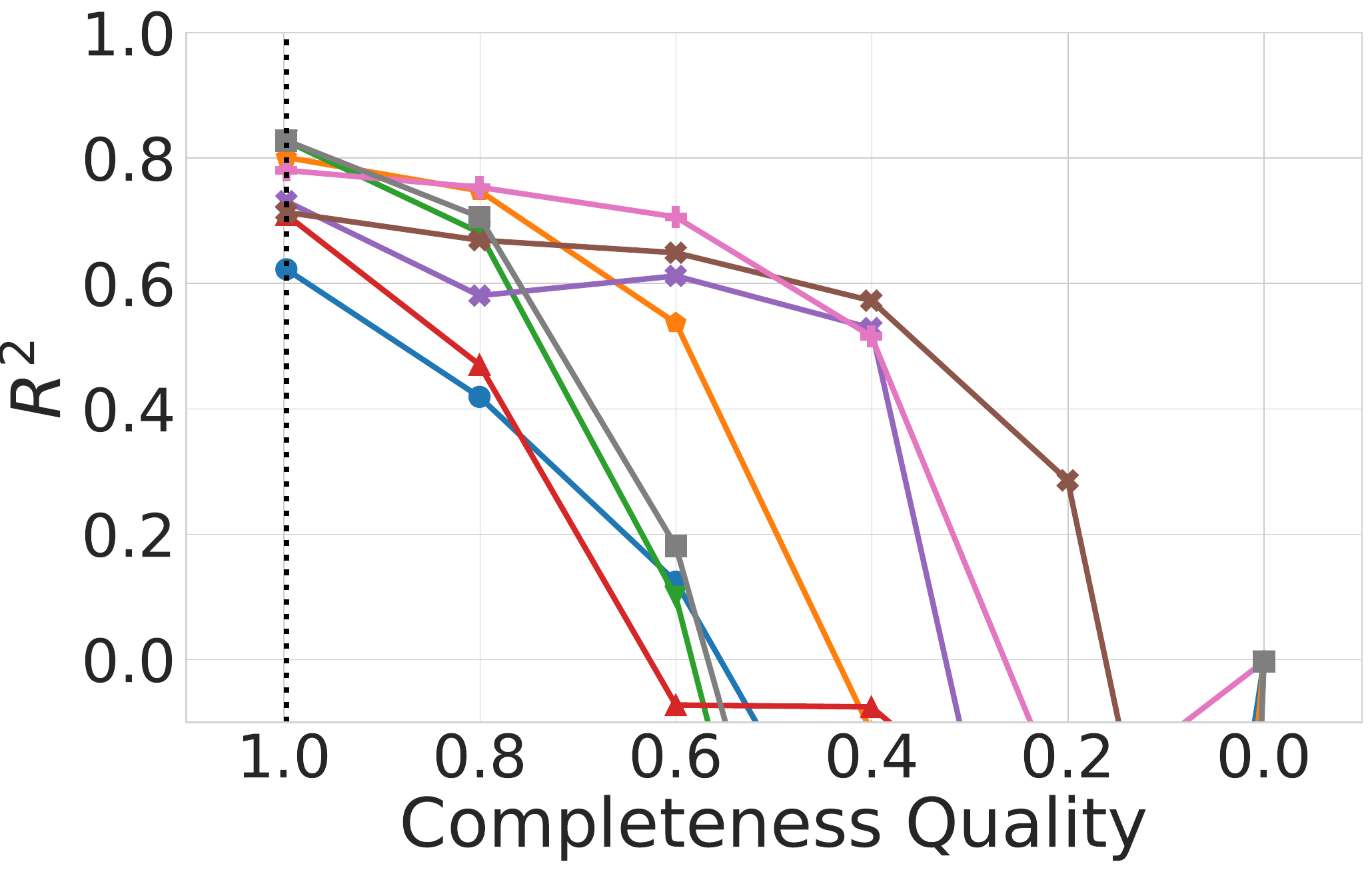}
        \caption{Completeness (Sce. 1)}
        \label{fig:regression-results-all-completeness-1-houses}
    \end{subfigure}
   \begin{subfigure}[b]{0.32\textwidth}
        \includegraphics[width=\textwidth]{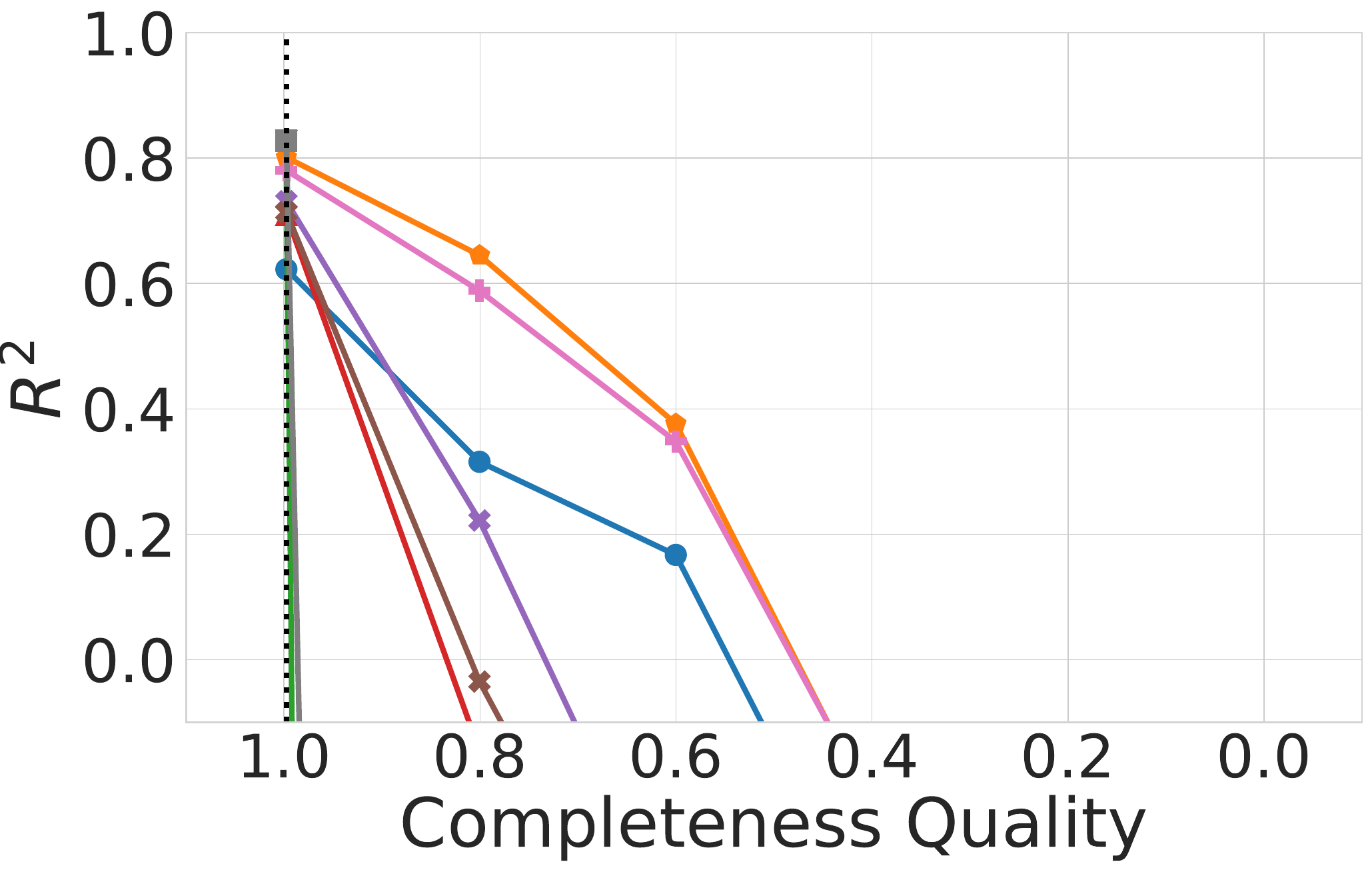}
        \caption{Completeness (Sce. 2)}
        \label{fig:regression-results-all-completeness-2-houses}
    \end{subfigure}
   \begin{subfigure}[b]{0.32\textwidth}
        \includegraphics[width=\textwidth]{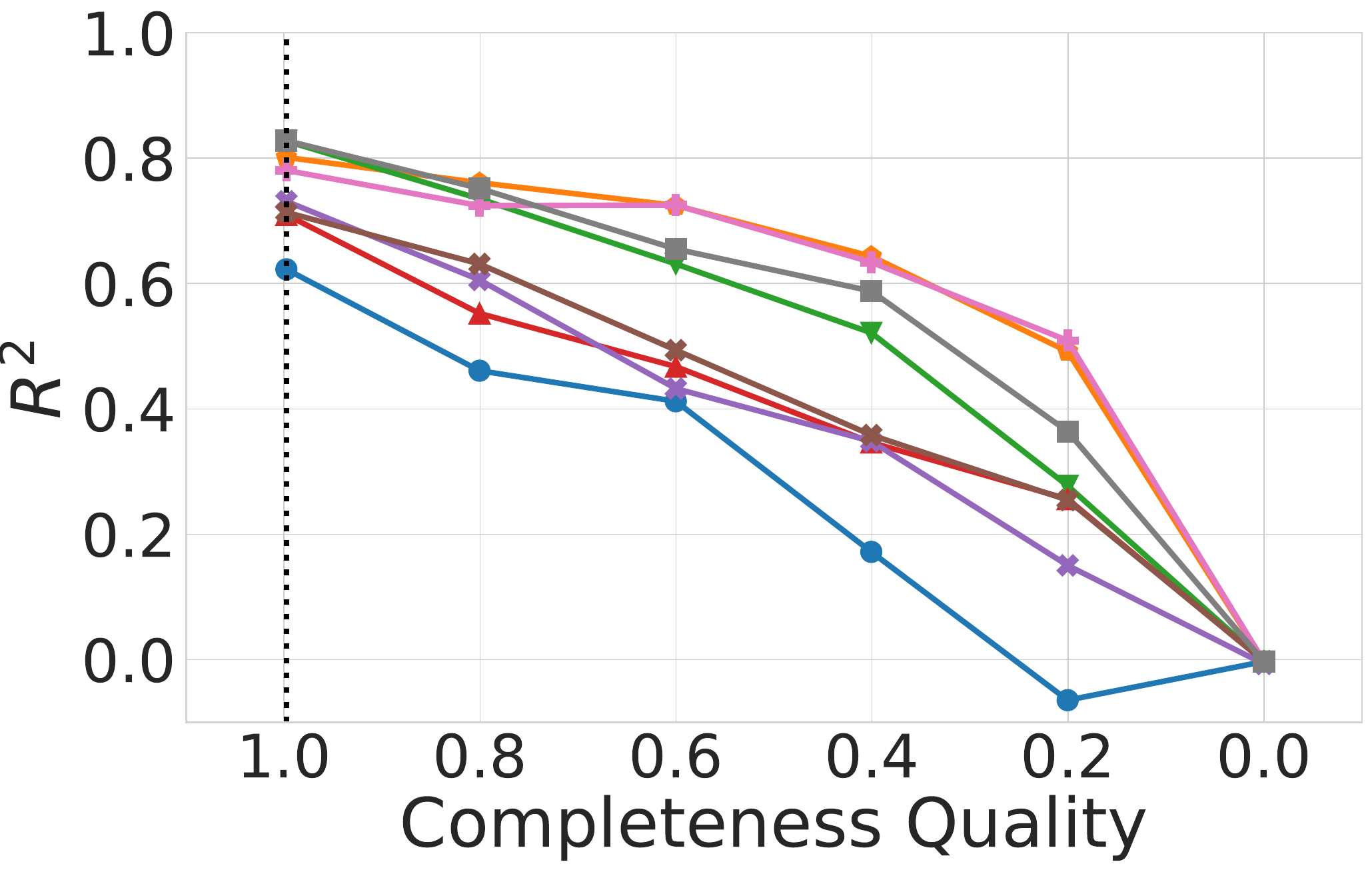}
        \caption{Completeness (Sce. 3)}
        \label{fig:regression-results-all-completeness-3-houses}
    \end{subfigure}  \\
    \begin{subfigure}[b]{0.32\textwidth}
        \includegraphics[width=\textwidth]{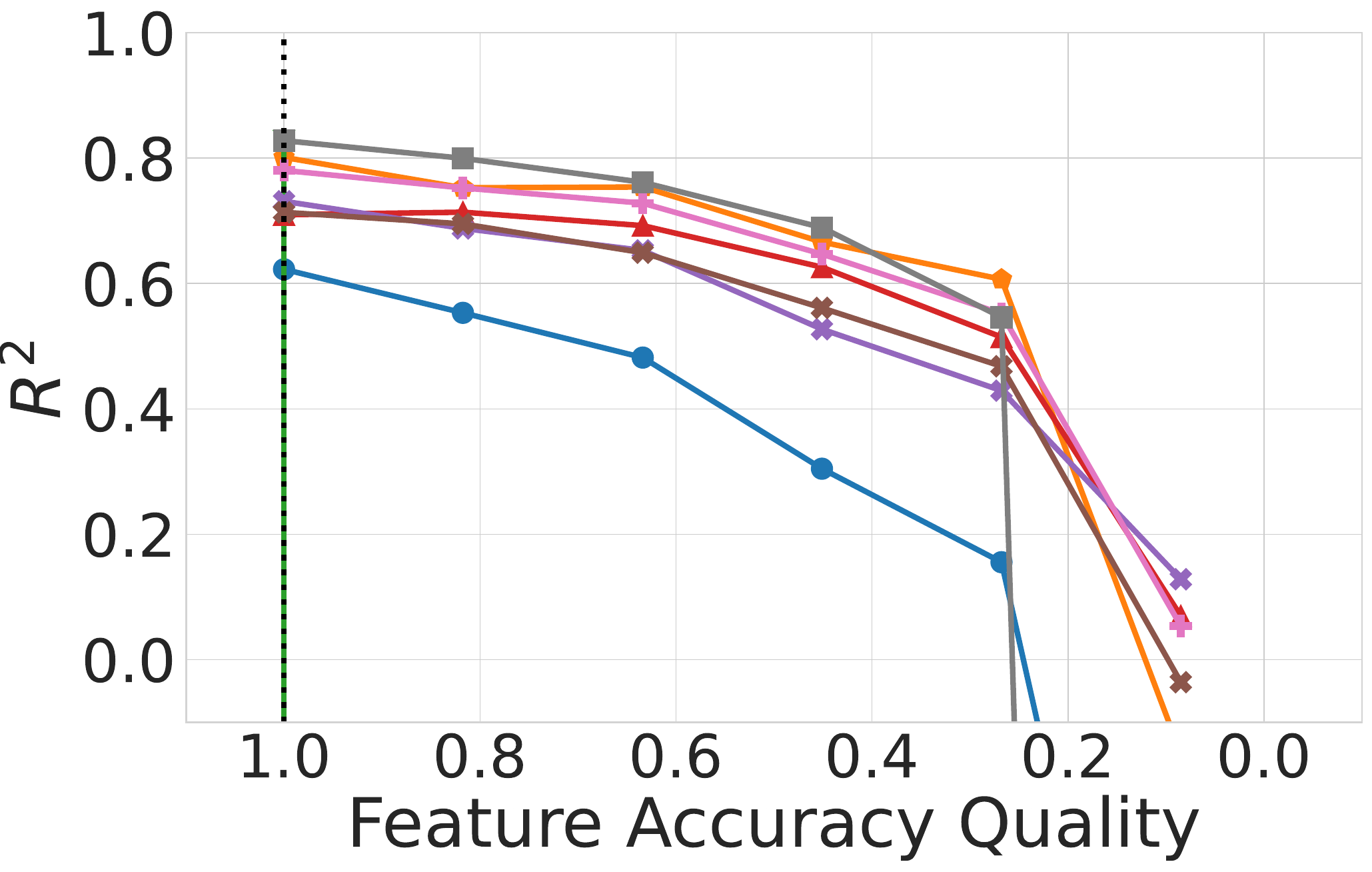}
        \caption{Feature Accuracy (Sce. 1)}
        \label{fig:regression-results-all-FeatureAccuracy-1-houses}
    \end{subfigure}
   \begin{subfigure}[b]{0.32\textwidth}
        \includegraphics[width=\textwidth]{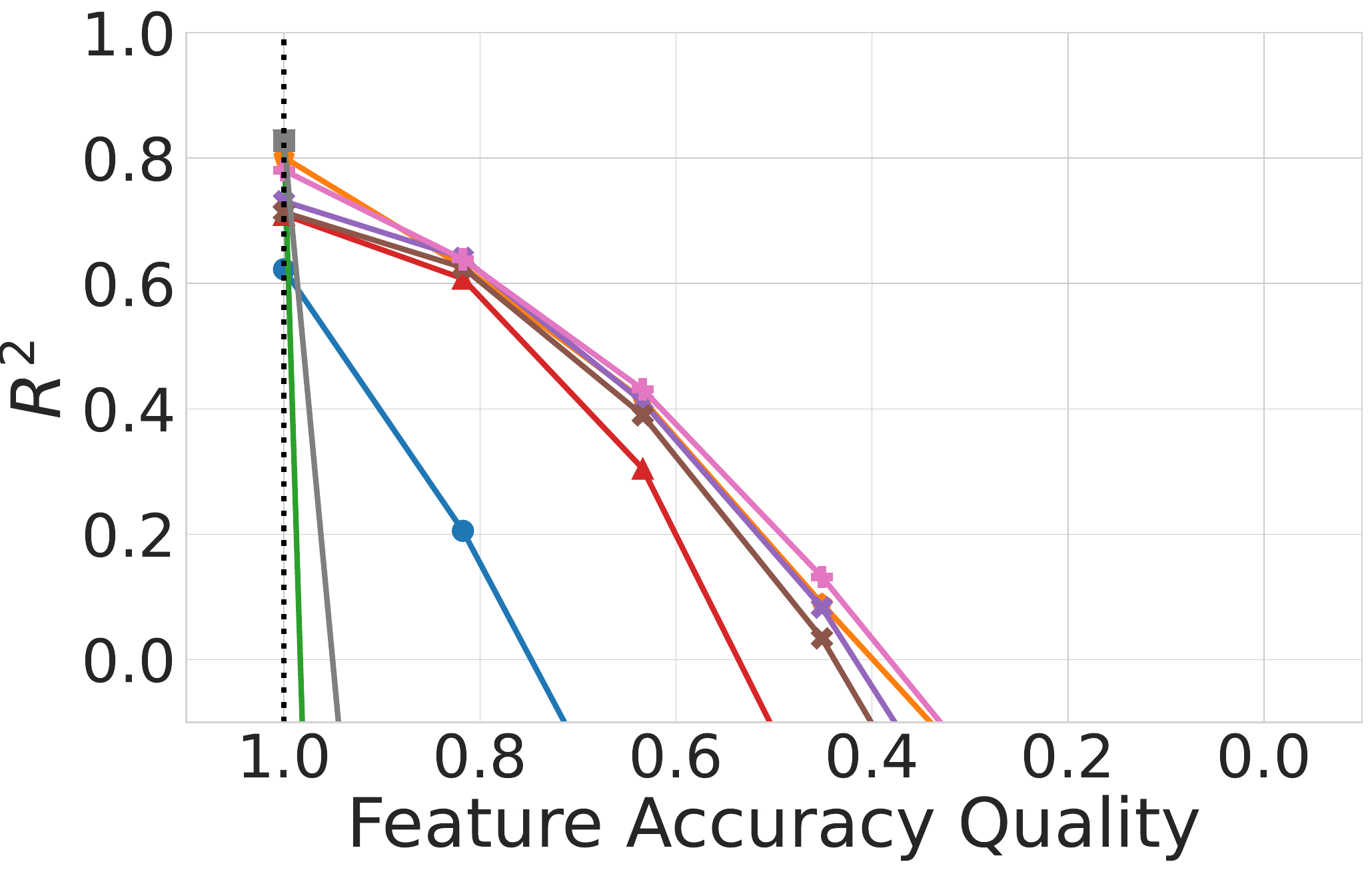}
        \caption{Feature Accuracy (Sce. 2)}
        \label{fig:regression-results-all-FeatureAccuracy-2-houses}
    \end{subfigure}
    \begin{subfigure}[b]{0.32\textwidth}
        \includegraphics[width=\textwidth]{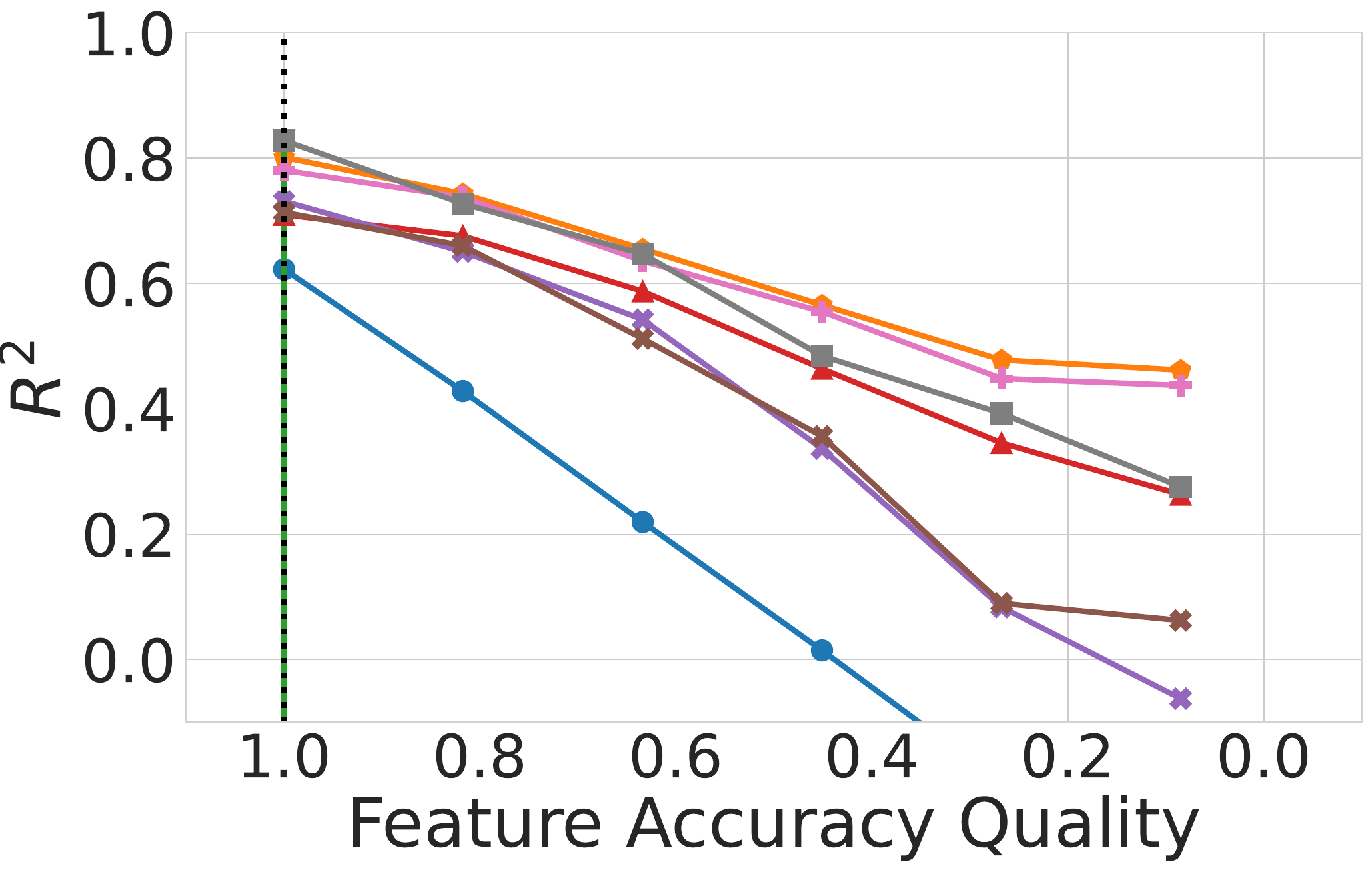}
        \caption{Feature Accuracy (Sce. 3)}
        \label{fig:regression-results-all-FeatureAccuracy-3-houses}
    \end{subfigure} \\
    \begin{subfigure}[b]{0.32\textwidth}
        \includegraphics[width=\textwidth]{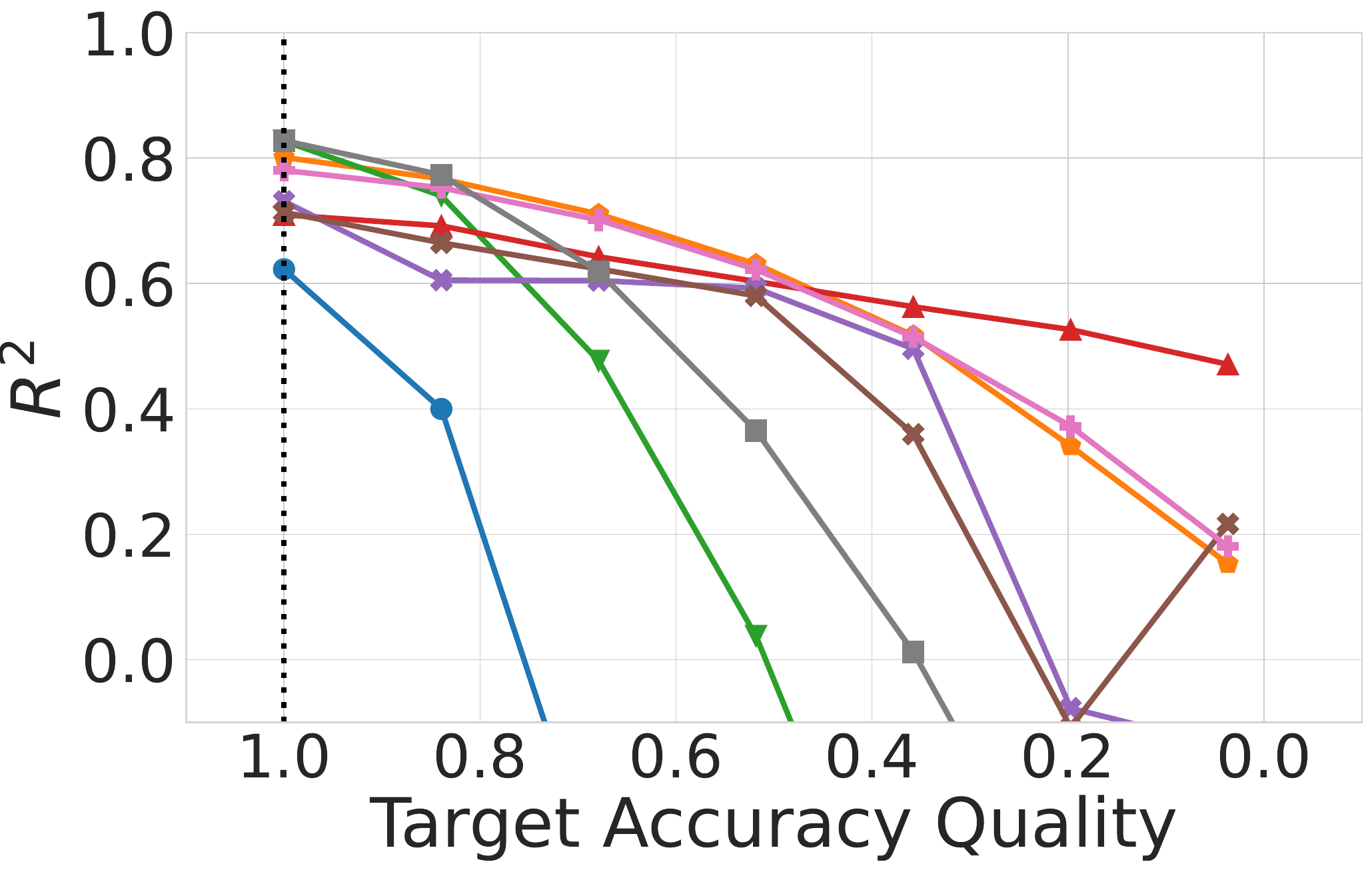}
        \caption{Target Accuracy}
        \label{fig:regression-results-all-TargetAccuracy-1-houses}
    \end{subfigure}
    \begin{subfigure}[b]{0.32\textwidth}
        \includegraphics[width=\textwidth]{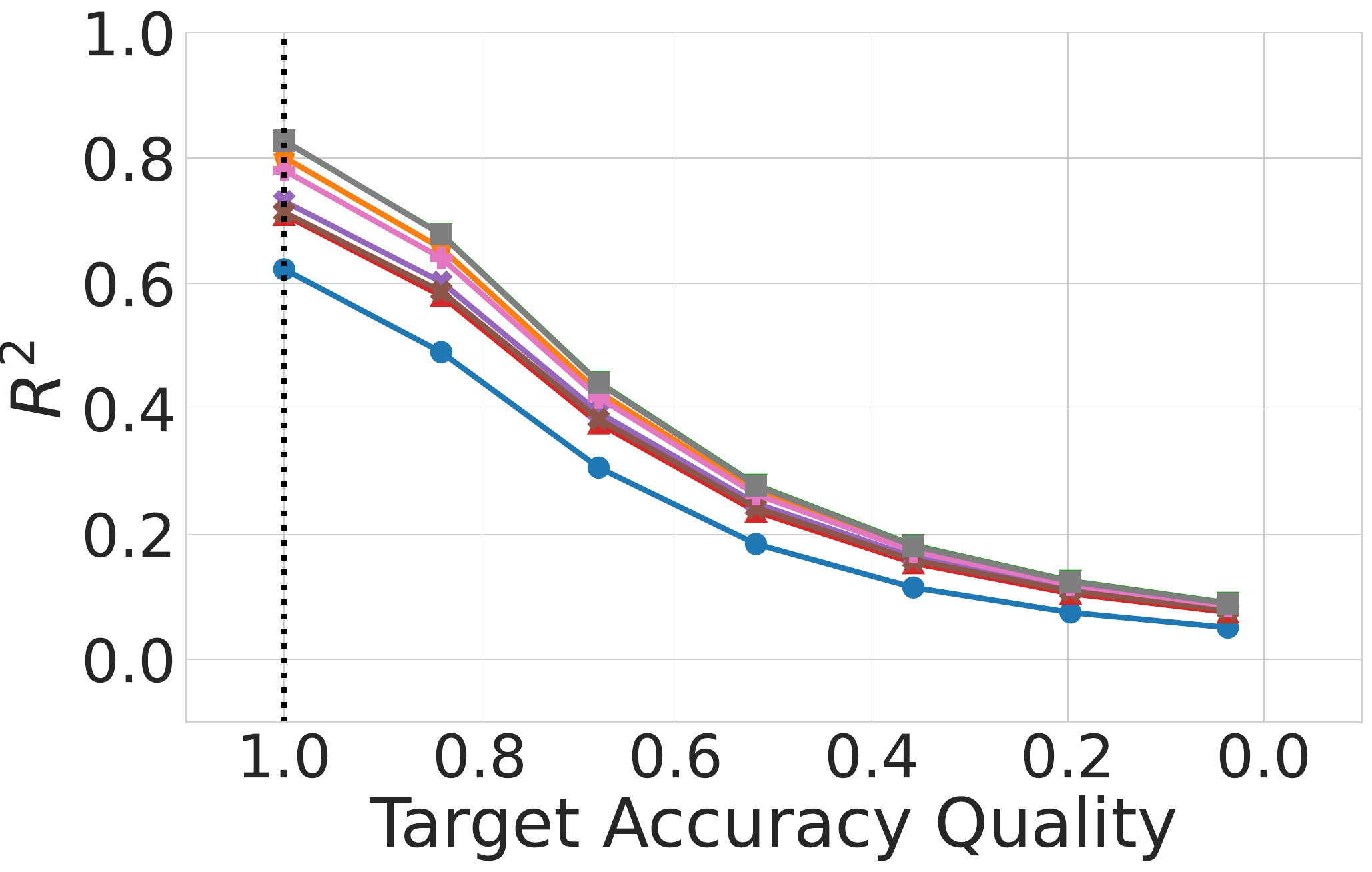}
        \caption{Target Accuracy}
        \label{fig:regression-results-all-TargetAccuracy-2-houses}
    \end{subfigure}
\begin{subfigure}[b]{0.32\textwidth}
        \includegraphics[width=\textwidth]{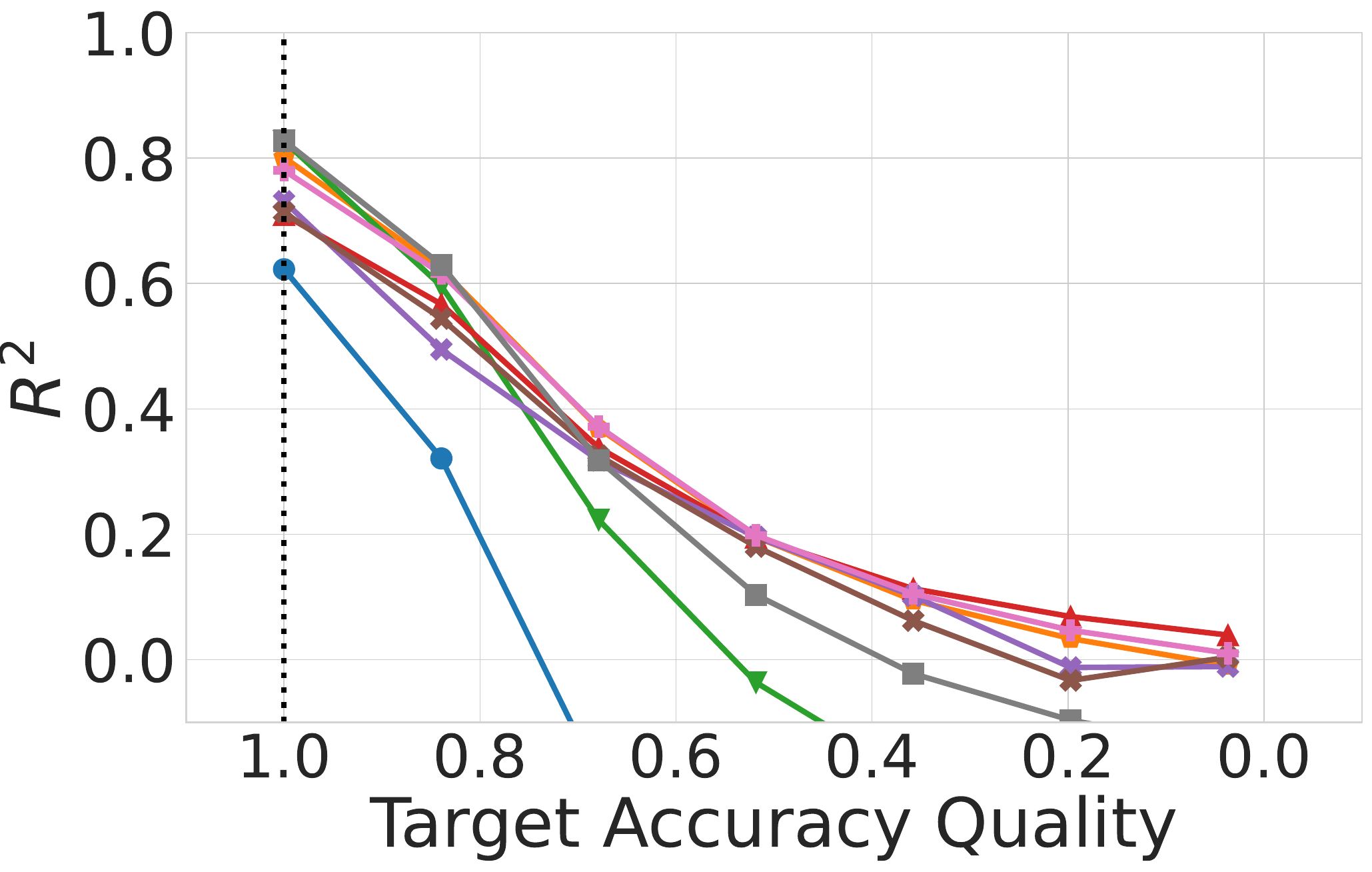}
        \caption{Target Accuracy}
        \label{fig:regression-results-all-TargetAccuracy-3-houses}
    \end{subfigure}  \\
    \begin{subfigure}[b]{0.32\textwidth}
        \includegraphics[width=\textwidth]{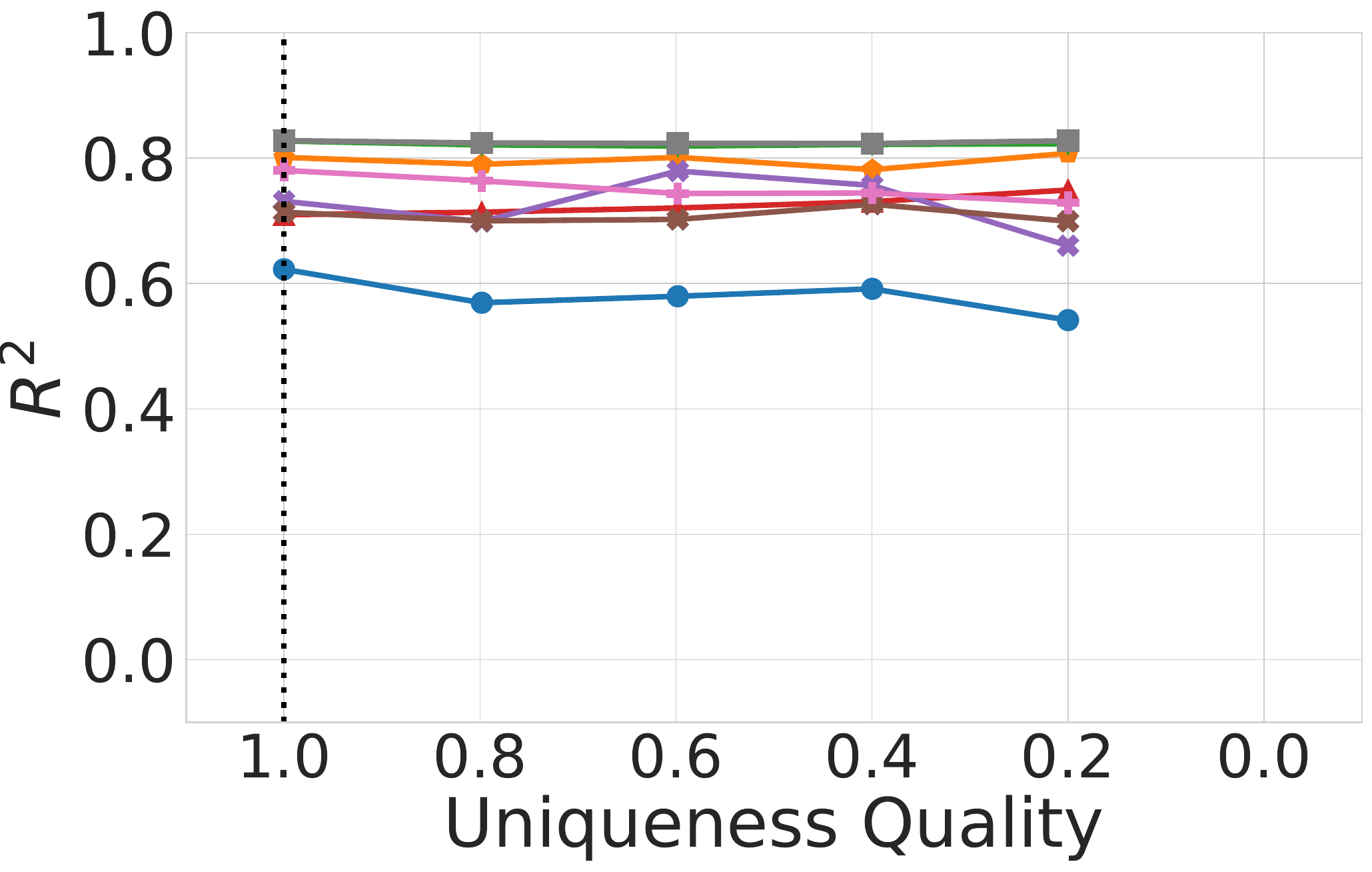}
        \caption{Uniqueness (single duplicate) (Sce. 1)}
        \label{fig:regression-results-all-Uniqueness_dc1-1-houses}
    \end{subfigure}   
    \begin{subfigure}[b]{0.32\textwidth}
        \includegraphics[width=\textwidth]{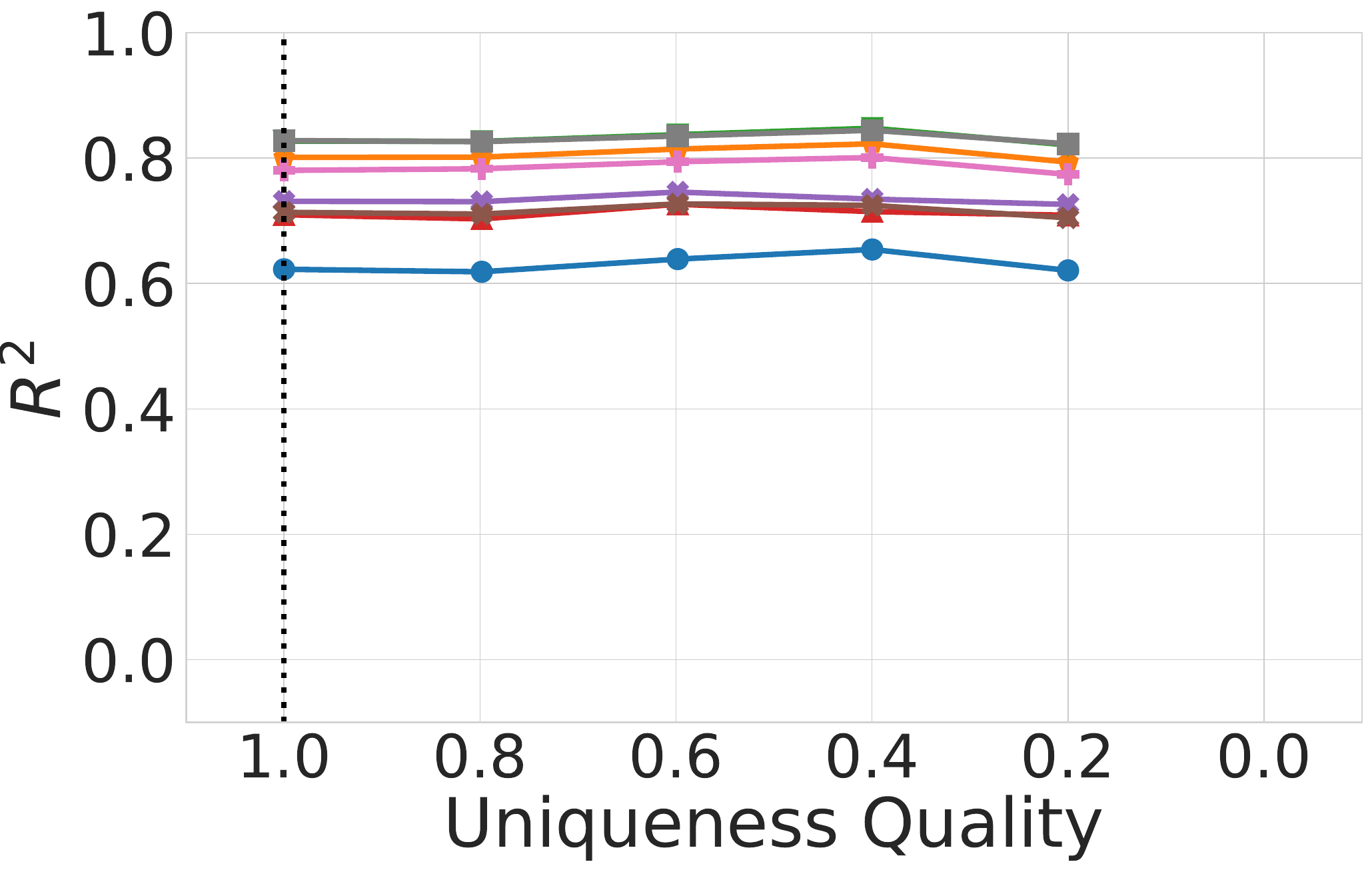}
        \caption{Uniqueness (single duplicate) (Sce. 2)}
        \label{fig:regression-results-all-Uniqueness_dc1-2-houses}
    \end{subfigure}  
    \begin{subfigure}[b]{0.32\textwidth}
        \includegraphics[width=\textwidth]{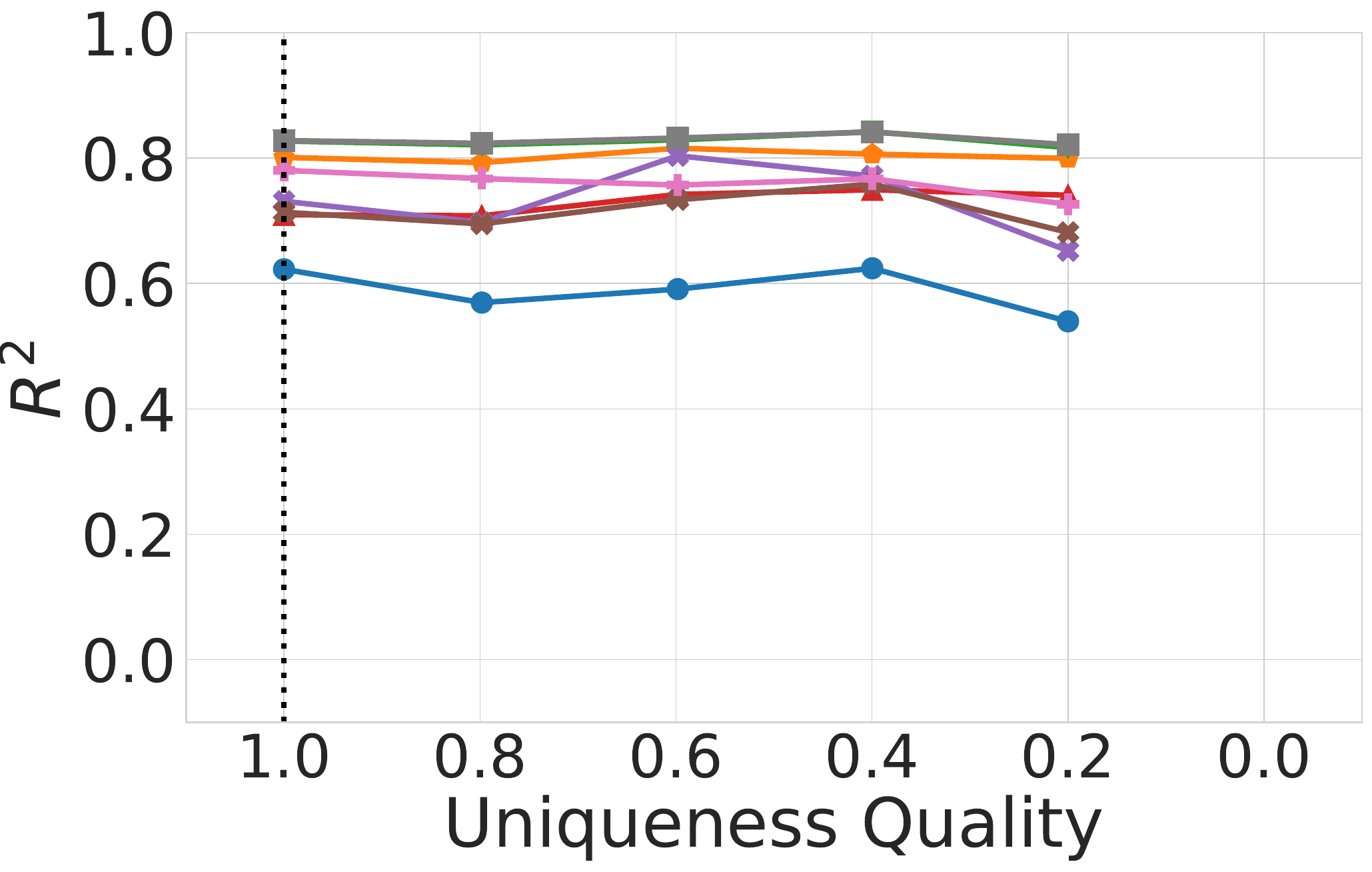}
        \caption{Uniqueness (single duplicate) (Sce. 3)}
        \label{fig:regression-results-all-Uniqueness_dc1-3-houses}
    \end{subfigure} \\
    \begin{subfigure}[b]{0.32\textwidth}
        \includegraphics[width=\textwidth]{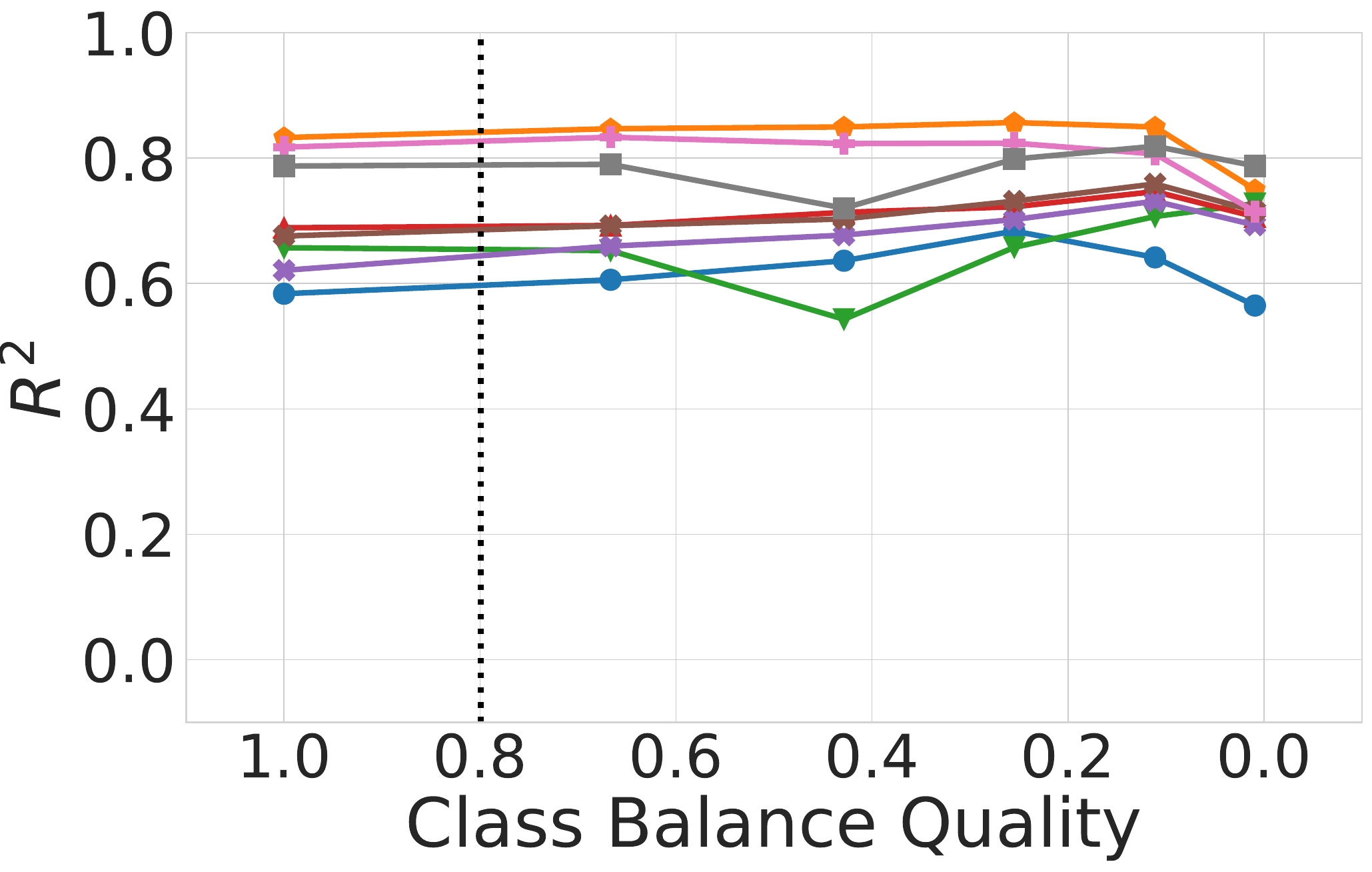}
        \caption{Class Balance (Sce. 1)}
        \label{fig:regression-results-all-ClassBalance-1-houses}
    \end{subfigure}
\begin{subfigure}[b]{0.32\textwidth}
        \includegraphics[width=\textwidth]{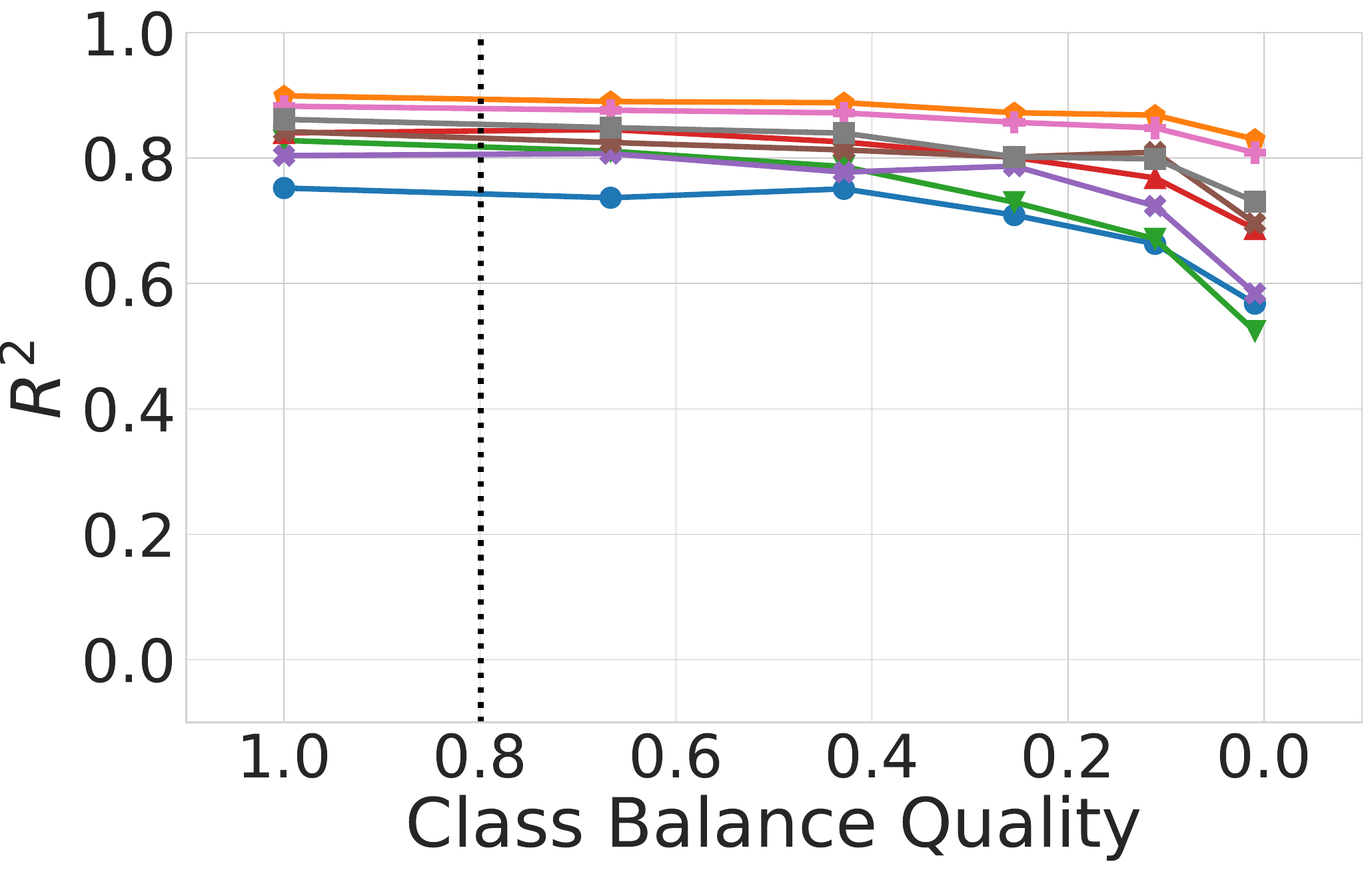}
        \caption{Class Balance (Sce. 2)}
        \label{fig:regression-results-all-ClassBalance-2-houses}
    \end{subfigure}
    \begin{subfigure}[b]{0.32\textwidth}
        \includegraphics[width=\textwidth]{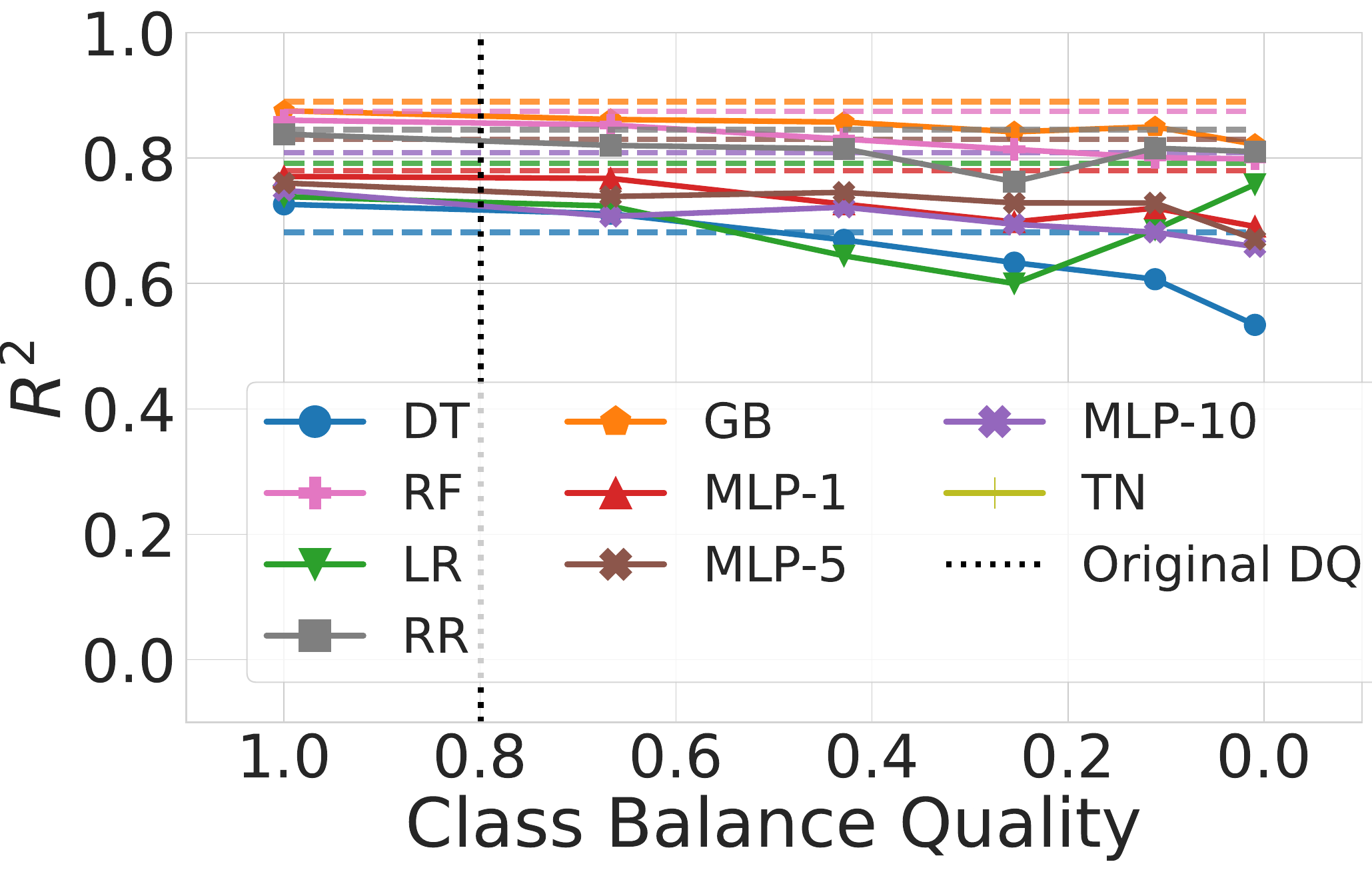}
        \caption{Class Balance (Sce. 3)}
        \label{fig:regression-results-all-ClassBalance-3-houses}
    \end{subfigure}
 \end{adjustbox} 
    \caption{\revision{$R^2$ of the regression algorithms for \textsf{Houses} dataset.}}
    \label{fig:regression-results-all-house}
\end{figure*}

\begin{figure*}[!htbp]
\captionsetup[subfigure]{aboveskip=-1pt,belowskip=-1pt}
    \centering
\begin{adjustbox}{minipage=\linewidth}
    \begin{subfigure}[b]{0.32\textwidth}
        \includegraphics[width=\textwidth]{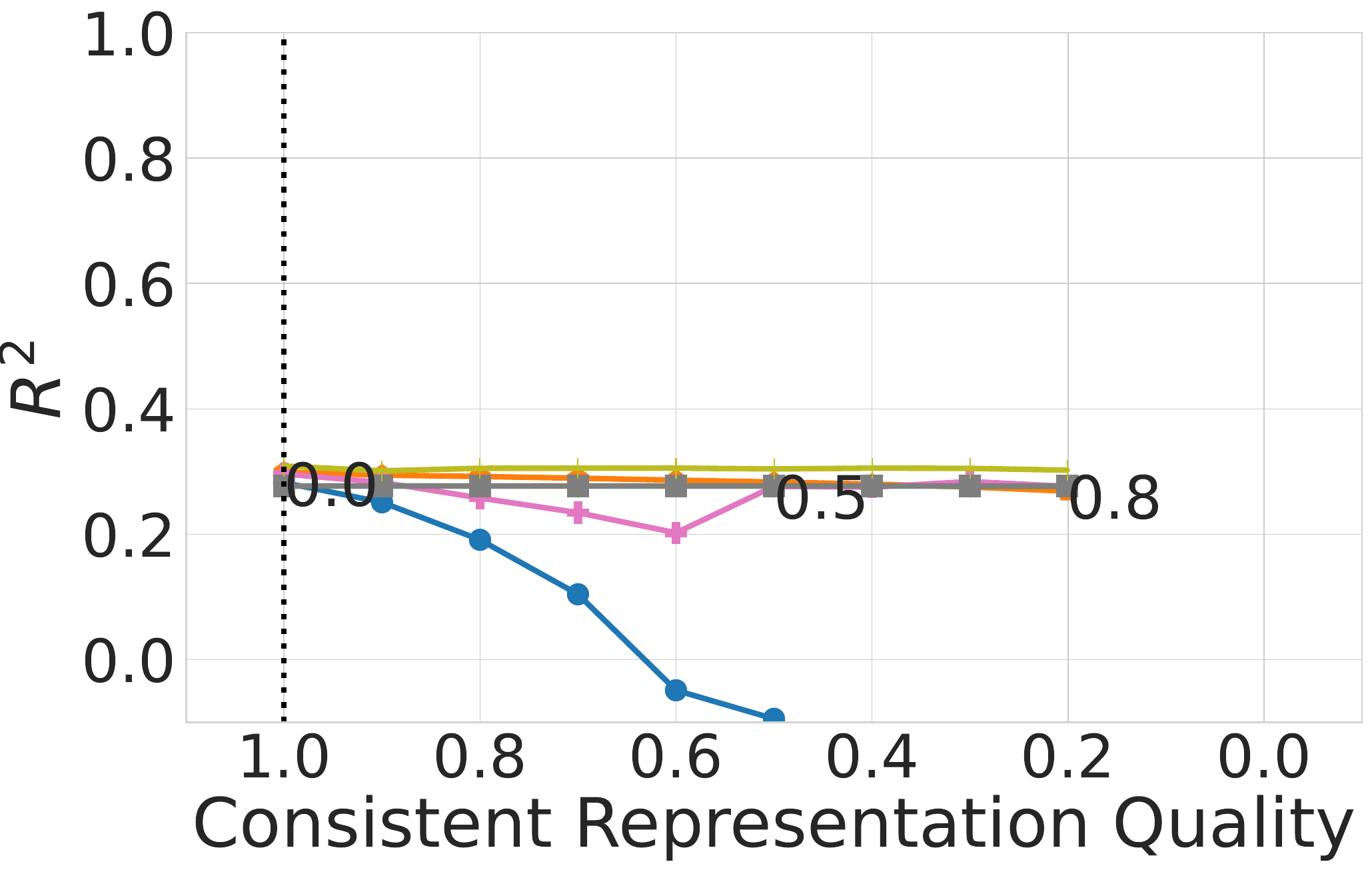}
        \caption{Consistency with $k_{v}=5$ (Sce. 1)}
        \label{fig:regression-results-all-ConsistentRepresentationk5-1-covid}
    \end{subfigure}
    \begin{subfigure}[b]{0.32\textwidth}
        \includegraphics[width=\textwidth]{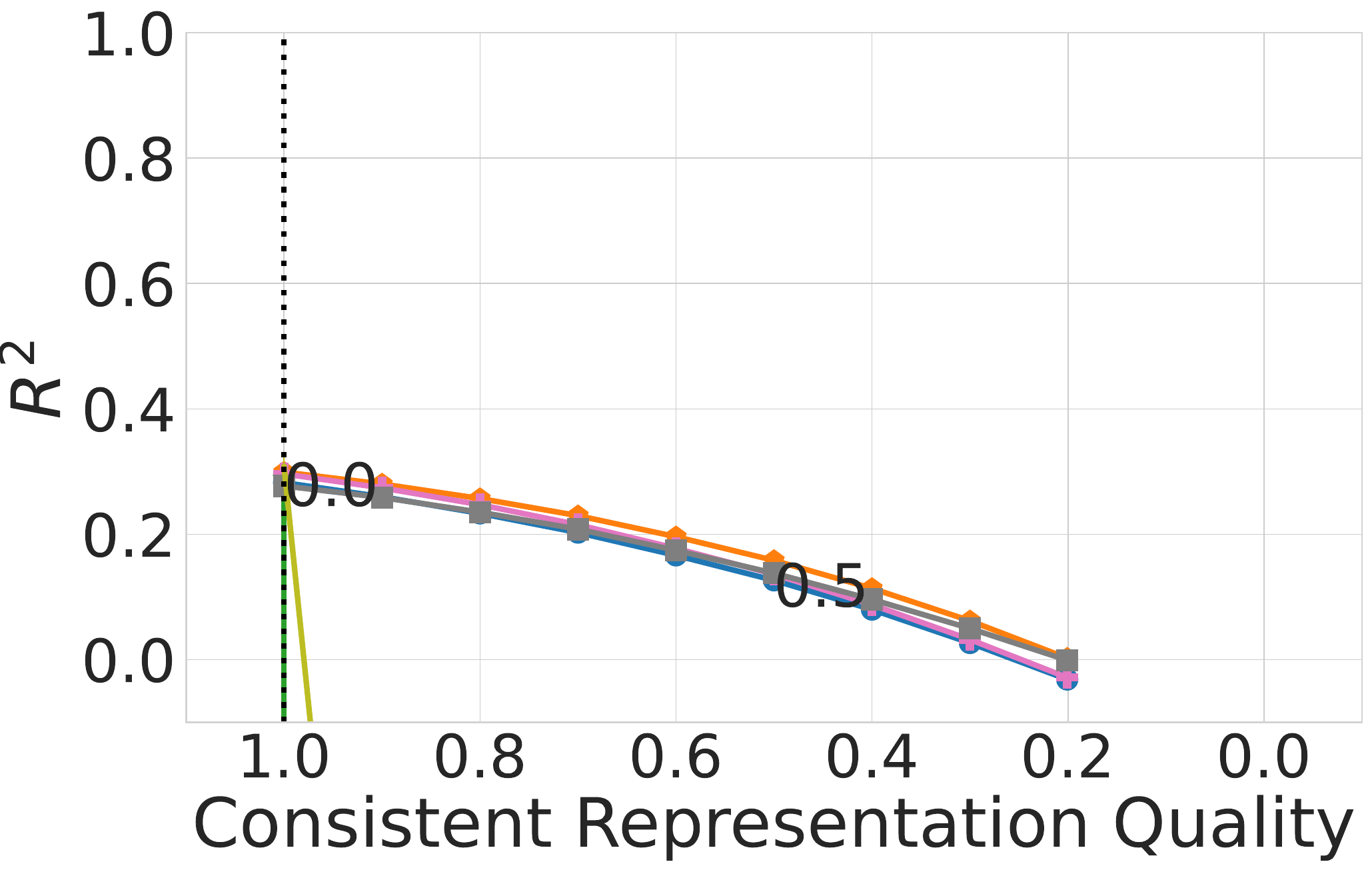}
        \caption{Consistency with $k_{v}=5$ (Sce. 2)}
        \label{fig:regression-results-all-ConsistentRepresentationk5-2-covid}
    \end{subfigure}
   \begin{subfigure}[b]{0.32\textwidth}
        \includegraphics[width=\textwidth]{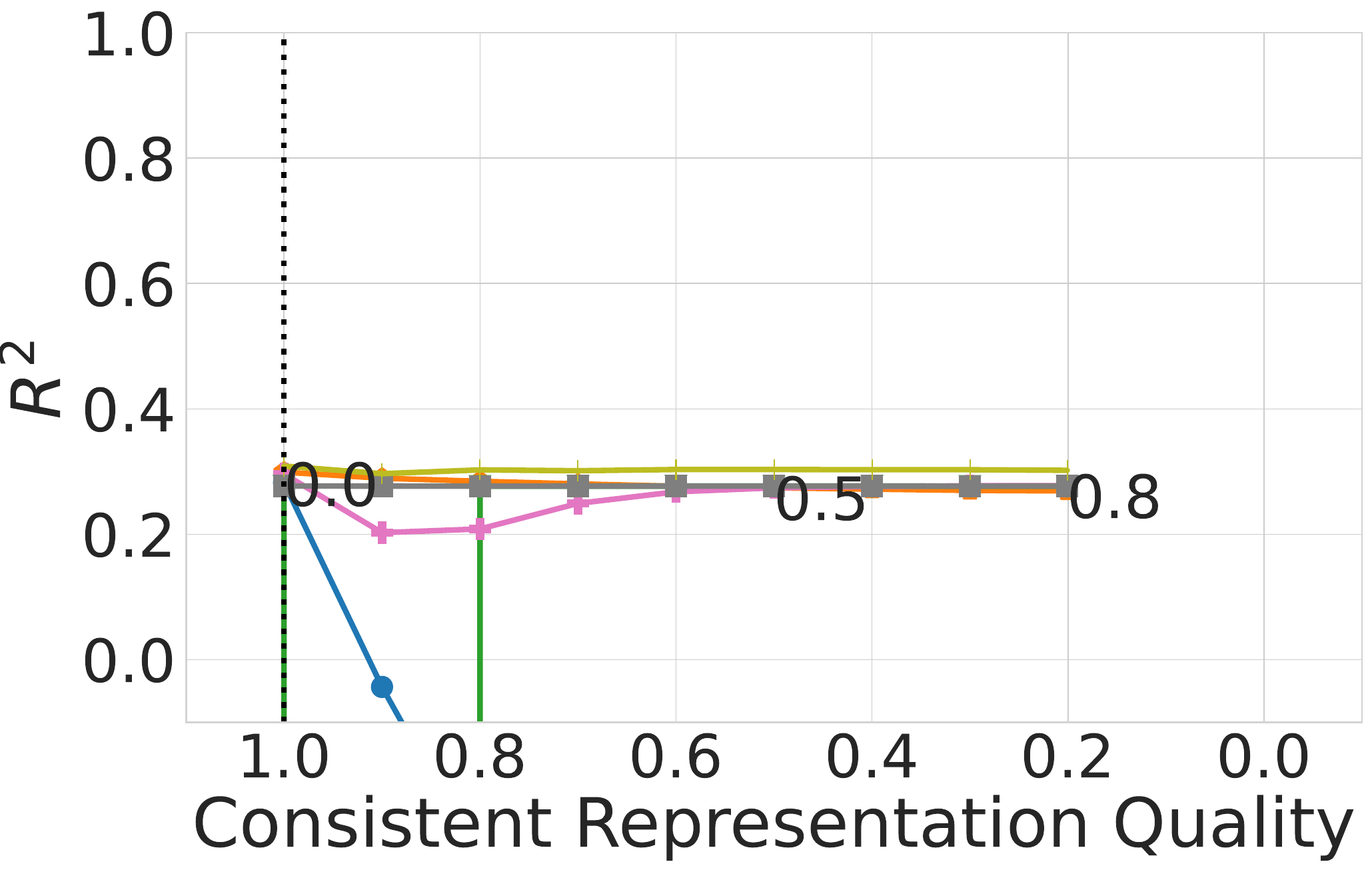}
        \caption{Consistency with $k_{v}=5$ (Sce. 3)}
        \label{fig:regression-results-all-ConsistentRepresentationk5-3-covid}
    \end{subfigure} \\
    \begin{subfigure}[b]{0.32\textwidth}
        \includegraphics[width=\textwidth]{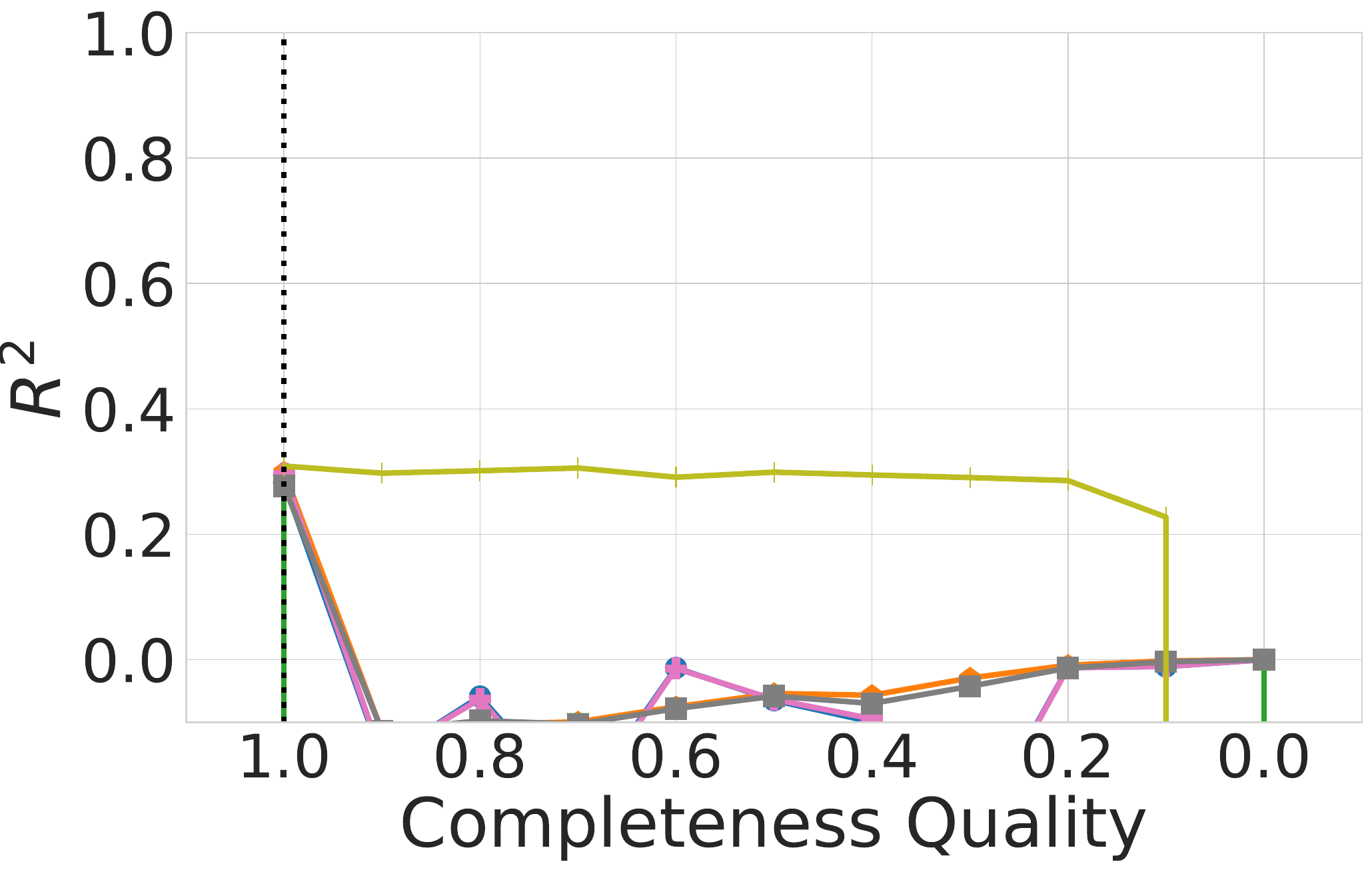}
        \caption{Completeness (Sce. 1)}
        \label{fig:regression-results-all-completeness-1-covid}
    \end{subfigure}
   \begin{subfigure}[b]{0.32\textwidth}
        \includegraphics[width=\textwidth]{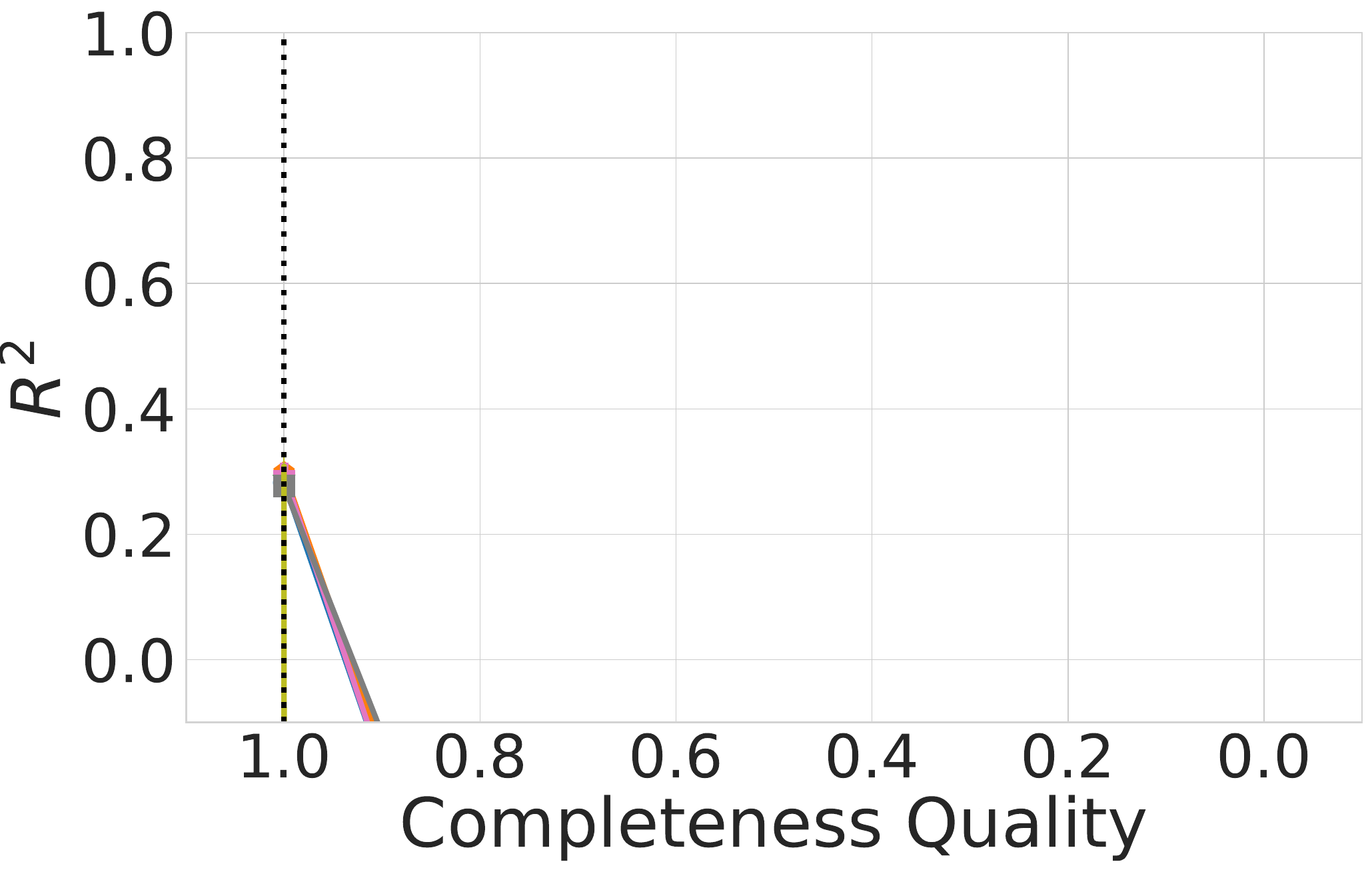}
        \caption{Completeness (Sce. 2)}
        \label{fig:regression-results-all-completeness-2-covid}
    \end{subfigure}
   \begin{subfigure}[b]{0.32\textwidth}
        \includegraphics[width=\textwidth]{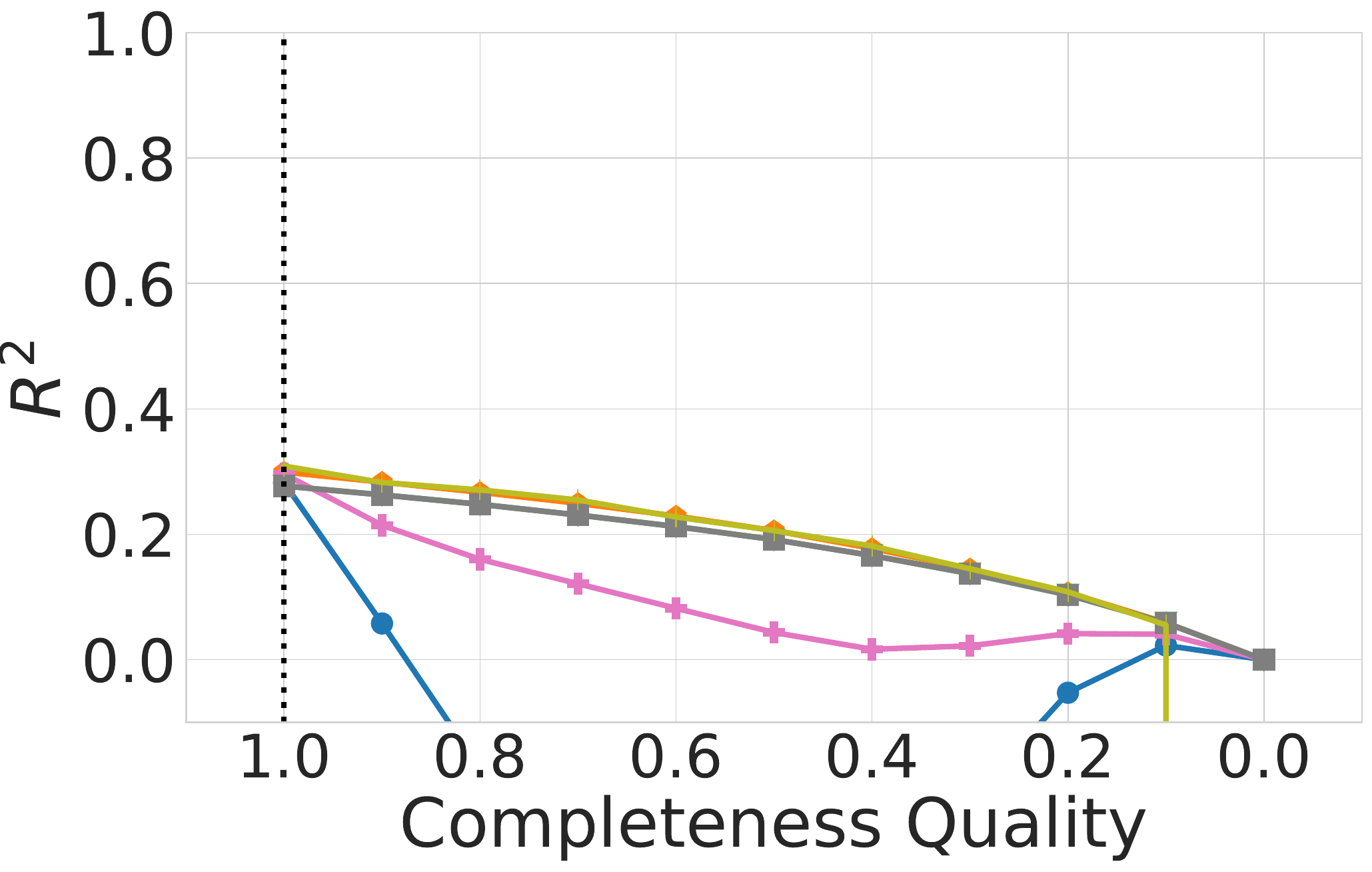}
        \caption{Completeness (Sce. 3)}
        \label{fig:regression-results-all-completeness-3-covid}
    \end{subfigure}  \\
    \begin{subfigure}[b]{0.32\textwidth}
        \includegraphics[width=\textwidth]{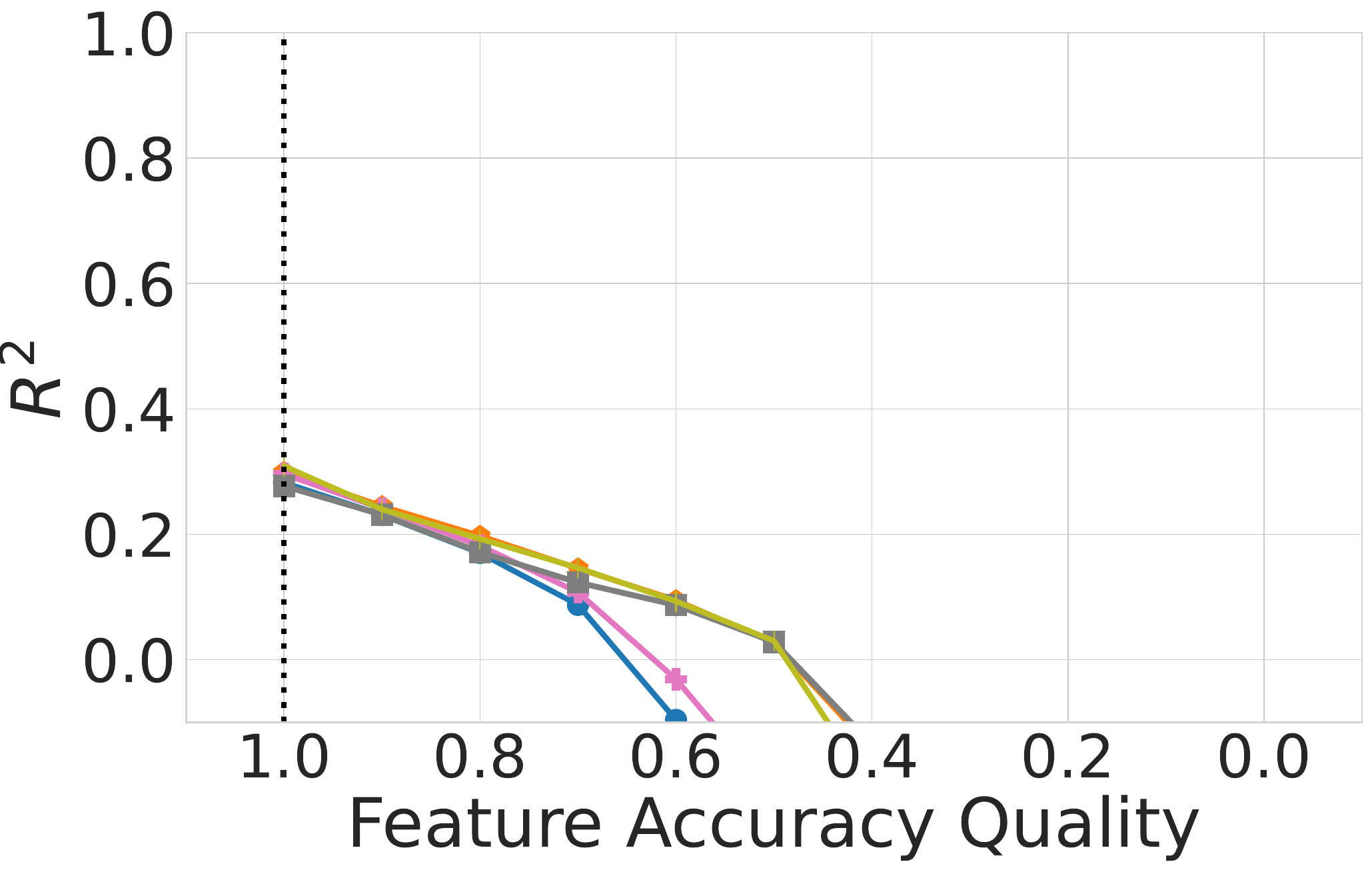}
        \caption{Feature Accuracy (Sce. 1)}
        \label{fig:regression-results-all-FeatureAccuracy-1-covid}
    \end{subfigure}
   \begin{subfigure}[b]{0.32\textwidth}
        \includegraphics[width=\textwidth]{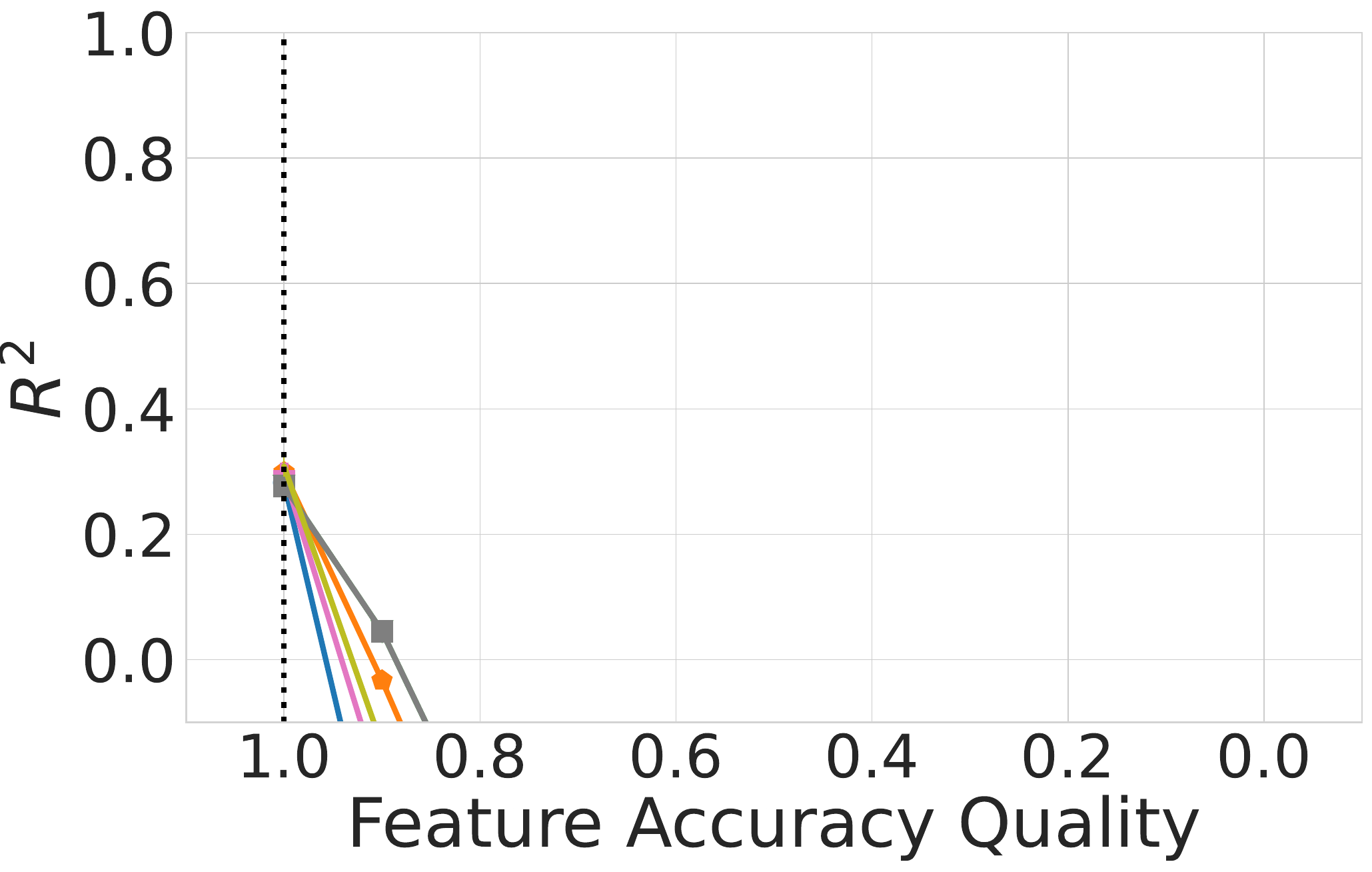}
        \caption{Feature Accuracy (Sce. 2)}
        \label{fig:regression-results-all-FeatureAccuracy-2-covid}
    \end{subfigure}
    \begin{subfigure}[b]{0.32\textwidth}
        \includegraphics[width=\textwidth]{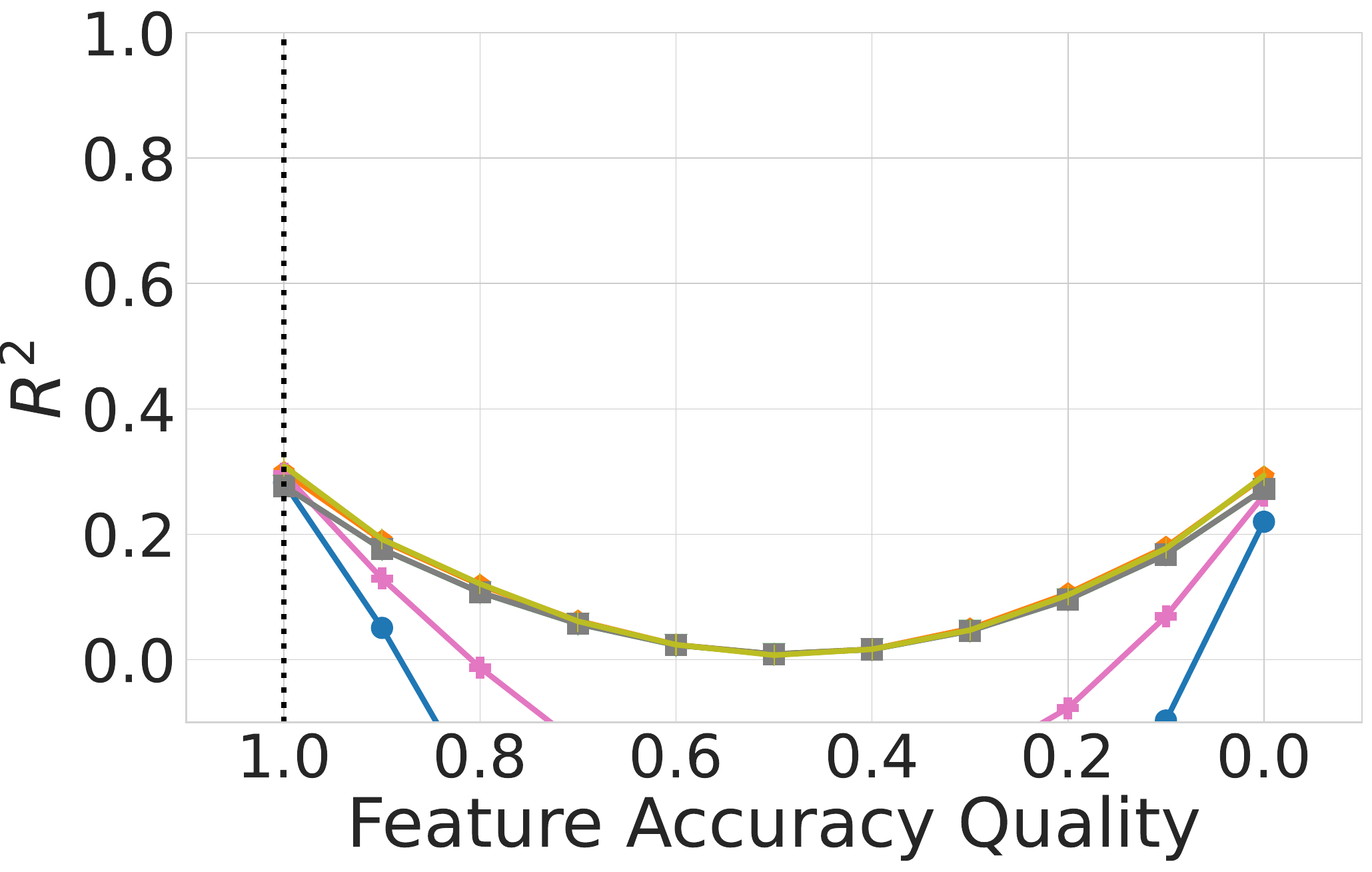}
        \caption{Feature Accuracy (Sce. 3)}
        \label{fig:regression-results-all-FeatureAccuracy-3-covid}
    \end{subfigure} \\
    \begin{subfigure}[b]{0.32\textwidth}
        \includegraphics[width=\textwidth]{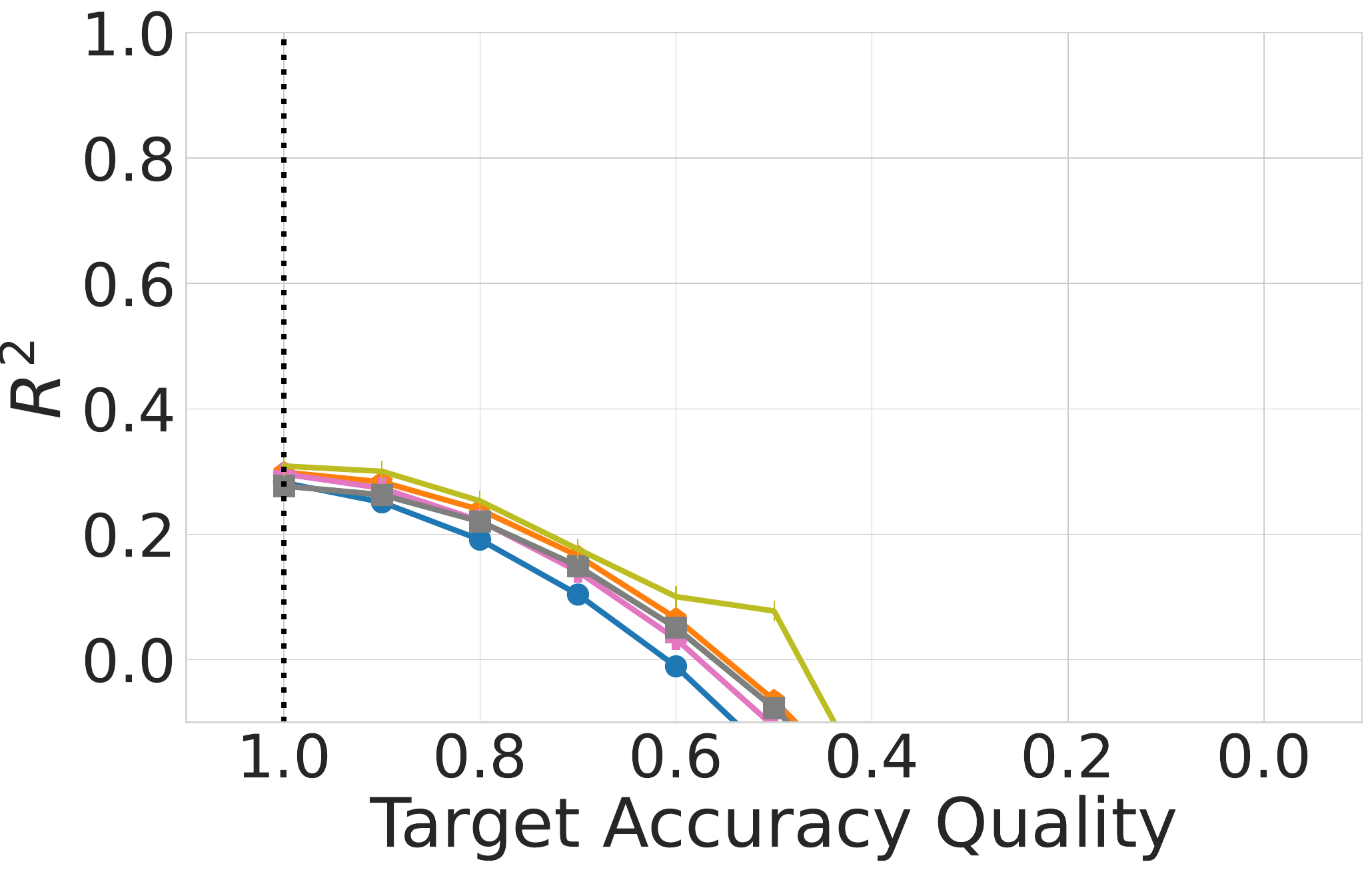}
        \caption{Target Accuracy (Sce. 1)}
        \label{fig:regression-results-all-TargetAccuracy-1-covid}
    \end{subfigure}
    \begin{subfigure}[b]{0.32\textwidth}
        \includegraphics[width=\textwidth]{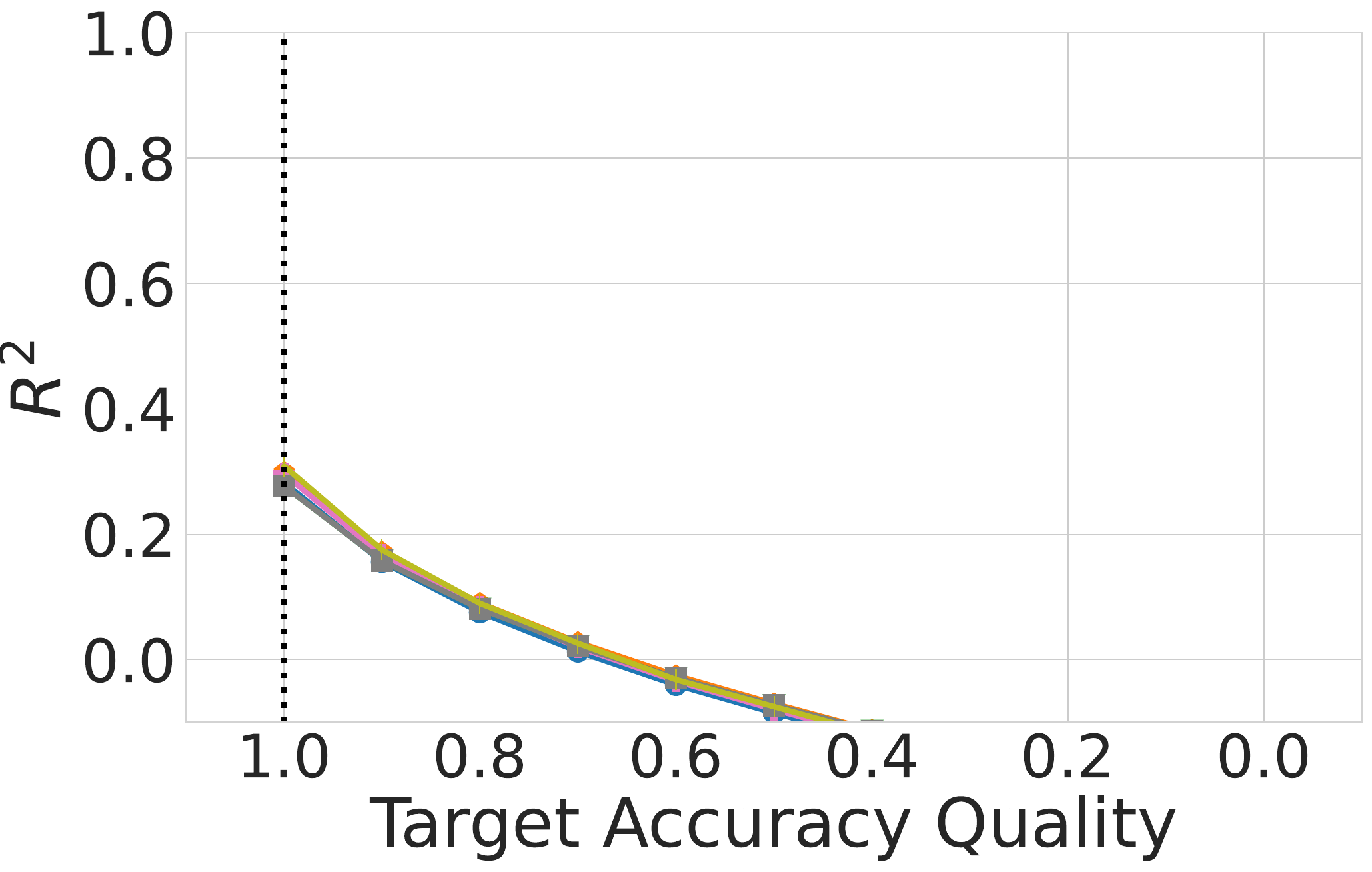}
        \caption{Target Accuracy (Sce. 2)}
        \label{fig:regression-results-all-TargetAccuracy-2-covid}
    \end{subfigure}
\begin{subfigure}[b]{0.32\textwidth}
        \includegraphics[width=\textwidth]{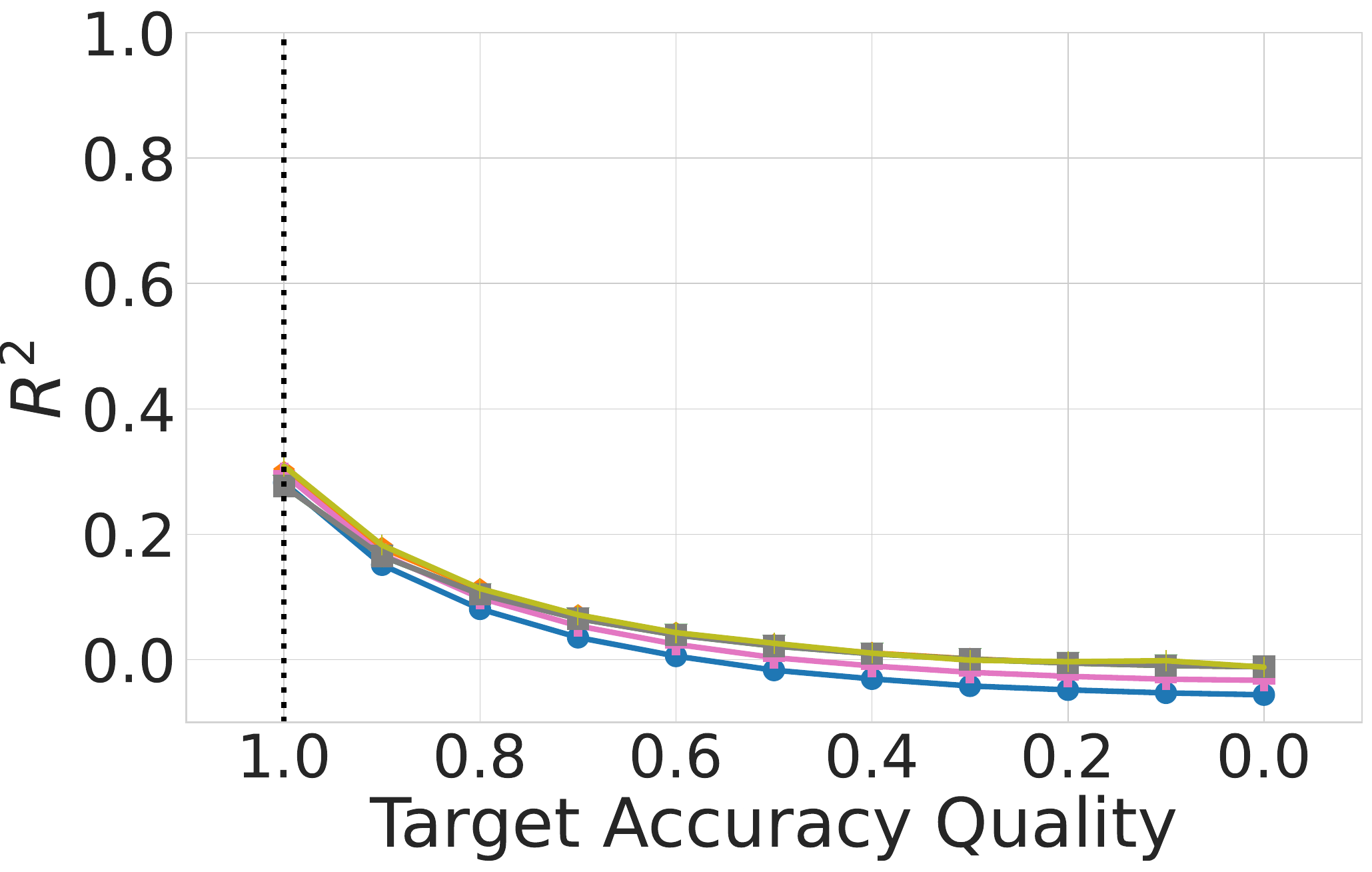}
        \caption{Target Accuracy (Sce. 3)}
        \label{fig:regression-results-all-TargetAccuracy-3-covid}
    \end{subfigure}  \\
    \begin{subfigure}[b]{0.32\textwidth}
        \includegraphics[width=\textwidth]{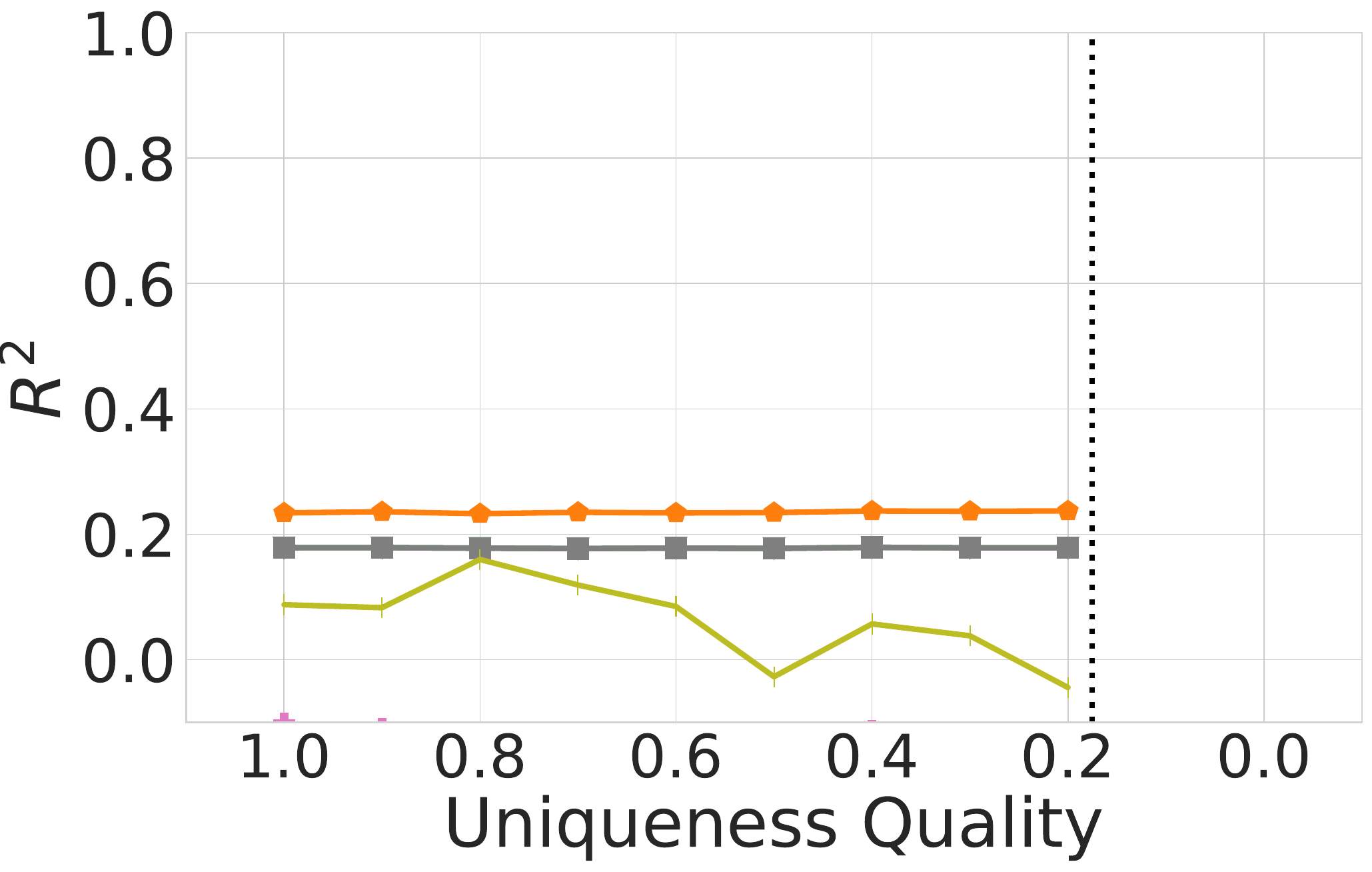}
        \caption{Uniqueness (single duplicate) (Sce. 1)}
        \label{fig:regression-results-all-Uniqueness_dc1-1-covid}
    \end{subfigure}   
    \begin{subfigure}[b]{0.32\textwidth}
        \includegraphics[width=\textwidth]{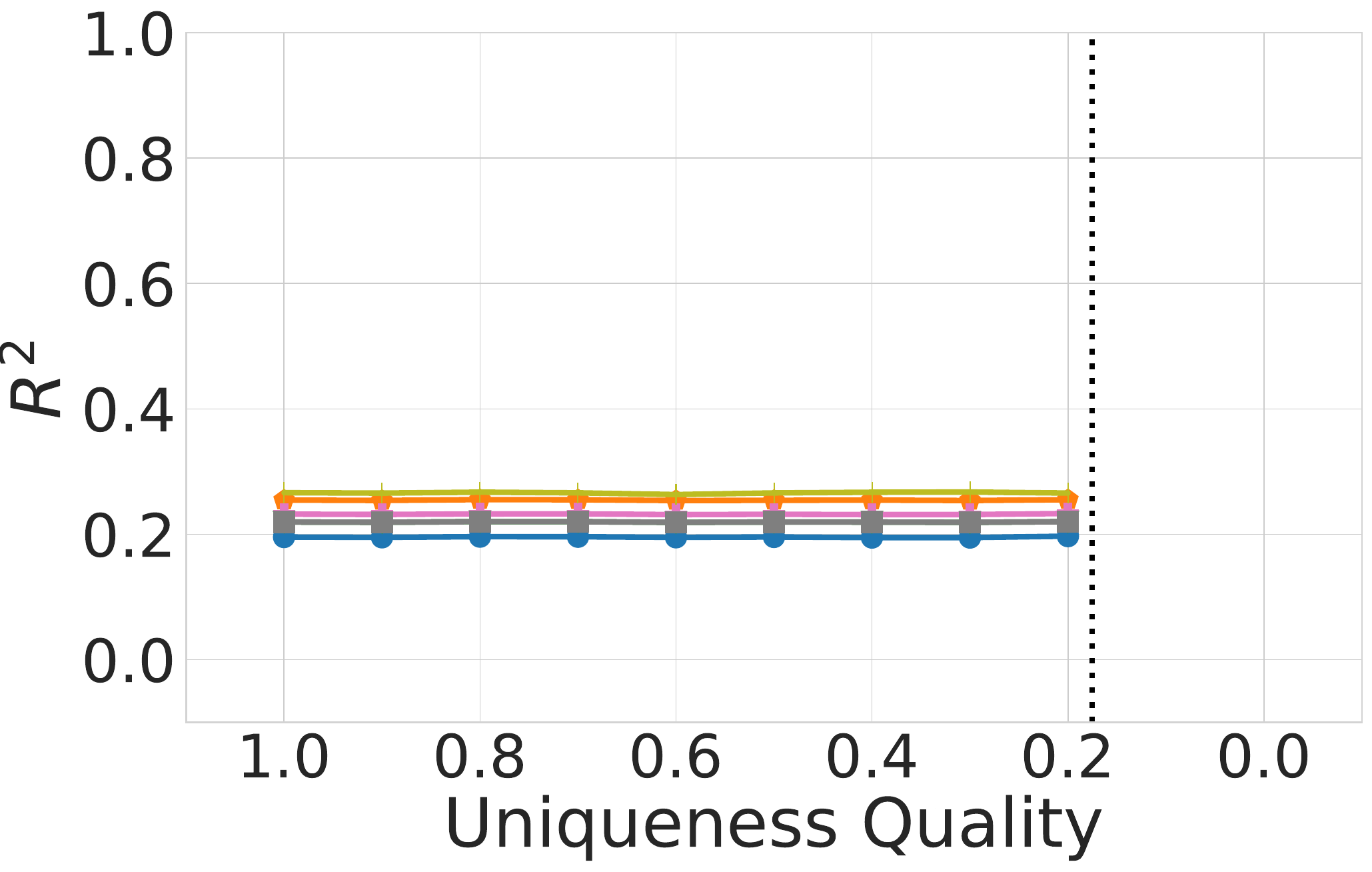}
        \caption{Uniqueness (single duplicate) (Sce. 2)}
        \label{fig:regression-results-all-Uniqueness_dc1-2-covid}
    \end{subfigure}  
    \begin{subfigure}[b]{0.32\textwidth}
        \includegraphics[width=\textwidth]{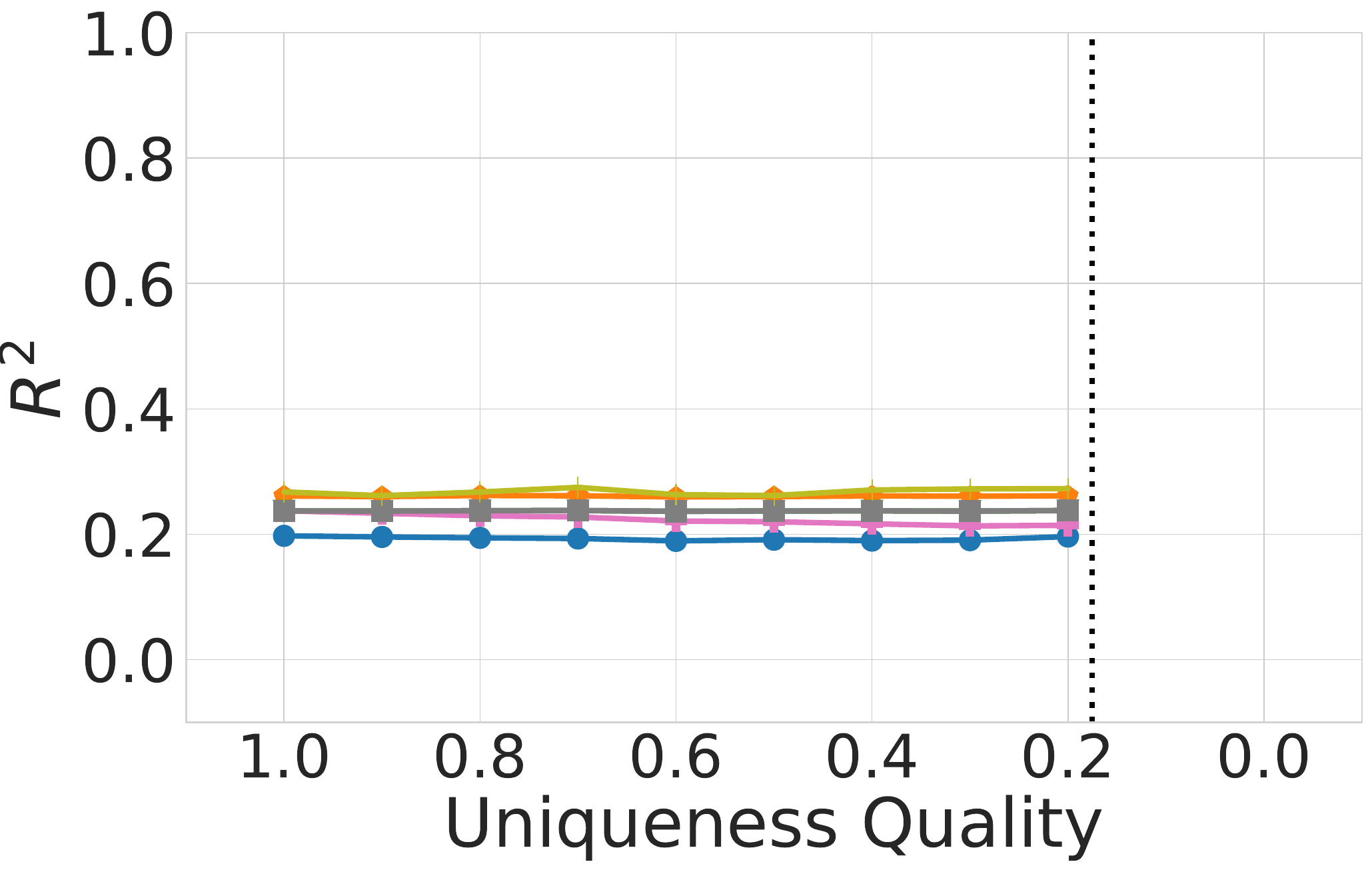}
        \caption{Uniqueness (single duplicate) (Sce. 3)}
        \label{fig:regression-results-all-Uniqueness_dc1-3-covid}
    \end{subfigure} \\
    \begin{subfigure}[b]{0.32\textwidth}
        \includegraphics[width=\textwidth]{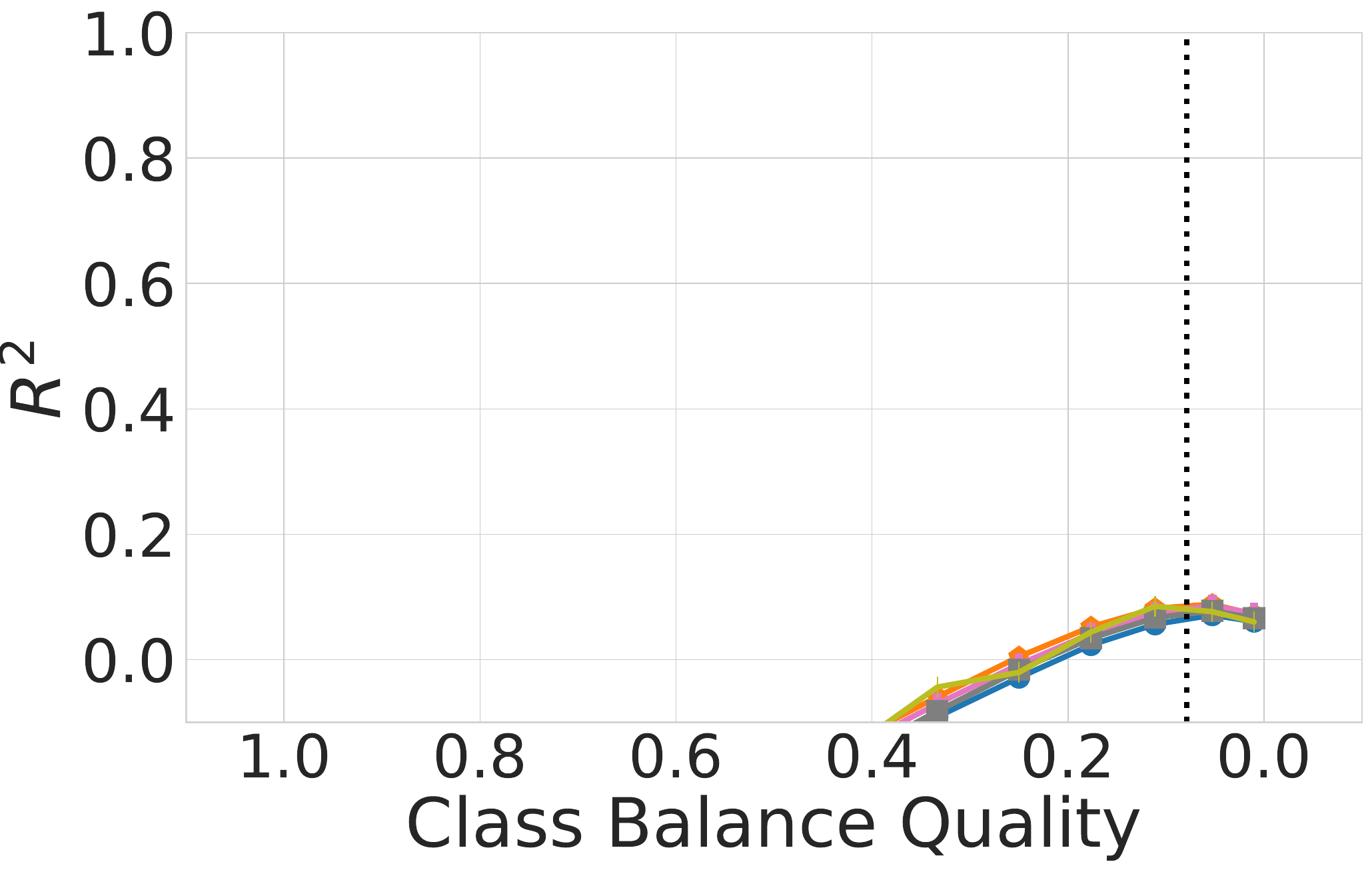}
        \caption{Class Balance (Sce. 1)}
        \label{fig:regression-results-all-ClassBalance-1-covid}
    \end{subfigure}
\begin{subfigure}[b]{0.32\textwidth}
        \includegraphics[width=\textwidth]{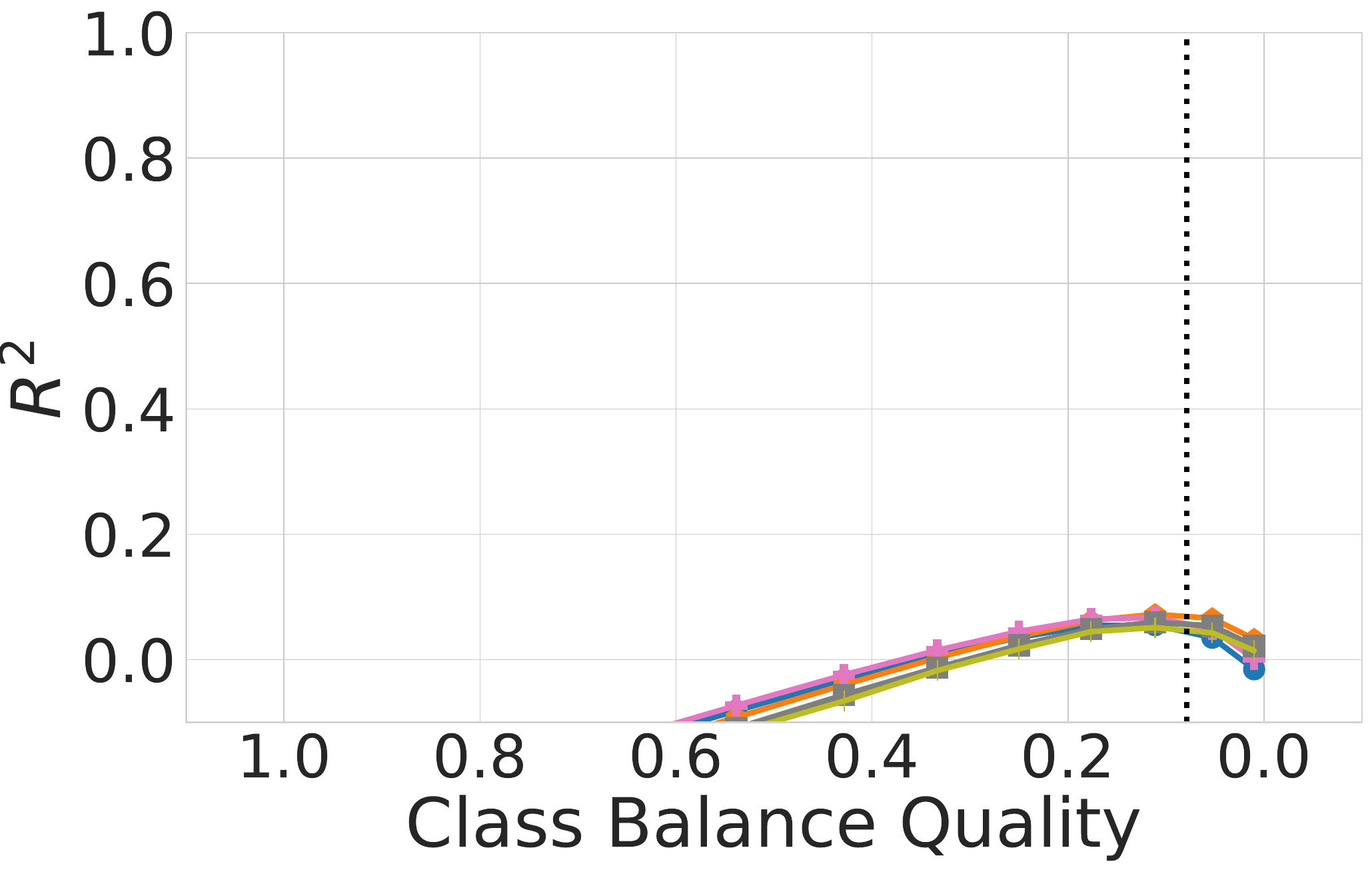}
        \caption{Class Balance (Sce. 2)}
        \label{fig:regression-results-all-ClassBalance-2-covid}
    \end{subfigure}
    \begin{subfigure}[b]{0.32\textwidth}
        \includegraphics[width=\textwidth]{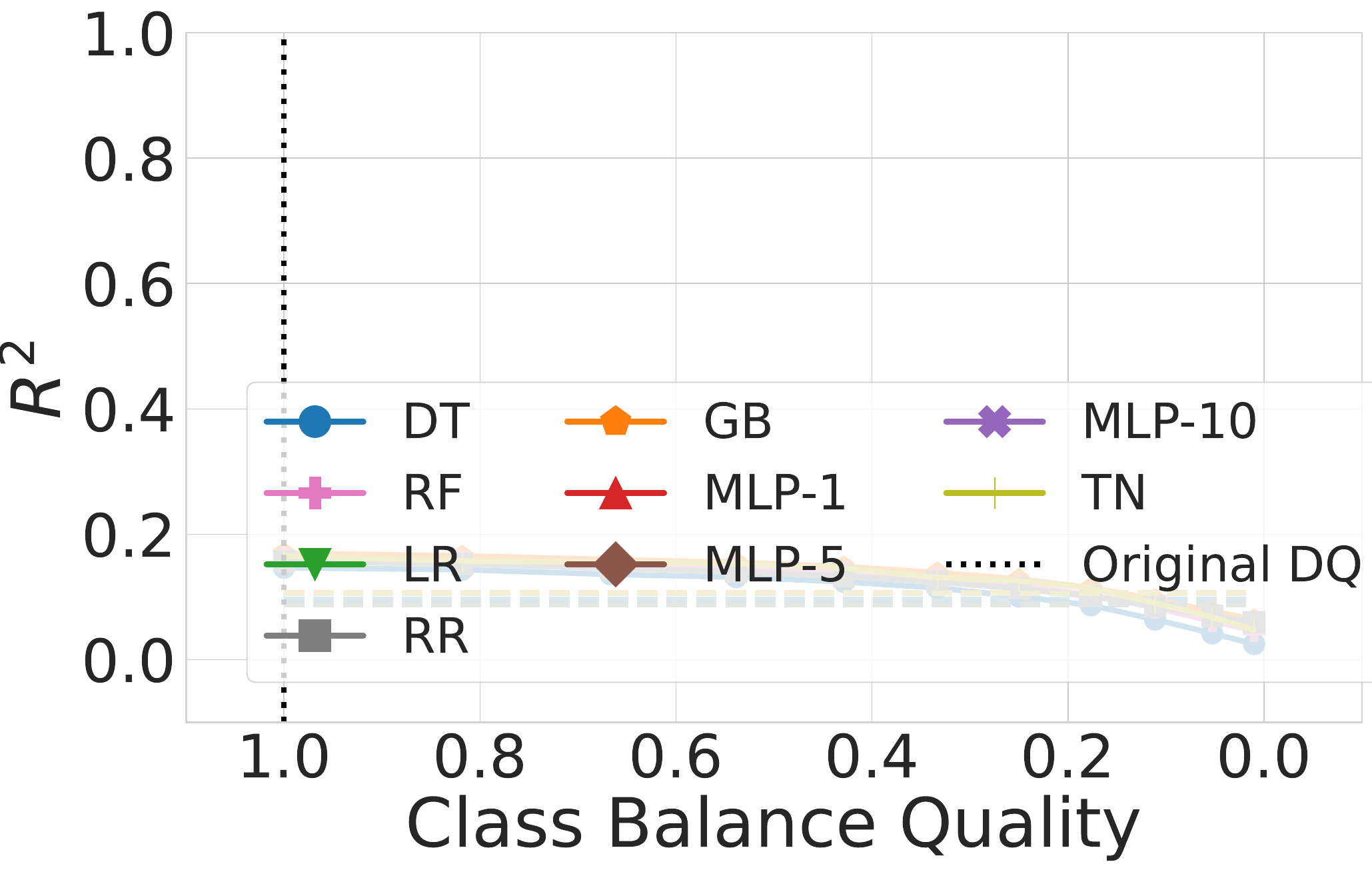}
        \caption{Class Balance(Sce. 3)}
        \label{fig:regression-results-all-ClassBalance-3-covid}
    \end{subfigure}
 \end{adjustbox} 
    \caption{\revision{$R^2$ of the regression algorithms for \textsf{COVID} dataset.}}
    \label{fig:regression-results-all-covid}
\end{figure*}

%% file: Latex_Figure/regression/Consistent_Representation_5.tex
\begin{figure*}[t]
    \centering
\raisebox{0.4\height}{\rotatebox{90}{Scenario 1}}\hspace{0.3em}
    \begin{subfigure}[b]{0.23\linewidth}
        \includegraphics[width=\linewidth]{figures/regression/Consistent_repr_k5/ConsistentRepresentation_4_house_prices_prepared_train_polluted_test_original.pdf}
        \caption{\textsf{Houses}}
        \label{fig:regression-results-all-ConsistentRepresentationk5-1-houses}
    \end{subfigure}
\begin{subfigure}[b]{0.23\linewidth}
        \includegraphics[width=\linewidth]{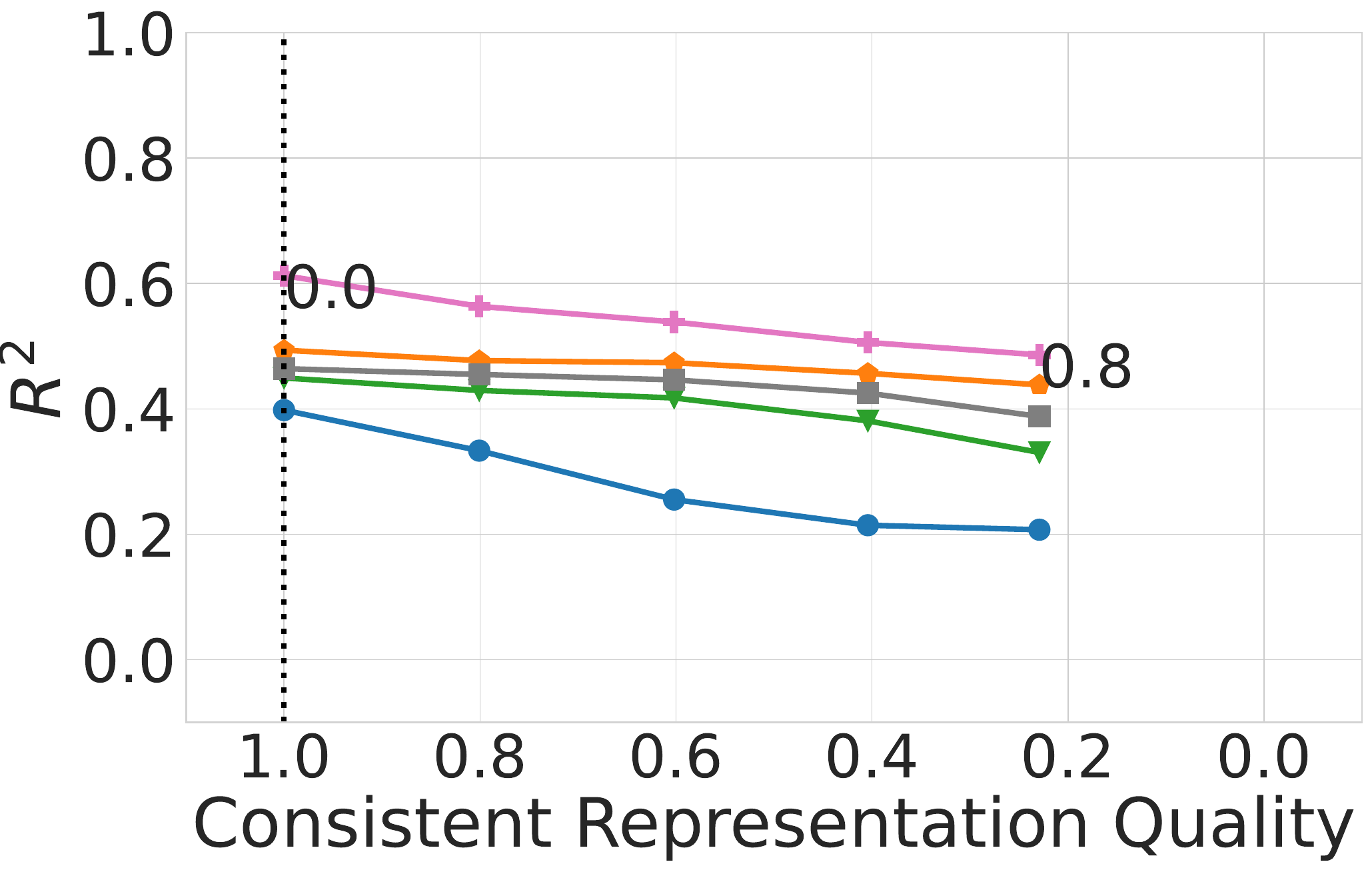}
        \caption{\textsf{IMDB}}
        \label{fig:regression-results-all-ConsistentRepresentationk5-1-imdb}
    \end{subfigure}
    \begin{subfigure}[b]{0.23\linewidth}
        \includegraphics[width=\linewidth]{figures/regression/Consistent_repr_k5/ConsistentRepresentation_4_covid_data_pre_processed_regression_train_polluted_test_original.pdf}
        \caption{\textsf{COVID}}
        \label{fig:regression-results-all-ConsistentRepresentationk5-1-covid}
    \end{subfigure}
    \begin{subfigure}[b]{0.23\linewidth}
        \includegraphics[width=\linewidth]{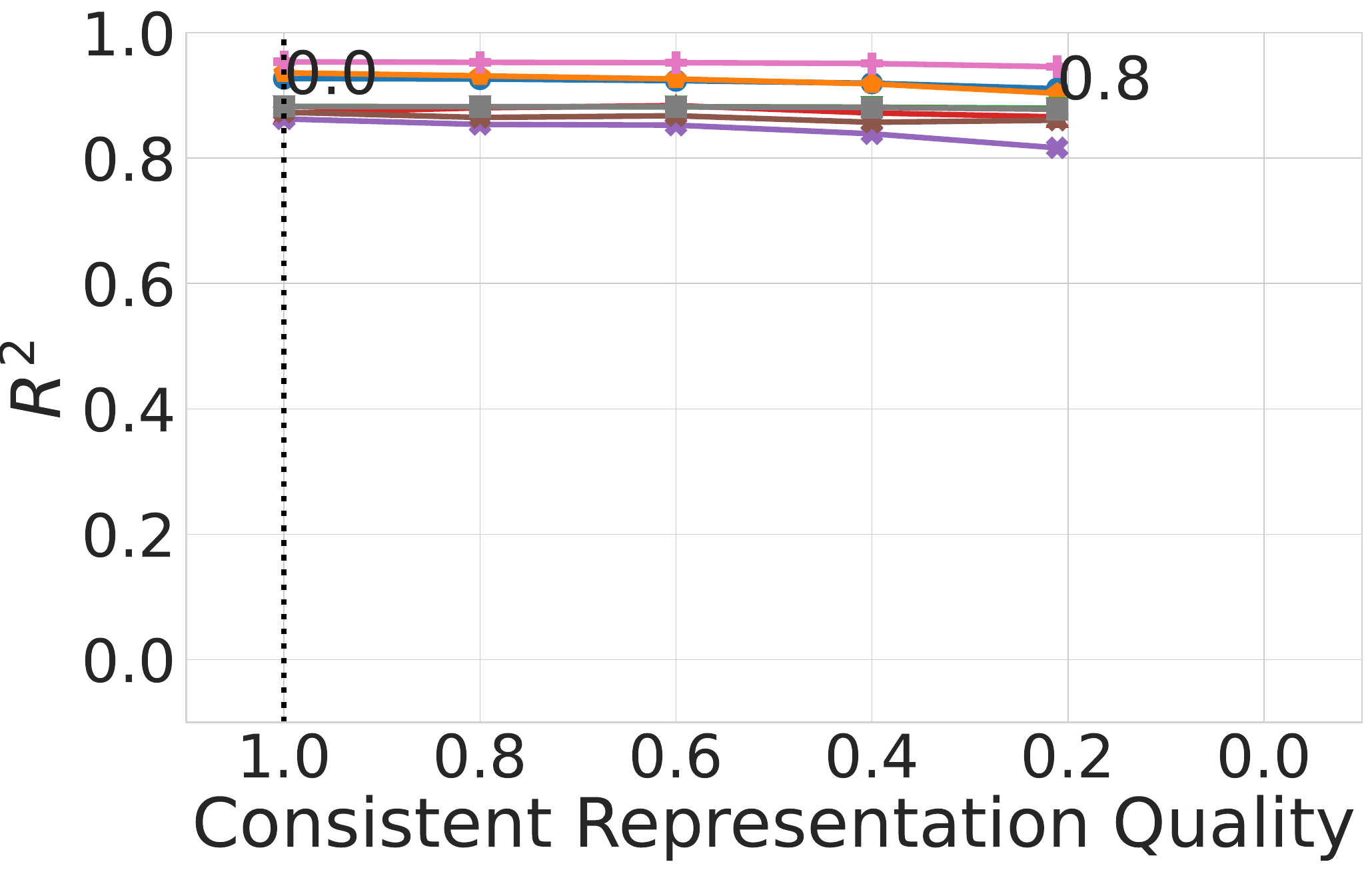}
        \caption{\textsf{Cars}}
        \label{fig:regression-results-all-ConsistentRepresentationk5-1-cars}
    \end{subfigure}

\raisebox{0.4\height}{\rotatebox{90}{Scenario 2}}\hspace{0.3em}
    \begin{subfigure}[b]{0.23\linewidth}
        \includegraphics[width=\linewidth]{figures/regression/Consistent_repr_k5/ConsistentRepresentation_4_house_prices_prepared_train_original_test_polluted.pdf}
        \caption{\textsf{Houses}}
        \label{fig:regression-results-all-ConsistentRepresentationk5-2-houses}
    \end{subfigure}
\begin{subfigure}[b]{0.23\linewidth}
        \includegraphics[width=\linewidth]{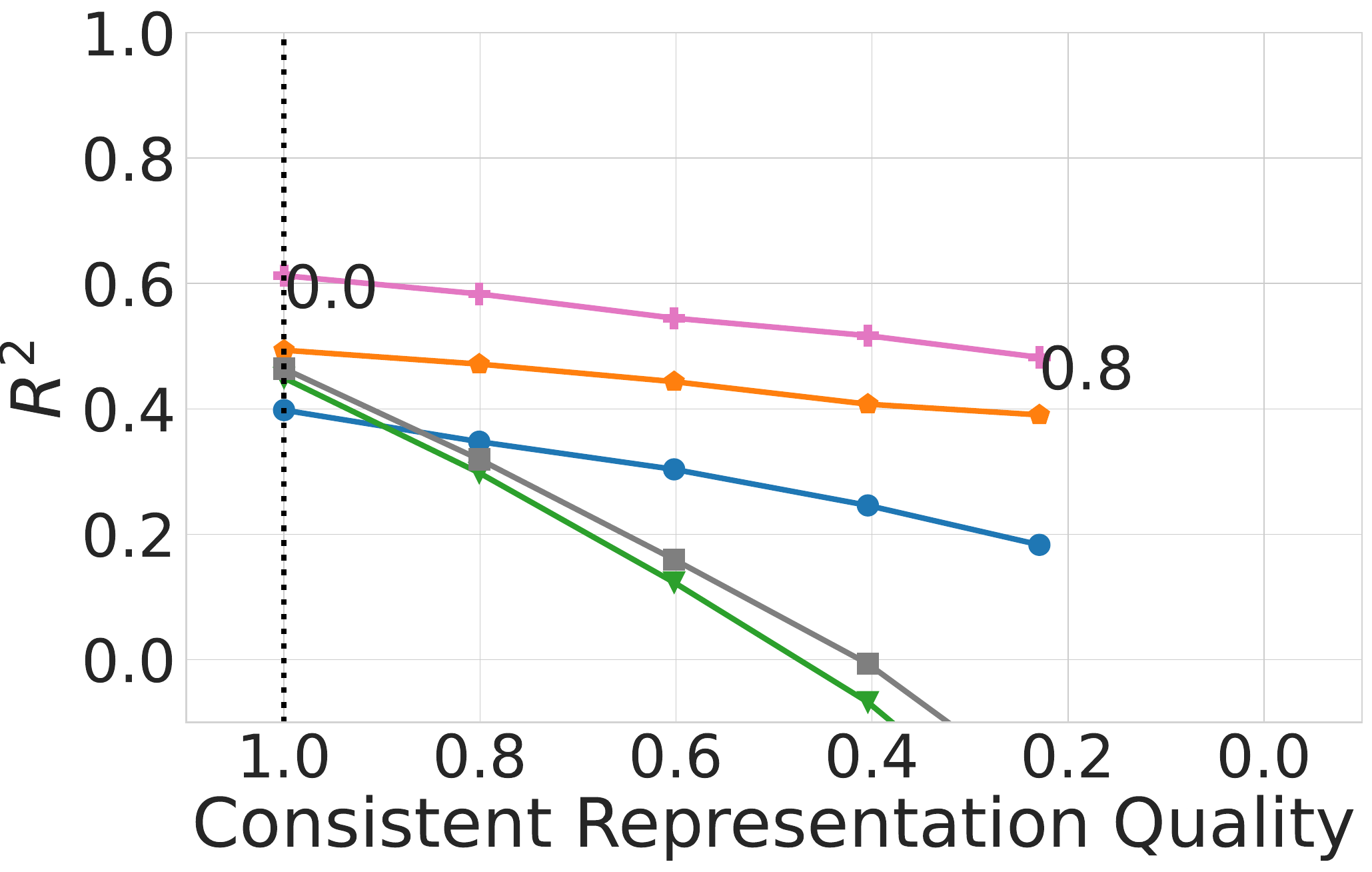}
        \caption{\textsf{IMDB}}
        \label{fig:regression-results-all-ConsistentRepresentationk5-2-imdb}
    \end{subfigure}
    \begin{subfigure}[b]{0.23\linewidth}
        \includegraphics[width=\linewidth]{figures/regression/Consistent_repr_k5/ConsistentRepresentation_4_covid_data_pre_processed_regression_train_original_test_polluted.pdf}
        \caption{\textsf{COVID}}
        \label{fig:regression-results-all-ConsistentRepresentationk5-2-covid}
    \end{subfigure}
    \begin{subfigure}[b]{0.23\linewidth}
        \includegraphics[width=\linewidth]{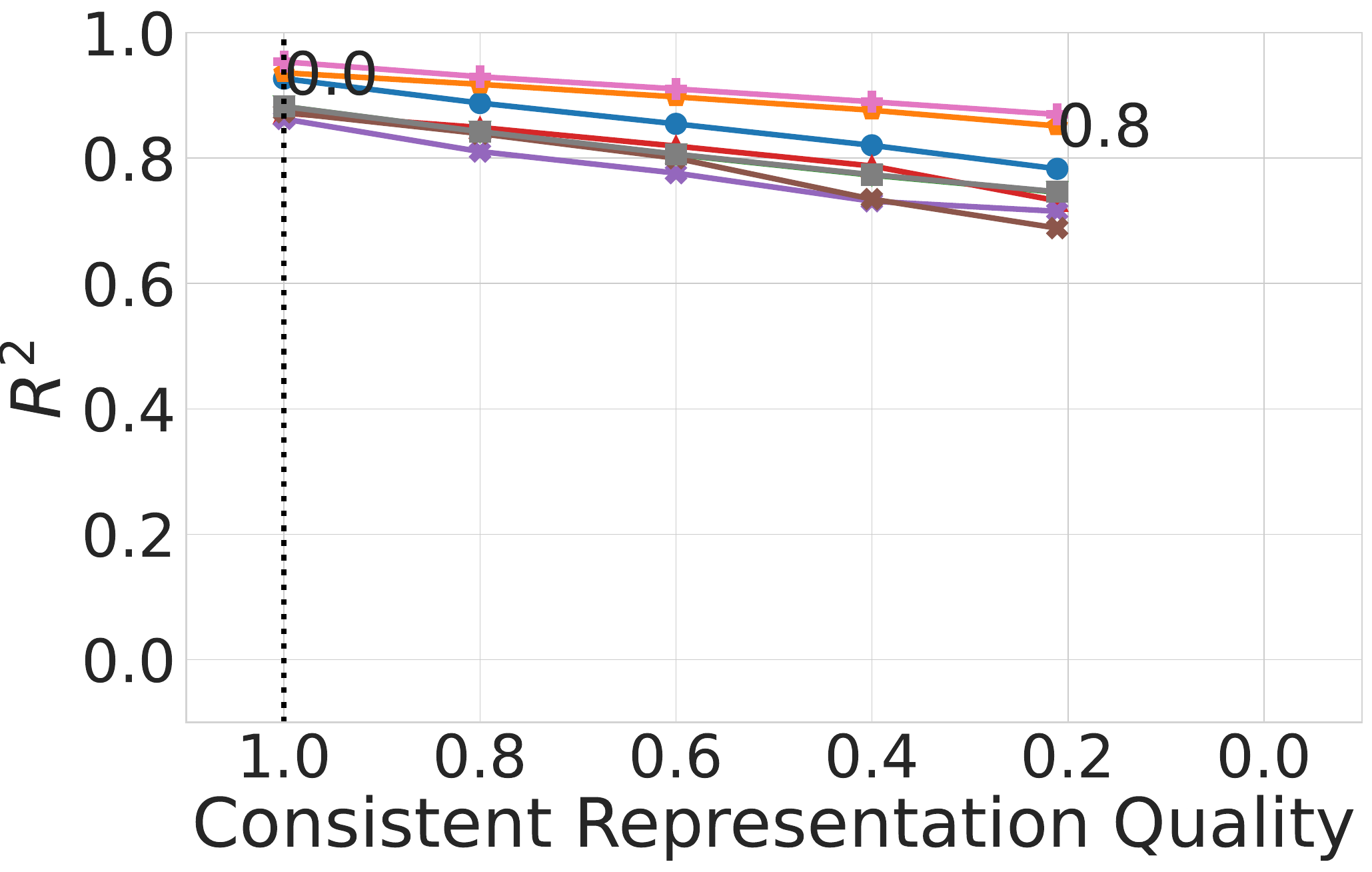}
        \caption{\textsf{Cars}}
        \label{fig:regression-results-all-ConsistentRepresentationk5-2-cars}
    \end{subfigure}

\raisebox{0.4\height}{\rotatebox{90}{Scenario 3}}\hspace{0.3em}
    \begin{subfigure}[b]{0.23\linewidth}
        \includegraphics[width=\linewidth]{figures/regression/Consistent_repr_k5/ConsistentRepresentation_4_house_prices_prepared_train_polluted_test_polluted.pdf}
        \caption{\textsf{Houses}}
        \label{fig:regression-results-all-ConsistentRepresentationk5-3-houses}
    \end{subfigure}
\begin{subfigure}[b]{0.23\linewidth}
        \includegraphics[width=\linewidth]{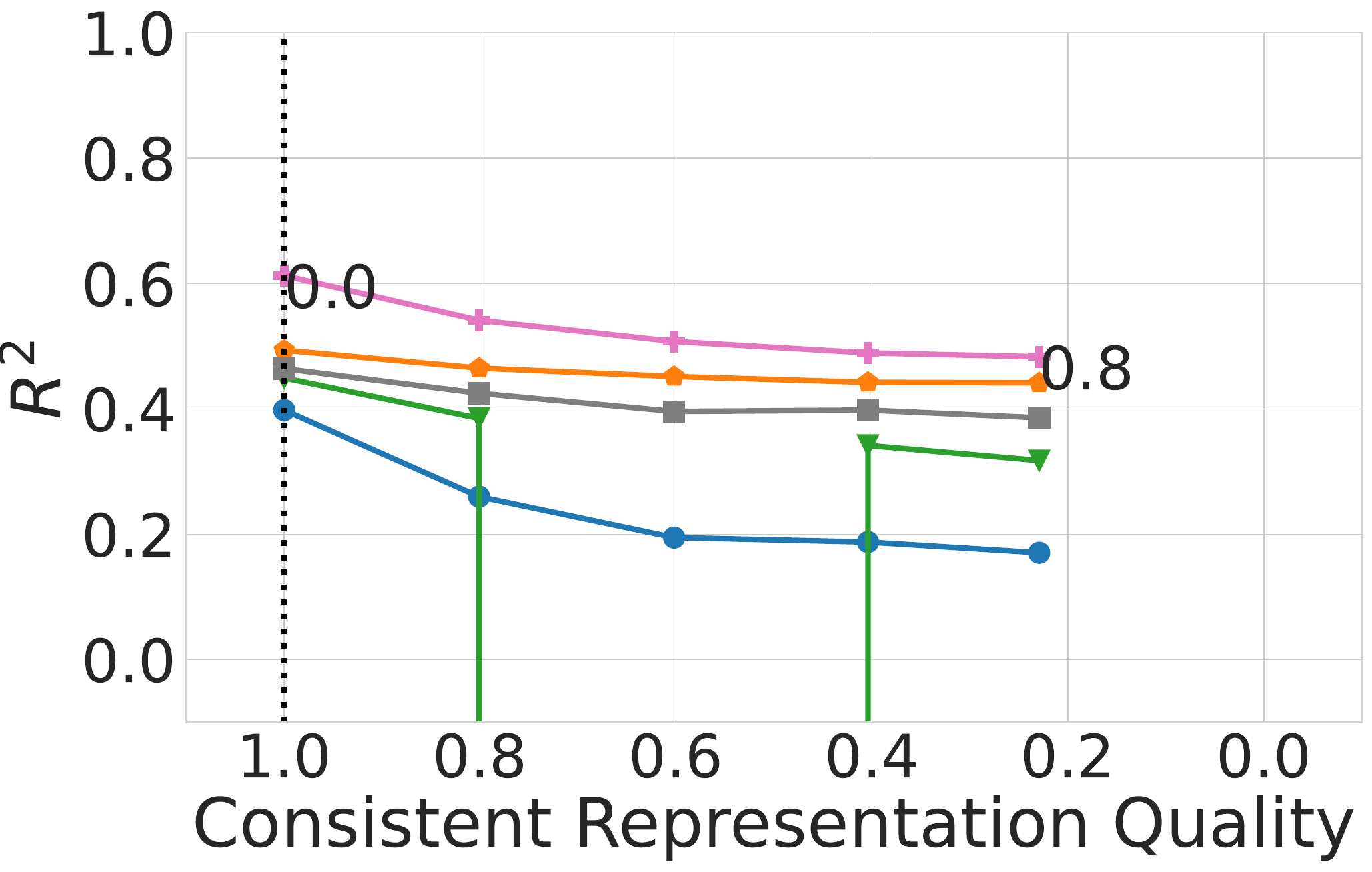}
        \caption{\textsf{IMDB}}
        \label{fig:regression-results-all-ConsistentRepresentationk5-3-imdb}
    \end{subfigure}
    \begin{subfigure}[b]{0.23\linewidth}
        \includegraphics[width=\linewidth]{figures/regression/Consistent_repr_k5/ConsistentRepresentation_4_covid_data_pre_processed_regression_train_polluted_test_polluted.pdf}
        \caption{\textsf{COVID}}
        \label{fig:regression-results-all-ConsistentRepresentationk5-3-covid}
    \end{subfigure}
    \begin{subfigure}[b]{0.23\linewidth}
        \includegraphics[width=\linewidth]{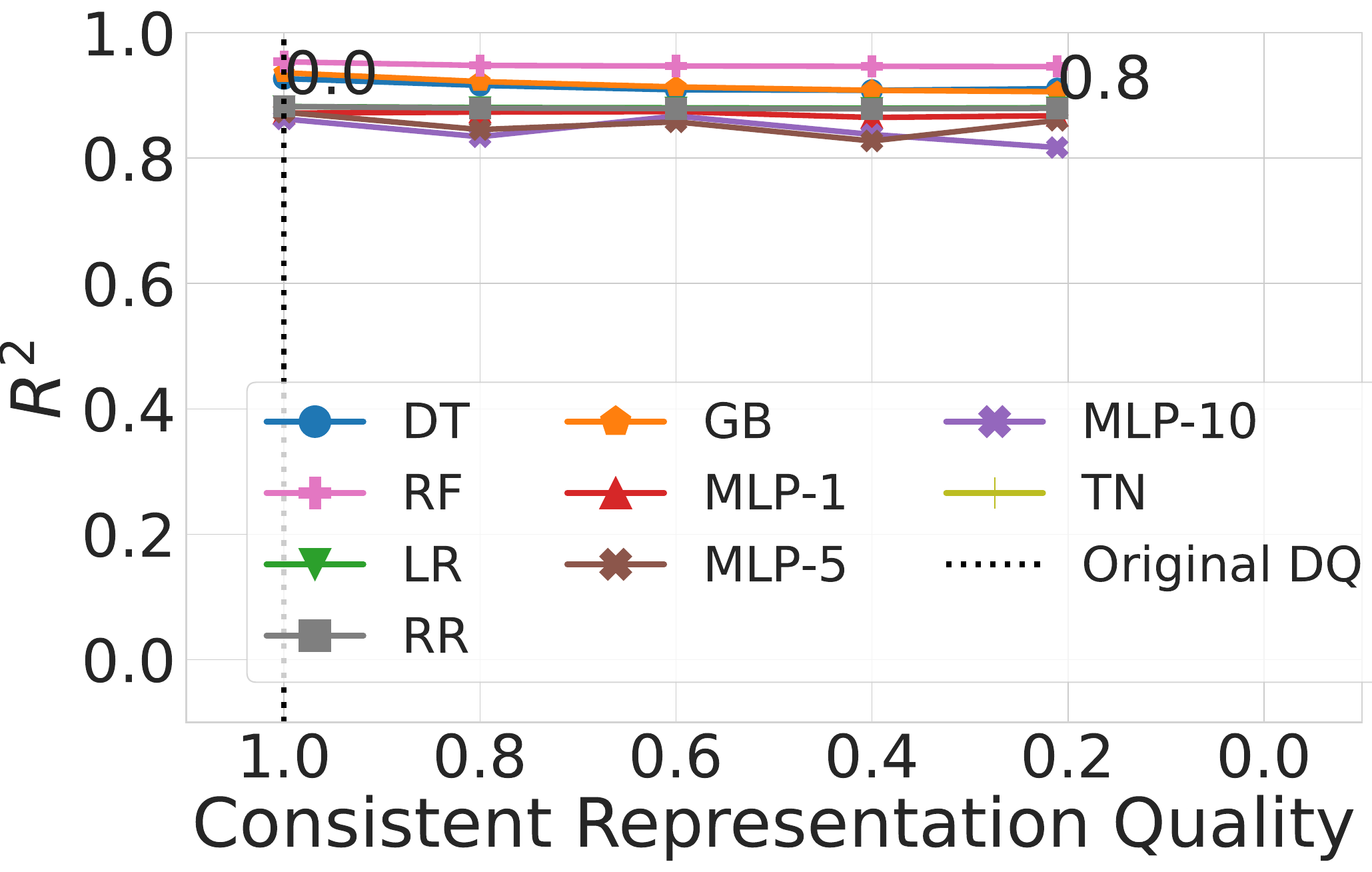}
        \caption{\textsf{Cars}}
        \label{fig:regression-results-all-ConsistentRepresentationk5-3-cars}
    \end{subfigure}
    \caption{$R^2$ of the regression algorithms for consistent representation with $k_v = 5$.}
    \label{fig:regression-results-all-ConsistentRepresentationk5}
\end{figure*}

%% file: Latex_Figure/regression/Completeness.tex
\begin{figure*}[t]
    \centering
\raisebox{0.4\height}{\rotatebox{90}{Scenario 1}}\hspace{0.3em}
    \begin{subfigure}[b]{0.23\linewidth}
        \includegraphics[width=\linewidth]{figures/regression/Completeness/Completeness_house_prices_prepared_train_polluted_test_original.pdf}
        \caption{\textsf{Houses}}
        \label{fig:regression-results-all-completeness-1-houses}
    \end{subfigure}
\begin{subfigure}[b]{0.23\linewidth}
        \includegraphics[width=\linewidth]{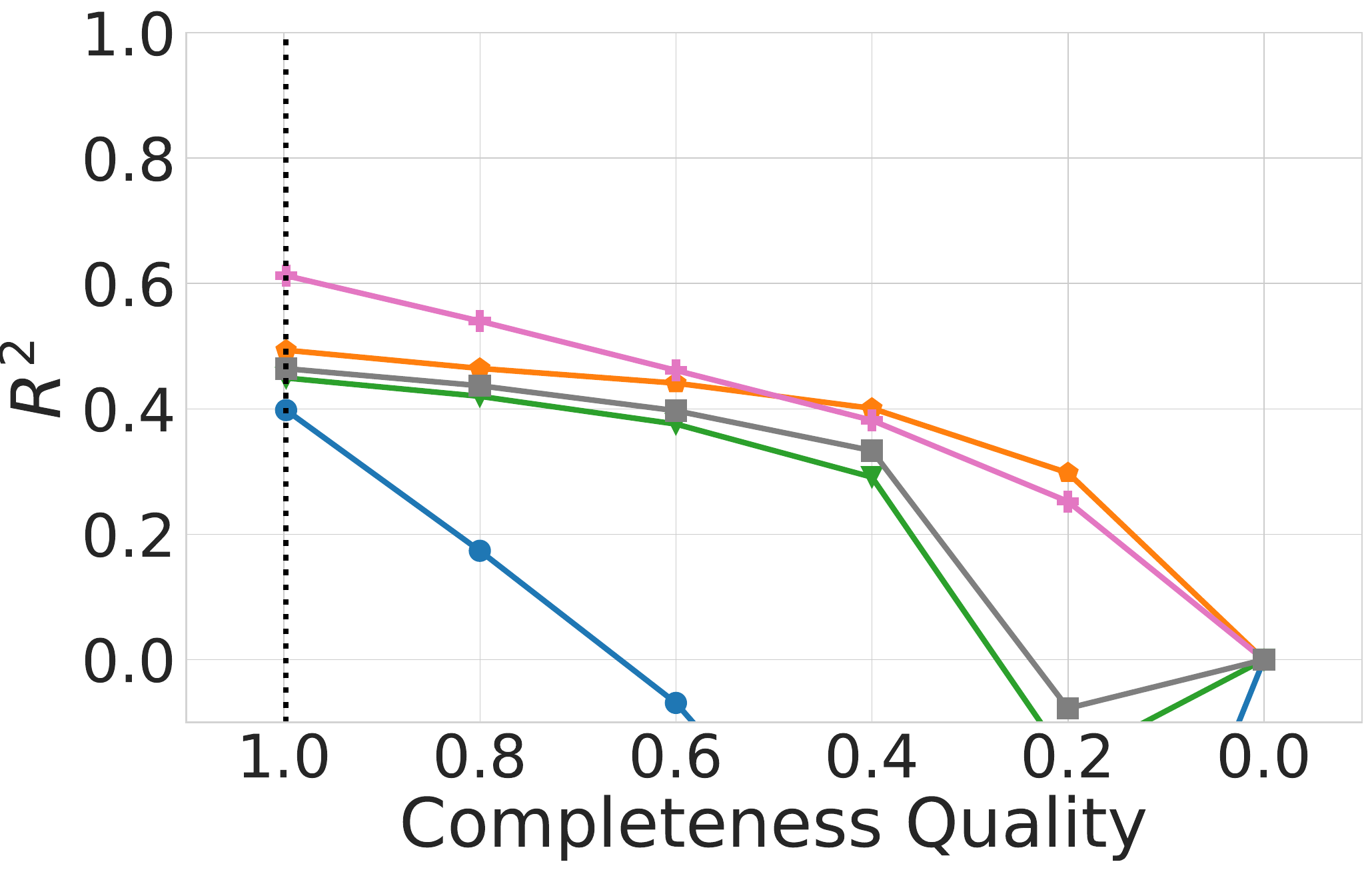}
        \caption{\textsf{IMDB}}
        \label{fig:regression-results-all-completeness-1-imdb}
    \end{subfigure}
    \begin{subfigure}[b]{0.23\linewidth}
        \includegraphics[width=\linewidth]{figures/regression/Completeness/Completeness_covid_data_pre_processed_regression_train_polluted_test_original.pdf}
        \caption{\textsf{COVID}}
        \label{fig:regression-results-all-completeness-1-covid}
    \end{subfigure}
    \begin{subfigure}[b]{0.23\linewidth}
        \includegraphics[width=\linewidth]{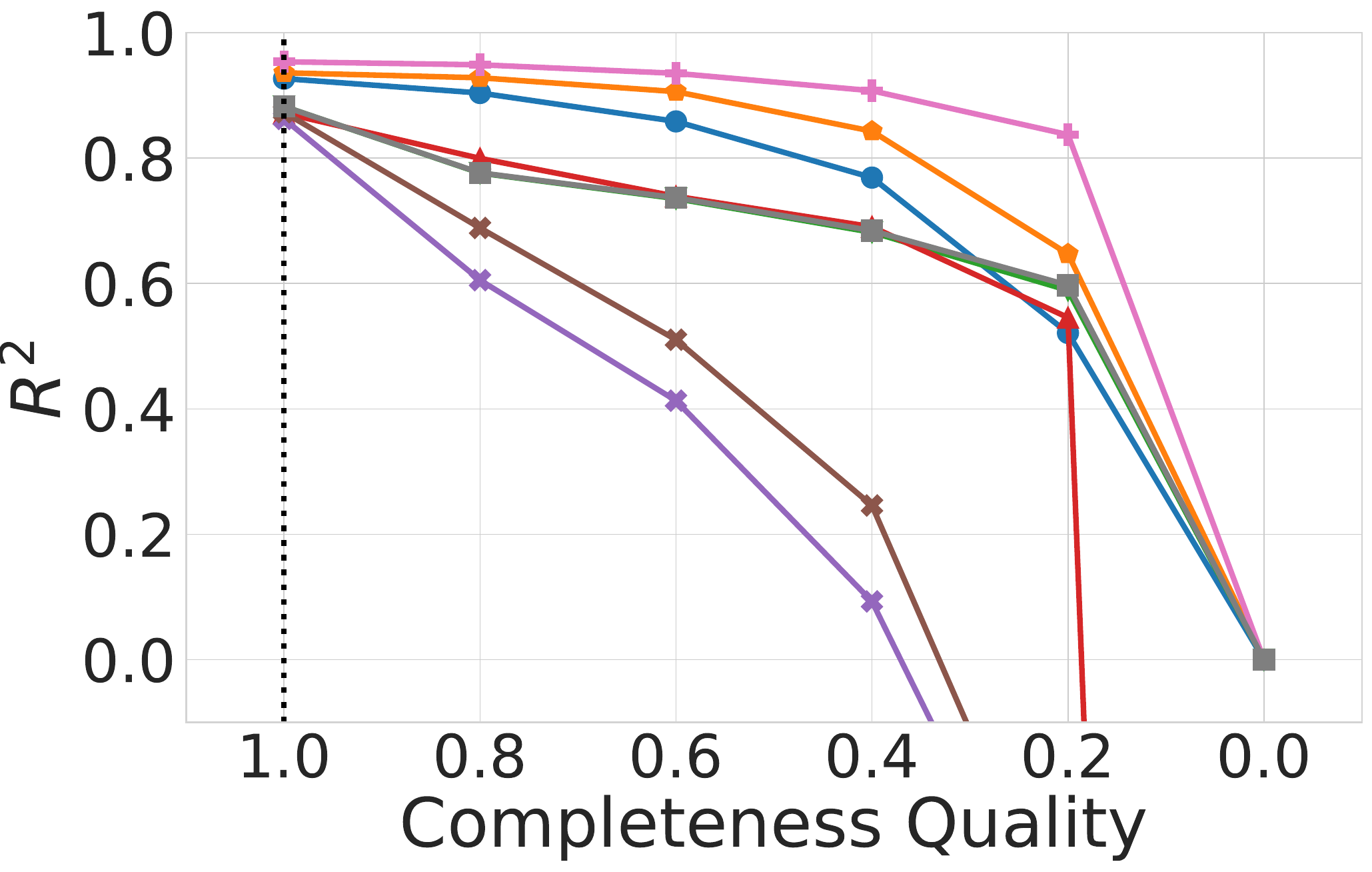}
        \caption{\textsf{Cars}}
        \label{fig:regression-results-all-completeness-1-cars}
    \end{subfigure}

\raisebox{0.4\height}{\rotatebox{90}{Scenario 2}}\hspace{0.3em}
    \begin{subfigure}[b]{0.23\linewidth}
        \includegraphics[width=\linewidth]{figures/regression/Completeness/Completeness_house_prices_prepared_train_original_test_polluted.pdf}
        \caption{\textsf{Houses}}
        \label{fig:regression-results-all-completeness-2-houses}
    \end{subfigure}
\begin{subfigure}[b]{0.23\linewidth}
        \includegraphics[width=\linewidth]{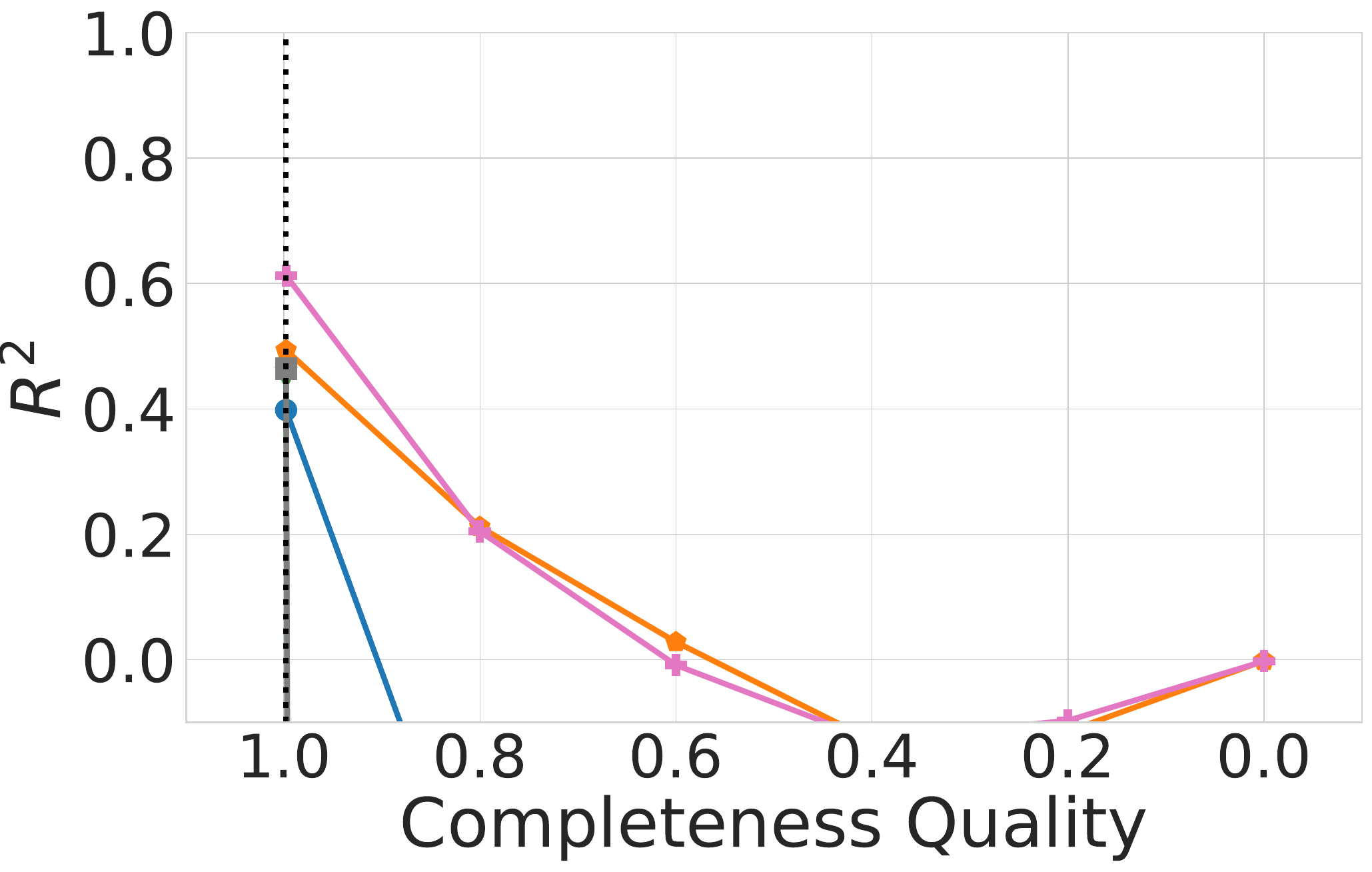}
        \caption{\textsf{IMDB}}
        \label{fig:regression-results-all-completeness-2-imdb}
    \end{subfigure}
    \begin{subfigure}[b]{0.23\linewidth}
        \includegraphics[width=\linewidth]{figures/regression/Completeness/Completeness_covid_data_pre_processed_regression_train_original_test_polluted.pdf}
        \caption{\textsf{COVID}}
        \label{fig:regression-results-all-completeness-2-covid}
    \end{subfigure}
    \begin{subfigure}[b]{0.23\linewidth}
        \includegraphics[width=\linewidth]{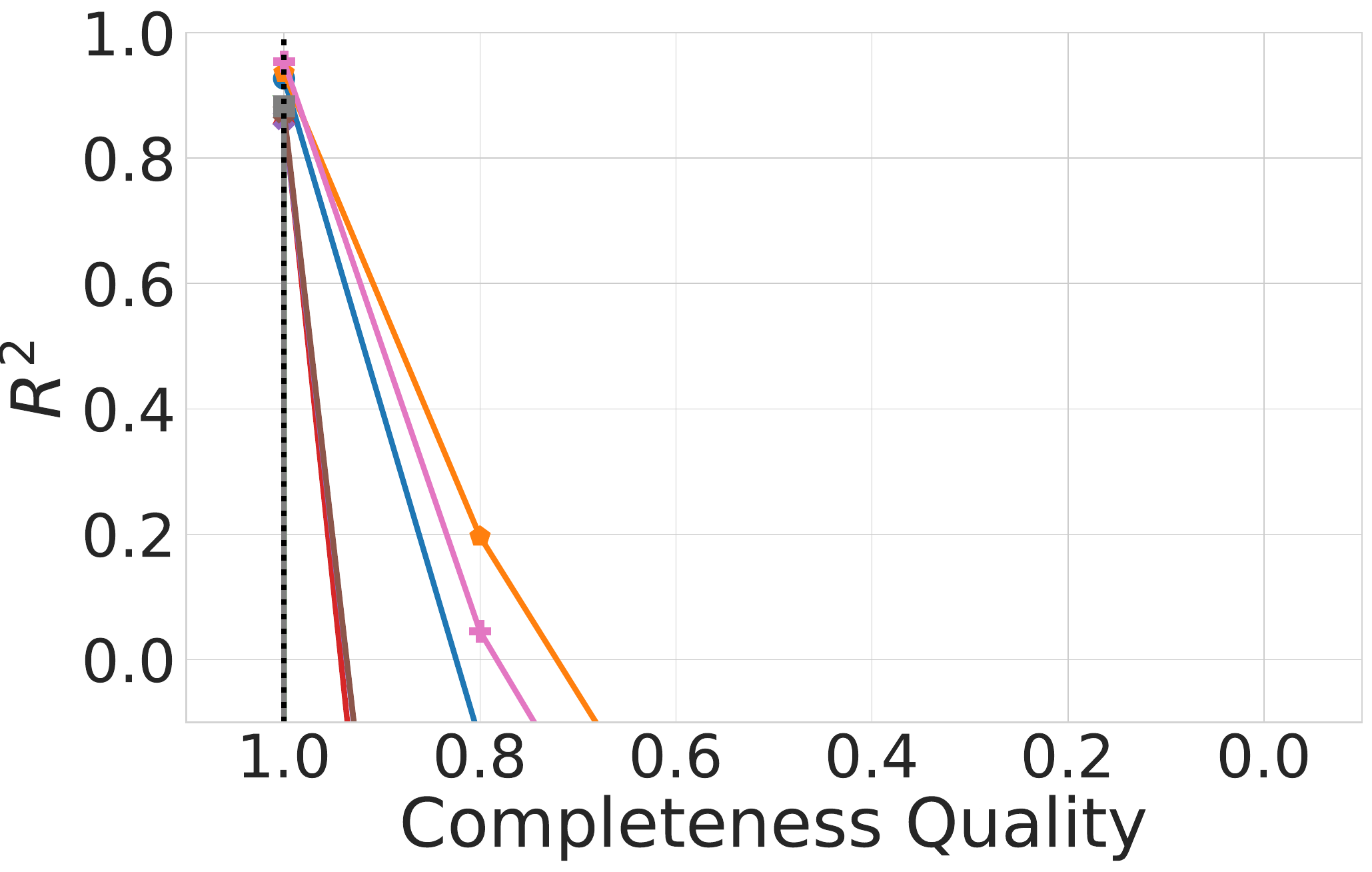}
        \caption{\textsf{Cars}}
        \label{fig:regression-results-all-completeness-2-cars}
    \end{subfigure}

\raisebox{0.4\height}{\rotatebox{90}{Scenario 3}}\hspace{0.3em}
    \begin{subfigure}[b]{0.23\linewidth}
        \includegraphics[width=\linewidth]{figures/regression/Completeness/Completeness_house_prices_prepared_train_polluted_test_polluted.pdf}
        \caption{\textsf{Houses}}
        \label{fig:regression-results-all-completeness-3-houses}
    \end{subfigure}
\begin{subfigure}[b]{0.23\linewidth}
        \includegraphics[width=\linewidth]{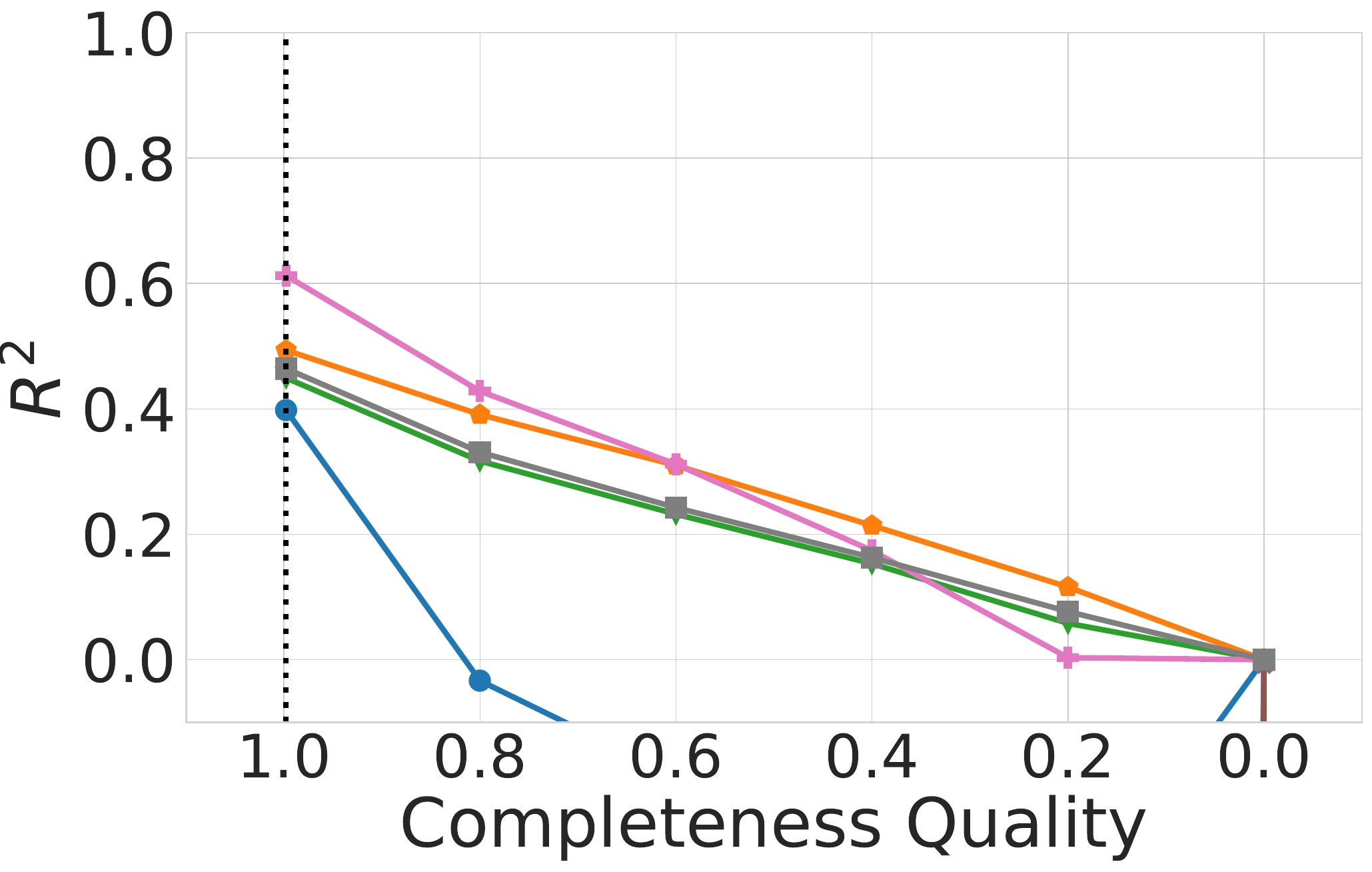}
        \caption{\textsf{IMDB}}
        \label{fig:regression-results-all-completeness-3-imdb}
    \end{subfigure}
    \begin{subfigure}[b]{0.23\linewidth}
        \includegraphics[width=\linewidth]{figures/regression/Completeness/Completeness_covid_data_pre_processed_regression_train_polluted_test_polluted.pdf}
        \caption{\textsf{COVID}}
        \label{fig:regression-results-all-completeness-3-covid}
    \end{subfigure}
    \begin{subfigure}[b]{0.23\linewidth}
        \includegraphics[width=\linewidth]{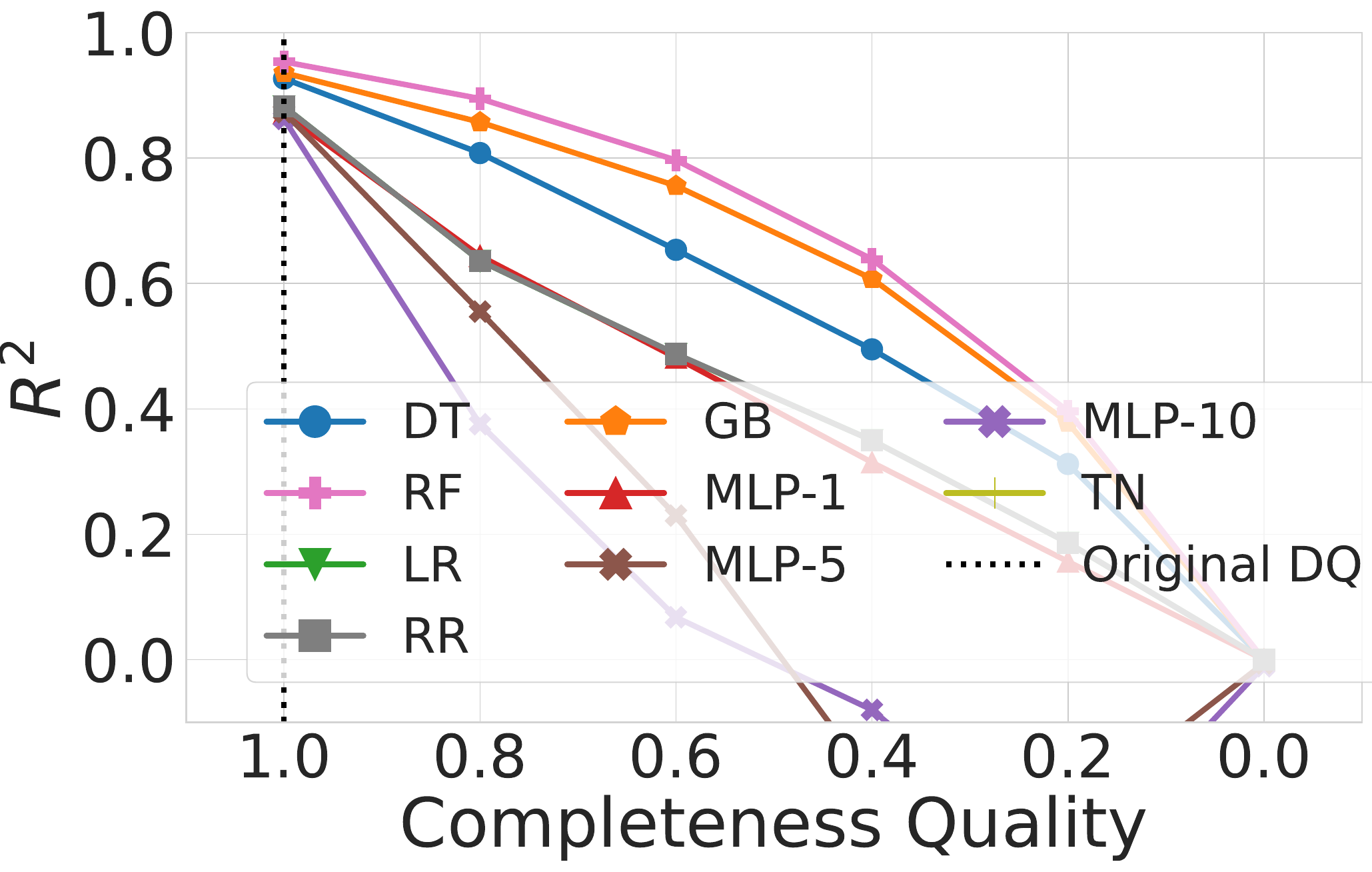}
        \caption{\textsf{Cars}}
        \label{fig:regression-results-all-completeness-3-cars}
    \end{subfigure}
    \caption{$R^2$ of the regression algorithms for completeness.}
    \label{fig:regression-results-all-completeness}
\end{figure*}

%% file: Latex_Figure/regression/Feature_Accurecy.tex
\begin{figure*}[t]
    \centering
\raisebox{0.4\height}{\rotatebox{90}{Scenario 1}}\hspace{0.3em}
    \begin{subfigure}[b]{0.23\linewidth}
        \includegraphics[width=\linewidth]{figures/regression/FeatureAccuracy/FeatureAccuracy_house_prices_prepared_train_polluted_test_original.pdf}
        \caption{\textsf{Houses}}
        \label{fig:regression-results-all-FeatureAccuracy-1-houses}
    \end{subfigure}
\begin{subfigure}[b]{0.23\linewidth}
        \includegraphics[width=\linewidth]{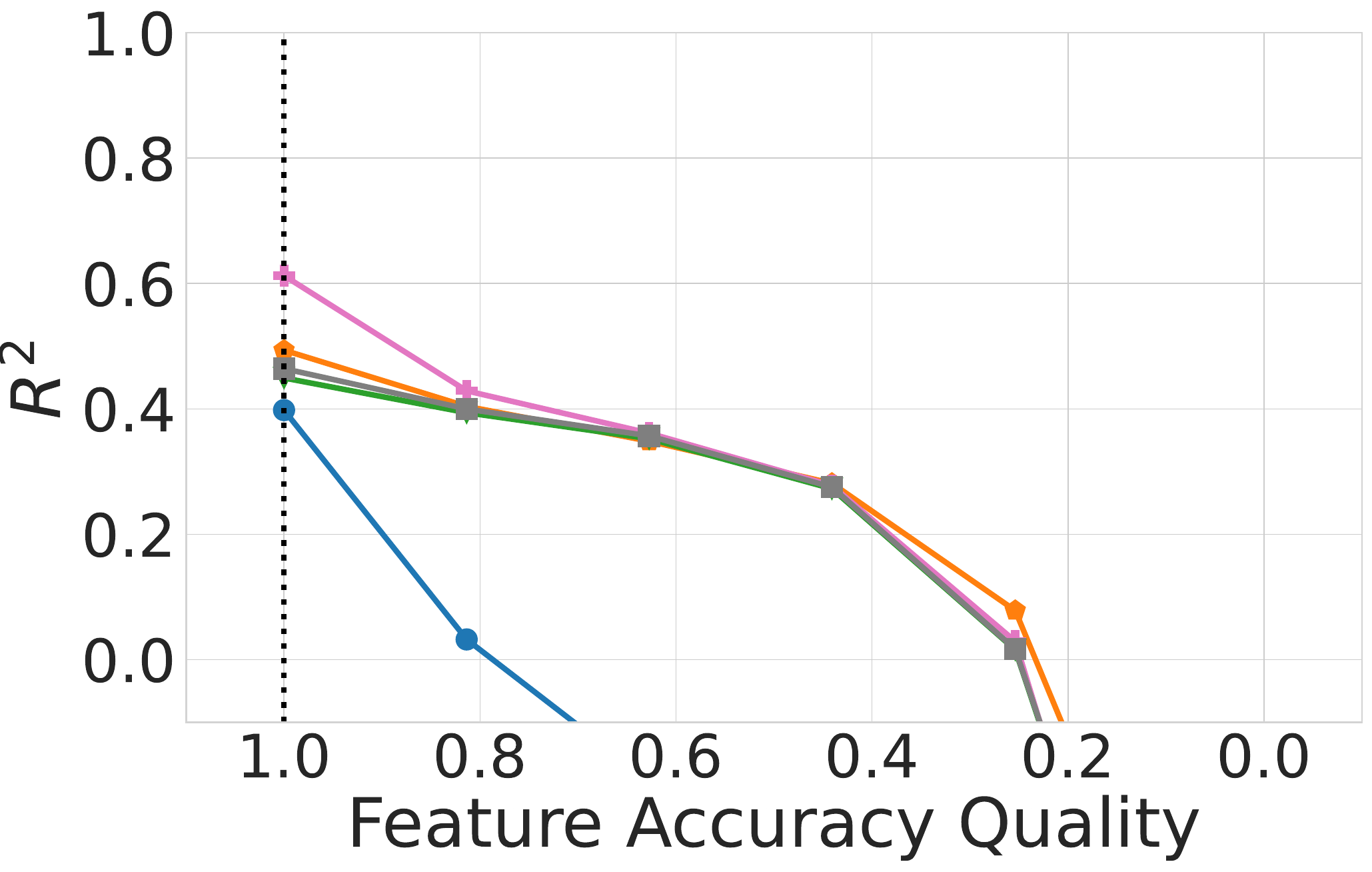}
        \caption{\textsf{IMDB}}
        \label{fig:regression-results-all-FeatureAccuracy-1-imdb}
    \end{subfigure}
    \begin{subfigure}[b]{0.23\linewidth}
        \includegraphics[width=\linewidth]{figures/regression/FeatureAccuracy/FeatureAccuracy_covid_data_pre_processed_regression_train_polluted_test_original.pdf}
        \caption{\textsf{COVID}}
        \label{fig:regression-results-all-FeatureAccuracy-1-covid}
    \end{subfigure}
    \begin{subfigure}[b]{0.23\linewidth}
        \includegraphics[width=\linewidth]{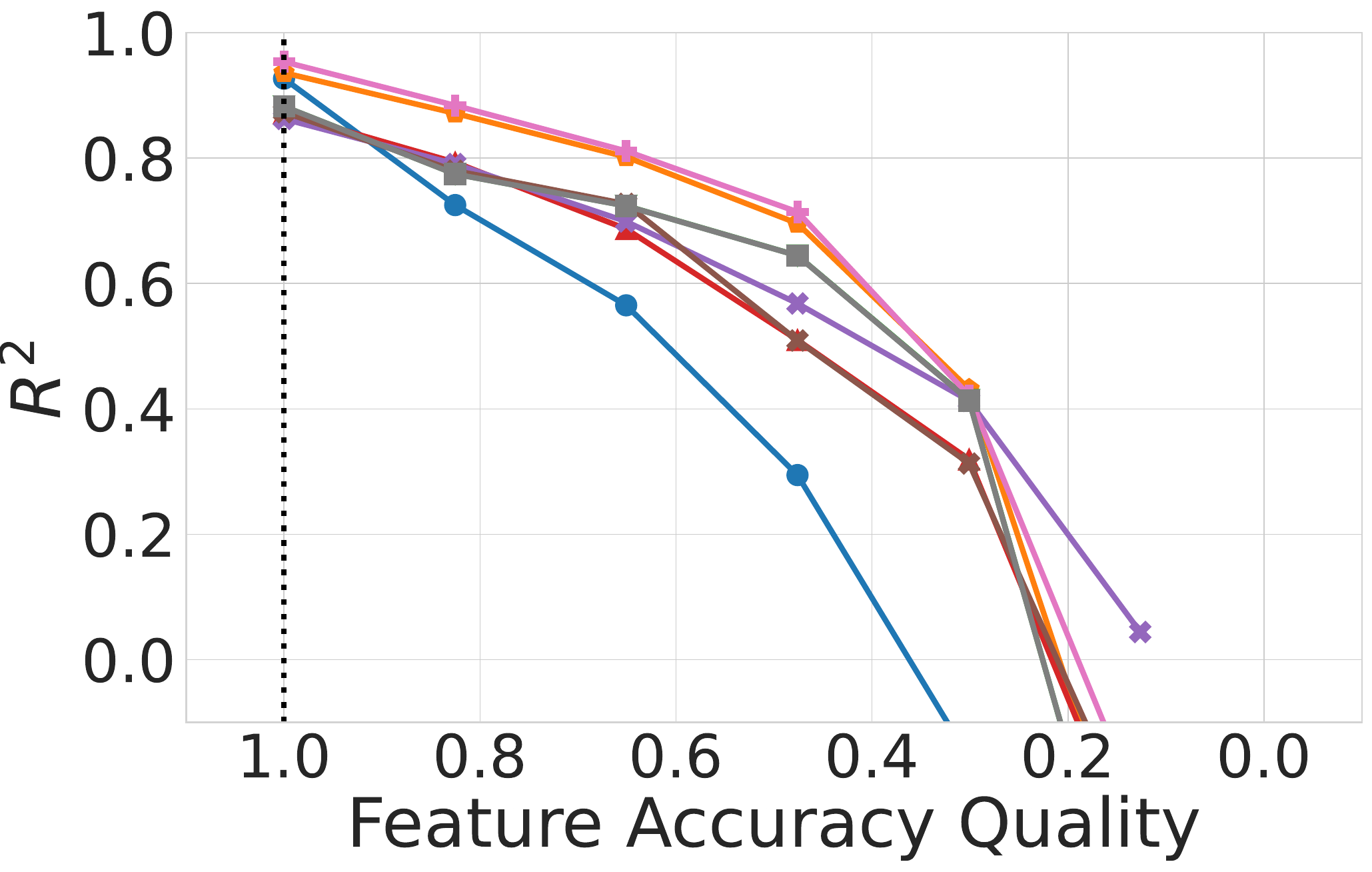}
        \caption{\textsf{Cars}}
        \label{fig:regression-results-all-FeatureAccuracy-1-cars}
    \end{subfigure}

\raisebox{0.4\height}{\rotatebox{90}{Scenario 2}}\hspace{0.3em}
    \begin{subfigure}[b]{0.23\linewidth}
        \includegraphics[width=\linewidth]{figures/regression/FeatureAccuracy/FeatureAccuracy_house_prices_prepared_train_original_test_polluted.pdf}
        \caption{\textsf{Houses}}
        \label{fig:regression-results-all-FeatureAccuracy-2-houses}
    \end{subfigure}
\begin{subfigure}[b]{0.23\linewidth}
        \includegraphics[width=\linewidth]{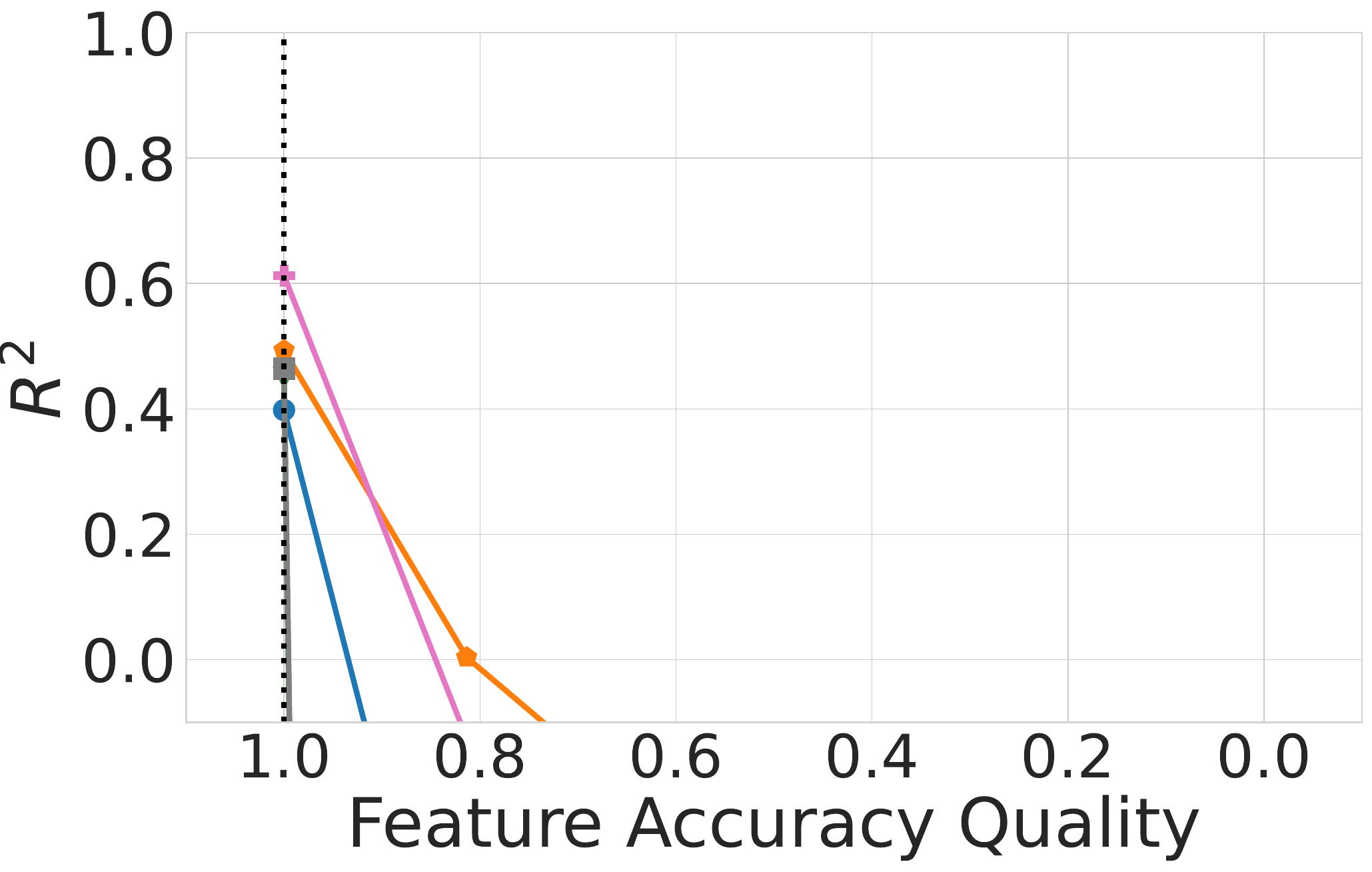}
        \caption{\textsf{IMDB}}
        \label{fig:regression-results-all-FeatureAccuracy-2-imdb}
    \end{subfigure}
    \begin{subfigure}[b]{0.23\linewidth}
        \includegraphics[width=\linewidth]{figures/regression/FeatureAccuracy/FeatureAccuracy_covid_data_pre_processed_regression_train_original_test_polluted.pdf}
        \caption{\textsf{COVID}}
        \label{fig:regression-results-all-FeatureAccuracy-2-covid}
    \end{subfigure}
    \begin{subfigure}[b]{0.23\linewidth}
        \includegraphics[width=\linewidth]{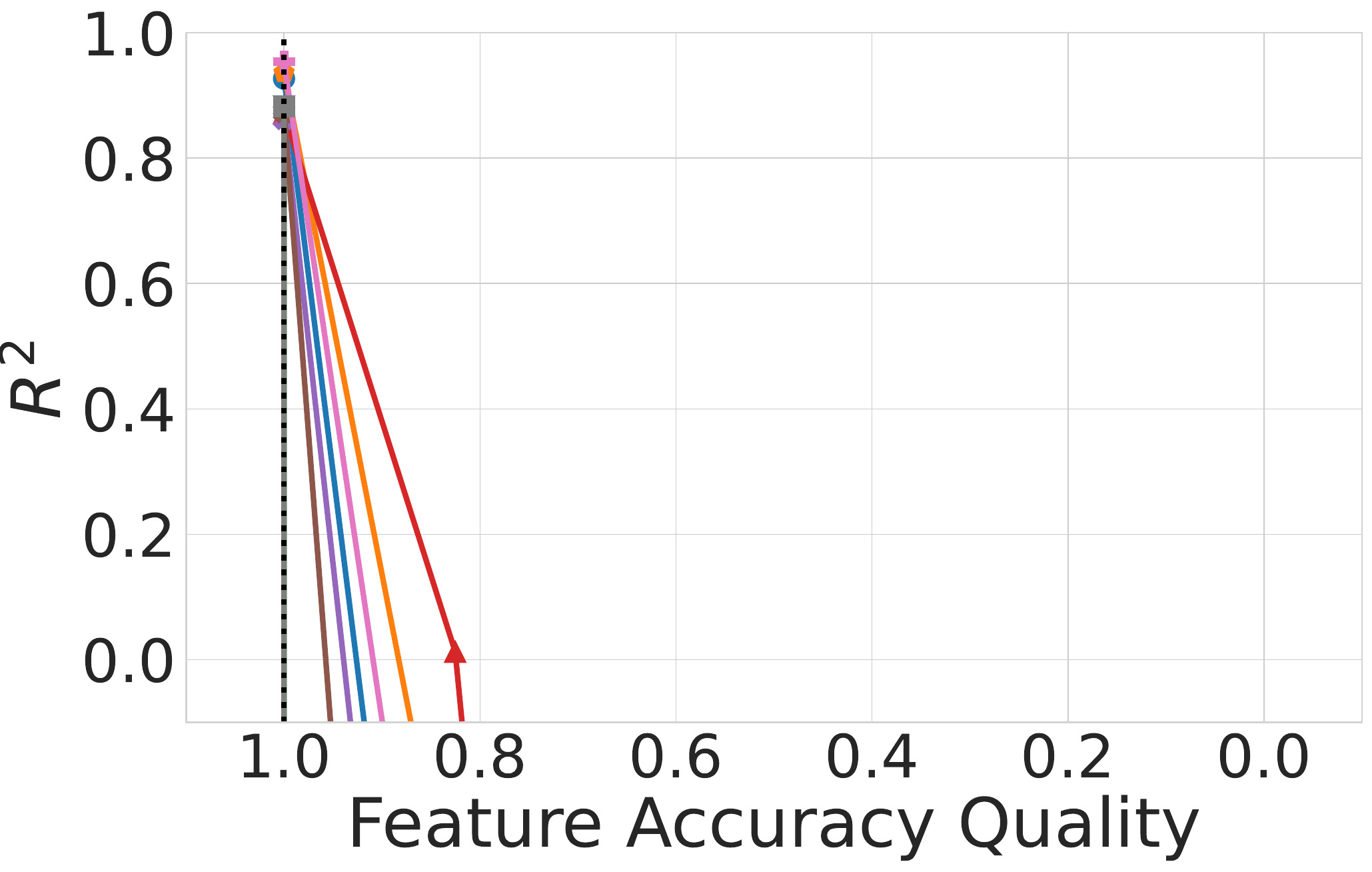}
        \caption{\textsf{Cars}}
        \label{fig:regression-results-all-FeatureAccuracy-2-cars}
    \end{subfigure}

\raisebox{0.4\height}{\rotatebox{90}{Scenario 3}}\hspace{0.3em}
    \begin{subfigure}[b]{0.23\linewidth}
        \includegraphics[width=\linewidth]{figures/regression/FeatureAccuracy/FeatureAccuracy_house_prices_prepared_train_polluted_test_polluted.pdf}
        \caption{\textsf{Houses}}
        \label{fig:regression-results-all-FeatureAccuracy-3-houses}
    \end{subfigure}
\begin{subfigure}[b]{0.23\linewidth}
        \includegraphics[width=\linewidth]{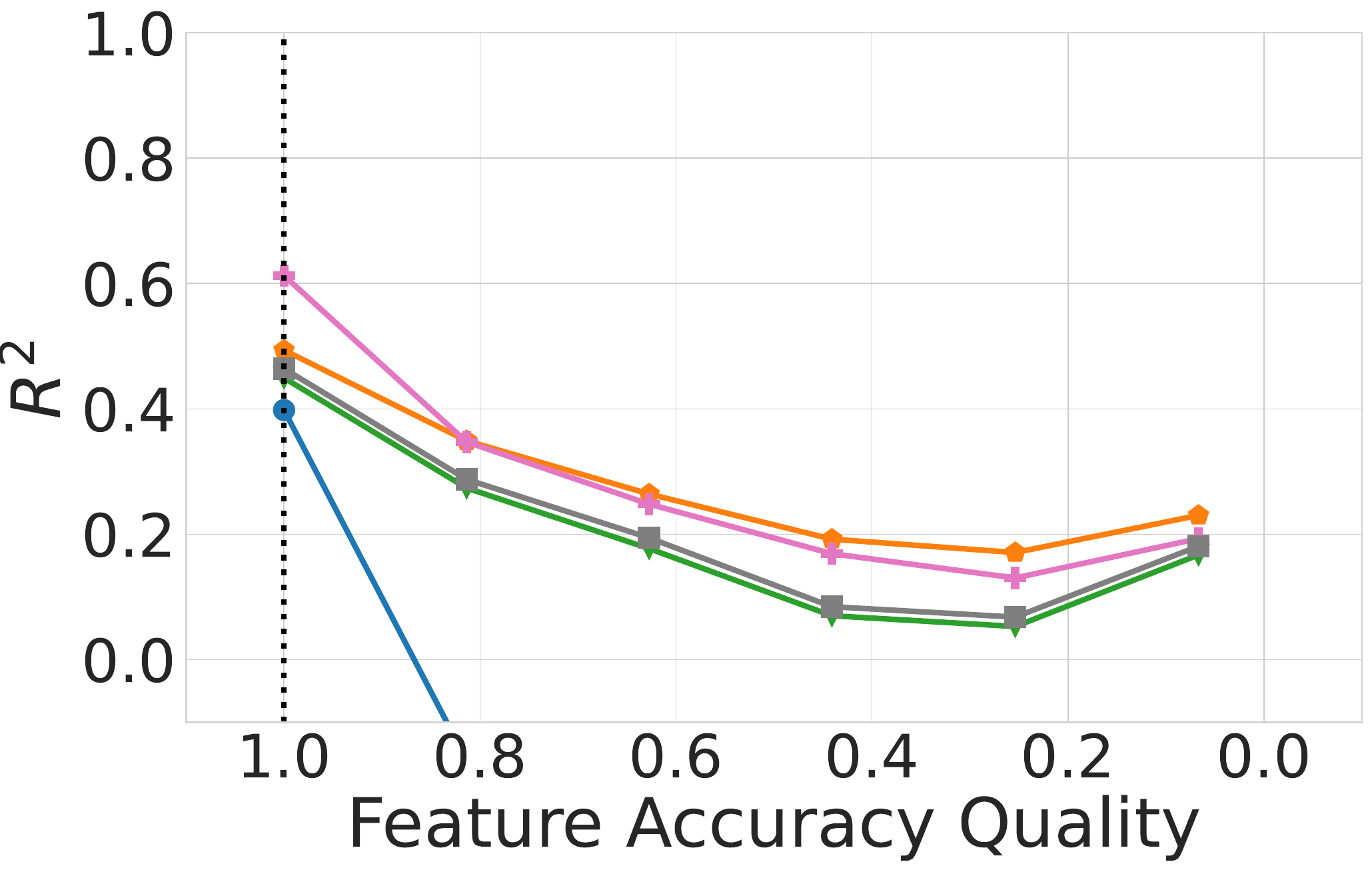}
        \caption{\textsf{IMDB}}
        \label{fig:regression-results-all-FeatureAccuracy-3-imdb}
    \end{subfigure}
    \begin{subfigure}[b]{0.23\linewidth}
        \includegraphics[width=\linewidth]{figures/regression/FeatureAccuracy/FeatureAccuracy_covid_data_pre_processed_regression_train_polluted_test_polluted.pdf}
        \caption{\textsf{COVID}}
        \label{fig:regression-results-all-FeatureAccuracy-3-covid}
    \end{subfigure}
    \begin{subfigure}[b]{0.23\linewidth}
        \includegraphics[width=\linewidth]{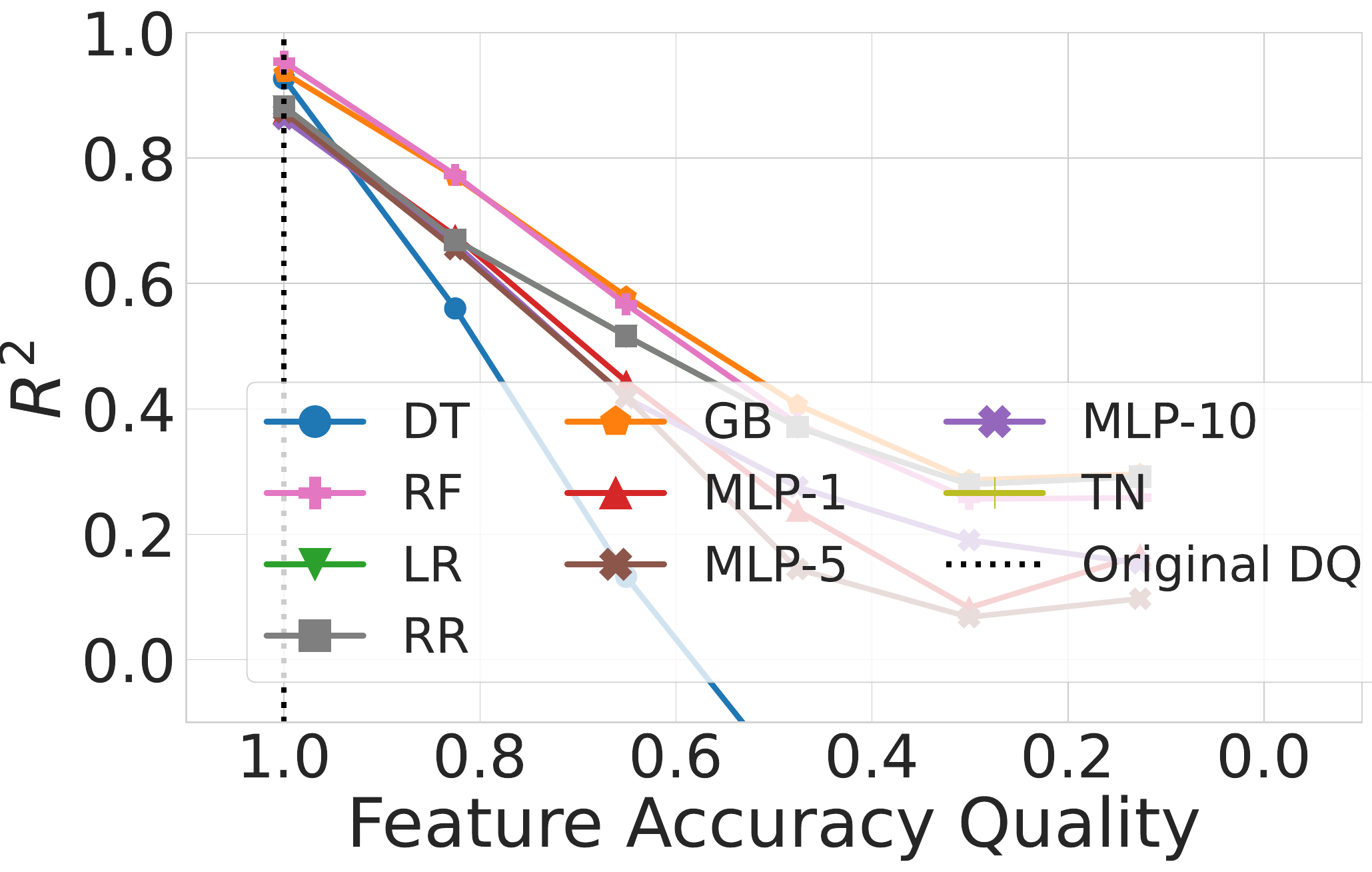}
        \caption{\textsf{Cars}}
        \label{fig:regression-results-all-FeatureAccuracy-3-cars}
    \end{subfigure}
    \caption{$R^2$ of the regression algorithms for feature accuracy.}
    \label{fig:regression-results-all-FeatureAccuracy}
\end{figure*}

%% file: Latex_Figure/regression/Target_Accurecy.tex
\begin{figure*}[t]
    \centering
\raisebox{0.4\height}{\rotatebox{90}{Scenario 1}}\hspace{0.3em}
    \begin{subfigure}[b]{0.23\linewidth}
        \includegraphics[width=\linewidth]{figures/regression/TargetAccuracy/TargetAccuracy_house_prices_prepared_train_polluted_test_original.pdf}
        \caption{\textsf{Houses}}
        \label{fig:regression-results-all-TargetAccuracy-1-houses}
    \end{subfigure}
\begin{subfigure}[b]{0.23\linewidth}
        \includegraphics[width=\linewidth]{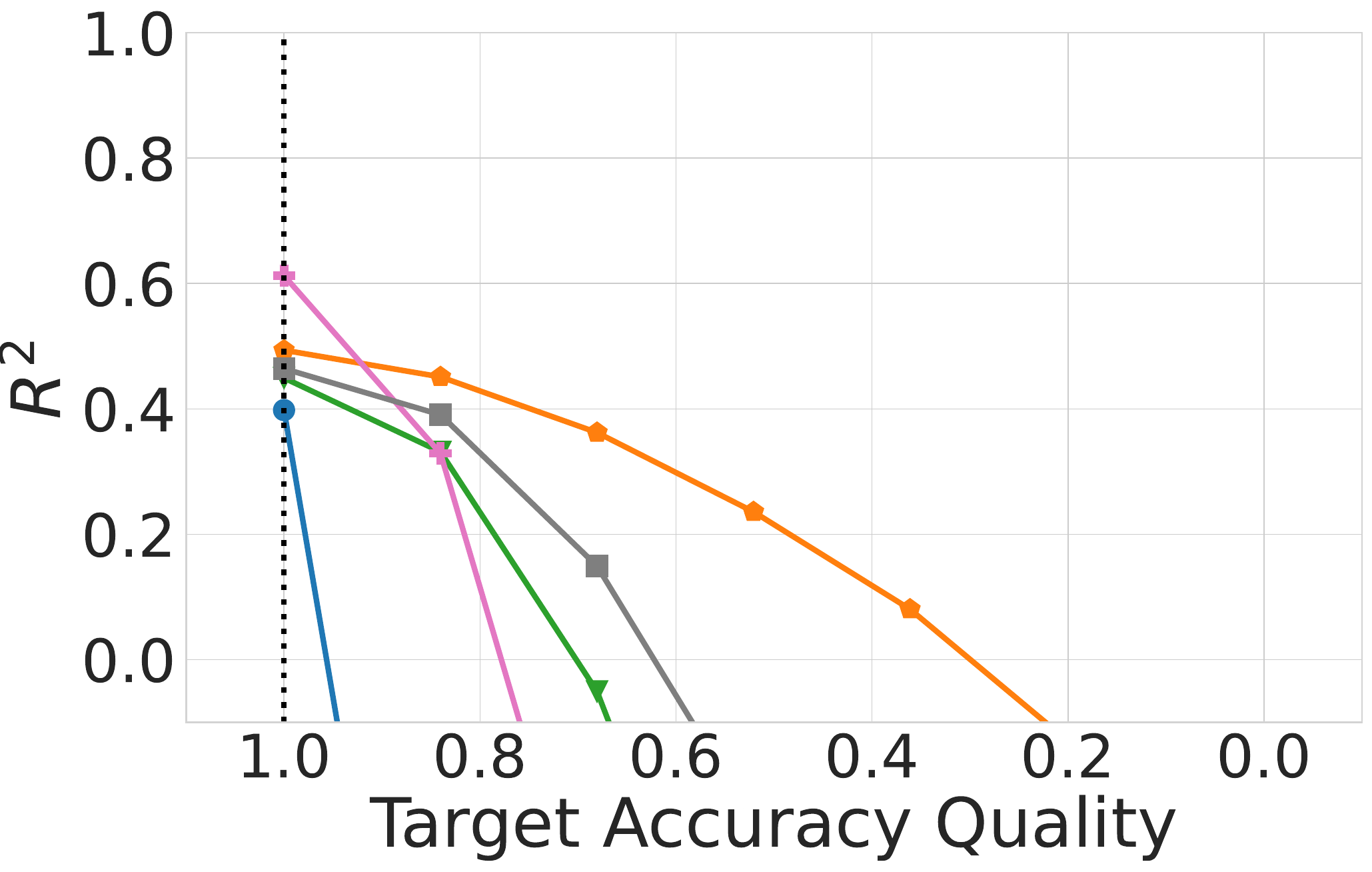}
        \caption{\textsf{IMDB}}
        \label{fig:regression-results-all-TargetAccuracy-1-imdb}
    \end{subfigure}
    \begin{subfigure}[b]{0.23\linewidth}
        \includegraphics[width=\linewidth]{figures/regression/TargetAccuracy/TargetAccuracy_covid_data_pre_processed_regression_train_polluted_test_original.pdf}
        \caption{\textsf{COVID}}
        \label{fig:regression-results-all-TargetAccuracy-1-covid}
    \end{subfigure}
    \begin{subfigure}[b]{0.23\linewidth}
        \includegraphics[width=\linewidth]{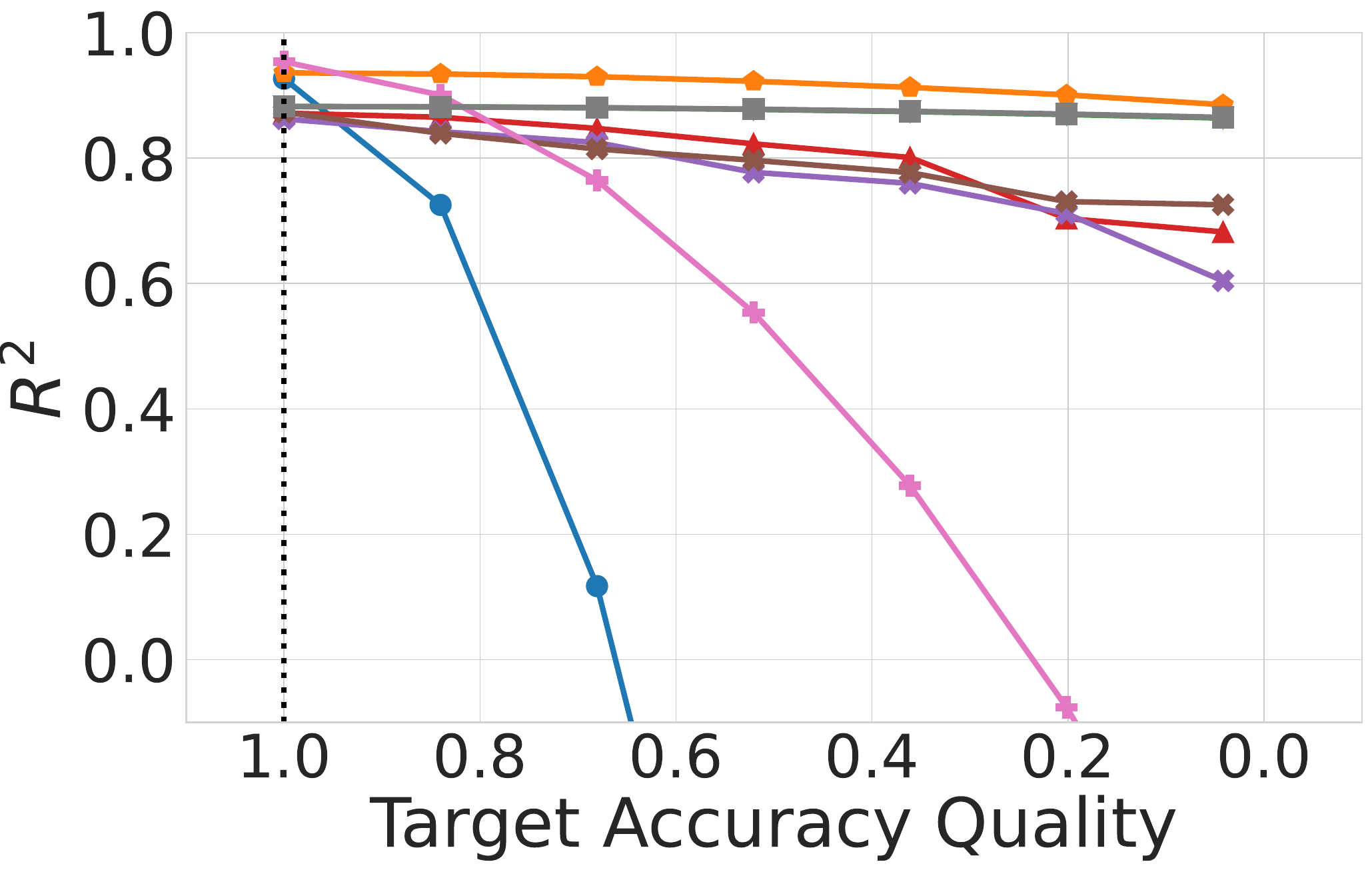}
        \caption{\textsf{Cars}}
        \label{fig:regression-results-all-TargetAccuracy-1-cars}
    \end{subfigure}

\raisebox{0.4\height}{\rotatebox{90}{Scenario 2}}\hspace{0.3em}
    \begin{subfigure}[b]{0.23\linewidth}
        \includegraphics[width=\linewidth]{figures/regression/TargetAccuracy/TargetAccuracy_house_prices_prepared_train_original_test_polluted.pdf}
        \caption{\textsf{Houses}}
        \label{fig:regression-results-all-TargetAccuracy-2-houses}
    \end{subfigure}
\begin{subfigure}[b]{0.23\linewidth}
        \includegraphics[width=\linewidth]{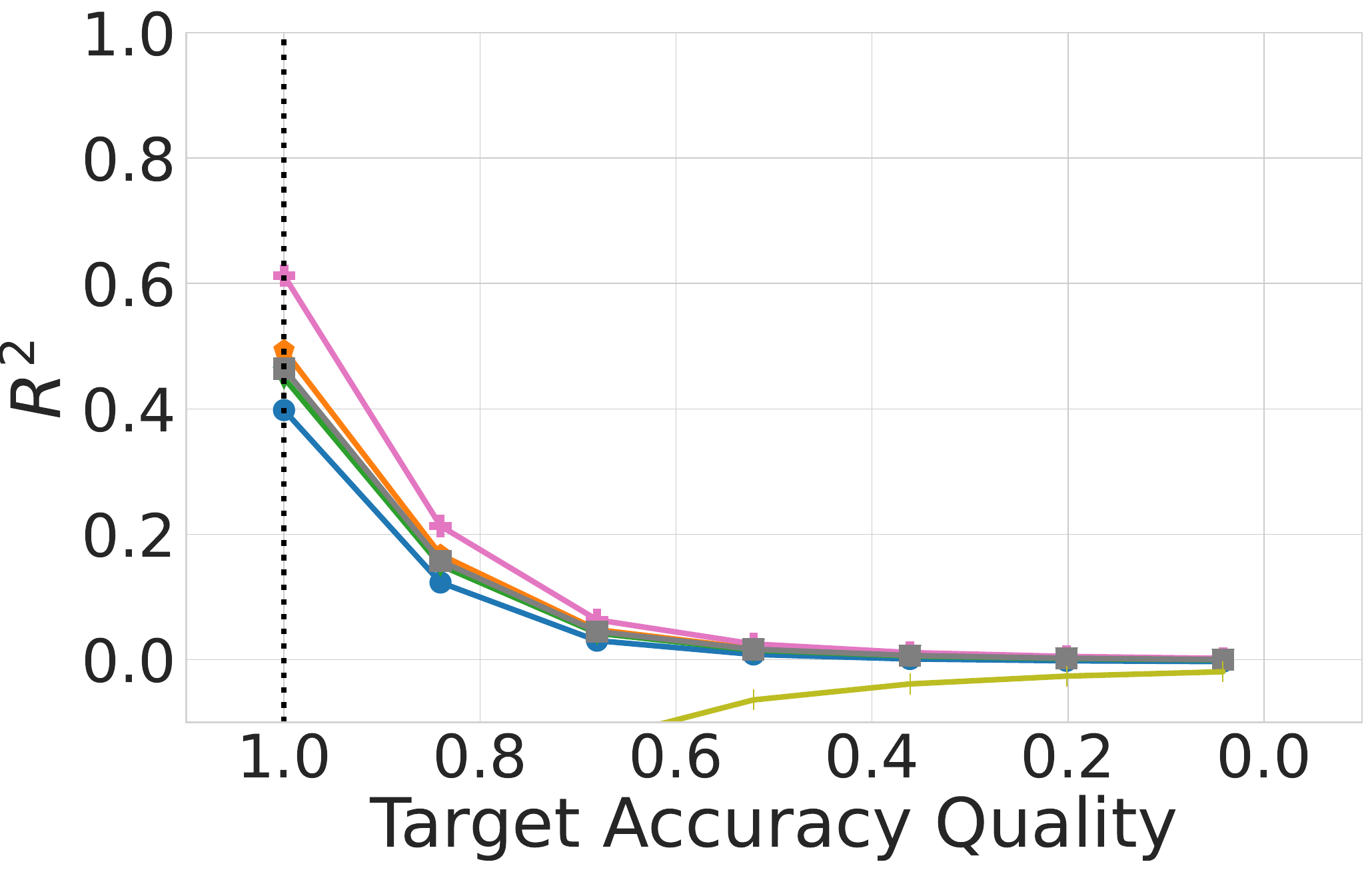}
        \caption{\textsf{IMDB}}
        \label{fig:regression-results-all-TargetAccuracy-2-imdb}
    \end{subfigure}
    \begin{subfigure}[b]{0.23\linewidth}
        \includegraphics[width=\linewidth]{figures/regression/TargetAccuracy/TargetAccuracy_covid_data_pre_processed_regression_train_original_test_polluted.pdf}
        \caption{\textsf{COVID}}
        \label{fig:regression-results-all-TargetAccuracy-2-covid}
    \end{subfigure}
    \begin{subfigure}[b]{0.23\linewidth}
        \includegraphics[width=\linewidth]{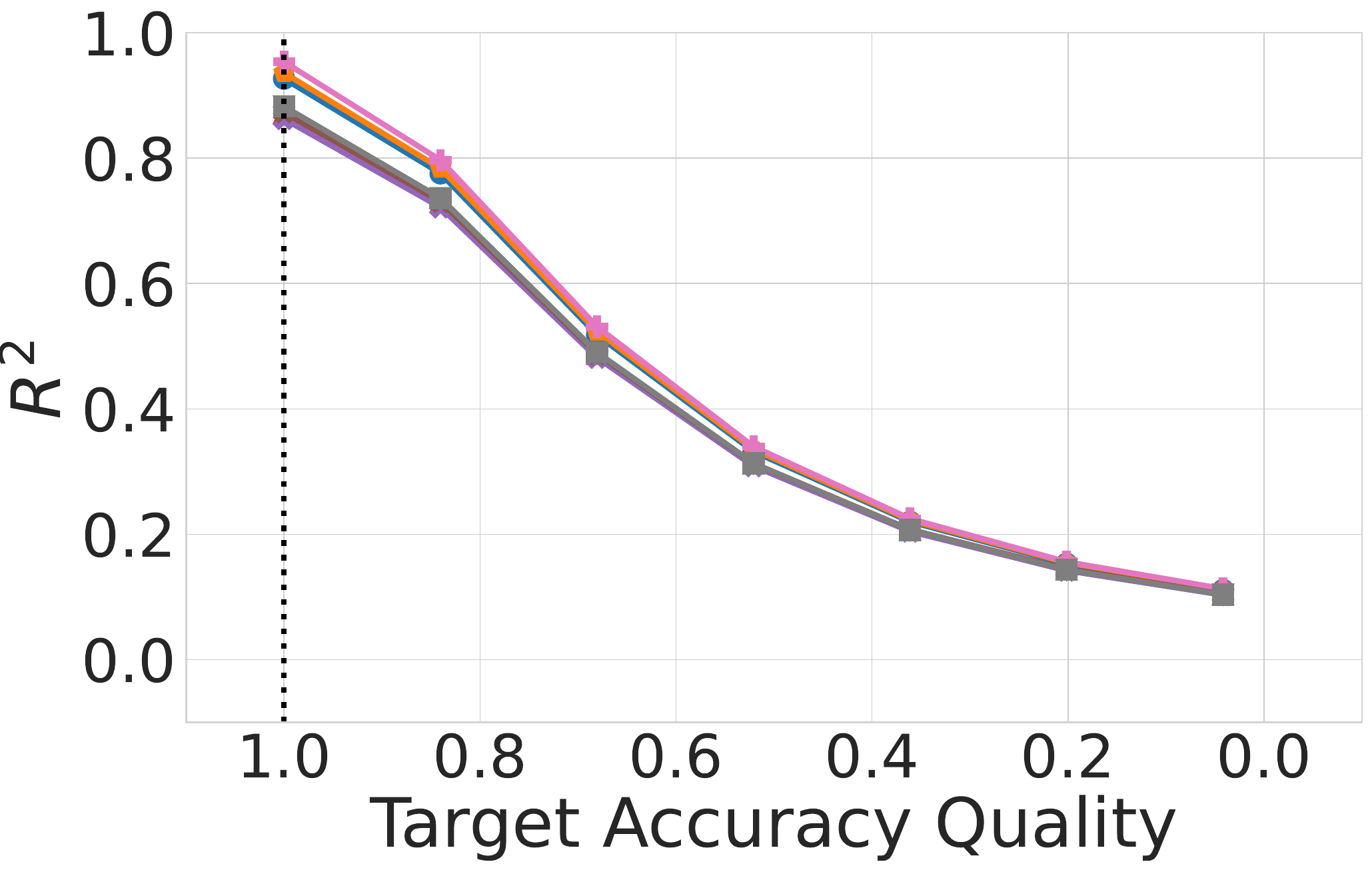}
        \caption{\textsf{Cars}}
        \label{fig:regression-results-all-TargetAccuracy-2-cars}
    \end{subfigure}

\raisebox{0.4\height}{\rotatebox{90}{Scenario 3}}\hspace{0.3em}
\begin{subfigure}[b]{0.23\linewidth}
        \includegraphics[width=\linewidth]{figures/regression/TargetAccuracy/TargetAccuracy_house_prices_prepared_train_polluted_test_polluted.pdf}
        \caption{\textsf{Houses}}
        \label{fig:regression-results-all-TargetAccuracy-3-houses}
    \end{subfigure}
\begin{subfigure}[b]{0.23\linewidth}
        \includegraphics[width=\linewidth]{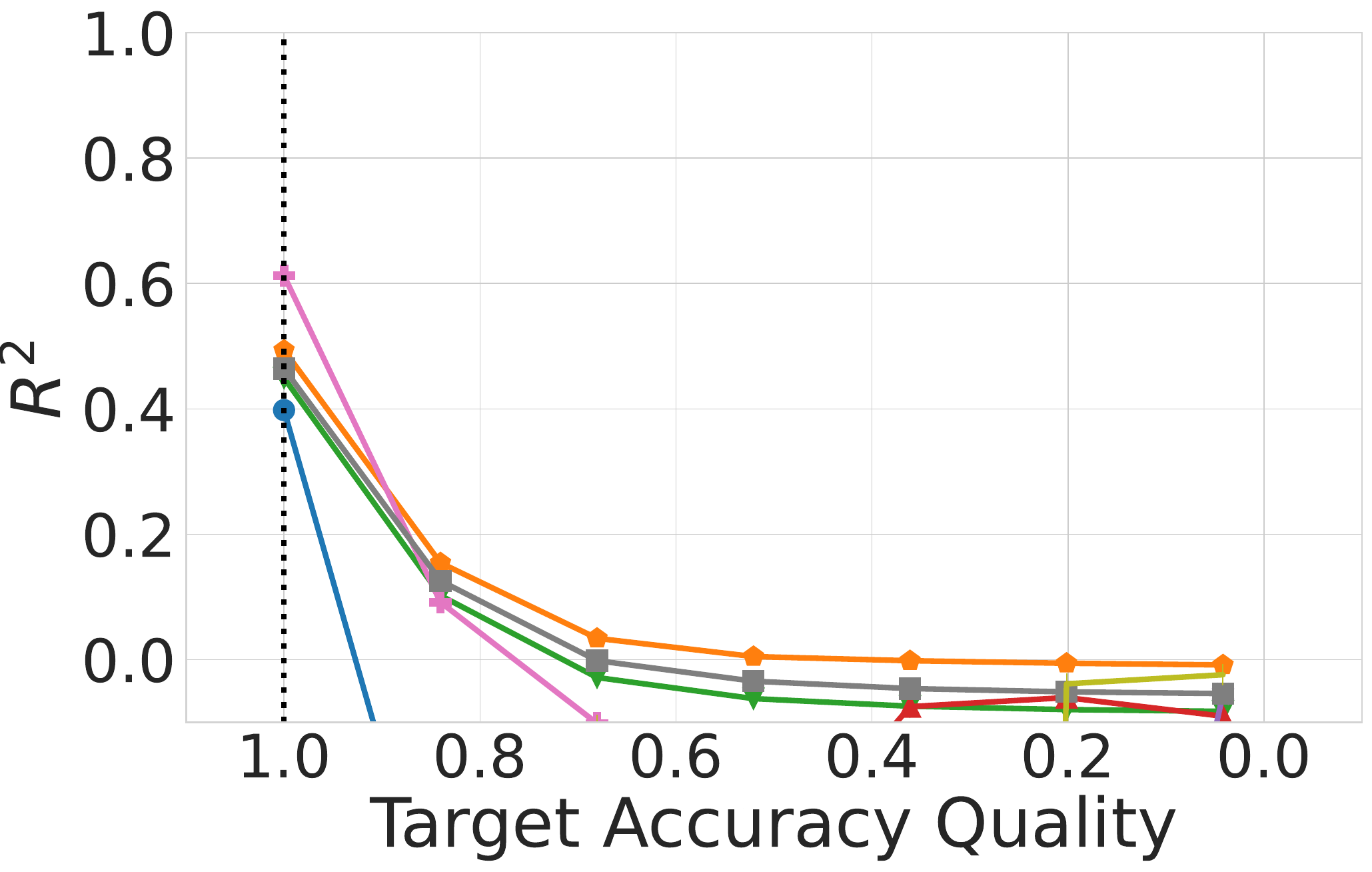}
        \caption{\textsf{IMDB}}
        \label{fig:regression-results-all-TargetAccuracy-3-imdb}
    \end{subfigure}
    \begin{subfigure}[b]{0.23\linewidth}
        \includegraphics[width=\linewidth]{figures/regression/TargetAccuracy/TargetAccuracy_covid_data_pre_processed_regression_train_polluted_test_polluted.pdf}
        \caption{\textsf{COVID}}
        \label{fig:regression-results-all-TargetAccuracy-3-covid}
    \end{subfigure}
    \begin{subfigure}[b]{0.23\linewidth}
        \includegraphics[width=\linewidth]{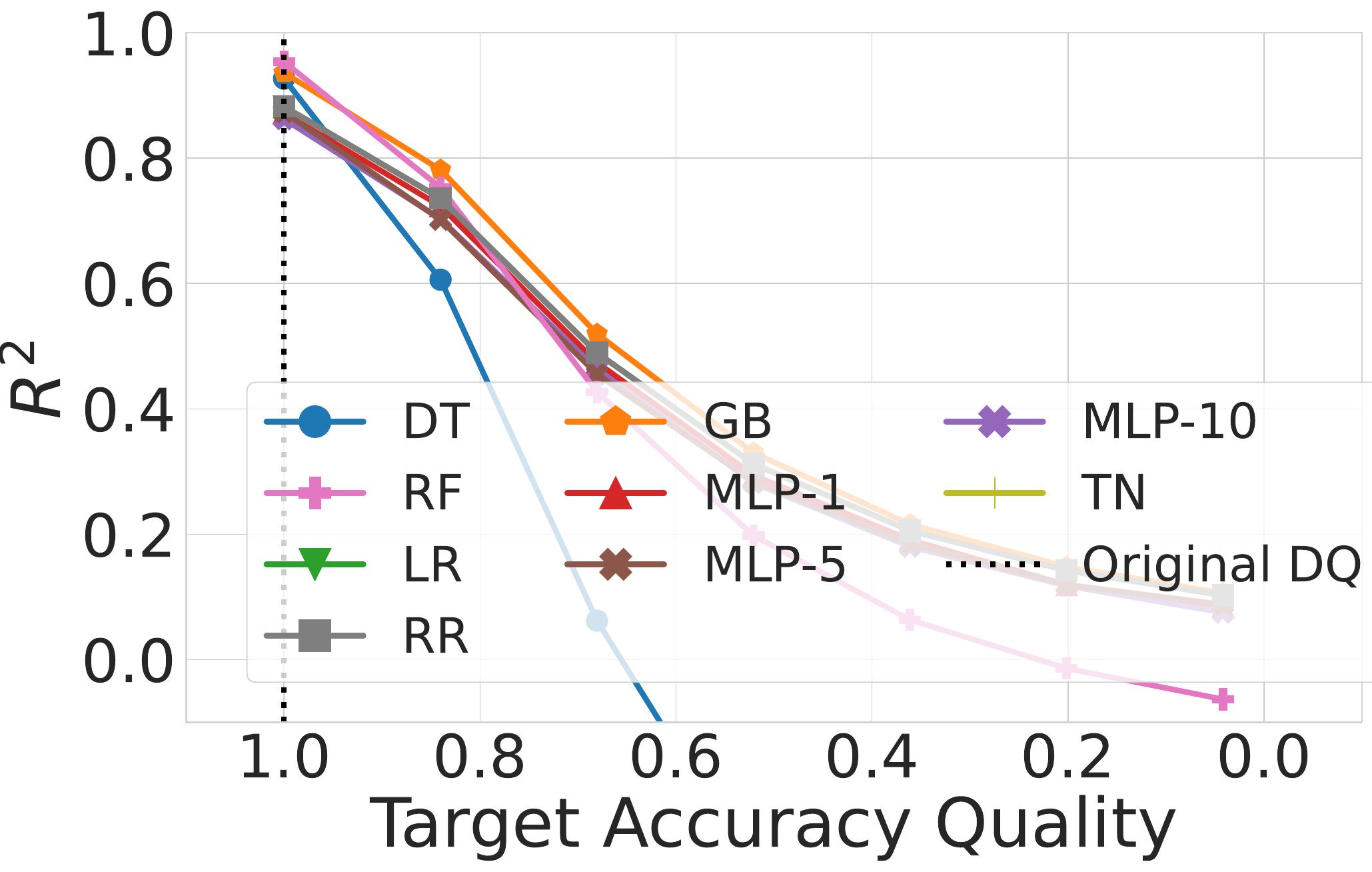}
        \caption{\textsf{Cars}}
        \label{fig:regression-results-all-TargetAccuracy-3-cars}
    \end{subfigure}
    \caption{$R^2$ of the regression algorithms for target accuracy.}
    \label{fig:regression-results-all-TargetAccuracy}
\end{figure*}

%% file: Latex_Figure/regression/Uniqueness_1.tex
\begin{figure*}[t]
    \centering
\raisebox{0.4\height}{\rotatebox{90}{Scenario 1}}\hspace{0.3em}
    \begin{subfigure}[b]{0.23\linewidth}
        \includegraphics[width=\linewidth]{figures/regression/Uniqueness_duplicatecount_1/Uniqueness_7_house_prices_prepared_train_polluted_test_original.pdf}
        \caption{\textsf{Houses}}
        \label{fig:regression-results-all-Uniqueness_dc1-1-houses}
    \end{subfigure}
\begin{subfigure}[b]{0.23\linewidth}
        \includegraphics[width=\linewidth]{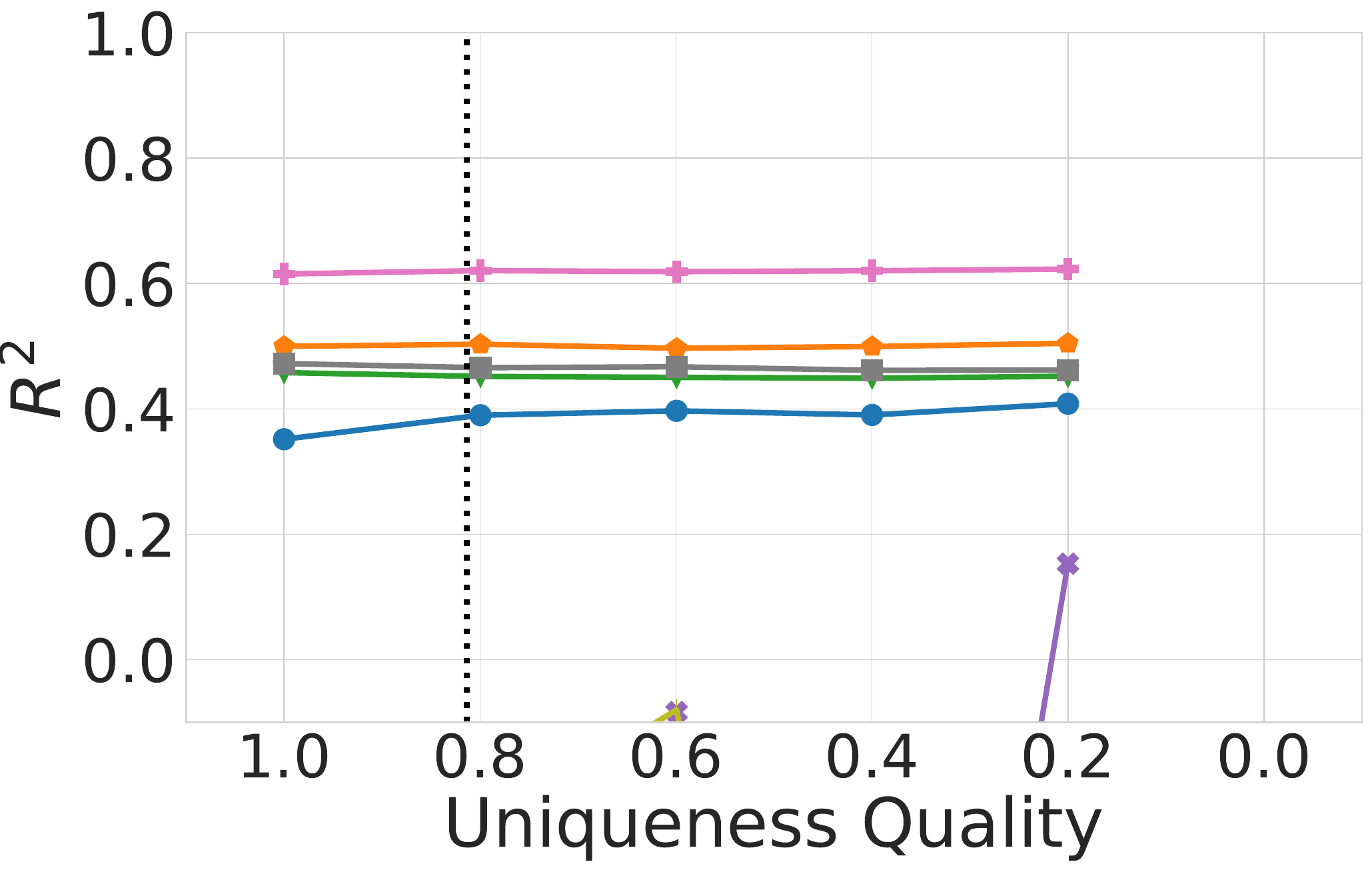}
        \caption{\textsf{IMDB}}
        \label{fig:regression-results-all-Uniqueness_dc1-1-imdb}
    \end{subfigure}
    \begin{subfigure}[b]{0.23\linewidth}
        \includegraphics[width=\linewidth]{figures/regression/Uniqueness_duplicatecount_1/Uniqueness_7_covid_data_pre_processed_regression_train_polluted_test_original.pdf}
        \caption{\textsf{COVID}}
        \label{fig:regression-results-all-Uniqueness_dc1-1-covid}
    \end{subfigure}
    \begin{subfigure}[b]{0.23\linewidth}
        \includegraphics[width=\linewidth]{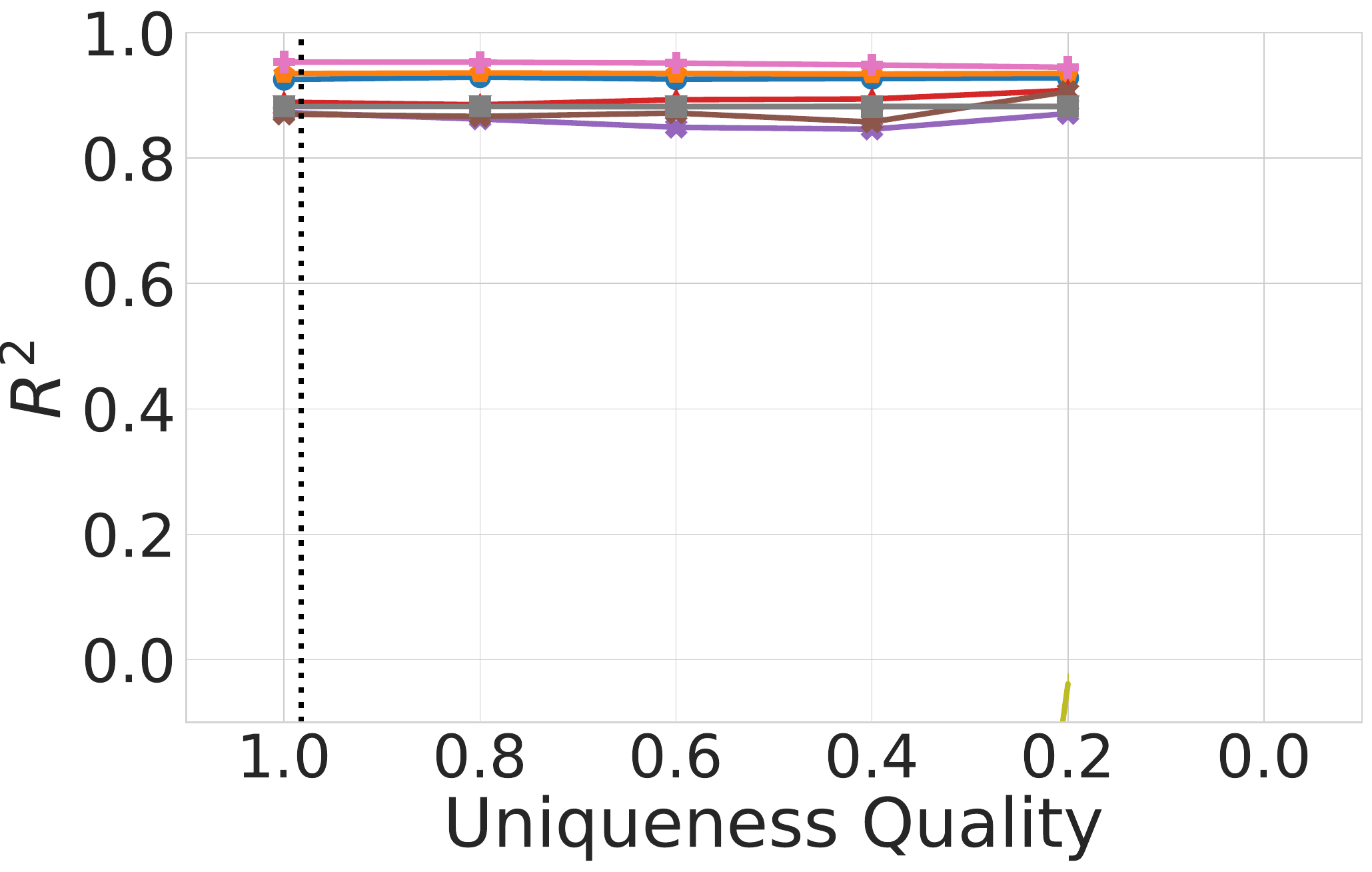}
        \caption{\textsf{Cars}}
        \label{fig:regression-results-all-Uniqueness_dc1-1-cars}
    \end{subfigure}

\raisebox{0.4\height}{\rotatebox{90}{Scenario 2}}\hspace{0.3em}
    \begin{subfigure}[b]{0.23\linewidth}
        \includegraphics[width=\linewidth]{figures/regression/Uniqueness_duplicatecount_1/Uniqueness_7_house_prices_prepared_train_original_test_polluted.pdf}
        \caption{\textsf{Houses}}
        \label{fig:regression-results-all-Uniqueness_dc1-2-houses}
    \end{subfigure}
\begin{subfigure}[b]{0.23\linewidth}
        \includegraphics[width=\linewidth]{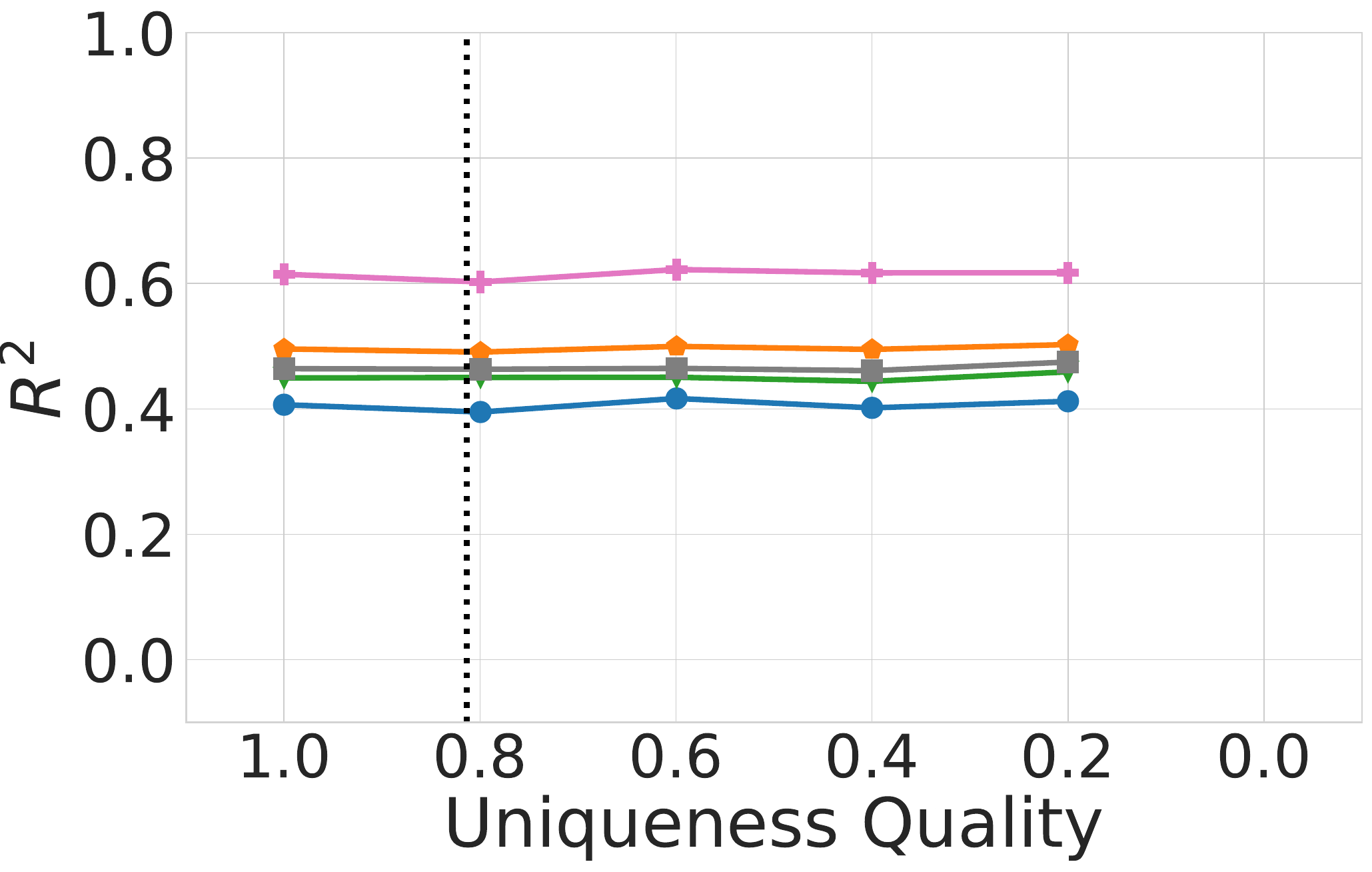}
        \caption{\textsf{IMDB}}
        \label{fig:regression-results-all-Uniqueness_dc1-2-imdb}
    \end{subfigure}
    \begin{subfigure}[b]{0.23\linewidth}
        \includegraphics[width=\linewidth]{figures/regression/Uniqueness_duplicatecount_1/Uniqueness_7_covid_data_pre_processed_regression_train_original_test_polluted.pdf}
        \caption{\textsf{COVID}}
        \label{fig:regression-results-all-Uniqueness_dc1-2-covid}
    \end{subfigure}
    \begin{subfigure}[b]{0.23\linewidth}
        \includegraphics[width=\linewidth]{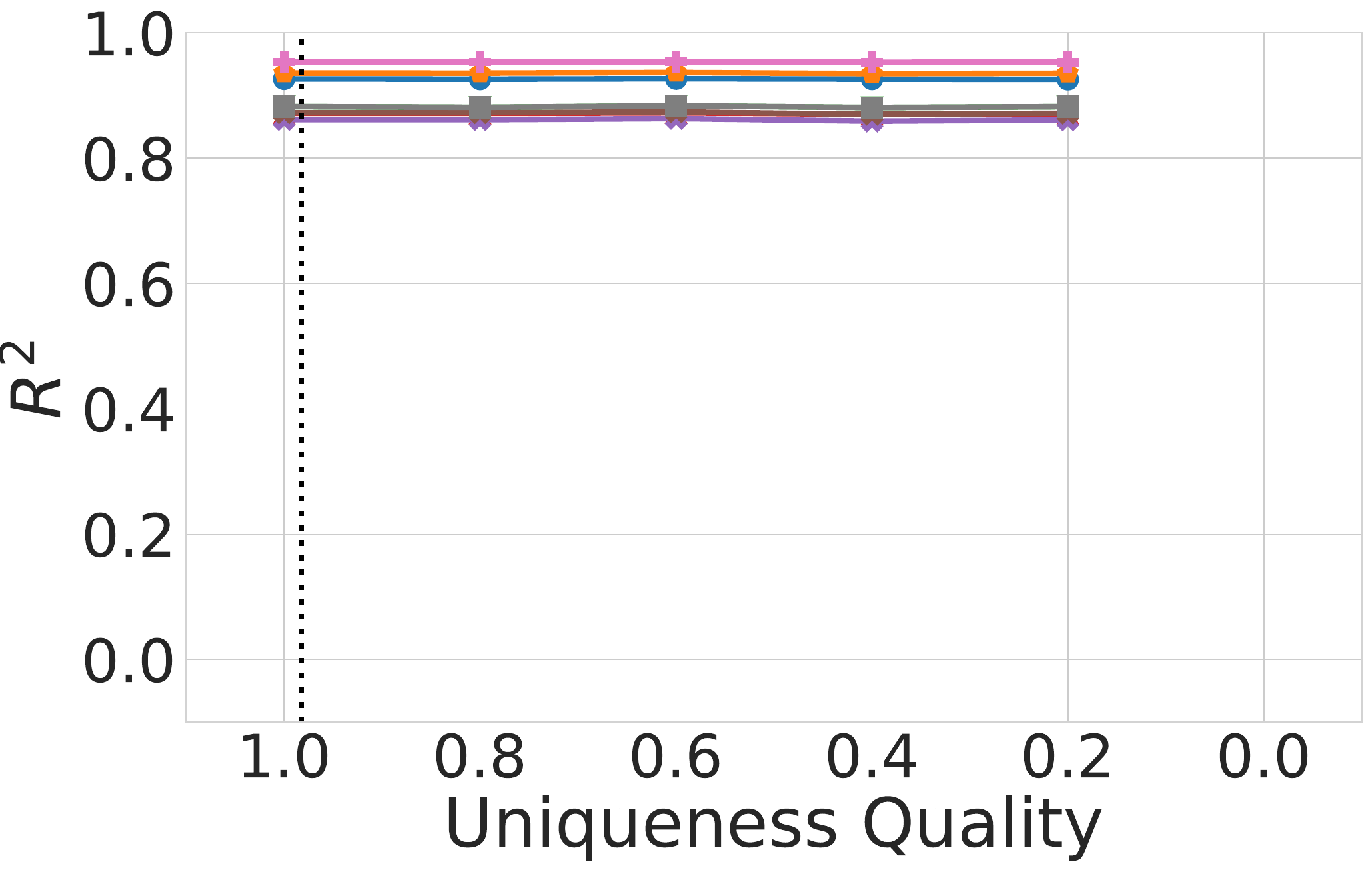}
        \caption{\textsf{Cars}}
        \label{fig:regression-results-all-Uniqueness_dc1-2-cars}
    \end{subfigure}

\raisebox{0.4\height}{\rotatebox{90}{Scenario 3}}\hspace{0.3em}
    \begin{subfigure}[b]{0.23\linewidth}
        \includegraphics[width=\linewidth]{figures/regression/Uniqueness_duplicatecount_1/Uniqueness_7_house_prices_prepared_train_polluted_test_polluted.pdf}
        \caption{\textsf{Houses}}
        \label{fig:regression-results-all-Uniqueness_dc1-3-houses}
    \end{subfigure}
\begin{subfigure}[b]{0.23\linewidth}
        \includegraphics[width=\linewidth]{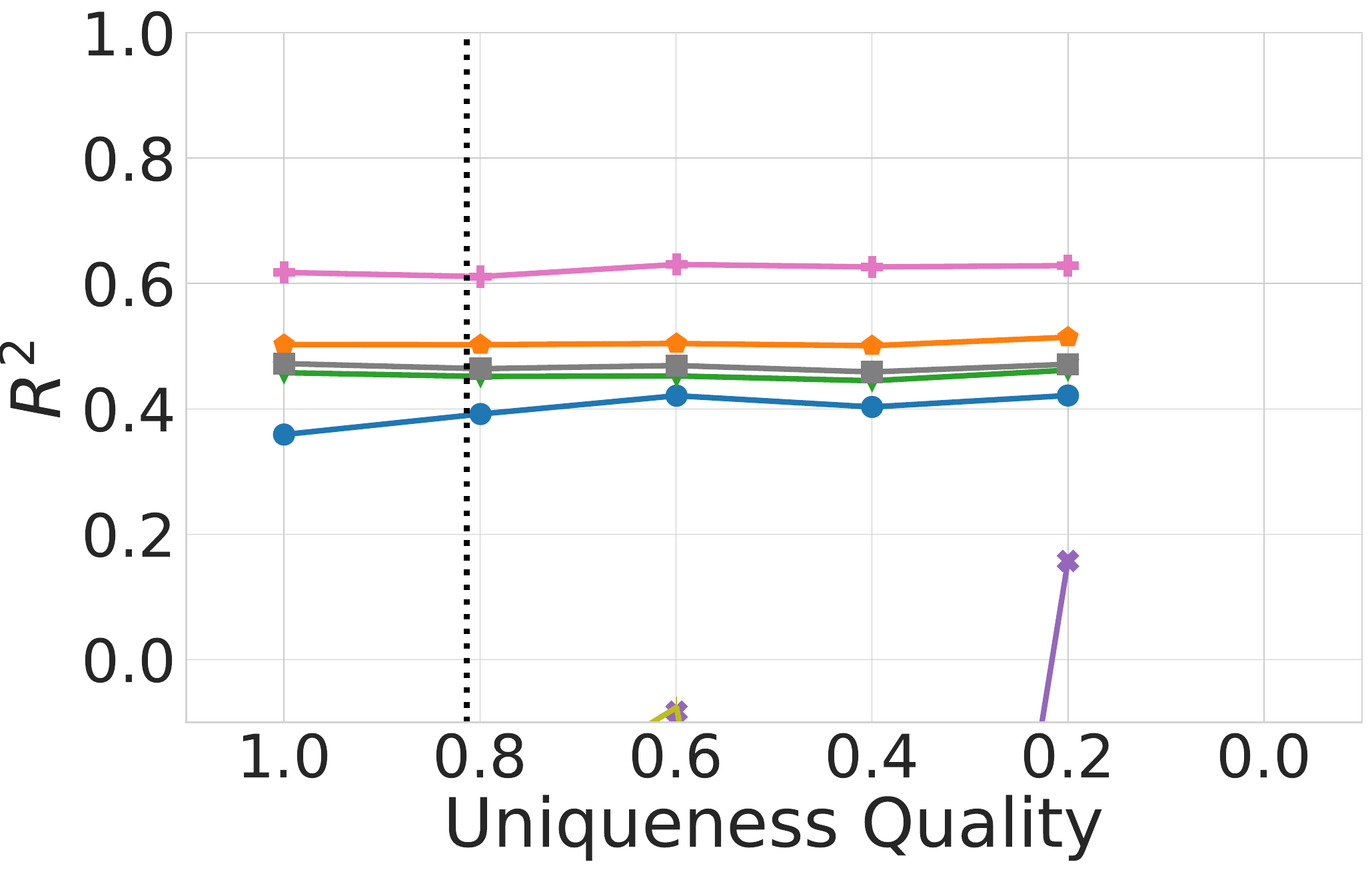}
        \caption{\textsf{IMDB}}
        \label{fig:regression-results-all-Uniqueness_dc1-3-imdb}
    \end{subfigure}
    \begin{subfigure}[b]{0.23\linewidth}
        \includegraphics[width=\linewidth]{figures/regression/Uniqueness_duplicatecount_1/Uniqueness_7_covid_data_pre_processed_regression_train_polluted_test_polluted.pdf}
        \caption{\textsf{COVID}}
        \label{fig:regression-results-all-Uniqueness_dc1-3-covid}
    \end{subfigure}
    \begin{subfigure}[b]{0.23\linewidth}
        \includegraphics[width=\linewidth]{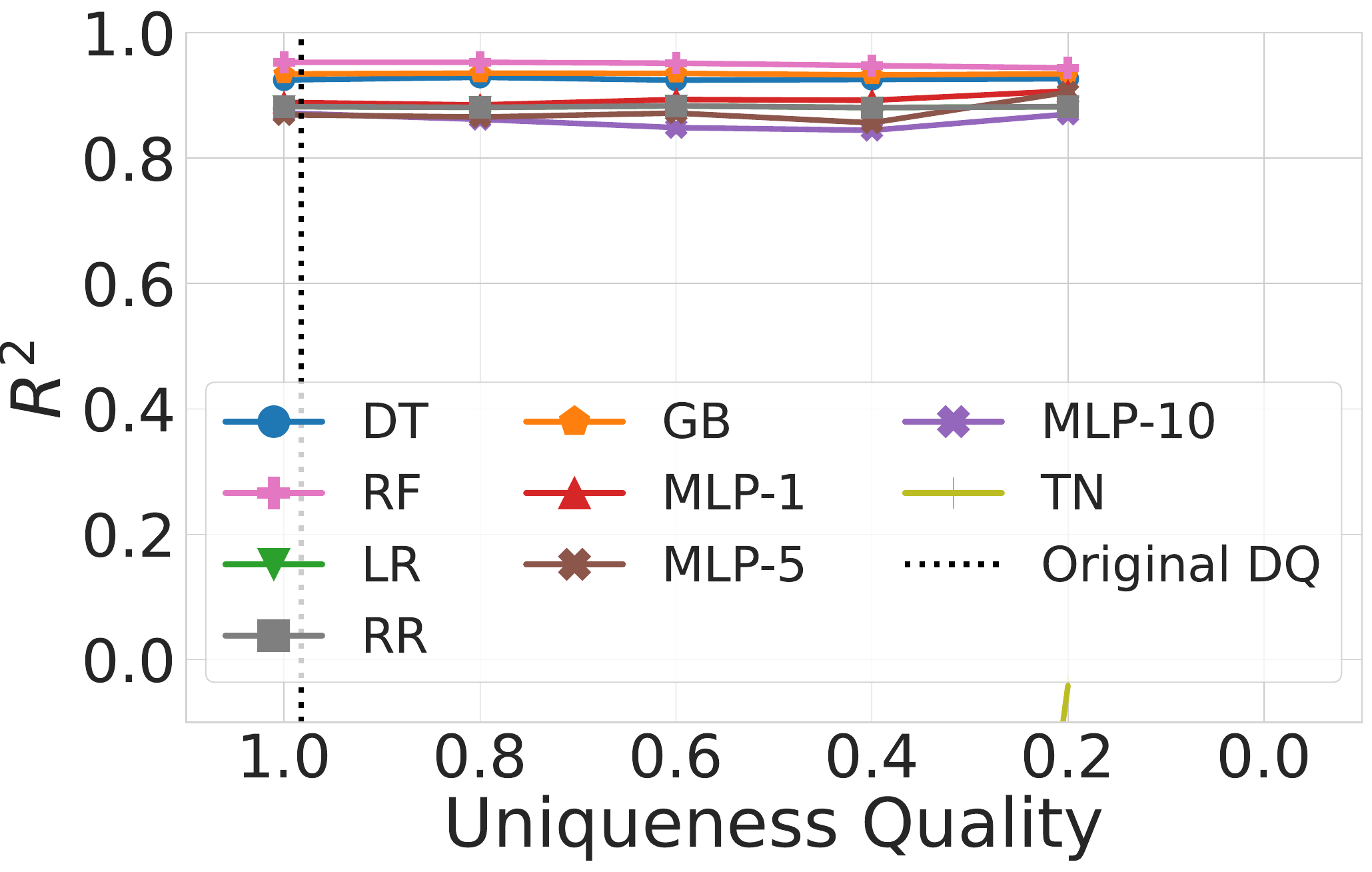}
        \caption{\textsf{Cars}}
        \label{fig:regression-results-all-Uniqueness_dc1-3-cars}
    \end{subfigure}
    \caption{$R^2$ of the regression algorithms for uniqueness with all rows having duplicate count of 1.}
    \label{fig:regression-results-all-Uniqueness_dc1}
\end{figure*}

%% file: Latex_Figure/regression/Class_Balance.tex
\begin{figure*}[t]
    \centering
\raisebox{0.4\height}{\rotatebox{90}{Scenario 1}}\hspace{0.3em}
\begin{subfigure}[b]{0.23\linewidth}
        \includegraphics[width=\linewidth]{figures/regression/ClassBalance/ClassBalance_house_prices_prepared_train_polluted_test_original.pdf}
        \caption{\textsf{Houses}}
        \label{fig:regression-results-all-ClassBalance-1-houses}
    \end{subfigure}
\begin{subfigure}[b]{0.23\linewidth}
        \includegraphics[width=\linewidth]{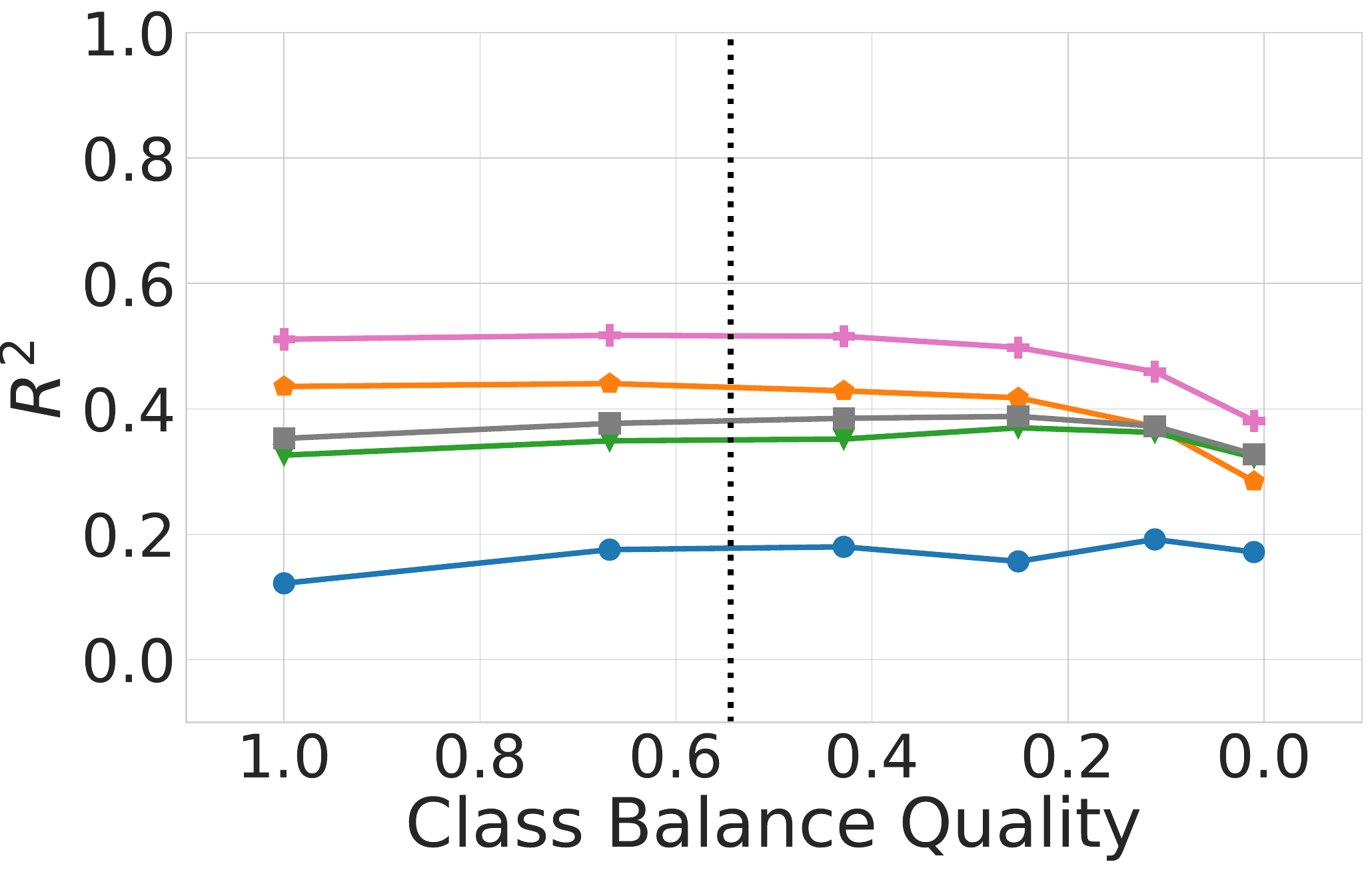}
        \caption{\textsf{IMDB}}
        \label{fig:regression-results-all-ClassBalance-1-imdb}
    \end{subfigure}
\begin{subfigure}[b]{0.23\linewidth}
        \includegraphics[width=\linewidth]{figures/regression/ClassBalance/ClassBalance_covid_data_pre_processed_regression_train_polluted_test_original.pdf}
        \caption{\textsf{COVID}}
        \label{fig:regression-results-all-ClassBalance-1-covid}
    \end{subfigure}
\begin{subfigure}[b]{0.23\linewidth}
        \includegraphics[width=\linewidth]{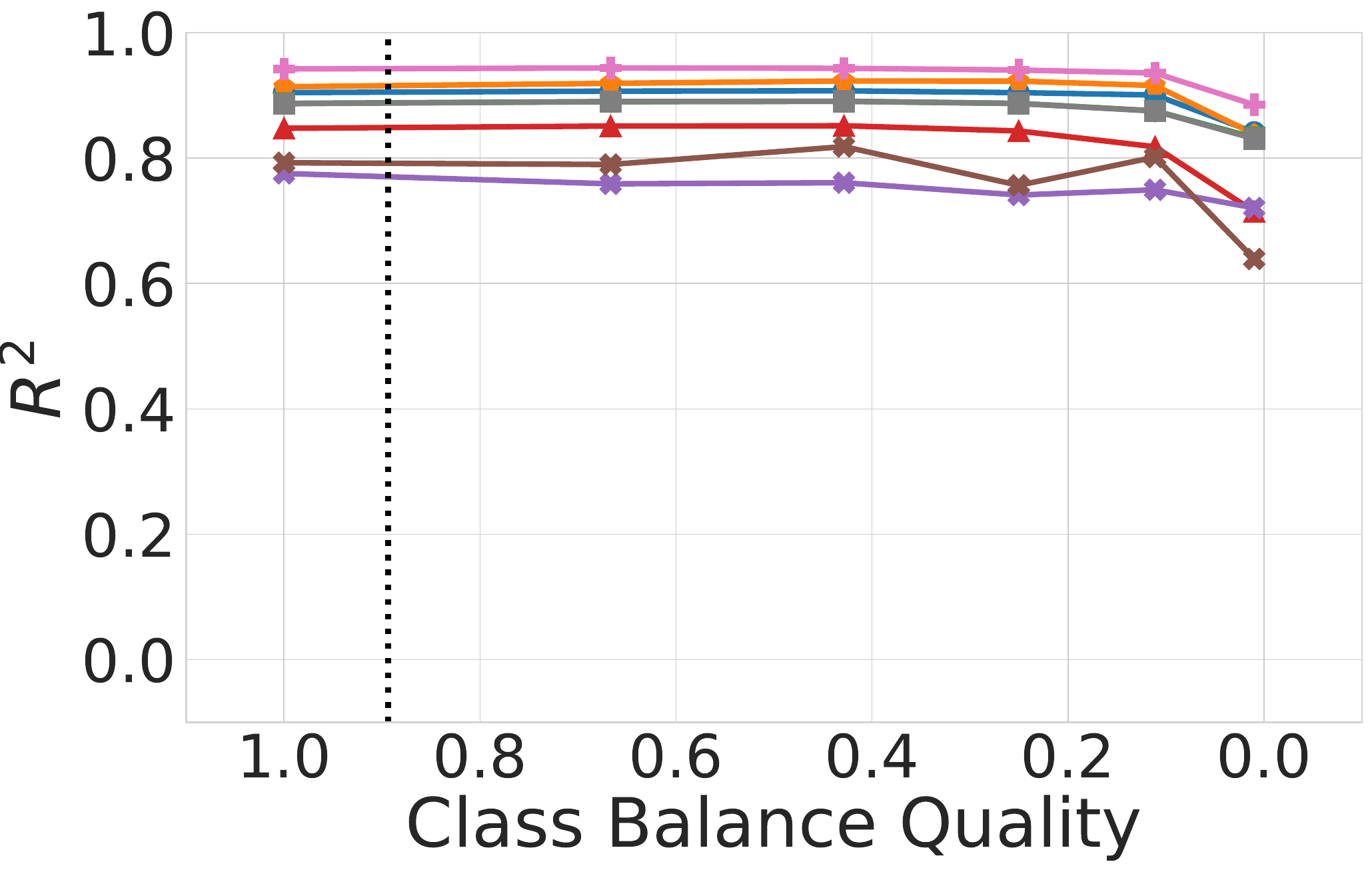}
        \caption{\textsf{Cars}}
        \label{fig:regression-results-all-ClassBalance-1-cars}
    \end{subfigure}

\raisebox{0.4\height}{\rotatebox{90}{Scenario 2}}\hspace{0.3em}
\begin{subfigure}[b]{0.23\linewidth}
        \includegraphics[width=\linewidth]{figures/regression/ClassBalance/ClassBalance_house_prices_prepared_train_original_test_polluted.pdf}
        \caption{\textsf{Houses}}
        \label{fig:regression-results-all-ClassBalance-2-houses}
    \end{subfigure}
\begin{subfigure}[b]{0.23\linewidth}
        \includegraphics[width=\linewidth]{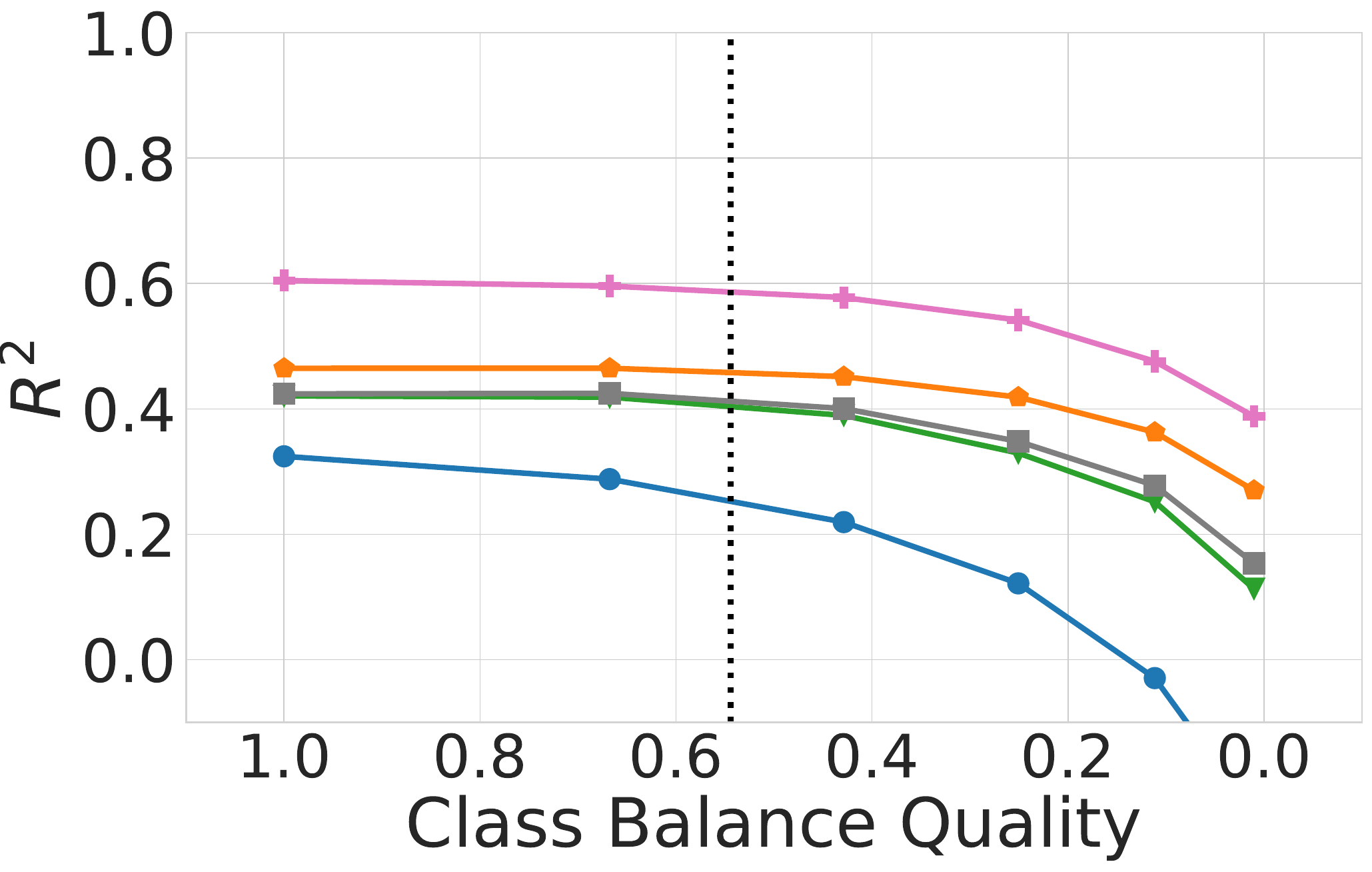}
        \caption{\textsf{IMDB}}
        \label{fig:regression-results-all-ClassBalance-2-imdb}
    \end{subfigure}
\begin{subfigure}[b]{0.23\linewidth}
        \includegraphics[width=\linewidth]{figures/regression/ClassBalance/ClassBalance_covid_data_pre_processed_regression_train_original_test_polluted.pdf}
        \caption{\textsf{COVID}}
        \label{fig:regression-results-all-ClassBalance-2-covid}
    \end{subfigure}
\begin{subfigure}[b]{0.23\linewidth}
        \includegraphics[width=\linewidth]{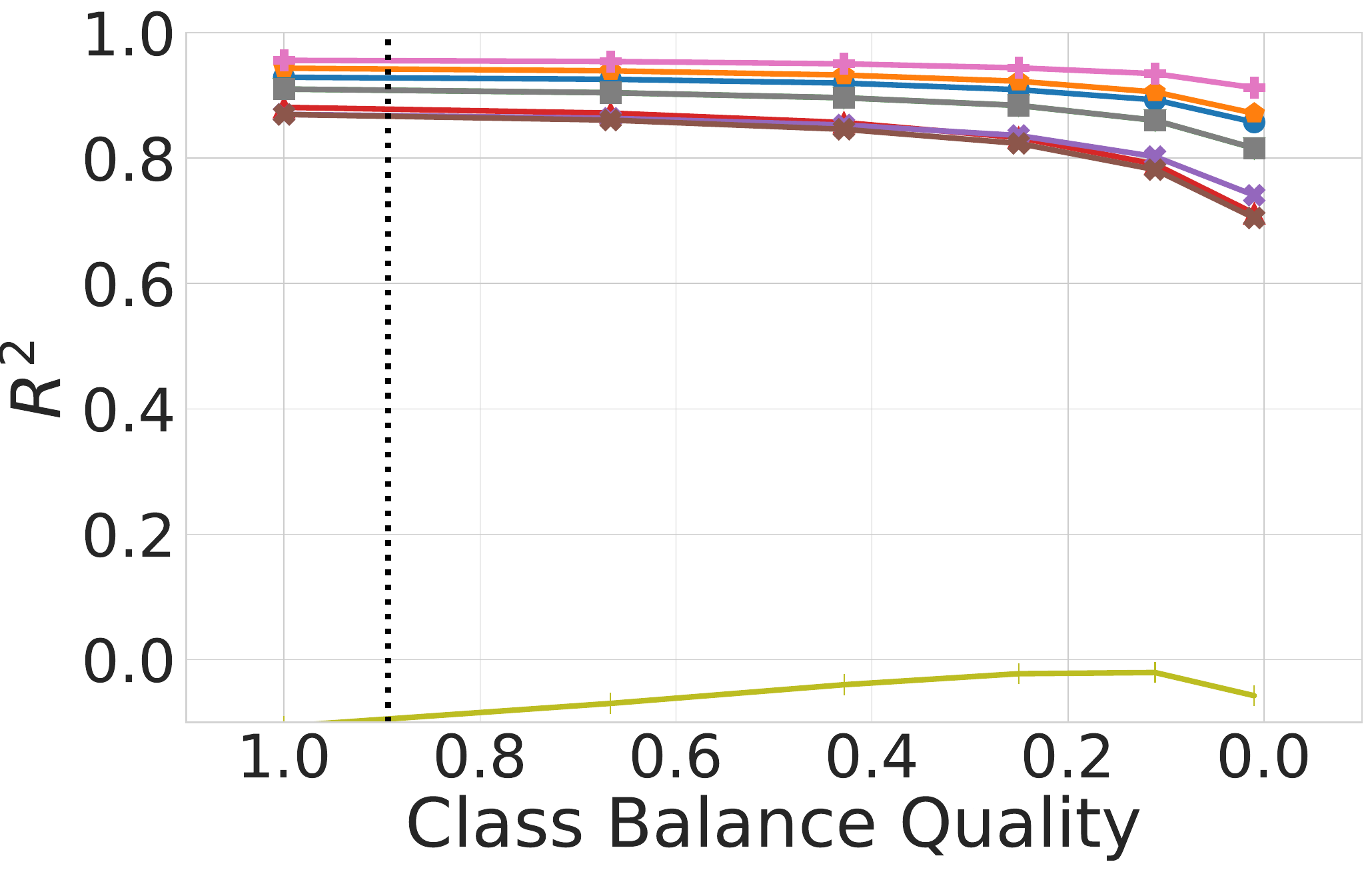}
        \caption{\textsf{Cars}}
        \label{fig:regression-results-all-ClassBalance-2-cars}
    \end{subfigure}

\raisebox{0.4\height}{\rotatebox{90}{Scenario 3}}\hspace{0.3em}
\begin{subfigure}[b]{0.23\linewidth}
        \includegraphics[width=\linewidth]{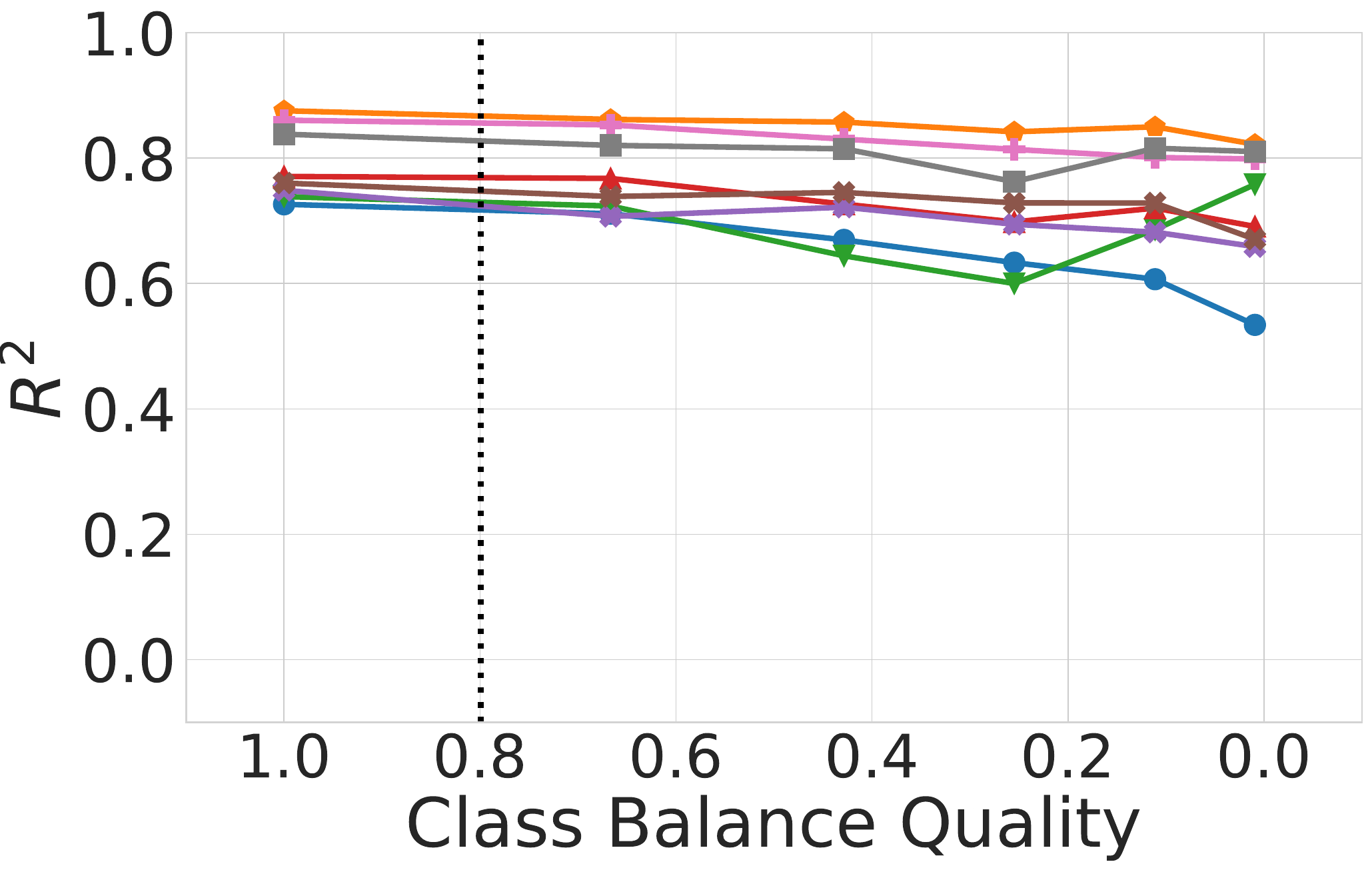}
        \caption{\textsf{Houses}}
        \label{fig:regression-results-all-ClassBalance-3-houses}
    \end{subfigure}
\begin{subfigure}[b]{0.23\linewidth}
        \includegraphics[width=\linewidth]{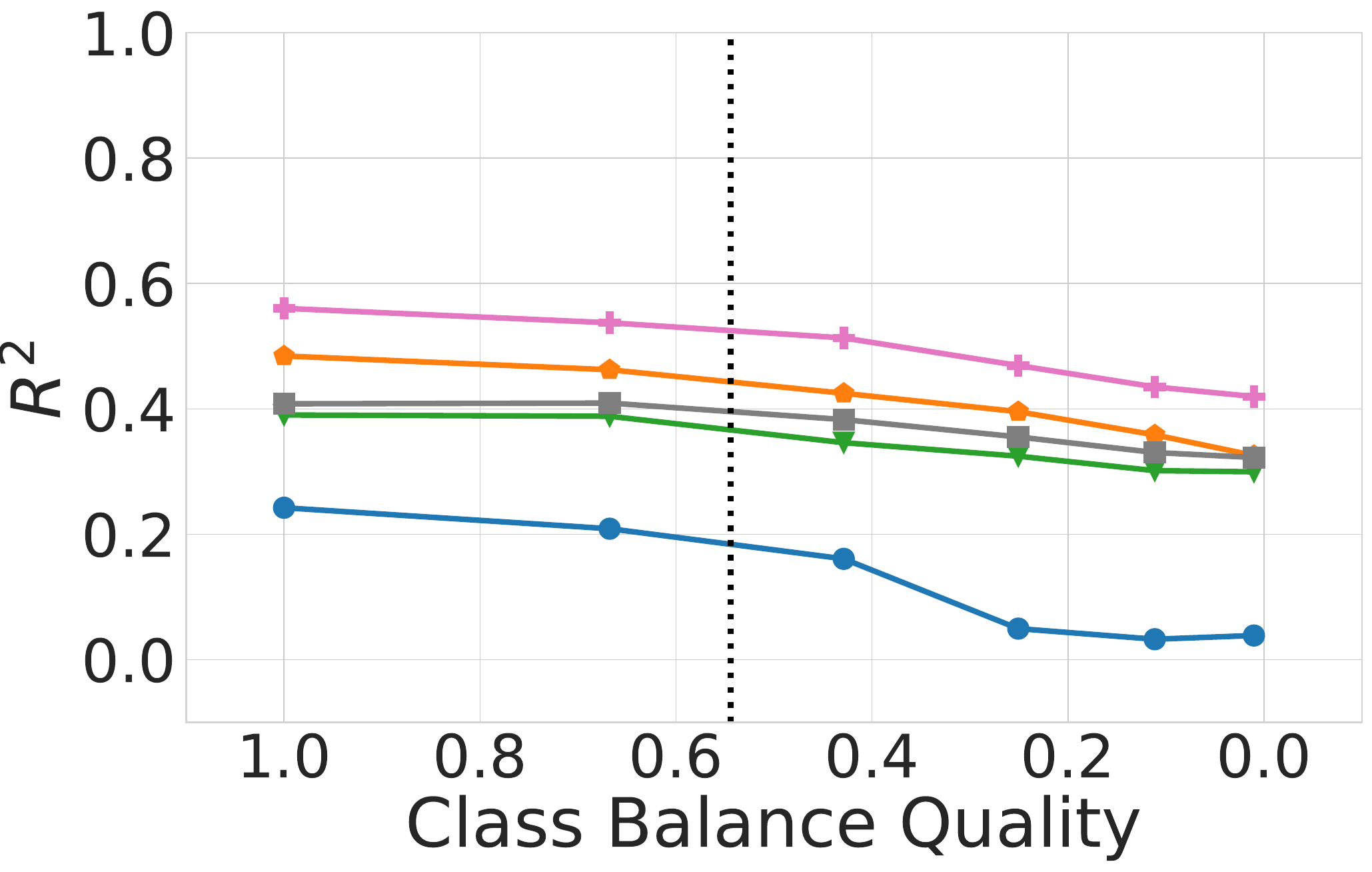}
        \caption{\textsf{IMDB}}
        \label{fig:regression-results-all-ClassBalance-3-imdb}
    \end{subfigure}
\begin{subfigure}[b]{0.23\linewidth}
        \includegraphics[width=\linewidth]{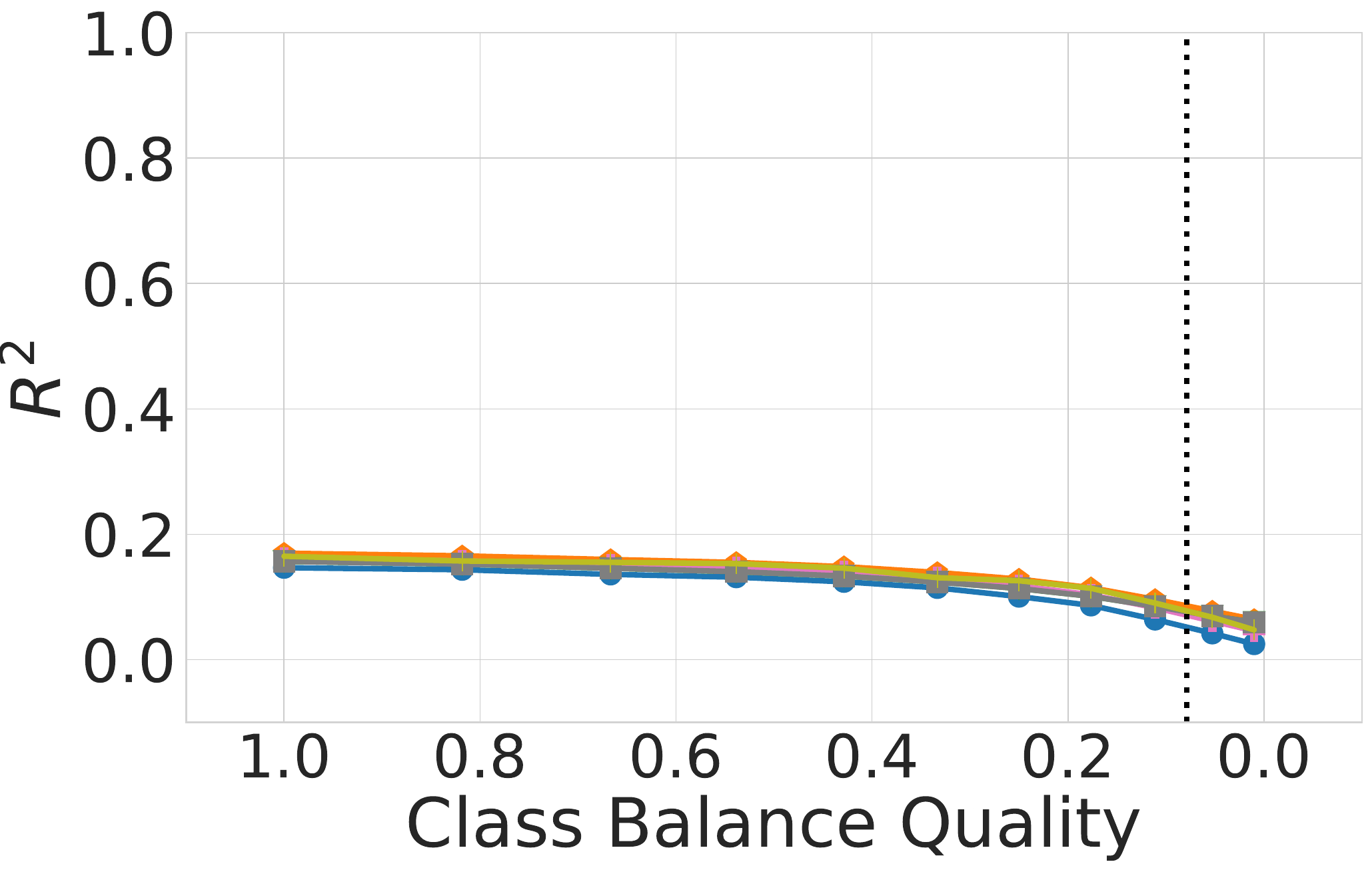}
        \caption{\textsf{COVID}}
        \label{fig:regression-results-all-ClassBalance-3-covid}
    \end{subfigure}
\begin{subfigure}[b]{0.23\linewidth}
        \includegraphics[width=\linewidth]{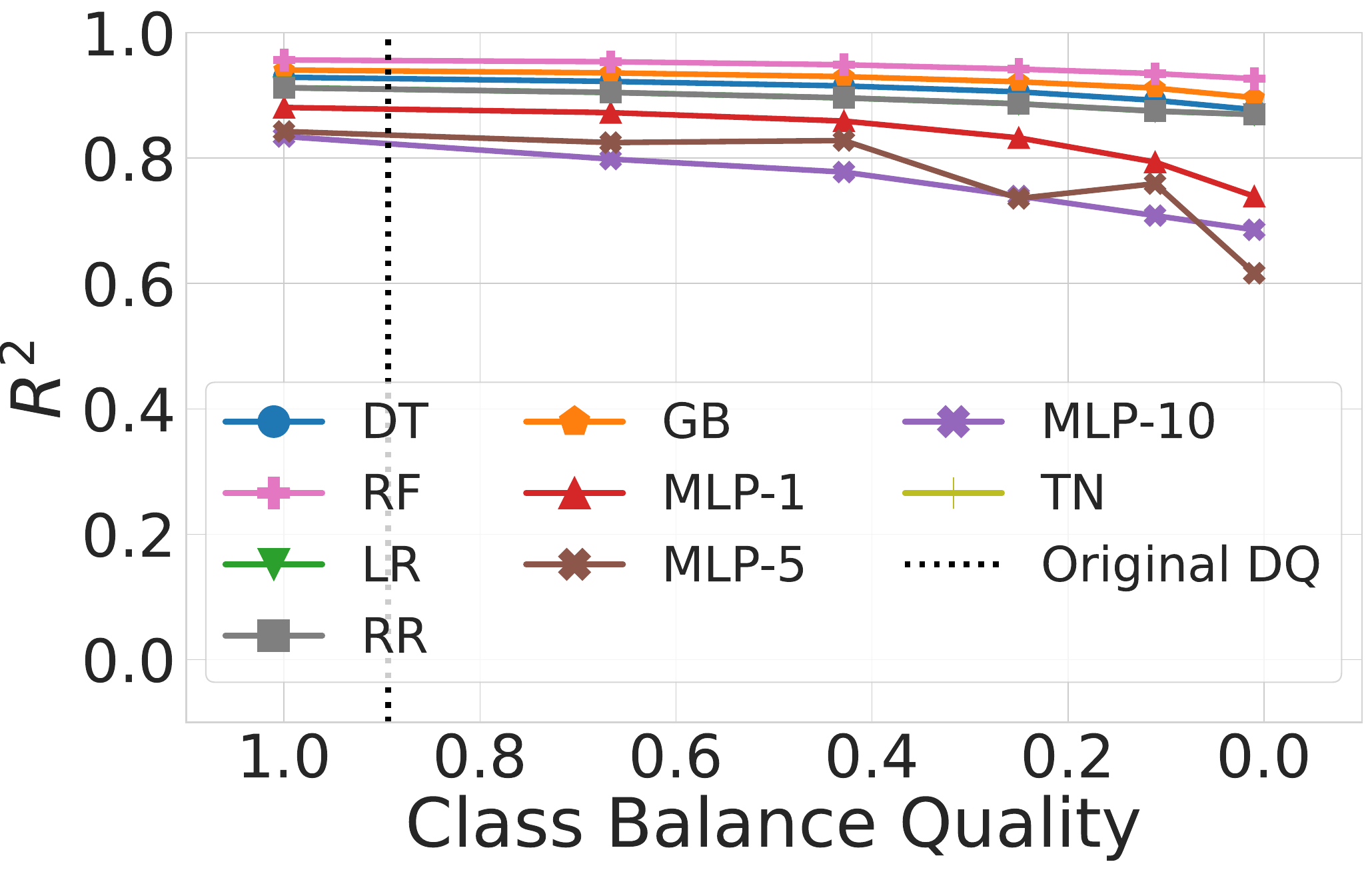}
        \caption{\textsf{Cars}}
        \label{fig:regression-results-all-ClassBalance-3-cars}
    \end{subfigure}
    \caption{$R^2$ of the regression algorithms for target class balance.}
    \label{fig:regression-results-all-ClassBalance}
\end{figure*}

%% file: 60-results/63-clustering.tex
\subsection{Clustering}
\label{subsec:clustering-results}
\PaperShort{\input{Latex_Figure/clustering/summary_convtype}}
Finally, we discuss \revision{in the following section} the effect of degrading the six data quality dimensions of \revision{four} datasets, namely: \textsf{Bank}, \textsf{Covertype}\revision{,} \textsf{Letter} \revision{and \textsf{COVID}} on the performance of five clustering algorithms, namely: Gaussian mixture clustering, $k$-means/$k$-prototypes, agglomerative clustering, OPTICS and autoencoder. 

\PaperLong{All plots shown are scaled to share the same y-axis. 
This comes at the downside of \textsf{Bank} plots having a lower readability, as the algorithms' performance is significantly worse on this dataset than on the others.}
\PaperShort{\revision{In Figure~\ref{fig:clustering-results-all-covertype}, we show the results of the experiments using the \textsf{Covertype} dataset.
We only show the y-axis up to an AMI score of 0.5, which is the highest archived score in our experiments.}}

Here, we start by some general notes and observations which are beneficial for the discussion in the rest of this section.
\textsf{Bank} is not originally designed for the clustering task, \revision{as it is} very low in dimensionality and contains a high amount of duplicated data points. 
Upon further analysis, we found that some combinations of the two categorical features dominate the data, with the combination of \textit{job = blue-collar} and \textit{marital = married} being found in about~$24.8\%$ of samples and the combinations' \textit{technician, married} and \textit{admin., married} together making up another approximately~$18.6\%$ of the samples. 
Therefore, we know that large parts of the data are not distinguishable by anything but the \texttt{age} feature. 
We question this feature's impact, as it is likely dominated due to the one-hot encoding, which adds more dimensions to the data.
As this encodes the categorical values in the data into one dimension per value, it distinguishes the impact of each individual dimension, which disproportionately affects the age feature which is not encoded in this way. 
Therefore, we are aware that the clustering quality on \textsf{Bank} is expected to be low. 
We argue that this is not an issue in most cases, as we can still derive some relative changes in clustering quality and argue about the impact of any given quality dimension regarding this dataset. 
However, we also believe that the findings on \textsf{Bank} are to be taken with a grain of salt and more weight should be put into findings on \textsf{Covertype}\revision{,} \textsf{Letter} \revision{and \textsf{COVID}.} 

\revision{All algorithms demonstrate significantly greater robustness to lower data quality in the \textsf{COVID} dataset. 
However, the AMI score is generally much lower compared to the other datasets.}

We start by a general observation:
We observed that the Gaussian mixture algorithm assumes data that follows a single distribution,~i.e., the random peaks and drops in its performance are caused by its sensitivity to the distribution of samples.
\revision{All algorithms demonstrate significantly greater robustness to lower data quality in the \textsf{COVID} dataset. 
However, the AMI score is generally much lower compared to the other datasets.}\revision{All algorithms demonstrate significantly greater robustness to lower data quality in the \textsf{COVID} dataset. 
However, the AMI score is generally much lower compared to the other datasets.}

\stitle{Consistent Representation}
\PaperLong{\input{Latex_Figure/clustering/Consistent_Representation}}
\PaperLong{As previously mentioned in Section~\ref{subsubsection:clustering-datasets}, \textsf{Letter} is the only dataset that the consistent representation polluter cannot be applied to, as it does not contain categorical features. 
This is why we only show three plots in Figure~\ref{fig:clustering-consistent-repr-results}.
We only consider those consistent representation pollution results that generated four additional representations per categorical value, not those adding only one new representation per value.}

\PaperShort{We do not report on \textsf{Letter} as it has no categorical features. Figure}\PaperLong{Figures~\ref{fig:clustering-consistent-repr-bank-results} and}~\ref{fig:clustering-consistent-repr-covertype-results} indicate\PaperShort{s} that the performance of the $k$-means/$k$-prototypes and OPTICS algorithms is not affected by the degrading consistency of \textsf{Covertype}\PaperLong{as well as \textsf{Bank}}.\PaperShort{The same observation applies to \textsf{Bank}.} 
In contrast, the agglomerative algorithm is strongly impaired even with only a small pollution degree. 
This susceptibility arises because the algorithm uses a Boolean distance matrix to represent differences between values for categorical features rather than a distance matrix with actual distances like for numerical features. 
This makes it easy to disturb the bottom-up combination process by altering the representation of the values.
\PaperLong{For example, with a quality measure of~$0.9$, the agglomerative clustering algorithm may encounter enough distant samples that are sufficiently removed from their original cluster samples to form their own subtrees.
These subtrees may not merge into the main cluster trees during the bottom-up process before the tree structure is cut off at the desired number of output clusters.
Additionally, this can cause clusters that were initially separated to be merged before these outlier subtrees are merged, degrading the quality of clusters that used to be correctly identified.}

For the autoencoder approach, we suspect that its observed random behavior is due to the randomness of the neural network training and its dependence on the sampled data for each of the five runs, which is amplified in datasets with many features like \textsf{Covertype}. 
The Gaussian mixture algorithm's performance slightly decreases on average with decreasing dataset quality. 
\PaperLong{This general reduction is in a similar range for both \textsf{Bank} and \textsf{Covertype}, with a performance difference of about~0.02 between the unpolluted~(baseline) and fully polluted variants of the datasets.
However, we see a similar behavior for \textsf{COVID}, where the performance first decreases, and with increasing pollution, the performance increases again.} 
We attribute the unexpected results~(initial decrease, then increase in AMI) to the observation made at the beginning of this section.

\stitle{Completeness}
\PaperLong{\input{Latex_Figure/clustering/Completeness}}
For all datasets, we notice\PaperLong{in Figure~\ref{fig:clustering-completeness-results}} a decline in average AMI for all algorithms as the completeness decreases.\PaperLong{While the different datasets show differing behavior, this general trend stays true among all of them.}

Figure~\PaperLong{\ref{fig:clustering-completeness-letter-results}}\PaperShort{\ref{fig:clustering-completeness-covertype-results}},\PaperLong{depicting the results on \textsf{Letter},} \revision{shows} that this decline is rapid and approaching zero in a negative exponential fashion.\PaperLong{Nevertheless, the autoencoder seems more robust against the insertion of default values. 
The OPTICS algorithm is most affected by a degradation of the dataset completeness, only identifying one cluster at a completeness quality of~$0.9$ already~(see Figure~\ref{fig:clustering-completeness-letter-nclusters}).
We believe this to be due to the shifting of data points from their original clusters to outlier clusters, where data points group due to their equality in the inserted default values.
This both decreases the density of existing clusters and creates a new, potentially high-density cluster, which may skew the OPTICS algorithm's perception of the expected density of clusters in the dataset.}
Yet, for \textsf{Covertype}\PaperLong{(see Figure~\ref{fig:clustering-completeness-covertype-results})}, the Gaussian mixture, autoencoder and $k$-prototypes algorithms performance degrades almost linearly and in a slightly slower fashion compared to \revision{\textsf{Bank} and \textsf{Letter}}.
This is because of the high number of categorical features that are more resilient against the insertion of default values.
\PaperLong{This hypothesis is strengthened by our findings on the low dimensional \textsf{Bank}\PaperLong{(see Figure~\ref{fig:clustering-completeness-bank-results})}, which also contains categorical features and is more resilient against clustering performance degradation than \textsf{Letter} for the completeness quality dimension.
For the OPTICS algorithm we observe an almost identical behavior to \textsf{Letter}, however, it does slightly recover some performance on low-quality datasets. 
This does line up with the peak of clusters identified by the algorithm~(see Figures~\ref{fig:clustering-completeness-covertype-nclusters} and \ref{fig:clustering-completeness-covid-nclusters} in Appendix), leading us to believe that this small increase in clustering quality is simply a coincidence with the algorithm having identified smaller clusters that, by chance, match up with parts of the original clusters in the data.}

As shown in Figure~\ref{fig:clustering-completeness-covertype-results}, the agglomerative algorithm drops almost to zero at a completeness of~$0.9$.
\PaperLong{We attribute this behavior to the default values added to represent missing values in the dataset.}
The added placeholders are not values found in \textsf{Covertype}\PaperLong{for either the numerical or categorical features, as there are no defaults for \texttt{empty} or \texttt{unknown} given in the original data}.
This can create small clusters of outliers far away from the rest of the data points. 
The effect is exacerbated by the one-hot encoding, where new dimensions are created for the placeholders\PaperLong{introduced into the data, aligning all datasets with inserted defaults along these dimensions. 
The outliers may be so far away from the original data points, that the agglomerative clustering method does not merge them into an existing cluster with original data points in it before it is forced to return the clusters found.
This has two effects:
Firstly, there are now clusters of outliers that likely do not share any resemblance to the original clusters in the algorithm's output. 
Secondly, clusters previously correctly identified in the original data are now merged as the number of clusters returned is fixed and are therefore no longer correct}.
\revision{As stated before, in \textsf{COVID}, as the largest dataset, the algorithms are generally more robust against any considered data quality issue, including the completeness.}

\PaperLong{For \textsf{Bank}, we can observe the overall decline in performance for all but the autoencoder and OPTICS clustering approaches~(see Figure~\ref{fig:clustering-completeness-bank-results} or Figure~\ref{fig:stretched-bank-results-completeness} in Appendix), generally matching our expectations derived from the previous two datasets. 
The OPTICS and autoencoder algorithms' clustering performance linearly correlates with the average number of clusters identified by the approaches~(see Figure~\ref{fig:clustering-completeness-covertype-nclusters} in Appendix). 
The behavior of the Gaussian mixtures as well as agglomerative algorithms is justified by the issues with \textsf{Bank} in general and the Gaussian mixtures algorithm's particular restrictions regarding the data distribution mentioned at the beginning of Section~\ref{subsec:clustering-results}.}

\stitle{Feature Accuracy}
\PaperLong{\input{Latex_Figure/clustering/Feature_Accurecy}}
For \textsf{Letter}, consisting of only numerical features, we observe a very uniform decrease in the AMI score when decreasing the dataset feature accuracy for almost all algorithms\PaperLong{(see Figure~\ref{fig:clustering-feature-accuracy-letter-results})}. 
This is due to the normally distributed noise we apply to numerical features\PaperLong{degrading their accuracy in a gradually increasing fashion, continuously decreasing the distinguishability of individual clusters by spreading out their points more and more. 
In addition, we can observe a slight improvement in performance for the Gaussian mixture algorithm in the first step of the pollution~(i.e., for an aggregated feature accuracy quality of~$\geq0.9$). 
We suspect this to be caused by the addition of noise, bringing the data closer to the assumed underlying mixture of Gaussian distributions. 
Moreover, we assume that the performance of OPTICS decreases at the first step of the pollution as the noise causes the density to approach a uniform level rapidly, reducing the cluster distinguishability and unifying them to a single cluster when viewed from only a density perspective~(see Figure~\ref{fig:clustering-feature-accuracy-letter-nclusters} in Appendix)}.

\PaperLong{\input{Latex_Figure/clustering/Target_Accurecy}}
In Figure~\ref{fig:clustering-feature-accuracy-covertype-results}, the Gaussian mixture algorithm\PaperLong{exhibits a special behavior} at the maximum pollution\PaperLong{: it} almost reaches the AMI score achieved on the clean dataset: \textsf{Covertype} consists mainly of binary features that have been completely inverted \PaperLong{in its fully polluted version. 
Therefore, the dataset contains mainly the same information as before the pollution}. 
The slight difference in performance comes from the ten remaining numerical features. \PaperLong{They do not share the same information at any two points of the pollution. 
To investigate why the performance of Gaussian mixture clustering on \textsf{Covertype} improves slightly around a feature accuracy quality of about~$0.6$ requires further research. 
Here, we expected a behavior more similar to that of the agglomerative clustering approach.} 
\revision{We see a similar behavior in the \textsf{COVID} dataset, including the slight difference in the performance between the fully clean and fully polluted state.}
\PaperLong{COVID has only one numerical attribute.}
The agglomerative algorithm behaves similarly \revision{to Gaussian mixture algorithm},\PaperLong{dropping in performance as the pollution starts and spiking again at full pollution,} but here the drop is again caused by the approach's sensibility to outliers created in the high-dimensional space.
\PaperLong{However, unlike Gaussian mixture, it shows a near-constant performance while it does not produce usable results on the polluted datasets, which is in line with our expectations.} 
Consequently, both the Gaussian mixture and the agglomerative algorithms cannot meaningfully interpret datasets that have mainly binary features and has low feature accuracy.
OPTICS behaves similarly on \textsf{Covertype} and \textsf{Letter}.
\PaperLong{In contrast to the explanation of OPTICS' behavior with \textsf{Letter}, here we assume for \textsf{Covertype} that OPTICS forms small clusters of very high density in a few dimensions, causing the majority of samples to become outliers~(see Figure~\ref{fig:clustering-feature-accuracy-covertype-nclusters} in Appendix).} 
\revision{OPTICS on \textsf{COVID} demonstrates robust performance, showing a stable AMI score regardless of the level of pollution.}
\revision{For all considered datasets t}he autoencoder and $k$-Prototypes show an almost linear decrease in AMI score, though this decrease is not as rapid as the rate of quality degradation\PaperLong{for \textsf{Letter} when its feature accuracy is polluted.

In Figure~\ref{fig:clustering-feature-accuracy-bank-results} (or focused Figure~\ref{fig:stretched-bank-results-feature-accuracy} in Appendix), we can see the effect of the feature accuracy polluter on \textsf{Bank}. 
The performance increases for the OPTICS and the autoencoder approaches in the first and second step of the pollution are striking. 
Looking more closely at the number of clusters OPTICS finds at a quality of~$1.0$~(see Figure~\ref{fig:clustering-feature-accuracy-bank-nclusters} in Appendix), we see that OPTICS found two clusters in the original dataset and~19 clusters after the first pollution step. 
Therefore, we assume that OPTICS profits from very dense clusters, created by the high number of duplicates in this dataset, being broken up in this first step of the pollution. 
This is due to the added noise and the higher variance of the data introduced by removing duplicates through pollution of some of their values.
With the autoencoder, on the other hand, you can see that it achieves an increasing AMI score as long as not all the searched six clusters have been found. 
From the point where the autoencoder has identified the desired six clusters, its performance decreases similarly to that of the $k$-means algorithm.
This descent is probably due to producing false clusters or more outliers. 
Like the $k$-means algorithm just mentioned, the agglomerative clustering and Gaussian mixture algorithms also experience a consistent degradation of their results when the feature accuracy is increasingly lowered. 
This decrease of the performance is less rapid for \textsf{Bank} than for \textsf{Letter}}.

\stitle{Target Accuracy}
\PaperLong{\input{Latex_Figure/clustering/Uniqueness}}
The pollution of the target accuracy should have no influence on any clustering algorithm, as they are unsupervised: a changed target class does not affect the clustering itself. 
However, the fact that we can still see a degrading curve in Figures\PaperLong{~\ref{fig:clustering-target-accuracy-letter-results}, \ref{fig:clustering-target-accuracy-covertype-results}, \ref{fig:clustering-target-accuracy-covid-results}, and }\ref{fig:clustering-target-accuracy-bank-results}\revision{\PaperShort{(and also for the other considered datasets)}} is due to us comparing the clustering result with the polluted target classes and not with the original target classes when calculating \revision{the} AMI \revision{score}.

\PaperLong{An interesting observation is the very slight but always existing increase in each algorithms' performance for \textsf{Letter} at very low dataset qualities. 
We assume that a high pollution level changes enough target labels that, by chance, enough of the same labels are again changed to the same label, recreating parts of their original clusters. 
This effect is more noticeable with datasets that have fewer classes~(i.e., \textsf{COVID}, \textsf{Bank} and \textsf{Covertype}) in their ground truth dataset, which strengthens our assumption.
}

\stitle{Uniqueness}
We used the polluter that inserts duplicates based on a normal distribution.
The Gaussian mixture, $k$-means/$k$-prototypes and agglomerative clustering algorithms are not affected by the pollution on the considered datasets as shown in Figures\PaperLong{\ref{fig:clustering-uniqueness-results}} \PaperShort{\ref{fig:clustering-uniqueness-covertype-results}}.
\revision{On \textsf{COVID}, Gaussian Mixture shows a slight performance increase of up to 5\pt after pollution\PaperLong{(see Figure~\ref{fig:clustering-uniqueness-covid-results})}}.
\revision{For \textsf{Covertype} and \textsf{Letter}}, the increase in duplicates helps the OPTICS approach to improve its performance: as a density-based clustering algorithm, it can benefit from somewhat evenly inserted duplicates as they increase the overall density of clusters and especially their cores without creating very dense cores consisting of only the same data point many times. 
The autoencoder approach also enhances its performance above a certain number of duplicates for \textsf{Letter}.\PaperLong{We assume that it is easier for the autoencoder to learn the encoding if there are more duplicates in \textsf{Letter}.} 
In contrast, we cannot see a clear behavior of the autoencoder on \textsf{Covertype}, which is probably due to\PaperLong{the additional time it would have needed to converge as well as} the generally random and sample-dependent nature of neural network training.
\revision{Also, for the \textsf{COVID} dataset, the autoencoder is robust against the inserted duplicates in its performance\PaperLong{(see Figure~\ref{fig:clustering-uniqueness-covid-results}}).}

\stitle{Target Class Balance}
We observe a difference in behavior between \textsf{Letter} with numeric-only features and high-class count and \textsf{Bank}, \textsf{Covertype} and \textsf{COVID} with fewer classes and features.
In \textsf{Letter}\PaperLong{, shown in Figure~\ref{fig:clustering-class-balance-letter-results}}, the imbalance in cluster sizes actually slightly improved the AMI score.
\PaperLong{This behavior is approximately linear across all algorithms except for the autoencoder-based clustering, which continues to improve over time but exhibits irregular behavior. 
We attribute this irregularity to the neural network training step, which is highly sensitive to the data it provides, which is changed as the class balance changes.
We propose two potential reasons for the slight improvement in clustering quality as the target class balance becomes more imbalanced.} 
For\PaperShort{example, for} OPTICS, we found that the number of clusters being identified on average increased as the dataset grew more imbalanced\PaperLong{(see Figure~\ref{fig:clustering-class-balance-letter-nclusters} in Appendix), likely since clusters became denser for those classes which grew due to the imbalancing process. 
This is also the reason behind the improvement of other algorithms' performance for \textsf{Letter}. 
As algorithms were never able to correctly identify all clusters, as evident by the AMI score being far from~$1.0$, the loss of information for some clusters may have been outweighed by the gain of information in other areas of the dataset, allowing algorithms to pick up on some clusters so much better, that it improved the overall clustering quality score. 
Additionally, we have to consider that smaller clusters also have less of an impact on the overall clustering quality, so their loss may not be reflected heavily in the overall adjusted mutual information score.}

\PaperLong{\input{Latex_Figure/clustering/Class_Balance}}
For \textsf{Bank}, \textsf{Covertype} and \textsf{COVID}~(see Figure\PaperLong{s~\ref{fig:clustering-class-balance-bank-results}, ~\ref{fig:clustering-class-balance-covid-results}, and} \ref{fig:clustering-class-balance-covertype-results})\PaperShort{, }\PaperLong{, we split the results analysis into three parts. 
First, we look at the $k$-Means/$k$-Prototypes and agglomerative clustering algorithms. T}\PaperShort{t}he $k$-means/$k$-prototypes algorithms' AMI declines slightly and linearly, which can also be observed for the agglomerative clustering on \textsf{Covertype}.
This decline is caused by the inability of these algorithms to identify the shrunken clusters reliably\PaperLong{, while they are also not majorly improving their recognition of the growing clusters}.\PaperLong{This may be the case in these datasets but not in \textsf{Letter}, as \textsf{Covertype}, \textsf{COVID} and \textsf{Bank} are harder to handle for the clustering algorithms in general, resulting in less improvement for the growing clusters. 
Additionally, there are fewer clusters overall to balance out or overtake the loss of quality coming with the difficulty to identify smaller clusters~(see Figure~\ref{fig:clustering-all-nclusters}).}
The irregular behavior of the agglomerative clustering on \textsf{Bank} hints at this algorithm's instability on this particular dataset.
\PaperLong{We can see similar behavior when comparing it to the clustering results achieved when polluting the uniqueness quality dimension~(see Figure~\ref{fig:clustering-uniqueness-bank-results}), which strengthens our assumption of the agglomerative clustering algorithm's issues with this relatively small dataset with many repeating entries.

Secondly, we analyze the results of the OPTICS and autoencoder clustering methods. 
Here, we can see very differing behavior between the three datasets, with OPTICS and autoencoder-based clustering performance being relatively constant in \textsf{Bank}~(the autoencoder has a score of~0 here), while they are experiencing drastic changes in quality for \textsf{Covertype} and \textsf{COVID}.
However, we still group these because the behavior of these algorithms matches the general behavior of the average number of clusters they were able to identify at each stage of pollution~(see Figures~\ref{fig:clustering-all-nclusters}). 
We can observe that the AMI score decreases if the number of clusters identified by the autoencoder or OPTICS algorithm decreases, with the AMI score being zero if only one cluster is identified. 
If the average number of clusters identified stays relatively constant, so does the algorithm performance, except a spike at a quality of about~$0.35$ in \textsf{Covertype} for the autoencoder clustering approach.
The two algorithms can identify any number of clusters; however, the closer they get to identifying the actual number of clusters, the more likely they are to also truly identify groupings of data points in the dataset, even if these are only the cores of existing clusters. 
Just as hypothesized for the OPTICS algorithm on \textsf{Letter}, we believe that the change in the number of identified clusters identified by these algorithms is rooted in the more or less dense regions in the polluted dataset compared to the original data.}

OPTICS and autoencoder show a relatively constant behavior for \textsf{Bank}, \textsf{COVID}, and \textsf{Covertype}.
The Gaussian mixture clustering shows different behavior in \textsf{Bank}, \textsf{COVID} and \textsf{Covertype}; however, in each case, it does not behave clearly due to the assumption of a specific distribution\PaperLong{, as mentioned at the beginning of Section~\ref{subsec:clustering-results}}. 
Therefore, we believe that altering the samples in the dataset provided to the Gaussian mixture clustering can have positive or negative effects, depending on whether this new sampling more or less resembles the assumed distribution of data.

%% file: Latex_Figure/clustering/summary_convtype.tex
\begin{figure*}[!htbp]
    \centering
    \begin{subfigure}[b]{0.32\textwidth}
        \includegraphics[width=\textwidth]{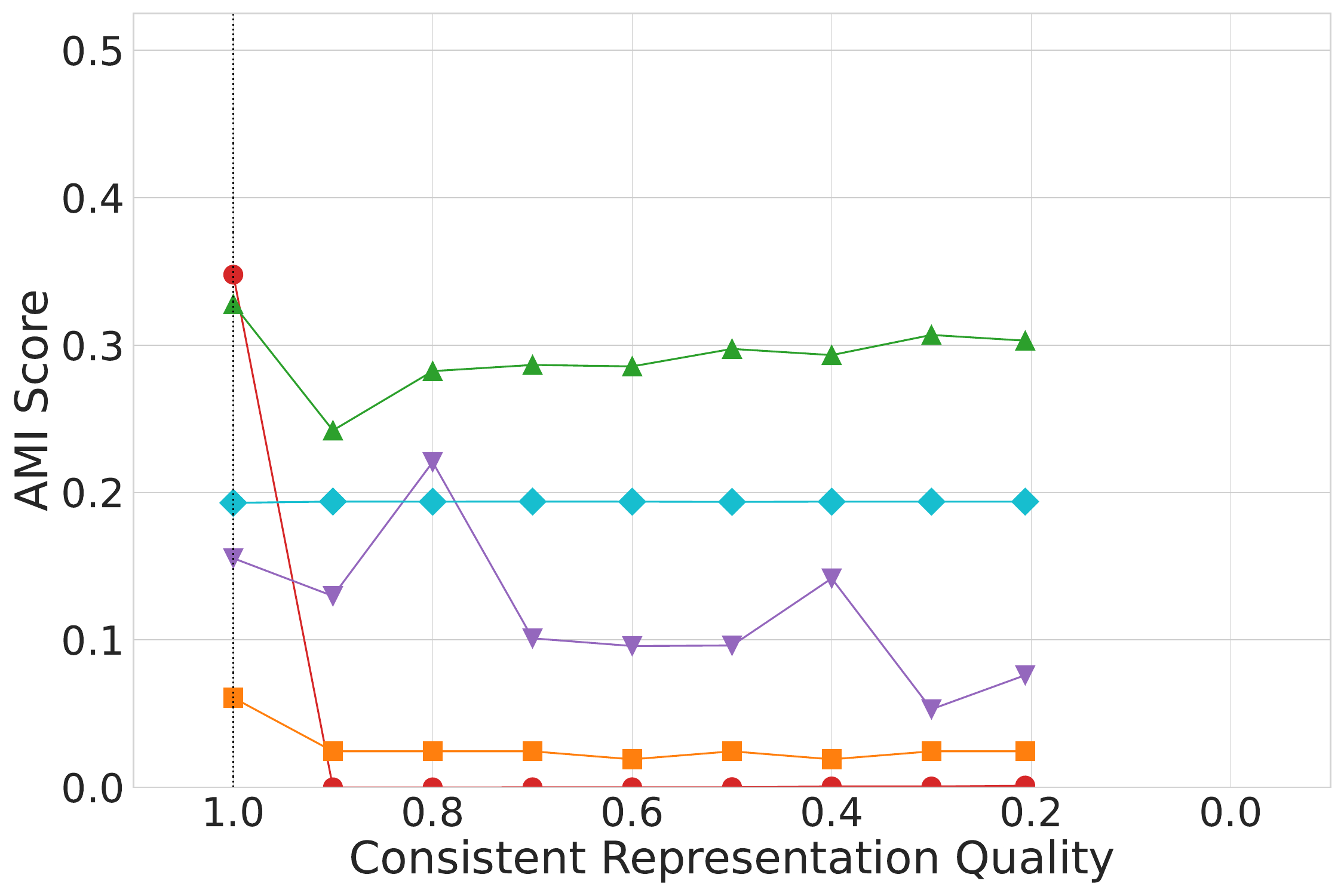}
        \caption{Consistency with $k_{v}=5$}
        \label{fig:clustering-consistent-repr-covertype-results}
    \end{subfigure}
    \begin{subfigure}[b]{0.32\textwidth}
        \includegraphics[width=\textwidth]{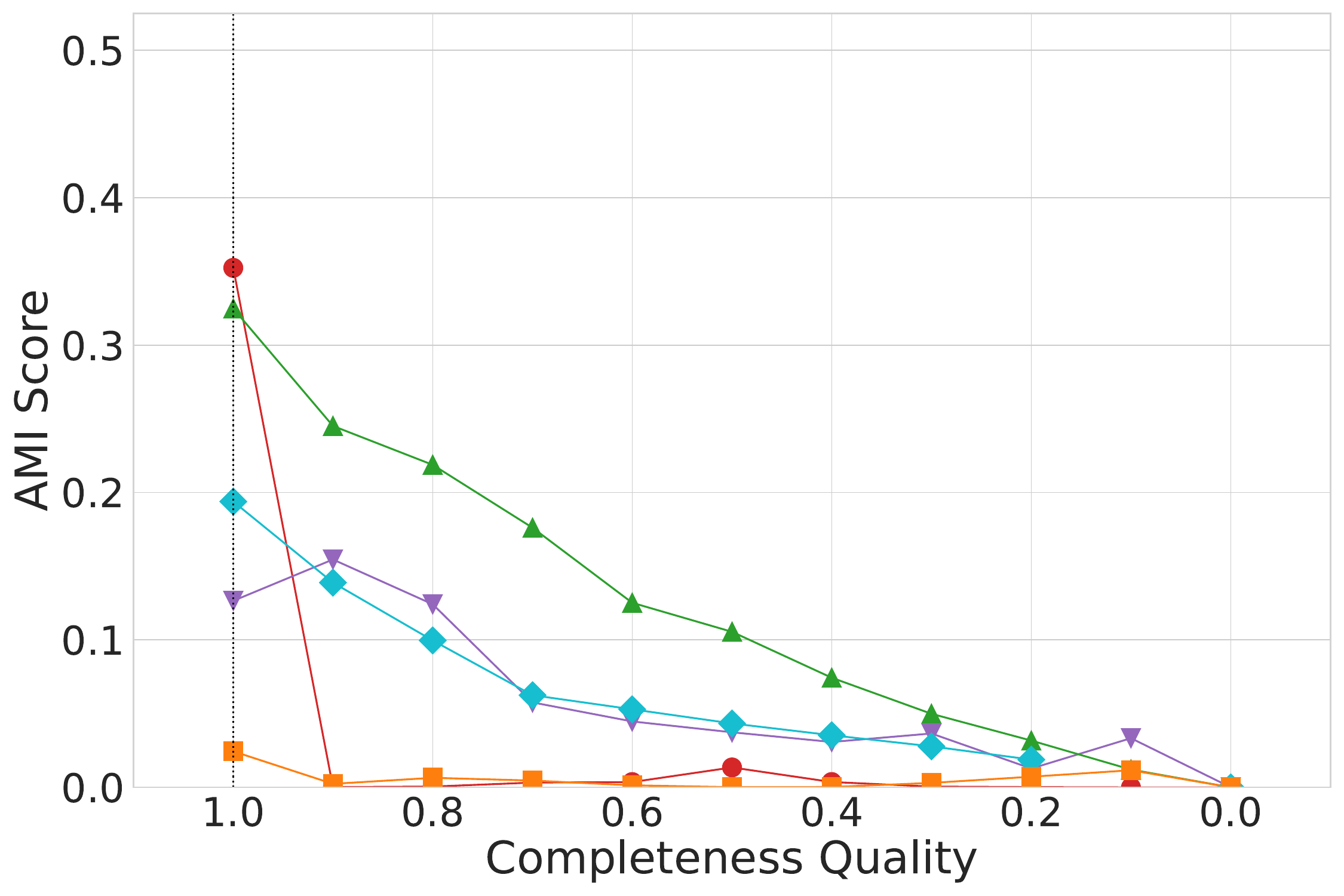}
        \caption{Completeness}
        \label{fig:clustering-completeness-covertype-results}
    \end{subfigure}
\begin{subfigure}[b]{0.32\textwidth}
        \includegraphics[width=\textwidth]{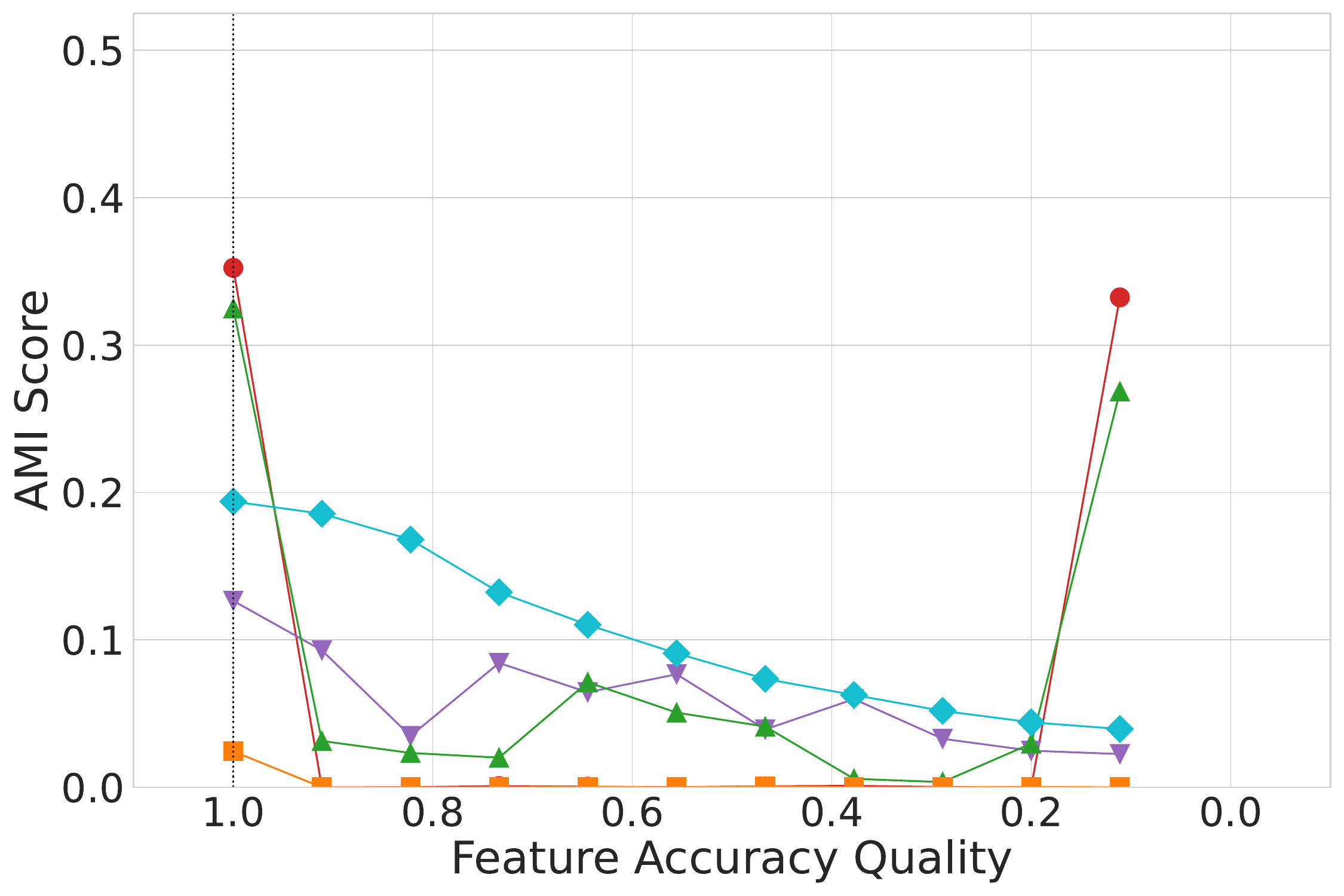}
        \caption{Feature Accuracy}
        \label{fig:clustering-feature-accuracy-covertype-results}
\end{subfigure}

    \begin{subfigure}[b]{0.32\textwidth}
        \includegraphics[width=\textwidth]{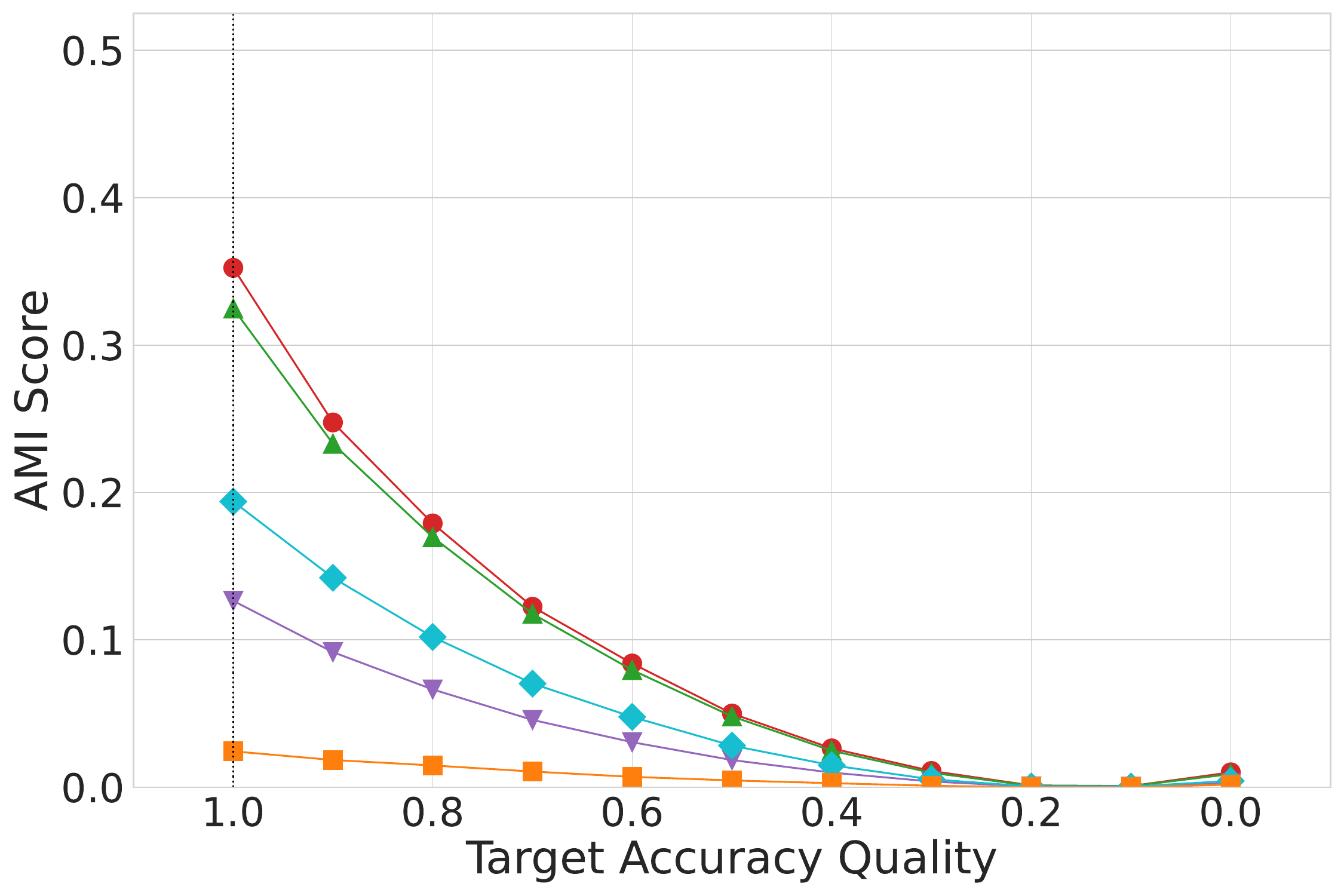}
        \caption{Target Accuracy}
        \label{fig:clustering-target-accuracy-covertype-results}
    \end{subfigure}
    \begin{subfigure}[b]{0.32\textwidth}
        \includegraphics[width=\textwidth]{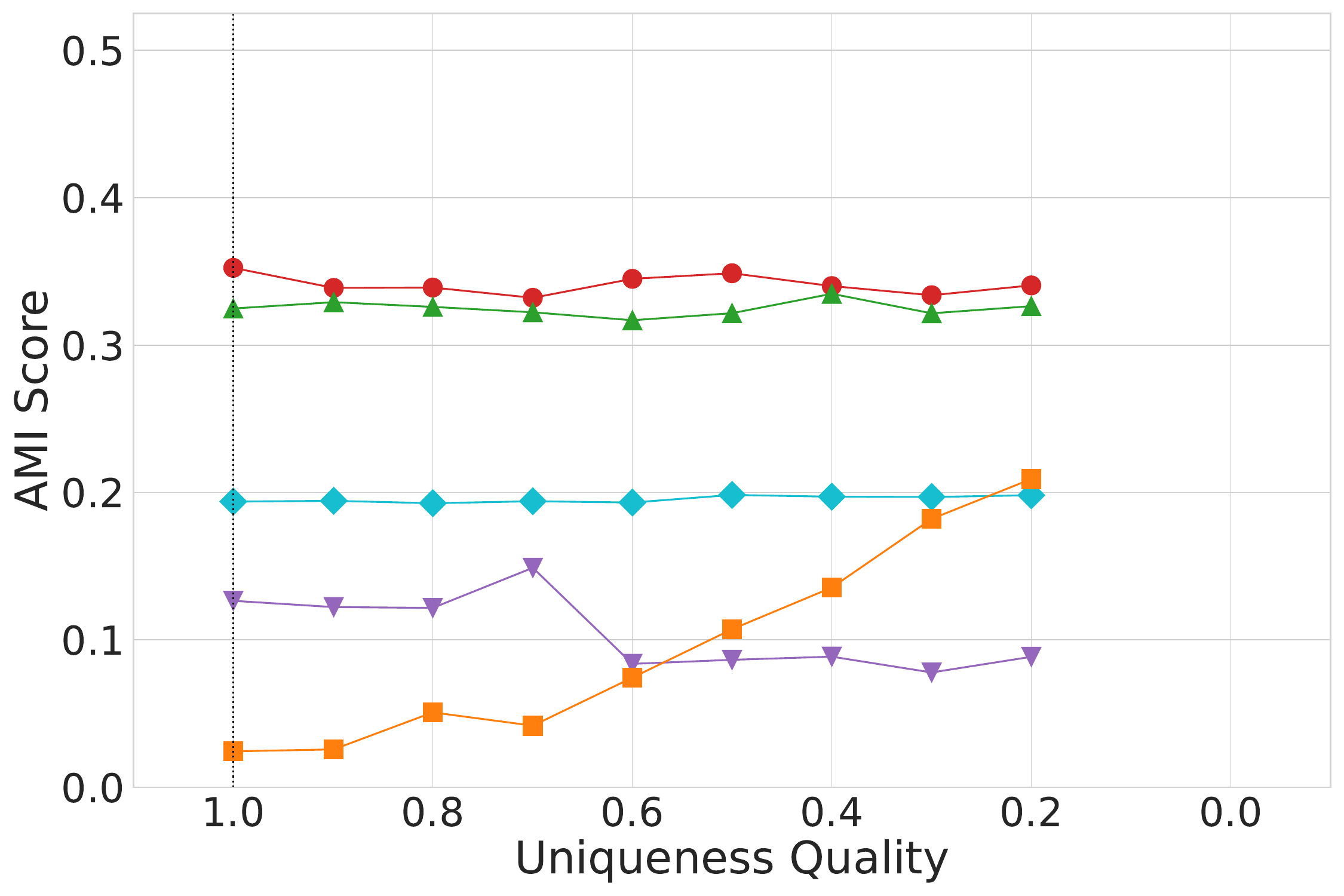}
        \caption{Uniqueness dataset 
        }
        \label{fig:clustering-uniqueness-covertype-results}
    \end{subfigure}
    \begin{subfigure}[b]{0.32\textwidth}
        \includegraphics[width=\textwidth]{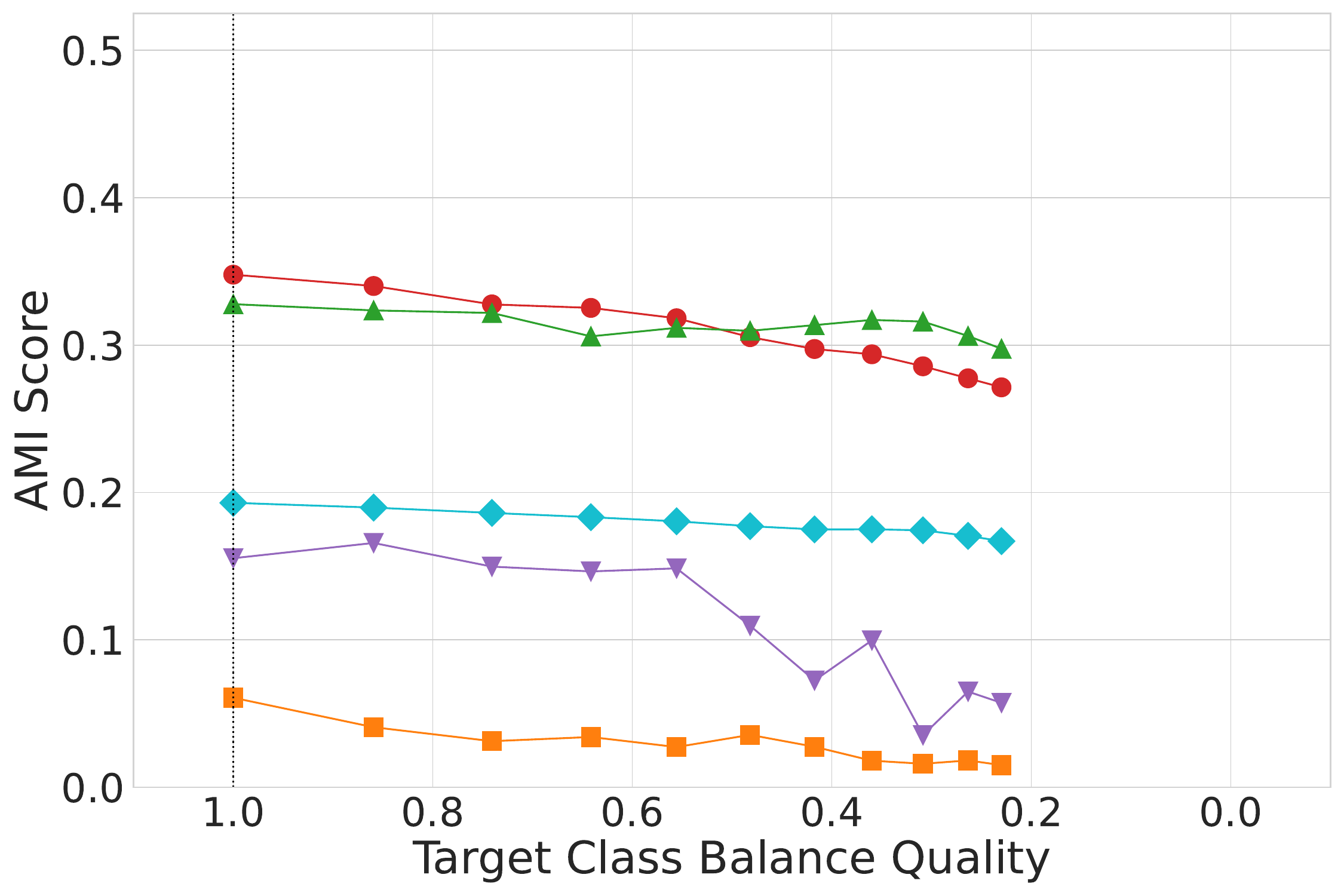}
        \caption{Class Balance}
        \label{fig:clustering-class-balance-covertype-results}
    \end{subfigure}    
 \caption{AMI score of the clustering algorithms for \textsf{Covertype}.}
 \label{fig:clustering-results-all-covertype}
\end{figure*}

%% file: Latex_Figure/clustering/Consistent_Representation.tex
\begin{figure*}[ht]
    \centering
\begin{subfigure}[b]{0.3\linewidth}
        \includegraphics[width=\linewidth]{figures/clustering/covtype/ConsistentRepresentation/adj_mut_info.pdf}
        \caption{Covertype}
        \label{fig:clustering-consistent-repr-covertype-results}
    \end{subfigure}
\begin{subfigure}[b]{0.3\linewidth}
        \includegraphics[width=\linewidth]{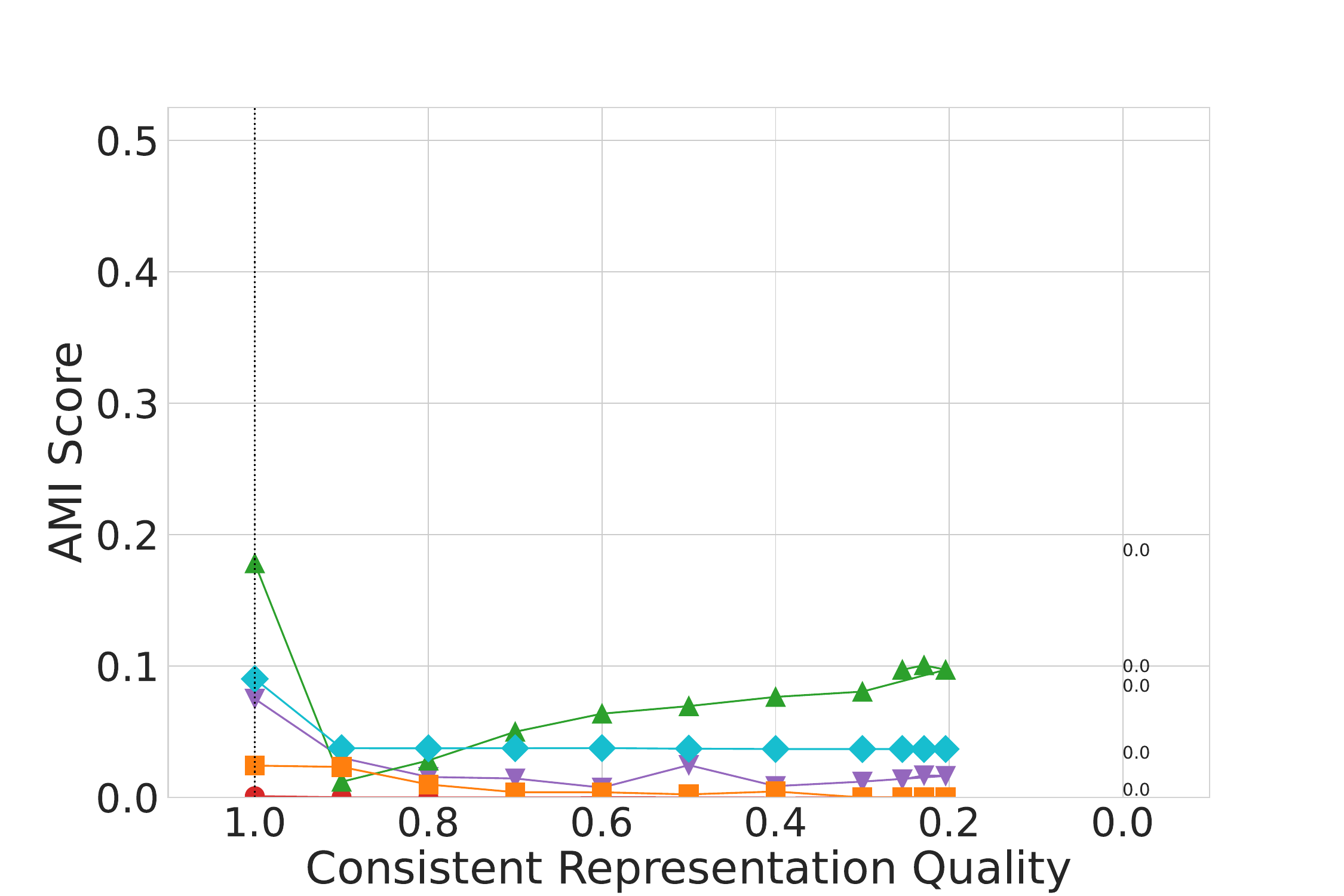}
        \caption{COVID}
        \label{fig:clustering-consistent-repr-covid-results}
    \end{subfigure}
    \begin{subfigure}[b]{0.3\linewidth}
        \includegraphics[width=\linewidth]{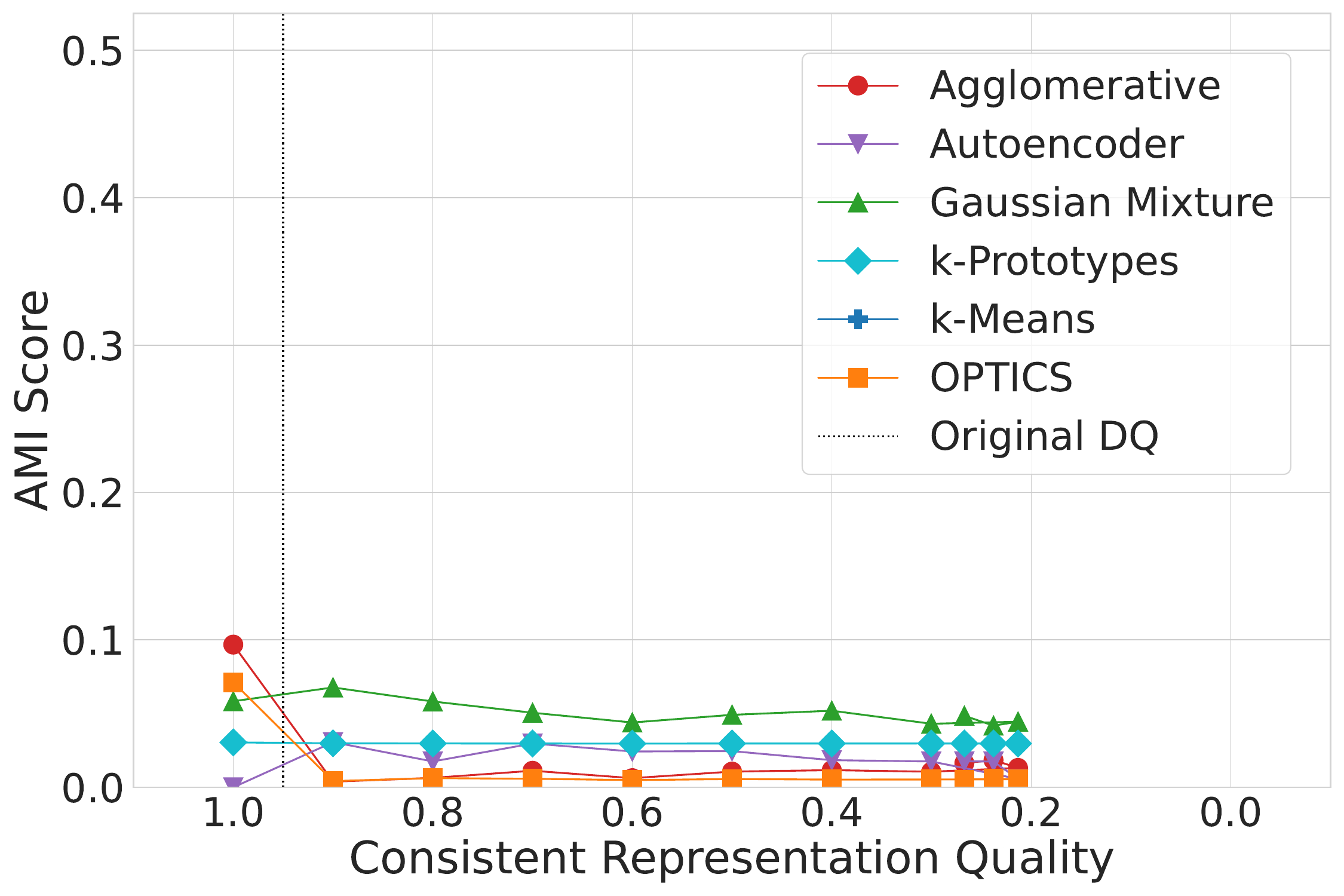}
        \caption{Bank}
        \label{fig:clustering-consistent-repr-bank-results}
    \end{subfigure}
    \caption{AMI score for consistent representation with $k_{v}=5$ dimension and clustering algorithms.}
    \label{fig:clustering-consistent-repr-results}
\end{figure*}

%% file: Latex_Figure/clustering/Completeness.tex
\begin{figure*}[htbp]
    \centering
\begin{subfigure}[b]{0.4\linewidth}
        \includegraphics[width=\linewidth]{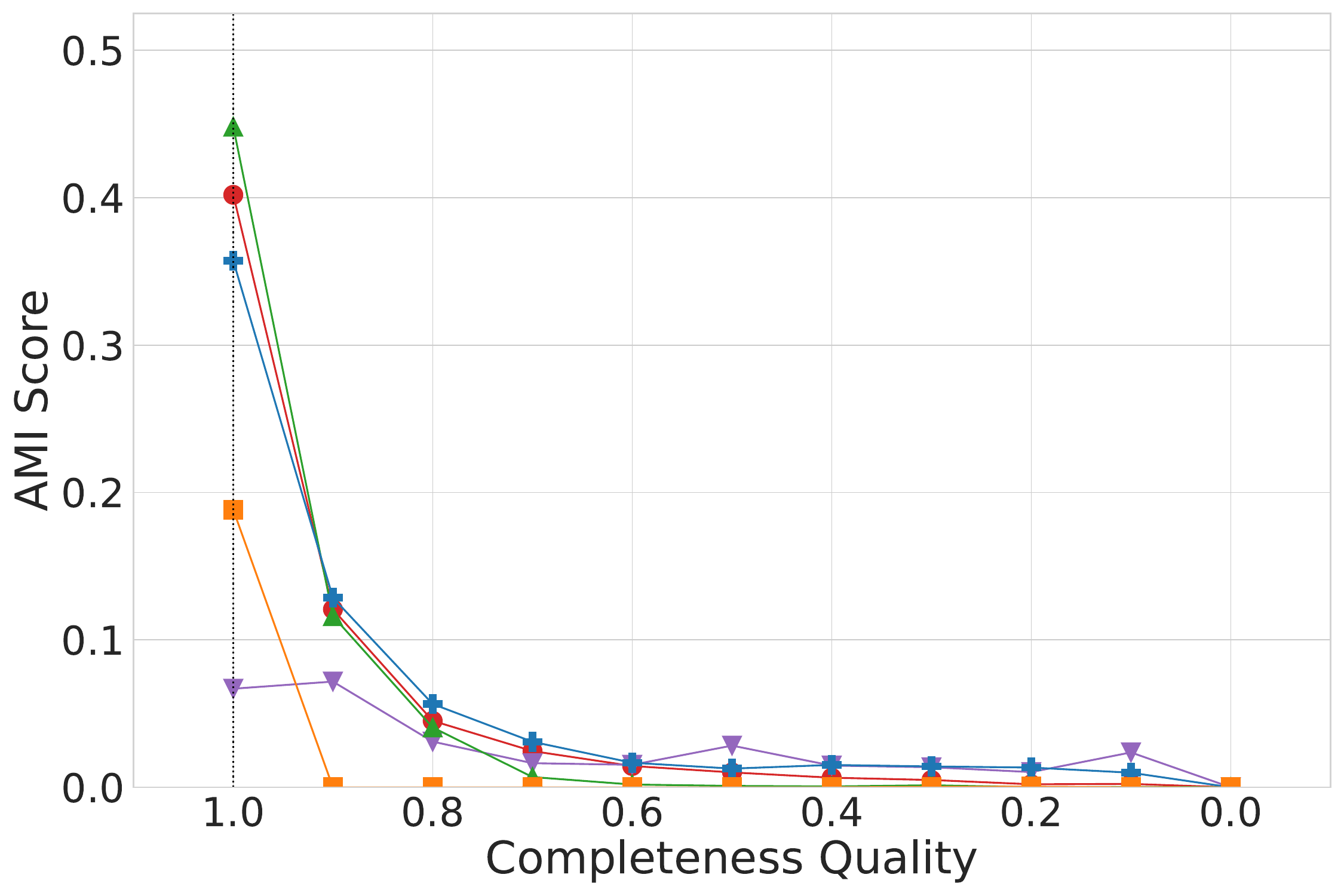}
        \caption{\textsf{Letter}}
        \label{fig:clustering-completeness-letter-results}
    \end{subfigure}
\begin{subfigure}[b]{0.4\linewidth}
        \includegraphics[width=\linewidth]{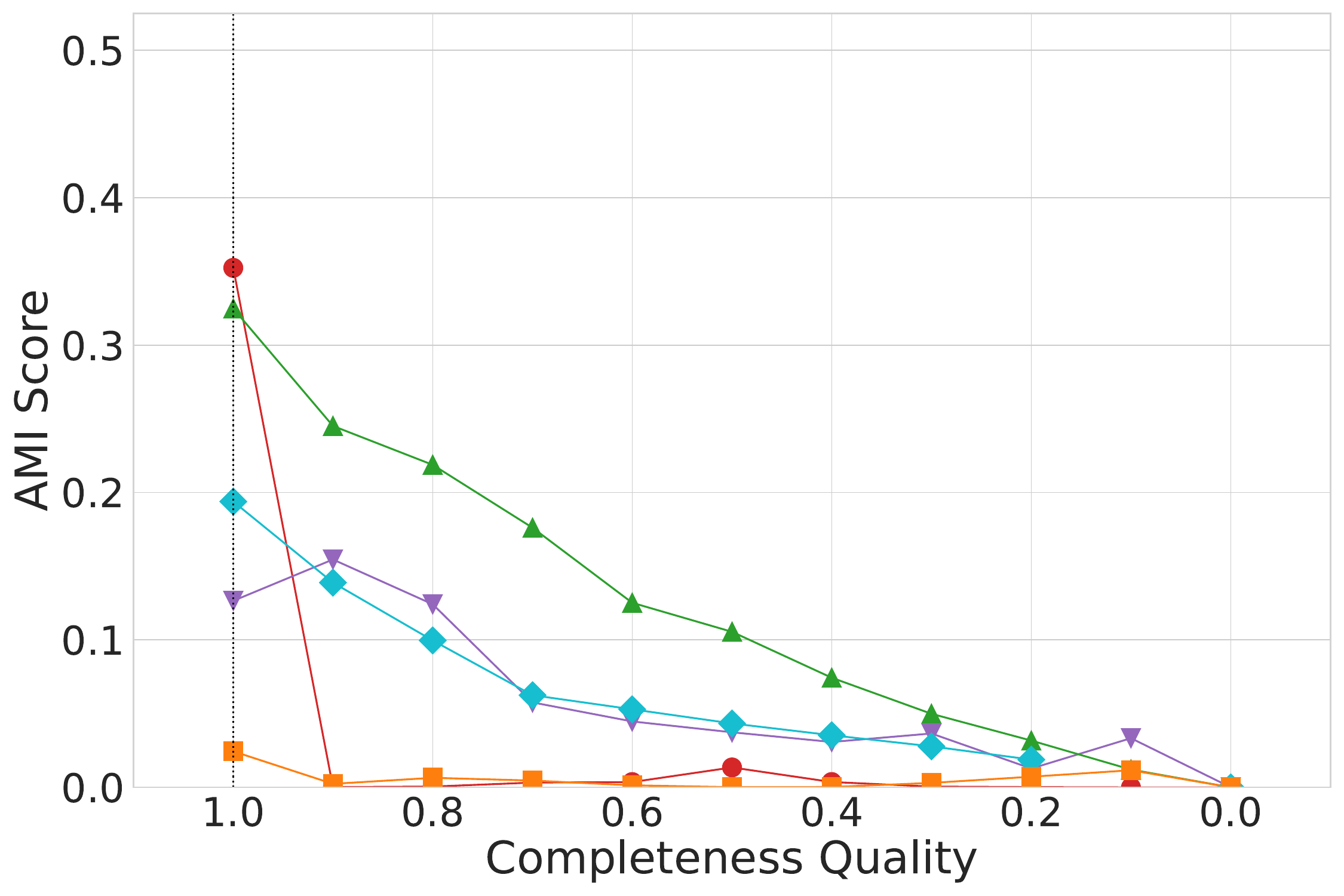}
        \caption{\textsf{Covertype}}
        \label{fig:clustering-completeness-covertype-results}
    \end{subfigure}
    
\begin{subfigure}[b]{0.4\linewidth}
        \includegraphics[width=\linewidth]{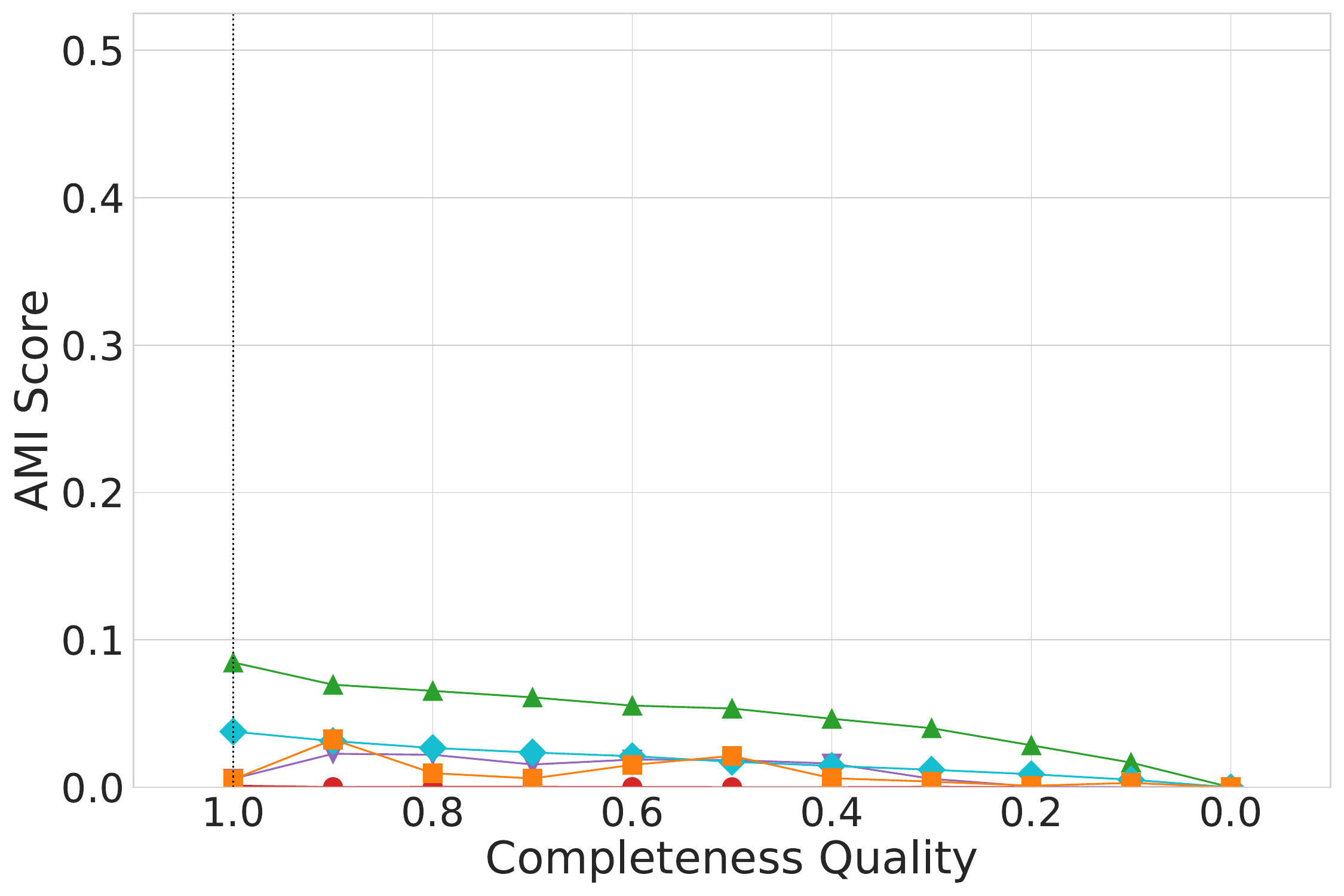}
        \caption{\textsf{COVID}}
        \label{fig:clustering-completeness-covid-results}
    \end{subfigure}
\begin{subfigure}[b]{0.4\linewidth}
        \includegraphics[width=\linewidth]{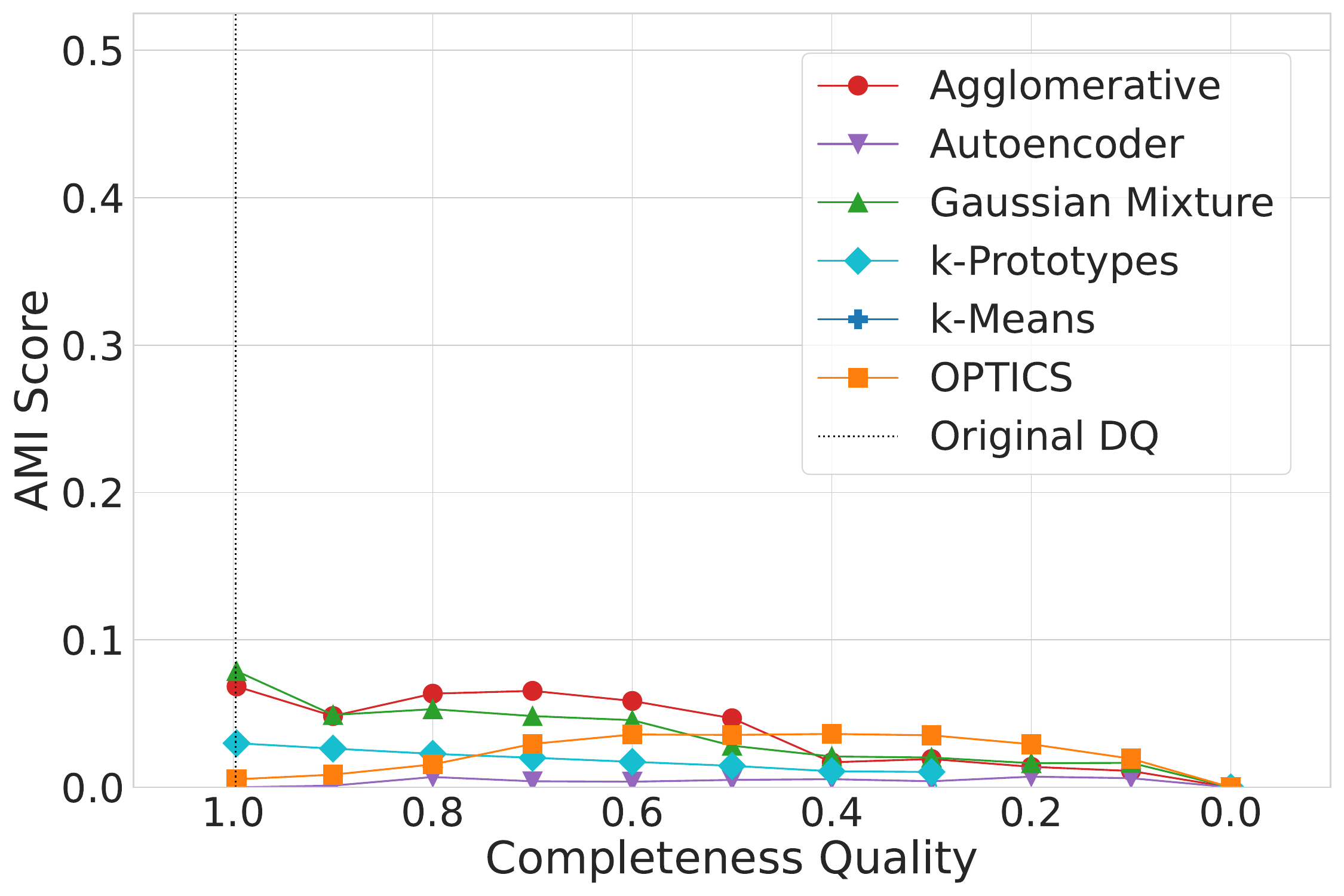}
        \caption{\textsf{Bank}}
        \label{fig:clustering-completeness-bank-results}
    \end{subfigure}
    \caption{AMI score for completeness dimension and clustering algorithms.}
    \label{fig:clustering-completeness-results}
\end{figure*}

%% file: Latex_Figure/clustering/Feature_Accurecy.tex
\begin{figure*}[htbp]
    \centering
\begin{subfigure}[b]{.4\linewidth}
        \includegraphics[width=\linewidth]{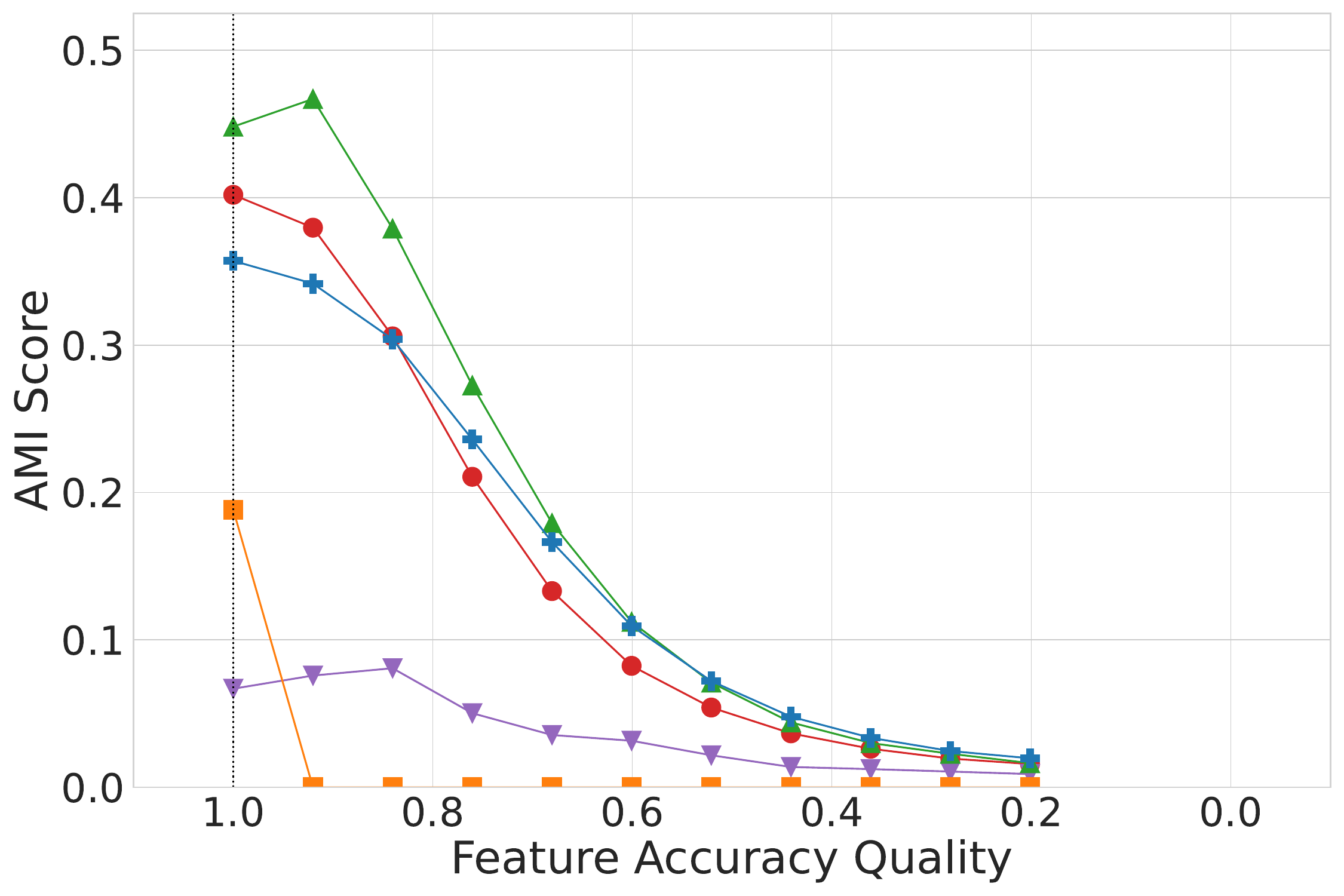}
        \caption{\textsf{Letter}}
        \label{fig:clustering-feature-accuracy-letter-results}
    \end{subfigure}
\begin{subfigure}[b]{.4\linewidth}
        \includegraphics[width=\linewidth]{figures/clustering/covtype/FeatureAccuracy/adj_mut_info.pdf}
        \caption{\textsf{Covertype}}
        \label{fig:clustering-feature-accuracy-covertype-results}
    \end{subfigure}
    
\begin{subfigure}[b]{.4\linewidth}
        \includegraphics[width=\linewidth]{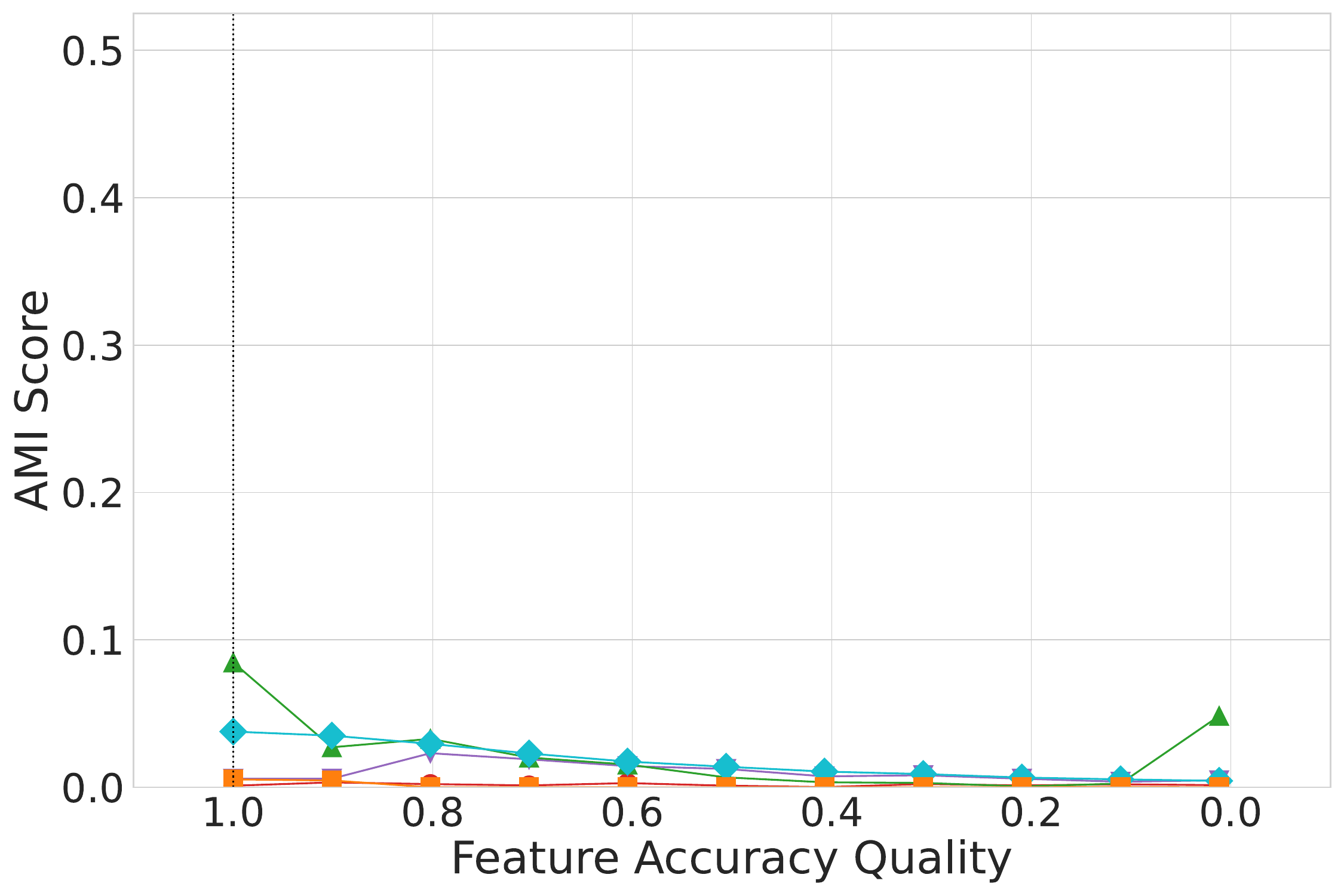}
        \caption{\textsf{COVID}}
        \label{fig:clustering-feature-accuracy-covid-results}
    \end{subfigure}
\begin{subfigure}[b]{.4\linewidth}
        \includegraphics[width=\linewidth]{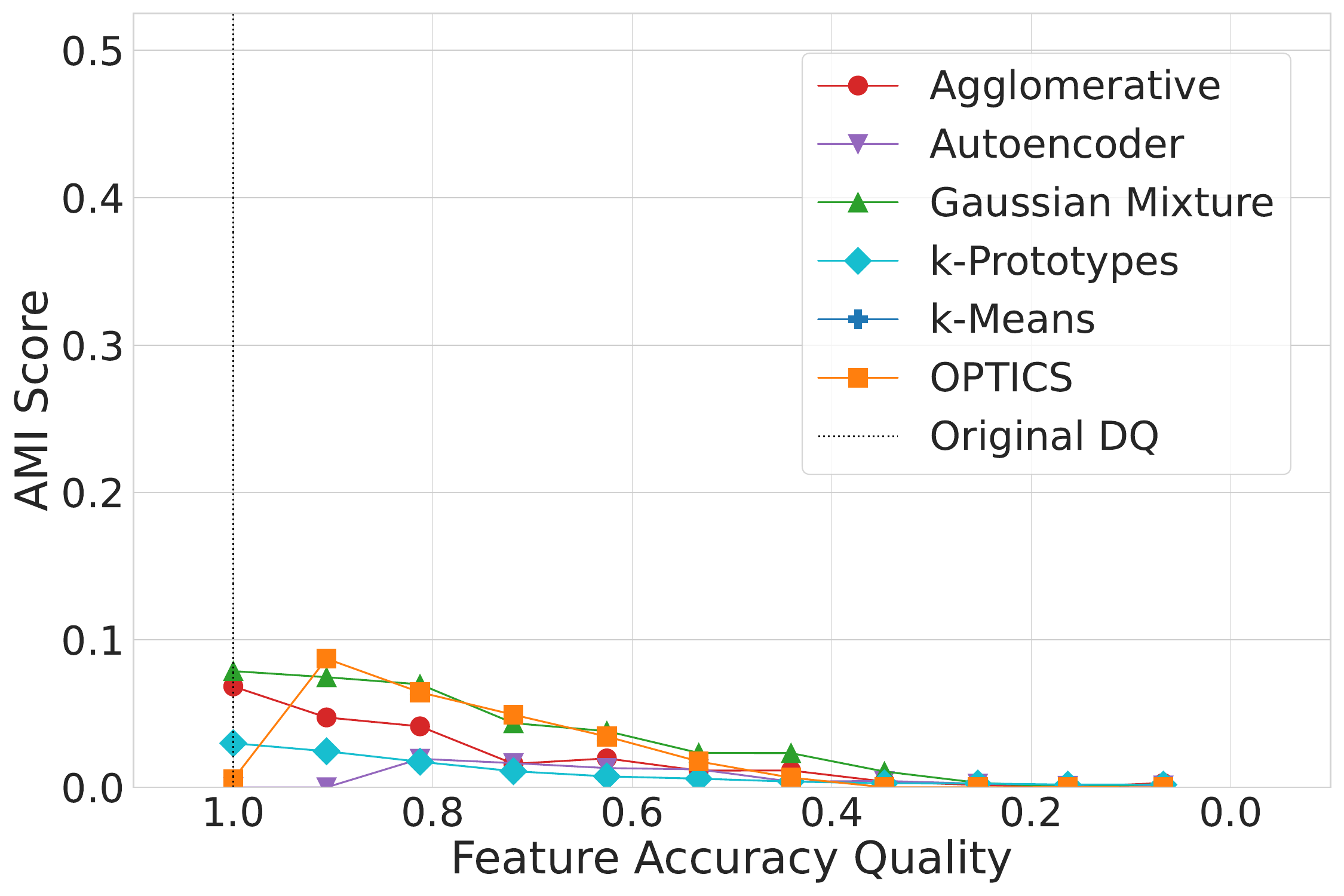}
        \caption{\textsf{Bank}}
        \label{fig:clustering-feature-accuracy-bank-results}
    \end{subfigure}
    \caption{AMI score for feature accuracy dimension and clustering algorithms.}
    \label{fig:clustering-feature-accuracy-results}
\end{figure*}

%% file: Latex_Figure/clustering/Target_Accurecy.tex
\begin{figure*}[htbp]
    \centering
\begin{subfigure}[b]{.4\linewidth}
        \includegraphics[width=\linewidth]{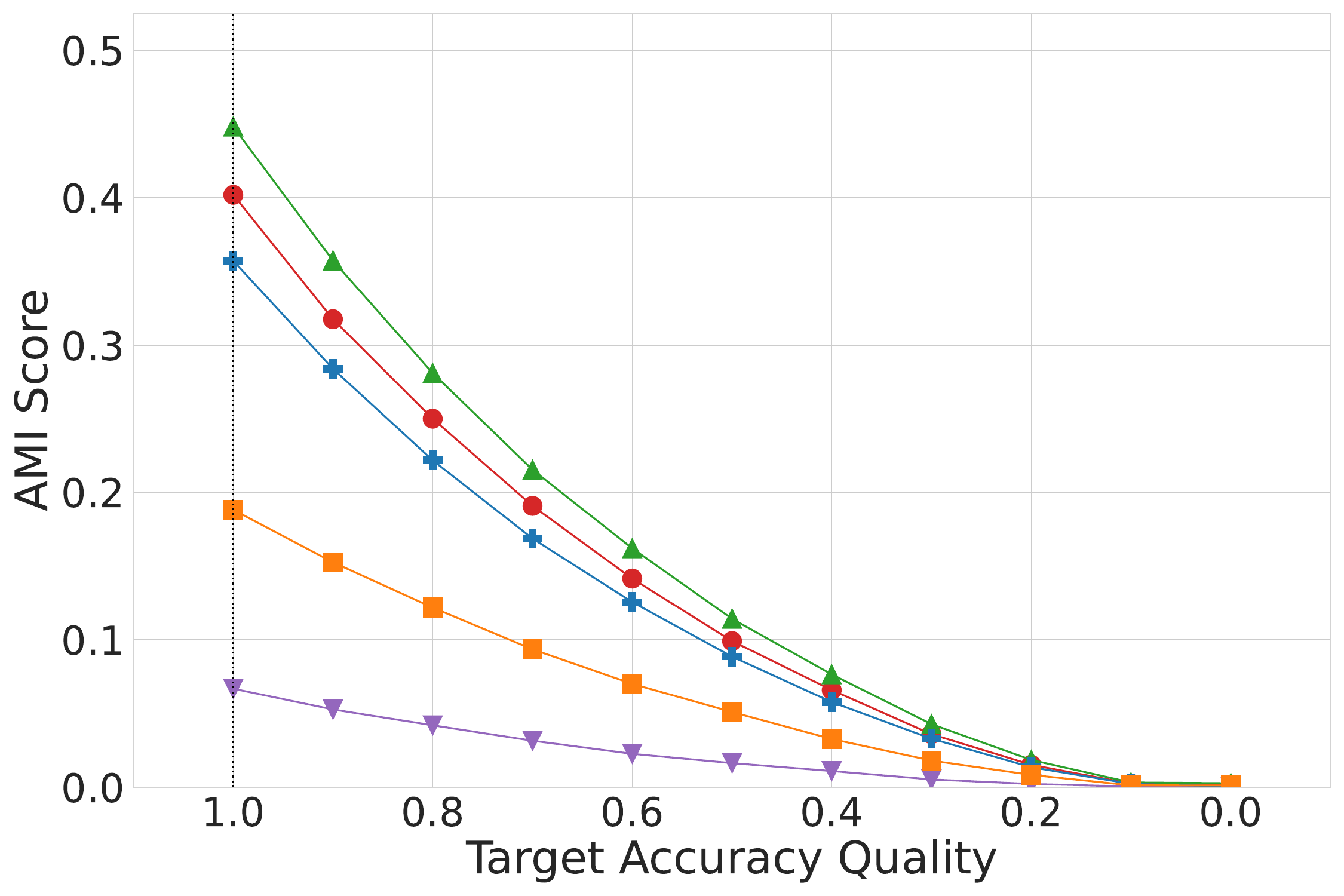}
        \caption{\textsf{Letter}}
        \label{fig:clustering-target-accuracy-letter-results}
    \end{subfigure}
\begin{subfigure}[b]{.4\linewidth}
        \includegraphics[width=\linewidth]{figures/clustering/covtype/TargetAccuracy/adj_mut_info.pdf}
        \caption{\textsf{Covertype}}
        \label{fig:clustering-target-accuracy-covertype-results}
    \end{subfigure}
    
\begin{subfigure}[b]{.4\linewidth}
        \includegraphics[width=\linewidth]{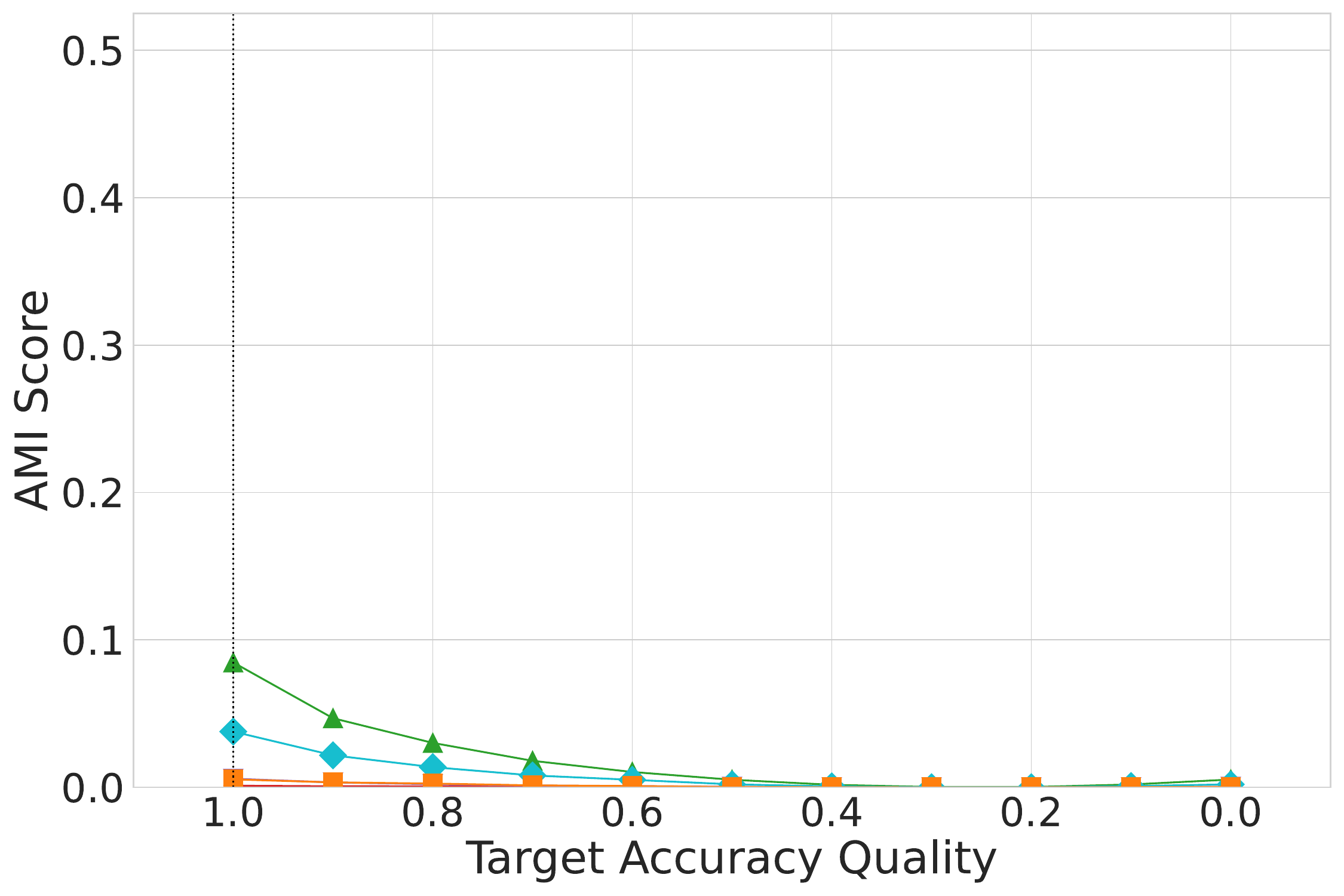}
        \caption{\textsf{COVID}}
        \label{fig:clustering-target-accuracy-covid-results}
    \end{subfigure}
\begin{subfigure}[b]{.4\linewidth}
        \includegraphics[width=\linewidth]{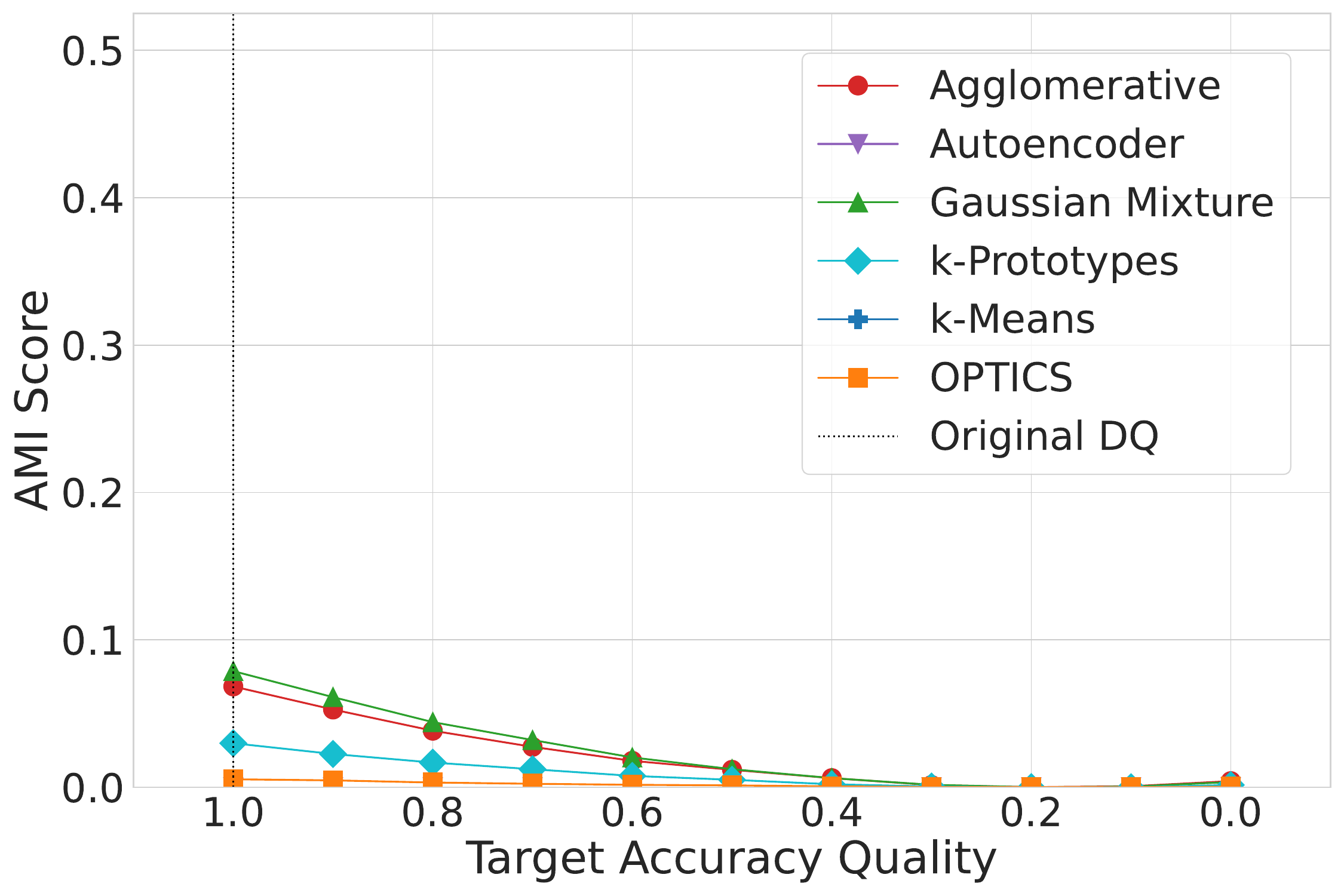}
        \caption{\textsf{Bank}}
        \label{fig:clustering-target-accuracy-bank-results}
    \end{subfigure}
    \caption{AMI score for target accuracy dimension and clustering algorithms.}
    \label{fig:clustering-target-accuracy-results}
\end{figure*}

%% file: Latex_Figure/clustering/Uniqueness.tex
\begin{figure*}[!htbp]
    \centering
\begin{subfigure}[b]{.4\linewidth}
        \includegraphics[width=\linewidth]{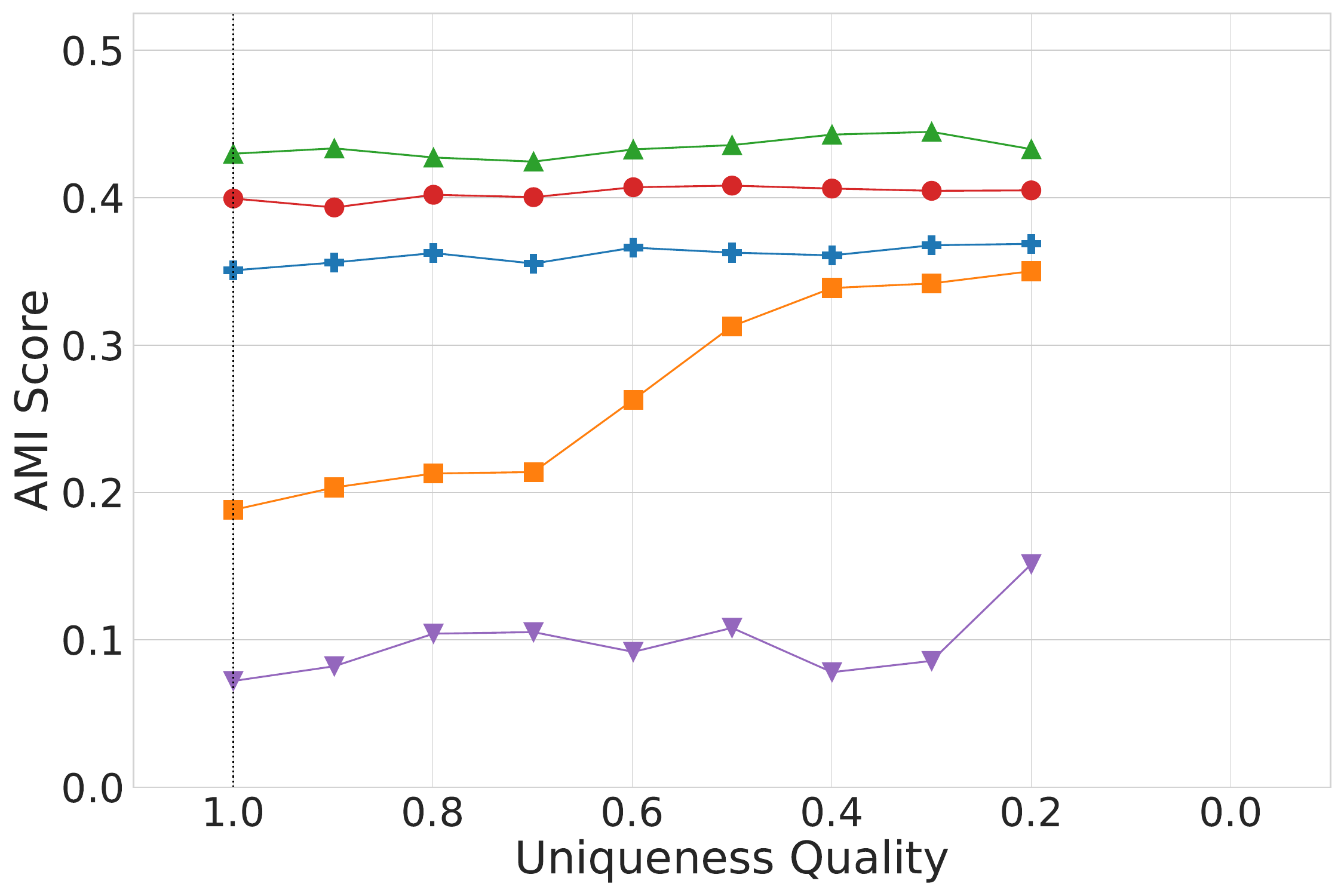}
        \caption{\textsf{Letter}}
        \label{fig:clustering-uniqueness-letter-results}
    \end{subfigure}
\begin{subfigure}[b]{.4\linewidth}
        \includegraphics[width=\linewidth]{figures/clustering/covtype/Uniqueness/adj_mut_info.pdf}
        \caption{\textsf{Covertype}}
        \label{fig:clustering-uniqueness-covertype-results}
    \end{subfigure}
    
\begin{subfigure}[b]{0.4\linewidth}
        \includegraphics[width=\linewidth]{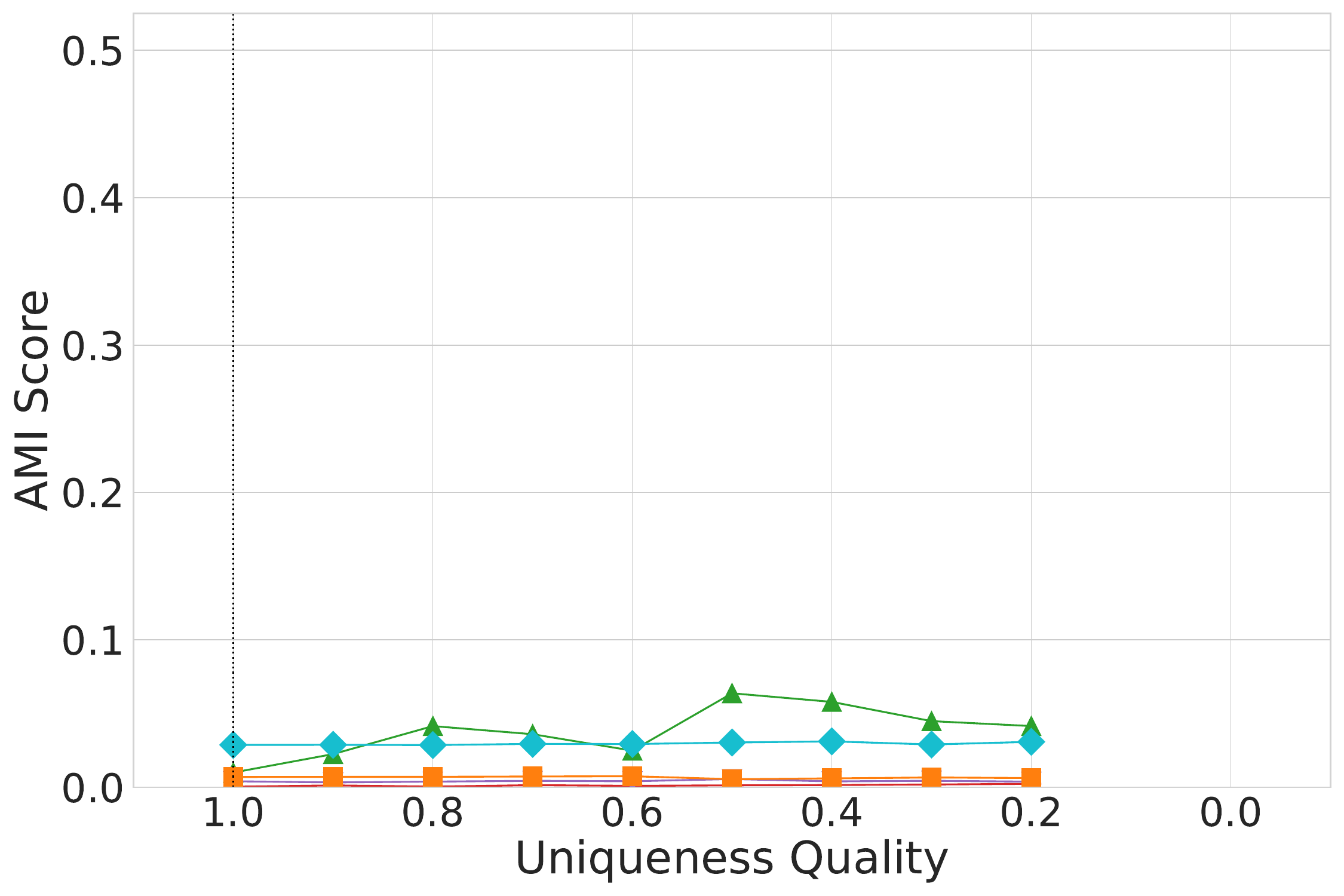}
        \caption{\textsf{COVID}}
        \label{fig:clustering-uniqueness-covid-results}
    \end{subfigure}
\begin{subfigure}[b]{0.4\linewidth}
        \includegraphics[width=\linewidth]{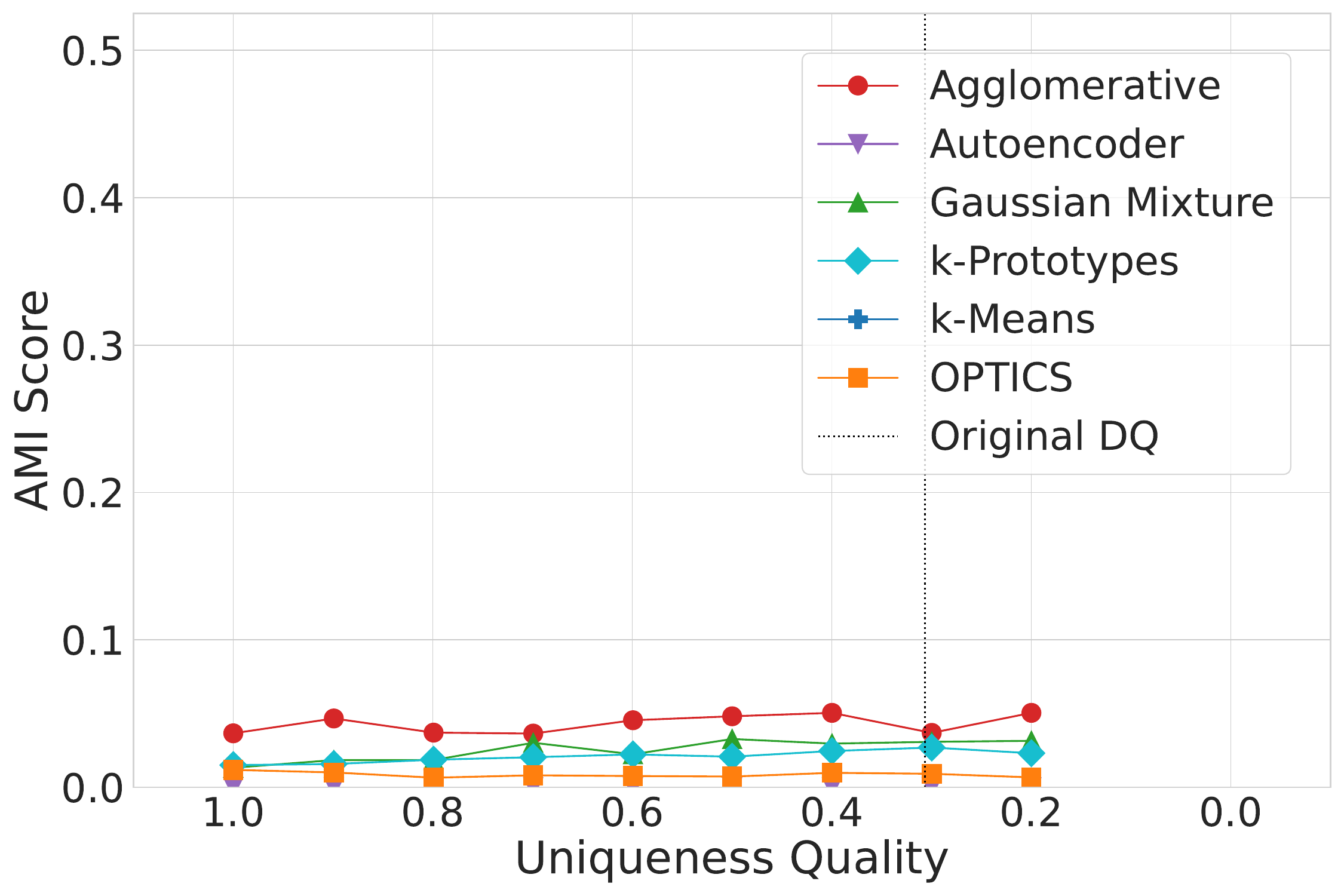}
        \caption{\textsf{Bank}}
        \label{fig:clustering-uniqueness-bank-results}
    \end{subfigure}
    \caption{AMI score for uniqueness dimension and clustering algorithms.}
    \label{fig:clustering-uniqueness-results}
\end{figure*}

%% file: Latex_Figure/clustering/Class_Balance.tex
\begin{figure*}[htbp]
    \centering
    \begin{subfigure}[b]{.4\linewidth}
        \includegraphics[width=\linewidth]{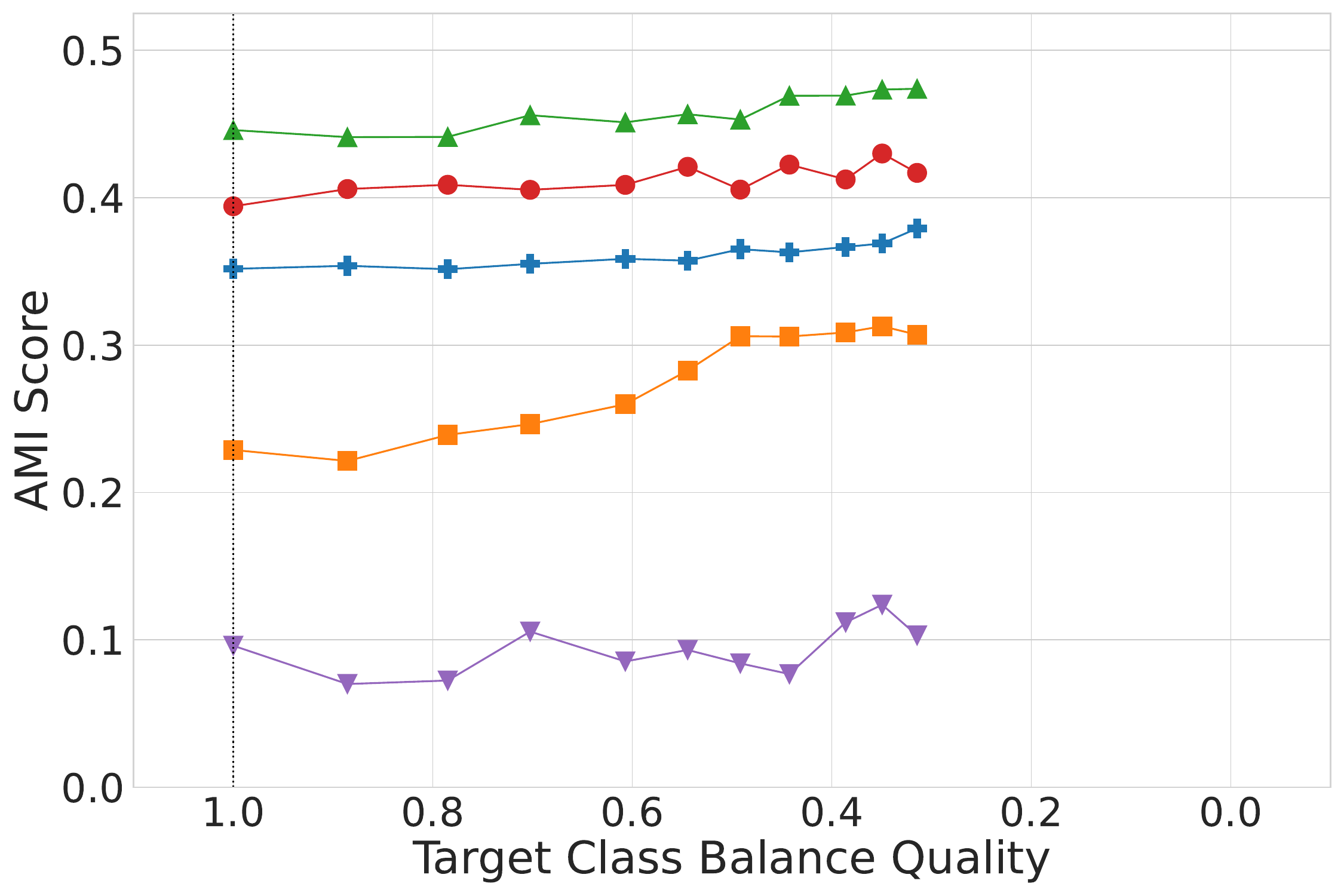}
        \caption{\textsf{Letter}}
        \label{fig:clustering-class-balance-letter-results}
    \end{subfigure}
    \begin{subfigure}[b]{.4\linewidth}
        \includegraphics[width=\linewidth]{figures/clustering/covtype/ClassBalance/adj_mut_info.pdf}
        \caption{\textsf{Covertype}}
        \label{fig:clustering-class-balance-covertype-results}
    \end{subfigure}
    
    \begin{subfigure}[b]{.4\linewidth}
        \includegraphics[width=\linewidth]{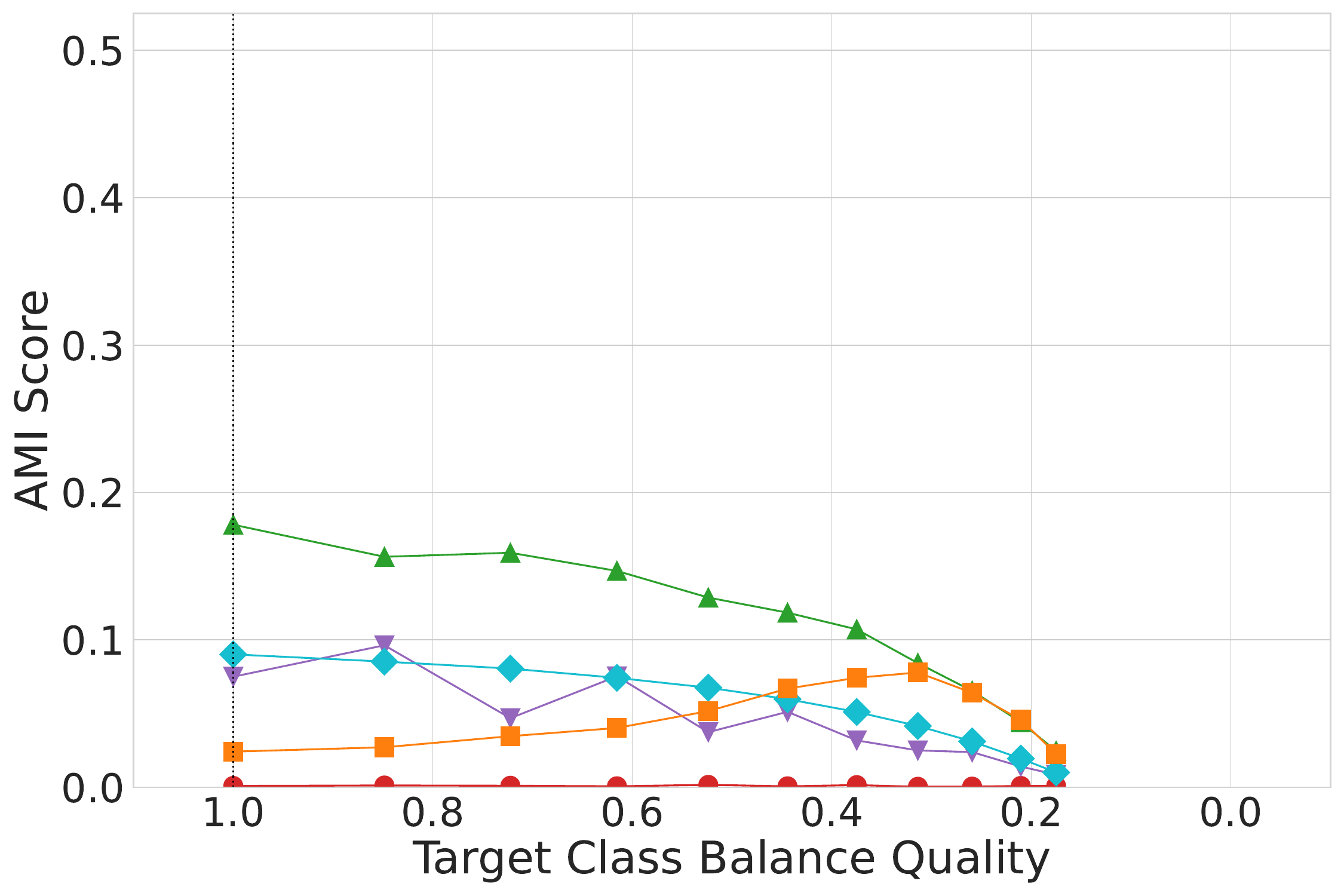}
        \caption{\textsf{COVID}}
        \label{fig:clustering-class-balance-covid-results}
    \end{subfigure}
    \begin{subfigure}[b]{.4\linewidth}
        \includegraphics[width=\linewidth]{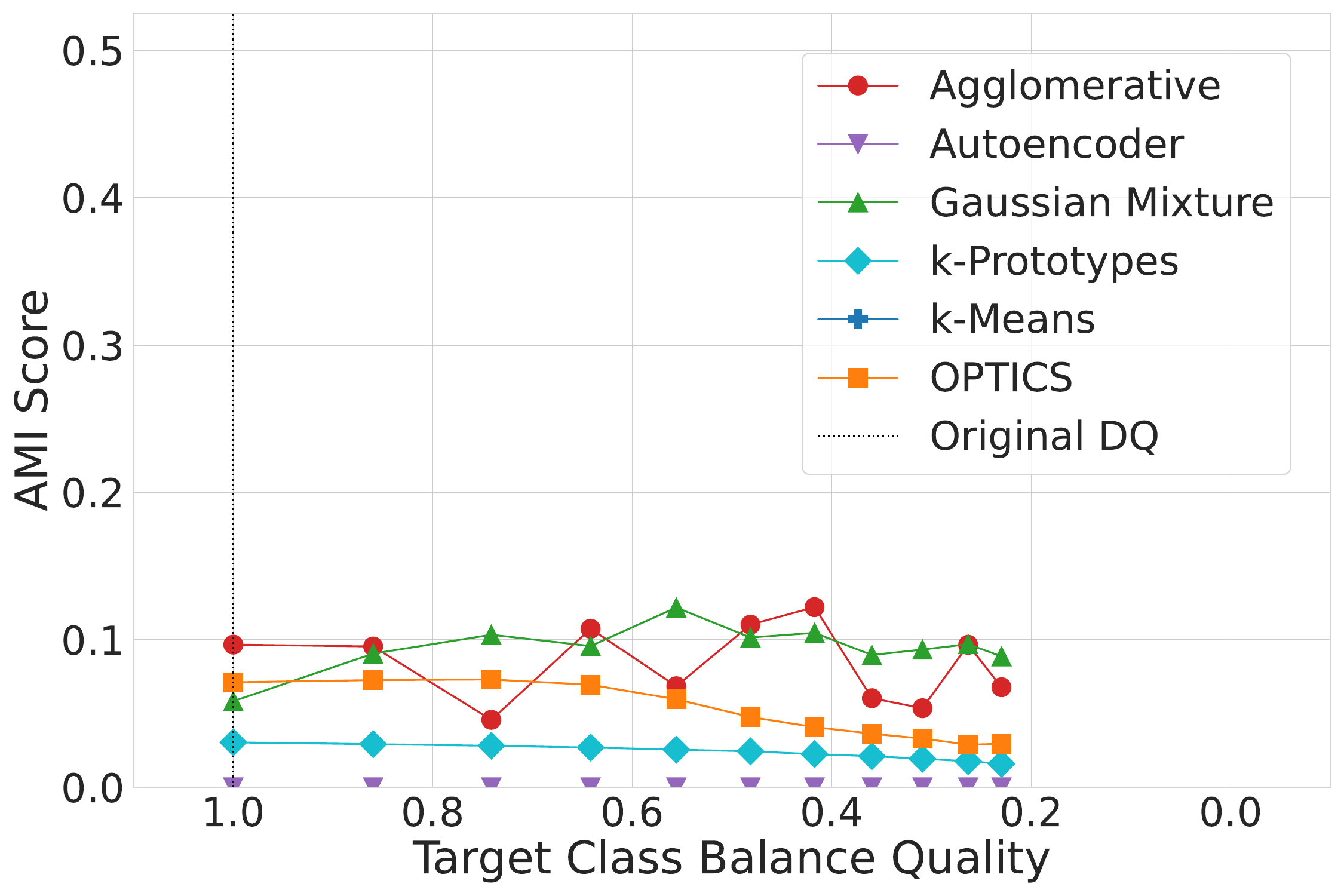}
        \caption{\textsf{Bank}}
        \label{fig:clustering-class-balance-bank-results}
    \end{subfigure}
    \caption{AMI score for target class balance dimension and clustering algorithms.}
    \label{fig:clustering-class-balance-results}
\end{figure*}

%% file: 70-conclusion.tex
\revision{\section{Discussion}}
\label{sec:discussion}

In this work, we experimentally studied the impact of six data quality dimensions on \revision{19} ML algorithms from three ML tasks: classification, regression and clustering. 
We ran many experiments using quality-degraded versions of nine real-world datasets. 
In addition, we distinguished between three scenarios that a data scientist could face while developing ML pipelines: (1)~polluted training set; (2)~polluted test set; (3)~polluted training and test sets.  
\revision{In the following Section~\ref{sec:qualitative_trends}, we derive high-level trends from the experiments, focusing on the interplay between the ML task and various data quality dimensions.
Subsequently, Section~\ref{sec:qualitative_examples} qualitatively examines selected  examples from the experiments where data pollution led to a concrete change in the predictions of the considered ML model.}\\

\subsection{Qualitative Trends} \label{sec:qualitative_trends}
Table~\ref{table:overview_effect_task} shows a high-level view of our findings, which we summarize along with recommendations in the following paragraphs. 
\revision{The thresholds for categorizing the effects in Table~\ref{table:overview_effect_task} were determined by observing the performance trends, i.e., the average relative change in model performance metrics compared to a baseline established using the clean data, across different levels of data pollution over all datasets and ML models. 
In particular, we analyzed the performance degradation between using clean data and data polluted to approximately 50\% quality. 
The thresholds were then defined based on these observations:} 
\revision{
\begin{itemize}
    \item Low Effect (\checkmark): If the degradation in performance was less than 5\pt relative to the baseline, we classified the impact as low. Models showing less than a 5\pt drop maintained a performance level that might be acceptable depending on the application context.
    \item Moderate Effect ($\bigcirc$): A performance degradation between 5\pt and 25\pt was considered a moderate effect. This range indicates that the data quality issue has a noticeable impact on model performance, potentially affecting the reliability of the results.
    \item High Effect (\ding{53}): Any degradation exceeding 25\pt was deemed a high effect. Such a significant decline suggests that the data quality issue critically undermines the model's effectiveness, necessitating immediate attention and remediation.
\end{itemize}
\stitle{Example} In classification, regression, and clustering tasks, we noticed that for missing values (especially exceeded 40\pt), the performance degradation on average crossed the 25\pt mark. Therefore, completeness was marked as having a high effect when missing data was substantial.}

\revision{While the results presented in Section~\ref{sec:results} and the following discussion suggest that certain data quality dimensions, such as uniqueness, consistent representation, and target class balance, may not significantly impact the performance of classification, regression, and clustering tasks in some scenarios, they emphasize that the overall goal should always be to maintain high data quality. 
In critical domains, such as medicine or autonomous systems, even minor data quality issues can have serious consequences.
For instance, in medical diagnosis systems, even each misclassification can lead to incorrect diagnoses: a model might misclassify a malignant tumor as benign due to poor target accuracy or missing values in the training data. 
Thus, relying on lower-quality data should be carefully considered against the potential risks in these settings, where incorrect model outputs could lead to severe outcomes.}
\begin{table}[ht]
\centering
\caption{\revision{The effect of data quality dimensions per ML task. \checkmark: low effect, $\bigcirc$: moderate effect, \ding{53}: high effect.}}
\label{table:overview_effect_task}
\begin{tabular}{r|l|l|l|l|l|l|}
\cline{2-7}
 &
  \multicolumn{1}{r|}{\rotatebox{90}{\hbox{\tabular{@{}l}Consistency\endtabular}}} &
  \multicolumn{1}{r|}{\rotatebox{90}{\hbox{\tabular{@{}l}Completeness\endtabular}}} &
  \multicolumn{1}{r|}{\rotatebox{90}{\hbox{\tabular{@{}l}Feat.-Accuracy\endtabular}}} &
  \multicolumn{1}{r|}{\rotatebox{90}{\hbox{\tabular{@{}l}Tar.-Accuracy\endtabular}}} &
  \multicolumn{1}{r|}{\rotatebox{90}{\hbox{\tabular{@{}l}Uniqueness\endtabular}}} &
  \rotatebox{90}{\hbox{\tabular{@{}l}Class Balance\endtabular}} \\ \hline
\multicolumn{1}{|l|}{Classification} 
& \checkmark & \ding{53} & \ding{53} & \ding{53} & \checkmark & $\bigcirc$ \\ \hline
\multicolumn{1}{|l|}{Regression}     
& \checkmark & \ding{53} & \ding{53} & \ding{53} & \revision{\checkmark} & $\bigcirc$ \\ \hline
\multicolumn{1}{|l|}{Clustering}     
& \checkmark & \ding{53}  & \ding{53} & \checkmark & \checkmark & \checkmark \\ \hline
\end{tabular}
\end{table}

\stitle{Classification}
We found that the quality dimensions with the least impact are \textit{uniqueness}, \textit{consistent representation} and \textit{target class balance} as long as the balance is not shifted towards a very extreme case. 
This suggests that with only a very low loss in classifier performance on tabular data, data scientists can skip pre-processing steps like exact deduplication, unifying inconsistent representations, and carefully balancing the target variable.

For \textit{completeness}, our findings suggest that missing values influence the performance of the classifiers the most when the classifier is not trained on data with missing values. 
The results also show that having less than~40\% missing values in the training phase does not significantly decrease the model's performance, which could be needed to increase robustness if no information is provided on how to impute the test data once the model reaches production.

Reducing the \textit{target accuracy} of the training data results in the degradation of model performance to below a majority classifier performance with target accuracy lower than~$1/|\textit{classes}|$.
However, the results suggest that if~80\% or more of the data are labeled correctly, one can train classification models on that noisy data without significant performance loss.
Our insights also reinforce the importance of labeling test data carefully, as labeling errors could then lead to a significant underestimation of the model's performance\PaperLong{(e.g., with~40\% label errors in~\textsf{Telco} test data, the perceived performance of the classifier is below a baseline model)}. 

A very low feature accuracy in the training data can also lead to a strong model performance degradation. 
However, similar to the target accuracy, there is a certain robustness to such errors, especially for linear models. 
Yet, the extent of this robustness is dataset-dependent. 
Reduced feature accuracy during serving time is also problematic but behaves more smoothly and in a linear trend, which makes performance estimation much easier if one can estimate the quality of the serving data. 
Moreover, even though the performance degrades constantly, it stays above baseline levels most of the time.

\stitle{Regression} We found that the quality dimensions with the
largest impact on regression performance are \textit{completeness}, \textit{feature accuracy} and \textit{target accuracy}. 
A decrease in quality in one of those dimensions leads to a worse-than-linear decrease in algorithm performance. 
Furthermore, missing values or inaccurate features in the test data without training the ML algorithm on such kind of data leads to even worse performance.
The dimensions \textit{uniqueness} and \textit{target class balance} show little impact. 
\revision{However}, we observed some degradation with small datasets and datasets with many features. 
We recommend not focusing on these two dimensions when using large training datasets with a few features. 
\textit{Consistent representation} has an impact as soon as the new representations outweigh the old ones.

For linear regression-based algorithms, ridge regression performs better or is equal to linear regression because regularization helps cope better with polluted data.
For tree-based algorithms, random forests always outperform decision trees.
The performance of multi-layer perceptron variants highly depend on both the quality dimension and the dataset characteristics. 
Among all considered algorithms, random forest performs best across most cases and thus is the most robust.
One exception is the target accuracy dimension.
This insight coincides with the findings by Aleryani et al.\ investigating imputation methods for classification~\cite{aleryani2020multiple}.
They found ensemble learning the best performing approach for imputation when dealing with missing data.

\stitle{Clustering}
We found the quality dimensions of \textit{completeness} and \textit{feature accuracy} to be the most impactful.
In both cases, a degradation of the dataset quality led to a decrease in clustering performance, which was linear for datasets with categorical features and seemingly exponential concerning \textsf{Letter} containing only numerical features. 
The density-based OPTICS algorithm was most affected by \textit{completeness} and \textit{feature accuracy} changes.
The $k$-means/$k$-prototypes, agglomerative clustering, and Gaussian mixture clustering generally behaved similarly.
The autoencoder approach was least affected by changes in these two quality dimensions.

Interestingly, the Gaussian mixture benefited from the slight noise added to numerical features in the first step of feature accuracy pollution.
Therefore, we recommend using this approach when dealing with datasets containing exclusively numerical features if one suspects inaccuracies in their feature values.
In the case of mixed-type datasets, we can recommend only the $k$-Prototypes algorithm, as its degradation to initial performance trade-off performs the best in our experiments.

Surprisingly, the \textit{consistent representation}, \textit{target accuracy}, \textit{target class balance}, and \textit{uniqueness} dimensions had a very low impact on the performance of the majority of the examined algorithms.
In contrast, adding some duplicate values to a completely duplicate-free dataset, decreasing its uniqueness, significantly improved OPTICS's performance.
We recommend refraining from using agglomerative clustering when dealing with a dataset with inconsistent representations for its categorical features, as it could not deal with this pollution.
Overall, the $k$-means/$k$-prototypes algorithms,~i.e., the centroid-based family, showed the most robustness regarding the six data quality dimensions.
Nevertheless, the agglomerative and Gaussian mixture clustering algorithms outperformed the $k$-means/$k$-prototypes approaches on clean data.
\PaperLong{
If the given data does not contain categorical and numerical features, where the features may take on erroneous values, we can also safely recommend using the Gaussian mixture clustering approach.
While it restricts usage to datasets that are approximately drawn from a mixture of Gaussian distributions, this approach generally performed better than most of the other clustering algorithms while being more resilient against data quality issues than the similarly well-performing agglomerative clustering algorithm.}

\subsection{\revision{Qualitative Examples}} \label{sec:qualitative_examples}

\begin{table}
    \centering
    \begin{tabular}{l|cclrr} \toprule
         & CombID & Model & \begin{tabular}[c]{@{}l@{}}Data Quality\\Dimension\end{tabular} & RecordID & \begin{tabular}[c]{@{}l@{}}Pol.\\Degree\end{tabular} \\ \midrule
        Bad Model & SC & SVM & Consistency & 162 & 5\%\\
         & GC & GB & Completeness & 1191 & 5\%\\
         & MT & MLP & Target Accuracy & 99 & 5\%\\\midrule
        Bad Data & MF & MLP & Feature Accuracy & 818 & 5\% \\
         & LT & LogR & Target accuracy & 976 & 5\%\\
         & KC & KNN & Completeness & 7284 & 5\% \\ \midrule
         Lucky Model& TC & TN & Consistency & 423 & 5\% \\
         & MC & MLP & Completeness & 358 & 5\% \\
         & DT & DT & Target accuracy & 1351 & 5\% \\\bottomrule
    \end{tabular}
    \caption{\revision{Exemplary records from the \textsf{Telco} dataset for selected combinations of ML algorithms and data quality dimensions in the classification task and Scenario 3, considering the cases \textit{bad model}, \textit{bad data} and \textit{lucky model}.}}
    \label{tbl:qualitative_examples}
\end{table}

\revision{In this section, we present selected examples where data pollution influenced the predictions of the considered ML model.
Specifically, these examples are records from the test data in which pollution has led to a change in the prediction made for them by the respective ML model.
Our focus is on Scenario~3, where both the training and test data are polluted and on the classification task.
This section complements our quantitative analysis by illustrating concrete cause-and-effect relationships where pollution changed model predictions. To better understand these relationships, we introduce three cases:}

\revision{\begin{description}
    \item[Bad model] When a model is trained on clean data, it makes a correct prediction for a record in the test data. However, if the model is trained on data with $x\%$ pollution, it makes an \emph{incorrect prediction}, even though the \emph{test record} itself remains \emph{unpolluted}.
    \item[Bad data] When a model is trained on clean data, it makes a \emph{correct prediction} for a record in the test data. However, if the model is trained on the data with $x\%$ pollution, it makes an \emph{incorrect prediction} for a \emph{test record} that is also \emph{polluted}.
    \item[Lucky model] When a model is trained on clean data, it makes an \emph{incorrect prediction} for a record in the test data. However, when the model is trained on the data with $x\%$ pollution, it (luckily) makes a \emph{correct prediction} for that test record while the \emph{record} itself remains \emph{unpolluted}.
\end{description}}

\revision{We regard the \textsf{Telco} dataset, considering various \emph{combinations} of data quality dimensions and ML algorithms.
The test data of \textsf{Telco} consists of \numprint{1407} records and \numprint{19} features.
For the record selection and their assignment to one of the introduced cases, we compare the values of the features and the label before the pollution with those after the pollution. 
As \textit{uniqueness} and \textit{target class balance} alter the size of the test data during pollution, they are unsuitable for a record wise comparison. Thus, we focus on \textit{consistency}, \textit{completeness}, \textit{feature accuracy}, and \textit{target accuracy} to showcase different behaviors. 
}

\revision{
For each of the three case, we selected three fixed combinations of ML algorithms and data quality dimensions, while ensuring that each classification algorithm is considered at least once.
For each selected combination, we selected a random record among those that first met the criteria for the respective case, i.e., at the lowest pollution level.
Table~\ref{tbl:qualitative_examples} presents the three cases along with their selected combinations (abbreviated under \textit{CombID}), where each combination has an associated record with the \textit{RecordID} from the test data at the pollution degree (\textit{Pol.\ degree}).
The pollution degree refers to the data as a whole, not specifically to the selected record. That is, a pollution degree of $x\%$ implies that $x\%$ of the data is polluted, but not necessarily the record itself.
Note that we added an extra pollution level of $5\%$ for a more fine grained analysis.}

\revision{
In the SC combination, when the SVM algorithm was trained on clean data, it made the correct prediction for this considered record, with the \textit{RecordID} $162$.
However, when the SVM algorithm is trained on data with $5\%$ pollution, the model misclassifies this record, despite the record itself remaining unpolluted.
This highlights a \textit{bad model} case, where the model's prediction is negatively affected by the data pollution elsewhere in the dataset.}

\revision{Since the test records are actually polluted in the \textit{bad data} case, we also list in Table~\ref{tbl:qualitative_examples_bad data} the original values alongside their respective polluted counterparts only for the polluted features out of the 19 features.
For example, for the KC combination at a $5\%$ pollution degree, the KNN model produces an incorrect prediction for the record with the \textit{RecordID} $284$. At this pollution degree, only 2 out of 19 features of the record are polluted. 
Since KNN relies on feature similarity for classification, the pollution of the features distorts its nearest-neighbor relationships, ultimately leading to misclassification.
}
\begin{table}[ht]
\centering
\begin{tabular}{crclrr}
\toprule
CombID & RecordID & \begin{tabular}[c]{@{}l@{}}Pol.\\Degree\end{tabular} & \begin{tabular}[c]{@{}l@{}}Polluted\\Feature\end{tabular}        & \begin{tabular}[c]{@{}l@{}}Original\\Value\end{tabular} & \begin{tabular}[c]{@{}l@{}}Polluted\\Value\end{tabular} \\ \midrule
MF    & 818 & 5\%                                              & \textsf{tenure}         & \texttt{72}             & \texttt{71.242689}      \\
     &  &                                                      & \textsf{PaperlessBill}  & \texttt{Yes}            & \texttt{No}             \\  
     &  &                                                      & \textsf{PayMethod}      & \texttt{Elec check}     & \texttt{Credit Card}    \\  
     &  &                                                      & \textsf{MonCharges}     & \texttt{110.75}         & \texttt{114.472121}     \\
     &  &                                                      & \textsf{TotCharges}     & \texttt{7751.7}         & \texttt{7761.316083}    \\ 
     \midrule
LT   & 976 & 5\%                                               & \textsf{Churn}          & \texttt{No}             & \texttt{Yes}            \\ \midrule
KC   & 284 & 5\%                                               & \textsf{MultLines}      & \texttt{Yes}            & \texttt{empty}          \\
     &  &                                                      & \textsf{TotCharges}     & \texttt{2688.45}        & \texttt{-1.0}           \\ 
\bottomrule
\end{tabular}
\caption{\revision{Exemplary records affected by the \textbf{bad data} case, including their original and polluted values \textbf{only for the polluted features} out of the 19 features per record.}}
\label{tbl:qualitative_examples_bad data}
\end{table}

\revision{Considering the TC combination, TN misclassified the record with the \textit{RecordID} 423 from the test data after training on clean data.
However, after introducing inconsistencies in $5\%$ of the training data and training TN on this data, the model correctly classified record 423, although the record itself was not part of the pollution.
This represents a \textit{lucky model} case, where the prediction for the considered record was positively influenced by data pollution.
However, as demonstrated in our various experiments and summarized in Table~\ref{table:overview_effect_task}, this is an exception rather than the rule. 
In general, data quality issues negatively impact the predictions of the underlying ML model.}


\revision{\subsection{Limitations} The experimental space of our analysis is enormous, leading to some decisions that limit the scope of the evaluation.
First, we abstained from optimizing hyperparameters for all algorithms. 
This lack is intentional, as we are interested in the impact of data quality on the algorithms' performance. 
However, this omission also means that the reported performance is not necessarily the best for any algorithm -- a not unrealistic situation in real/world scenarios.
Another ``hyperparameter'' for the implementation of the completeness polluter is our choice of placeholder values.
Comparing different values is a potential area for future work.
Next, in the clustering algorithms' evaluation, we implemented the autoencoder using a basic neural network that had not yet been optimized for its particular task.
Improvements could include using network components other than linear layers or incorporating the clustering performance on the code space into the loss function.
Additionally, we did not account for the information loss caused by encoding high-dimensional data into a two-dimensional embedded space.
}

\revision{\section{Conclusion and Future Work}
\label{sec:conclusion}
In the field of artificial intelligence, the importance of data quality is increasingly being recognized, leading to a shift from a model-centric approach to a data-centric perspective.}

\revision{This study empirically examined the impact of six key data quality dimensions on machine learning performance across three tasks (classification, regression, and clustering) using 19 widely used machine learning algorithms and ten diverse datasets. We conducted the evaluation under various scenarios of high and low training and test data quality.
Our results highlight how different data quality issues affect machine learning performance in distinct ways, offering valuable insights for data scientists when developing machine learning pipelines in the presence of data quality issues.
}

\revision{Several avenues for future work emerge.}
In future work, it is desirable to evaluate the clustering algorithms using not only adjusted mutual information score but also metrics such as the absolute size of overlap between the original and generated data clustering and the average and variance of the cluster sizes in the algorithm output.
We also plan to expand our evaluation in two directions: adding more data quality dimensions and conducting a deeper evaluation per machine learning model.

%% file: appendix/appendix.tex
\section{Additional plots}
This appendix provides some additional plots from the conducted experiments that were discussed in the report, but moved to the appendix to improve the readability of the paper.

We show in Figure~\ref{fig:classification-results-all-ConsistentRepresentation}, the performance of the classification algorithms when we decrease the data consistency with~$k_v = 2$. The same for clustering algorithm is shown in Figure~\ref{fig:regression-results-all-ConsistentRepresentation}. 
We also included in Figure~\ref{fig:regression-results-all-Uniqueness_dctnormal} the plots from the regression experiments where the uniqueness polluter added duplicates following a normal distribution.

For better understanding some clustering results, it is beneficial to see the correlation between the number of clusters identified by the OPTICS and Autoencoder clustering approaches to their performance. 
Please note that the Agglomerative and $k$-Means/$k$-Prototypes algorithms are always returning the expected number of clusters as this is one of their input parameters. 
The Gaussian Mixtures algorithm is also coerced to do this; however, it is not guaranteed to return the correct number of clusters. 
Therefore, these three lines commonly overlap and only one of them is visible in the plots.
The number of clusters shown in Figure~\ref{fig:clustering-all-nclusters} is averaged over the~5 different runs of each algorithm on a given dataset, and can therefore also take on floating-point values.

Finally, as the line plots shown in the clustering result chapter~(see Section~\ref{subsec:clustering-results}) for the \textsf{Bank} dataset are difficult to read because of the shared y-axis among the different datasets, we decided to also add a focused version of these plots in Figure~\ref{fig:stretched-bank-results}.

\input{Latex_Figure/classification/Consistent_Representation_2}
\input{Latex_Figure/regression/Consistent_Representation_2}
\input{Latex_Figure/regression/Uniqueness_norm}
\input{Latex_Figure/clustering/n_cluster}
\input{Latex_Figure/clustering/bank_focused}

%% file: Latex_Figure/classification/Consistent_Representation_2.tex
\begin{figure*}[b]
    \centering 
\raisebox{0.4\height}{\rotatebox{90}{Scenario 1}}\hspace{0.1em}
\begin{subfigure}[b]{0.23\linewidth}
        \includegraphics[width=\linewidth]{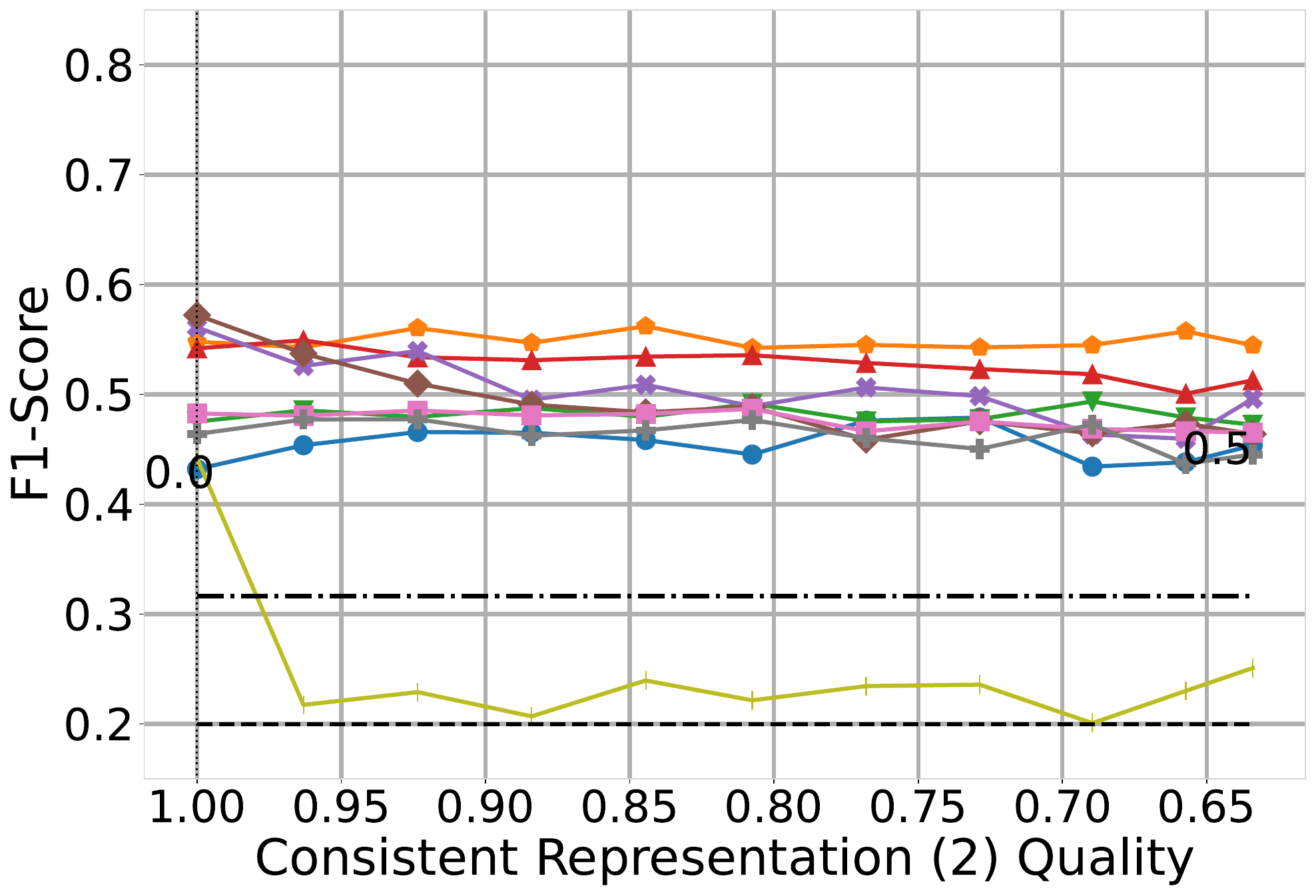}
        \caption{\textsf{Contraceptive}}
        \label{fig:classification-results-all-ConsistentRepresentation-1-contra}
    \end{subfigure}
\begin{subfigure}[b]{0.23\linewidth}
        \includegraphics[width=\linewidth]{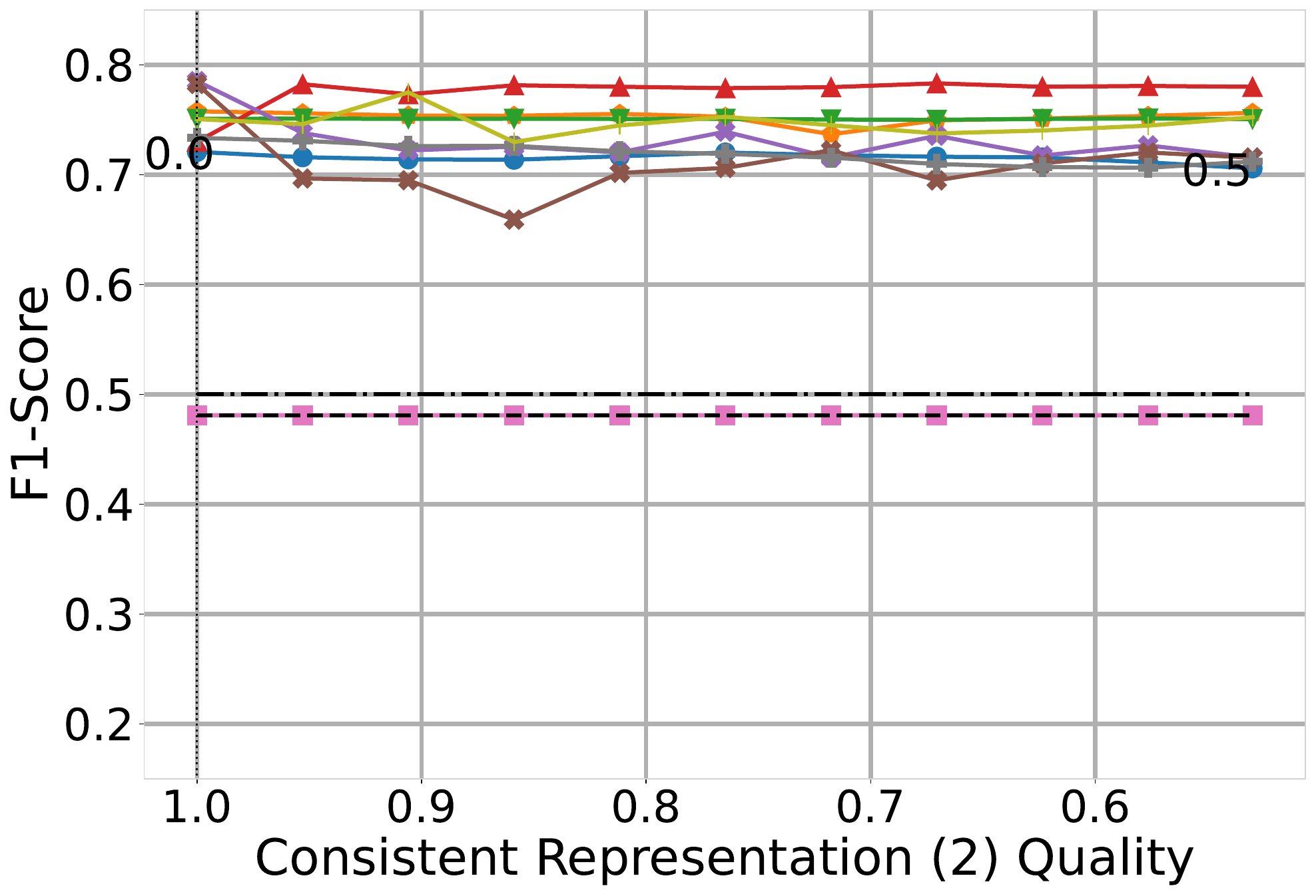}
        \caption{\textsf{COVID}}
        \label{fig:classification-results-all-ConsistentRepresentation-1-covid}
    \end{subfigure}
\begin{subfigure}[b]{0.23\linewidth}
        \includegraphics[width=\linewidth]{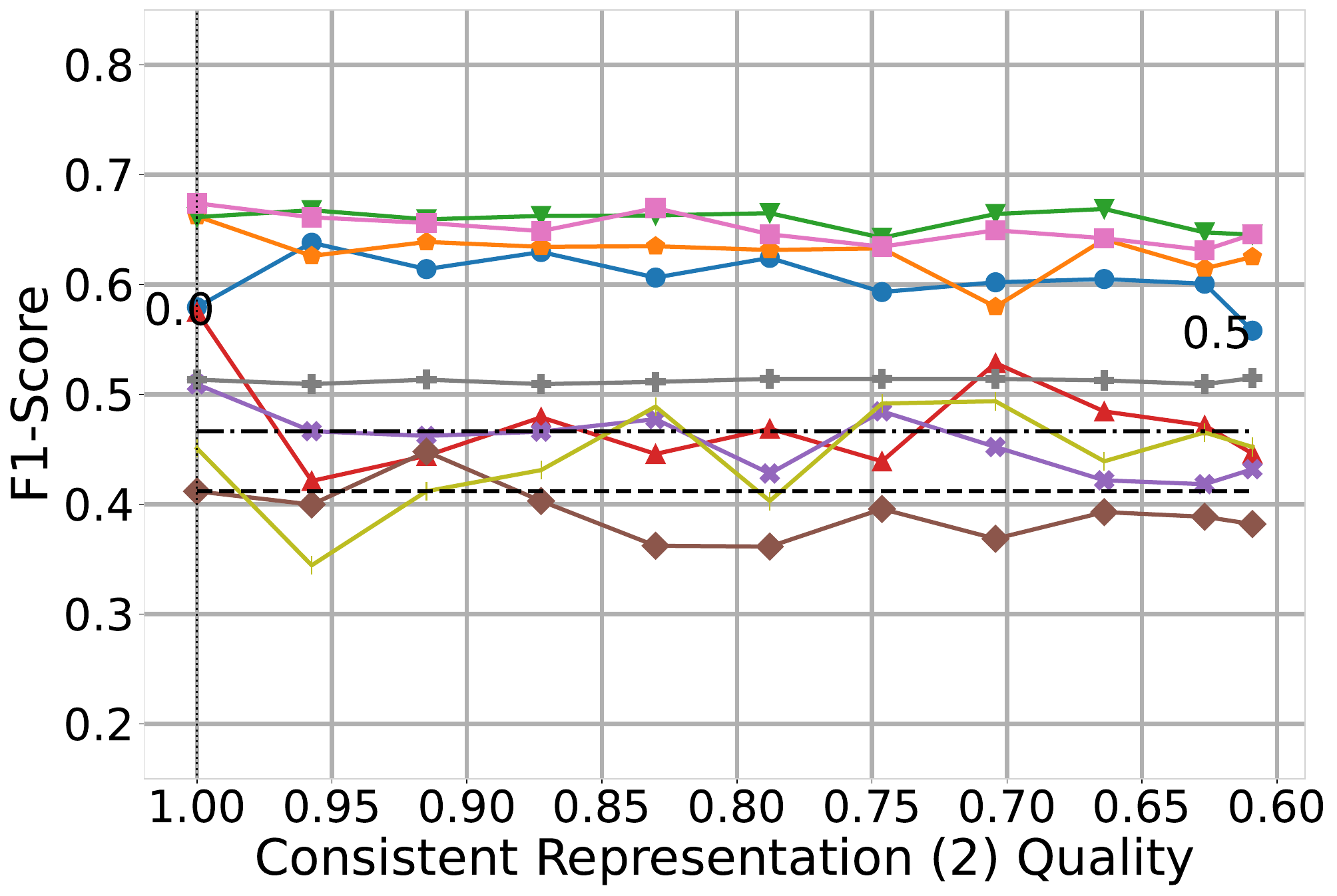}
        \caption{\textsf{Credit}}
        \label{fig:classification-results-all-ConsistentRepresentation-1-credit}
    \end{subfigure}
\begin{subfigure}[b]{0.23\linewidth}
        \includegraphics[width=\linewidth]{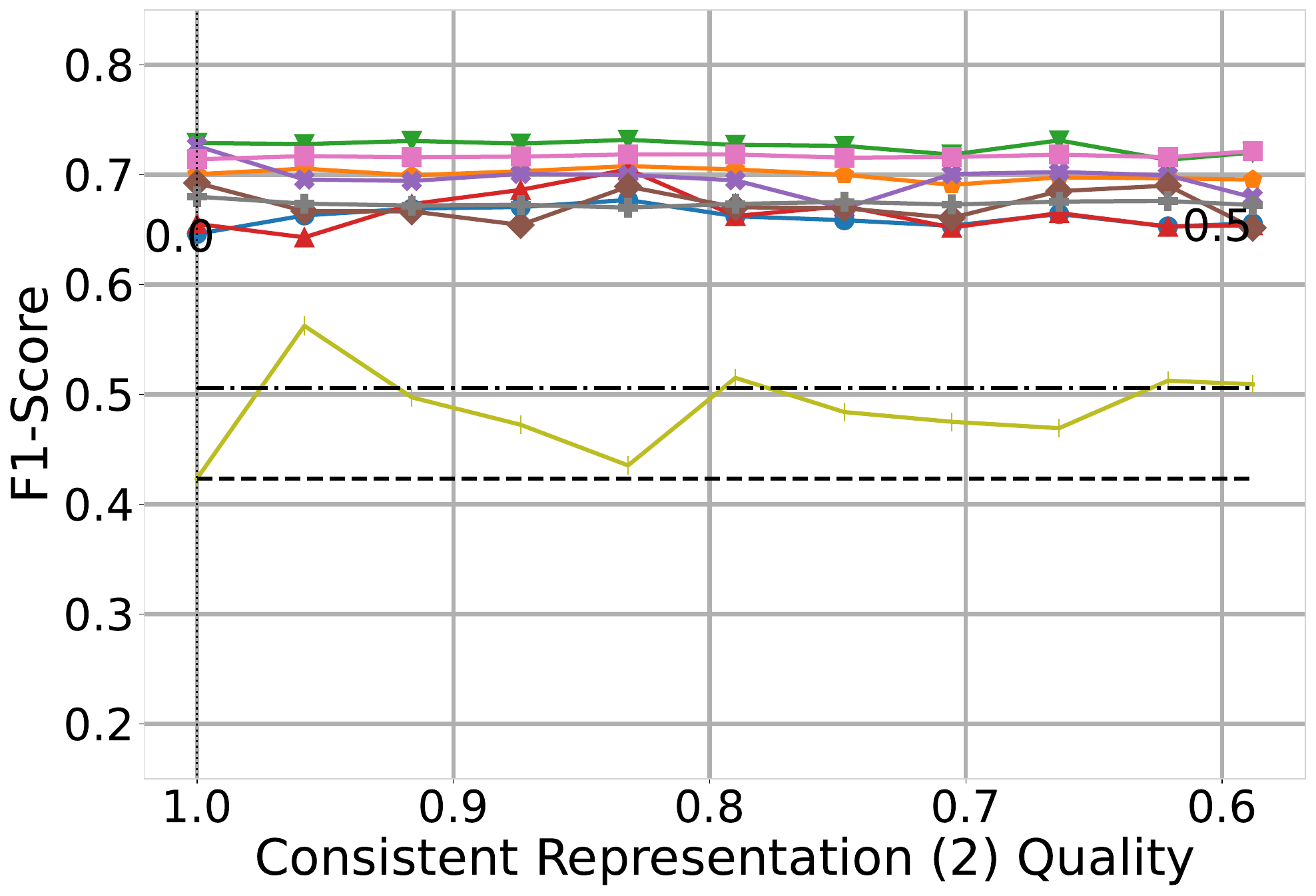}
        \caption{\textsf{Telco}}
        \label{fig:classification-results-all-ConsistentRepresentation-1-telco}
    \end{subfigure}

\raisebox{0.4\height}{\rotatebox{90}{Scenario 2}}\hspace{0.1em}
\begin{subfigure}[b]{0.23\linewidth}
        \includegraphics[width=\linewidth]{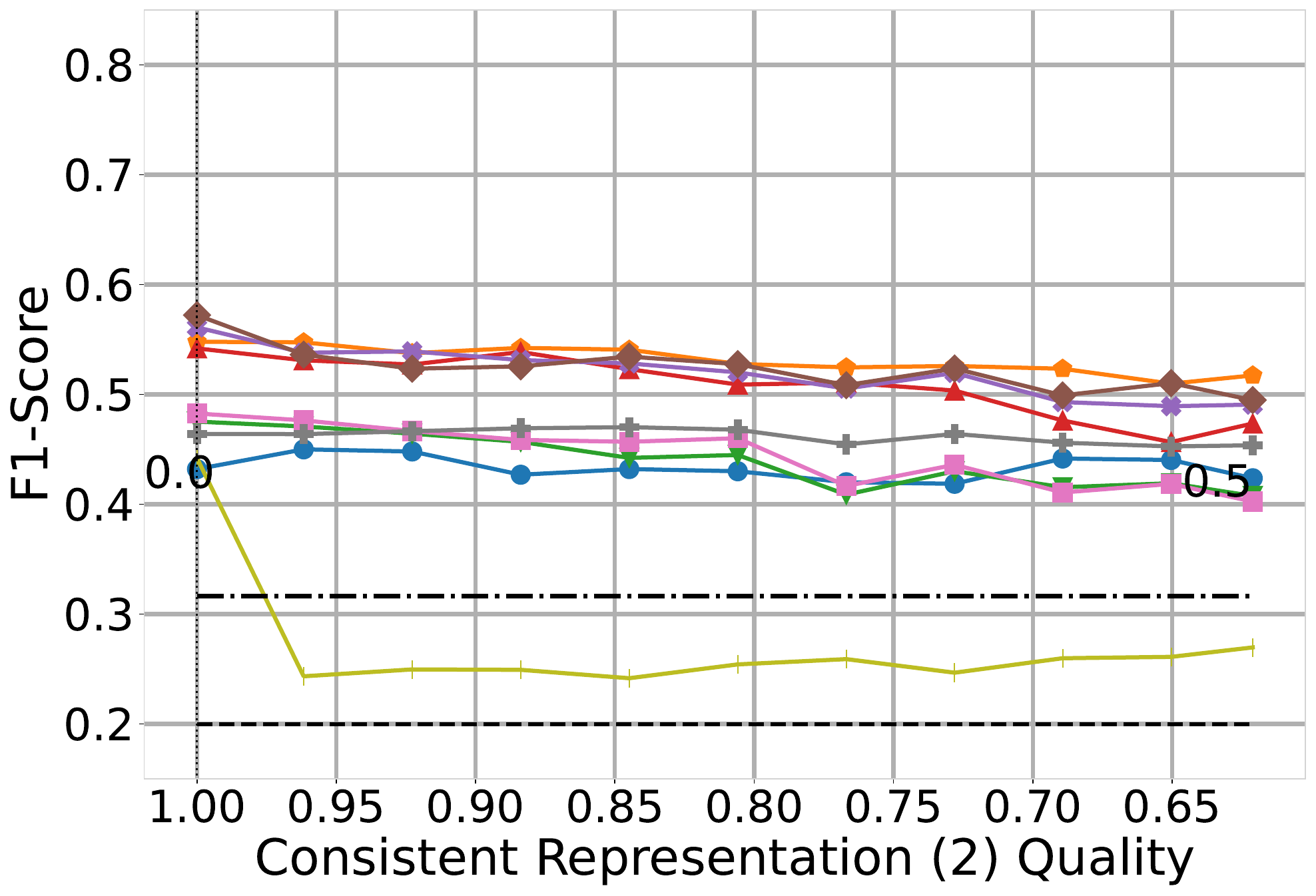}
        \caption{\textsf{Contraceptive}}
        \label{fig:classification-results-all-ConsistentRepresentation-2-contra}
    \end{subfigure}
\begin{subfigure}[b]{0.23\linewidth}
        \includegraphics[width=\linewidth]{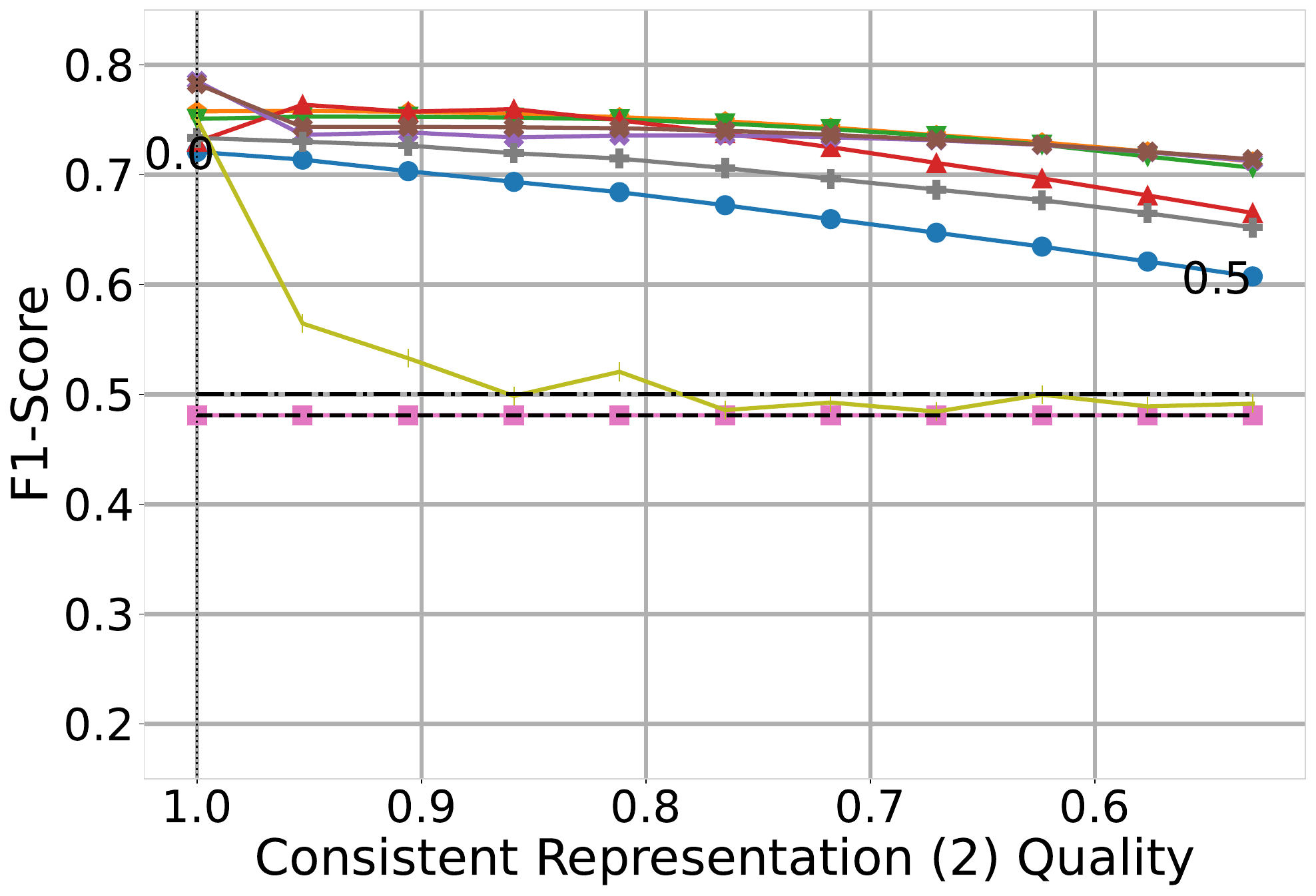}
        \caption{\textsf{COVID}}
        \label{fig:classification-results-all-ConsistentRepresentation-2-covid}
    \end{subfigure}
\begin{subfigure}[b]{0.23\linewidth}
        \includegraphics[width=\linewidth]{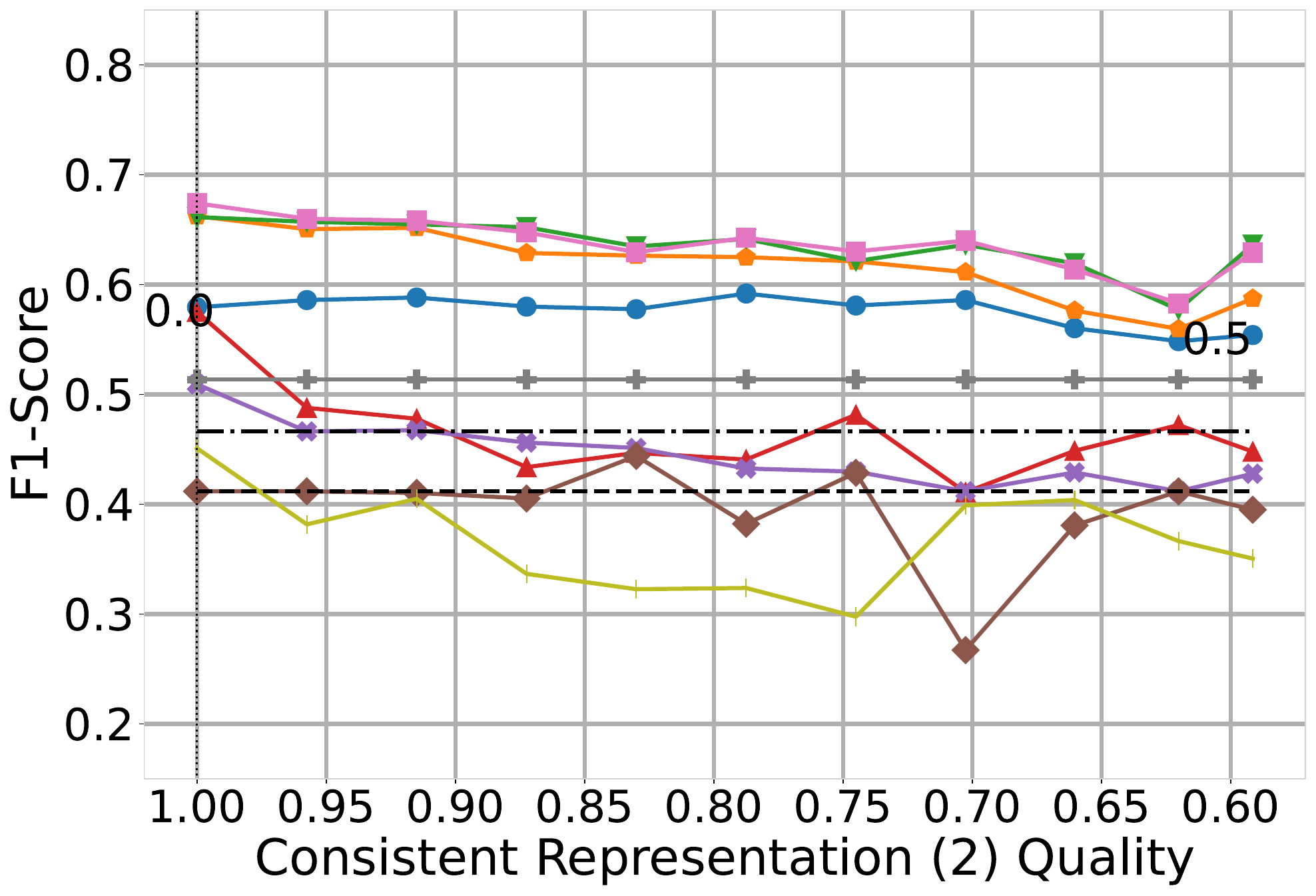}
        \caption{\textsf{Credit}}
        \label{fig:classification-results-all-ConsistentRepresentation-2-credit}
    \end{subfigure}
\begin{subfigure}[b]{0.23\linewidth}
        \includegraphics[width=\linewidth]{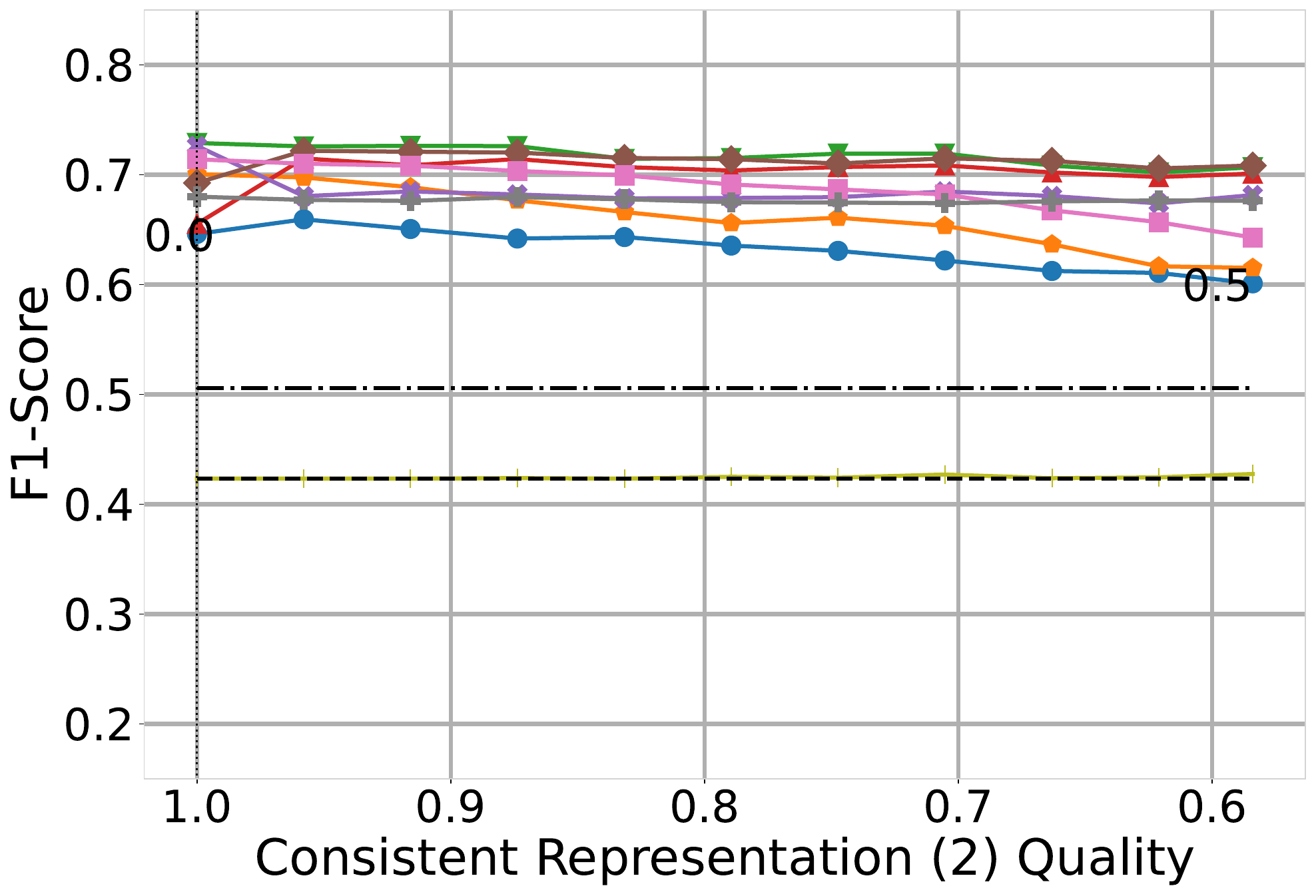}
        \caption{\textsf{Telco}}
        \label{fig:classification-results-all-ConsistentRepresentation-2-telco}
    \end{subfigure}

\raisebox{0.4\height}{\rotatebox{90}{Scenario 3}}\hspace{0.1em}
\begin{subfigure}[b]{0.23\linewidth}
        \includegraphics[width=\linewidth]{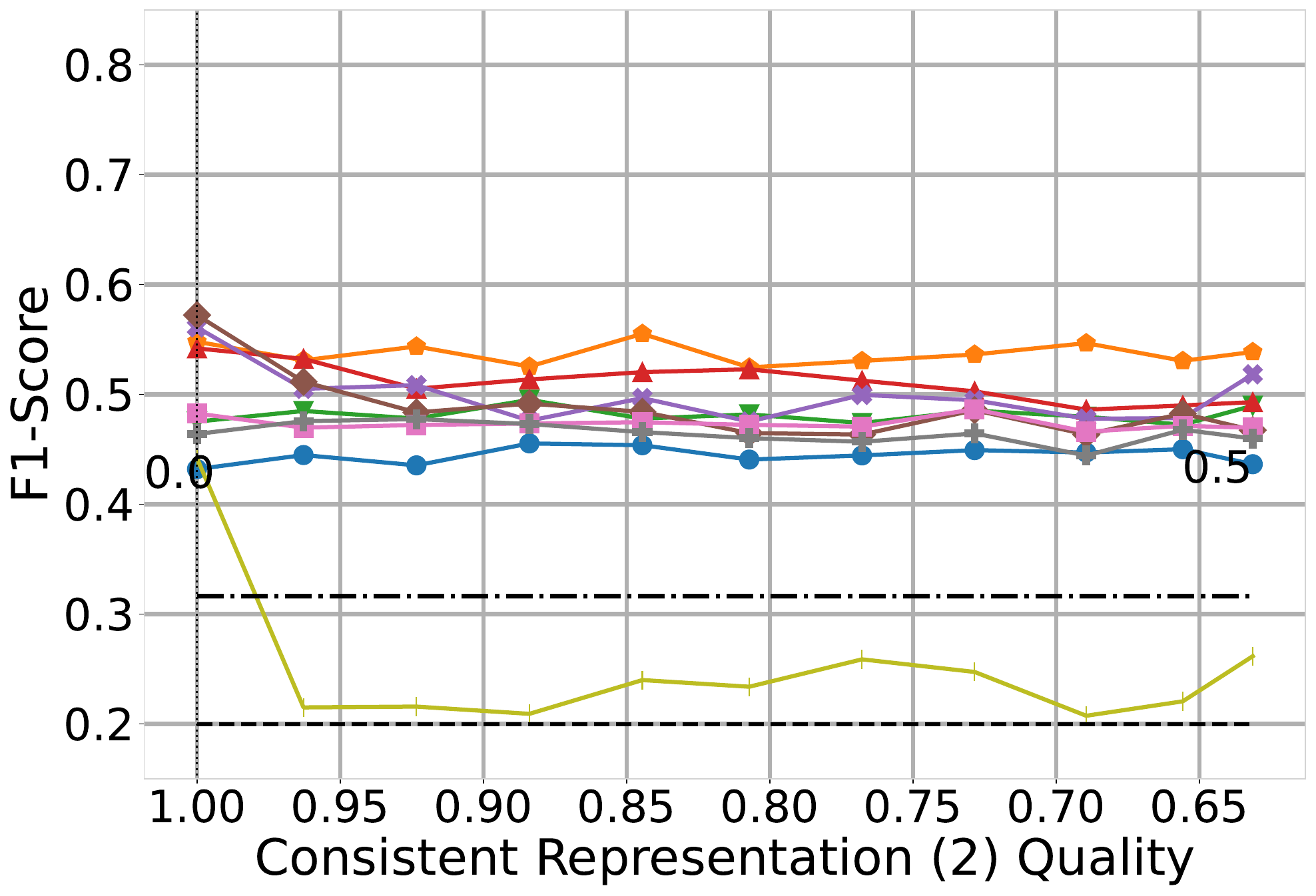}
        \caption{\textsf{Contraceptive}}
        \label{fig:classification-results-all-ConsistentRepresentation-3-contra}
    \end{subfigure}
\begin{subfigure}[b]{0.23\linewidth}
        \includegraphics[width=\linewidth]{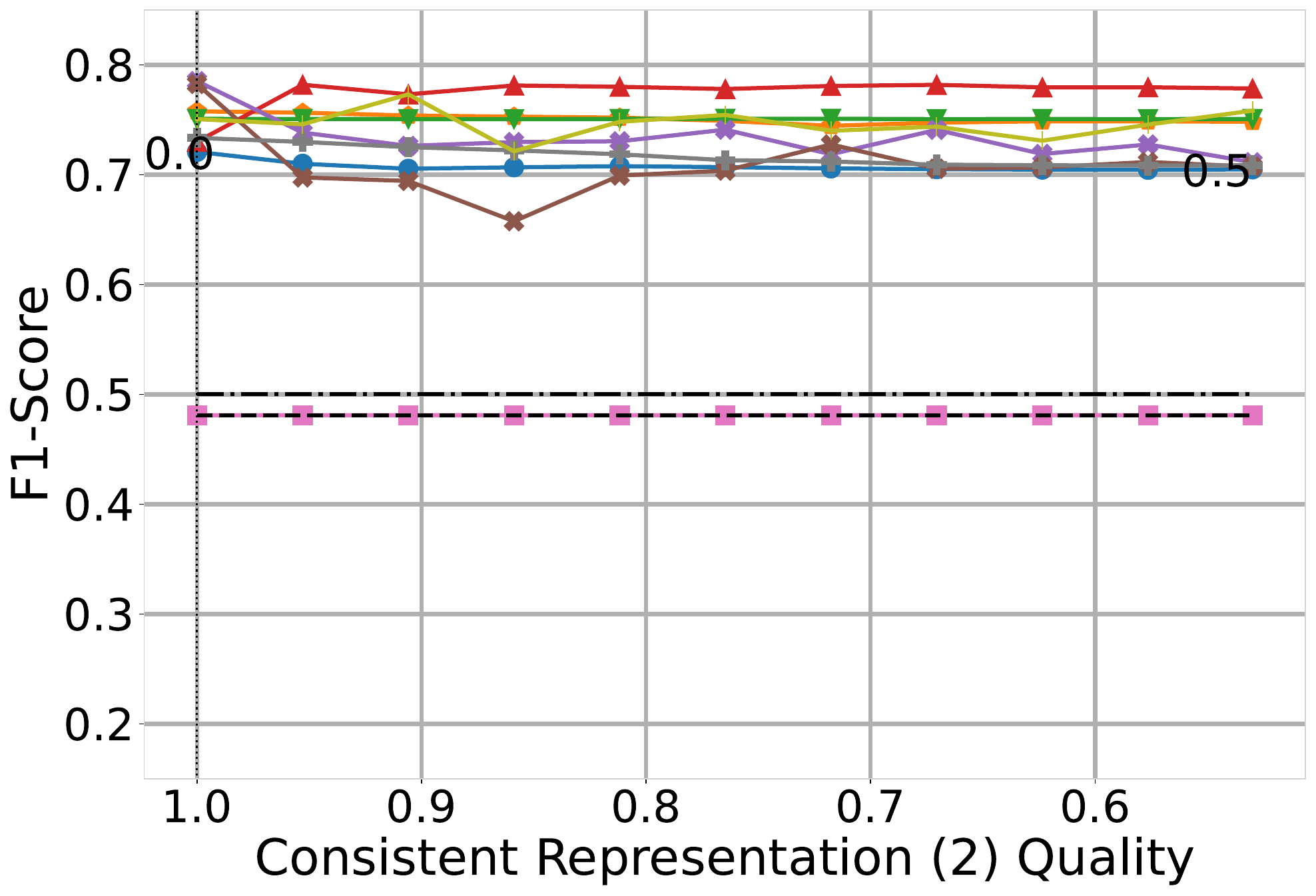}
        \caption{\textsf{COVID}}
        \label{fig:classification-results-all-ConsistentRepresentation-3-covid}
    \end{subfigure}
\begin{subfigure}[b]{0.23\linewidth}
        \includegraphics[width=\linewidth]{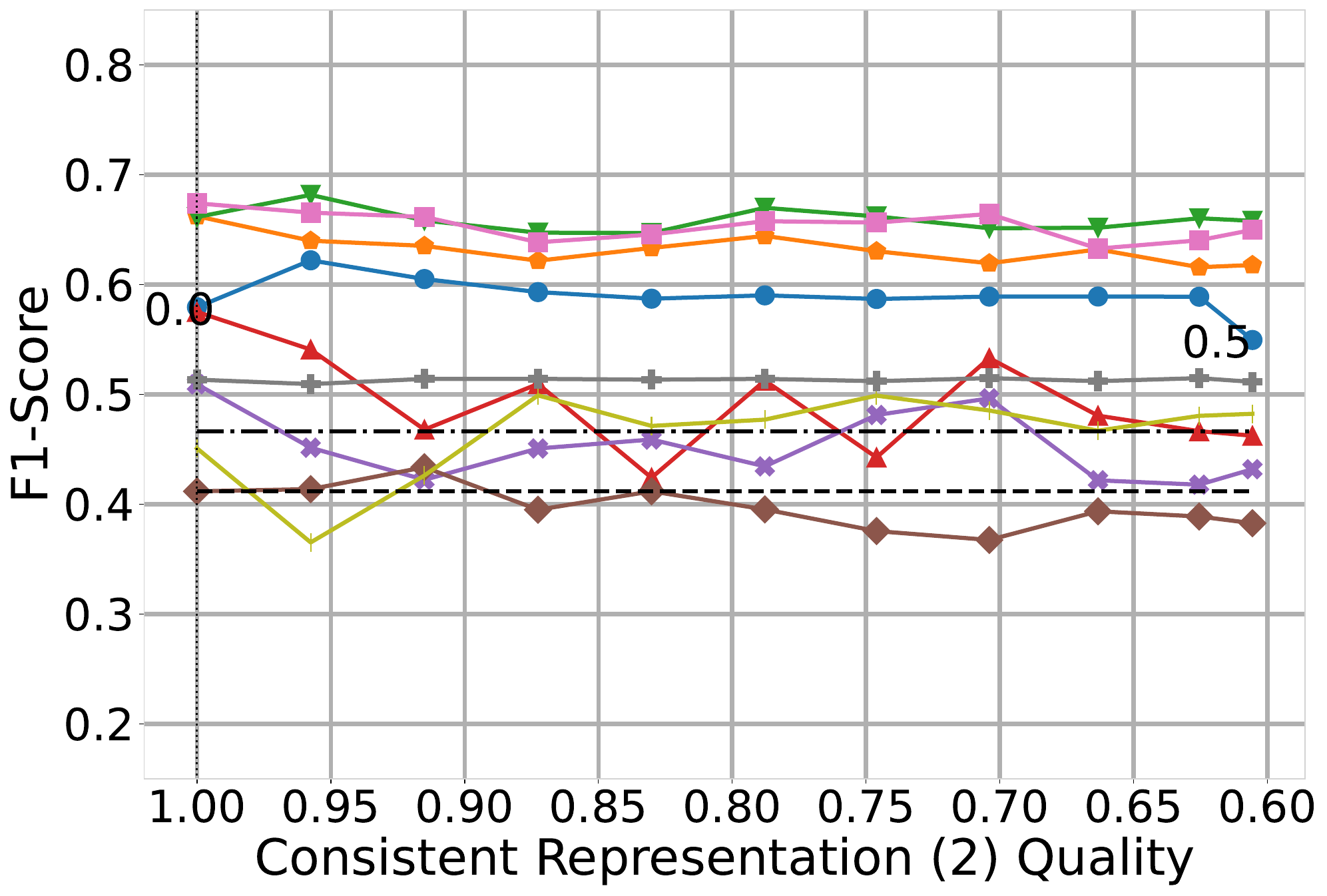}
        \caption{\textsf{Credit}}
        \label{fig:classification-results-all-ConsistentRepresentation-3-credit}
    \end{subfigure}
\begin{subfigure}[b]{0.23\linewidth}
        \includegraphics[width=\linewidth]{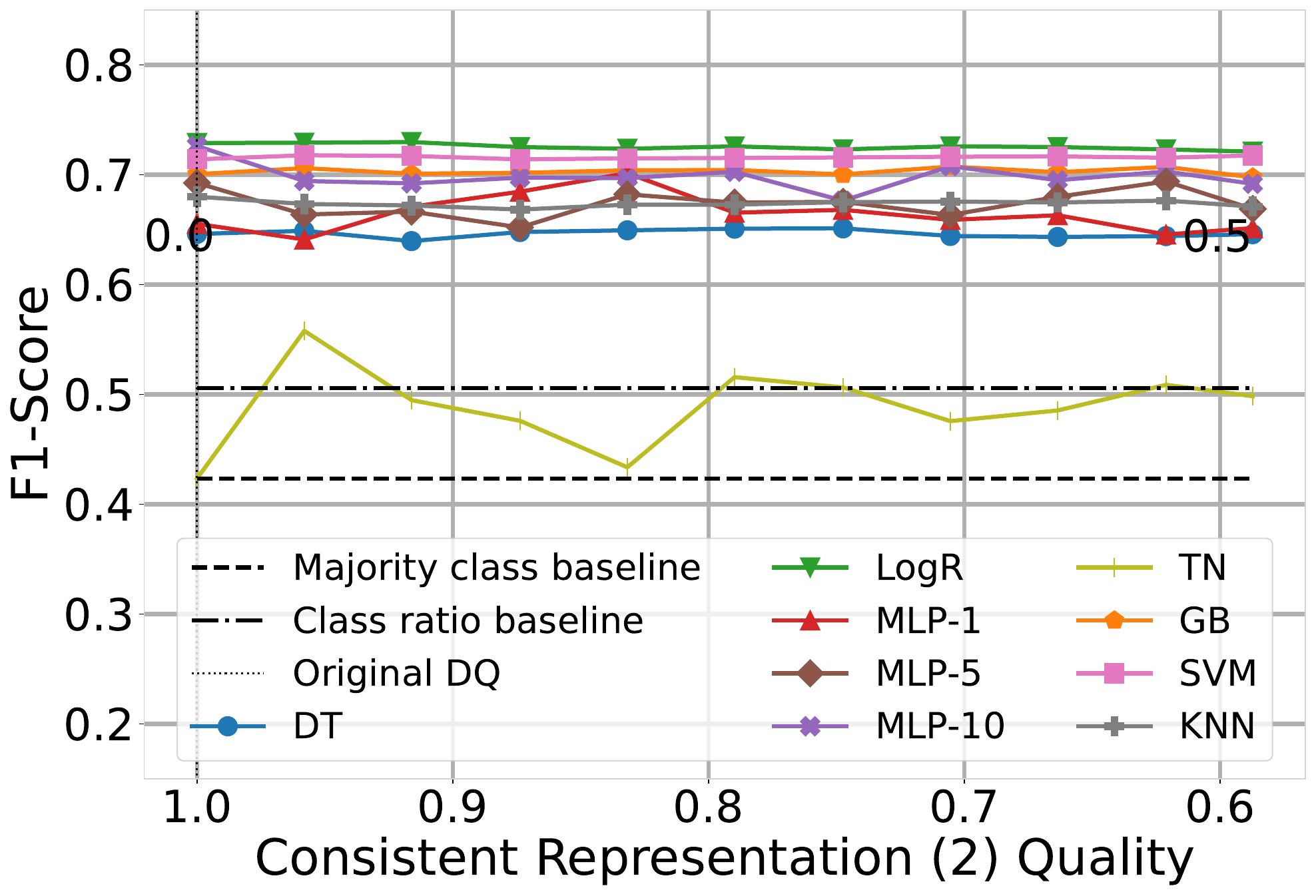}
        \caption{\textsf{Telco}}
        \label{fig:classification-results-all-ConsistentRepresentation-3-telco}
    \end{subfigure}
    \caption{$F_1$-scores of the classification algorithms for consistent representation with $k_v = 2$.}
    \label{fig:classification-results-all-ConsistentRepresentation}
\end{figure*}

%% file: Latex_Figure/regression/Consistent_Representation_2.tex
\begin{figure*}[!htbp]
    \centering
\raisebox{0.4\height}{\rotatebox{90}{Scenario 1}}\hspace{0.3em}
    \begin{subfigure}[b]{0.23\linewidth}
        \includegraphics[width=\linewidth]{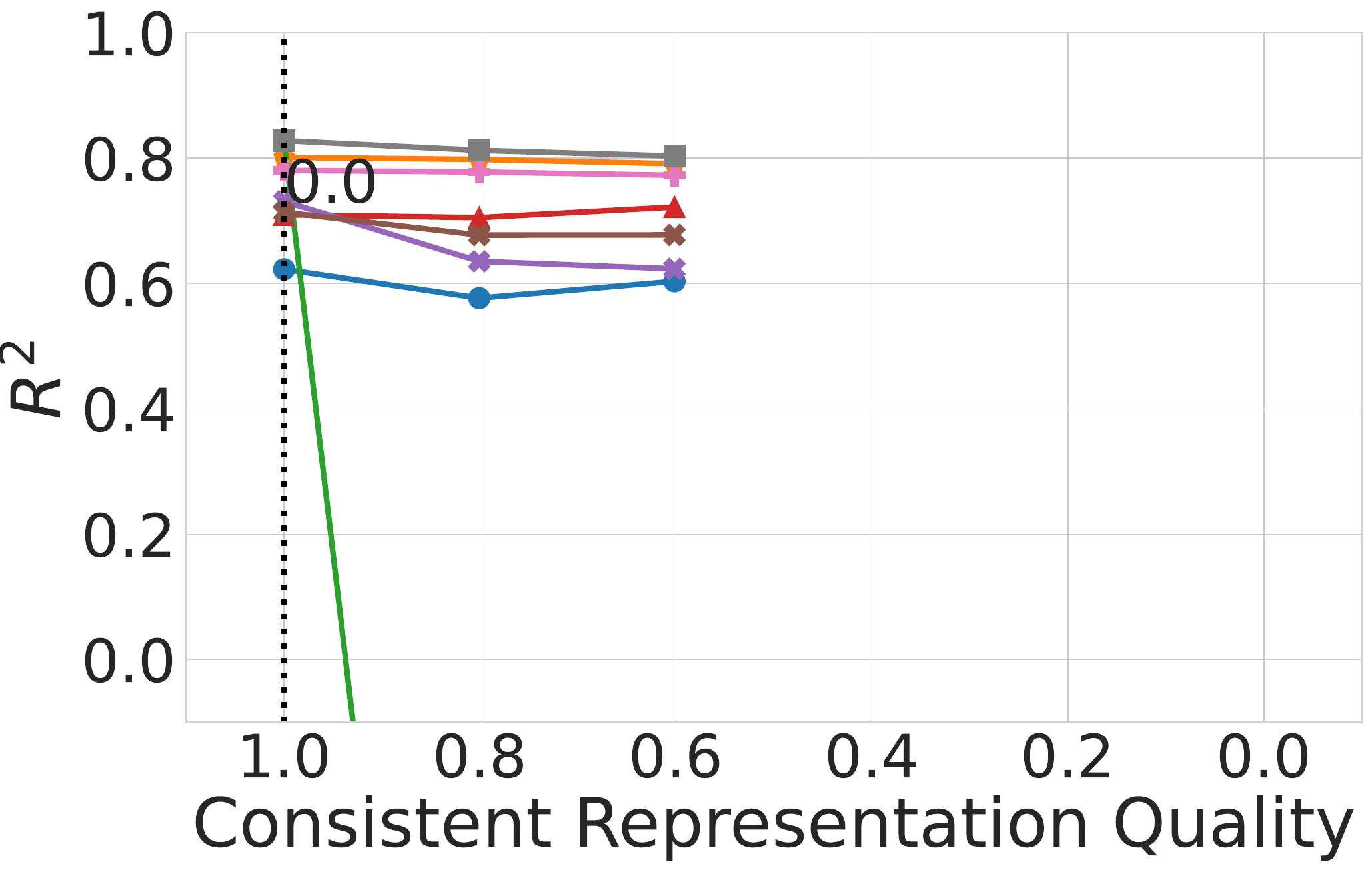}
        \caption{\textsf{Houses}}
        \label{fig:regression-results-all-ConsistentRepresentation-1-houses}
    \end{subfigure}
    \begin{subfigure}[b]{0.23\linewidth}
        \includegraphics[width=\linewidth]{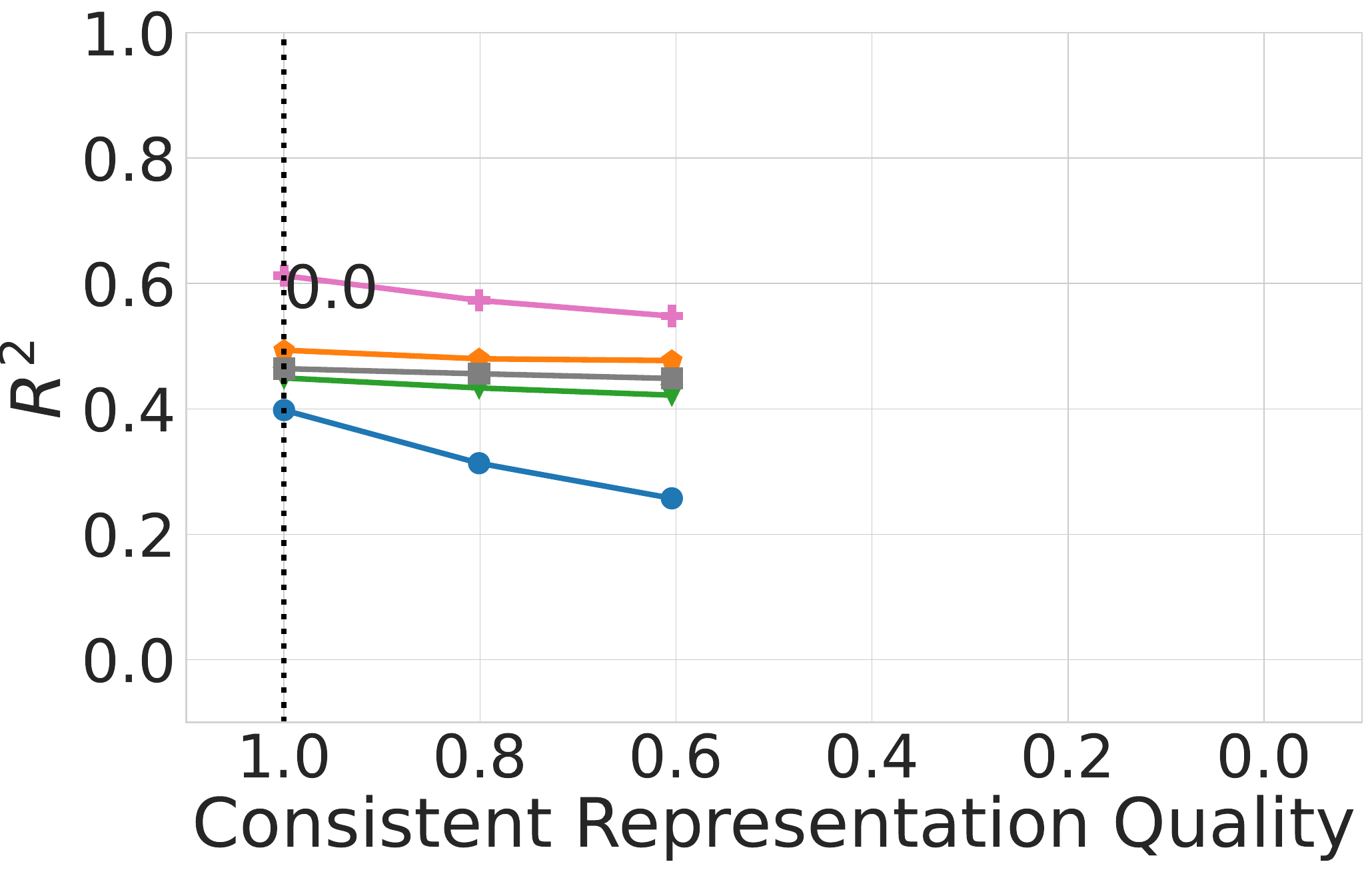}
        \caption{\textsf{IMDB}}
        \label{fig:regression-results-all-ConsistentRepresentation-1-imdb}
    \end{subfigure}
    \begin{subfigure}[b]{0.23\linewidth}
        \includegraphics[width=\linewidth]{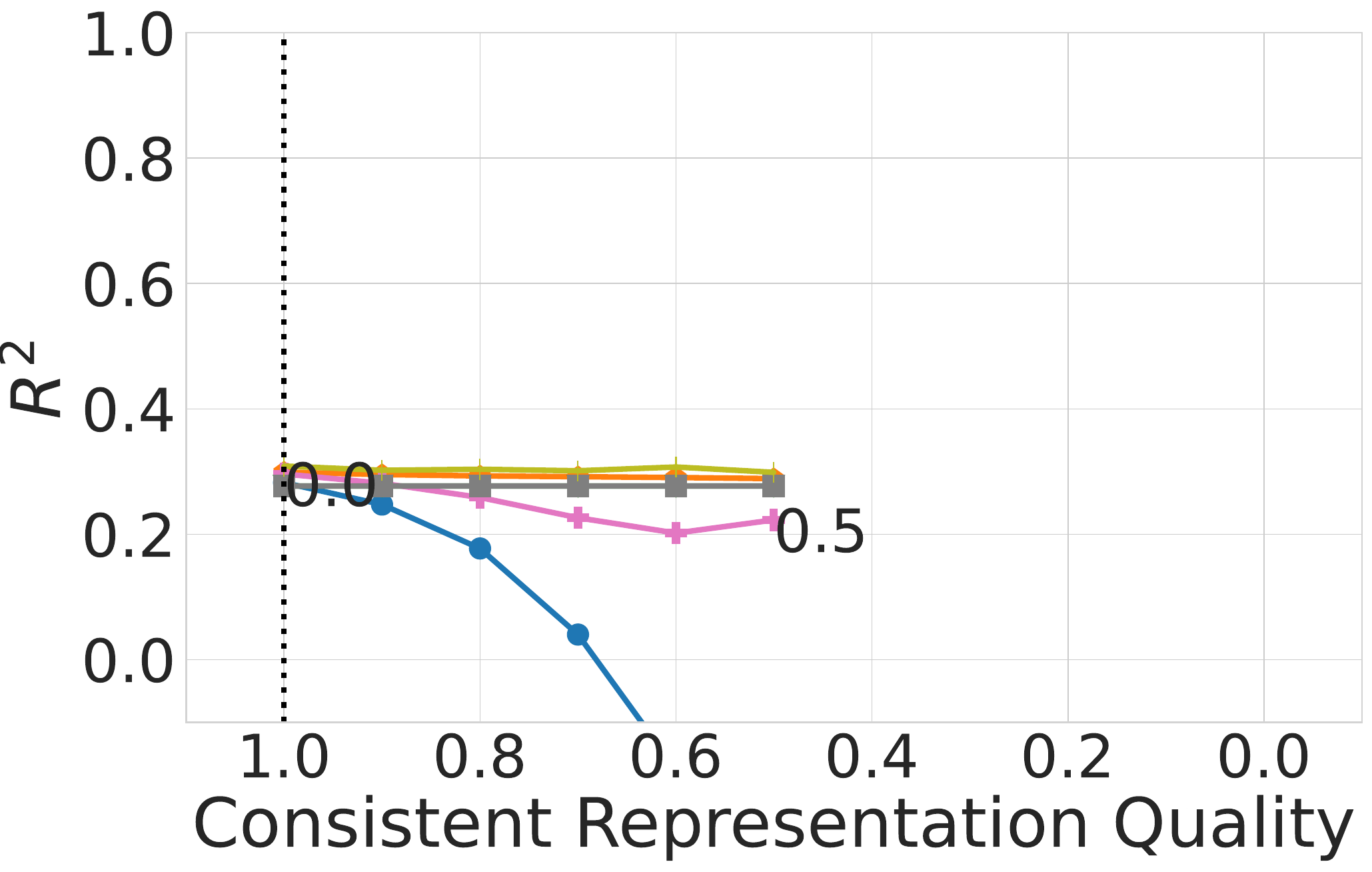}
        \caption{\textsf{COVID}}
        \label{fig:regression-results-all-ConsistentRepresentation-1-covid}
    \end{subfigure}
    \begin{subfigure}[b]{0.23\linewidth}
        \includegraphics[width=\linewidth]{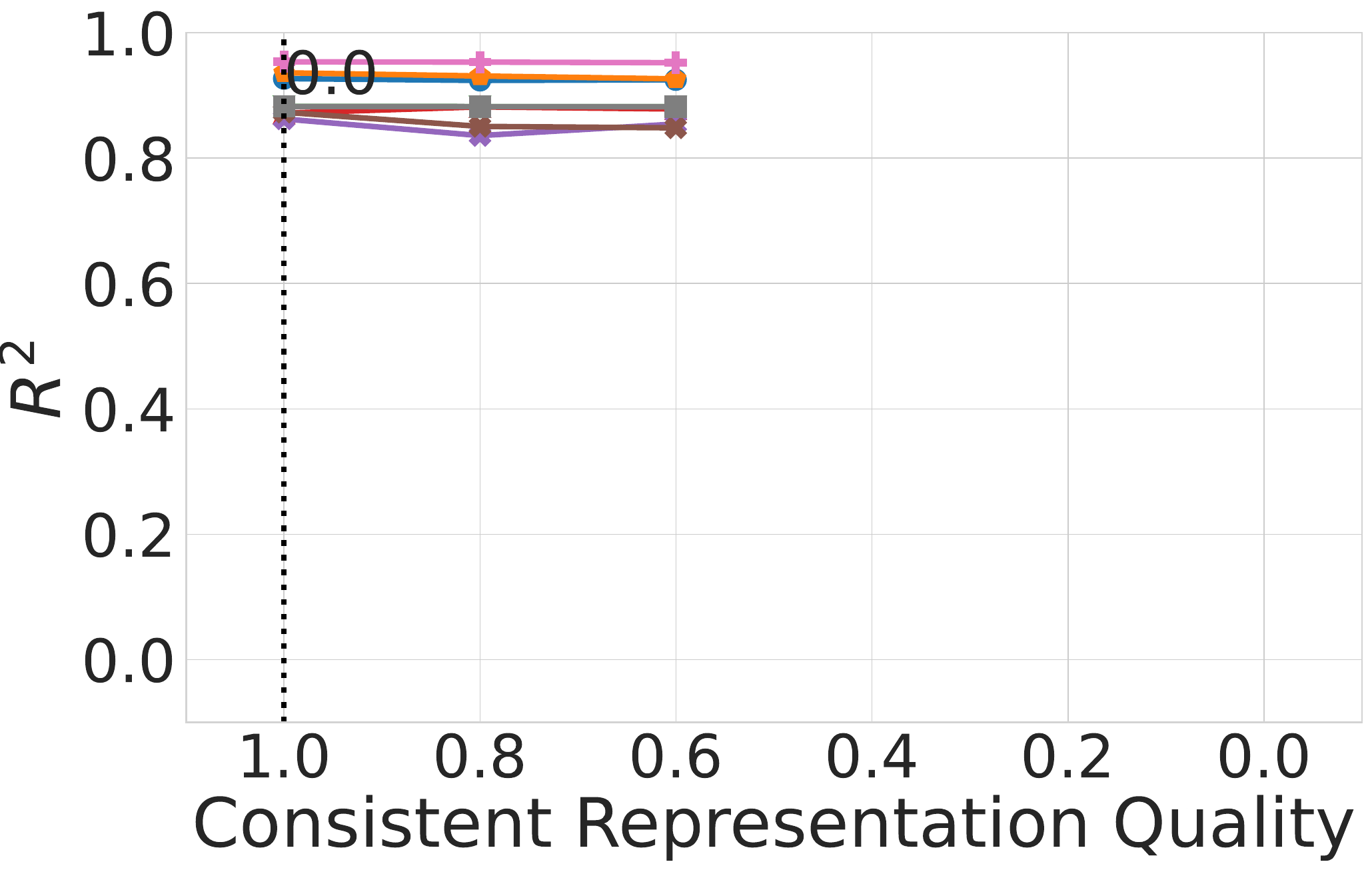}
        \caption{\textsf{Cars}}
        \label{fig:regression-results-all-ConsistentRepresentation-1-cars}
    \end{subfigure}

\raisebox{0.4\height}{\rotatebox{90}{Scenario 2}}\hspace{0.3em}
    \begin{subfigure}[b]{0.23\linewidth}
        \includegraphics[width=\linewidth]{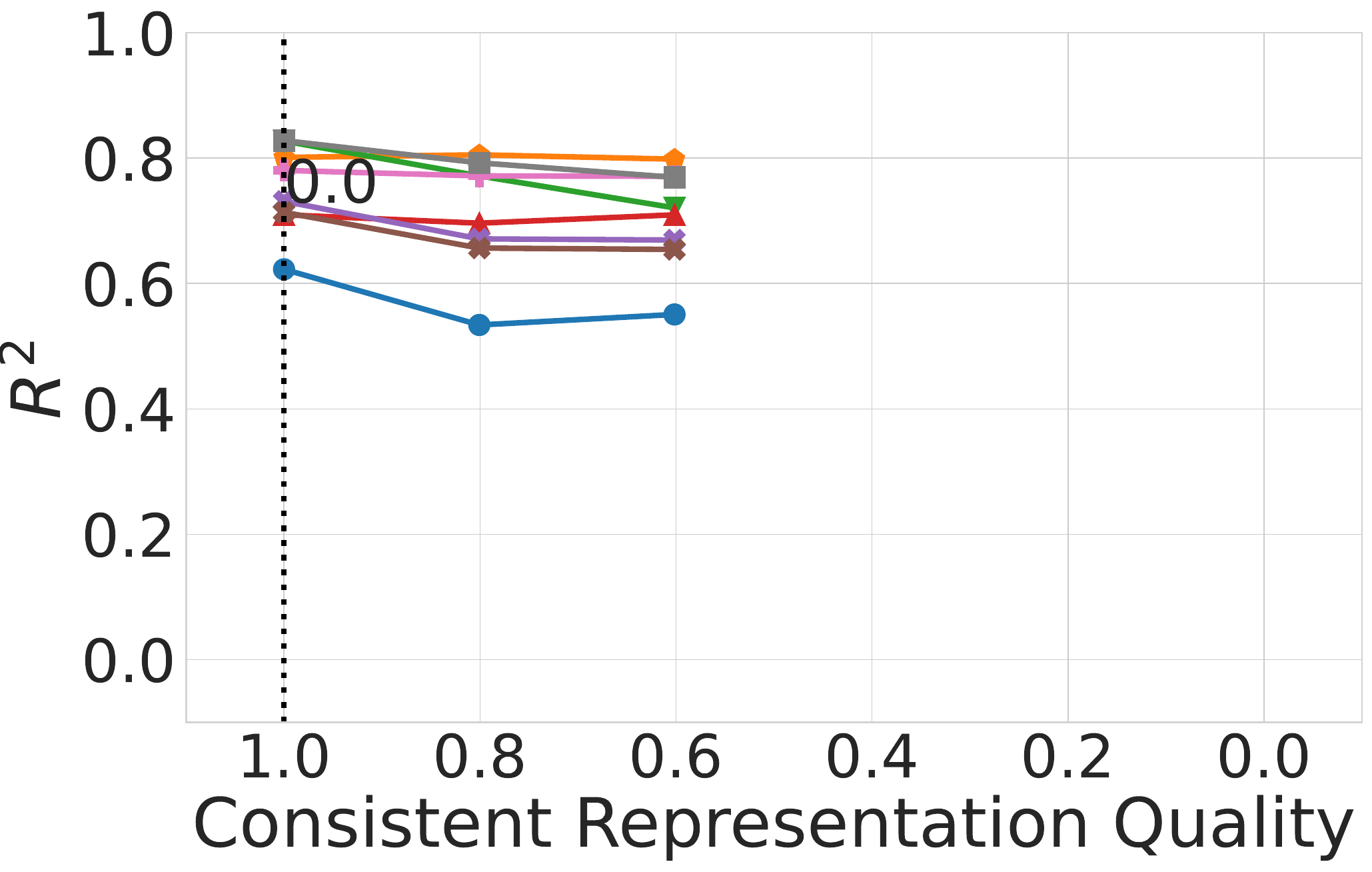}
        \caption{\textsf{Houses}}
        \label{fig:regression-results-all-ConsistentRepresentation-2-houses}
    \end{subfigure}
    \begin{subfigure}[b]{0.23\linewidth}
        \includegraphics[width=\linewidth]{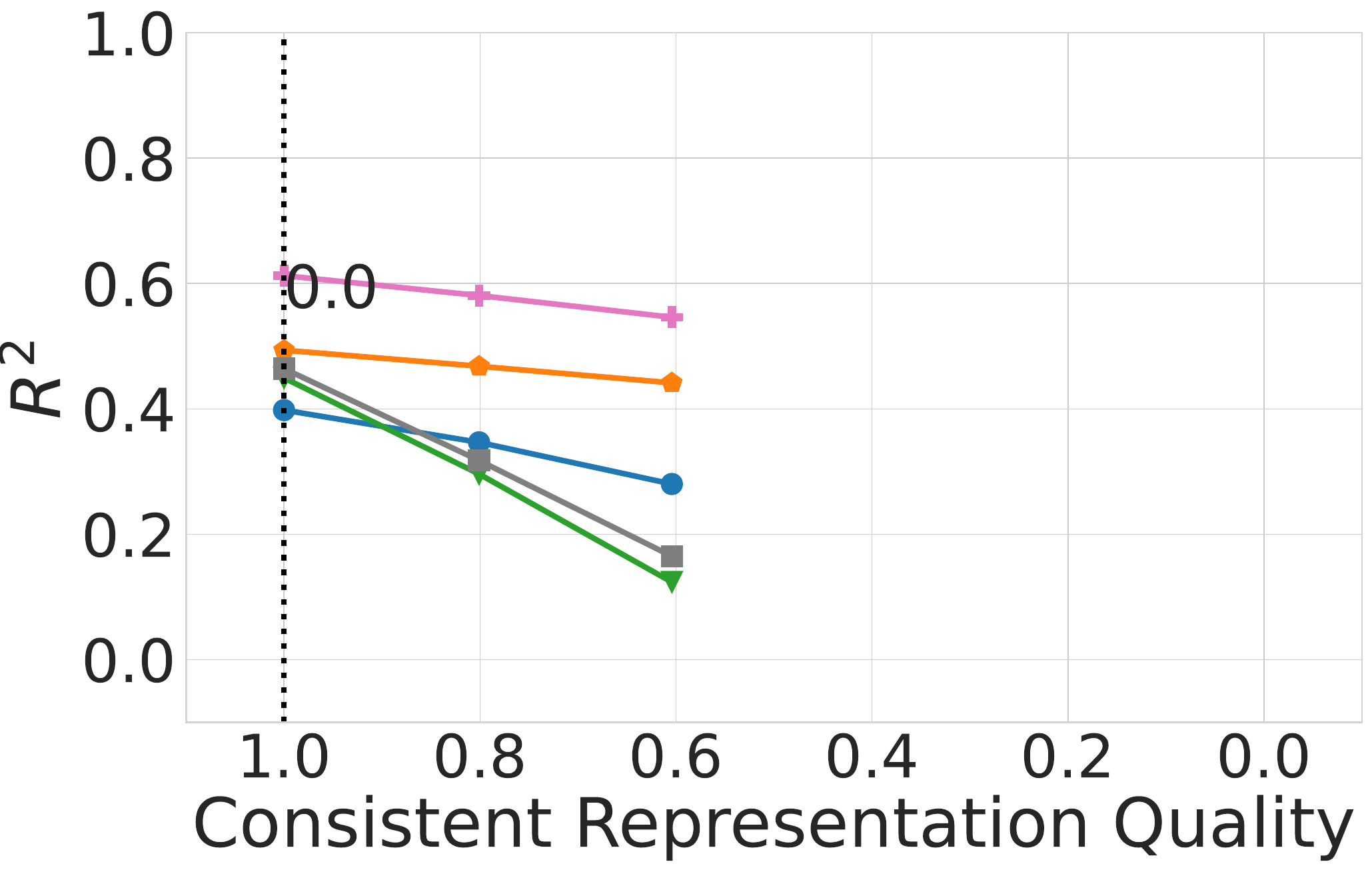}
        \caption{\textsf{IMDB}}
        \label{fig:regression-results-all-ConsistentRepresentation-2-imdb}
    \end{subfigure}
    \begin{subfigure}[b]{0.23\linewidth}
        \includegraphics[width=\linewidth]{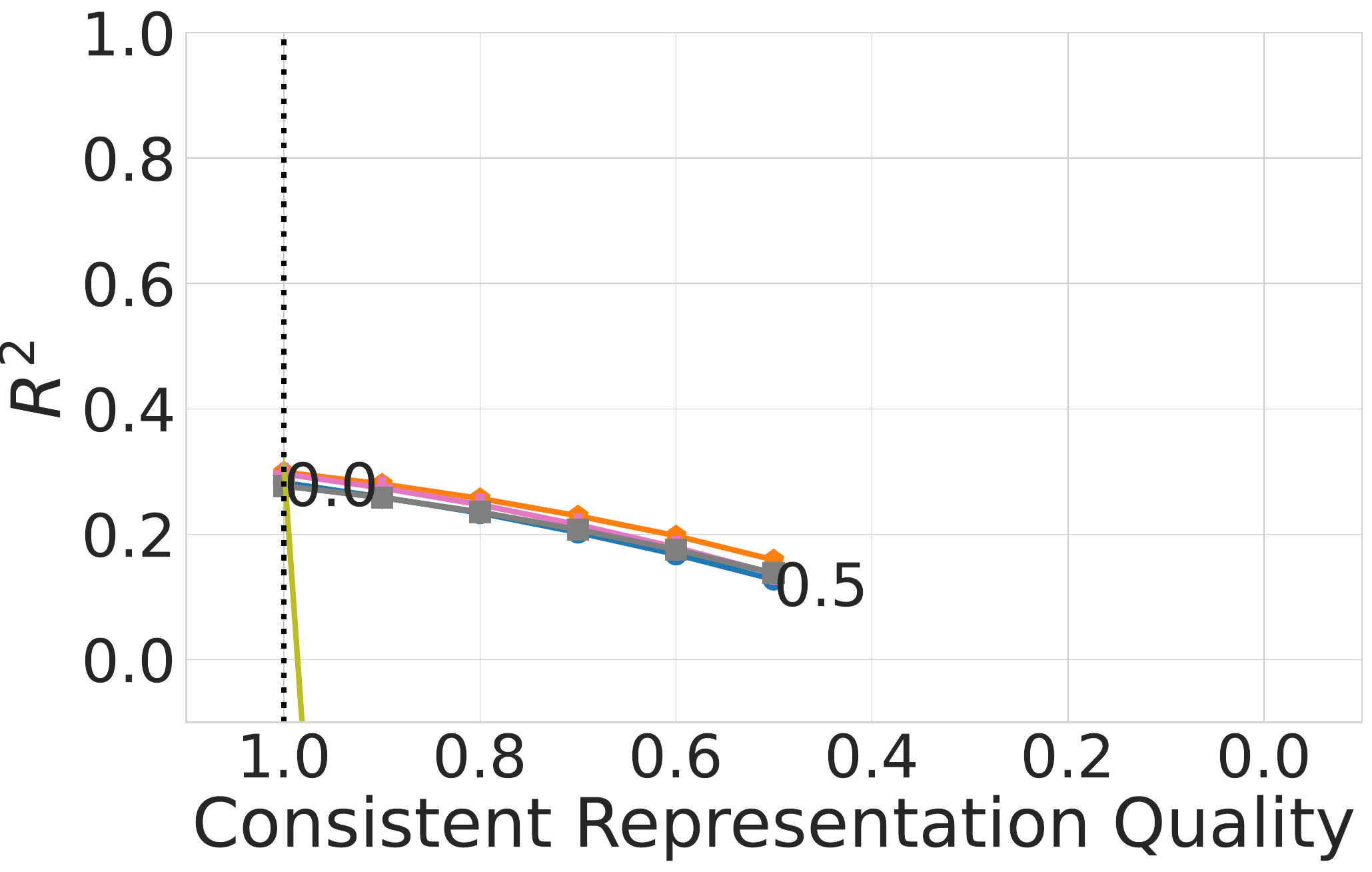}
        \caption{\textsf{COVID}}
        \label{fig:regression-results-all-ConsistentRepresentation-2-covid}
    \end{subfigure}
    \begin{subfigure}[b]{0.23\linewidth}
        \includegraphics[width=\linewidth]{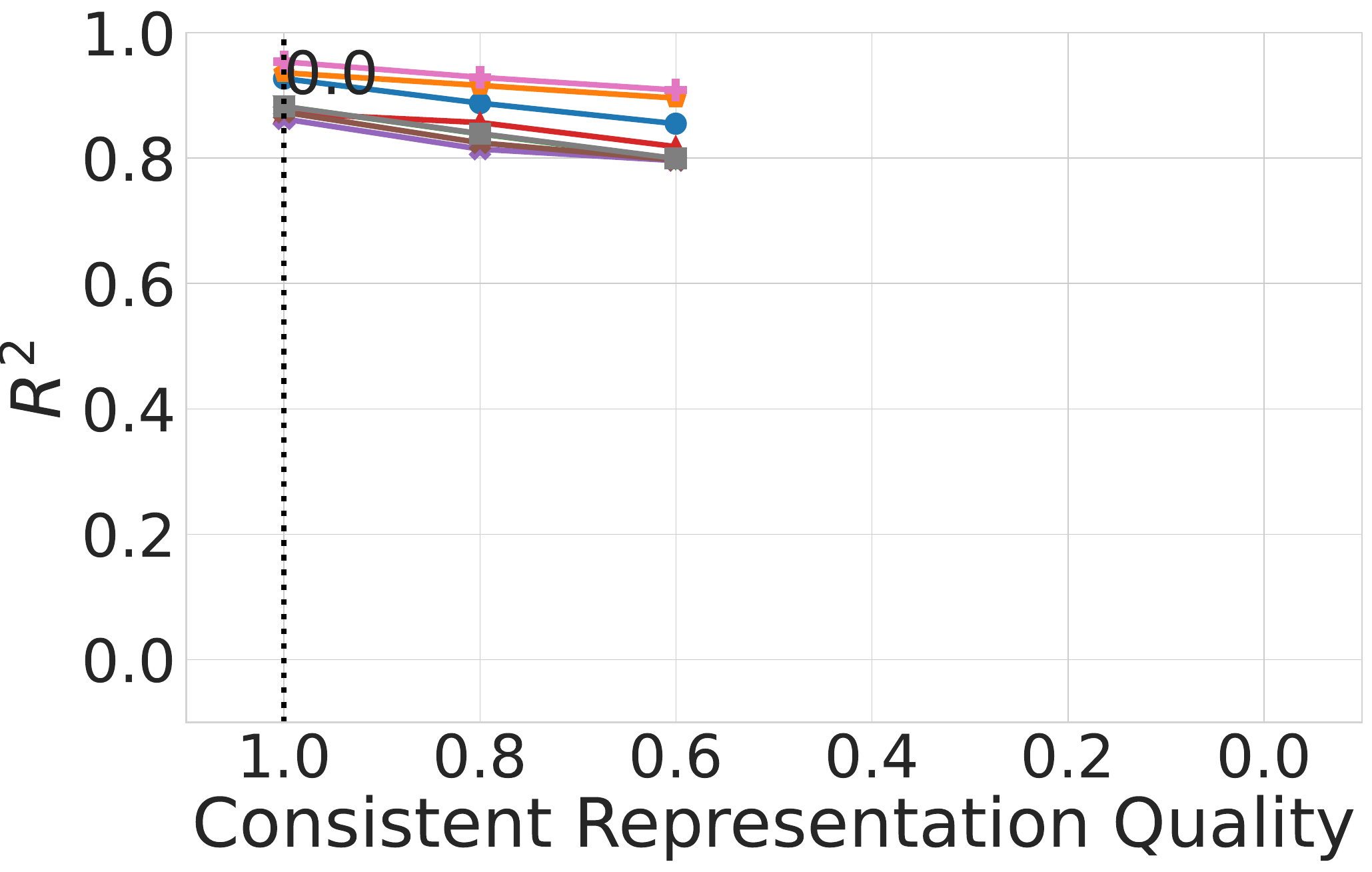}
        \caption{\textsf{Cars}}
        \label{fig:regression-results-all-ConsistentRepresentation-2-cars}
    \end{subfigure}

\raisebox{0.4\height}{\rotatebox{90}{Scenario 3}}\hspace{0.3em}
    \begin{subfigure}[b]{0.23\linewidth}
        \includegraphics[width=\linewidth]{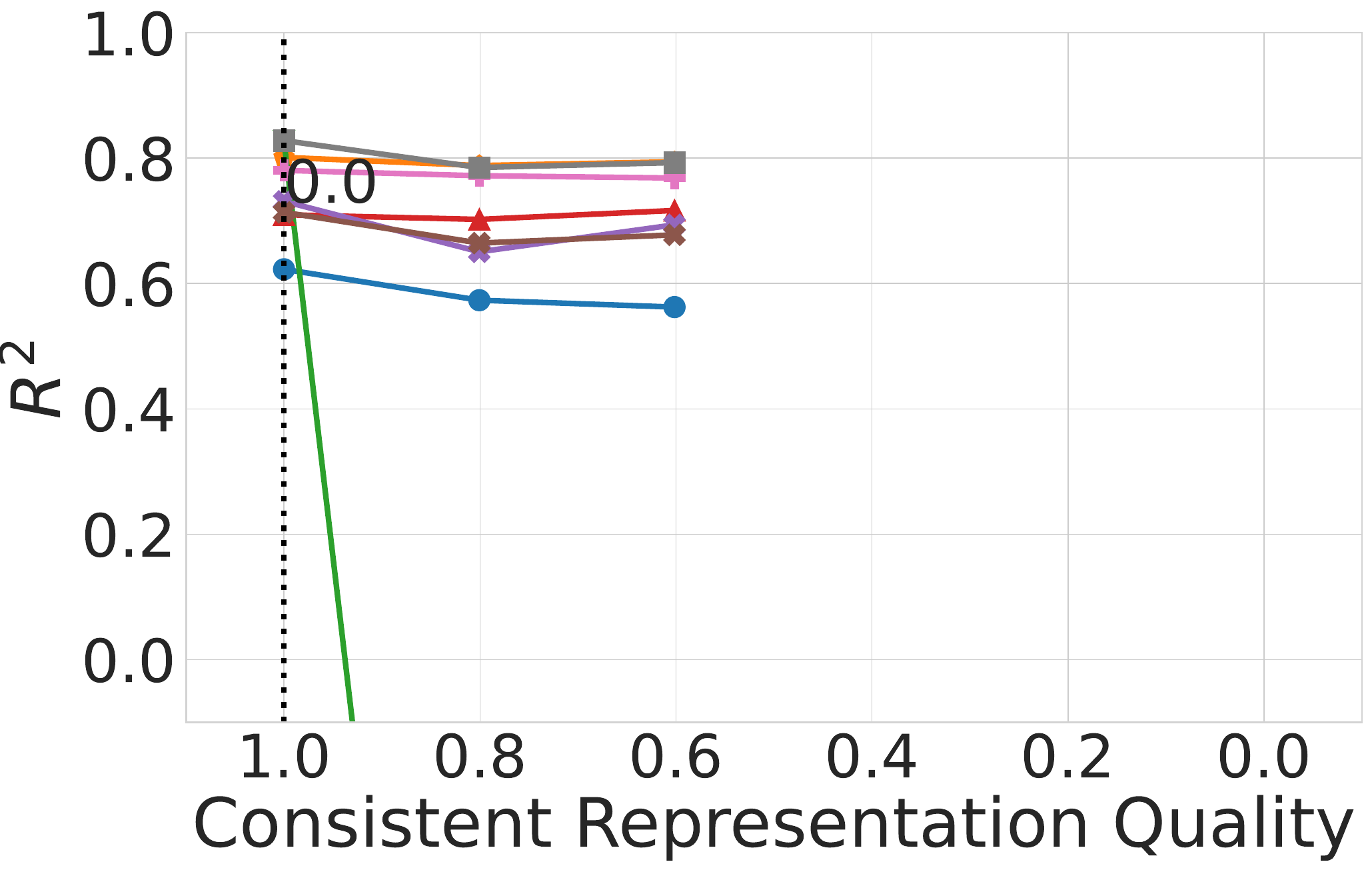}
        \caption{\textsf{Houses}}
        \label{fig:regression-results-all-ConsistentRepresentation-3-houses}
    \end{subfigure}
    \begin{subfigure}[b]{0.23\linewidth}
        \includegraphics[width=\linewidth]{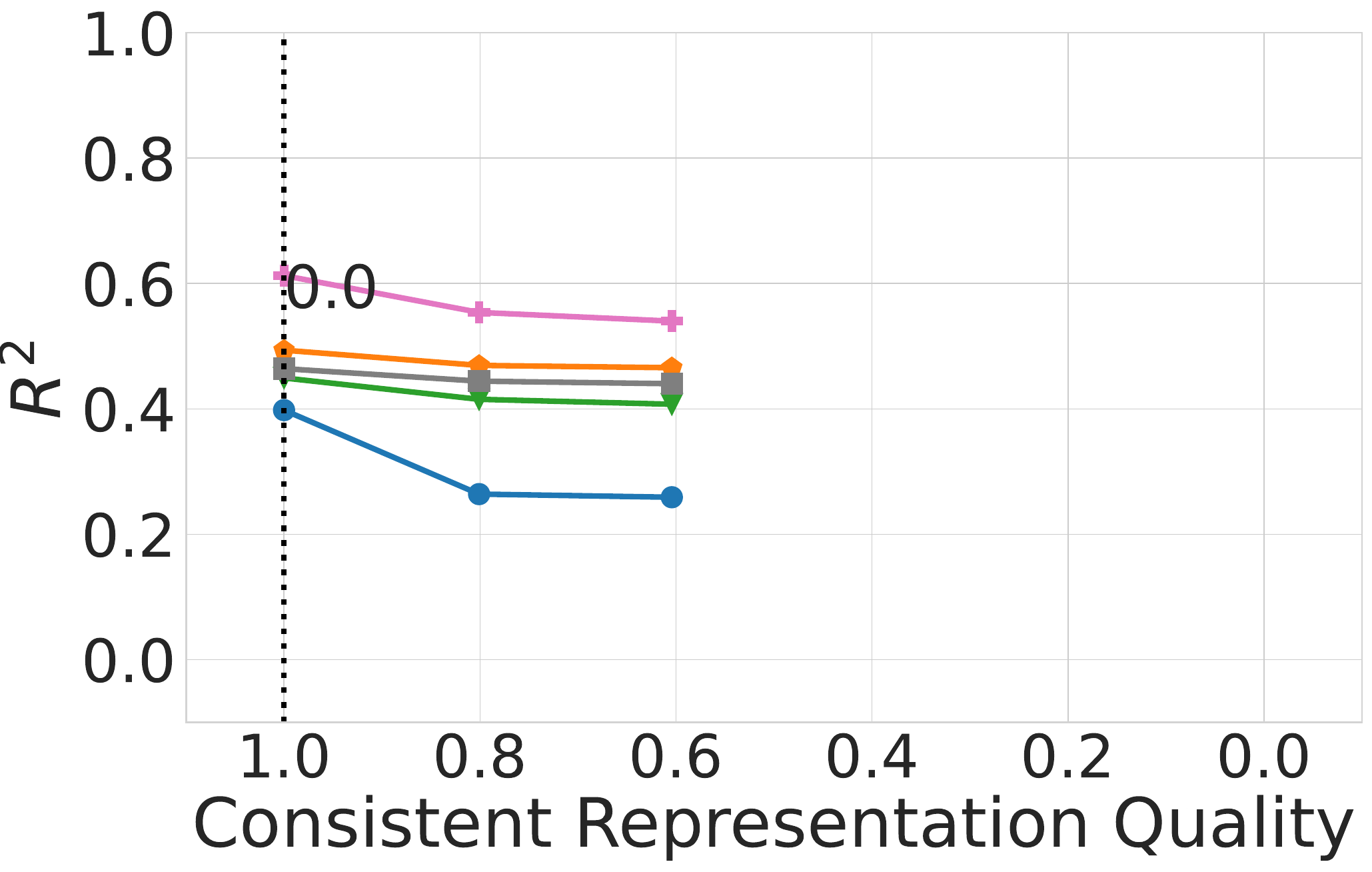}
        \caption{\textsf{IMDB}}
        \label{fig:regression-results-all-ConsistentRepresentation-3-imdb}
    \end{subfigure}
    \begin{subfigure}[b]{0.23\linewidth}
        \includegraphics[width=\linewidth]{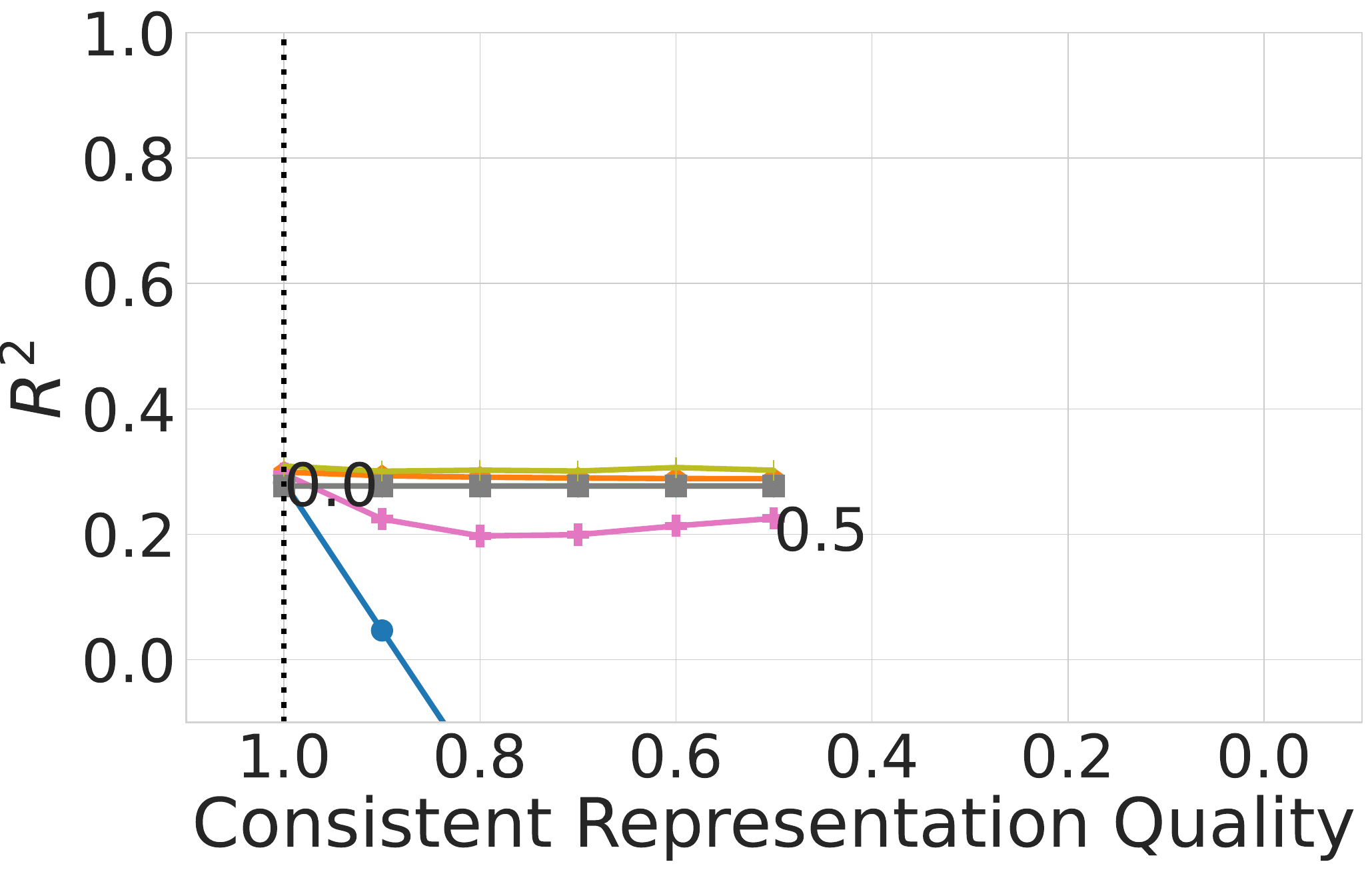}
        \caption{\textsf{COVID}}
        \label{fig:regression-results-all-ConsistentRepresentation-3-covid}
    \end{subfigure}
    \begin{subfigure}[b]{0.23\linewidth}
        \includegraphics[width=\linewidth]{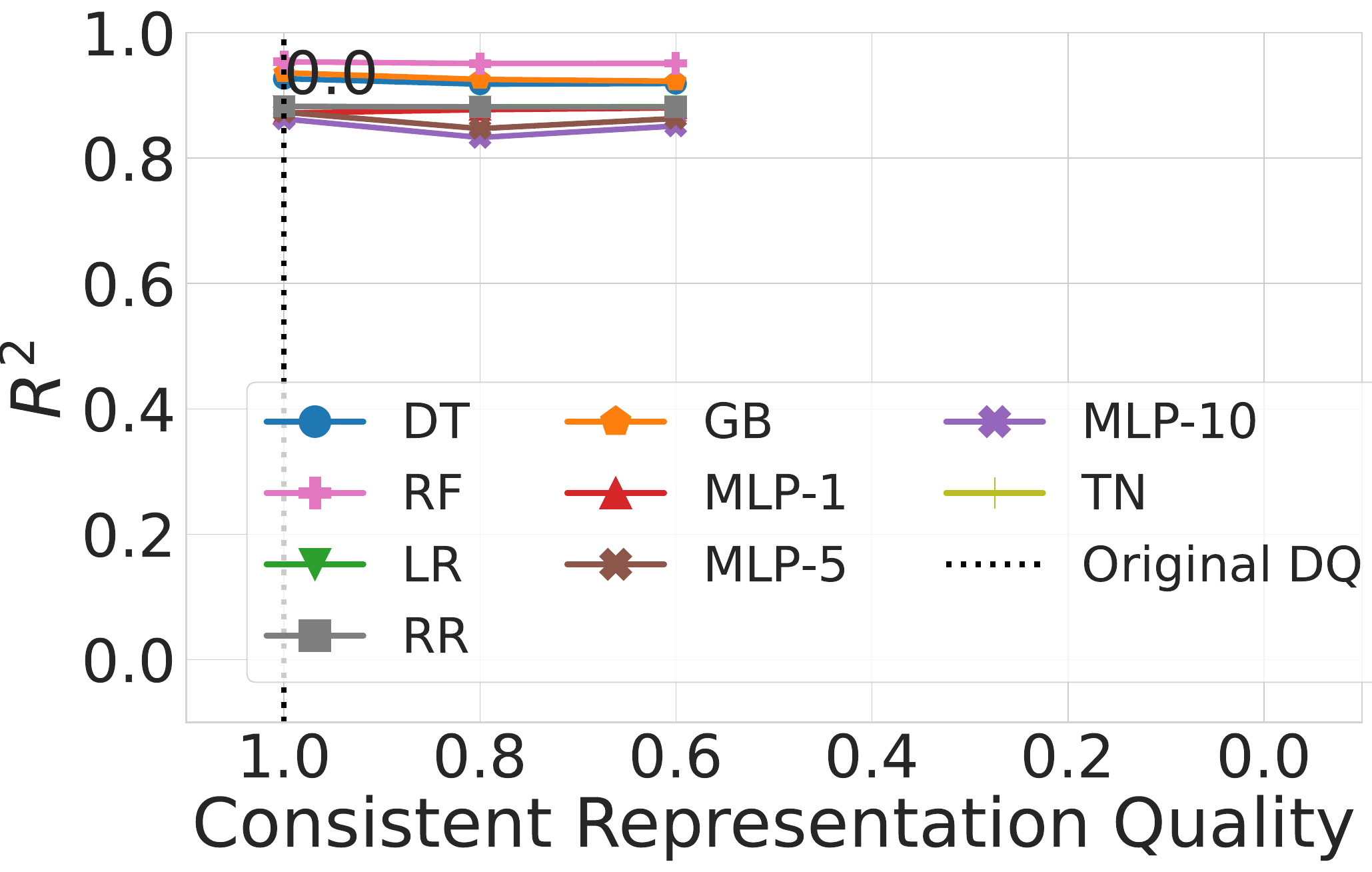}
        \caption{\textsf{Cars}}
        \label{fig:regression-results-all-ConsistentRepresentation-3-cars}
    \end{subfigure}
    \caption{$R^2$ of the regression algorithms for consistent representation with $k_v = 2$.}
    \label{fig:regression-results-all-ConsistentRepresentation}
\end{figure*}

%% file: Latex_Figure/regression/Uniqueness_norm.tex
\begin{figure*}[!htbp]
    \centering
\raisebox{0.4\height}{\rotatebox{90}{Scenario 1}}\hspace{0.3em}
    \begin{subfigure}[b]{0.23\linewidth}
        \includegraphics[width=\linewidth]{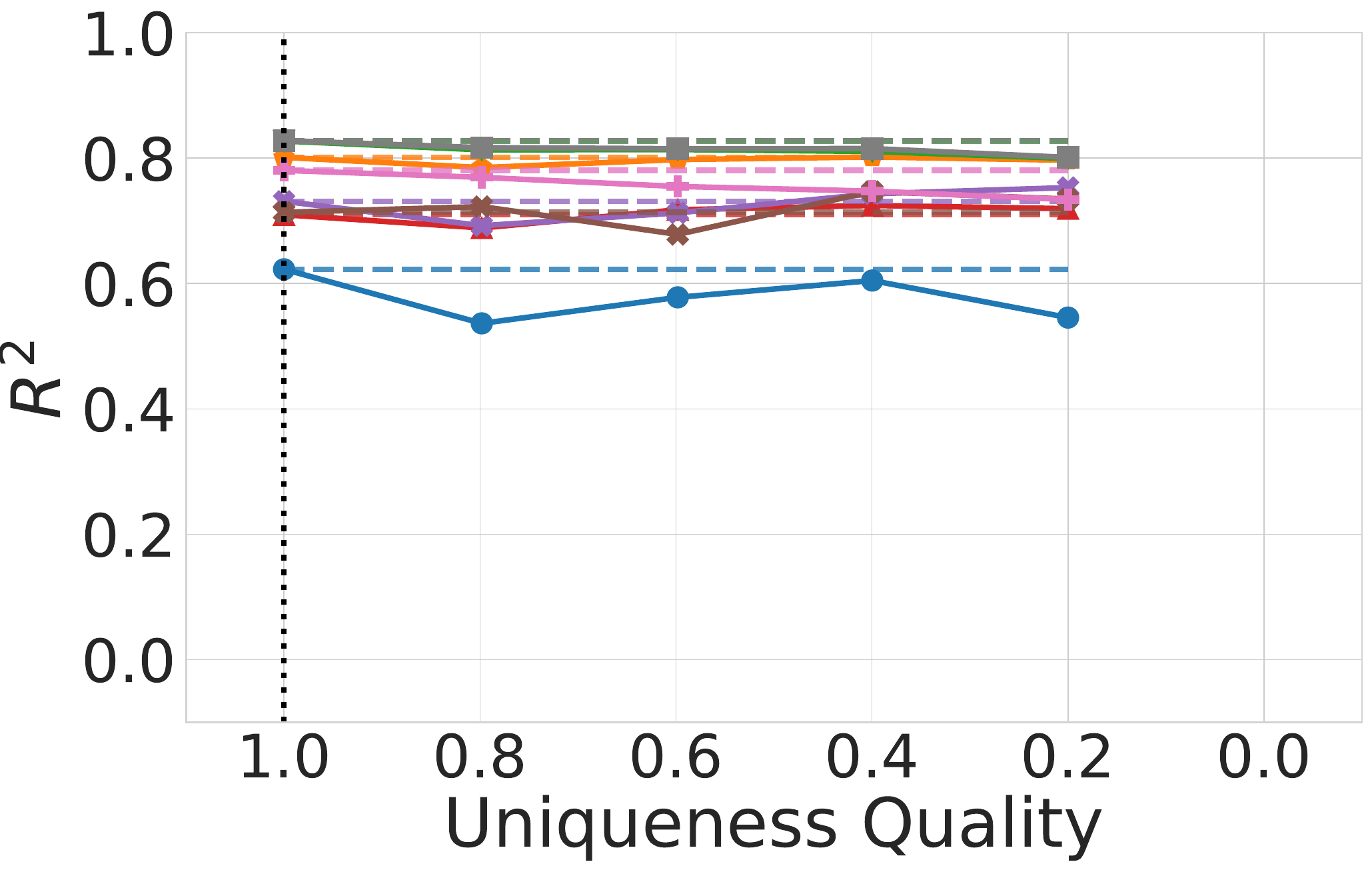}
        \caption{\textsf{Houses}}
        \label{fig:regression-results-all-Uniqueness_dctnormal-1-houses}
    \end{subfigure}
    \begin{subfigure}[b]{0.23\linewidth}
        \includegraphics[width=\linewidth]{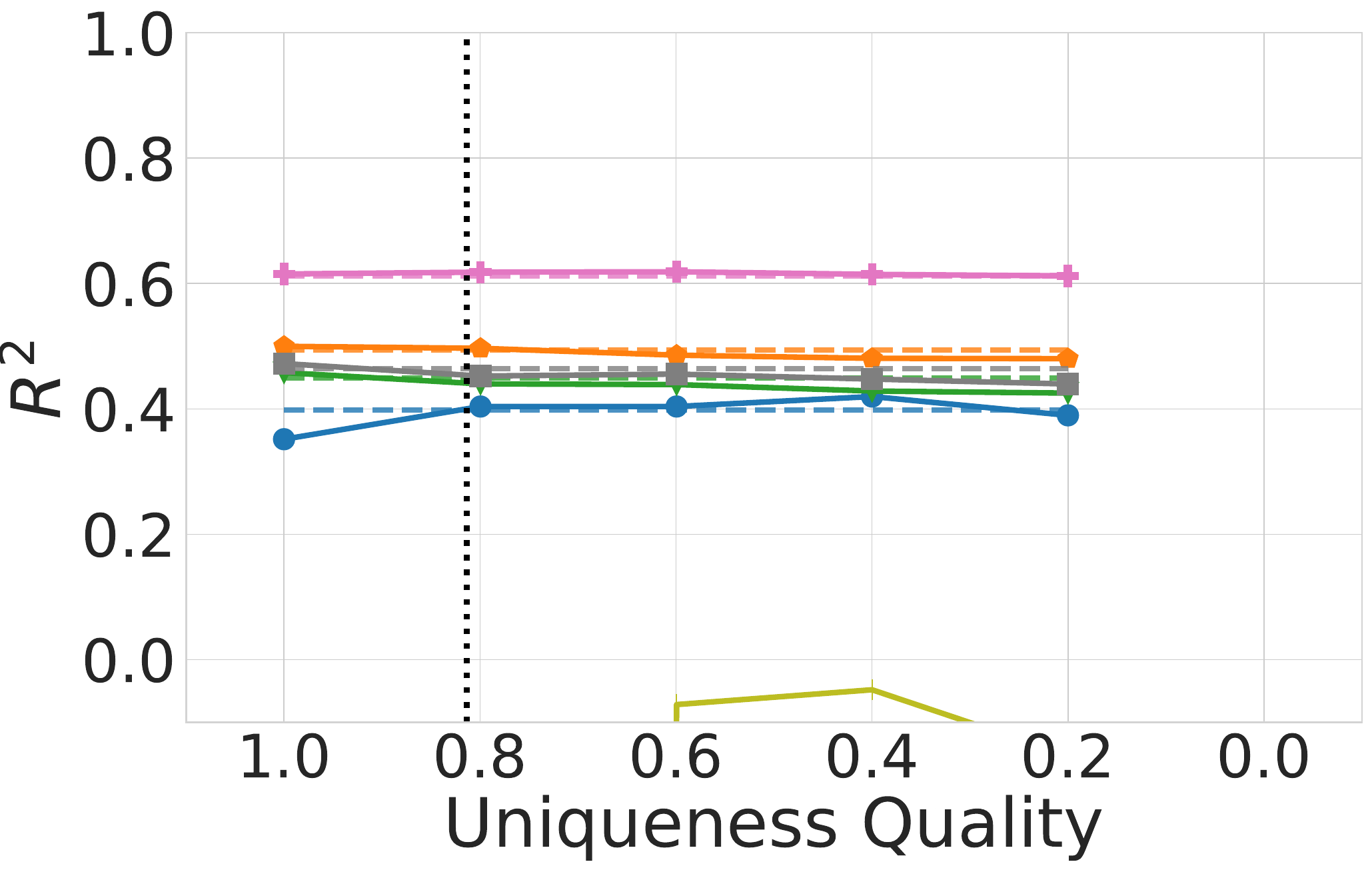}
        \caption{\textsf{IMDB}}
        \label{fig:regression-results-all-Uniqueness_dctnormal-1-imdb}
    \end{subfigure}
    \begin{subfigure}[b]{0.23\linewidth}
        \includegraphics[width=\linewidth]{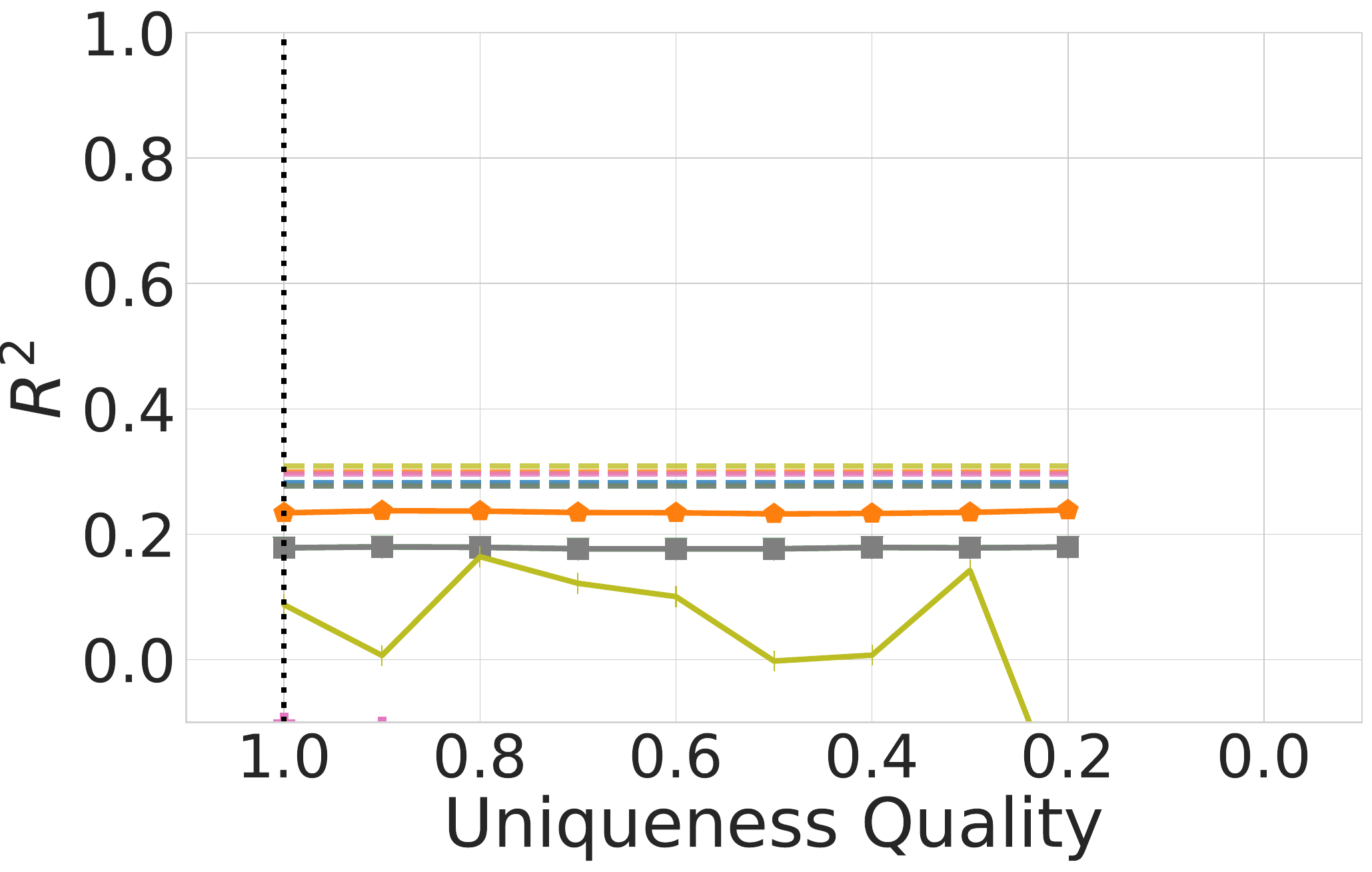}
        \caption{\textsf{COVID}}
        \label{fig:regression-results-all-Uniqueness_dctnormal-1-covid}
    \end{subfigure}
    \begin{subfigure}[b]{0.23\linewidth}
        \includegraphics[width=\linewidth]{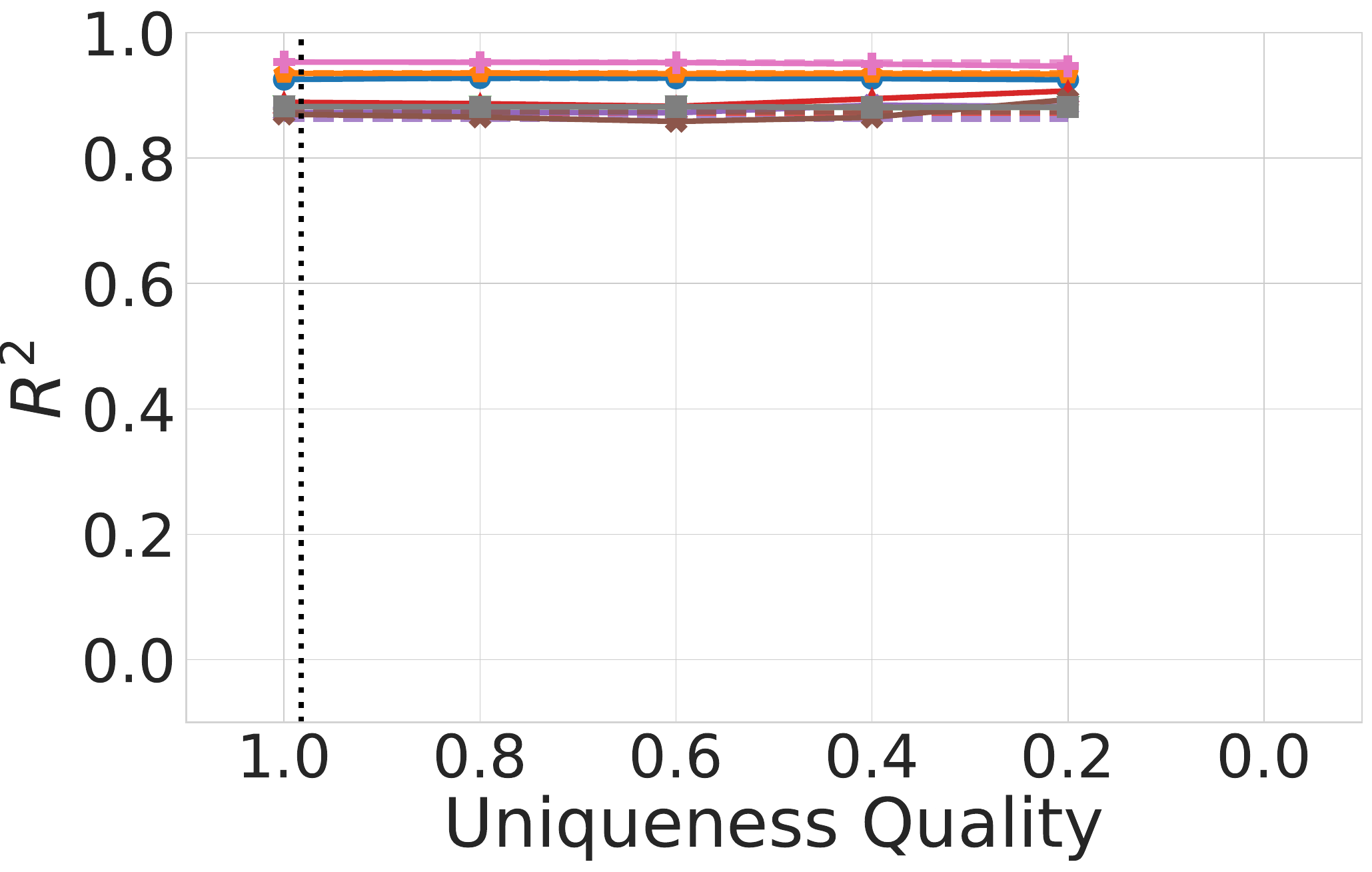}
        \caption{\textsf{Cars}}
        \label{fig:regression-results-all-Uniqueness_dctnormal-1-cars}
    \end{subfigure}

\raisebox{0.4\height}{\rotatebox{90}{Scenario 2}}\hspace{0.3em}
    \begin{subfigure}[b]{0.23\linewidth}
        \includegraphics[width=\linewidth]{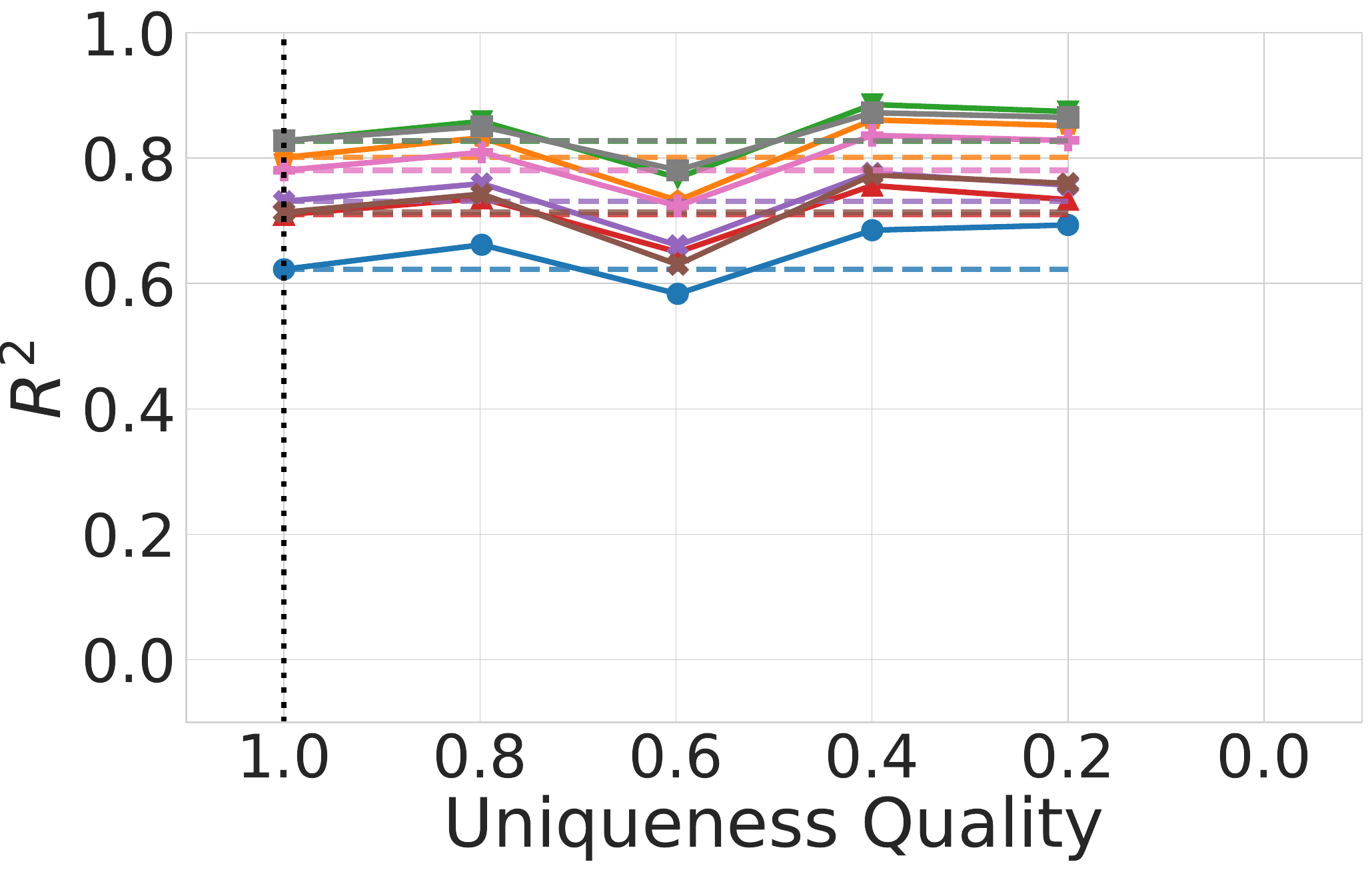}
        \caption{\textsf{Houses}}
        \label{fig:regression-results-all-Uniqueness_dctnormal-2-houses}
    \end{subfigure}
    \begin{subfigure}[b]{0.23\linewidth}
        \includegraphics[width=\linewidth]{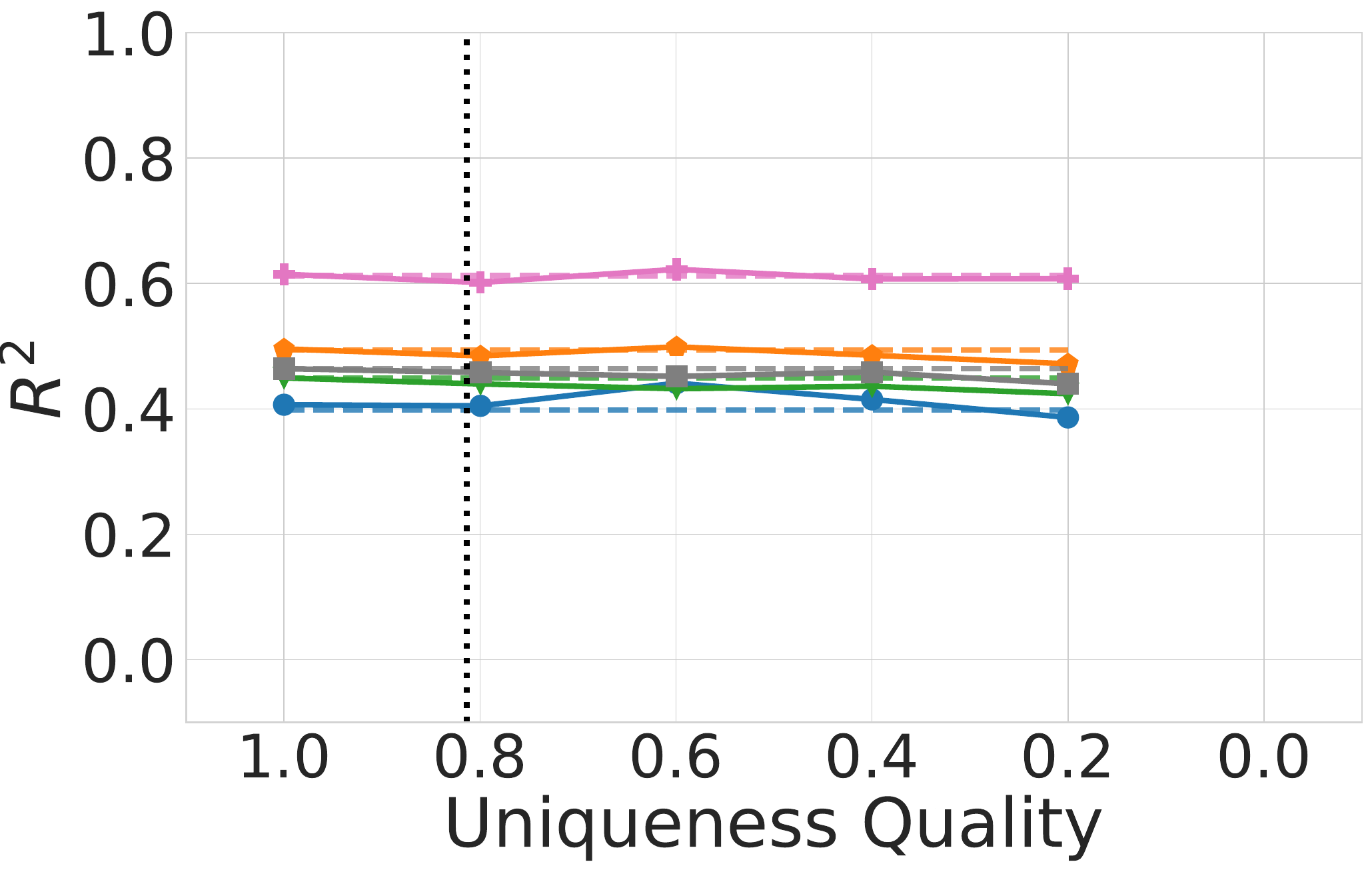}
        \caption{\textsf{IMDB}}
        \label{fig:regression-results-all-Uniqueness_dctnormal-2-imdb}
    \end{subfigure}
    \begin{subfigure}[b]{0.23\linewidth}
        \includegraphics[width=\linewidth]{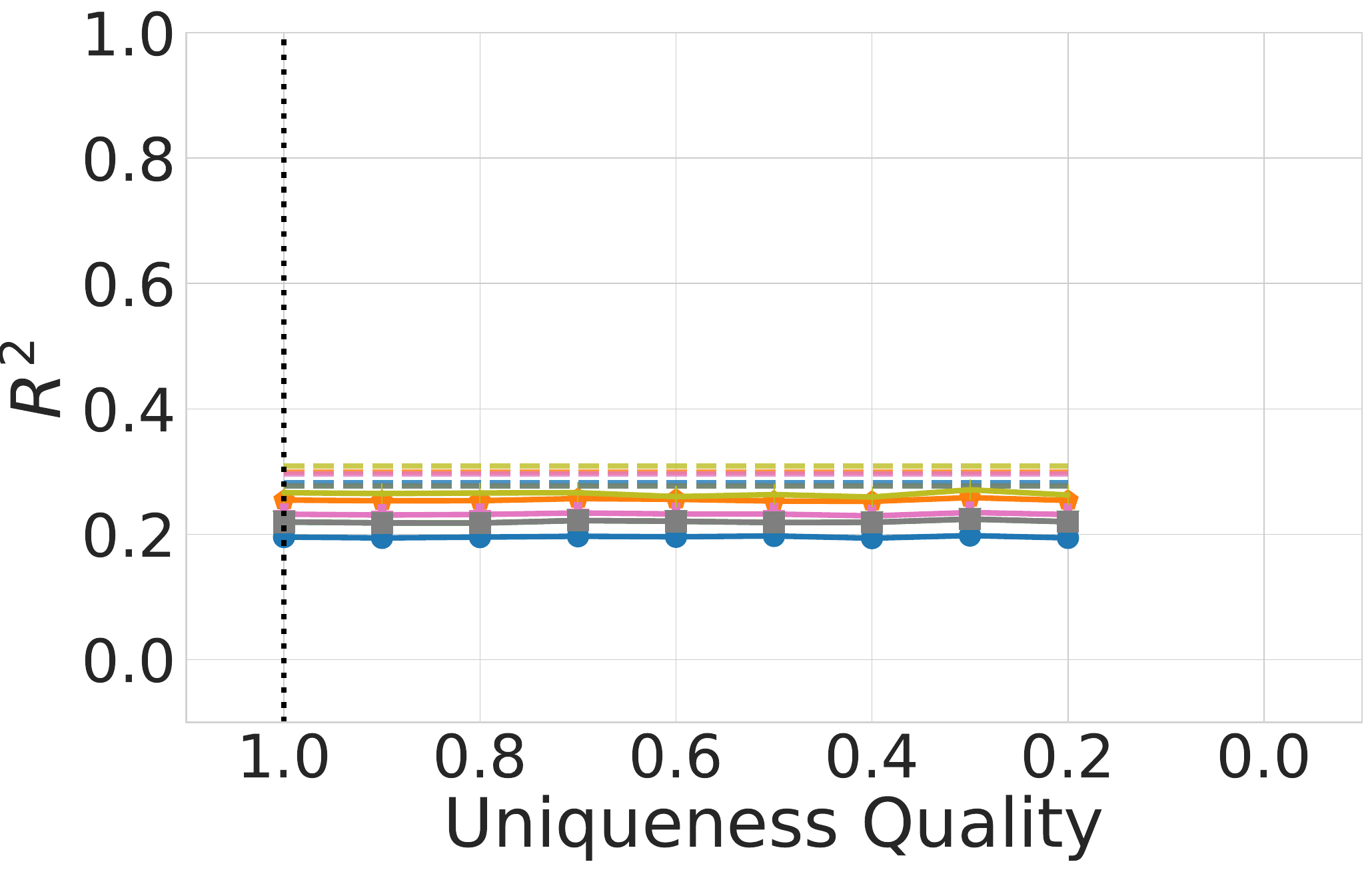}
        \caption{\textsf{COVID}}
        \label{fig:regression-results-all-Uniqueness_dctnormal-2-covid}
    \end{subfigure}
    \begin{subfigure}[b]{0.23\linewidth}
        \includegraphics[width=\linewidth]{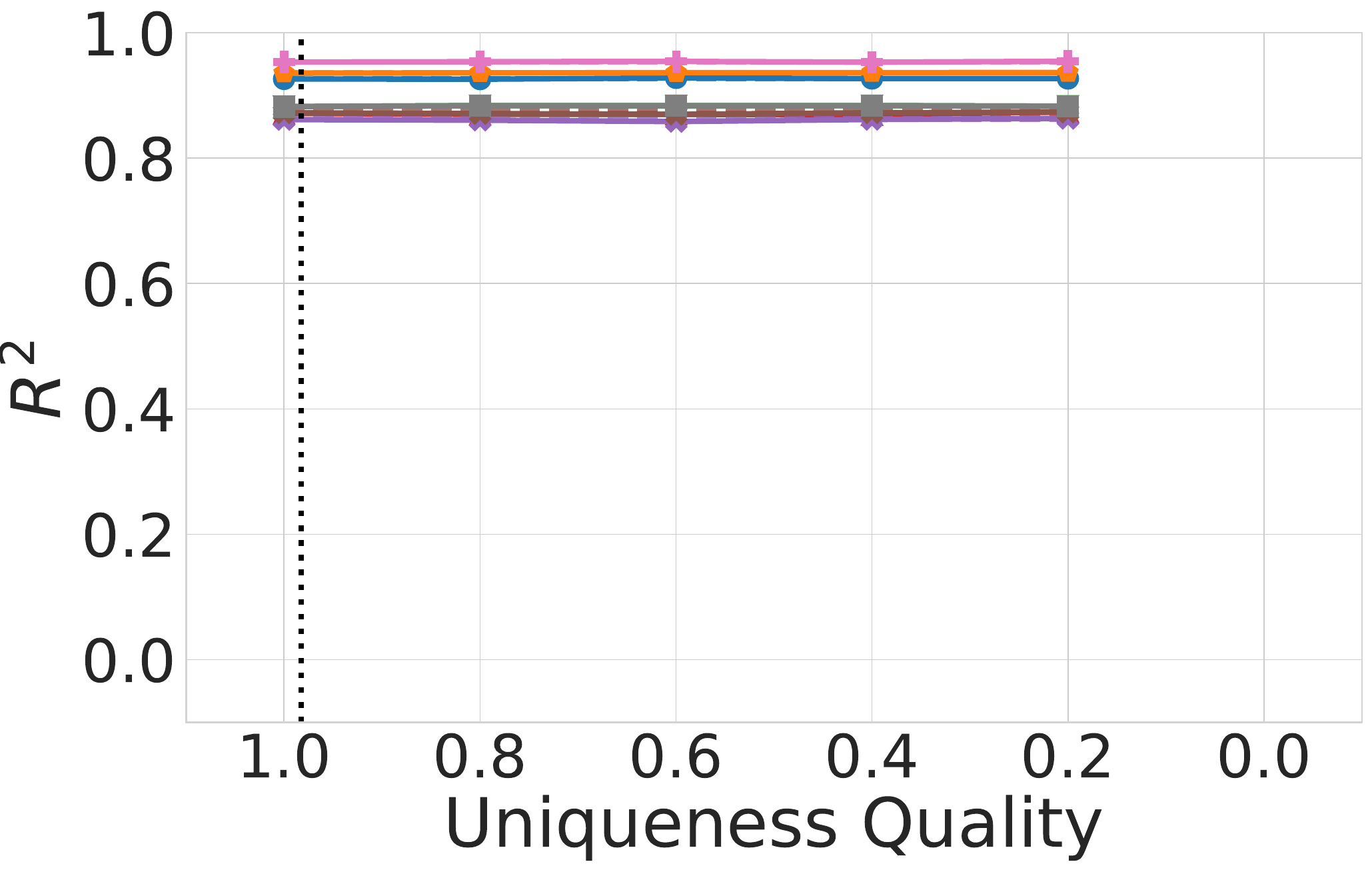}
        \caption{\textsf{Cars}}
        \label{fig:regression-results-all-Uniqueness_dctnormal-2-cars}
    \end{subfigure}

\raisebox{0.4\height}{\rotatebox{90}{Scenario 3}}\hspace{0.3em}
    \begin{subfigure}[b]{0.23\linewidth}
        \includegraphics[width=\linewidth]{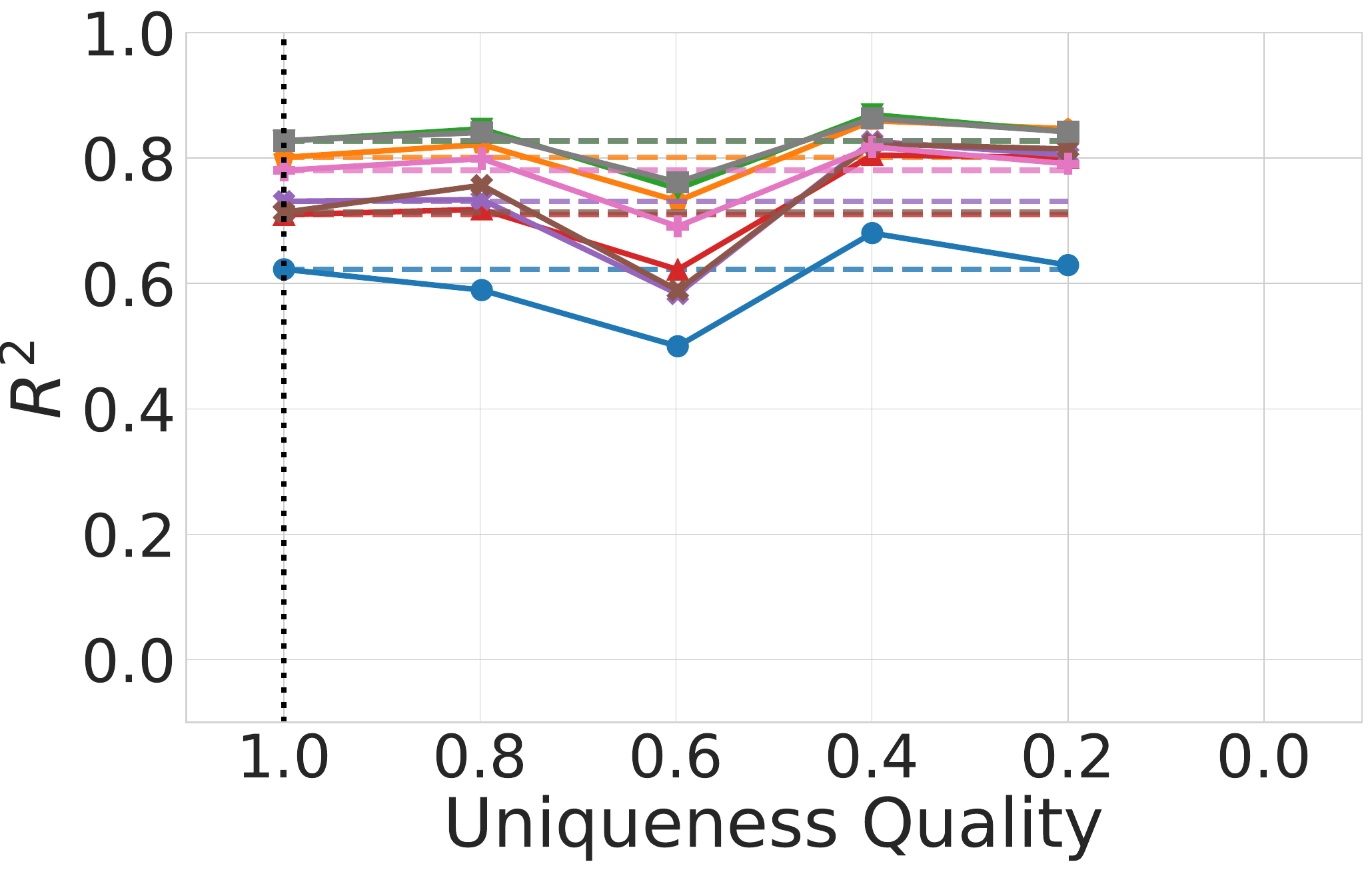}
        \caption{\textsf{Houses}}
        \label{fig:regression-results-all-Uniqueness_dctnormal-3-houses}
    \end{subfigure}
    \begin{subfigure}[b]{0.23\linewidth}
        \includegraphics[width=\linewidth]{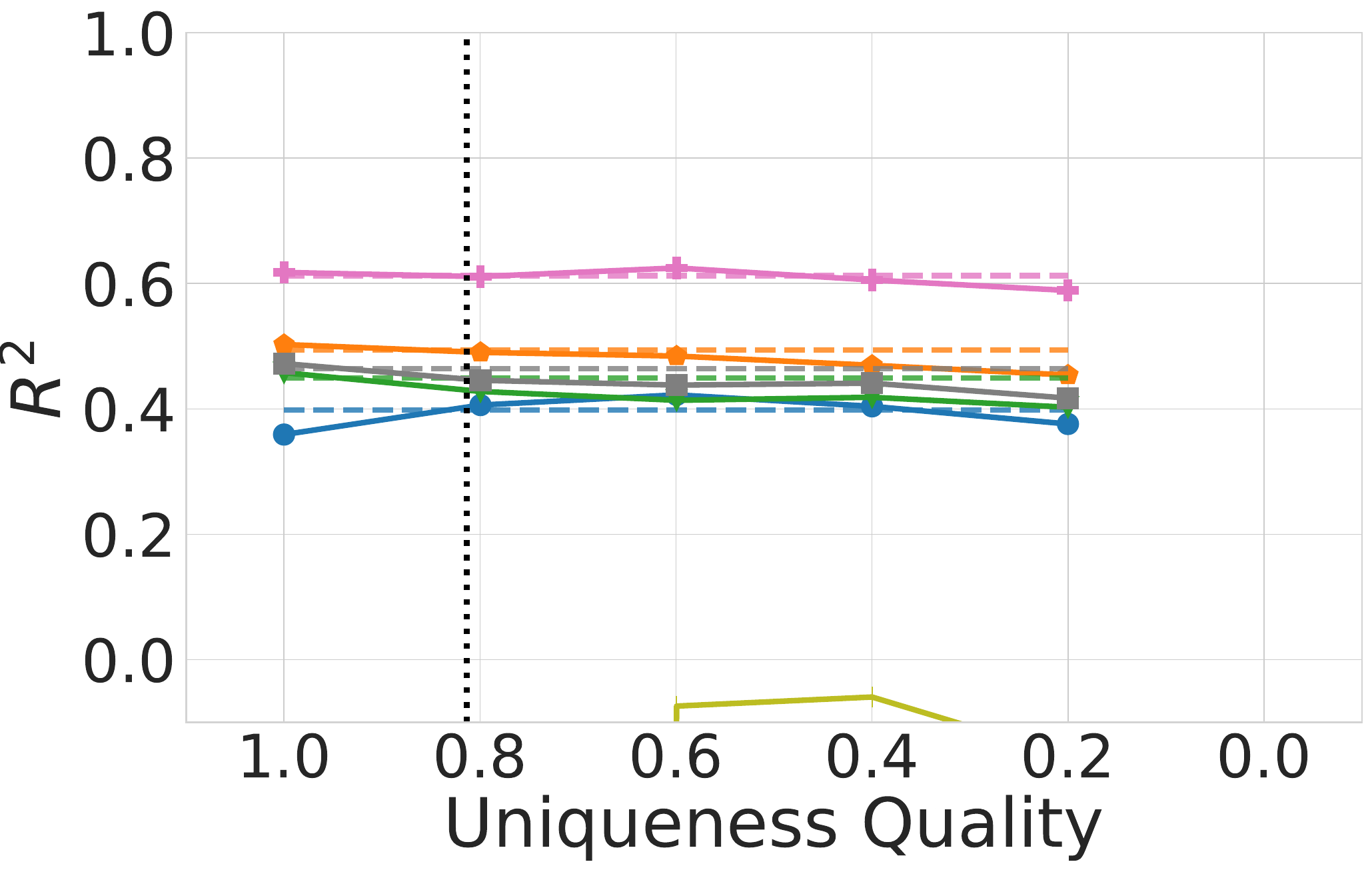}
        \caption{\textsf{IMDB}}
        \label{fig:regression-results-all-Uniqueness_dctnormal-3-imdb}
    \end{subfigure}
    \begin{subfigure}[b]{0.23\linewidth}
        \includegraphics[width=\linewidth]{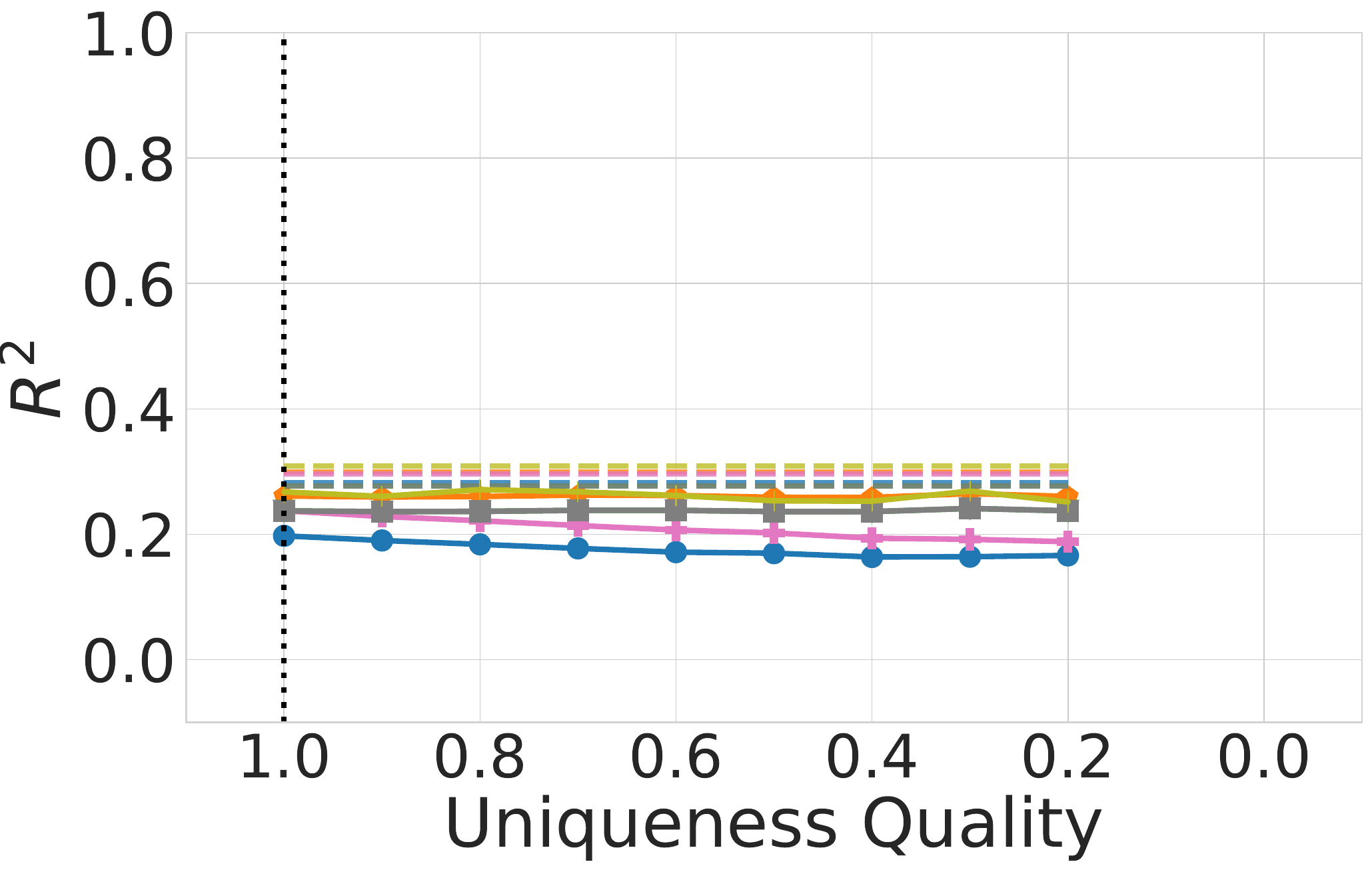}
        \caption{\textsf{COVID}}
        \label{fig:regression-results-all-Uniqueness_dctnormal-3-covid}
    \end{subfigure}
    \begin{subfigure}[b]{0.23\linewidth}
        \includegraphics[width=\linewidth]{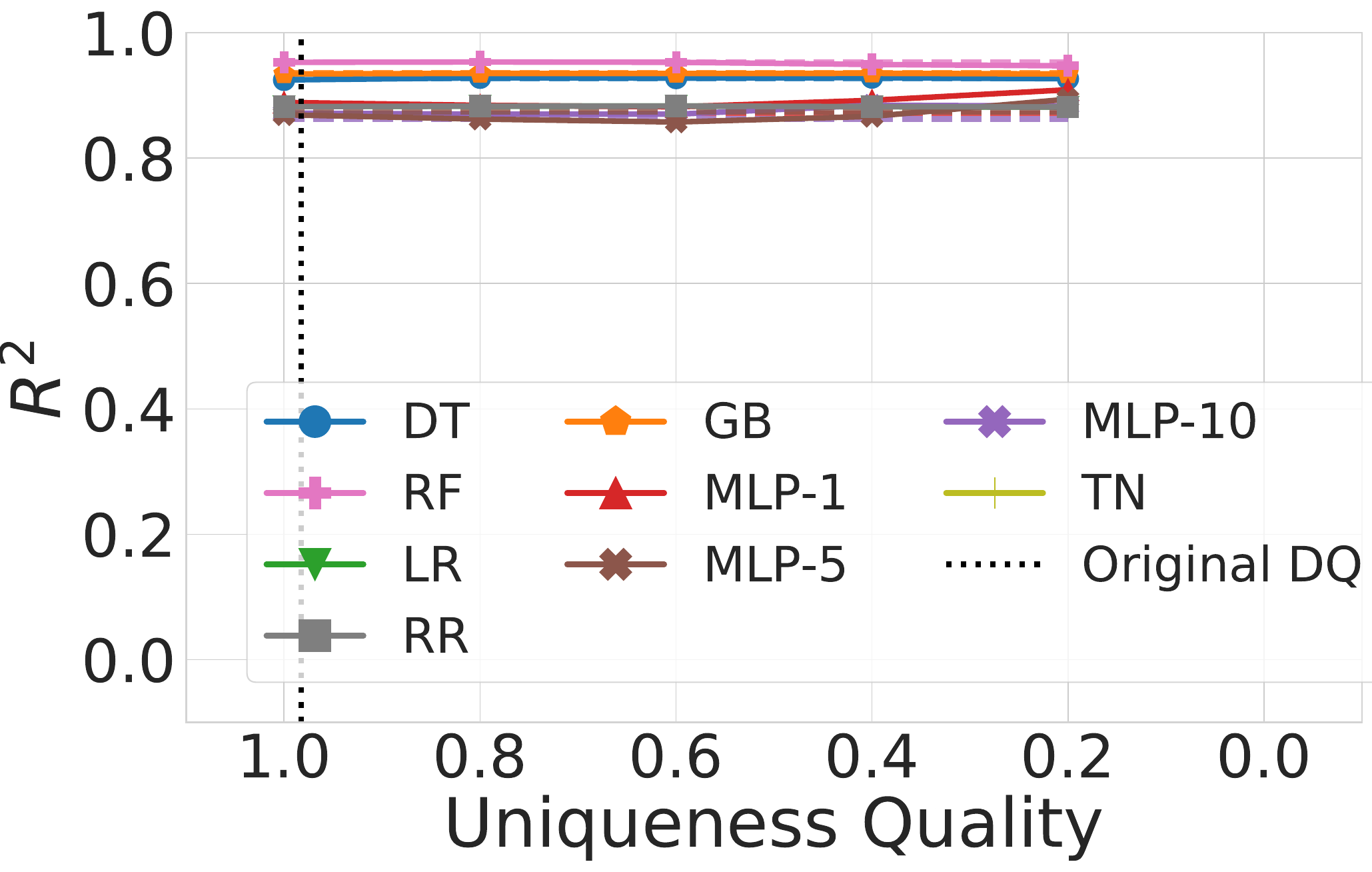}
        \caption{\textsf{Cars}}
        \label{fig:regression-results-all-Uniqueness_dctnormal-3-cars}
    \end{subfigure}
    \caption{$R^2$ of the regression algorithms for uniqueness with duplicate count sampled by normal distribution.}
    \label{fig:regression-results-all-Uniqueness_dctnormal}
\end{figure*}

%% file: Latex_Figure/clustering/n_cluster.tex
\begin{figure*}[!hbp]
    \centering
    \begin{subfigure}[b]{0.23\linewidth}
        \includegraphics[width=\linewidth]{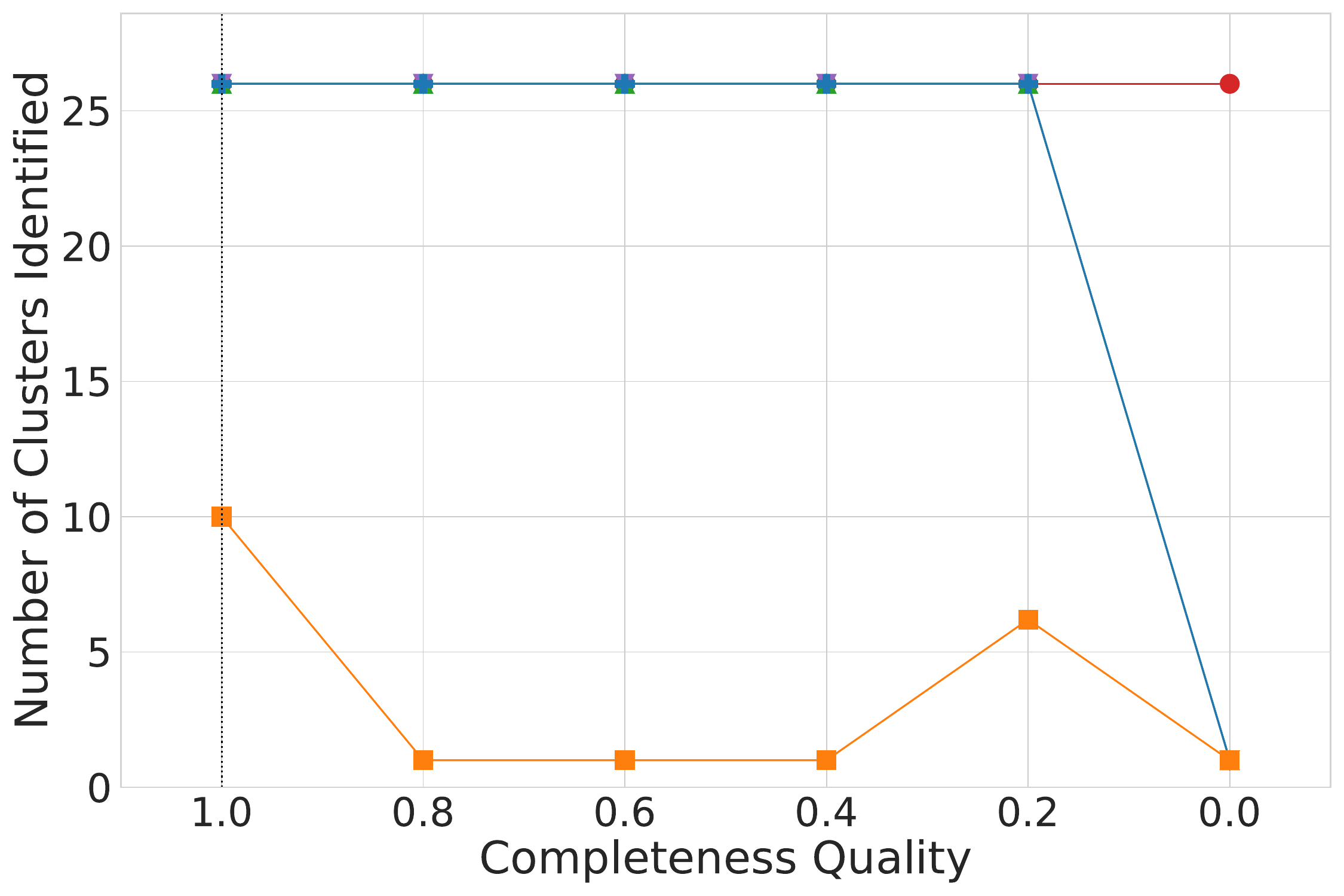}
        \caption{\textsf{Letter}, expected 26 clusters}
        \label{fig:clustering-completeness-letter-nclusters}
    \end{subfigure}
    \begin{subfigure}[b]{0.23\linewidth}
        \includegraphics[width=\linewidth]{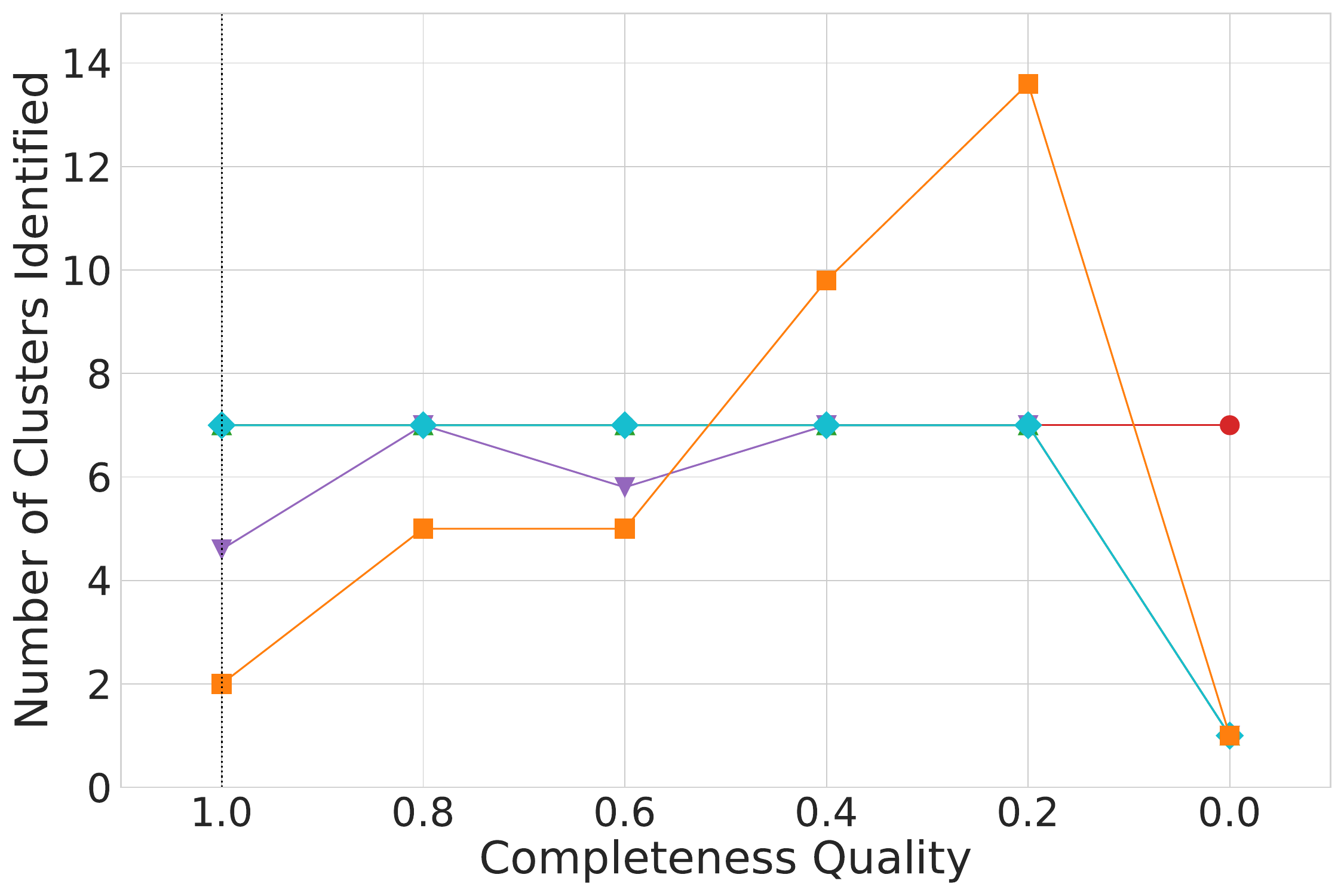}
        \caption{\textsf{Covertype}, expected 7 clusters}
        \label{fig:clustering-completeness-covertype-nclusters}
    \end{subfigure}
    \begin{subfigure}[b]{0.23\linewidth}
        \includegraphics[width=\linewidth]{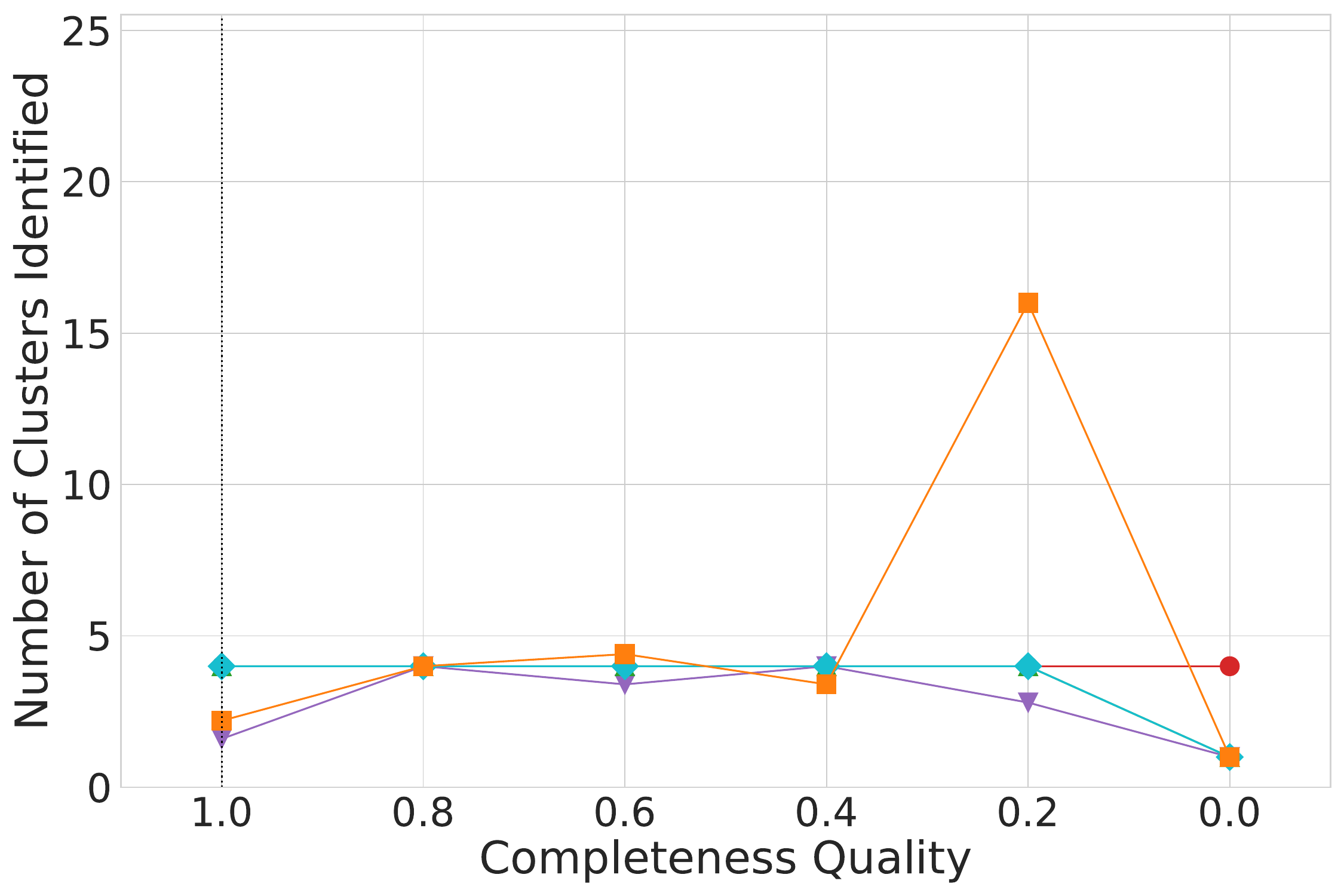}
        \caption{\textsf{COVID}, expected 4 clusters}
        \label{fig:clustering-completeness-covid-nclusters}
    \end{subfigure}
    \begin{subfigure}[b]{0.23\linewidth}
        \includegraphics[width=\linewidth]{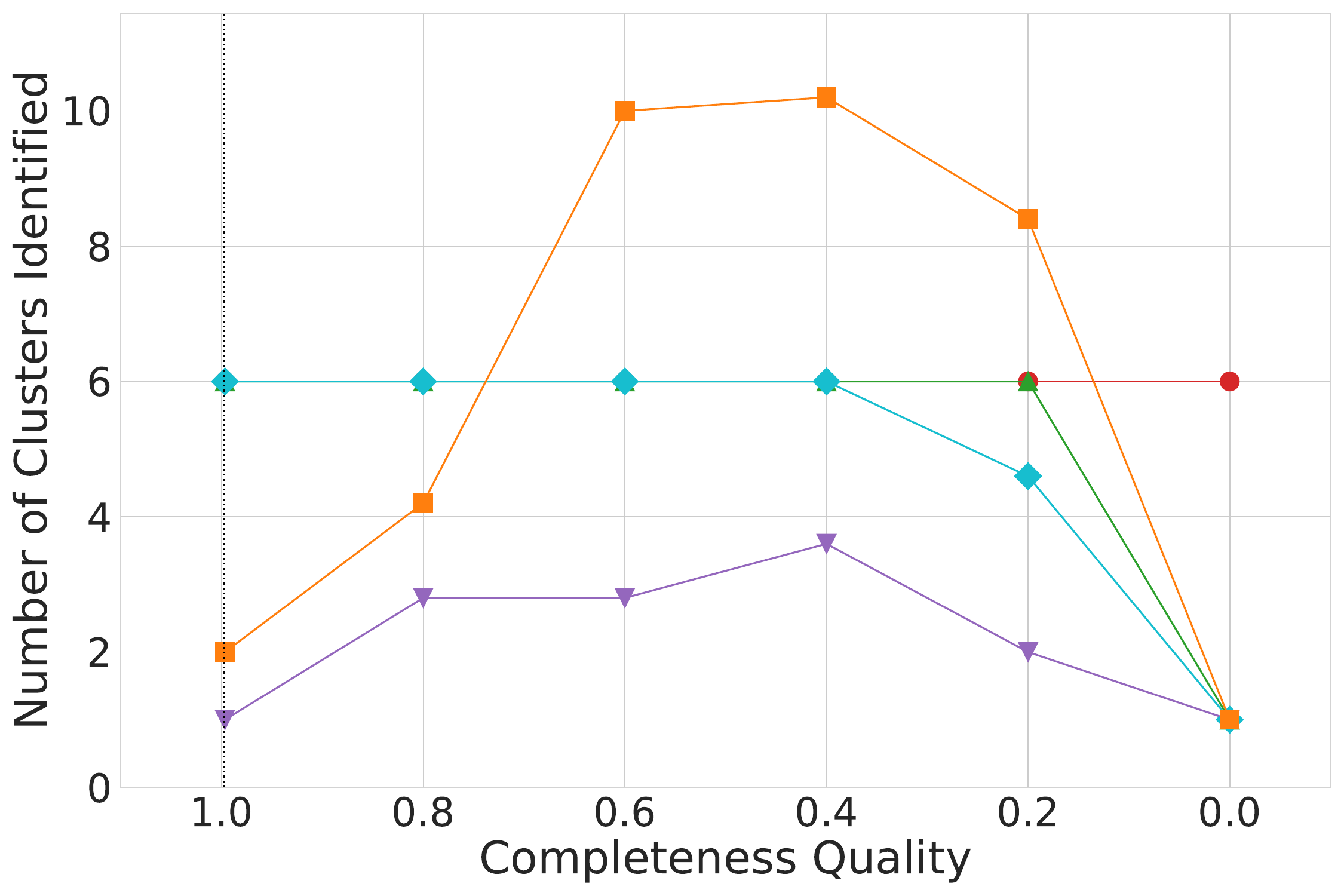}
        \caption{\textsf{Bank}, expected 6 clusters}
        \label{fig:clustering-completeness-bank-nclusters}
    \end{subfigure}

    \begin{subfigure}[b]{0.23\linewidth}
        \includegraphics[width=\linewidth]{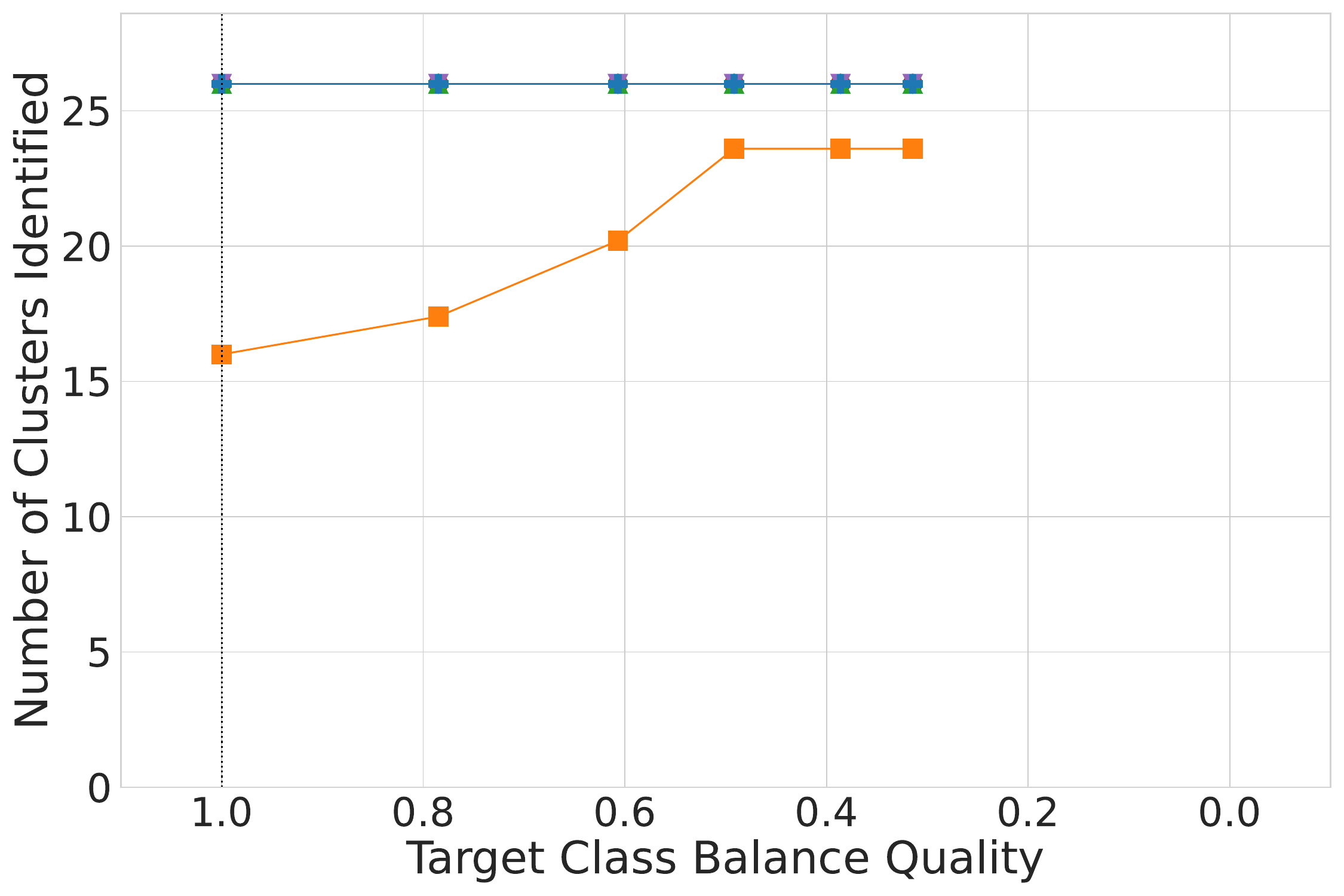}
        \caption{\textsf{Letter}, expected 26 clusters}
        \label{fig:clustering-class-balance-letter-nclusters}
    \end{subfigure}
    \begin{subfigure}[b]{0.23\linewidth}
        \includegraphics[width=\linewidth]{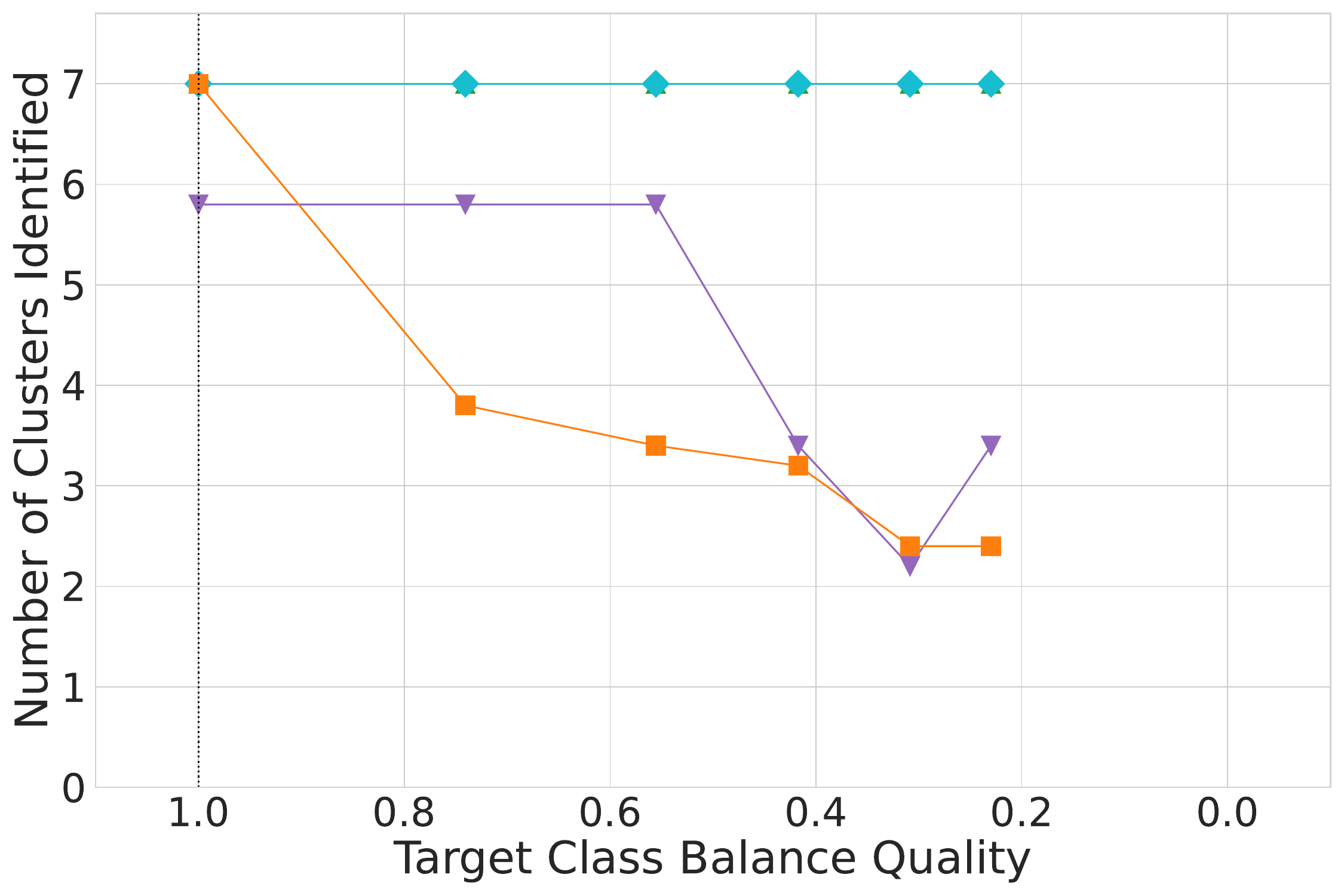}
        \caption{\textsf{Covertype}, expected 7 clusters}
        \label{fig:clustering-class-balance-covertype-nclusters}
    \end{subfigure}
    \begin{subfigure}[b]{0.23\linewidth}
        \includegraphics[width=\linewidth]{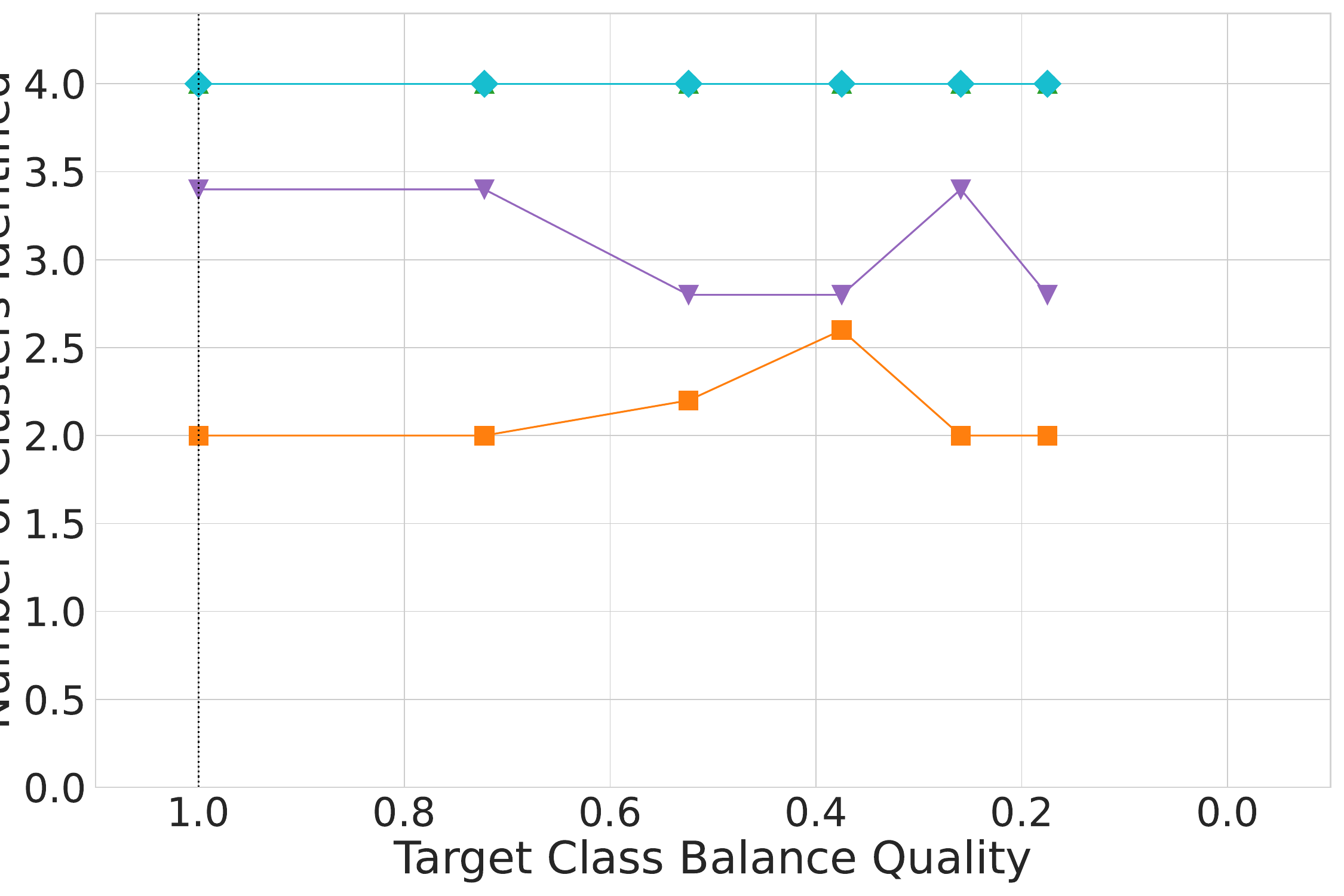}
        \caption{\textsf{COVID}, expected 4 clusters}
        \label{fig:clustering-class-balance-covid-nclusters}
    \end{subfigure}
    \begin{subfigure}[b]{0.23\linewidth}
        \includegraphics[width=\linewidth]{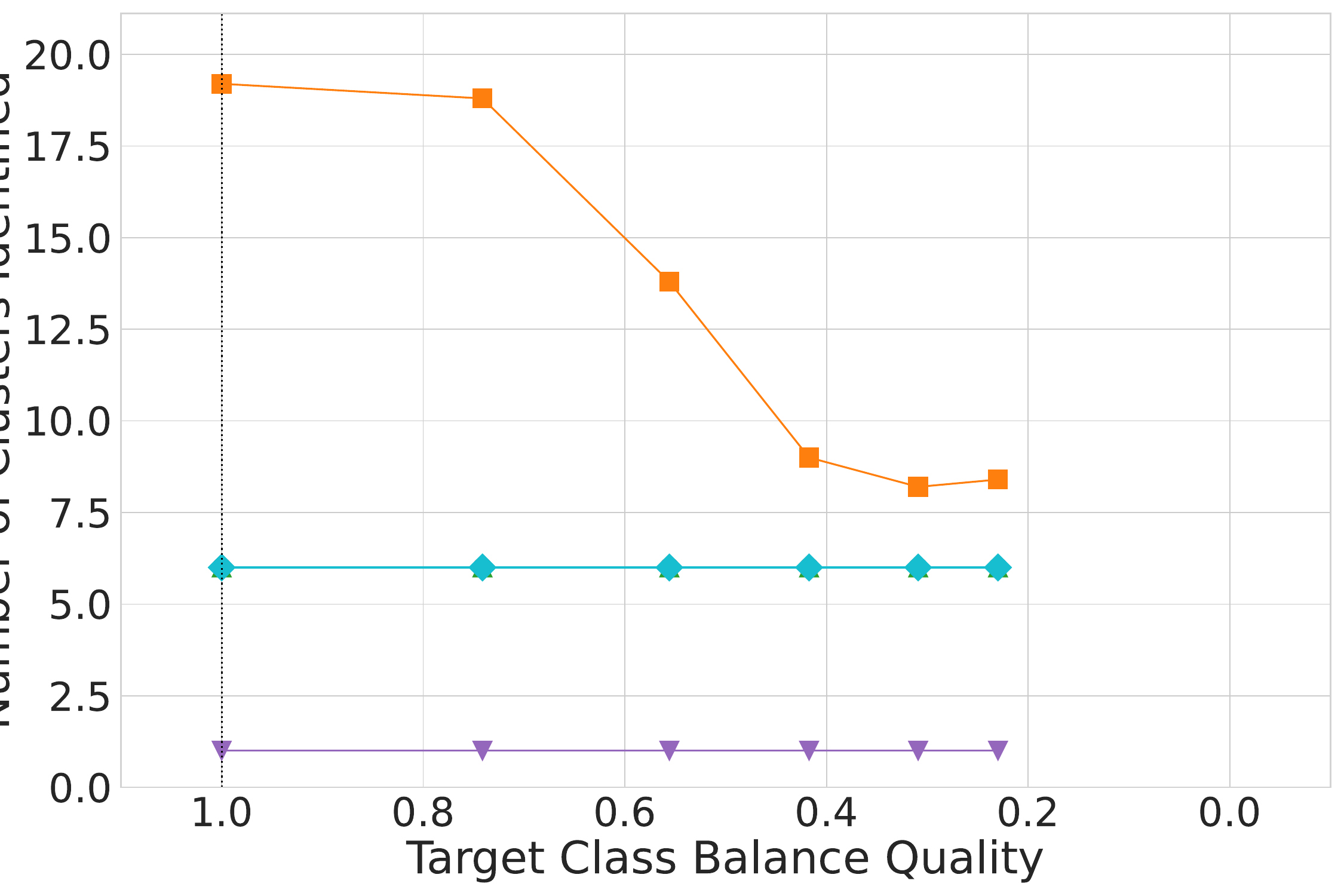}
        \caption{\textsf{Bank}, expected 6 clusters}
        \label{fig:clustering-class-balance-bank-nclusters}
    \end{subfigure}

   \begin{subfigure}[b]{0.23\linewidth}
        \includegraphics[width=\linewidth]{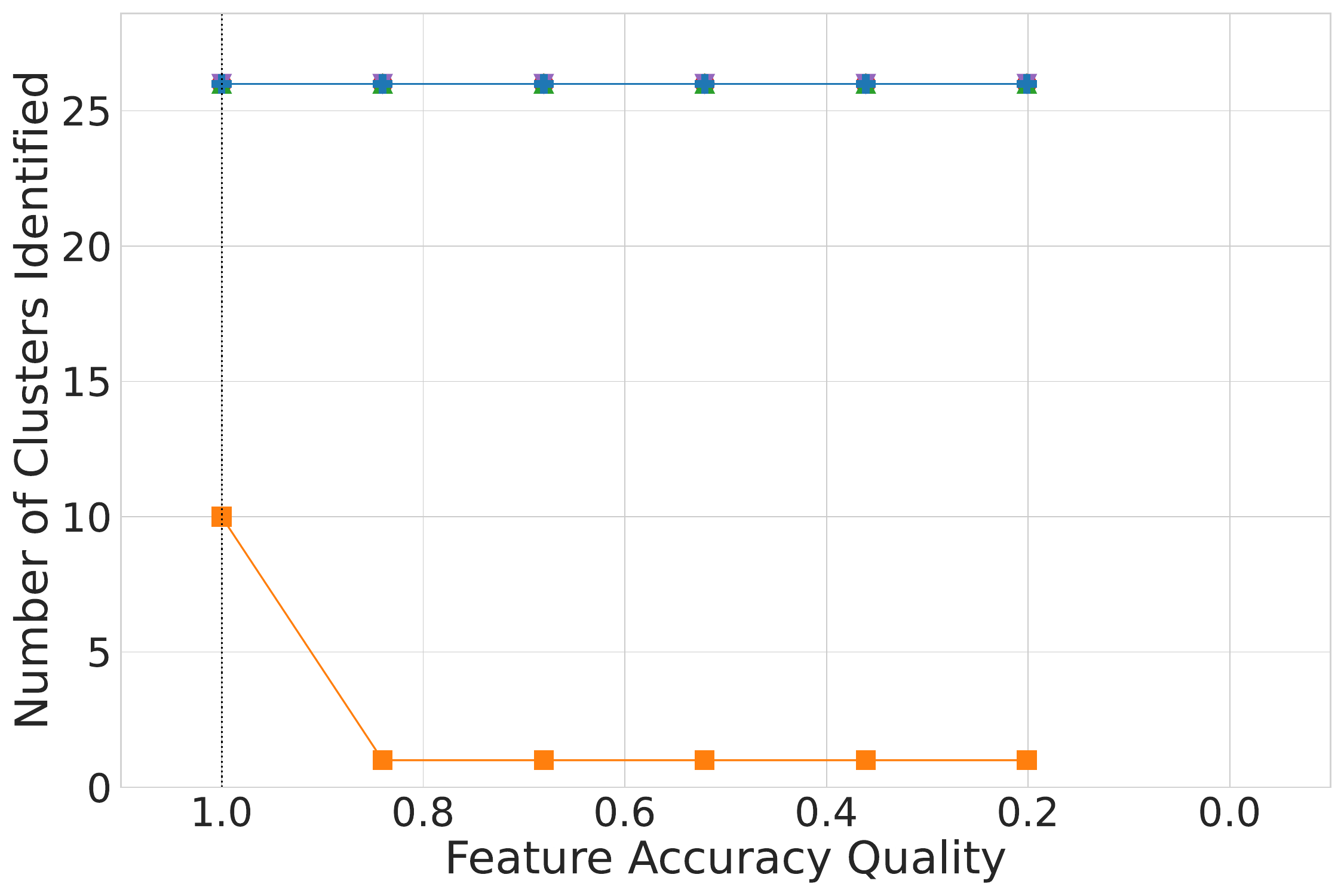}
        \caption{\textsf{Letter}, expected 26 clusters}
        \label{fig:clustering-feature-accuracy-letter-nclusters}
    \end{subfigure}
    \begin{subfigure}[b]{0.23\linewidth}
        \includegraphics[width=\linewidth]{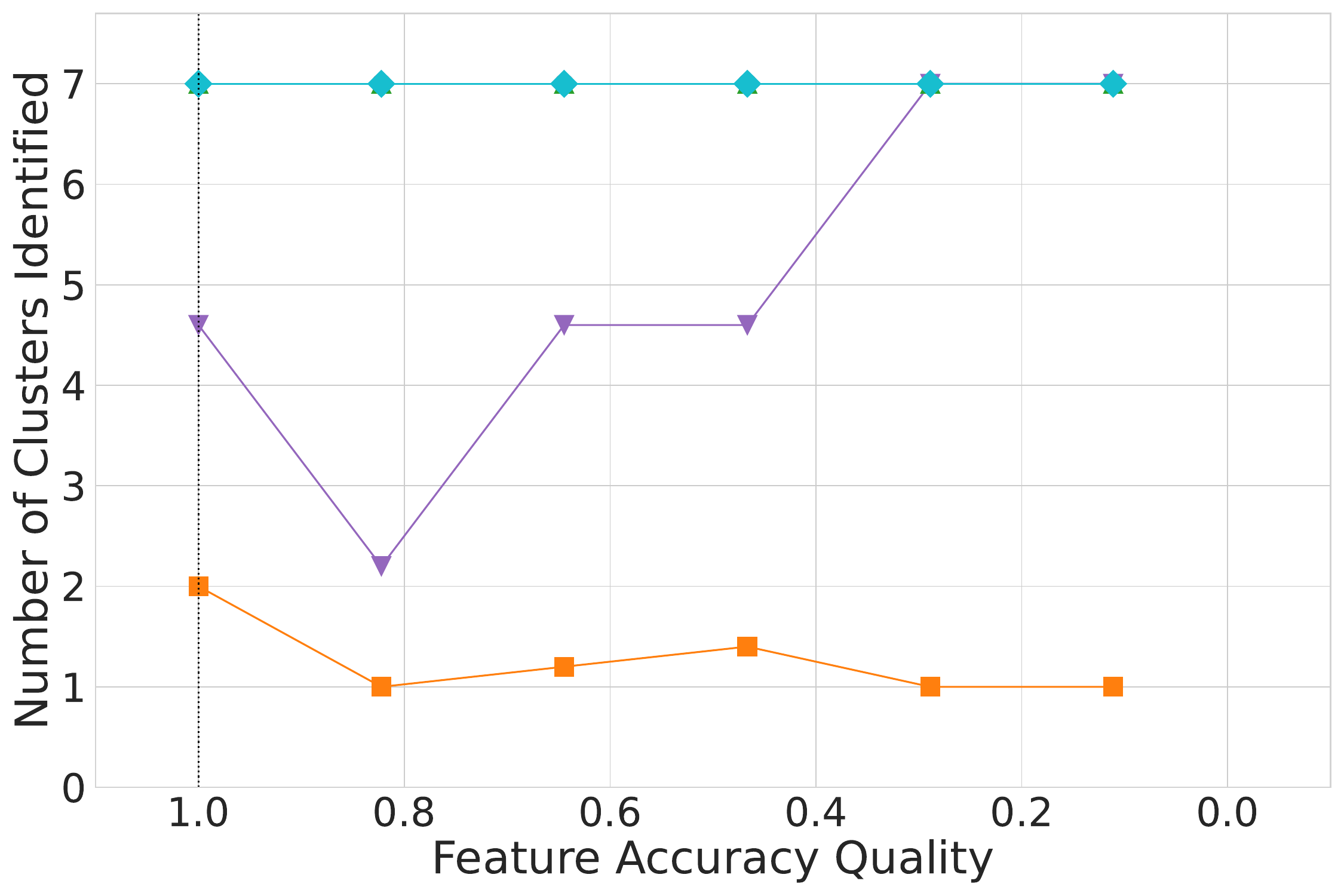}
        \caption{\textsf{Covertype}, expected 7 clusters}
        \label{fig:clustering-feature-accuracy-covertype-nclusters}
    \end{subfigure}
    \begin{subfigure}[b]{0.23\linewidth}
        \includegraphics[width=\linewidth]{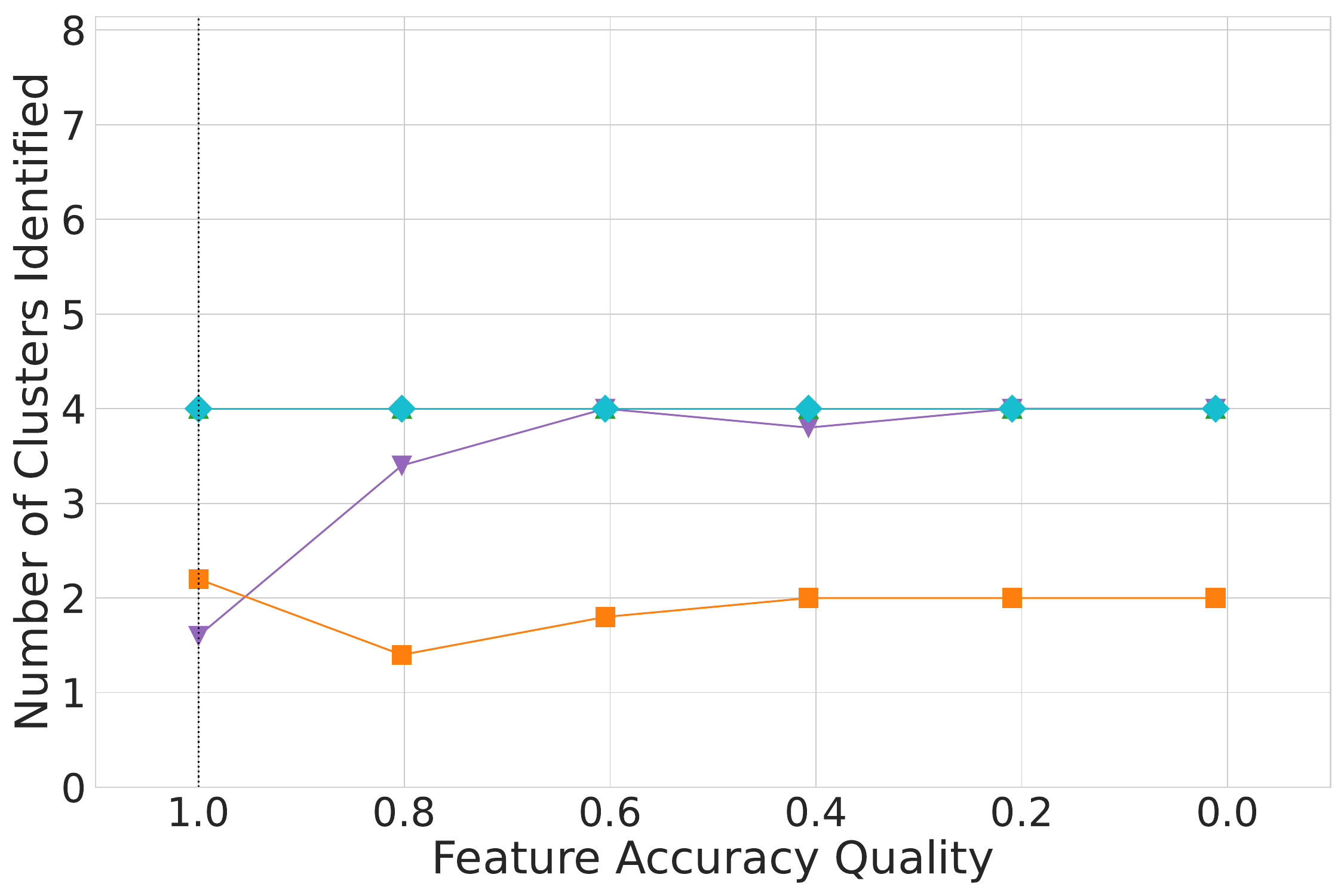}
        \caption{\textsf{COVID}, expected 4 clusters}
        \label{fig:clustering-feature-accuracy-covid-nclusters}
    \end{subfigure}
    \begin{subfigure}[b]{0.23\linewidth}
        \includegraphics[width=\linewidth]{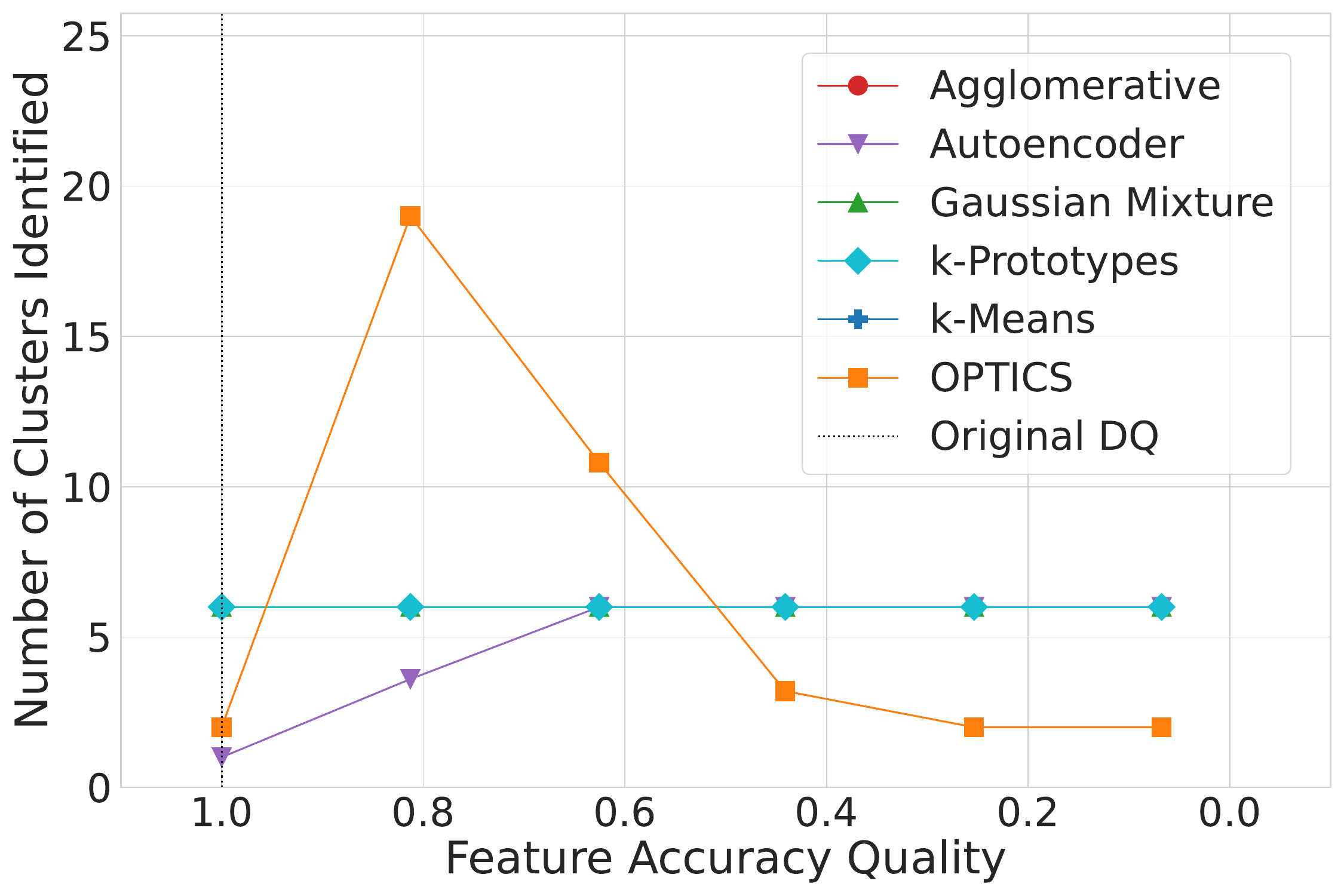}
        \caption{\textsf{Bank}, expected 6 clusters}
        \label{fig:clustering-feature-accuracy-bank-nclusters}
    \end{subfigure}
    \caption{Average number of clusters identified for completeness, target class balance, and feature accuracy dimension.}
    \label{fig:clustering-all-nclusters}
 \end{figure*}

%% file: Latex_Figure/clustering/bank_focused.tex
\begin{figure*}[!hbp]
    \centering
    \begin{subfigure}[b]{0.32\textwidth}
        \includegraphics[width=\textwidth]{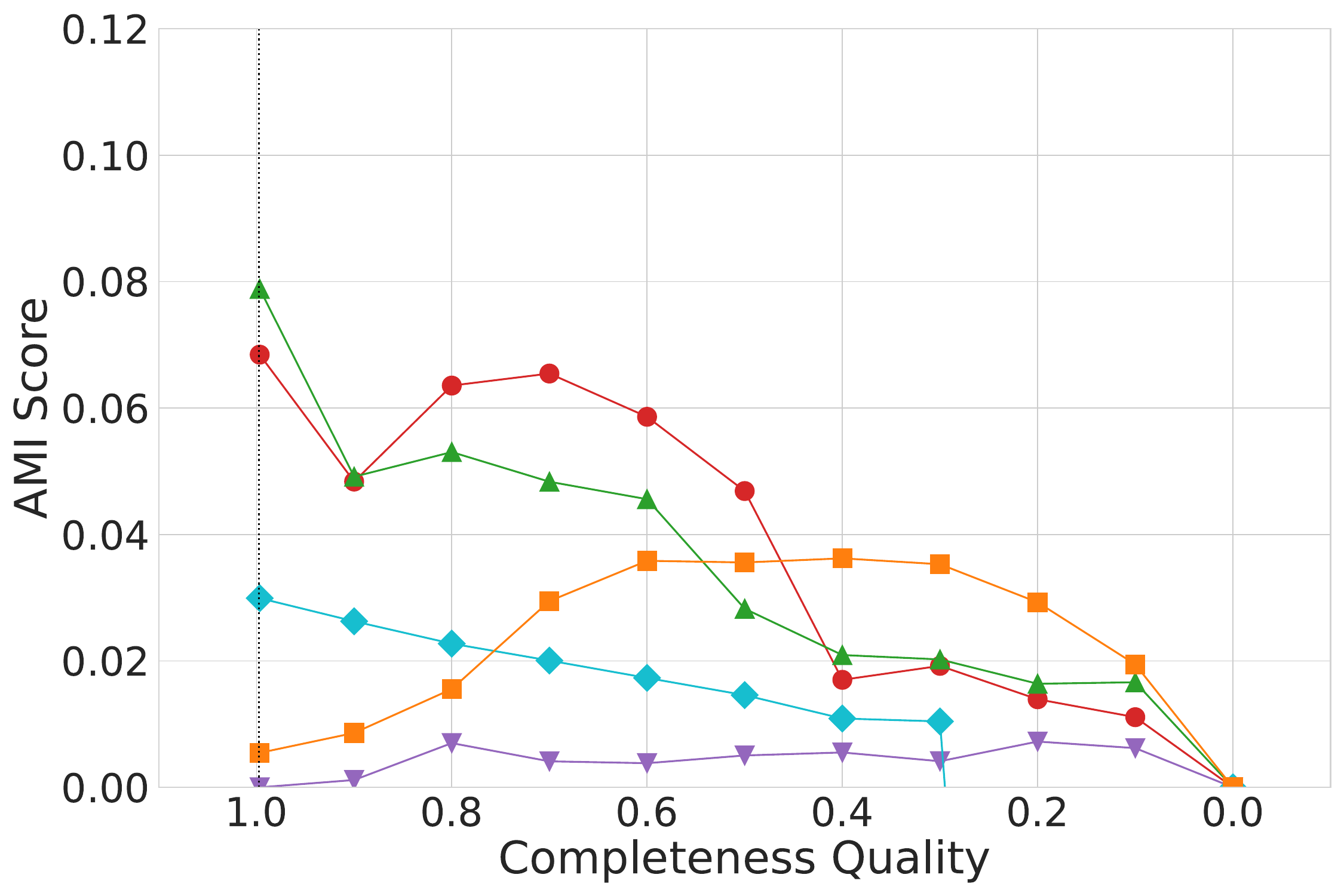}
        \caption{Completeness}
        \label{fig:stretched-bank-results-completeness}
    \end{subfigure}
    \begin{subfigure}[b]{0.32\textwidth}
        \includegraphics[width=\textwidth]{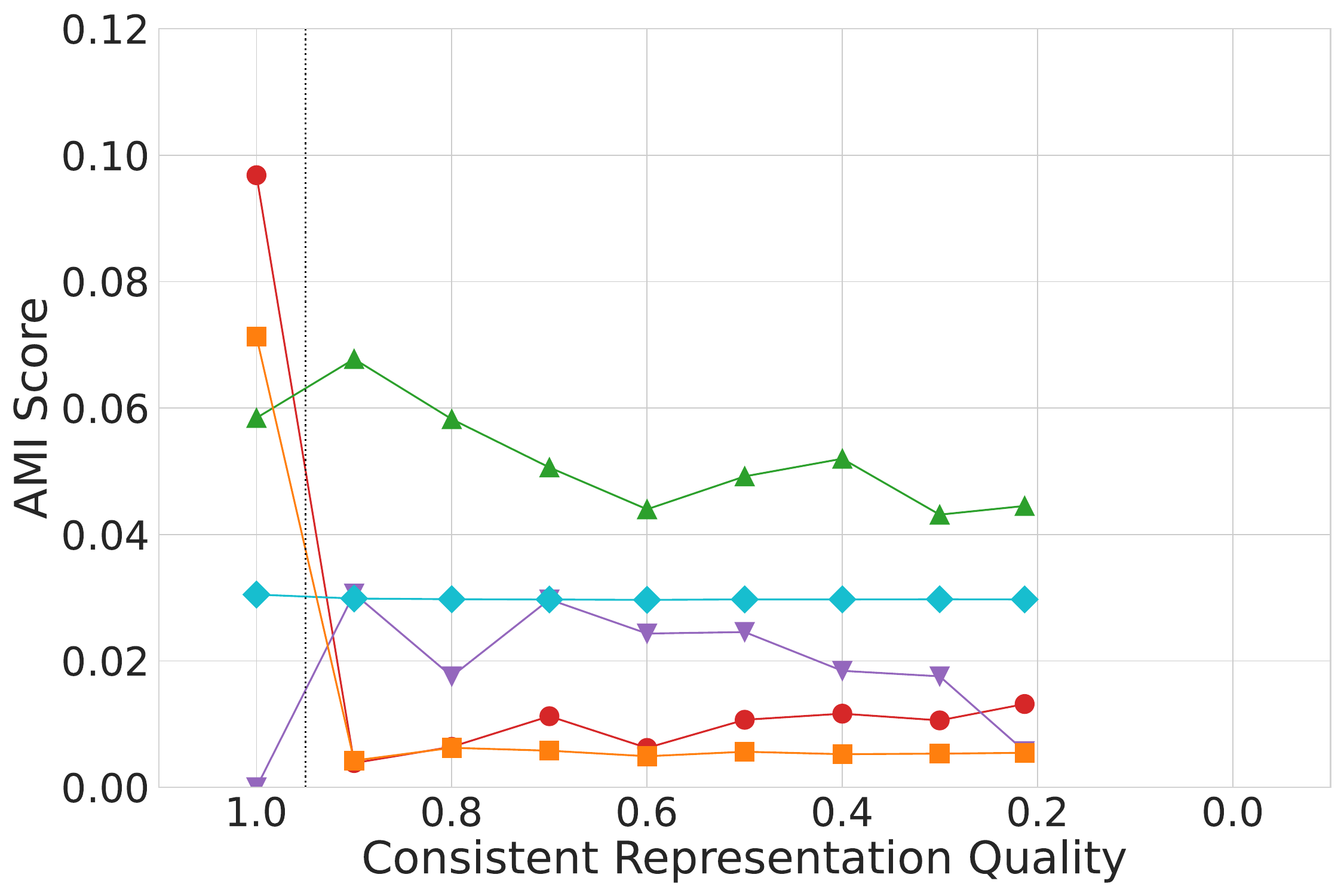}
        \caption{Consistent Representation}
        \label{fig:stretched-bank-results-consistent-representation}
    \end{subfigure}
    \begin{subfigure}[b]{0.32\textwidth}
        \includegraphics[width=\textwidth]{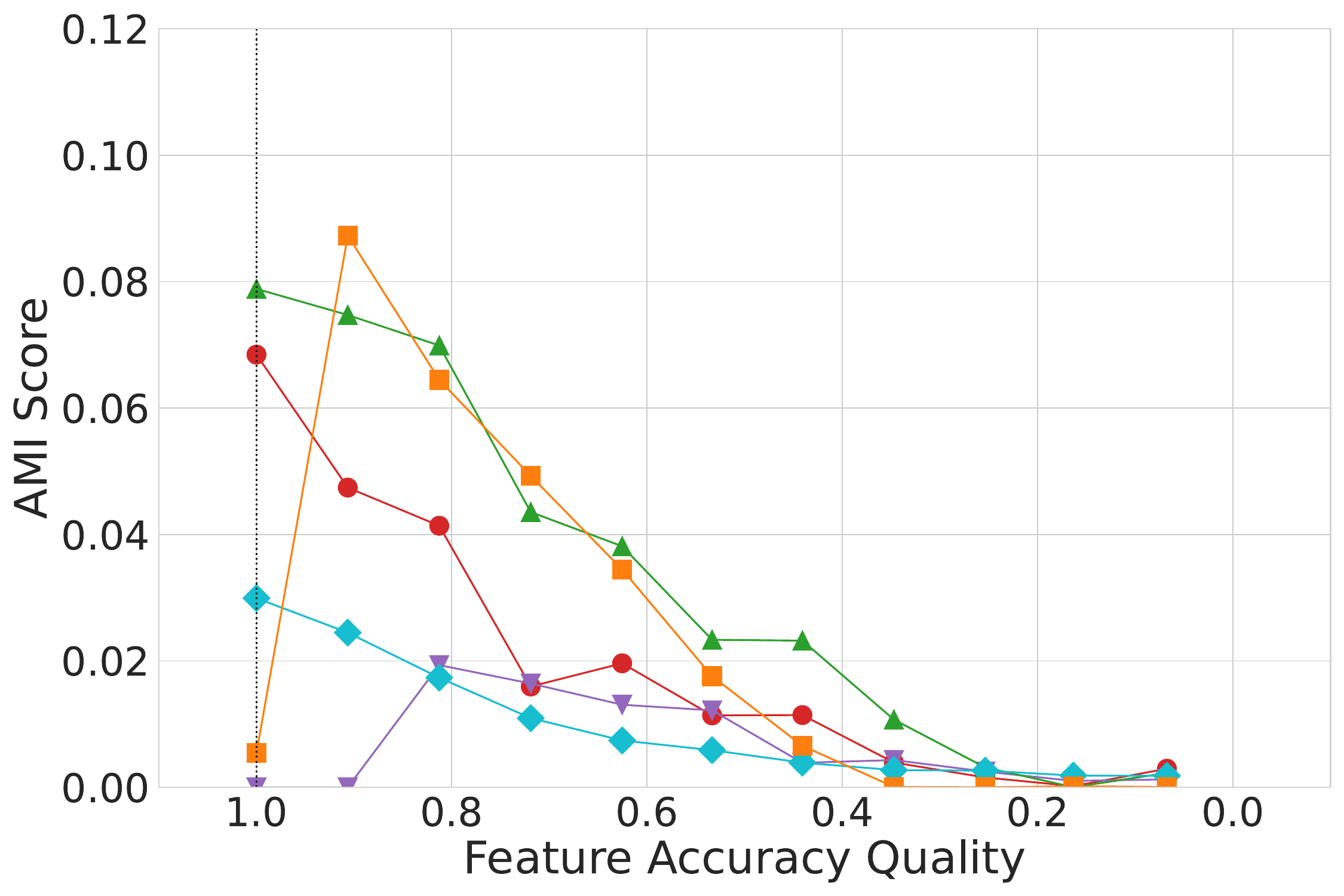}
        \caption{Feature Accuracy}
        \label{fig:stretched-bank-results-feature-accuracy}
    \end{subfigure}
    \begin{subfigure}[b]{0.32\textwidth}
        \includegraphics[width=\textwidth]{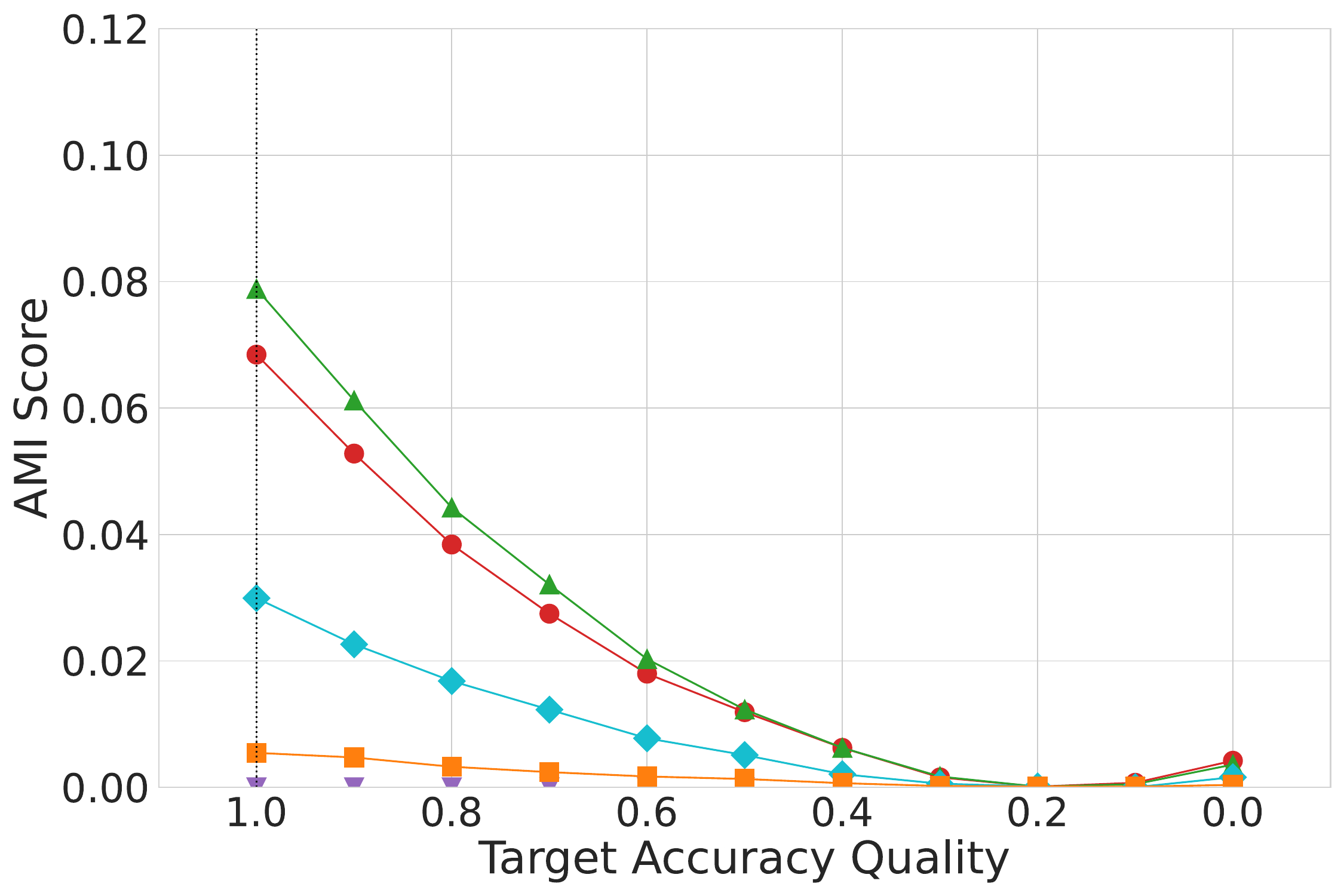}
        \caption{Target Accuracy}
        \label{fig:stretched-bank-results-target-accuracy}
    \end{subfigure}
    \begin{subfigure}[b]{0.32\textwidth}
        \includegraphics[width=\textwidth]{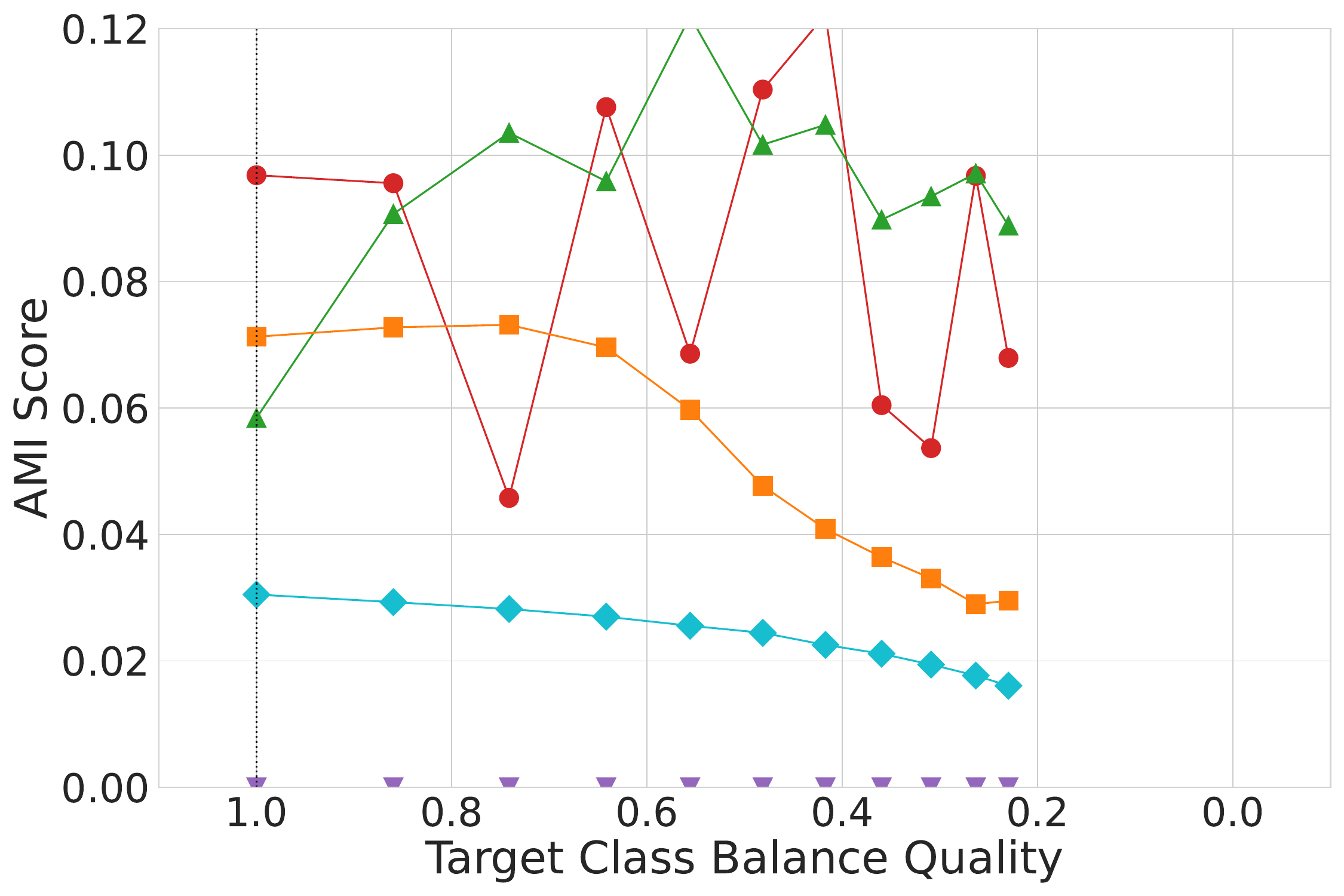}
        \caption{Class Balance}
        \label{fig:stretched-bank-results-class-balance}
    \end{subfigure}
    \begin{subfigure}[b]{0.32\textwidth}
        \includegraphics[width=\textwidth]{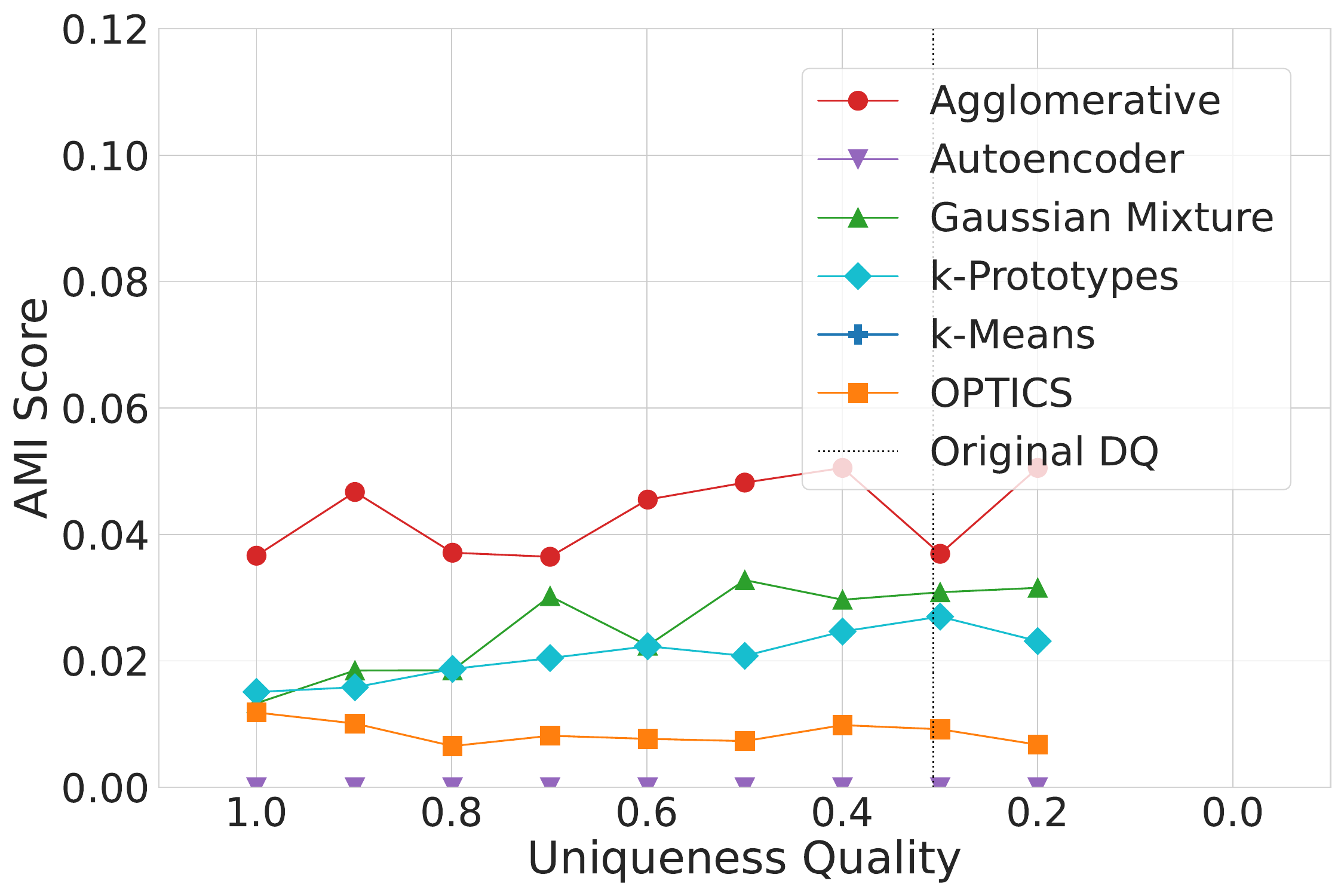}
        \caption{Uniqueness}
        \label{fig:stretched-bank-results-uniqueness}
    \end{subfigure}
    \caption{AMI score of the clustering algorithms for \textsf{Bank}.}
    \label{fig:stretched-bank-results}
\end{figure*}